\documentclass[12pt]{memoir}
\usepackage[left=1in, right=1in]{geometry}
\usepackage{hyperref}
\usepackage[utf8]{inputenc}
\usepackage{physics}
\usepackage{url}
\usepackage{amsmath, amsfonts}
\usepackage{mathtools}
\usepackage{stmaryrd}
\usepackage{graphicx, float}
\usepackage{subcaption}
\usepackage{setspace}
\usepackage{subcaption}
\usepackage{hyperref}
\usepackage{titling,lipsum}
\usepackage{amsthm}
\usepackage{amssymb}
\usepackage[table]{xcolor}
\usepackage{mathrsfs}
\usepackage{amsfonts}
\usepackage{tensor}
\usepackage{enumitem}   
\usepackage{multirow}
\usepackage{dsfont}
\usepackage[all]{xy}
\usepackage{enumitem}
\usepackage{fancyhdr}
\usepackage{framed} 
\usepackage{tikz}
\usepackage{tikz-cd}
\usepackage{tikz-3dplot}
\usepackage{tcolorbox}
\usepackage{subcaption} 
\usepackage{youngtab}
\usepackage{scalerel}
\usepackage{caption,subcaption,array,makecell,longtable,lscape,wrapfig,comment,placeins}

\colorlet{shadecolor}{gray!15}

\newcommand{\defeq}{\vcentcolon=}
\newcommand{\eqdef}{=\vcentcolon}

\usepackage[T1]{fontenc}
\usepackage{xcolor,calc, blindtext}

\usetikzlibrary{positioning, fit}

\usepackage{bbm}
\usepackage{float}

    \setSingleSpace{1.1}
    \SingleSpacing
    \definecolor{chaptercolor}{gray}{0.8}
    \newcommand\numindent{\kern37pt}
    \newlength\chaptertitleboxheight
    \makechapterstyle{hansen}{

      \setlength\midchapskip{0pt}
      \setlength\beforechapskip{0.5\baselineskip}
      \setlength{\afterchapskip}{3\baselineskip}
      
      \renewcommand\chaptitlefont{%
        \normalfont%
        \huge
        \bfseries%
        \raggedleft%
      }%
      \settototalheight\chaptertitleboxheight{%
        \parbox{\textwidth}{\chaptitlefont \strut bg\\bg\strut}}
      }
    \chapterstyle{hansen}
    \aliaspagestyle{chapter}{empty} 

\theoremstyle{definition}
\newtheorem{definition}{Definition}[section] 

\newtheorem*{remark}{Remark}

\newtheorem{theorem}{Theorem}[section] 
\newtheorem{corollary}{Corollary}[theorem]

\newtheorem{proposition}{Proposition}[section]

\newcommand{\C}{\mathbb{C}}
\newcommand{\R}{\mathbb{R}}

\numberwithin{equation}{section}

\DeclareFontFamily{T1}{calligra}{}
\DeclareFontShape{T1}{calligra}{m}{n}{<->s*[1.44]callig15}{}
\DeclareMathAlphabet\mathcalligra   {T1}{calligra} {m} {n}
\DeclareMathAlphabet\mathzapf       {T1}{pzc} {mb} {it}
\DeclareMathAlphabet\mathchorus     {T1}{qzc} {m} {n}
\DeclareMathAlphabet\mathrsfso      {U}{rsfso}{m}{n}

\def\eg{{\textit{e.g.}}}
\def\ie{{\textit{i.e.}}}

\def\SU{\mathrm{SU}}
\def\USp{\mathrm{USp}}
\def\SO{\mathrm{SO}}
\def\Spin{\mathrm{Spin}}
\def\Sp{\mathrm{Sp}}

\def\su{\mathfrak{su}}

\def\DWeight#1#2#3{\bigl(\raise2.5pt\hbox{${}_{#1}$}{}^{#2}_{#3}\bigr)}

\def\AAWeight#1#2{\bigl(\raise0pt\hbox{${}^{#1}_{#2}$}\bigr)}

\def\so{{\mathfrak{so}}}

\def\nn{\nonumber}

\def\so{\mathfrak{so}}

\def\*{\partial}

\def\transpose{\intercal}

\def\II{{\mathscr I}}

\def\RR{{\mathbb R}}
\def\HH{{\mathbb H}}
\def\ZZ{{\mathbb Z}}

\DeclareMathOperator{\atantwo}{arctan2}

\def\tr{\hbox{tr}}

\def\Re{\hbox{Re}}
\def\Im{\hbox{Im}}

\def\SU{\hbox{SU}}
\def\USp{\hbox{USp}}
\def\SO{\hbox{SO}}
\def\Spin{\hbox{Spin}}
\def\Sp{\hbox{Sp}}

\def\invsq#1{{1\over\sqrt{1+|#1|^2}}}

\newcommand{\Diff}{\operatorname{Diff}}
\newcommand{\Homeo}{\operatorname{Homeo}}
\newcommand{\Diffp}{\Diff^{+}}
\newcommand{\Homeop}{\Homeo^{+}}
\newcommand{\Sm}{C^{\infty}}

\def\Re{\mathrm{Re}\hskip1pt}
\def\Im{\mathrm{Im}\hskip1pt}

\def\SU{\mathrm{SU}}
\def\USp{\mathrm{USp}}
\def\SO{\mathrm{SO}}
\def\Spin{\mathrm{Spin}}
\def\Sp{\mathrm{Sp}}

\def\so{{\mathfrak{so}}}

\def\epsilonzero{\overset{\scriptscriptstyle 0}\varepsilon}
\def\omegazero{\overset{\scriptscriptstyle 0}\omega}
\def\Omegazero{\overset{\scriptscriptstyle 0}\Omega}

\def\Rzero{\overset{\scriptscriptstyle 0}R}

\def\Vol{\hbox{Vol}}

\def\nn{\nonumber}

\def\so{\mathfrak{so}}

\def\*{\partial}

\def\transpose{\intercal}

\def\sF{{\mathscr F}}

\def\RR{{\mathbb R}}
\def\HH{{\mathbb H}}
\def\OO{{\mathbb O}}
\def\ZZ{{\mathbb Z}}

\def\tr{\hbox{tr}}

\def\Re{\mathrm{Re}\hskip1pt}
\def\Im{\mathrm{Im}\hskip1pt}

\def\SU{\mathrm{SU}}
\def\USp{\mathrm{USp}}
\def\SO{\mathrm{SO}}
\def\Spin{\mathrm{Spin}}
\def\Sp{\mathrm{Sp}}

\def\su{\mathfrak{su}}

\def\invsq#1{{1\over\sqrt{1+|#1|^2}}}

\def\II{{\mathscr I}}

\newcommand{\paral}{\mathbin{\!/\mkern-5mu/\!}}

\def\Ppar{P_{\scaleto{\paral}{6pt}}}
\def\Pperp{P_{\scaleto{\perp}{5pt}}}

\newcommand{\g}{\mathfrak{g}}

\usepackage{pgfplots}
 
\tdplotsetmaincoords{60}{115}
\pgfplotsset{compat=newest}

\setsecnumdepth{subsection}      
\settocdepth{section}

\begin{document}

\begin{titlingpage}
    
\centering
{\scshape\Huge  \textbf{A Physicist's Visit \\ to \\
Exotic Spheres}\par}

\vspace{1.5cm}
{Tancredi Schettini Gherardini}

\vspace{1.5cm}

{PhD Thesis \par}

\vspace{1.5cm}

{Supervised by\par
    Professor David S. Berman}
	
\vspace{1.5cm}

{Submitted to Queen Mary University of London for the degree of Doctor of Philosophy on 11/07/2025.}

\vspace{1cm}

{Revised Version, April 2026}

\vspace{2.7cm}

\begin{figure}[h]
    \centering
    \includegraphics[width=0.4\linewidth]{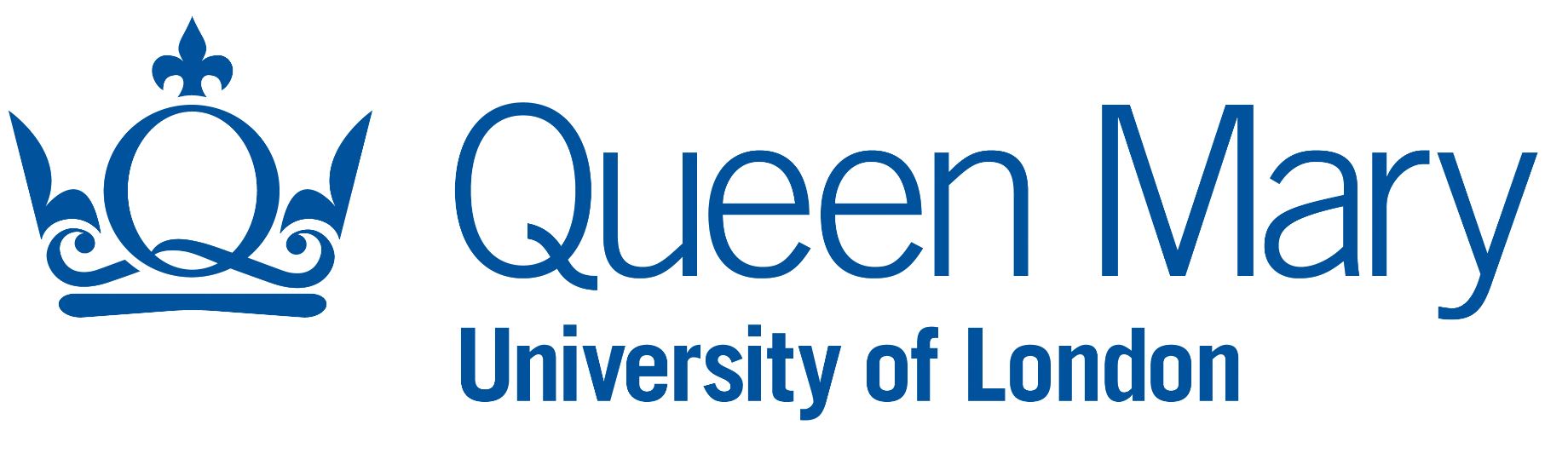}
    
\end{figure}

\end{titlingpage}

\newpage

\
\vfil
\hfil 
\begin{align}
 Ai \,\,\,\, miei \,\,\,\, genitori \,\,\,\, e \,\,\,\, ad \,\,\,\, Irene \nonumber \\ \nonumber \\
 To \,\,\,\, my \,\,\,\, parents \,\,\,\, and \,\,\,\, to \,\,\,\, Irene \nonumber
\end{align}
\hfil
\vfil




\newpage

\

\newpage

\vfil

{\centering \textbf{Abstract} \vspace{0.75cm} \\ 

This thesis discusses exotic $7$--spheres, i.e.~manifolds that are homeomorphic but not diffeomorphic to the ordinary $7$--sphere, using a set of analytical and computational tools from theoretical physics. The theory of fibre bundles and instantons, together with their relation to Yang--Mills theory, are reviewed, before presenting a generalisation of self-duality to \textit{twisted self-duality}. The formalism required to derive and geometrically interpret some solutions to twisted-self-duality is relevant to the main subject of this thesis: investigating the geometry of the Gromoll--Meyer sphere. Through a Kaluza--Klein ansatz, motivated by bundle-theoretic arguments, an analytic expression for a family of Riemannian metrics on the Gromoll--Meyer sphere is derived. After a detailed study of its geometric constituents, recast as quaternionic-valued objects, the metric with maximal isometry is identified. Its curvature properties are also studied and the associated energy conditions are assessed. Then, an up-to-date and broader overview on the current work concerning exotic spheres and exotic manifolds in general is offered, before focusing again on the Gromoll--Meyer sphere, but this time under the lens of differential topology. Some explicit realisations of the homeomorphism between an exotic $7$-sphere and an ordinary one are discussed, together with their possible interpretations in the context of general relativity. Finally, a numerical algorithm for finding Riemannian Einstein metrics on arbitrary manifolds is presented; it is based on machine learning, and highly generalisable in many directions. The current work on implementing its application to exotic spheres is also discussed. The thesis ends with an ample discussion of possible future directions.}

\vfil

\chapter*{Preface}

\section*{Motivation} 
The intertwining of physics and mathematics goes back to the Babylonian-Greek period (\cite{Neugebauer1957,Kline1972}). While mathematics was almost immediately recognised as a discipline of its own, it took many more centuries for physics to gain its independence and leave the \textit{natural phylosophy} umbrella. During those centuries, it was almost always the case that facts and results from mathematics were used to explain physical phenomena. In other words, until the end of the Medieval age, all that ``physics'' did was borrowing some abstract mathematical knowledge to describe some observations in the real word (\cite{Duhem1969}).\footnote{By \textit{physics} here we mean the discipline that was actually performing the quantitative, predictive work we associate with physics today.} During the 16\textsuperscript{th} and 17\textsuperscript{th} centuries, however, something begun to change: physics became a driving force of mathematical discovery. Besides the famous case of calculus being invented to model the rate of a change of some physical quantity, there are a number of examples which show how questions from physics forced mathematicians to invent new concepts, structures and entire fields - see \cite{Boyer1987}. Predicting ship trajectories led to the formalisation of certain projections and spherical functions, predicting projectile trajectories led to further study of certain differential equations, and understanding the vibrations of strings in musical instruments launched entire branches of analysis and functional expansion theory (\cite{Truesdell1960}). In summary, a transition period happened, during which physics went from being a mere application of already-existing mathematical results to natural phenomena, to being a cornucopia of open questions seeking the development of new mathematical frameworks in order to be answered, as described in \cite{Cohen1981}. This shift of paradigm, which also coincides with the birth of physics in its modern meaning, is one of the turning points of the human's progress in the quantitative realm, according to \cite{ Popper1975, Kuhn1962}. The one just described is just one side of the \textit{scientific revolution} which took place after the Medieval age. Another side of it, which is more relevant to this thesis, is the birth of mathematical theory-building about the physical world. Instead of restricting themselves to observed phenomena, scientists started to make mathematically justified speculations about new ones. Using Popper's language, the ``hypothesis formation'' includes implications beyond the simple explanation of a known phenomenon (\cite{Popper1934, Popper1959}). It predicts the existence of new ones, which need to be verified in order to corroborate the underlying theory. This new science, dynamical and speculative, although with a robust and rigorous underlying method, had a fundamentally new aspect: the creation of (falsifiable) theories. While the scrutiny of a theory via experiments follows relatively structured methods, there is no universal algorithm for coming up with a new theory. Citing Popper once again: ``...the act of conceiving or inventing a theory, seems to me to be neither susceptible of logical analysis nor of rational justification.'' (\cite{Popper1959}). This is, in essence, one of the most fundamental aspects underlying theoretical high energy physics, which lies at the boundary between rational analysis and creative imagination. In response to this feature, theoretical physicists have often followed a somewhat vague principle of mathematical simplicity and elegance, when formulating new theories. Although being often associated with stronger predictive power and fewer ad-hoc assumptions, mathematical beauty cannot be quantified nor correlated with a theory's success. However, it has happened that new theories were created following some elegant mathematical ideas more than some empirical observations. It is the case of Dirac’s relativistic wave equation for the electron, which led to the prediction of antimatter before any experimental hint of its existence (\cite{Dirac1928,Dirac1931}), Schr\"odinger equation (\cite{Schrodinger1926a}), and many others. In these circumstances, the source of inspiration consisted of some pattern within the mathematical framework, rather data coming from a given experiment. This motivation, which we might suggestively call ``listening mathematics' whispers'', is behind many breakthroughs of the previous century.\footnote{These whispers, of course, are not always right; for instance, see Weil's attempt at unifying gravity, electromagnetism in \cite{Weyl1918}, or Eddington's fundamental theory \cite{Eddington1946}, or the Bohr–Kramers–Slater (BKS) theory developed in \cite{Bohr1924, BotheGeiger1925}.}\footnote{During the last few decades, the opposite has also happened: ``listening to physics' whispers'' has led to significant advances in mathematics; for example, the case of mirror symmetry (\cite{GreenePlesser1990, CandelasEtAl1991}).} The reason behind the study of exotic spheres from the perspective of a physicist, which is the subject of this thesis, is exactly to focus on one of these whispers. Differential geometry is the mathematical language underlying the most successful classical theories of fundamental physics: general relativity, electromagnetism (gauge theory), weak and strong interactions (Yang--Mills theory). The existence of exotic differentiable structures is a very curious result in differential geometry, which has no clear physical implications on these theories. Exotic spheres are the first and simplest examples of manifolds carrying exotic differentiable structures, and therefore they seem the best candidates to start assessing such implications. It might be the case that the consequences of this mathematical fact are irrelevant in terms of our understanding of the universe. If this were true, it would still be a valuable scientific result. The aim of this thesis is to provide some insights into how to address such an investigation.

\section*{Statement of Originality and Publications}

I, Tancredi Schettini Gherardini, certify that, except where explicit
acknowledgement is made, the work presented in this thesis is
entirely my own and has not been submitted for any other degree or
qualification at this or any other institution.

The following papers (peer reviewed or undergoing review) form the basis of Chapters \ref{chap:2}, \ref{chap:3}, \ref{chap:4}, \ref{chap:5} of this thesis.

\begin{itemize}
  \item  Chapter \ref{chap:2} is based on ``Twisted Self-duality'', published in \hyperlink{https://www.worldscientific.com/doi/abs/10.1142/S0217751X23500859?srsltid=AfmBOopLrfXzOvC2UZ16y-QAwsHsLFMvunOsVOutVJ4Rvh1XYHrha8JF&journalCode=ijmpa}{Int. J. Mod. Phys. A, Vol. 38 (2023)}, \hyperlink{https://arxiv.org/abs/2208.09891
}{arXiv:2208.09891}, with David Berman.
  
  \item The first part of Chapter \ref{chap:3} is based on ``Exotic Spheres' Metrics and Solutions via Kaluza--Klein Techniques'', published in \hyperlink{https://link.springer.com/article/10.1007/JHEP12(2023)100}{J. High Energ. Phys. 2023, 100}, \hyperlink{https://arxiv.org/abs/2309.01703
}{arXiv:2309.01703}; the second part of Chapter \ref{chap:3} is based on ``Curvature of an Exotic 7-sphere'', published in \hyperlink{https://www.sciencedirect.com/science/article/pii/S0393044025001743}{J. Geom. Phys. Vol. 216 (2025)}, \hyperlink{https://arxiv.org/abs/2410.01909
}{arXiv:2410.01909}.

  \item Chapter \ref{chap:5} is based on ``AInstein: Numerical Einstein Metrics via Machine Learning'', published in \hyperlink{https://iopscience.iop.org/article/10.1088/3050-287X/ae1117}{AI for Science, Vol. 1, No. 2 (2025)}, \hyperlink{https://arxiv.org/abs/2502.13043}{arXiv:2502.13043}.\footnote{As another application of the \textit{AInstein} neural network, the pre-print \cite{cortes2026machinelearningapproachnirenberg} appeared after the submission of this thesis.} 
\end{itemize}

Some unpublished material appearing for the first time in this thesis is contained in Chapter \ref{chap:3} and Chapter \ref{chap:4}.

\section*{A Note on the Title}
The title of this thesis is a reference to a presentation, later turned into a paper, by Jean-Pierre Bourguignon, named ``A Mathematician's Visit to Kaluza--Klein Theory'' (\cite{bourguignon1989mathematicians}). It is a rigorous bundle-theoretic overview on Kaluza--Klein theories. In his words: ``Our little venture into physics will give us the opportunity of comparing the attitudes of mathematicians and physicists in this type of problems, and hopefully to propose some guidelines for later developments.'' 

The idea behind the study of exotic spheres is of similar nature. It is a topic originated from and researched by (almost) only mathematicians, and we hope that applying some theoretical physics techniques and suggesting some physical interpretations might hint at new interesting directions. The title was chosen accordingly.

\section*{Changes from the Defended Version}
This version differs from the thesis defended in September 2025 only in the following minor respects: corrections of typos and small numerical factors in a few computations, minor revisions to the literature overviews, and the inclusion of a detailed derivation in Chapter $\ref{chap:4}$, Section \ref{subsec:Tamura_chap4}, that was only sketched in the original version.

\newpage

\tableofcontents

\chapter{Introduction, Structure and Conventions}

\section{Introduction}
Milnor's discovery of exotic spheres in \cite{10.2307/1969983} contributed to the birth of a completely new branch of mathematics: differential topology. Before then, the topological structure and the differentiable structure seemed inextricably connected, with the former uniquely specifying the latter. Exotic $7$--spheres, homeomorphic to the ordinary $7$--sphere but not diffeomorphic to it, showed for the first time how these two structures can be decoupled. Since then, a very large number of other manifolds with the same property, i.e.~\textit{exotic}, have been discovered, receiving large attention from the mathematical community, leading to several breakthroughs such as \cite{KervaireMilnor1963,  Brieskorn66, Freedman82, Taubes1987}, just to cite some of the most influential works. Despite many new exotic manifolds have been found since Milnor's seminar paper, exotic $7$-spheres remain among the few ones admitting (relatively) simple realisations; one might argue that this is one of the reasons why many of their properties have been understood and formalised (see \cite{GroveZiller00, nuimeprn10073}). Nevertheless, they remain a subject of active research mainly because of two facts. The first one, more appealing to mathematicians, is that there are many open questions about their geometric features, such as those stated in \cite{nuimeprn10073}. The second one, more appealing to physicists, is that the role of exotic spheres - and manifolds in general - within theories of (quantum) gravity have not been gauged in depth (see \cite{book}). The latter motivation is what led many of the investigations presented in this thesis, but a number of results of mathematical relevance were also collected during the process. We now review the relevant background to put these studies into context. 

Milnor's original realisation of exotic spheres was as $S^3$ bundles over $S^4$. He constructed a family of fibre bundles, including exotic spheres as well as the ordinary $7-$sphere, realised as a quaternionic Hopf fibration. As it is well-known, the key bundle-theoretic element that specifies the $7$-sphere written as $S^3 \hookrightarrow S^7 \xrightarrow{} S^4$ is the \textit{instanton}. This (anti-)self-dual connection first appeared as a finite-action solution of $\mathrm{SU(2)}$ Yang--Mills theory in the BPST paper \cite{BELAVIN197585}, and its relation to $S^7$ was only later recognised in \cite{Trautman:1977im}. Since then, self-duality equations have always played a prominent role in many physical theories, as well as in a number of mathematical results. With the study of supergravity theories, self-duality relations combined with additional non-trivial algebraic constraints begun to appear: \cite{Cremmer_1998, Cremmer_1998_2, Tseytlin:1990nb,Tseytlin:1990va}. This motivates the first investigation presented in this thesis, based on \cite{Berman:2022dpj}: a natural generalisation of the study in \cite{https://doi.org/10.48550/arxiv.1412.2768}, focusing on the \textit{twisted self-duality} relation for $\text{SO(4)}$ Yang--Mills theory and deriving two natural solutions. Their bundle-theoretic interpretation naturally leads to discussing total spaces that are intimately related to exotic spheres.

A well-known machinery to describe the geometry of a total space respecting the underlying bundle structure is given by Kaluza--Klein formalism - see the seminal papers \cite{Kaluza1921, Klein1926}, as well as the later developments in \cite{10.1063/1.522434, Salam:1981xd, DUFF19861}. Although the physicists' interpretation of Kaluza--Klein theory is associated with reducing a higher-dimensional theory to a lower-dimensional one, the reverse direction is just as legitimate (\cite{bourguignon1989mathematicians}); the metric ansatz can be used as a prescription for uplifting lower-dimensional geometrical objects to a higher-dimensional Riemannian metric on the total space. It is the latter viewpoint which is adopted in this thesis: following \cite{Gherardini:2023uyx, berman2024curvatureexotic7sphere}, we derive an analytic expression for a maximally isometric geometry on one of the exotic spheres - the Gromoll--Meyer sphere. As noted in those papers, we could only find two (very brief) discussions pointing in this direction in \cite{FREUND1985263, YAMAGISHI198447}. The fact that such a construction was never explicitly realised and studied in detail was one of the main motivations behind our studies, which are presented in the central part of this thesis.  Our view is that obtaining a concrete and manageable coordinate description for a metric, together with the derivation of its main curvature properties, is a key step towards trying to embed it as a solution of higher-dimensional theories of (super)gravity. The results of \cite{Gherardini:2023uyx, berman2024curvatureexotic7sphere} in this direction also provide a tool for studying some key mathematical properties associated with open questions, such as the study of sectional curvature.

The geometric structure mentioned above is a consequence of the inequivalent differentiable structure that is carried by an exotic sphere. Very little is known about the ``obstruction'' that prevents such an atlas from being smoothly deformed into the one of the ordinary sphere. This is the subject of the third investigation presented, which discusses some realisations of (continuous but non-differentiable) maps relating exotic spheres to ordinary ones, and some possible physical implications. We note that the differential-topological properties of exotic spheres have been shown to arise in string theory in \cite{Witten:1985xe, 10.1063/1.529078}. Some interpretations of exotic spaces in cosmology and (quantum) gravity were also put forward in \cite{Brans:1992mj, Asselmeyer:1996bh, Schleich_1999}. Our discussion aims at suggesting possible consequences of inequivalent differentiable structures inside the framework of general relativity, in a way that is neither formal nor exhaustive. \\

One of the main difficulties in studying exotic spheres and exotic manifolds in general, is the absence of symmetries. Exotic $7$--spheres, for instance, are not homogeneous spaces (\cite{3c71f429-d750-3ab8-af97-277a9e5ed9e8}). The maximal symmetry available is also much smaller than that of the standard geometries of \textit{non}-exotic manifolds. As a consequence, one is naturally led to consider numerical tools, to go beyond what can be done analytically (see \cite{Headrick:2005ch, Douglas2006, Gentle2004} for pioneering works in this direction). This is the subject of the last investigation presented, which is based on machine learning techniques. The use of such tools for tackling open problems in theoretical physics and differential geometry has led to new insights in the context of Calabi--Yau metrics, presented in \cite{Ashmore:2019wzb, Douglas:2020hpv, Anderson:2020hux, Jejjala:2020wcc,Larfors:2021pbb}, as well as in other scenarios involving the approximation of metrics on non-trivial settings studied in \cite{deluca2024, Li:2023, chen2024}. We discuss a numerical scheme, first presented in \cite{hirst2025ainsteinnumericaleinsteinmetrics}, which approximates Riemannian Einstein metrics on arbitrary manifolds using a neural networks; its effectiveness is corroborated by recovering the usual round metrics on spheres of various dimensions. The generalisation of this method to the case of lens spaces and exotic spheres is discussed.

\section{Structure and Content}
This thesis is structured as follows. \\
Chapter \ref{chap:2} reviews Yang--Mills theory and the theory of fibre bundles. It gives an overview on how these two topics are two faces of the same coin, before specialising to self-dual connections/gauge fields in $\text{SU(2)}$ Yang--Mills theory. Then, twisted self-duality equations are presented, alongside with two solutions; their geometric interpretation is discussed, and naturally leads to considering bundles associated to the ones constructed by Milnor. The results presented in this chapter, beyond the standard literature review, appeared in \cite{Berman:2022dpj}. \\
Chapter \ref{chap:3} is the main chapter of the thesis. It introduces the concept of inequivalent differentiable structures, and presents its first (and arguably most natural) realisation through exotic $7$--spheres. A thorough discussion of its geometry, in terms of the Kaluza--Klein formalism, is presented, by reviewing the results of \cite{Gherardini:2023uyx} and \cite{berman2024curvatureexotic7sphere}. \\
Chapter \ref{chap:4} summarises some facts about the differential-geometric and differentia-topological properties of exotic spheres and of some other exotic manifolds. Moreover, it discusses some features of the homeomorphic maps between ordinary spheres and exotic ones, together with some possible interpretations in the context of general relativity. \\
Chapter \ref{chap:5} reviews a recent numerical method, based on machine learning, for approximating Riemannian metrics on arbitrary manifolds (\cite{hirst2025ainsteinnumericaleinsteinmetrics}). We present its application to the case of ordinary spheres and discuss the route to generalising it to exotic ones.

\section{Notation and Conventions}
\label{sec:Notation_and_conventions_ch1}
Since most of the work in this thesis lies somewhere between theoretical physics and pure mathematics, a number of formal and stylistic choices were not trivial to make. Below we summarise some of the notational and conventional aspects that might lead to ambiguities, due to different definitions being present in the literature.

\subsection*{Typography}
A word in \emph{italics} signals either the \emph{first appearance} of a technical term,
or a deliberate emphasis — most often to stress negation, as in ``the bundle is \emph{not} trivial.''  Italics are used sparingly; no other font variants are employed for emphasis. 
Inverted commas are used to denote a ``handwavy'' use of a concept, in a slighlty inappropriate and not rigorous way.


\subsection*{Relations}
We use $\cong_{\mathrm{diff}}$ to denote diffeomorphism and $\cong_{\mathrm{hom}}$ for homeomorphism. Equivalence relations are denoted by $\sim$, which is also used to mean ``roughly'' or ``behaves like''; this will be evident from the context.

\subsection*{Indices}
Throughout the thesis, we denote four-dimensional indices with Greek letters from the second half of the alphabet: $\mu , \nu, \rho, \sigma, \cdots$. The corresponding ``flat'' indices (see \ref{sec:VielbeinFormalism} for an overview of the vielbein formalism) are denoted by Latin letters from the first half of the alphabet: $a,b,c,d, \cdots $. These run from $1$ to $4$ (Chapter \ref{chap:2} and first half of Chapter \ref{chap:3}) or from $0$ to $3$ (second half of chapter \ref{chap:3}). This is due to different literature conventions for instantons, coming from the physics and mathematics literature, respectively; the choice adopted in each part of this thesis is clearly stated in the relevant sections. \\
The only instances of a letter subscript which does not denote an index are in Section \ref{subsec_twisted_solutions_chap2} (where ``L'' stands for \textit{Lorentzian} and ``E'' for \textit{Euclidean}), in Section \ref{sec:Quaternionc_chapB} (where ``L'' stands for \textit{left}) and in Section \ref{subsec:Kaluza_Klein_on_lens_chap3} (where subscripts are used to label different charts); the context should make the notation clear in each of these cases. \\
There are a number of other circumstances where indices are used in relation to a coordinate basis on a given manifold (not four-dimensional) or to a Lie algebra basis. For those cases, we reserve ourselves the freedom to adopt different conventions on a case-by-case basis; the range and meaning of each index choice is clearly specified when needed. \\
Einstein summation convention is also used; this means that when two indices are repeated in the same expression, they are being summed over:
\begin{align}
    x_{\mu} x_{\mu} \quad \mathrm{means} \quad \sum_{\mu} x_{\mu } x_{\mu} \, .
\end{align}
Apart from Section \ref{subsec_twisted_solutions_chap2}, no distinction is made between indices up or down, because the Kronecker delta is involved in their raising and lowering. \\
For the special case of three-vectors, arrows are sometimes used to denote them:
\begin{align}
    \vec{A} \quad \mathrm{means} \quad (A_1, A_2, A_3) \, .
\end{align}

\subsection*{Curvature(s)}
The Christoffel symbols and Riemann tensor in our notation read
\begin{align}
\begin{aligned}
&\Gamma^\mu{}_{\nu \sigma}  :=\frac{1}{2} g^{\mu \rho}\left(\partial_\nu g_{\sigma \rho}+\partial_\sigma g_{\nu \rho}-\partial_\rho g_{\nu \sigma}\right) \, , \\
&R^\rho{}_{\sigma \mu \nu}=\partial_\mu \Gamma_{\nu \sigma}^\rho-\partial_\nu \Gamma_{\mu \sigma}^\rho+\Gamma_{\mu \lambda}^\rho \Gamma_{\nu \sigma}^\lambda-\Gamma_{\nu \lambda}^\rho \Gamma_{\mu \sigma}^\lambda \, .
\end{aligned}
\end{align}
As usual, the components of the Ricci tensor are denoted as $R_{\mu \nu}$ (and obtained via contracting two indices of the Riemann tensor); in coordinate-free notation, we use $Ric(g)$ to refer to the Ricci tensor associated with $g$. The scalar curvature is $R$, with appropriate subscripts where it might be confused with the radius of some sphere. We are not be concerned with sectional or Gaussian curvature in what follows.

\subsection*{Special Tensors}
This work deals almost exclusively with Riemannian geometry. The only appearance of a Lorentzian metric, in Chapter \ref{chap:2}, is due to a mere coincidence: the conjugation of a quaternion, in components, involves the Minkowski metric. We only use the ``mostly-plus'' signature, and two forms of Minkowski metric can be found: $\eta = \mathrm{diag}(1,1,1,-1)$ (in Chapter \ref{chap:2}) and $^o\eta=\mathrm{diag}(-1,1,1,1)$ (in Chapter \ref{chap:Mathematica_definitions_and_theorems}). \\
As for the epsilon tensor, we follow usual convention $\varepsilon_{1234}=+1$, $\varepsilon_{0123}=+1$. 

\subsection*{Pauli Matrices and 't Hooft Symbols}
Throughout this thesis, we always refer to the standard Pauli matrices
\begin{equation}
  \sigma^{1}= \begin{pmatrix} 0 & 1 \\ 1 & 0 \end{pmatrix}, 
  \qquad
  \sigma^{2}= \begin{pmatrix} 0 & -\,\mathrm{i} \\ \mathrm{i} & 0 \end{pmatrix},
  \qquad
  \sigma^{3}= \begin{pmatrix} 1 & 0 \\ 0 & -1 \end{pmatrix},
\end{equation}
which obey
\begin{equation}
  \sigma^{a}\sigma^{b}= \delta^{ab}\,I+ \mathrm{i}\,\varepsilon^{abc}\sigma^{c},
  \qquad
  \operatorname{Tr}(\sigma^{a}\sigma^{b}) = 2\,\delta^{ab} \,
\end{equation}
where $I$ is the $2 \times 2$ identity matrix.
Different generators will be used depending on the section. \\
Throughout Chapter \ref{chap:2} and for the first half of Chapter \ref{chap:3}, we adopt the following quaternionic basis:
\begin{align}
\sigma^{\mu} =(\vec{\sigma}, i I) = (\sigma^1, \sigma^2, \sigma^3, i I) \, , \quad 
\bar{\sigma}{^\mu}  =(\vec{\sigma},-i I) = (\sigma^1, \sigma^2, \sigma^3, - iI) \, ,
\end{align}
where $I$ is the $2 \times 2$ identity matrix. We define
\begin{align}
    \sigma^{\mu \nu}=\frac{1}{2}\left(\sigma^\mu \bar{\sigma}^\nu-\sigma^\nu \bar{\sigma}^\mu\right) \, , \quad \bar{\sigma}^{\mu \nu}=\frac{1}{2}\left(\bar{\sigma}^\mu \sigma^\nu-\bar{\sigma}^\nu \sigma^\mu\right) \, ,
\end{align}
satisfying\footnote{Two comments. For mathematicians: this is the first instance of the Einstein summation convention in use. For physicists: we raise and lower indices freely since we are in Euclidean signature.} 
\begin{align}
    \sigma_{\mu \nu}=-\frac{1}{2} \epsilon_{\mu \nu \rho \sigma} \sigma_{\rho \sigma} \, , \quad \bar{\sigma}_{\mu \nu}=\frac{1}{2} \epsilon_{\mu \nu \rho \sigma} \bar{\sigma}_{\rho \sigma} \, .
\end{align}
The former is referred as anti-self-duality, and the latter as self-duality. These symbols naturally lead to the 't Hooft symbols as
\begin{align}
    \bar{\sigma}_{\mu \nu}=i \eta_{a \mu \nu} \sigma^a, \quad \sigma_{\mu \nu}=i \bar{\eta}_{a \mu \nu} \sigma^a \, .
\end{align}
This recovers the conventional component-wise definition:
\begin{align}
  \;
  \eta^{a}_{\mu\nu}
    = \varepsilon_{a\mu\nu 4}
      + \delta_{a\mu}\,\delta_{\nu 4}
      - \delta_{a\nu}\,\delta_{\mu 4}, \qquad
  \;
  \bar\eta^{a}_{\mu\nu}
    = \varepsilon_{a\mu\nu 4}
      - \delta_{a\mu}\,\delta_{\nu 4}
      + \delta_{a\nu}\,\delta_{\mu 4},
  \label{eq:tHooftDef}
\end{align}
where \(a=1,2,3\).
They satisfy
\begin{align}
  &\eta^{a}_{\mu\nu}= +\tfrac12\,\varepsilon_{\mu\nu\rho\sigma}\,\eta^{a}_{\rho\sigma} \, ,
  \quad \bar\eta^{a}_{\mu\nu}= -\tfrac12\,\varepsilon_{\mu\nu\rho\sigma}\,\bar\eta^{a}_{\rho\sigma} \, ,
  \label{eq:SelfDuality}
\end{align}
so that \(\eta^{a}_{\mu\nu}\) (\(\bar\eta^{a}_{\mu\nu}\)) are the (anti-)self-dual 't Hooft symbols.
More properties can be found in Section \ref{subsec:Some_Relations_in_Components}. These are the most common definitions throughout the literature on instantons in theoretical physics.

\subsection*{Quaternions}
The second part of Chapter \ref{chap:3} is a based on the use of purely quaternionic objects, and hence we adopt the more natural choice
\begin{align}
      e_{c} =  ( I, - i \Vec{\sigma}) \, , \quad \bar{ e}_{c} = ( I, i \Vec{\sigma}) \,  ,
     \label{eq:Defn_e}
\end{align}
which associates the first component with the real part of the quaternion. In these conventions, the multiplication (associative but non-commutative) is specified by the relation:
$e_ie_j=-\delta_{ij}+\epsilon_{ijk}e_k$. 
Conjugation is defined by $1\mapsto1$, $e_i\mapsto-e_i$, so with $x=x^ae_a=x^0+x^ie_i$ the conjugated element is denoted as
$\bar x=x^0-x^ie_i$. It satisfies $\overline{xy}=\bar y\bar x$.
The real part, considered as a real number, reads $\Re x={1\over2}(x+\bar x)=x^0$, and the imaginary part is
$\Im x={1\over2}(x-\bar x)=x^ie_i$.
With these definitions, a component $x^a$ is extracted from $x\in\HH$ as $x^a=\Re(x\bar e^a)$.
We write the modulus as $|x|^2=x^ax^a=x\bar x=\bar x x$; it is, of course, multiplicative: $|xy|=|x||y|$.
Any non-zero quaternion has a unique inverse $x^{-1}={\bar x\over|x|^2}$.
A useful ``sigma matrix identity'' is $x\bar y+y\bar x=2\Re(x\bar y)$. Finally, we note that, when extracting the components of naturally antisymmetric quaternionic objects, the \textit{reversed} 't Hooft symbols appear (see Section \ref{subsec:'tHooft_notation_and_quaternions_chap3} and Section \ref{sec:Quaternionc_chapB}):
\begin{align}
& ^o\eta_{i m n} =\epsilon_{i m n 0} -\delta_{i m} \delta_{n 0} + \delta_{i n} \delta_{m 0} \, , \nonumber \\
& ^o\bar{\eta}_{i m n}=\epsilon_{i m n 0} + \delta_{i m} \delta_{n 0} - \delta_{i n} \delta_{m 0} \, ,
\label{eq:'tHooft_zero}
\end{align}
where $^o\eta_{i m n}$ is selfdual and $^o\bar{\eta}_{i m n}$ is anti-selfdual.

\chapter{Fibre Bundles, Instantons and \\ 
(Twisted) Self-duality \,\,\,\,\,\,\,\,\,\,\,\,\,\,\,\,\,\,\,\,\,}
\label{chap:2}
The elegant description of Maxwell and Yang--Mills theories using the language of fibre bundles is arguably one of the most fruitful intersections between pure mathematics and theoretical physics. This chapter reviews this topic, with a special focus on $\text{SU(2)}$ Yang--Mills theory, (twisted) self-duality equations, (twisted) instantons and their geometric interpretation (\cite{Berman:2022dpj}).

\section{Introduction, Overview and Structure}
The idea that a \emph{potential} rather than a field strength is the primary dynamical object appears already in Maxwell’s synthesis of electricity and magnetism, where the four–potential is defined only up to a gradient (\cite{Maxwell1865}).  Quantum mechanics sharpened this ambiguity: Fock showed in $1932$ that the phase of a wave-function can be changed independently at each point of configuration space without affecting observables, a first glimpse of \emph{local} (gauge) symmetry - see for instance \cite{Fock1932}. Dirac’s analysis of the magnetic monopole made the lesson topological: a globally defined potential need not exist at all, and the resulting charge quantisation hints at non-trivial ``gluing data’’ for potentials defined on overlapping patches, as discussed in \cite{Dirac1931}. The Aharonov–Bohm effect turned this from speculation into measurable fact, demonstrating that such global issues really matter for quantum interference (\cite{AharonovBohm1959}).

While these clues accumulated, pure mathematicians were building a rigorous and precise framework in which the facts above are naturally accommodated. Whitney’s construction of smooth sphere–bundles provided the first systematic examples of spaces whose fibres vary smoothly over a base manifold - see the seminal papers \cite{Whitney1940} and \cite{doi:10.1073/pnas.21.7.464}, as well as the first systematic treatment of the topic in \cite{7a97cef4-8443-3a57-aced-2037f84b9e06}. A decade later, Ehresmann supplied the missing geometric ingredient by defining a \emph{connection} on an arbitrary bundle, thereby formalising the notion of parallel transport and curvature (\cite{Ehresmann1951}).

The decisive step that unified the two threads was taken in the mid-twentieth century. In \cite{YangMills1954}, Yang and Mills generalised Maxwell’s $U(1)$ symmetry to non-Abelian isospin rotations and discovered that insisting on local symmetry inevitably introduces a new field that transforms like an Ehresmann connection. Trautman was the first to state this equivalence explicitly: electromagnetic and Yang–Mills potentials are local representatives of a connection on a principal bundle, while the physical fields are its curvature forms - see \cite{Trautman1970}.\footnote{To be precise, \cite{Trautman1970} is based on a series of lectures that Trautman gave at King's College London in $1967$, three years prior to the publication. Between $1967$ and $1970$, mimeographed versions of these lecture notes were circulating among physicists, as \cite{Kerner1968} and \cite{10.1063/PT.3.2799} testify.} Wu and Yang then elaborated on the ``dictionary’’ that matches gauge-theoretic notions with their bundle-theoretic counterpart in  \cite{WuYang1975}.  Atiyah’s lectures distilled the picture: all classical gauge theories live naturally on principal fibre bundles, their dynamics governed by curvature and their topological sectors by characteristic classes, as summarised in \cite{Atiyah1979}. \\
While this geometric unification was being formalised, physicists discovered finite-action \emph{self-dual} solutions of the Euclidean Yang–Mills equations.  The pioneering work of Belavin, Polyakov, Schwarz and Tyupkin introduced these \textit{instantons} as exact solutions \cite{BELAVIN197585}, and ’t~Hooft soon revealed their profound quantum significance \cite{tHooft:1976snw}.  Mathematicians quickly recognised their utility: the Atiyah–Drinfeld–Hitchin–Manin (ADHM) construction provided a complete algebraic-geometric classification of instantons on $\mathbb{R}^{4}$ \cite{ATIYAH1978185}, cementing the role of gauge theory in modern differential and algebraic geometry. \\
Thus, Maxwell's electromagnetic potential, the non-Abelian fields of Yang and Mills, and the mathematical formalism developed of Whitney and Ehresmann are now recognised as facets of one geometric object: a principal fibre bundle endowed with a connection, whose self-dual instanton sectors link physics, topology and geometry in a remarkably unified framework. Such a beautiful interplay between mathematics and physics is summarised in \cite{Nakahara:2003nw} and \cite{RevModPhys.52.175} (among many others), which will be followed when reviewing this material in the first part of the chapter. 

The second part covers some of the results of \cite{Berman:2022dpj}, which is motivated by a central mathematical fact: in four dimensions with Euclidean signature, the Hodge Star acting on a two form curvature squares to one. This allows to think of the Hodge star as an operator on the two form curvature with two eigenvalues, $+1$ and $-1$. The self-dual solutions correspond to the positive eigenvalue and the anti-self-dual solutions to the negative eigenvalue. When one has a similar non-trivial operator acting on the internal Lie algebra, i.e.~an operator that squares to one, one may then also decompose the Lie algebra in this way, according to eigenvectors of this involution. Combining the two operators of Hodge star and the internal involution allows the construction of a \textit{twisted self-duality}, a term coined in \cite{Cremmer_1998, Cremmer_1998_2}. In fact, this sort of twisted self-duality has appeared in various formulations of supergravity and in world sheet duality symmetric formulations of the string theory \cite{Tseytlin:1990nb,Tseytlin:1990va,Berman:2007xn,Berman:2013eva,Alfonsi:2021bot}.\footnote{For more treatments of duality invariant reformulations of string theory and generalised (exceptional) geometry, the reader is referred to the (non-exhaustive set of) seminal papers \cite{Hull:2007gg, PachecoWaldram:2008egg, Hillmann:2009e7, HullZwiebach:2009dft, HohmHullZwiebach:2010bia, HohmHullZwiebach:2010gmetric, Berman:2011gg, Berman:2011so55, Berman:2011jh, Berman:2012uy,  Berman_2020}.} A very natural setting for studying the twisted self-duality equations and its solutions is that of $\text{SO(4)}$ Yang--Mills theory. This is the subject of the study presented in \cite{Berman:2022dpj}, and summarised in the next sections.
The structure of the chapter is the following. \\
In the first section, \ref{sec:Yang_Mills_and_bundles_chap2}, we provide an overview on Yang--Mills theory and principal fibre bundles, as independent frameworks. Then, we outline the formulation of the physical theory in rigorous bundle-theoretic language. \\
In Section \ref{sec:Self_duality_chap2}, we introduce self-duality equations in Yang--Mills theories and their most famous solutions: the $\text{SU(2)}$ (BPST) instanton. The geometric interpretation of such a solutions as a connection on the $S^3$-bundle over $S^4$ whose total space is $S^7$ is provided. Then, we discuss a modification of the self-duality equations by the inclusion of a ``twist'', in Section \ref{sec:Twisted_self_duality_chap2}. We comment on some of their properties, present a natural solution for $\mathrm{su}(2) \times \mathrm{su}(2) = \mathrm{so}(4)$-valued connections, both in Lorentzian and Euclidean signatures. We also comment on the geometric interpretation of the latter, which leads us to consider bundles that are intimately related to those constructed by Milnor when he discovered exotic spheres. The discussion of Milnor's work is postponed until the next chapter. \\
Section \ref{sec:Conclusions_chap2} provides a summary of the results and some comments on natural further investigations that naturally follow from them.

\section{Yang--Mills Theory and Bundle Theory}
\label{sec:Yang_Mills_and_bundles_chap2}
Classically, a gauge field is introduced as a Lie-algebra--valued vector potential
$A_\mu(x)$, but its natural home is differential geometry. Similarly, the field strength admits a very elegant geometric interpretation.
In the following, we present (part of) the dictionary between Yang--Mills physics and principal bundles.

\subsection{Yang--Mills Theory}
\label{subsec:Yang_Mills_chap2}

Let $G$ be a compact Lie group with Lie algebra $\mathfrak g$ and basis $\{T^a\}$ normalized by  
\(\operatorname{tr}(T^aT^b)=\tfrac12\delta^{ab}\).  
Spacetime usually assumed to be a $4$‐dimensional Lorentzian manifold $(M,g_{\mu\nu})$ with Hodge star $*$. \\
The \emph{gauge potential} is a Lie–algebra–valued 1–form
\begin{align}
A \;=\; A_\mu^a\,T^a\,dx^\mu \;\in\; \Omega^{1}(M,\mathfrak g),
\end{align}
acting on matter fields through the gauge‐covariant derivative
\begin{align}
D_\mu = \partial_\mu + A_\mu,
\qquad
D = d + A .
\end{align}
Its associated \emph{field–strength} 2–form is  
\begin{align}
F \;=\; dA + A\wedge A 
   \;=\; \tfrac12\,F_{\mu\nu}^a\,T^a\,dx^\mu\wedge dx^\nu,
\quad
F_{\mu\nu}^a
     = \partial_\mu A_\nu^a - \partial_\nu A_\mu^a
       + f^{abc}A_\mu^bA_\nu^c .
\label{eq:Field_strength_physics}
\end{align}
The structural identity $dF + A\wedge F - F\wedge A = 0$ becomes, in components,
\(D_{[\mu}F_{\nu\rho]} = 0\).\\

The Yang–Mills action appears identically in form and component notation:
\begin{align}
S_{\text{YM}}
  = -\frac{1}{2g^2}\int_M \operatorname{tr}\bigl(F\wedge *F\bigr)
  = -\frac{1}{4g^2}\int_M d^dx\,\sqrt{-g}\;F_{\mu\nu}^aF^{\mu\nu\,a}.
\end{align}
Variation with respect to $A$ yields the equations of motion
\begin{align}
D*F \;=\; 0,
\qquad\Longleftrightarrow\qquad
D_\mu F^{\mu\nu}=0,\;
\text{i.e. }\
\partial_\mu F^{\mu\nu\,a}
      + f^{abc}A_\mu^bF^{\mu\nu\,c}=0 .
\end{align}
To conclude our brief overview of Yang--Mills theory, let us turn to gauge transformations. For a smooth map $g:M\!\to\! G$, they read:
\begin{align}
A \mapsto gAg^{-1}+gdg^{-1},
\quad
F \mapsto gFg^{-1},
\label{eq:Gauge_transf_physics}
\end{align}
Infinitesimally, i.e.~with infinitesimal parameter 
$\epsilon=\epsilon^aT^a$, one has \(\delta_\epsilon A_\mu^a = D_\mu\epsilon^a\).
These transformations leave both the action and the equations of motion invariant.
The dual presentation above—Lie–algebra components for explicit calculations, and differential forms for geometric clarity—will be employed interchangeably in what follows.

\subsection{(Principal) Fibre Bundles}
\label{subsec:Princ_fibre_bundles_chap2}
Let $M$ be a smooth manifold of dimension $n$. Let $M$ and $E$ be smooth manifolds and let $\pi: E \to M$ be a smooth surjective map. Then the triple $(E, \pi, M)$ is called a \emph{bundle}, which we also denote as $E \xrightarrow{\pi} M$. The manifold $E$ is called the \emph{total space}, $M$ the \emph{base space}, and $\pi$ the \emph{projection}. A bundle $E' \xrightarrow{\pi'} M'$ is called a \emph{subbundle} of $E \xrightarrow{\pi} M$ if $E' \subseteq E$ and $M' \subseteq M$ are submanifolds and $\pi' = \pi|_{E'}$. Given a point $p \in M$, the pre-image $F_p = \pi^{-1}(p)$ is called the \emph{fibre} at $p$.

Let $G$ be a Lie group acting effectively on a smooth manifold $F$ from the left.\footnote{To be pedantic, one would first need to discuss coordinate bundles and then present a fibre bundle as an equivalence class of coordinate bundles. We are not pedantic; see \cite{Nakahara:2003nw}.} The bundle $(E, \pi, M)$ is promoted to a \emph{(differentiable) fibre bundle}\label{th:Fibre_bundle}, denoted by the collection $(E, \pi, M, F, G)$, if the following properties hold:
\begin{itemize}
    \item Every fibre is diffeomorphic to the \emph{typical fibre}: $F_p = \pi^{-1}(p) \cong F$ for all $p \in M$.
    \item $M$ is equipped with an open cover $\{U_i\}_{i \in I}$ together with diffeomorphisms
    \begin{align}
        \phi_i: U_i \times F \longrightarrow \pi^{-1}(U_i),
    \end{align}
    satisfying $\pi \circ \phi_i(p, f) = p$ for all $p \in U_i$ and $f \in F$. The inverse map $\phi_i^{-1}: \pi^{-1}(U_i) \to U_i \times F$ is called a \emph{local trivialisation}.
    \item For each $p \in U_i$, the restricted map $\phi_{i,p}: F \to F_p$ defined by $\phi_{i,p}(f) = \phi_i(p, f)$ is a diffeomorphism. On any non-empty overlap $U_i \cap U_j$, the composition
    \begin{align}
        t_{ij}(p) = \phi_{i,p}^{-1} \circ \phi_{j,p}: F \longrightarrow F
    \end{align}
    is an element of the \emph{structure group} $G$. The maps $t_{ij}: U_i \cap U_j \to G$ are called \emph{transition functions}, and they satisfy $\phi_i(p,\, t_{ij}(p) \cdot f) = \phi_j(p, f)$.
\end{itemize}

The transition functions are smooth and satisfy the \emph{cocycle conditions}:
\begin{enumerate}
    \item On $U_i \cap U_j \cap U_k$: $t_{ik}(p) = t_{ij}(p)\, t_{jk}(p)$.
    \item On $U_i \cap U_j$: $t_{ij}(p) = (t_{ji}(p))^{-1}$.
    \item On $U_i$: $t_{ii}(p) = e$.
\end{enumerate}
Conversely, given an open cover $\{U_i\}$ of $M$ and a collection of smooth maps $\{t_{ij}: U_i \cap U_j \to G\}$ satisfying these cocycle conditions, one can construct a fibre bundle with structure group $G$ and typical fibre $F$ (see \cite{Nakahara:2003nw}, for instance).

If we choose a different set of local trivialisations $\{\phi'_i\}$, we obtain new transition functions $\{t'_{ij}\}$ related to the old ones by
\begin{align}
    t'_{ij}(x) = h_i(x)\, t_{ij}(x)\, (h_j(x))^{-1},
\end{align}
where $h_i: U_i \to G$ are smooth maps. Two bundles are isomorphic if their transition functions are related in this way.

A \emph{section} of a bundle $\pi: E \to M$ is a smooth map $s: M \to E$ such that $\pi \circ s = \mathrm{id}_M$. Note that while many general fibre bundles (such as vector bundles) inherently possess global sections without being trivial, the existence of a global section is a much stronger condition in the specific case of a \emph{principal} bundle, which we now introduce.

\bigskip

A fibre bundle of particular importance arises when the typical fibre $F$ is the structure group $G$ itself. This is called a \emph{principal fibre bundle} (or \emph{principal $G$-bundle}): it is a fibre bundle $(P, \pi, M, G, G)$ in which the typical fibre coincides with the structure group, together with a free right action $R: P \times G \to P$, written $(p, g) \mapsto R_g(p) = p \cdot g$, satisfying:
\begin{enumerate}
    \item $(p \cdot g_1) \cdot g_2 = p \cdot (g_1 g_2)$ for all $p \in P$ and $g_1, g_2 \in G$.
    \item $p \cdot e = p$, where $e$ is the identity element of $G$.
\end{enumerate}
We often denote the bundle by the quadruple $(P, M, \pi, G)$ or simply $P(M,G)$.

Concretely, the fibre bundle structure and the right action are linked as follows. In the general fibre bundle definition, $G$ acts on $F$ from the left via the transition functions. When $F = G$, this left action is simply left multiplication: $t_{ij}(p)$ acts on $g \in G$ by $g \mapsto t_{ij}(p)\, g$. The additional right action, on the other hand, acts on the total space $P$ from the right and is free and transitive on each fibre $\pi^{-1}(x) \cong G$. It is this right action that provides the principal bundle with its characteristic rigidity: once a single point in a fibre is specified, the entire fibre is determined by the group action.

This structure must satisfy the following axioms:
\begin{enumerate}[label=(\roman*)]
    \item The projection $\pi$ is a smooth, surjective map.
    \item The right action of $G$ on $P$ is \emph{free}, meaning that if $p \cdot g = p$ for some $p \in P$, then $g = e$.
    \item The fibres of the bundle are the orbits of the $G$-action. That is, for any $x \in M$, the fibre $\pi^{-1}(x)$ is an orbit of $G$. This implies that $\pi(p \cdot g) = \pi(p)$ for all $p \in P$, $g \in G$, and that the action is transitive on each fibre.
    \item The bundle is \emph{locally trivial}. For any $x \in M$, there exists an open neighbourhood $U \subset M$ of $x$ and a diffeomorphism
    \begin{align}
        \psi_U: \pi^{-1}(U) \to U \times G,
    \end{align}
    called a \emph{local trivialisation}, with the following properties:
    \begin{itemize}
        \item $\psi_U(p) = (\pi(p), \phi_U(p))$, where $\phi_U: \pi^{-1}(U) \to G$ is a smooth map. This ensures that the first component of the map is the projection, i.e., $\mathrm{pr}_1 \circ \psi_U = \pi$.
        \item The map is $G$-equivariant in the sense that
        \begin{align}
            \phi_U(p \cdot g) = \phi_U(p)\, g,
        \end{align}
        which implies $\psi_U(p \cdot g) = (\pi(p),\, \phi_U(p)\, g)$.
    \end{itemize}
\end{enumerate}
Each fibre $\pi^{-1}(x)$ is diffeomorphic to the structure group $G$. Note that the local trivialisation conventions we chose differ slightly in direction between the general fibre bundle and the principal bundle setting. In the former, a local trivialisation is a diffeomorphism $\phi_i: U_i \times F \to \pi^{-1}(U_i)$, mapping \emph{from} the product \emph{to} the total space. For a principal bundle, it is common to write the local trivialisation in the opposite direction, as a diffeomorphism $\psi_U: \pi^{-1}(U) \to U \times G$, mapping \emph{from} the total space \emph{to} the product. The two conventions are simply related by $\psi_U = \phi_U^{-1}$, and the transition functions agree: if $\psi_i(p) = (\pi(p), \phi_i(p))$ and $\psi_j(p) = (\pi(p), \phi_j(p))$ for the principal bundle, then
\begin{align}
    \phi_i(p) = t_{ij}(\pi(p))\, \phi_j(p),
\end{align}
which is the same relation $t_{ij}(x) = \phi_{i,x}^{-1} \circ \phi_{j,x}$ from the fibre bundle definition above, now expressed in terms of left multiplication in $G$.

\bigskip

A principal bundle is \emph{trivial} if and only if it admits a global section $s: M \to P$ (i.e., $\pi \circ s = \mathrm{id}_M$). If such a section exists, one can define a global trivialisation $\Psi: M \times G \to P$ by $\Psi(x, g) = s(x) \cdot g$.

Let $H$ be a Lie subgroup of $G$. A \emph{reduction of the structure group} from $G$ to $H$ is a submanifold $P_H \subset P$ such that $(P_H, M, \pi|_{P_H}, H)$ is a principal $H$-bundle. A reduction exists if and only if the transition functions $\{t_{ij}\}$ of $P(M,G)$ can be chosen to take their values in the subgroup $H \subset G$.

\bigskip

Let us now go through some canonical examples.

\begin{itemize}
    \item \textbf{Trivial Bundle:} The simplest principal bundle is the product manifold $P = M \times G$. The projection is $\pi = \mathrm{pr}_1: M \times G \to M$, so $\pi(x,h) = x$. The right action of $g \in G$ on a point $(x,h) \in M \times G$ is defined as $(x,h) \cdot g = (x, hg)$.

    \item \textbf{Frame Bundle:} Let $E$ be a vector bundle of rank $k$ over $M$. For each point $x \in M$, the fibre $E_x$ is a $k$-dimensional vector space. A \emph{frame} at $x$ is an ordered basis $\{e_1, \dots, e_k\}$ of $E_x$. The \emph{frame bundle} of $E$, denoted $F(E)$, is the set of all frames at all points of $M$:
    \begin{align}
        F(E) = \bigcup_{x \in M} \{ \text{frames in } E_x \}.
    \end{align}
    The projection $\pi: F(E) \to M$ maps a frame at $x$ to the point $x$. The structure group is the general linear group $\mathrm{GL}(k, \mathbb{R})$. The right action of an element $A = (A^i_j) \in \mathrm{GL}(k, \mathbb{R})$ on a frame $u = (e_1, \dots, e_k)$ is given by
    \begin{align}
        u \cdot A = u' = (e'_1, \dots, e'_k), \quad \text{where} \quad e'_j = \sum_{i=1}^k e_i A^i_j.
    \end{align}
    This defines a principal bundle $F(E)(M, \mathrm{GL}(k, \mathbb{R}))$.

    A particularly important case is when $E = TM$, the tangent bundle of $M$. The corresponding frame bundle $F(TM)$ is called the \emph{linear frame bundle} of $M$, often denoted $F(M)$. Its structure group is $\mathrm{GL}(n, \mathbb{R})$. If $M$ is equipped with a Riemannian metric, we can consider only orthonormal frames. This constitutes a reduction of the structure group from $\mathrm{GL}(n, \mathbb{R})$ to the orthogonal group $\mathrm{O}(n)$. If, in addition, $M$ is orientable, we can restrict to positively-oriented orthonormal frames, which reduces the structure group further to the special orthogonal group $\mathrm{SO}(n)$.

    \item \textbf{Complex Hopf fibration:} $S^{3} \to S^{2}$ with fibre $S^{1}$, a principal $U(1)$-bundle whose non-triviality is measured by the first Chern class $c_{1} = 1 \in H^{2}(S^{2}, \mathbb{Z})$. This will be mentioned in Chapter~\ref{chap:3}, Section~\ref{sec:lens_spaces_chap3}.

    \item \textbf{Quaternionic Hopf fibration:} $S^{7} \to S^{4}$ with fibre $S^{3}$, a principal $\operatorname{Sp}(1) \cong \mathrm{SU}(2)$-bundle whose second Chern class $c_{2} = 1$ generates $H^{4}(S^{4}, \mathbb{Z})$. This will be outlined in Section~\ref{subsec:geometric_instaton_chap2}, and described in more detail in Chapter~\ref{chap:3}, Section~\ref{sec:ordinary_sphere_chap3}.
\end{itemize}

Principal bundles thus provide a global framework for keeping track of smoothly varying group data over manifolds. Their topological character is encoded in the transition functions, while geometric refinements (connections, curvature, characteristic classes) will be introduced in later subsections. Note that this is just a quick overview on the topic, and very detailed presentations of fibre bundles can be found in many classic textbooks and reviews. Two introductory expositions, very close to the author's heart, are provided in \cite{Nakahara:2003nw} and \cite{RevModPhys.52.175}; they also cover a good portion of the material presented in the rest of this section.

\subsection{The Fibre Bundle Structure Underlying Yang--Mills}
\label{subsec:Fibre_bundles_and_Yang_Mills_chap2}

We now introduce the concept of a connection on a principal bundle $P(M,G)$, which provides a way to differentiate sections of associated vector bundles and defines the notion of parallel transport. Let $\g$ be the Lie algebra of the structure group $G$.

A connection on $P$ is a specification of a ``horizontal'' direction at every point $p \in P$. The tangent space $T_p P$ at any point $p \in P$ naturally contains a \emph{vertical subspace} $V_p P$, which consists of vectors tangent to the fibre passing through $p$. Formally, $V_p P = \ker(\pi_*: T_p P \to T_{\pi(p)} M)$. The vertical subspace $V_p P$ is isomorphic to the Lie algebra $\g$. This isomorphism is established by the \emph{fundamental vector field} $A^*$ on $P$ associated with an element $A \in \g$, defined as
\begin{align}
    A^*_p = \left. \frac{d}{dt} \right|_{t=0} (p \cdot \exp(tA)).
\end{align}
A \emph{connection} is a choice of a complementary \emph{horizontal subspace} $H_p P \subset T_p P$ at each $p \in P$, such that
\begin{align}
    T_p P = V_p P \oplus H_p P.
\end{align}
This choice must be smooth and compatible with the group action, meaning $(R_g)_* H_p = H_{p \cdot g}$ for all $g \in G$.

This geometric definition is elegantly captured by a \emph{connection 1-form} $\omega$, which is a $\g$-valued 1-form on the total space $P$. The form $\omega$ is defined by the following two properties:
\begin{enumerate}
    \item It maps fundamental vector fields back to their corresponding Lie algebra elements: $\omega(A^*_p) = A$ for all $A \in \g$.
    \item It is equivariant under the right action of $G$: $R_g^* \omega = \mathrm{Ad}(g^{-1}) \omega$, where $\mathrm{Ad}(g^{-1})$ is the adjoint representation of $G$ on its Lie algebra $\g$.
\end{enumerate}
The horizontal subspace at $p$ is then simply the kernel of $\omega_p$: $H_p P = \ker(\omega_p)$.

The \emph{curvature} of the connection $\omega$ is a $\g$-valued 2-form $\Omega$ on $P$, which measures the failure of the horizontal distribution to be integrable. It is defined by the \emph{Cartan structure equation}:
\begin{align}
    \Omega = d\omega + \frac{1}{2}[\omega, \omega],
\end{align}
where the bracket is the wedge product of forms combined with the Lie bracket in $\g$. The curvature $\Omega$ is a horizontal form, meaning $\Omega(X,Y) = 0$ if either $X$ or $Y$ is a vertical vector. It also transforms equivariantly: $R_g^* \Omega = \mathrm{Ad}(g^{-1}) \Omega$.

\bigskip

The language of connections and curvature finds its most direct physical application in Yang--Mills gauge theory. The abstract objects $\omega$ and $\Omega$ on $P$ are related to the familiar fields on the base space $M$ by choosing a local section. Let $\{U_i\}$ be an open cover of $M$ and let $s_i: U_i \to P$ be a local section over each patch.

The \emph{gauge potential} (or \emph{gauge field}) on $U_i$ is the $\g$-valued 1-form $A_i$ obtained by pulling back the connection 1-form $\omega$ via the section $s_i$:
\begin{align}
    A_i = s_i^* \omega.
\end{align}
Similarly, the \emph{field strength} on $U_i$ is the $\g$-valued 2-form $F_i$ obtained by pulling back the curvature 2-form $\Omega$:
\begin{align}
    F_i = s_i^* \Omega.
\end{align}
By pulling back the Cartan structure equation, we find the relationship between the gauge field and its field strength:
\begin{align}
    F_i = s_i^*(d\omega + \frac{1}{2}[\omega, \omega]) = d(s_i^*\omega) + \frac{1}{2}[s_i^*\omega, s_i^*\omega] = dA_i + \frac{1}{2}[A_i, A_i] \, .
    \label{eq:Field_strength_maths}
\end{align}
On an overlap region $U_i \cap U_j$, the local sections are related by the transition function $t_{ij}: U_i \cap U_j \to G$, such that $s_j(x) = s_i(x) \cdot t_{ij}(x)$. This change of local section induces a transformation on the gauge potential. Using the properties of $\omega$, one finds that the gauge potentials $A_i$ and $A_j$ are related by
\begin{align}
    A_j = \mathrm{Ad}(t_{ij}^{-1}) A_i + t_{ij}^{-1} dt_{ij} \, .
    \label{eq:Gauge_transf_maths}
\end{align}
This is precisely the \emph{gauge transformation} law for a non-abelian gauge field from \ref{eq:Gauge_transf_physics}, with $g$ being identified with $t_{ij}^{-1}$. The function $t_{ij}(x)$ plays the role of the local gauge transformation. The field strength, in turn, transforms more simply:
\begin{align}
    F_j = \mathrm{Ad}(t_{ij}^{-1}) F_i \, .
\end{align}
This shows that the field strength transforms covariantly under a gauge transformation. The Yang--Mills action is constructed from the gauge--invariant quantity $\mathrm{Tr}(F \wedge *F)$, where the trace is taken in the Lie algebra.

\section{Self-duality}
\label{sec:Self_duality_chap2}
This section discusses the self-duality equations as a first-order system of partial differential equations which implies minimisation of the Yang--Mills action, and its most famous solutions: $\mathrm{SU(2)}$ instantons. The geometric interpretation of these solutions as connections on $S^7$ (i.e.~the quaternionic Hopf fibration) is also given.

\subsection{The Equations}
\label{subsec:self_dual_equations_chap2}

Let us consider four-dimensional \textit{Euclidean} Yang–Mills theory. As discussed, one treats the gauge potential 
\(A_\mu(x)\,dx^\mu\in\Omega^{1}(M,\mathfrak g)\) as the dynamical field, which can be identified with the connection
on a principal $G$–bundle $P\!\to\!M$ and takes values in the Lie algebra $\mathfrak g$
of the compact gauge group \(G\).  
Its field strength is given by \eqref{eq:Field_strength_maths} (or \eqref{eq:Field_strength_physics}). Placing the theory on a Riemannian four-manifold \((M,g)\) and using the Hodge dual
\(*:\Omega^{2}\!\to\!\Omega^{2}\) with \(*^{2}=+1\), the Euclidean Yang–Mills action is  
\begin{align}
S[A] \;=\;- \frac{1}{2g^{2}}\int_{M}\!\operatorname{tr}\!\bigl(F\wedge* F\bigr)
         =- \frac{1}{4g^{2}}\!\int_{M}\!d^{4}x\,\operatorname{tr}(F_{\mu\nu}F^{\mu\nu}),
\end{align}
where \(g\) is the Yang–Mills coupling and \(\operatorname{tr}\) is a negative-definite Killing form, \(\operatorname{tr}(T_aT_b)=-\tfrac12\delta_{ab}\). This action is manifestly invariant under gauge transformations and, from a mathematical perspective, it provides a canonical choice of connection on the bundle, i.e.~that which extremises the functional above.
According to the discussion above, varying \(A_\mu\) gives the (second-order) Yang–Mills equations  
\begin{align}
D^\mu F_{\mu\nu}=0
\quad \Longleftrightarrow\quad D\!* F = 0 \, .
\end{align}
The 4-form \(\operatorname{tr}(F\wedge F)\) is closed and its integral
\begin{align}
k \;=\; \frac{1}{8\pi^{2}}\int_{M}\!\operatorname{tr}\!\bigl(F\wedge F\bigr)\in\mathbb Z
\end{align}
measures the second Chern number: physically, the number of times the gauge field wraps the group manifold as one covers space-time.  In the path integral this topological sector label $k$ appears as an instanton number.

One can rewrite the action by completing the square:
\begin{align}
\operatorname{tr}(F\wedge* F)
   = \tfrac12\,\operatorname{tr}\!\bigl((F\mp* F)\wedge*(F\mp* F)\bigr)
     \;\pm\;\operatorname{tr}(F\wedge F).
\end{align}
After integration, this gives the Bogomol'nyi bound
\begin{align}
S[A] \;=\; \frac{1}{4g^{2}}\!\int_{M}\!\operatorname{tr}\!\bigl((F\mp* F)\wedge*(F\mp* F)\bigr)
           \;\pm\;\frac{8\pi^{2}}{g^{2}}\,k
           \;\;\ge\;\; \frac{8\pi^{2}}{g^{2}}\lvert k\rvert,
\end{align}
since the first term is non-negative.  
Saturation occurs precisely when that square vanishes, yielding the self-duality or anti-self-duality equations
\begin{align}
\,F = \pm\,* F\, .
\end{align}
These are first-order but automatically solve the full Yang–-Mills equations because $D\!* F=0$ follows from the Bianchi identity $DF=0$ when $F=\pm* F$.

Physically, such self-dual configurations are instantons (for $+$) or anti-instantons ($-$):
localised lumps of colour-magnetic field in Euclidean time that tunnel between degenerate classical vacua labelled by different $k$.
Each carries the minimal Euclidean action
\begin{align}
S_\text{instanton} = \frac{8\pi^{2}}{g^{2}}\lvert k\rvert,
\end{align}
so their exponential weight \(\exp(-S)\) dominates non-perturbative amplitudes.

\subsection{The SU(2) Solutions}
\label{subsec:instantons_chap2}
Belavin, Polyakov, Schwartz and Tyupkin (\cite{BELAVIN197585}) discovered the first explicit
solution of the self–duality equation \(F=* F\) in pure Yang–Mills theory,
working on \(\mathbb{R}^{4}\) with gauge group \(\text{SU(2)}\). The gauge potential can be elegantly written by employing the self-dual ’t Hooft symbols
\(\eta^{a}_{\mu\nu}\) one has\footnote{See Conventions Section \ref{sec:Notation_and_conventions_ch1} for details on the 't Hooft symbols.}
\begin{align}
    A_{\mu}(x)=
\frac{2\,\eta^{a}_{\mu\nu}(x-x_{0})^{\nu}}
     {(x-x_{0})^{2}+\rho^{2}}\;T_{a} \, ,
\label{eq:Insta_def}
\end{align}
where $(x-x_{0})^{2} \equiv \bigl|x^{\mu}-x_{0}^{\mu}\bigr|^{2}$, $
\rho>0$ is the \textit{size},
$x_{0}^{\mu}\in\mathbb R^{4}$ is the \textit{centre} and
$T_{a}=\tfrac{\sigma_{a}}{2\, i}$.
In addition to self-duality, which guarantees that this gauge field satisfies the Yang--Mills equations, there is another key property of this expression: it tends to a \textit{pure gauge} configuration as one approaches infinity:
\begin{align}
    A_\mu \stackrel{|x|^2 \rightarrow \infty}{=} U^{-1} \partial_\mu U \, ,
\end{align}
with $U$ given by (see Section \ref{sec:Notation_and_conventions_ch1} for definition of $\sigma, \bar{\sigma}$):
\begin{align}
    U(x)  = i x_\mu \sigma_\mu / \sqrt{x^2}, \quad U^{-1}(x)=-i x_\mu \bar{\sigma}_\mu / \sqrt{x^2} \, .
\end{align}
The associated field strength reads
\begin{align}
F_{\mu\nu}(x)=
\frac{4\,\rho^{2}}{\bigl[(x-x_{0})^{2}+\rho^{2}\bigr]^{2}}
\;\eta^{a}_{\mu\nu}\,T_{a} \, .
\end{align}
Since \(\eta^{a}_{\mu\nu}=\tfrac12\varepsilon_{\mu\nu\rho\sigma}\eta^{a}_{\rho\sigma}\),
the two-form \(F\) satisfies
\( F = * F
\), as anticipated,
showing that the configuration is self-dual.
The Euclidean Lagrangian density is
\begin{align}
\mathcal L(x)=
\frac{1}{4g^{2}}\operatorname{tr}\bigl(F_{\mu\nu}F^{\mu\nu}\bigr)
            =\frac{96\,\rho^{4}/g^{2}}
                   {\bigl[(x-x_{0})^{2}+\rho^{2}\bigr]^{4}}.
\end{align}
Integrating over \(\mathbb R^{4}\) yields
\begin{align}
S_{\text{inst}}=\int d^{4}x\,\mathcal L(x)=\frac{8\pi^{2}}{g^{2}},
\qquad
k = \frac{1}{8\pi^{2}}\!\int\!\operatorname{tr}(F\wedge F)=1 \, ,
\label{eq:Instanton_number_and_winding}
\end{align}
saturating the Bogomol'nyi bound in the \(k=1\) sector.
The solution possesses
\begin{align}
4 \text{ (translations)} + 1 \text{ (scale)} + 3 \text{ (global } \text{SU(2)}\text{ orientation)} = 8
\end{align}
collective coordinates.  Therefore, the one-instanton moduli space is (see \cite{tong2005tasilecturessolitons}, for instance)
\begin{align}
\mathcal M_{k=1}^{SU(2)}
   \cong \mathbb R^{4}_{x_{0}} \times \mathbb R^{+}_{\rho} \times \text{SU(2)}_{U}/\mathbb Z_{2}.
\end{align}
Note that the anti-instanton solution only differs by replacing the self-dual 't Hooft symbols $\eta^{a}_{\mu\nu}$ with the anti-self-dual symbols $\bar{\eta}^{a}_{\mu\nu}$  (again, we adopt the conventional definition spelled out in Section \ref{sec:Notation_and_conventions_ch1}). This produces a minus sing in the second integral in in \eqref{eq:Instanton_number_and_winding}. 

The coordinate expression for the instanton presented above is standard in both the physics and mathematical literature - although the quaternionic description is more common in the latter  (\cite{BELAVIN197585,Atiyah1979}); the rephrasing instantons in terms of quaternionic-valued objects will be reviewed in Chapter \ref{chap:3}, Section \ref{sec:Quat_inst_and_spheres_chap3}. Another expression, which is common in the physics literature as it arises from a natural ansatz when applying the (anti-)self-duality constraint, is the so-called ``singular gauge'' - see \cite{vandoren2008lectures}. In practice, a gauge transformation with respect to $U(x)$ defined above
gives the \emph{singular gauge} potential (see Section \ref{sec:Instantons_in_Components} for some explicit calculations)
\begin{align}
A_{\mu}^{\mathrm{Sing}}(x)=
\frac{2\,\rho^{2}}{(x-x_{0})^{2}\bigl[(x-x_{0})^{2}+\rho^{2}\bigr]}
\;\bar{\eta}^{a}_{\mu\nu}(x-x_{0})^{\nu}\,T_{a} \, .
\label{eq:Sing_gauge_inst}
\end{align}
This expression is singular at $|x - x_0| = 0$, and near $x = x_0$ it is dominated by the pure gauge configuration:
\begin{align}
    A_\mu \stackrel{x \rightarrow x_0}{\simeq} U \partial_\mu U^{-1} \, .
\end{align}
Hence, the behaviour which characterised the ``regular'' solution \eqref{eq:Insta_def} at infinity can now be found close to the origin. This is a manifestation of the fact that the two expressions are related by a space-time inversion transformation, which is nothing but a change of stereographic projection from one pole to the other (\cite{vandoren2008lectures}); we comment more on this in the next subsection, alongside the description of the rigorous geometric picture. The first multi-instanton solutions, i.e.~solutions with instanton number larger than one, appeared in the singular gauge: \cite{Witten1977, thooftunp, PhysRevD.15.1642}, with the last one being recognised as the most general ansatz.\footnote{The unpublished paper \cite{thooftunp} is cited both in \cite{EGUCHI1980213} and \cite{RevModPhys.51.461}.} According to \cite{Giambiagi:1977yg}, the multi-instanton configuration can be written as
\begin{align}
A_\mu(x)=\sum_{i=1}^N f_i(x) g_i^{-1}(x) \partial_\mu g_i(x) \, ,
\end{align}
where 
\begin{align}
f_i(x)=\frac{\rho_i^2}{y_i^2}\left(1+\sum_{j=1}^N \frac{\rho_j^2}{y_j^2}\right)^{-1} \, , \quad (y_i)^{\mu} = x^{\mu} - \lambda_i^{\mu} \, ,
\end{align}
and
\begin{align}
   g_i =\frac{y^{\mu} \sigma_{\mu} }{\sqrt{y_i^2}} \,\,\,\, \implies \,\,\,\,  g_i^{-1}(x) \partial_\mu g_i(x) =  - \sigma_{\mu \nu} \frac{y_{i \nu}}{y_i^2} \, , \quad g_i(x) \partial_\mu g_i^{-1}(x) =  - \bar{\sigma}_{\mu \nu} \frac{y_{i \nu}}{y_i^2} \, .
   \label{eq:Pure_gauges}
\end{align}
The paper above also presents a constructive prescription for ``gauging away the singularity'', i.e.~removing through a suitable gauge transformation, the pathological behaviour at $a = \lambda_i$. For the limiting case of $N = 1$, this reduces to the simple transformation presented above, whose geometrical interpretation is provided below. For $N>1$, this becomes quite non-trivial, and the case $N=2$ is treated in Chapter \ref{chap:3}, Section \ref{subsec:k2_chap3}.

In summary, using physics jargon, the BPST (anti-)instanton represents a finite-action bubble of self-locked colour-magnetic flux in Euclidean space-time. In other words, it is a non-trivial solution to the (anti-)self-duality equation, therefore solving the Yang--Mills equations, and it yields a finite action. In Minkowski signature it mediates
tunnelling between classical Yang-–Mills vacua whose Chern–Simons numbers differ
by one unit. Regular and singular gauges correspond to two different descriptions of the same potential, related by a singular gauge transformation that we now identify with a stereographic change of coordinates. \\
So far we have mainly focused on reviewing instantons through the physics narrative, but we now turn to a more rigorous and mathematically-oriented description of these objects.

\subsection{The Geometric Interpretation of the Solution}
\label{subsec:geometric_instaton_chap2}
As first noted in \cite{Trautman:1977im} (according to \cite{10.1063/PT.3.2799}), the BPST solution has a deep geometric interpretation. Although it was originally written as living on \(\mathbb{R}^{4}\),
the bundle picture emerges once one \textit{compactifies}
\(\mathbb{R}^{4}\) to the four–sphere
\(S^{4}=\mathbb{R}^{4}\cup\{\infty\}\).\footnote{This procedure has nothing to do with the physics meaning of \textit{compactification}, and a short discussion can be found in Section \ref{subsec:One_point_compactifications}.} The idea of adding a ``point at infinity'', is naturally motivated by a simple physical fact, which we already mentioned: in order to have a finite action, the gauge field must become pure gauge as $r \xrightarrow{} \infty$, where \(r^{2}=x_{\mu}x_{\mu}\). This constrains the behaviour as one approaches infinity, hinting to the fact that the instanton might have a natural support on $S^4$ rather than $\mathbb{R}^4$. The mathematical reason for this involves $S^7$ written as an $S^3$-bundle over $S^4$, and will be mentioned later. We begin by considering once again the map
\begin{align}
U(x)\;=\; i \frac{\sigma_{\mu} x_{\mu}}{r} = \frac{- I x_4 + ix_1 \sigma^1 + i x_2 \sigma^2 + i x_3 \sigma^3}{\sqrt{x_1^2 + x_2^2 + x_3^2 + x_4^2}}  
\, ,
\end{align}
which, as $r \xrightarrow{} \infty$, should be interpreted as the identity map $U:\,S^{3}_{\infty}\longrightarrow SU(2)\simeq S^{3}$. One can verify that, for a fixed $r$, this map is indeed the identity from $S^3$ (which can be thought as spacial infinity) to $SU(2)$ (thought as a group manifold);
consequently, its degree is \(\deg(U)=1\).
The associated pure gauge field (which is nothing but the Maurer–Cartan form) reads
\begin{align}
\Theta\;:=\;U^{-1}\!dU
   \;=\;\frac{1}{r^{2}}\Bigl(x_{4}\,dx_{j}-x_{j}\,dx_{4}
   -\varepsilon_{jk\ell}x_{k}\,dx_{\ell}\Bigr)\;\frac{\mathrm i\sigma_{j}}{2}
   \;=\;\frac{2}{r^{2}}\,
        \eta^{a}_{\mu\nu}\,x^{\nu}\,T_{a}\,dx^{\mu} \, .
\end{align}
See Section \ref{sec:Instantons_in_Components} for detailed calculations leading to this result and to the analogous one for $U(x)^{-1}$. To keep this behaviour at infinity, needed to guarantee finiteness of the action, but ensure regularity at the origin, one can consider a scalar profile of the form
\begin{align}
f(r)=\frac{r^{2}}{r^{2}+\rho^{2}}
\quad(0\le f<1),
\end{align}
and set
\begin{align}
A^{\text{reg}}
   \;=\;f(r)\,\Theta
   \;=\;\frac{2\,r^{2}}{r^{2}+\rho^{2}}\,
         \eta^{a}_{\mu\nu}\frac{x^{\nu}}{r^{2}}\,T_{a}\,dx^{\mu}
   \;=\;\frac{2\,\eta^{a}_{\mu\nu}x^{\nu}}{r^{2}+\rho^{2}}\,
         T_{a}\,dx^{\mu}.
\end{align}
This is the usual BPST potential that is nonsingular at the origin and
behaves as a pure gauge (\(\Theta\)) at infinity.
Its curvature is
\begin{align}
F^{\text{reg}}
   =dA^{\text{reg}}+A^{\text{reg}}\wedge A^{\text{reg}}
   =\frac{4\,\rho^{2}}{(r^{2}+\rho^{2})^{2}}\,
     \eta^{a}_{\mu\nu}T_{a}\,\frac12\,dx^{\mu}\wedge dx^{\nu} \, ,
\end{align}
with $F^{\text{reg}}=* F^{\text{reg}}$, of course.
In order to complete the global picture, it is useful to find the expression for the connection on the other local trivialisation. Since we are dealing with fibre bundles, this is a two-step process. Firstly, one has to express the connection in the coordinates of the other local trivialisation (i.e.~pull it back), and then the apply the gauge transformation in \eqref{eq:Gauge_transf_maths} (or \eqref{eq:Gauge_transf_physics}). Let us proceed with the first step. Assuming that the coordinates $\{x^{\mu} \}$ correspond to those obtained via the stereographic projection from one pole, we label as $y^{\mu}$ those corresponding to the projection from the other one - see Section \ref{subsec:Working_definitions_chapA1} for more details. According to our definition therein, the two coordinates are related through $y_\mu = \frac{\rho^2 x_\mu}{x^2}$.
In order to compute
\begin{align}
A'_\mu(y) = \frac{\partial x_\nu}{\partial y_\mu} A_\nu(x(y)) \, ,
\end{align}
one needs the Jacobian matrix of the inverse transformation $\frac{\partial x_\nu}{\partial y_\mu}$. Again, this can be found in Section \ref{subsec:Working_definitions_chapA1}, and it reads
\begin{align}
\frac{\partial x_\nu}{\partial y_\mu}   = \frac{\rho^2}{y^2} \left( \delta_{\nu\mu} - \frac{2y_\nu y_\mu}{y^2} \right) \, .
\end{align}

The functional substitution $A_\mu^a(x(y))$ simplifies to just substituting $x \xrightarrow{} y$, and this is a consequence of the conformal symmetry of the problem - see \cite{PhysRevD.14.517}. These results amount to:
\begin{align*}
A'{}^a_\mu(y) =  \left[ \frac{\rho^2}{y^2} \left( \delta_{\nu\mu} - \frac{2y_\nu y_\mu}{y^2} \right) \right] \left[ \frac{2 \eta_{a\nu\rho} y_\rho}{y^2 + \rho^2} \right] = \frac{2\rho^2 \eta_{a\mu\rho} y_\rho}{y^2(y^2 + \rho^2)} \, .
\end{align*}
This is the anti-instanton gauge field written in singular gauge. If we had started with a regular anti-instanton, then we would have found \eqref{eq:Sing_gauge_inst}. This shows that the so-called ``singular gauge'' is a result of an incomplete transformation, or equivalently of an inappropriate description using the coordinate of one patch for describing the connection in the other local trivialisation. To complete the transformation, the missing step is to implement the appropriate gauge transformation, spelled out in \eqref{eq:Gauge_transf_maths}, or equivalently in \eqref{eq:Gauge_transf_physics}; analogously to what we discussed before for the case of the instanton, gauge‐transforming (with $g = U(x)$, according to \eqref{eq:Gauge_transf_physics}, or $t_{ij} = U(x)^{-1}$, according to \eqref{eq:Gauge_transf_maths}), yields
\begin{align}
    A''_{\mu} (y) = \frac{2\,\bar{\eta}^{a}_{\mu\nu}x^{\nu}}{r^{2}+\rho^{2}}\,
         T_{a}\,dx^{\mu} \, ,
\end{align}
which is indeed regular. Note that there is still a (somewhat disturbing) asymmetry in finding that the same global object has different self-duality properties in the two local descriptions. This happens because the standard stereographic projection is orientation-reversing, and a change of orientation switches self-dual objects with anti-self-dual ones (and viceversa). The stereographic projection can be modified to achieve the orientation-preserving change of coordinates $y_\mu = \frac{-\rho^2 \eta_{\mu \nu} x^\nu}{x^2}$, with $\eta_{\mu \nu} = \mathrm{diag}(1,1,1,-1)$. The details of this transformation and the corresponding atlas are discussed in \ref{subsec:Working_definitions_chapA1}, where the only difference lies in the use of $^o\eta = (-1,1,1,1)$ instead of $\eta$. Although this definition might seem a bit artificial at this stage, it naturally arises when adopting the quaternionic description (see \ref{subsec:Working_definitions_chapA1} and \ref{subsec:'tHooft_notation_and_quaternions_chap3}). The Jacobian associated with this change of coordinates reads:
\begin{align}
    \frac{\partial x_{\mu}}{\partial y_{\nu}}
   =\frac{- \rho^{2}\,\eta_{\mu \sigma}}{y^2}
     \bigl[\,\delta_{\sigma\nu}- \frac{2 y_{\sigma}y_{\nu}}{\rho^2}\bigr].
\end{align}
With such a choice, one finds that
\begin{align}
    A'{}^a_\mu(y)= \frac{2\rho^2 \bar{\eta}_{a\mu\rho} y_\rho}{y^2(y^2 + \rho^2)} \,
\end{align}
which, under a suitable gauge transformation ($g = U(x)^{-1}$, according to \eqref{eq:Gauge_transf_physics}, or $t_{ij} = U(x)$, according to \eqref{eq:Gauge_transf_maths}) yields:
\begin{align}
    A''_{\mu} (y) = \frac{2\,\eta^{a}_{\mu\nu}x^{\nu}}{r^{2}+\rho^{2}}\,
         T_{a}\,dx^{\mu} \, .
\end{align}
Hence, using an appropriate orientation-preserving atlas on $S^4$, 
the two local expressions for the connection are identical. 

Let us now provide the geometric interpretation of this global picture. The map \(U:S^{3}_{\infty}\to \mathrm{SU(2)}\), viewed as \(U:S^{3} \to S^3 \), has homotopy $1$ and can be interpreted as a map ``gluing'' the equator of $S^4$. Recall that the transition functions of principal fibre bundles are maps from the overlap between the two patches on the base $U_{\alpha} \cap U_{\beta}$ to the fibre group manifold $G$. If we let $S^4$ be the base of some fibre bundle, and $S^3 \sim \mathrm{SU(2)}$ be the fibre, then one can identify the map above with the transition function that glues the northern and southern trivial bundles over \(S^{4}\) into a non‐trivial principal bundle. The total space of this principal $S^3$--bundle over $S^4$, with \(c_{2}=1\), is clearly $7$-dimensional. A claim that will be justified in the next chapter is that this is the seven-sphere. Therefore, in the light of the connection between fibre bundles and Yang--Mills theory outlined in the previous section, it follows that the instanton is the connection defined on the degree-one \(\mathrm{SU(2)}\) bundle over \(S^{4}\) whose total space is $S^7$. Of course, if the map $U$ had been trivial ($U(x) = 1$), then the associated gauge field/connection would have also been trivial and the corresponding bundle would have been simply $S^4 \times S^3$.

\section{Twisted Self-duality}
\label{sec:Twisted_self_duality_chap2}
This section presents the two realisations of twisted self-duality equations studied in \cite{Berman:2022dpj}, together with their natural solutions. The latter are also given a geometric interpretation.

\subsection{The Equations}
\label{subsec:twisted_equations_chap2}
Let us denote with $\mathcal{F}^a$ the curvature two-form that takes values in some Lie algebra $\mathfrak{g}$. The upper index, $a=1,...,\mathrm{dim}(\mathfrak{g})$ is a vector index in a particular basis for the algebra, $\mathfrak{g}$.
Let $\mathcal{J}$ be the operator that acts on the vector space such that 
\begin{align}
    (\mathcal{J})^2=1 \, .
    \label{eq:Squares_to_one}
\end{align}
Twisted self-duality is then when the Hodge star is combined with $\mathcal{J}$, so that the curvature obeys the equation:
\begin{align}
    \mathcal{F}=*\mathcal{J}\mathcal{F} \, .
    \label{eq:TSDE_first_time}
\end{align}
Solutions of this equation are discussed in the next section, under the obvious assumption (necessary for consistency) that the combined operator satisfies
\begin{align}
    (*\mathcal{J})^2=1  \, .
    \label{eq:Squares_to_one}
\end{align}
This is trivially true in the case described above since both the Hodge star and $\mathcal{J}$ are involutions. But it does also beg the fascinating question of just demanding (\ref{eq:Squares_to_one}) while allowing $*^2=-1$ and $\mathcal{J}^2=-1$.
In four dimensional Lorentzian space, $*^2=-1$, which rules out real self-dual solutions. Twisted self-dual solutions, however, are not ruled out provided one has a twist operator such that $\mathcal{J}^2=-1$.
Although it would be interesting to classify possible operators $\mathcal{J}$ for any given gauge algebra $\mathfrak{g}$, this is not discussed here. Instead, in order to be as simple and constructive as possible, we focus on a specific realisation of the twisted self-dual equation. Bypassing the abelian case (which was explored as part of \cite{https://doi.org/10.48550/arxiv.1412.2768} was one of the motivations for this investigation), the simplest non-abelian gauge group is $su(2)$, and so a natural way to construct an algebra with an operator $\mathcal{J}$ is simply to have two copies of $\text{su(2)}$ and have $\mathcal{J}$ map between them.
And so we consider the case of ${\mathfrak{g}}=\text{su(2)} \oplus \text{su(2)}$ and
\begin{align}
\mathcal{J}=\begin{pmatrix}
0 & 1 \\
1 & 0
\end{pmatrix} \, ,
\label{eq:J_matrix}
\end{align}
for Euclidean signature. For the Lorentzian case it is sufficient to modify the above operator by introducing a minus sign in one of the two off-diagonal entries.
Of course, one can then use the isomorphism $\text{so(4)}=\text{su(2)} \oplus \text{su(2)}$ and think of this as an $\text{so(4)}$ gauge theory.

\subsection{The SO(4) Solutions}
\label{subsec_twisted_solutions_chap2}
\subsubsection{Euclidean}
Explicitly, we seek a solution to the equation
\begin{align}
    \left( \begin{array}{c}
      F  \\
     \bar{F} \\
\end{array}
\right) = * \begin{pmatrix}
0 & 1 \\
1 & 0
\end{pmatrix}
\left( \begin{array}{c}
      F  \\
     \bar{F} \\
\end{array}
\right),
\label{Twisted_self_dual_su2su2_explicit}
\end{align}
with $F,\bar{F}$ being $\text{su(2)}$ curvature forms associated with $\text{su(2)}$ connections $A,\bar{A}$, respectively.
We start by using some very basic techniques and concepts from matrix algebra: eigenvectors and eigenvalues. Let us think of $F,\bar{F}$ in terms of their colour indices, so that each of them is a three-vector in colour space. Then, we perform the following decomposition:
\begin{align}
     \left( \begin{array}{c}
      F  \\
     \Bar{F} \\
\end{array}
\right)= \left( \begin{array}{c}
      F^1 \\
      \, F^{1} \\
\end{array}
\right) + 
\left( \begin{array}{c}
      F^2  \\
      \, -F^{2} \\
\end{array}
\right),
\label{eq:Field_str_evectors_decomp}
\end{align}
where it should be clear that $F^1, F^2 $ are also three-vectors. The two vectors on the RHS are linearly independent. What we are doing is nothing but a decomposition into eigenvectors of the matrix $\begin{pmatrix}
0 & 1 \\
1 & 0
\end{pmatrix}  $.

And so, we are decomposing the $\text{su(2)} \oplus \text{su(2)}$ field strength into components that live in the two eigenspaces defined by the almost product structure above. Then, we obtain:
\begin{align}
     \left( \begin{array}{c}
      F^1 \\
      \, F^{1} \\
\end{array}
\right) + 
\left( \begin{array}{c}
      F^2  \\
     - \, F^{2} \\
\end{array}
\right) = * \begin{pmatrix}
0 & 1 \\
1 & 0
\end{pmatrix} \Bigg[  \left( \begin{array}{c}
      F^1 \\
      \, F^{1} \\
\end{array}
\right) + 
\left( \begin{array}{c}
      F^2  \\
     - \, F^{2} \\
\end{array}
\right)  \Bigg] = * \Bigg[     \left( \begin{array}{c}
      F^1 \\
      \, F^{1} \\
\end{array}
\right) -
\left( \begin{array}{c}
      F^2  \\
     - \, F^{2} \\
\end{array}
\right)   \Bigg].
\label{eq:Euclidean decomposition}
\end{align}
This gives the following two equations (which are in fact six):
\begin{align}
   \, F^1= * \, F^1  \nonumber \\
   - \, F^2 = * \, F^2.
   \label{eq:Eigenvectors_eqns}
\end{align}
Hence, we obtained two self-dual equations, which are now untwisted. We know the solution to those in terms of $F^1$ and $F^2$: they must be proportional to the 't Hooft symbols. Now, we would be tempted to complete the job by solving for the gauge fields $A^1$ and $A^2$ that correspond to $F^1$ and $F^2$, respectively. Such solutions are to the usual instanton and anti-instanton. However, the \textit{real} field strengths in our theory are $F$ and $\bar{F}$, and it is them that need to be written in terms of gauge fields $A$ and $\bar{A}$. \footnote{Clearly, if we look at the Abelian case, such consideration is superfluous. However, in the non-Abelian case, the relation between $A^{1,2}$ and $A, \bar{A}$ is highly non-trivial.}

A very simple solution can be found when we restrict only to one of the two eigenspaces. Specifically, setting $F^2=0$ yields the usual $\text{su(2)}$ self-dual instanton:
\begin{align}
    A_{\mu }^a=\bar{A}_{\mu }^a= A^{1}_{\mu }{}^a = 2 \frac{\eta_{\mu \nu}^{a}\left(x-x_{0}\right)_{\nu}}{\left(x-x_{0}\right)^{2}+\rho^{2}}, \nonumber \\
     F_{\mu \nu }^a=\bar{F}_{\mu \nu}^a= F^{1}_{\mu \nu}{}^a = -4 \frac{\eta_{\mu \nu}^{a}\rho^2}{[\left(x-x_{0}\right)^{2}+\rho^{2}]^2}.
     \label{eq:Two_instantons_twisted}
\end{align}
This solution has clearly a ``instanton number'' of 2, since each of the two $\text{su(2)}$ blocks contributes with instanton number 1, and the moduli space is the same as the one for a single instanton. \\
If we restrict to the other eigenspace, things radically change. We need to solve simultaneously:
\begin{align}
    \begin{cases}
     -F^{2}=* F^{2} \\
     F^{2}= d A - [A,A] \\
     -F^{2}= d \bar{A} - [\bar{A},\bar{A}] 
    \end{cases}
\end{align}
The first two equations are satisfied by the anti-instanton solution (i.e.~instanton with instanton number $-1$) and the final equation is for a field strength with the opposite sign of the anti-instanton. Note that since the field strength is non-linear in potentials it is not possible to generate the opposite field strength by scaling the potential by $-1$. We have not be able to construct solutions for this choice.\\
Finally, we are left with the most general case of both eigenmodes contributing to the field strength. We found that the most natural solution for this case emerges with a Lorentzian background, and it is the subject of the next subsection.


\subsubsection{Lorentzian}
Let us now take a brief detour, and focus on Lorentzian signature for the only time in this thesis. We present an interesting solution to the twisted self-dual equation with $\text{su(2)} \oplus \text{su(2)}$ gauge group in four-dimensional \textit{Minkowski spacetime}, with metric $\eta = \mathrm{diag}(1,1,1,-1)$. As we mentioned, we can take the same twist matrix $\mathcal{J}$ that appeared in the Euclidean setting and introduce a minus sign in one of the non-zero entries.\\
Thus, we obtain the equation:
\begin{align}
    \left( \begin{array}{c}
      F  \\
     \bar{F} \\
\end{array}
\right) = *_L \begin{pmatrix}
0 & -1 \\
1 & 0
\end{pmatrix}
 \left( \begin{array}{c}
      F  \\
     \bar{F} \\
\end{array}
\right),
\label{eq:Twisted_Lor}
\end{align}
where the subscript L emphasizes that now the Hodge dual is in an (unusual) Lorentzian spacetime and therefore it involves Minkowski metrics. As previously, $F,\bar{F}$ are $\text{su(2)}$ curvature forms associated with $\text{su(2)}$ connections $A,\bar{A}$, respectively. As before, we start by decomposing the field strength into eigenvectors of the twist matrix, which are now complex:
\begin{align}
     \left( \begin{array}{c}
      F  \\
     \bar{F} \\
\end{array}
\right)= \left( \begin{array}{c}
      F^1 \\
     i \, F^{1} \\
\end{array}
\right) + 
\left( \begin{array}{c}
      F^2  \\
     -i \, F^{2} \\
\end{array}
\right),
\label{eq:Field_str_evectors_decomp_Lorentz}
\end{align}
Using the above decomposition, we obtain from \ref{eq:Twisted_Lor}:
\begin{align}
     \left( \begin{array}{c}
      F^1 \\
     i \, F^{1} \\
\end{array}
\right) + 
\left( \begin{array}{c}
      F^2  \\
     -i \, F^{2} \\
\end{array}
\right) = *_L \begin{pmatrix}
0 & -1 \\
1 & 0
\end{pmatrix} \Bigg[  \left( \begin{array}{c}
      F^1 \\
     i \, F^{1} \\
\end{array}
\right) + 
\left( \begin{array}{c}
      F^2  \\
     -i \, F^{2} \\
\end{array}
\right)  \Bigg] = \nonumber \\
*_L \Bigg[   -i  \left( \begin{array}{c}
      F^1 \\
     i \, F^{1} \\
\end{array}
\right) +  i
\left( \begin{array}{c}
      F^2  \\
     -i \, F^{2} \\
\end{array}
\right)   \Bigg].
\end{align}
Hence, we need to solve
\begin{align}
   i \, F^1= *_L \, F^1  \nonumber \\
   -i \, F^2 = *_L \, F^2.
   \label{eq:Eigenvectors_eqns}
\end{align}
They are solved by the Wick-rotated version of the 't Hooft symbols, which we will denote by $\omega$ and $\bar{\omega}$, respectively. They are given by\footnote{We stress that, since we are now dealing with a Minkowski background, one must be careful when raising or lowering the indices with $\eta$.}:
\begin{align}
    \omega^{a \,\mu \nu} = \begin{cases}
    \omega^{a \, 44}=0 \\
    \omega^{a \, 4 \nu} =  i \delta^{a \nu} \hspace{0.4cm} \textrm{for} \,\,\, \nu=1,2,3 \\
    \omega^{a \, \mu 4} =  -i \delta^{a \mu} \hspace{0.4cm} \textrm{for} \,\,\, \mu=1,2,3 \\
     \omega^{a \, \mu \nu} =  -\epsilon^{a \mu \nu} \hspace{0.4cm} \textrm{for} \,\,\, \mu,\nu=1,2,3,
    \end{cases}
    \nonumber \\
    \bar{\omega}^{a \,\mu \nu} = \begin{cases}
    \bar{\omega}^{a \, 44}=0 \\
    \bar{\omega}^{a \, 4 \nu} =  -i \delta^{a \nu} \hspace{0.4cm} \textrm{for} \,\,\, \nu=1,2,3 \\
    \bar{\omega}^{a \, \mu 4} =  i \delta^{a \mu} \hspace{0.4cm} \textrm{for} \,\,\, \mu=1,2,3 \\
     \bar{\omega}^{a \, \mu \nu} =  -\epsilon^{a \mu \nu} \hspace{0.4cm} \textrm{for} \,\,\, \mu,\nu=1,2,3,
    \end{cases}
\end{align}
and satisfy
\begin{align}
    \omega=-i *_L \omega \quad \quad \textrm{and} \quad \quad \bar{\omega}=i *_L \bar{\omega}.
\end{align}
We see that $F^1$ must be proportional to $\omega$ and $F^2$ to $\bar{\omega}$.
We choose:
\begin{align}
(F^1)^{a \, \mu \nu}=\frac{1}{2}(-1+i)\omega^{a \, \mu \nu} f(x^{\mu})=\Big( \frac{1}{2} \bar{\eta}_L^{a \, \mu \nu} - \frac{i}{2} \eta_L^{a \, \mu \nu} \Big) f(x^{\mu}), \nonumber \\
(F^2)^{a \, \mu \nu}=\frac{1}{2}(-1-i)\bar{\omega}^{a \, \mu \nu} f(x^{\mu})=\Big( \frac{1}{2} \bar{\eta}_L^{a \, \mu \nu} + \frac{i}{2} \eta_L^{a \, \mu \nu} \Big)f(x^{\mu}) ,
\end{align}
where $\eta_L^{a \,\mu \nu}$ and $\bar{\eta}_L^{a \,\mu \nu}$ are defined as:
\begin{align}
    \eta_L^{a \,\mu \nu} = \begin{cases}
    \eta_L^{a \, 44}=0 \\
    \eta_L^{a \, 4 \nu} =   \delta^{a \nu} \hspace{0.4cm} \textrm{for} \,\,\, \nu=1,2,3 \\
    \eta_L^{a \, \mu 4} =  -\delta^{a \mu} \hspace{0.4cm} \textrm{for} \,\,\, \mu=1,2,3 \\
     \eta_L^{a \, \mu \nu} =  \epsilon^{a \mu \nu} \hspace{0.4cm} \textrm{for} \,\,\, \mu,\nu=1,2,3 \,\,\, ,
    \end{cases} \nonumber \\
    \bar{\eta}_L^{a \,\mu \nu} = \begin{cases}
    \eta_L^{a \, 44}=0 \\
    \eta_L^{a \, 4 \nu} =  - \delta^{a \nu} \hspace{0.4cm} \textrm{for} \,\,\, \nu=1,2,3 \\
    \eta_L^{a \, \mu 4} =  \delta^{a \mu} \hspace{0.4cm} \textrm{for} \,\,\, \mu=1,2,3 \\
     \eta_L^{a \, \mu \nu} =  \epsilon^{a \mu \nu} \hspace{0.4cm} \textrm{for} \,\,\, \mu,\nu=1,2,3 \,.
    \end{cases}
    \label{eq:'tHooft_tensors}
\end{align}
They match the standard t' Hooft symbols when the indices are downstairs, i.e.~$\eta_L^{\,\,a}{}_{ \mu \nu}=\eta^{a}_{ \mu \nu}$ and $\bar{\eta}_L^{\,\,a}{}_{ \mu \nu}=\bar{\eta}^{a}_{ \mu \nu}$.
We added the subscript ``L'' to remind that in this context the position of the indices matters, since raising and lowering happens through the Minkowski metric $\eta$.
This way, we have that, according to \eqref{eq:Field_str_evectors_decomp_Lorentz}:
\begin{align}
     F^{a \, \mu \nu} = \bar{\eta}_L^{a \, \mu \nu} f(x^{\mu}) \,\,\,\,\,\,\,\, \textrm{and} \,\,\,\,\,\,\,\, \bar{F}^{a \, \mu \nu} = \eta_L^{a \, \mu \nu} f(x^{\mu}).
\end{align}
With the usual instanton ansatz in mind, we can set $f$ to be
\begin{align}
f(x^{\mu})=-4 \frac{\rho^2}{[(x-x_0)^2_E + \rho^2]^2} = -4\frac{\rho^2}{[(x-x_0)^{\gamma} (x-x_0)^{\delta} \delta_{\gamma \delta} + \rho^2]^2},
\end{align}
where we implicitly used the subscript E to denote the Euclidean norm. As for the gauge fields associated to these field strengths, we propose the expressions:
\begin{align}
    A_{\mu} = 2 \frac{\bar{\eta}_{\mu \nu}^{a}\left(x-x_{0}\right)^{\nu}}{\left(x-x_{0}\right)^{2}_E+\rho^{2}}, \nonumber \\
    \bar{A}_{\mu} = 2 \frac{\eta_{\mu \nu}^{a}\left(x-x_{0}\right)^{\nu}}{\left(x-x_{0}\right)^{2}_E+\rho^{2}}.
    \label{eq:Weird_Solution}
\end{align}
To verify that they give the field strengths above, we perform the usual calculation leading to the regular instanton, which gives:
\begin{align}
    \bar{F}_{\mu \nu}^{a}= 
     - 4\eta_{\mu \nu}^{a} \frac{\rho^2 }{[\left(x-x_{0}\right)^{2}_E + \rho^{2}]^2}.
\end{align}
We used:
\begin{align}
    \epsilon_{a b c} \eta_{b \mu \nu} \eta_{c \rho \sigma}=\delta_{\mu \rho} \eta_{a \nu \sigma}+\delta_{\nu \sigma} \eta_{a \mu \rho}-\delta_{\mu \sigma} \eta_{a \nu \rho}-\delta_{\nu \rho} \eta_{a \mu \sigma},
\end{align}
which is a well known result from the early studies of the $\text{SU(2)}$ instanton (see \ref{subsec:Some_Relations_in_Components}). Since the same relation holds for $\bar{\eta}$, we can verify that, analogously, the gauge field $A$ has field strength
\begin{align}
    F^a_{\mu \nu} = - 4\bar{\eta}_{\mu \nu}^{a} \frac{\rho^2 }{[\left(x-x_{0}\right)^{2}_E + \rho^{2}]^2}.
\end{align}
This is a very curious solution, since it mixes Lorentzian and Euclidean features. It satisfies the Lorentzian twisted self-dual equation, but it involves the use of the Euclidean norm. It should be noted that, since this solution only preserves the 3D rotational symmetry from the usual Euclidean instantons, i.e.~it is not invariant under Lorentz boosts. One should then think of this solution as sponteneously breaking the Lorentz symmetry. Consequently, the associated Lorentz boost moduli can appear in the solution, as can be seen by performing the analogous calculation with the coordinates transformed; this yields  
\begin{align}
\bar{F}^a_{\mu \nu} = - 4\eta_{\mu \nu}^{a} \frac{\rho^2 }{[\left(x-x_{0}\right)^{' \, 2}_E + \rho^{2}]^2},
\end{align}
with the prime labelling the boosted coordinates, and similarly for $F$. \\

This is a solution of the twisted self duality equation in 4d Lorentzian space but it is important to realise that the Bogomoln’yi argument does not apply in this instance (the Lorentzian action is not positive definite due to the presence of the time-like direction). So although we have successfully constructed a solution to the Lorentzian twisted self-duality equations these solutions do not have to satisfy the Yang--Mills equations of motion and indeed one can check that they do not. These solutions would be relevant for a first order theory which imposes the twisted self-duality as a constraint such as a twisted BF type theory.

\subsection{The Geometric Interpretation of the Solution}
\label{subsec:geometric_twisted_chap2}
Let us begin from the Lorentzian solution. We believe the existence of such a solution to be intimately related with the discussion in Section \ref{subsec:geometric_instaton_chap2}. Specifically, the appearance of the Minkowski metric in the context of instantons is due to purely geometrical reasons, and it follows from the requirement of the transition functions to preserve the orientation (or, spelled out in Section \ref{subsec:Working_definitions_chapA1}, to preserve the underlying quaternionic structure). Hence, although it was not noted at the time of writing \cite{Berman:2022dpj}, the interplay between the Lorentzian and Euclidean metrics which arises in the Lorentzian solution that we discussed could admit an elegant reformulation on geometrical grounds.

Coming to the Euclidean solutions presented in the previous section, they are more ``standard'', and should be interpreted as being defined on an underlying $S^3 \times S^3$ bundle over $S^4$. The total space of this manifold, however, is not a ``nice'' 10-dimensional manifold. Similarly to how the $\text{SU(2)}$-bundle described in Section \ref{subsec:geometric_instaton_chap2} is labelled by the instanton number, the same idea applies here for this type of bundles, which can accommodate two $\text{SU(2)}$ gauge fields. Their instanton numbers, which, as we discussed, are intimately related to the homotopy class of the transition functions, allow to write a (discrete) classification; concretely, this follows from: 
\begin{align}
\pi_3(S^3 \times S^3) = \mathbb{Z} \times \mathbb{Z} \, .
\label{eq:pi3_s3_times_s3}
\end{align}
Since $\pi_3(S^3 \times S^3) \simeq \pi_3(\text{SO(4)})$ (see \cite{McEnroe2016MILNORSCO}), one can equivalently consider $\text{SO(4)}$ principal bundles, which is what is done in \cite{Rigas1978}; following their notation, the bundles can be labelled as $P_{m,n}$, where $m,n$ specify the homotopy class in \eqref{eq:pi3_s3_times_s3}. In the conventions of \cite{Rigas1978}, the solution \eqref{eq:Two_instantons_twisted} corresponds to the (double cover of the) principal $\text{SO(4)}$-bundle $P_{(0,1)}$; one can find some of its properties it by looking at \textit{case e.} in Section 1 of that paper. We do not insist on this analysis further, as it is one of the subjects of the next chapter (Section \ref{subsec:Milnor_construction_chap3}), where bundles \textit{associated} to $P_{m,n}$ are discussed in detail. We limit ourselves to observing that, just by considering the addition of a second $\text{SU(2)}$ gauge fields, one enters the mathematical set-up where Milnor discovered the first exotic $7$--spheres.

\section{Summary and Outlook}
\label{sec:Conclusions_chap2}

In this chapter, we introduced Yang--Mills theory in its bundle-theoretic formulation and reviewed the geometric interpretation of the BPST (as well as other) instanton solutions. By satisfying the self-duality equation, they minimise the action, making them of great relevance for physics; they also carry a very rich underlying mathematical structure, which is the reason behind their importance in mathematics. After reviewing these concepts, we specialised to a modified version of the self-duality equations, called \textit{twisted self-duality}. We outlined the origin and significance of these relations, and then we proceeded to study their incarnation in $\text{SO(4)}$ Yang--Mills theory, both in Euclidean and Lorentzian spacetime. We presented one solution for each signature, and provided their geometric interpretations. 

There is a clear set of remaining questions to be answered that are posed by the discussion in this chapter. One of them is whether there is a more interesting group that admits an appropriate involution, and, if the answer is positive, what are the solutions to the twisted self-dual equation in such a setting. From the perspective of exceptional field theory, which was not reviewed in this chapter, but is one of the main motivations for the study of twisted self-duality, a number of questions naturally arise. The first one being: can one find the full solutions to $E_7$ exceptional field theory once gravity is coupled, including both the internal generalised metric and the external metric? This seems very plausible, which leads to the second one: can certain dimensional reductions be useful for constructing solutions in lower dimensional exceptional field theories? 
Beside this set of questions inspired by duality-invariant theories, it is interesting to consider other theories where twisted self-duality plays a key role. Various constructions come to mind for topological theories or generalised BF type of theories. There is much in this direction left to explore.

\chapter{Exotic Spheres, Kaluza--Klein \\ 
and Quaternions \,\,\,\,\,\,\,\,\,\,\,\,\,\,\,\,\,\,\,\,\,\,\,\,\,\,\,\,\,\,\,}
\label{chap:3}

This is the main chapter of this thesis. It introduces the concept of inequivalent differentiable structures, and presents its first (and arguably most natural) realisation through exotic $7$--spheres. A thorough discussion of its geometry, in terms of the Kaluza--Klein formalism, is presented, based on \cite{Gherardini:2023uyx} and \cite{berman2024curvatureexotic7sphere}.

\section{Introduction, Overview and Structure}
\label{sec:Intro_chap3}

The construction of exotic spheres by Milnor represents one of the major results in modern differential
geometry (see the seminal work \cite{10.2307/1969983}). These manifolds constitute a family of seven-dimensional spaces which are homeomorphic but not 
diffeomorphic to $S^7$, first discovered as total spaces of non-trivial 3-sphere bundles over $S^4$. \\
Since the advent of general relativity, essentially all areas of theoretical physics have been permeated by differential
geometry to some extent. Motivated by the crucial role played by the metric in any physical theory, the
geometry of a manifold is usually what physicists tend to focus on. The topological characterisation, 
despite its spreading in the physics literature happened slightly later and with less rapidity, is 
also very present in current research. What seems to be missing is the intermediate layer between the two:
differentiable structures. \\
The appearance of exotic spheres in the physics literature is very rare. Shortly after their discovery,
they have been discussed by \cite{FREUND1985263} and \cite{YAMAGISHI198447} in the context of supergravity, although the discussion is very brief. Since then, they were considered in the
context of gravitational instantons in \cite{Witten:1985xe} and \cite{10.1063/1.529078}, and their interpretation as topological defects was discussed in \cite{Rohm:1988yz}. The same is true for exotic manifolds (manifolds that are pairwise homeomorphic but not diffeomorphic) in general. The first consistent efforts in exploring the role of differentiable structures in physics came from Brans, who focused on how they
might be a source for gravity (\cite{Brans:1992mj}). His steps were followed by Asselmeyer-Maluga and Król, who also conducted similar investigations (see \cite{Asselmeyer-Maluga:2017tbn}, for instance). Implications of metric on exotic manifolds in cosmology were put forward also in \cite{Duston2011, Duston2022}. Finally, in \cite{Schleich_1999}, the authors considered a specific family 
of exotic manifolds, within the context of
gravitational path integrals. A very peculiar fact is the (almost total) absence of exotic spheres
from the string theory literature, where numerous families of seven dimensional spaces and the possible
geometries on them have been
studied - starting from the classic dimensional reductions in the 70's and 80's (a detailed list of Freund-Rubin compactifications can be found in \cite{CASTELLANI1984429}), up to the more recent
AdS-CFT investigations (\cite{Aharony_2000, acharya1999branes, Fr__1999}, to mention a few). It is interesting to note that different geometries on the seven-sphere, with the standard differentiable structure, have been investigated in these two contexts, as can be seen in \cite{Awada:1982pk}, \cite{POPE1985352} and \cite{Klebanov_2009}, respectively. However, the same is not true for different differentiable structures on the topological seven-sphere. This absence was noted for instance in \cite{Coquereaux:1983kj} (comment 2 therein), and, according to \cite{book}, it is due to the lack of explicit coordinates
for exotic spheres, which prevents coordinate expressions for geometrical objects such as the metric.
This explaination seems plausible by looking at the recent mathematical literature on exotic spheres.
Several studies regarding metrics on such spaces exist, but they often consist of existence results (see \cite{boyer2004einstein} and \cite{boyer2003einstein}), and even the few constructive ones are very formal, such as \cite{10.2307/1971078}, for instance. \\
In this chapter, we aim at bridging this gap by providing a general expression for a natural metric on
exotic spheres, with the specific case of the Gromoll-Meyer sphere worked out in full detail.\footnote{To be precise, our construction applies to those exotic spheres which appear among the family of bundles considered by Milnor, i.e. ten out of fourteen (ignoring orientation). The Gromoll--Meyer sphere is among those. For a more complete discussion, the reader is referred to \cite{nuimeprn10073}.} We obtain it by considering the original construction by Milnor, and realising a bundle metric via the Kaluza--Klein techniques developed in \cite{10.1063/1.525753}, which we now introduce.

Although Kaluza--Klein theories originated as a mechanism of dimensional reduction (from Kaluza’s proposal to embed four–dimensional general relativity in a five–dimensional theory, in \cite{Kaluza1921}), the underlying mathematical framework was yet to be uncovered. After he showed that the off-diagonal components of a higher-dimensional metric behave like the electromagnetic four-potential on space-time, Klein completed the picture by suggesting that the extra dimension is a microscopic circle in \cite{Klein1926}; at that point, however the notion of fibre bundle had not been developed. But even after it had, the Kaluza--Klein ansatz was not immediately recognised as describing a natural geometry on an abelian fibre bundle. It is only a few years after the publication of Yang and Mill's famous paper, while generalising the Kaluza--Klein ansatz to the non-abelian case was in progress (see \cite{zbMATH03272259}, problem 77), that it was recognised to be intimately related with bundle-theoretic objects, in \cite{Kerner1968} and \cite{Trautman1970}.\footnote{Once again, we emphasize that the publishing of \cite{Trautman1970} happened three years after the original lecture notes.} These works showed how the Kaluza--Klein metric is simply a very natural choice for a geometry respecting the underlying bundle structure, as neatly presented in \cite{10.1063/1.522434}. Since then, many different versions of Kaluza--Klein formalisms appeared; a number of them will be mentioned in this chapter, but \cite{10.1063/1.525753} is the most relevant one for our purposes. The mathematical feature underlying a generic Kaluza--Klein ansatz is that it specifies a \textit{Riemannian submersion}, i.e.~the total space of the bundle carries a Riemannian metric for which the fibres are orthogonal to the base. The relevance of such an ansatz was later recognised by mathematicians (see \cite{bourguignon1989mathematicians}, as well as \cite{Betounes:1992qz} for a more formal exposition), and used as a tool for constructing metrics on the total spaces of fibre bundles. Metrics that are built using this technique are known in the mathematical literature as \textit{Kaluza--Klein metrics}, or \textit{connection metrics} (constructions of this type for exotic spheres appear in \cite{10.2307/1999745} and \cite{Duran2001}, for instance). They are sometimes referred as \textit{inverse} Kaluza--Klein metrics in the physics literature, such as in \cite{DUFF19861}, to emphasise the uplift from a lower- to a
higher-dimensional metric, contrary to Kaluza and Klein's original spirit. This chapter discusses a Kaluza--Klein metric on $S^7$ viewed as the Hopf bundle, which appears in \cite{DUFF19861}, as a preliminary application of the bundle-theoretic formalism of \cite{10.1063/1.525753}. Then, the same prescription is used to obtain a Riemannian metric on one of the exotic $7$-spheres: the Gromoll--Meyer one. Its interpretation as a genuine reduction of Einstein theory in seven-dimensions down to four is also discussed, before performing a thorough study of its main properties (such as isometry, Ricci curvature and scalar curvature). We compare our findings to the existing results in the mathematical literature and comment on their physical implications in formulating static solutions of general relativity in eight dimensions. Our detailed coordinate expressions could be taken as a starting point for a number of further investigations within the context of supergravity, but also for a careful mathematical investigation of the sectional curvature of the metric on the Milnor bundle. \\

Let us outline the structure of the chapter. We begin by reviewing the notion of (exotic) differentiable structure in Section \ref{sec:Exotic_diff_struct_chap3}. The original (abelian) Kaluza--Klein theory is outlined in Section \ref{sec:lens_spaces_chap3}, both from the physics perspective and from the abstract bundle-theoretic one. Lens spaces are also introduced, and the special case of the Hopf bundle serves as an exemplification of the Kaluza--Klein ansatz. Then, it is the turn of the non-abelian Kaluza--Klein formalism, and its application on the $7$-sphere realised as a quaternionic Hopf bundle, discussed in Section \ref{sec:ordinary_sphere_chap3}. Section \ref{sec:Milnor_spheres_and_GM} introduces the generalisation of Kaluza--Klein to the case of associated bundles and the original construction of exotic spheres by Milnor. An explicit coordinate form of the metric on the Gromoll--Meyer sphere is obtained, and the significance of the the associated dimensional reduction from $7$ dimensions discussed. Section \ref{sec:Interlude_chap3} summarises the first part of the chapter and comments on what questions naturally follow from the results just presented. This leads to the introduction of an elegant computational tool: quaternions. Section \ref{sec:Quat_inst_and_spheres_chap3}, introduces the quaternionic notation adopted in this chapter, and illustrates how all of the geometrical quantities that appear in the Kaluza--Klein metric admit a natural description in terms of quaternionic-valued objects. In Section \ref{sec:Curvature_chap3}, we derive the general expression for the Ricci curvature and Ricci scalar associated to the Kaluza--Klein ansatz. Section \ref{sec:KKModuli_vs_Instanton_moduli_chap3} is devoted to an explicit construction of the $k=1$ and $k=2$ $\text{SU(2)}$ instantons, mainly focussing on their moduli space and on how to switch between the singular/regular gauge expressions; we also discuss the relation between the instantons' moduli space and the Kaluza--Klein metric's moduli space. This analysis motivates a special choice of moduli for the $k=1$ and $k=2$ instantons, assumed throughout Section \ref{sec:CalcRicciSec}, where we show that the corresponding Kaluza--Klein metric has maximal isometry, \ie, $\text{SO(3)}\times \text{O(2)}$ (\cite{10.2307/1971078}). Moreover, the Ricci tensor is explicitly computed and a condition for it to be positive is found. In Section \ref{sec:Energy_conditions_chap3}, this result is used to assess the energy conditions on the simplest choice of space-time involving the Gromoll--Meyer sphere $\Sigma$: an $8$-dimensional static space-time whose space-like part is $\Sigma$. We end with Section \ref{sec:Conclusions_chap3}, which contains a summary and a discussion about future directions.

\section{(Exotic) Differentiable Structures}
\label{sec:Exotic_diff_struct_chap3}
Let us review the definitions of differentiable structure and exotic manifold, motivating them with a simple example (which can be found in \cite{book}, for instance). If one wishes to skip the preamble and go straight to the definitions, they can be found at the end of the section. 

Building an atlas on a topological space allows us to do calculus on it, through its coordinates. It goes without saying that for a generic topological space the choice of atlas is far from unique. Does this mean that we have many \textit{distinct} ways to define calculus on a manifold, one for each atlas? \\
One would hope that this is not the case, otherwise we would have a profound ambiguity which seems hard to resolve, i.e.~why should one atlas be more fundamental than another. Fortunately, it turns out that almost always two atlases on a manifold ``contain the same information'', in the sense that they describe the same calculus. We shall now make this statement more rigorous. Let $M$ be a topological manifold, and let $\{(U_i, \psi_i) \}$, $\{ (V_i, \phi_i ) \}$ be two atlases on it. If their union is still an atlas, they are said to be \textit{compatible}. What this means is that the change of coordinates between any patch in the first atlas and any patch in the second one is $C^{\infty}$, so that one can smoothly move between the two descriptions without any issue. In this sense, the two atlases specify the same calculus on the manifold. Compatibility is an equivalence relation, and it seems natural to identify the differentiable structure of a differentiable manifold $(M,A=\{(U_i, \psi_i) \})$ with the equivalence class of the atlases compatible with $A$. \\
Unfortunately, this definition of differentiable structure is too restrictive, and the following example shows why. \\
Consider two differentiable manifolds with the same underlying topological space $M=\R$ with the usual topology. They are given by: 
\begin{align}
  M_1 = (M=\R, (U, \psi)) \quad  \textrm{and} \quad M_2=(M=\R, (V, \phi)),
\end{align}
where $U = V = \R$, and
\begin{align}
    \psi: U &\rightarrow \R \nonumber \\
    p &\mapsto x=p \nonumber \\
    \phi: V &\rightarrow \R \nonumber \\
    p &\mapsto y=p^{3}.
\end{align}
$U$ and $V$ clearly cover $M$ (separately), and $\psi$, $\phi$ are both homeos into an open set of $\R$ (which is $\R$ itself). \\
Since these are two atlases for the simplest non-trivial topological space, one would assume that they are compatible. To confirm this, let us look at their union: $\{(U,\psi), (V, \phi) \}$. The change of coordinates $\psi \circ \phi^{-1}$ gives $x=\sqrt[3]{y}$. This is not differentiable at the origin, hence it is not $C^{\infty}$. In other words, $\{(U,\psi), (V, \phi) \}$ is not an atlas, showing that $\{(U,\psi)\}$ and $ \{(V, \phi) \}$ are not compatible. At this point one might think that calculus is not uniquely defined on the real line, and that this might constitute an issue.
The resolution to this problem is that the two differentiable manifolds $M_1$ and $M_2$ are actually diffeomorphic. Let us consider the following map:
\begin{align}
    h: M_1 &\rightarrow M_2 \nonumber \\
    p &\mapsto p^{1/3}.
\end{align}
It is a homeomorphism. Seen as a map from $M_1$ to $M_2$, we see that its coordinate expression given by $ \phi \circ h \circ \psi^{-1}$, i.e. it reads $x \mapsto y=x$. It is smooth, its inverse is also smooth, so that $h$ defines a diffeomorphism. Hence, although the two atlases are not compatible, they are related via diffeomorphism, so that we can consider them to be equivalent. 

This leads to identify a \textit{differentiable structure} of a manifold $(M,A)$ as all the atlases compatible with $A$ \textit{and} all the atlases $A'$ such that $(M,A) \cong_{\mathrm{diff}} (M,A')$. 

The definition of \textit{exotic manifolds} follows quite naturally. Given two manifolds $(M,A)$ and $(M',A')$, they form an exotic pair if $M \cong_{\mathrm{top}} M'$ (they are topologically equivalent, i.e.~homeomorphic), but $(M,A) \ncong_{\mathrm{diff}} (M',A')$ (there does not exist a diffeomorphism between the two).

\section{Lens Spaces: A Prelude to Exotic Spheres}
\label{sec:lens_spaces_chap3}
Exotic spheres are a straightforward generalisation of lens spaces from the division algebra of complex numbers to the ones of quaternions. The word ``straightforward'' can misleading in that an important property emerges when we move from complex numbers to quaternions: non-commutativity. The existence of exotic differentiable structures is intimately related to the failure of commutativity, as we should describe soon. In this section, we introduce lens spaces, the lower-dimensional analogue of exotic spheres.\footnote{Note that lens spaces, and the special case of the Hopf fibration, are not just a mathematical construction without applications in physics (see \cite{URBANTKE2003125}); they are also the subject of very recent studies such as \cite{harada2025exactvacuumsolutionhopf}.}

\subsection{Definition (top-down)}
\label{subsec:lens_top_down_chap3}
We begin with the simplest non-trivial lens space, also known as the three-sphere, $S^3$, or the Hopf fibration; we follow the conventions outlined in Section \ref{subsec:Princ_fibre_bundles_chap2}. \\
Let $E=S^3$ be the unit three sphere, let $M=S^2$ with the complex stereographic atlas, and let $F=S^1 \cong U(1)$ (see Section \ref{subsec:Working_definitions_chapA1} for details on the stereographic projection). \\
We now show that, given a suitable projection map $\pi$ and appropriate local trivialisations on $M$, $(E, \pi, M, F)$ is a well defined principal fibre bundle.

Let $E=S^3$ be defined as usual by $(x_1)^2 +  (x_2)^2 +  (x_3)^2 +  (x_4)^2 =1$. Equivalently, if we define
\begin{align}
 z_0 \defeq x_1 + i x_2 \, , \quad  z_1 \defeq x_3 + i x_4 ,
\end{align}
then $E=S^3$ is specified by $|z_0|^2 + |z_1|^2 = 1$.
Similarly, we let $M=S^2$ be defined as $(y_1)^2 +  (y_2)^2 +  (y_3)^2 =1$. Note that there is not a natural restatement of this relation with complex numbers. We define the projection map (known as \textit{Hopf map} in the literature) as:
\begin{align}
    \pi : S^3 &\rightarrow S^2 \nonumber \\
    (x_1, x_2, x_3, x_4) &\mapsto \pi (x_1, x_2, x_3, x_4)= (y_1, y_2, y_3) = \nonumber \\ &( 2x_1 x_3 + 2 x_2 x_4 ,  2 x_2 x_3 - 2 x_1 x_4 , (x_1)^2 +  (x_2)^2 -  (x_3)^2 -  (x_4)^2 ).
    \label{eq:Hopf_projection}
\end{align}
It is easy to show from the above values of $y_{1,2,3}$ that $(y_1)^2 +  (y_2)^2 +  (y_3)^2 =1$ (given that $(x_1)^2 +  (x_2)^2 +  (x_3)^2 +  (x_4)^2 =1$). Also, we can quickly check that $\pi$ is onto. Hence, the map is well-defined. \\
Now, to construct the local trivialisations, we need to consider the complex stereographic atlas on $M=S^2$, as defined in Section \ref{subsec:Working_definitions_chapA1}; explicitly, one has:
\begin{align}
    Z=Y_1 - i Y_2 = \frac{y_1 - iy_2}{1 + y_3} = \frac{x_3 + ix_4}{x_1 + i x_2} = \frac{z_1}{z_0}, \quad \quad \textrm{with} \,\,\, (y_1,y_2,y_3) \in U_1,
    \label{eq:Coordinates_of_northern_patch_complex}
\end{align}
and
\begin{align}
    Z'=Y'_1 + i Y'_2 = \frac{y^1 + iy^2}{1 - y^3} = \frac{x_1 + i x_2}{ x_3 + ix_4} = \frac{z_0}{z_1}, \quad \quad \textrm{with} \,\,\, (y_1,y_2,y_3) \in U_2.
    \label{eq:Coordinates_of_southern_patch_complex}
\end{align}
In the third equality in the above lines, we are rewriting the point on (a chart of) $M=S^2$ in terms of a point of $E=S^3$. In doing so, we immediately see that a $U(1)$ ambiguity emerges, since $Z,Z'$ are invariant under $(z_0, z_1) \mapsto \lambda (z_0, z_1)$, with $\lambda \in U(1)=S^1$. And, clearly, $ \lambda (z_0, z_1)$ is still a point of $E=S^3$. \\
We start to see that $S^2$ is obtained by $S^3$ ``up to $S^1$'', and we now make this statement more rigorous by introducing the appropriate local trivialisations. We define:
\begin{align}
    \phi_1^{-1}: \pi^{-1}(U_1) &\rightarrow U_1 \cross S^1 \nonumber \\
    (z_0, z_1) &\mapsto \phi_1^{-1}  (z_0, z_1)  = (z_1 / z_0, z_0 / |z_0|),
\end{align}
and
\begin{align}
    \phi_2^{-1}: \pi^{-1}(U_2) &\rightarrow U_2 \cross S^1 \nonumber \\
    (z_0, z_1) &\mapsto \phi_1^{-1}  (z_0, z_1)  = (z_0 / z_1, z_1 / |z_1|).
\end{align}
A few comments are in order. $z_0 \neq 0$ on $U_1$ and $z_1 \neq 0$ on $U_2$, so the maps are non-singular. $z_1 /z_0$ spans $\phi_1(U_1)$ (according to \ref{eq:Coordinates_of_northern_patch_complex}), which can be mapped back to $U_1$ by acting $\phi_{1}^{-1}$. The same holds for the other map. Also, $z_0 / |z_0|$ for $z_0$ corresponding to a point in $U_1$ spans $S^1$, and so does  $z_1 / |z_1|$ for the other patch. Finally, the maps are invariant under $(z_0,z_1) \mapsto \lambda (z_0, z_1)$, as they should for consistency. Hence, considering that everything is smooth and invertible, the maps above qualify for the job of local trivialisations.\\
What happens in the overlap determines the transition function. Let $p$ be any point other than the poles. Then, we have that $
t_{12, p} = \phi_{1,p}^{-1} \circ \phi_{2,p}: S^1 \rightarrow S^1 $ specifies the transition function (see Section \ref{subsec:Princ_fibre_bundles_chap2}). Suppose that $\lambda=e^{i \phi} \in S^1$, and that $\psi_2(p)=Z'=he^{i\omega}$. Then, $\phi_{2,p}(\lambda) = \phi_2(Z', \lambda)=(z_0, z_1)$, with $z_0=h \sqrt{1 / (1+h^2)}e^{i (\phi + \omega )}$ and $z_1= \sqrt{1 / (1+h^2)}e^{i \phi }$.\footnote{This comes from solving $\begin{cases}
    Z' = z_0/ z_1 \\
    \lambda = z_1 / |z_1|
    \end{cases}$ subject to the constraint $|z_0|^2 + |z_1|^2 =1$.}\\

Then, we have that $\phi_1^{-1}(z_0, z_1)=(1/Z', Z'/|Z'| \lambda) $ . Noting that $1/Z'$ in $U_1$ is actually the same point as $Z'$ in $U_2$ (as we would expect), the transition function is just a pointwise map from the fibre to itself:
\begin{align}
    t_{12}(p)(\lambda)= \textrm{arg}(\psi_2(p)) \,  \lambda.
    \label{eq:Trans_func_Hopf}
\end{align}
We end by summarising the form of the local trivialisations in the other direction (which we implicitly used to derive the transition function):
\begin{align}
    \phi_1: U_1 \times S^1 &\rightarrow \pi^{-1}(U_1) \nonumber \\
    (Z, \lambda) &\mapsto \frac{1}{\sqrt{1 + |Z|^2}} \lambda (1 , Z) \, ,
\end{align}
and
\begin{align}
    \phi_1: U_2 \times S^1 &\rightarrow \pi^{-1}(U_2) \nonumber \\
    (Z, \lambda) &\mapsto \frac{1}{\sqrt{1 + |Z|^2}} \lambda (Z , 1) \, .
\end{align}
The constructive proof we just presented is called \textit{Hopf fibration}. As for all fibre bundles, the Hopf one is classified by $\pi_1 (S^1)$, which counts how many times the map \eqref{eq:Trans_func_Hopf} ``winds'' around the equator $S^1$. For the case above, this quantity is $1$. Some lens spaces, but not all, are obtained when allowing different values of $\pi_1 (S^1)$.\footnote{Similarly, some exotic spheres, but not all, are obtained via Milnor's construction. We focus on these lens spaces because their description closely mimics the one of exotic spheres as sphere bundles over spheres.} While $0$ corresponds simply to the product manifold $S^2 \times S^1$ and $-1$ to an orientation-reversed $S^3$, all other integers yield new non-trivial fibre bundles. If we let $\pi_1 (S^1 )$ of $t_{12}$ be $n$ (integer), then the resulting manifold is usually denoted as $L(1,n)$ in the literature - see Section \ref{subsec:Lens_app} for more details on the general definition of $L(m,n)$ that we are referring to, and \cite{Watkins1990} for a very insightful survey. Before proceeding to present the quaternionic version of the construction above, let us comment on why exotic spheres are promising candidates for detecting exotic differentiable structures, based on the properties of their lower-dimensional cousins. Lens spaces furnished the first explicit instances where the \emph{usual} algebraic invariants, i.e.~fundamental group and (co)homology, fail to
distinguish $3$-manifolds.  Already in 1919 J.~W.~Alexander showed that the two
spaces\footnote{We adopt the modern notation $L(p,q)=S^{3}/\!\sim$, where
$(z_{1},z_{2})\sim(e^{2\pi i/p}z_{1},e^{2\pi iq/p}z_{2})$, see Section \ref{subsec:Lens_app}.} $L(5,1)$ and $L(5,2)$ possess isomorphic fundamental groups
$\pi_{1}\cong\Bbb Z_{5}$ and identical homology
($H_{1} \cong \mathbb{Z}_5$ and $H_{2}=0$), yet are \emph{not} homeomorphic.
Even more striking are the pairs
$L(7,1)$ and $ L(7,2)$,
which share \emph{all} homotopy-theoretic data
($\pi_{1}$, $H_{\ast}$, and ordinary homotopy type), yet differ in their simple homotopy type - the ``intermediate layer’’ between algebraic topology
and full geometric topology.  Lens spaces thus revealed a hierarchy of
equivalences:
\begin{align}
\text{homeomorphic}\;\Longrightarrow\;
\text{simple-homotopy equivalent}\;\Longrightarrow\;
\text{homotopy equivalent} \, .
\end{align}
While the first two notions coincide for Lens spaces, the strictness of the second implication proved that extra geometric data is required to distinguish spaces that look identical from a purely algebraic perspective. This hierarchy motivated the development of
torsion invariants (Reidemeister, Whitehead) and, more broadly, the birth of geometric topology: a discipline concerned with classifying manifolds up to homeomorphism or diffeomorphism rather than merely up to homotopy. With this is mind, it is not surprising (a posteriori almost nothing is surprising!) that exotic spheres revealed the decoupling of two seemingly coinciding structures. In this case, the intermediate layer is not between algebraic topology and full geometric topology, but between topological structure and differentiable structure. \\

We postpone the discussion of $S^3$ bundles over $S^4$ to next section, where we describe a concrete manifestation of inequivalent differentiable structures; it is sometimes said that there is not enough room for exotic structures to exist on $S^1$ bundles over $S^2$, because of the presence of commutativity. We proceed to get rid of it in a few pages.

\subsection{Another Perspective (bottom-up)}
\label{subsec_lens_bottom_up_chap3}

In the previous discussion, we took the manifolds $E,M,F,G$ as the starting point. From those, we constructed projection $\pi$, local trivialisations maps $\phi_i$ and determined the transition functions $t_{ij}$, in order to specify a fibre bundle. In what follows, we take an alternative route: we will start with $M,F,G,t_{ij}$, and use them to obtain $E, \pi, \phi_i$ which give a fibre bundle. This approach is called bundle reconstruction (\cite{Nakahara:2003nw}), and it involves ``creating'' a new manifold, $E$, from the three ingredients (two if the bundle is principal) $E,F,G$ ($E,F$). \\
It is often said that ``a bundle describes a manifold that locally looks like a product of two spaces''. Such statement clearly refers to the total space $E$, which locally it looks like $M \cross F$; sometimes, however, it is not evident how $E$ can be built out of $M$ and $F$, which we now clarify by discussing a prototypical example. \\
Our claim is that the base space manifold $M$ together with an atlas $\{ (U_i, \psi_i ) \}$, the fibre manifold $F$ and the transition functions $t_{ij}: U_i \cap U_j \rightarrow F$ are enough to specify a principal fibre bundle uniquely. To show this, we start by constructing a new topological space $X$ as:
\begin{align}
    X= \cup_i X_i,
\end{align}
where $X_i= U_i \times F$. $X_i$ are differentiable manifolds (since they are product of two differentiable manifolds). It follows that $X$ is a differentiable manifold as well, because it is a \textit{disjoint} union of differentiable manifolds.\\
Now, we introduce an equivalence relation on elements of $X$:
\begin{align}
    X_i \ni (p,f) \sim (q,g) \in X_j \quad \textrm{iff} \quad p=q \in U_i \cap U_j \,\,\, \textrm{and} \,\,\, g=t_{ij}(p)f. 
\end{align}
Using such relation, we can define a new space (which will be our total space):
\begin{align}
    E = X/ \sim .
\end{align}
We also define the projection map simply as
\begin{align}
    \pi : E &\rightarrow M \nonumber \\
    [(p,f)] &\mapsto p
\end{align}
and the local trivialisations,
\begin{align}
    \phi_i : U_i \times F &\rightarrow \pi^{-1}(U_i) \nonumber \\
    (p,f) &\mapsto [(p,f)].
\end{align}
Now, by definition, $E$ must have a differentiable structure (and, consequently, we demand $\phi_i$ to be diffeomorphisms). We now build such structure, starting from the underlying topology. We declare that $O \subset E$ is open if 
\begin{align}
    \phi_i^{-1} (O \cap \pi^{-1}(U_i)) \subset U_i \times F \quad \textrm{is open} \quad \forall \, U_i.
\end{align}
We are pulling back the topology of $X$ via the local trivialisations - see Section \ref{subsec:Quotient_topology} for a discussion on this point. This automatically makes the $\phi_i$ homeomorphisms. The same we do for the additional differentiable structure: we again pull it back from $X$. Given an atlas $\{ V_i, \varphi_i \}$ for $F$, we define:
\begin{align}
    W_{ij} = \phi_i (U_i \times V_j) \subset E \\
    \chi_{ij} = (\psi_i, \varphi_j) \circ \phi_i^{-1} : E &\rightarrow \R.
\end{align}
Then, the Atlas $\{ (W_{ij}, \chi_{ij}) \}$ makes $E$ a differentiable manifold, and the smoothness of coordinates changes is ensured by the smoothness of $t_{ij}$. \\
Let us check that $\phi_i$ are really diffeomorphisms. Let $x \in \R$ be the coordinates of a point in $U_i \times F$. Then, the coordinates of its image (with respect to $\phi_i$) are given by:
\begin{align}
    x'(x)= \chi_{ij} \circ \phi_i \circ (\psi_i, \varphi_j)^{-1} x =  (\psi_i, \varphi_j) \circ \phi_i^{-1} \circ \phi_i \circ (\psi_i, \varphi_j)^{-1} x = x \, .
\end{align}
Again, a remark on the quotient topology can be found in \ref{subsec:Quotient_topology}. Hence, the objects just described, i.e.~the triple $M$ (with explicit atlas),$F$, $t_{ij}$ is actually the minimal information required to construct a (principal) fibre bundle. 

Let us apply this machinery to recover $S^3$ from the minimal data corresponding to the Hopf fibration. Let us consider $S^2$ with the complex stereographic atlas and identify $S^1$ with the fibre. Concerning the gluing data, treating an element of $S^1$ (globally) as $e^{i \theta}$, we specify $t_{12}(e^{i \theta}) = e^{i ( \theta + \arctan (y,x) )} $. Under these assumptions, then the bottom-up description of $S^3$ gives
\begin{align}
    (\mathbb{R}^2 \times S^1 \cup \mathbb{R}^2 \times S^1 ) / \sim \, , 
\end{align}
where $(x,y,e^{i  \theta}) \sim (\frac{x}{x^2 + y^2}, \frac{-y}{x^2 + y^2} , e^{i ( \theta + \arctan (y,x) )} )$. We revisit this construction, with an explicit realisation of the complete bundle atlas (including the atlas for $S^1$, which is usually neglected) in a few pages (Section \ref{subsec:Kaluza_Klein_on_lens_chap3}). 

\subsection{Abelian Kaluza--Klein}
\label{subsec:abelian_Kaluza_Klein_chap3}

Let us now discuss the (abelian) Kaluza--Klein formalism, both from a physics perspective and as a mathematical formalism for describing metrics on the total space of fibre bundles. In the next section, it will be exemplified by focusing on $S^3$ realised as the Hopf fibration.

\subsubsection{Dimensional-reduction perspective (physics)}

The seminal idea of Kaluza \cite{Kaluza1921} and Klein \cite{Klein1926} is to embed four-dimensional spacetime \(M^{4}\) in a five-dimensional manifold \(\mathcal{M}_{5}=M^{4}\times S^{1}\) equipped with the metric
\begin{align}
g_{AB}=
\begin{pmatrix}
g_{\mu\nu}(x)+\kappa^{2}\,\phi^{2}(x)A_{\mu}(x)A_{\nu}(x) & \kappa\,\phi^{2}(x)A_{\mu}(x)\\
\kappa\,\phi^{2}(x)A_{\nu}(x) & \phi^{2}(x)
\end{pmatrix},
\qquad A,B=0,\dots,4,\;\mu,\nu=0,\dots,3,
\label{eq:original_kaluza_klein_physics}
\end{align}
where \(y := \!x^{4}\sim y+2\pi R\) parametrises the circular extra dimension and \(\kappa\) is a coupling normalisation.  The \emph{cylindricity condition} \(\partial_{y} g_{AB}=0\) ensures that the Killing vector \(\partial_{y}\) generates an isometry corresponding to electromagnetism’s \(\mathrm{U}(1)\).

Expanding any five-dimensional field \(\Phi(x,y)\) in Fourier modes \(\mathrm{e}^{in\,y/R}\) shows that the zero-mode sector reproduces ordinary four-dimensional general relativity for \(g_{\mu\nu}\) together with a Maxwell field \(A_\mu\) and a scalar \(\phi\) (the radion).  The Einstein–Hilbert action
\begin{align}
S_{5}=\frac{1}{2\kappa_{5}^{2}}\int_{\mathcal{M}_{5}}\!\mathrm{d}^{5}x\,\sqrt{-g^{(5)}}\,R^{(5)}
\end{align}
dimensionally reduces to
\begin{align}
S_{4}=\frac{1}{2\kappa_{4}^{2}}\int_{M^{4}}\!\mathrm{d}^{4}x\,\sqrt{-g}\,
\bigl(R^{(4)}-\tfrac{\kappa^{2}\phi^{3}}{4}F_{\mu\nu}F^{\mu\nu}-\tfrac{3}{2}\,(\partial\ln\phi)^{2}\bigr)
\end{align}
after integrating over \(S^{1}\), performing a Weyl rescaling of the four-dimensional metric to pass to the Einstein frame, and appropriately rescaling Newton’s constant (\cite{Overduin_1997}).  The massive Kaluza–Klein tower has masses \(m_{n}=|n|/R\); if the compactification radius \(R\) is small enough, these modes lie beyond experimental reach, leaving the massless sector as an effective low-energy theory. Generalisations with higher-dimensional internal spaces and non-abelian isometry groups lead to Yang--Mills fields in four dimensions, as we shall shortly discuss, making the Kaluza–-Klein mechanism a prototype for modern supergravity and string compactifications. This mechanism, however, can be run ``backwards'', in what physicists call \textit{inverse Kaluza--Klein} prescription. As we mentioned, this appears in \cite{DUFF19861}, for instance, and it consists of building a higher-dimensional metric from lower-dimensional ingredients. Such an interpretation of the Kaluza--Klein formalism, which is usually more common among mathematicians, is now reviewed; of course, among the many references on the topic, we cannot help but referring the reader to \cite{bourguignon1989mathematicians} in particular, which motivated the title of this thesis.

\subsubsection{Bundle-metric perspective (mathematics)}
Let \(\pi:P\to M\) be a principal \(\mathrm{U}(1)\)-bundle over the Lorentzian four-manifold \((M,g)\) with connection one-form \(\omega\). Choosing a positive constant \(\kappa\) (eventually fixed by matching to Newton's constant) defines a \emph{Kaluza--Klein metric} on the total space \(P\) by
\begin{align}
G=\pi^{*}g+\kappa^{2}\,\omega\otimes\omega .
\end{align}
The horizontal distribution \(\ker\omega\) is orthogonal to the vertical fibres generated by the fundamental vector field \(\xi\) of the \(\mathrm{U}(1)\) action, and \(\|\xi\|^{2}_{G}= \kappa^{2}\). Because \(G\) is invariant under the right action of \(\mathrm{U}(1)\), the quotient \((P,G)/\mathrm{U}(1)\) recovers \((M,g)\).

Writing the connection locally as \(\omega=\mathrm{d}\theta+\pi^{*}A\) shows that the curvature two-form \(\Omega = \mathrm{d}\omega = \pi^{*}F\) encodes the electromagnetic field strength \(F = \mathrm{d}A\). The Levi-Civita connection of \(G\) splits along horizontal and vertical directions; in particular, the Ricci tensor satisfies
\begin{align}
\operatorname{Ric}_{G}\bigl(X,Y\bigr)
  =\operatorname{Ric}_{g}\!\bigl(\pi_{*}X,\pi_{*}Y\bigr)
  -\tfrac{\kappa^{2}}{2}\,(F\!\cdot\!F)\!\bigl(\pi_{*}X,\pi_{*}Y\bigr)
  +\cdots,
\end{align}
where the ellipsis denotes components with at least one vertical argument and \(F\!\cdot\!F\) is the symmetric tensor \(F_{\mu\lambda}F_{\nu}{}^{\lambda}\) pulled back to \(P\) (\cite{Bleecker1981}). 

Crucially, one cannot simply impose the five-dimensional vacuum Einstein equations \(\operatorname{Ric}_{G}=0\), as the purely vertical components would overconstrain the system to \(F_{\mu\nu}F^{\mu\nu}=0\). Instead, one computes the scalar curvature \(R_{G} = R_{g} - \frac{\kappa^{2}}{4}F_{\mu\nu}F^{\mu\nu}\) and invokes the variational principle. Varying the Einstein--Hilbert action of \(G\) on the total space \(P\) with respect to the base metric \(g\) and connection \(A\) yields the general Einstein--Maxwell system on \(M\):
\begin{align}
\operatorname{Ric}_{g}-\tfrac{1}{2}Rg = \tfrac{\kappa^{2}}{2}\,T^{\text{EM}}, 
\qquad \nabla^{\mu}F_{\mu\nu}=0 .
\end{align}
Because \(\mathrm{U}(1)\) bundles over \(M\) are classified by the first Chern class \(c_{1}(P)\in H^{2}(M;\mathbb{Z})\), magnetic charge quantisation \(\int_{S^{2}}F\in2\pi\mathbb{Z}\) appears naturally, and topologically non-trivial field configurations such as Dirac monopoles find a geometric home, as we shall describe later. Moreover, the formalism extends seamlessly to non-abelian structure groups, which will also be discussed in Section \ref{subsec:principal_Kaluza_Klein_chap3} and Section \ref{subsec:associated_Kaluza_Klein_chap3}. This viewpoint, more widespread among mathematicians, frames the Kaluza--Klein ansatz as a neat prescription for assembling a Riemannian metric on the total space of a bundle from the lower-dimensional constituents of the bundle, i.e.~the metric on the base, the metric on the fibre, and the connection. We proceed to apply this formalism to the case of lens spaces.

\subsection{Kaluza--Klein Metrics on Lens Spaces}
\label{subsec:Kaluza_Klein_on_lens_chap3}
In preparation for exotic spheres, let us describe lens spaces with a rigorous treatment. Let the images of the four patches be:
\begin{align}
    U_{A \alpha} = \R^2_A \times \R_{A ,\alpha} , \, \,   U_{A \beta} = \R^2_A \times \R_{A,\beta} , \, \, U_{B \alpha} = \R^2_B \times \R_{B,\alpha} , \, \, U_{B \beta} = \R^2_B \times \R_{B,\beta}, \, \,
\end{align}
where the subscripts are used to keep track of different coordinates; Latin letters refer to those of $S^2$ (the base) and Greek ones to the coordinates of $S^1$ (the fibre). This treatment differs from the standard one (which can be found in \cite{Nakahara:2003nw}, for instance) mainly because the $S^1$ part is treated using two patches here; as opposed to dealing with it globally. The change between $\R_{A , \alpha}$ and $\R_{A , \beta} $ depends on the choice of coordinates on the circle only. We choose to work with the one-dimensional stereographic projection discussed above, which we now summarise. Suppose we work with $S^1_A$, for concreteness. According to the stereographic projection atlas, spelled out for the general $S^n$ case in Section \ref{subsec:Working_definitions_chapA1}, one of the two charts takes the form:
\begin{align}
    \psi_{\alpha} : U_{\alpha} = S^1_A \backslash (0, 1) &\xrightarrow{} \R_{A , \alpha} \nonumber \\
    (x,y) &\mapsto u = \frac{x}{1-y} \, ,
\end{align}
Similarly, the other chart is associated with the projection from the other pole, which we denote here as $\psi_{\beta}$. If we let the new coordinate be denoted by $v$, then the change of coordinates reads $u = 1/v$.\footnote{The other option would be to use the angle parametrisation, where we have $\theta \in (- \pi, \pi)$ and $\theta' \in (  0, 2 \pi)$, with the transition function being $    \theta' = \begin{cases}
        \theta  \quad \quad \quad \, \, if \,\,\, \theta > 0 \\
        \theta + 2 \pi  \quad if \,\,\,  \theta < 0 \,
    \end{cases}$.}
Clearly, the corresponding one-dimensional Jacobian is given by $-1/v^2$. And, of course, the exact same thing can be done for $S^1_B$.\\
The one between $U_{A \sigma_1}$ and $U_{B \sigma_2} $ ($\sigma_{1,2}$ can be $\alpha$ or $\beta$) is a stereographic projection on the base combined with a twist on the fibre. Regarding the former, we denote the coordinates in one patch as $X=(x_1, x_2)$, and the usual transformation
\begin{align}
    y_1 = \frac{x_1}{x_1^2 + x_2^2} \, , \quad y_2 = \frac{x_2}{x_1^2 + x_2^2} \, 
\label{eq:2d_stereo_change}
\end{align}
as $1/ \bar{X}$ (the meaning of this notation is clarified in \ref{subsec:Working_definitions_chapA1}).
Just to give an example of one change of coordinates between the charts, let us provide the expression for $\sigma_1 = \sigma_2 = \alpha$:
\begin{align}
    (X, u) \mapsto (Y, u') = (1/\bar{X}, \frac{\cos (\Theta)}{1-\sin(\Theta)} ) \, ,
    \label{eq:Tra_alpha_alpha}
\end{align}
where $\Theta$ is given by $\Theta = k \atantwo(x_2 , x_1) + \atantwo( \frac{u^2 -1}{u^2 + 1} , \frac{2 u }{u^2 + 1 }) $. Note that $\atantwo(x_2 , x_1)$ just gives the angle of $X$ with respect to the horizontal axis, and that the inverse of the above transformation, i.e.~from $U_{B \alpha}$ to $U_{A \alpha}$, reads the same but with $ \atantwo(x_2 , x_1) \xrightarrow[]{} -  \atantwo(x_2 , x_1)$, as expected. The change of coordinates for the other patches are analogous, and they are provided in Section \ref{sec:Lens_chap5}, where an exhaustive description of the coordinate changes and their Jacobians is needed. \\
Now, given this explicit description of the bundle structure, we can discuss possible geometries on lens spaces. According to the Kaluza--Klein prescription \eqref{eq:original_kaluza_klein_physics}, the metric on the total space of these bundles reads:
\begin{align}
    ds^2 = d \Omega_2 + (d \Omega_1 + A )^2 \, ,
\end{align}
where $\Omega_2$ is a metric on the base manifold $S^2$, $\Omega^1$ is a metric on the fibre manifold $S^1$, and $A$ is a connection defined on the bundle, i.e.~a $\text{u(1)}-$valued gauge field living on $S^2$. The simplest non-trivial bundle of this family corresponds to $k= \pm 1$, which is nothing but the ordinary $3-$sphere. It is well known that the round metric on $S^3$ in ``angular coordinates'' reads:
\begin{align}
    ds^2 = \frac{1}{4} (d \psi^2 + d \theta^2 + d \phi^2 - 2 \cos \theta d \phi d \psi ) \, .
\end{align}
It is not a coincidence that this can be re-written as\footnote{Note that, according to the Section \ref{sec:Notation_and_conventions_ch1}, this metric satisfies $Ric(g) = 2 g = ( 3 -1) g$.}
\begin{align}
    d s^2=\frac{1}{4} \Big[ d \theta^2+\sin ^2 \theta d \phi^2 + (d \psi-\cos \theta d \phi)^2 \Big]\, ,
\end{align}
where $d \theta^2+\sin ^2 \theta d \phi^2 $ is the round metric on $S^2$ of radius $1/2$, and $\frac{1}{2} d \psi$ is the canonical metric on the fibre $S^1$. This points to the fact that $A = -\frac{1}{2} \cos \theta d \phi$ is the connection on the Hopf bundle. The simplest consistency check comes by calculating the first Chern number associated with $A$. This can be done in a few lines:
\begin{align}
    F = dA= \frac{1}{2} \sin \theta d \theta \wedge d \phi \, , \nonumber \\
    c_1=\frac{1}{2 \pi} \int_{S^2} F=\frac{1}{2 \pi} \int_0^{2 \pi} \int_0^\pi\left(\frac{1}{2} \sin \theta\right) d \theta d \phi =1 \, .
\end{align}
By re-writing this metric in the coordinates defined above, one finds that the round metric on $S^3$ (radius $1$), for $k=1$ (which is $S^3$ itself) reads:
\begin{align}
ds^2 = \frac{dx_1^2+dx_2^2}{(1+x_1^2+x_2^2)^2} + \frac{1}{4}\left( \frac{4du}{u^2+1} + \frac{(x_1^2+x_2^2)-1}{(x_1^2+x_2^2)+1} \frac{x_1 dx_2 - x_2 dx_1}{x_1^2+x_2^2} \right)^2 \, .
\end{align}
This result is clearly ill-defined at the origin, highlighting a very important issue to do with gauge fields: their definition always relies on some underlying choice of chart. This was already discussed in Section \ref{subsec:geometric_instaton_chap2}, and it will be crucial when discussing the construction of a metric on the exotic sphere; this toy-model example gives us the chance to review how to handle such a situation.

The singular term is
\begin{align}
\frac{x_{1}\,dx_{2}-x_{2}\,dx_{1}}{x_{1}^{2}+x_{2}^{2}} .
\end{align}

The standard connection on the Hopf fibration is
\begin{align}
A \propto \cos\theta_{b}\,d\phi_{b},
\end{align}
which is perfectly regular when expressed in \((\theta_{b},\phi_{b})\), appears singular in Cartesian coordinates because
\(d\phi_{b}\) itself is singular there. To show that the metric on \(S^{3}\) is
(of course) smooth, and the apparent divergence is purely
coordinate artefact, we choose instead the gauge that is regular at the North pole
(\(r_{s}=0,\;\theta_{b}=0\)):
\begin{align}
A_{N}
 \propto (1 - \cos \theta_b ) \, .
\end{align}
If the base \(S^{2}\) has radius \(R_{0}=1/2\) (so that
\(r_{s}^{2}=x_{1}^{2}+x_{2}^{2}\)), one may rewrite
\(\cos\theta_{b}=\tfrac{1-4r_{s}^{2}}{1+4r_{s}^{2}}\) and obtain
\begin{align}
A_{N}
=\frac{k}{2}\bigl(1-\cos\theta_{b}\bigr)\,d\phi_{b}
=k\,\frac{4\bigl(x_{1}\,dx_{2}-x_{2}\,dx_{1}\bigr)}
         {1+4\bigl(x_{1}^{2}+x_{2}^{2}\bigr)}.
\end{align}

The two potentials differ by a (singular) gauge transformation:
\begin{align}
A
=
A_{N}
-\frac{k}{2}\,d\phi_{b},
\end{align}
so each form is regular on its own patch, and together they describe the same smooth connection on the Hopf bundle. With this choice, the metric in the coordinates defined in above reads:
\begin{align}
ds^{2}
= 4 \frac{dx_{1}^{2}+dx_{2}^{2}}{\bigl(1+4r_{s}^{2}\bigr)^{2}}
  +\Bigl(
      \tfrac{2\,du}{u^{2}+1}
      +\tfrac{4\,(x_{1}\,dx_{2}-x_{2}\,dx_{1})}
             {1+4r_{s}^{2}}
    \Bigr)^{2} \, ,
\end{align}
where the components are manifestly finite at $x_{1}=x_{2}=0$. \\
This concludes our detour to lens spaces, which served as an introduction to a concrete example of a fibre bundles that shares many features with exotic spheres; the special case of the Hopf fibration allowed us to illustrate a neat example of Kaluza--Klein geometry.

\section{Ordinary $7-$spheres}
\label{sec:ordinary_sphere_chap3}

In this section, we generalise the previous discussions about the Hopf fibration and Kaluza--Klein formalism to higher dimensions. We do this by reviewing $S^7$ realised as a quaternionic Hopf fibration and considering non-abelian Kaluza--Klein theory on principal (non-abelian) fibre bundles.

\subsection{Quaternionic Hopf Fibration}
\label{subsec:quaternionic_hopf_chap3}

The expression for the Hopf map \eqref{eq:Hopf_projection} is specific for the Hopf fibration of $S^3$, and hides the generality of the construction, which is in one-to-one correspondence with the division algebras. The more general expression reads
\begin{align}
    \begin{aligned} h: S^{2 n-1} & \rightarrow S^n \\ \left(z_0, z_1\right) & \mapsto\left(2 z_0 \overline{z_1},\left|z_0\right|^2-\left|z_1\right|^2\right)\end{aligned},
\end{align}
for $n \in \{1,2,4,8 \}$ and the bar denotes conjugation in the appropriate division algebra. \\
For the following discussion, let us focus on $n=4$, which specifies the quaternionic Hopf fibration defining $S^7$, i.e.~the simplest non-trivial bundle in the family constructed by Milnor. We can again introduce quaternionic stereographic coordinates on the base, which in this case is $S^4$, analogously to \eqref{eq:Coordinates_of_northern_patch_complex} and \eqref{eq:Coordinates_of_southern_patch_complex}:
\begin{align}
    &Z'=Y_1' + i Y_2' +j Y_3' + k Y_4' = \frac{y_1 + iy_2 + jy_3 + ky_4}{1 - y_5} =\frac{2 z_0 \bar{z_1}}{1-\left|z_0\right|^2 + \left|z_1\right|^2} = \frac{z_0 \bar{z_1}}{\left|z_1\right|^2} = \frac{z_0}{z_1} \, ,\nonumber \\ &\textrm{with} \quad (y_1,y_2,y_3, y_4, y_5) \in U_2 \, ,
\label{eq:Coordinates_of_northern_patch_quaternion}
\end{align}
and similarly for the quaternionic stereographic projection from the other pole\footnote{Note that, similarly to the case of the ordinary Hopf fibration, and as discussed in Section \ref{subsec:Working_definitions_chapA1}, the minus signs in projection from the other pole ensure that $Z' = 1/Z$ instead of $1/\bar{Z}$}. The local trivialisations also take the same form:
\begin{align}
    \phi_1^{-1}: \pi^{-1}(U_1) &\rightarrow U_1 \cross S^3 \nonumber \\
    (z_0, z_1) &\mapsto \phi_1^{-1}  (z_0, z_1)  = (z_1 / z_0, z_0 / |z_0|)\, ,
\end{align}
and
\begin{align}
    \phi_2^{-1}: \pi^{-1}(U_2) &\rightarrow U_2 \cross S^3 \nonumber \\
    (z_0, z_1) &\mapsto \phi_1^{-1}  (z_0, z_1)  = (z_0 / z_1, z_1 / |z_1|) \, ,
\end{align}
with the only difference that $z_0$ and $z_1$ are quaternions now. This implies that, when finding the inverse of the local trivialisations, non-commutativity must be taken into account. For chart $U_2$, the problem amounts to solving $q=z_0 / z_1$ and $p = z_1/|z_1|$ simultaneously, which yields:
\begin{align}
    \phi_2: U_2 \cross S^3 &\rightarrow  \pi^{-1}(U_2) \nonumber \\
    (q, p) &\mapsto \left(z_0= \frac{qp}{\sqrt{1+|q|^2}}, z_1 = \frac{p}{\sqrt{1 + |q|^2}} \right) \, .
\end{align}
A similar result follows for the chart $U_1$.

\subsection{Non-Abelian (Principal) Kaluza--Klein}
\label{subsec:principal_Kaluza_Klein_chap3}

\subsubsection{Dimensional-reduction perspective (physics)}

After Kaluza and Klein, a natural generalisation is to start with a \((4+n)\)-dimensional manifold locally of the form
\(\mathcal{M}_{4+n}=M^{4}\times K^{n}\) in which the compact internal space
\(K^{n}\) possesses the isometry group \(G\) one wishes to gauge.  The simplest
choice is to take \(K^{n}=G\), i.e.~restricting to a principal bundle, itself equipped with its bi-invariant metric
\(h_{\alpha \beta}\) (\(\alpha,\beta=1,\dots,n=\dim G\)). Denote with $x$ the coordinates on $M^4$ and with $e_\mu{}^a(x)$ the vielbein on it (see Section \ref{sec:VielbeinFormalism} for details on how the vielbein is defined). Then, the Kaluza–Klein ansatz for the higher-dimensional metric reads\footnote{We find this to be the most common ansatz among the early supergravity community; it can be found in \cite{DUFF198490, Bailin:1987jd, Salam:1981xd}. Note that plus and minus signs are sensitive to choices of metric signature and of Lie algebra basis. Additionally, the metric components can be presented in bundle-adapted form or coordinate adapted form - see Section \ref{sec:Bundle_adapted_vs_coordinate_adapted}; both are commonly used in the literature, and they yield slightly different descriptions of the same metric.}
\begin{align}
    d s^2=\left(d x^\mu e_\mu{}^a(x)\right)^2 + \left(\left(d y^\alpha-d x^\mu \kappa A_\mu{}^\beta(x) K_\beta{}^\alpha(y)\right) e_\alpha{}^i(y)\right)^2 \, ,
    \label{eq:Kaluza-Klein_non-abelian_only_physics}
\end{align}
where $e_\alpha{}^i(y)$ is the vielbein on $G$, and $K_\beta{}^\alpha(y)$ are Killing vectors of $G$. Note that a generic Lie group $G$ has \textit{at least} $\mathrm{dim}(G)$ Killing vectors - those associated with the natural choice of a left-invariant metric (or equivalently right-invariant). But this a lower bound: $S^3 \sim \mathrm{SU(2)}$ with its bi-invariant metric has $6$ of them. Discussing how to deal with cases where one considers more Killing vector fields than the dimension of the internal space naturally leads to relaxing the assumption $K^n \sim G$ to $K^n \sim G / H$; this, in turn, leads to considering associated bundles, and we deal with it in the next section. For now, we restrict to group manifolds and left-invariant metrics on it, which is implicitly assumed in \eqref{eq:Kaluza-Klein_non-abelian_only_physics} by using indices $\alpha, \beta$ of the same kind. For this case, one can also identify $dy^{\alpha} e_\alpha{}^i = \theta^{i}$ with the left-invariant Maurer-Cartan one-forms
on \(G\) satisfying \(\mathrm{d}\theta^{i}=
-\tfrac{1}{2}f^{i}{}_{j k}\,\theta^{j}\wedge\theta^{k}\). Harmonic-expanding the
\((4+n)\)-dimensional metric fluctuations in the normal modes of the Laplacian
on \(K^{n}\) and keeping only the singlets yields a four-dimensional graviton
\(g_{\mu\nu}\), Yang–Mills gauge fields \(A^{\alpha}_{\mu}\) taking values in
\(\mathfrak{g}\), and a finite set of scalar fields (typically moduli
parametrising the size and shape of \(K^{n}\))
(\cite{10.1063/1.522434, 10.1063/1.525753}).  Inserting the ansatz into the Einstein–Hilbert
action and integrating over \(K^{n}\) produces the effective theory
\begin{align}
S_{4}= \frac{1}{2\kappa_{4}^{2}}\int_{M^{4}}\!\mathrm{d}^{4}x\,\sqrt{-g}\,
\Bigl(
  R^{(4)}
  -\frac{\kappa^{2}}{4}\,h_{\alpha \beta}F^{\alpha}_{\mu\nu}F^{\beta\,\mu\nu}
  -V_{\text{scal}}(\varphi)
\Bigr),
\end{align}
where \(F^{\alpha}_{\mu\nu}= \partial_{\mu}A^{\alpha}_{\nu}-\partial_{\nu}A^{\alpha}_{\mu}
+f^{\alpha}{}_{\beta \gamma}A^{\beta}_{\mu}A^{\gamma}_{\nu}\) and \(V_{\text{scal}}\) encodes the
curvature of the internal space.  Massive excitations appear as towers with
masses set by the eigenvalues of the Laplacian on \(K^{n}\), typically
\(m\sim 1/R\) if \(R\) is the characteristic radius of \(K^{n}\).
Historically, such models—sometimes called
\emph{compactified pure gravity}—provided the first geometric unification of
gravity with \(\mathrm{SU}(2)\) or \(\mathrm{SU}(3)\) Yang–Mills sectors, as commented in
\cite{DUFF19861}, and remain a benchmark for modern flux
compactifications in string theory.

\subsubsection{Bundle-metric perspective (mathematics)}

Let \(G\) be a compact, semisimple Lie group with Lie algebra \(\mathfrak{g}\) and let \(\pi:P\to(M^{4},g)\) be a principal \(G\)-bundle. Fix a bi-invariant inner product \(h\) on \(\mathfrak{g}\) (minus the Cartan--Killing form). A \emph{Kaluza--Klein metric} \(\hat{g}\) on the total space \(P\) is constructed by choosing a principal connection one-form \(\omega\in\Omega^{1}(P; \mathfrak{g})\) and setting
\begin{align}
    \hat{g} = \pi^{*}g + \kappa^{2}\,h_{ab}\,\omega^{a}\otimes\omega^{b}.
\end{align}
The vertical fibres are orbits of the right action \(R_{u}:P\to P\) (for \(u \in G\)), and the fundamental vector fields \(\xi_{a}\) obey \(\hat{g}(\xi_{a},\xi_{b})=\kappa^{2}h_{ab}\). Because the Lie derivative \(\mathcal{L}_{\xi_{a}}\hat{g}=0\), each \(\xi_{a}\) generates an isometry, meaning the Riemannian submersion \(\pi: (P,\hat{g}) \to (M^{4},g)\) has totally geodesic fibres isometric to \(G\).

Locally, via a trivialising section \(\sigma: U \subseteq M^4 \to P\), the connection pulls back to a Yang--Mills gauge potential \(A = \sigma^*\omega \in \Omega^1(U; \mathfrak{g})\). On this local patch \(U \times G\), the connection form is rigorously expressed as \(\omega_{(x,u)} = \operatorname{Ad}_{u^{-1}}(\pi^* A) + \theta\), where \(\theta = u^{-1}\mathrm{d}u\) is the left-invariant Maurer--Cartan form on \(G\). The curvature \(\Omega = \mathrm{d}\omega + \frac{1}{2}[\omega, \omega]\) projects to the Yang--Mills field strength \(F^{a} = \mathrm{d}A^{a}+ \frac{1}{2}f^{a}{}_{bc}A^{b}\wedge A^{c}\) on the base, while the Maurer--Cartan equations \(\mathrm{d}\theta^{a}= -\tfrac{1}{2}f^{a}{}_{bc}\theta^{b}\wedge\theta^{c}\) dictate the intrinsic geometry of the vertical group manifold.

While the principal connection \(\omega\) defines a horizontal/vertical splitting of the tangent bundle \(TP\), the Levi-Civita connection \(\hat{\nabla}\) of the metric \(\hat{g}\) does not preserve it. In the language of O'Neill's Riemannian submersions, this failure is measured exactly by the field strength \(F\). By evaluating O'Neill's equations, the horizontal-horizontal projection of the Ricci tensor of \(P\), \(\operatorname{Ric}_{\hat{g}}(X,Y)\), evaluates to
\begin{align}
    \operatorname{Ric}_{g}(X,Y) -\tfrac{\kappa^{2}}{2}\,h_{ab}\langle\,F^{a}(X, \cdot), F^{b}(Y, \cdot)\rangle_g \, .
\end{align}
Tracing this tensor (and adding the vertical trace) yields the total scalar curvature \(\hat{R} = R^{(4)} + R_{G} - \frac{\kappa^2}{4}h_{ab}F^a_{\mu\nu}F^{b\,\mu\nu}\), where \(R_G\) is the constant positive scalar curvature of the group manifold. Integrating \(\hat{R}\) over the fibres to form an action, and subsequently varying it, yields the Einstein--Yang--Mills equations on \(M^4\). (Note that if one allows the internal metric \(h_{ab}\) to fluctuate as a function of \(M^4\), this procedure also naturally generates scalar moduli fields and a corresponding potential). 

Finally, because principal \(G\)-bundles are classified topologically by characteristic classes (such as the second Chern class for \(G=\mathrm{SU}(N)\)), instanton number and magnetic charge quantisation emerge naturally as integrality conditions on \(\int_{S^{4}}\operatorname{tr}(F\wedge F)\) or \(\int_{S^{2}}\operatorname{tr}(F)\). 

Thus, in the formal geometric language, the non-abelian gauge potential and its field strength are strictly the principal connection and its curvature, while the unified metric \(\hat{g}\) packages the gravitational and gauge degrees of freedom into a single higher-dimensional Riemannian object.

\subsubsection{Combining both perspectives}
As a brief comment, these two treatments of Kaluza--Klein formalism for principal bundles were unified in a language common to both physicists and mathematicians, such as \cite{10.1063/1.522434, Kerner1968, PhysRevD.13.235}. We refer the reader to them for explicit derivations of the ansatz and its curvature components, through a rigorous bundle-theoretic treatment. The ultimate result is the same as \eqref{eq:Kaluza-Klein_non-abelian_only_physics}, up to a minus sign due to conventions, as they find
\begin{align}
    g_{a b}=\left(\begin{array}{c|c}
g_{\mu \nu}+g_{i k} B_\mu^i B_\nu^k & B_\mu^i g_{i k} \\
\hline g_{i k} B_\nu^k & g_{i k}
\end{array}\right) \, .
\label{eq:Principal_Kaluza_Klein_mathphys}
\end{align}

\subsection{Metric on $S^7$ via Inverse Kaluza--Klein}
\label{subsec:Kaluza_Klein_on_s7_chap3}

\subsubsection{Construction à la \cite{DUFF19861}}
We now briefly review the construction of the round metric on $S^7$ viewed as a principal bundle via an inverse Kaluza--Klein process, which might be familiar to physicists. To do so, we repeat the steps of the famous supergravity review by Duff, Nilsson and Pope, \cite{DUFF19861}, but with right-invariant objects instead of left-invariant ones. The equivalence between these two choices is discussed in the appendix, section \ref{Sec:Left_inv_vs_right_inv}.\\
We can write the line element defined by the usual round metric on $S^4$ as
\begin{align}
    ds^2(S^4)= d\mu^2 + \frac{1}{4}\sin^2\mu ( \bar{\Sigma}_i \bar{\Sigma}_i ),
\label{eq:Metric_on_S4_right_inv}
\end{align}
where $\bar{\Sigma}_i$ form a set of right-invariant one-forms on $S^3$ and $0 \leq \mu \leq \pi $. Clearly, this is the metric on the base space, and it is natural to also express $\bar{\Sigma}_i$ with a set of (three) angular coordinates. This is achieved by parametrising $U\in \mathrm{SU(2)}$ with Euler angles, computing the Maurer-Cartan form $dU U^{-1} $ and decomposing it in some basis for $\mathrm{su(2)}$. Some details on how this is done in \cite{DUFF19861} are summarised in section \ref{Sec:Left_inv_vs_right_inv}.  We use $\bar{\Sigma}_i$ and $\bar{\sigma}_i$ only when referring to the right-invariant objects in the conventions of \cite{DUFF19861}. We reserve other symbols for the right/left invariant objects defined in our conventions, which appear in the following sections. To avoid confusion of any sort, we also use a different set of Greek letters for the Euler angles: $\{ \alpha, \beta, \gamma \}$ are employed for the objects defined in our conventions, while $ \{ \phi, \theta, \psi \}$ are the letters employed in \cite{DUFF19861}.\\
Regardless of the details of the three implicit coordinates in \ref{eq:Metric_on_S4_right_inv}, the standard gauge potential for the $k=-1$ $\mathrm{SU(2)}$ instanton centered at the origin takes the form
\begin{align}
    A^i= \cos^2(\frac{1}{2} \mu) \bar{\Sigma}_i.
    \label{eq:Instanton_gauge}
\end{align}
By \textit{standard} we mean that the gauge field is valued in $\mathrm{su(2)}$ with generators $T_i$ conventionally chosen so that $[T_i, T_j ] = \epsilon_{ijk} T_k$. This is also discussed in the appendix (see \ref{sec:Appendix_instanton}), together with the derivation of the above equation. It will shortly become evident that
\ref{eq:Instanton_gauge} is the local expression for the connection of the principal bundle defined by the Hopf fibration. In fact, by choosing the vielbein on the fibre to be another set of right-invariant one-forms $\bar{\sigma}_i$ on $S^3$, we can construct a metric on the total space as
\begin{align}
    ds^2(S^7) = d\mu^2 + \frac{1}{4} \sin^2\mu ( \bar{\Sigma}_i \bar{\Sigma}_i ) + (\bar{\sigma}_i - A^i)^2.
    \label{eq:Metric_on_S7_right_inv}
\end{align}
By the same argument of \cite{DUFF19861} for the left-invariant case, this is exactly the usual round metric on $S^7$, as anticipated. When $A^i = 0$, this the usual round metric on $S^4$ with unit radius times a  round metric $S^3$ with radius $2$. \\
The coordinates employed here are useful for the case of a single anti-instanton that we are considering, but we find them to be unsuitable for the generalisation to $|k|>1$ instantons. Hence, we report here the same quantities in stereographic coordinates:
\begin{align}
    ds^2(S^4) = \frac{4 dx^\mu dx^\mu}{(1 + x^{\mu} x^{\mu})^2} \quad \mathrm{and} \quad A^i =  \frac{1}{x^\mu x^\mu +1 } \bar{\Sigma}_i.
\label{eq:Metric_and_instanton_in_stereographic}
\end{align}
From this expression we immediately see that the moduli of the instanton have been chosen to be trivial: the centre is at the origin and the size is set to one. \\
The right-invariant one-forms $\bar{\sigma_i}$ and $\bar{\Sigma}_i$ also admit different (but equivalent) descriptions. We can define them globally when $S^3$ is thought as embedded in $\R^4$, or locally by using Euler angles as coordinates on $S^3$. While the latter is the one used in \cite{DUFF19861}, we find the former more natural in the formalism outlined in Section \ref{subsec:associated_Kaluza_Klein_chap3}; more details on their definitions can be found in Section \ref{sec:Invariant_objects_appendix}. In what follows, we will use both of them and comment on the connection between the two.

\section{Milnor's Exotic Spheres and the Gromoll--Meyer Sphere}
\label{sec:Milnor_spheres_and_GM}
In this section, we introduce exotic spheres according to Milnor's original construction and review their associated bundles' structure in detail. Then, we review the Kaluza--Klein ansatz for the most general case of bundles with homogeneous fibres; in other words, non-principal non-abelian bundles. We then combine these two discussions by presenting a concrete coordinate expression for a metric on the ``simplest'' exotic sphere: the Gromoll--Meyer sphere.

\subsection{Milnor's Bundles}
\label{subsec:Milnor_construction_chap3}

\subsubsection{The Original Construction}

In this section, we outline Milnor's construction of exotic spheres. Some reviews on the topic are: ~\cite{McEnroe2016MILNORSCO, Bognat2018MILNORSES,Exot_world}. \\
Exotic spheres are defined as the total space of an $S^3$ (non-principal) bundle over $S^4$ with transition functions given living in $\text{SO(4)}$. By the bundle reconstruction theorem (see Section \ref{subsec_lens_bottom_up_chap3} and \cite{Nakahara:2003nw}), the minimal set of information necessary to uniquely specify the total space of a bundle consists of: $M$ (the base), $\{U_i \}$ (an atlas for the base), $F$ (the fibre), $t_{ij}$ (the transition functions, which also define the structure group). To reproduce Milnor's construction of exotic spheres we employ the isomorphism between $\mathbb{R}^4$ and quaternions, which clearly induces the one between $S^3$ and unit quaternions. With this in mind, we let
\begin{align}
    M=S^4  \, , \quad \{U_i \} = \{ (U_N, \psi_N), (U_S, \psi_S) \}, \quad F=S^3 \simeq  \mathbb{H}_* = \{z \in \mathbb{H} \, | \, \norm{z}^2=1 \},
\end{align}
where $U_N = S^4 \, \backslash \, \textrm{North Pole} $  and $\psi_N$ is the usual stereographic projection from $U_N$ to $\mathbb{R}^4 \simeq \mathbb{H}$, and analogously for $U_S$, $\psi_S$. The last piece of information consists of the transition functions, which in this case is just a single one, since there are two patches on the base. We define a family of transition functions, labelled by two integers, which will in general define inequivalent bundles:
\begin{align}
f_{h,l}: \psi_N(U_N \cap U_S) \times S^3 &\xrightarrow{} \psi_S(U_N \cap U_S) \times S^3 \nonumber \\
    (z,y) &\mapsto (\frac{1}{z}, \frac{z^h y z^l}{||z||^{h+l}}),
\end{align}
where $z \in \mathbb{H}$, $y \in \mathbb{H}_*$ and quaternion multiplication is understood as juxtaposition.\footnote{Note that a completely analogous construction exists for octonions, rather than quaternions, defining a $15$-dimensional total space; this is treated in \cite{Shimada_1957} and \cite{10.2969/jmsj/01010029}, for instance, but will not be discussed here.}
The case $h=1$, $l=0$ reduces to the usual Hopf fibration (as does $h=0$, $l=1$, just with different conventions). For generic $h$ and $l$, the bundle defined this way is not a principal one, since left and right multiplication combined produce an element of $\text{SO(4)}$. What Milnor was able to prove is that for $h+l=1$, the total space of the bundle, which we will denote by $E_{h,l}$, is homeomorphic to $S^7$. In addition to that, he showed that $E_{h,l}$ and $S^7$ cannot be diffeomorphic if $(h-l)^2 \neq 1(\bmod 7)$, implying that some total spaces are exotic manifolds.\\
To conclude this section, we make a few comments on the above transition functions. By associativity of quaternions, it immediately follows that the right action and the left action commute, and we have that
\begin{align}
    \frac{z^h y z^l}{||z||^{h+l}} = \frac{1}{||z||^{h+l}} L_{z^h} R_{z^l} y = \frac{1}{||z||^{h+l}} R_{z^l} L_{z^h} y.
\end{align}
By linearity of the quaternionic multiplication, we can assign a matrix form to the operations above. For $u=a + b \textbf{i} + c\textbf{j} + d\textbf{k}$ and $x=a' + b' \textbf{i} + c'\textbf{j} + d'\textbf{k}$,
\begin{align}
    L_u x=\left(\begin{array}{cccc}
a & -b & -c & -d \\
b & a & -d & c \\
c & d & a & -b \\
d & -c & b & a
\end{array}\right)\left(\begin{array}{c}
a^{\prime} \\
b^{\prime} \\
c^{\prime} \\
d^{\prime}
\end{array}\right) \, ,
\label{eq:Left_action_quat_SO(4)}
\\
 R_u x=\left(\begin{array}{cccc}
a & -b & -c & -d \\
b & a & d & -c \\
c & -d & a & b \\
d & c & -b & a
\end{array}\right)\left(\begin{array}{c}
a^{\prime} \\
b^{\prime} \\
c^{\prime} \\
d^{\prime}
\end{array}\right) \, .
\end{align} 
If $u$ is a unit quaternion, then both matrices belong to $\text{SO(4)}$, and one can verify that the they indeed commute in general.
Specifically, $\{ L_u | u \in \mathbb{H}_* \}$ is referred as the subgroup of \textit{left-isoclinic rotations}, which we will label as $\mathrm{SU(2)}_L$. Similarly, $\{ R_u | u \in \mathbb{H}_* \}$ are known as the \textit{right-isoclinic rotations} and this subgroup will be indicated by $\mathrm{SU(2)}_R$. These two subgroups are not disjoint: they share the identity and the central inversion. The group homomorphism:
\begin{align}
    S^3 \times S^3 &\xrightarrow{} \text{SO(4)} \nonumber \\
    (y,z) &\mapsto L_y R_{z^{-1}} \, ,
    \label{eq:From_SU(2)_times_SU(2)_to_SO(4)}
\end{align}
whose kernel consists of the two elements just mentioned, illustrates that $\mathrm{SU(2)} \times \mathrm{SU(2)}$ double covers $\text{SO(4)}$, as it is well known.

We now wish to explicitly show how this family of manifolds is associated to principal $\text{SO(4)}$-bundles over $S^4$, as mentioned in Section \ref{subsec:geometric_twisted_chap2}.\footnote{Note that exotic spheres constructed as associated bundles to principal $\text{SO(4)}-$bundles already appeared in \cite{Rigas1978} and \cite{BOUWKNEGT201546}, for instance.} The latter are classified by the ``winding'' of their transition functions, given by $\pi_3(\text{SO(4)})$. As before, we can invoke the lifting theorem (\cite{McEnroe2016MILNORSCO}) which yields $\pi_3(SO(4)) = \pi(S^3 \times S^3 ) = \mathbb{Z}\times \mathbb{Z}$. The two integers correspond to $h,l$ if we choose the transition functions to be:
\begin{align}
    t_{h,l}: \psi_N(U_N \cap U_S) \times \text{SO(4)} &\xrightarrow{} \psi_S(U_N \cap U_S) \times SO(4) \nonumber \\
    (z,g) &\mapsto (\frac{1}{z}, \frac{1}{||z||^{h+l}} L_{z^h} R_{z^l} g).
\end{align}
Note that this is essentially the same choice of transition functions for the Milnor's construction, but for the object that they acting on. In other words, the only difference is the fibre. \\
It is known that the associated fibre bundle has the same transition functions of the principal one, which means that if we consider for $E$ we will have that
\begin{align}
    t_{h,l}: \psi_N(U_N \cap U_S) \times \text{SO(4)}/\text{SO(3)} &\xrightarrow{} \psi_S(U_N \cap U_S) \times \text{SO(4)}/\text{SO(3)} \nonumber \\
    (z,a) &\mapsto (\frac{1}{z}, \frac{1}{||z||^{h+l}} L_{z^h} R_{z^l} a),
\end{align}
where $a$ is some element of the quotient $\text{SO(4)}/\text{SO(3)}$; this is further clarified in Section \ref{subsec:metric_on_Gromoll_Meyer_chap3}. Moreover, the action of $\text{SO(4)}$ elements preserves the isomorphism between $\text{SO(4)}/\text{SO(3)}$ and $S^3$. This shows that the transition functions above indeed describe Milnor's bundles.

\subsubsection{A Note on Different Conventions and Equivalences}
So far we have followed the original set-up, concerning both content and conventions; the only minor notational change, compared to Milnor's original paper, is that we use ``$l$'' to denote the exponent of the right-multiplying quaternion, instead of ``$j$''. There is, however, a different set of conventions that is quite widespread in the literature, which amounts to a different choice of labelling. Instead of defining the transition functions as $u^h v u^l$, they can be defined as\footnote{Note that this comes from a the following choice of generators of $\pi_3(\text{SO(4)})$: $\rho(u)v=uvu^{-1}$ and $\sigma(u) v = uv$ (\cite{10.2969/jmsj/01010029}).} 
\begin{align}
    u^{m+n} v u^{-m} \, .
    \label{eq:mn_conventions}
\end{align}
The relation with the original choice is immediate: 
\begin{align}
    h = m+ n \, , \quad l = -m \, ,
    \label{eq:mn_to_hl}
\end{align}
which will be useful for the following discussion on diffeomorphisms, based on \cite{CROWLEY2003363}.
This notation also appears in Chapter \ref{chap:4}, Section \ref{subsec:Tamura_chap4}, where we build upon the work of \cite{10.2969/jmsj/01010029} using these $m,n$ conventions. As a final word of caution, we point out that some authors switch $m$ and $n$ in their conventions; it is case of \cite{Rigas1978}, for instance, which we referred to in Section \ref{subsec:geometric_twisted_chap2} when discussing principal $SO(4)-$bundles. 

Let us now quickly comment on some relations among members of this family of bundles. Let the total space of the bundle defined by \eqref{eq:mn_conventions} be denoted by $F_{m,n}$; thus, $F_{m,n}$ defines the exact same object as $E_{h,l}$ when the relations in \eqref{eq:mn_to_hl} hold. Then, the following diffeomorphisms exist (\cite{CROWLEY2003363}):
\begin{align}
    &F_{m,n} \simeq F_{m+n , -n} \quad \mathrm{(change \,\, of \,\, orientation \,\, on \,\, the \,\, fibre)} \, , \nonumber \\
    &F_{m,n} \simeq F_{-m , -n} \quad \mathrm{(change \,\, of \,\, orientation \,\, on \,\, the \,\, base)} \, .
\end{align}
In our $h,l$ notation, this yields:
\begin{align}
    &E_{h,l} \simeq E_{-l , -h} \quad \mathrm{(change \,\, of \,\, orientation \,\, on \,\, the \,\, fibre)} \, , \nonumber \\
    &E_{h,l} \simeq E_{-h , -l} \quad \mathrm{(change \,\, of \,\, orientation \,\, on \,\, the \,\, base)} \, .
\end{align}
Combining the two one gets the diffeomorphism corresponding to an overall change of orientation:
\begin{align}
    E_{h,l} \simeq E_{l , h} \, .
\end{align}
These relations are derived more explicitly and reviewed in more detail in Section \ref{subsec:left_and_right_bundles_chap3}.

\subsection{Non-Abelian Non-Principal Kaluza--Klein}
\label{subsec:associated_Kaluza_Klein_chap3}

\subsubsection{Conventions and Basic Notions}
Here we summarise the conventions used and the main results about Lie groups and quotient spaces that will be needed in the following. For a more complete discussion, we refer the reader to \cite{10.1063/1.525753}. \\
Let $G$ be a Lie group and $X=G/H$ be the coset space obtained by quotienting $G$ by the subgroup $H$.
We can define a basis $\{T_i \}$ of the Lie algebra $\mathfrak{g}$ of G, with $i=1,...,\textrm{dim}(G)$ where the first $\textrm{dim}(H)$ indices span the Lie sub-algebra $\mathfrak{h} \subseteq \mathfrak{g}$ of $H$. We will denote those indices as $\hat{i}, \hat{j},...=1,...,\textrm{dim}(H)$. For the remaining indices, spanning the complementary space $\mathfrak{b} = \mathfrak{g} - \mathfrak{h}$, will use $\alpha, \beta, ...$. 
\\
A key property which we will assume is that the coset space is reductive, i.e.
\begin{align}
\textrm{Ad}_{\mathfrak{g}}(H) \mathfrak{b} \subseteq \mathfrak{b} \quad , \quad \textrm{i.e.} \quad h \mathfrak{b} h^{-1} \subseteq \mathfrak{b} \quad \forall h \in H.
\end{align}
This implies that $C_{\hat{i} \alpha}{}^{\hat{j}} = 0$, and both conditions are always satisfied if $H$ is compact. \\
Let us now make a few considerations at the level of the group (manifold).
Clearly we have a left action of $G$ on itself, given by $L_g h = g \cdot h$, and similarly a right action given by $R_h g = g \cdot h$. Moreover, $G$ has a transitive and effective left action on $X$ that we will denote by $\bar{L}: G\times X \xrightarrow{} X$. \\
Let us recall that any Lie group admits a set of left-invariant vector fields (generating right-translations), i.e.
\begin{align}
     e_i^{R}(g)=\left.\frac{d}{d t} R_{\exp \left(t T_i\right)}(g)\right|_{t=0},
\end{align}
and a set of right-invariant vector fields (generating left-translations), i.e.
\begin{align}
     e_i^{L}(g)=\left.\frac{d}{d t} L_{\exp \left(t T_i\right)}(g)\right|_{t=0}.
\end{align}
They satisfy 
\begin{align}
    \begin{aligned}
& {\left[e_i^{\mathrm{R}}, e_j^{\mathrm{R}}\right]=C_{i j}{ }^k e_k^{\mathrm{R}},} \\
& {\left[e_i^{\mathrm{L}}, e_j^{\mathrm{L}}\right]=-C_{i j}{ }^k e_k^{\mathrm{L}}.}
\end{aligned}
\end{align}
We can push forward the vector fields onto $X$ to obtain (given some coordinates $y$ on $X$):
\begin{align}
    K_i(y)=\left.\frac{d}{d t} \bar{L}_{\exp \left(t T_i\right)}(y)\right|_{t=0},
    \label{eq:K_i_definition}
\end{align}
which then obey
\begin{align}
    [K_i,K_j]=-C_{ij}{}^k K_k.
\end{align}
The $K_{\alpha}$'s form a basis on the neighbourhood of the origin $O$ of the coset space. Thus, in a neighbourhood of $O$, we have that
\begin{align}
    [K_{\alpha},K_{\beta}]=-C_{\alpha \beta}{}^i K_i{}^{\gamma}(y) K_{\gamma},
    \label{eq:K_matrix_commutator}
\end{align}
where $K_i{}^{\gamma}$ are the components of $K_i$ in the basis given by $\{K_{\alpha}\}$. Clearly, $K_{\alpha}{}^{\gamma}= \delta_{\alpha}{}^{\gamma}$, while $K_{\hat{i}}{}^{\gamma}(y)$ will be more general functions of $y$.

\subsubsection{Notions on Principal and Associated Bundles}
This section is short and technical, with the aim of summarising some aspects of the relation between the principal bundle and the associated one. The interested reader can find more complete discussions in \cite{10.1063/1.525753} and \cite{zbMATH03194988}, while the reader that prefers to avoid technicalities should skip this section. \\
Let us consider a principal bundle $P$ with projection $\pi$, base $M$ and fibre $G$. We will refer to the coordinates on the base as $x$. Let us denote a local trivialisation by $\psi: U \times G \xrightarrow{} \pi^{-1}(U)$. We define $\phi_x : G \xrightarrow{} \pi^{-1}(U)$ as $\phi_x(g)= \psi(x,g)$. \\
We now briefly outline how the bundle $E$ associated with $P$, with fiber $G/H$, is constructed. We denote the projection map on $E$ as $\eta$, and the local trivialisation as $\bar{\psi}: U \times G/H \xrightarrow{} \eta^{-1}(U)$. For more details on these maps, we refer the reader to \cite{10.1063/1.525753}, or to the classic reference \cite{zbMATH03194988}. In the latter, it is proven that $E$ is the quotient of $P$ by the right action of $H$, and we refer to this quotient map as $\tau: P \xrightarrow{} E$.
Finally, if we define $\bar{\phi}_x$ analogously to $\phi_x$, we obtain the following commuting diagram
\begin{align}
    \begin{array}{ccc}
P \supseteq \pi^{-1}(x) & \stackrel{\phi_x}{\longleftarrow} & G \\
\tau \downarrow & & \downarrow \mu \\
E \supseteq \eta^{-1}(x) & \stackrel{\bar{\phi}_x}{\longleftarrow} & G / H ,
\end{array}
\label{eq:Commuting_diagram}
\end{align}
where clearly $\mu$ is the quotient map on $G$.
This diagram summarises our construction and the relation between the principal and associated bundles. \\
Finally, we can obtain a basis for vertical vectors in $E$ by considering
\begin{align}
    \bar{e}_i(w)=(\bar{\phi}_x)_* K_i(y),
    \end{align}
where $w=\bar{\psi}(x,y)$ and we are using $y$ for the coordinates on the fibre.
As before, the subset $\{\bar{e}_{\alpha} \}$ can be chosen as the set of basis vectors, satisfying:
\begin{align}
    [\bar{e}_{\alpha},\bar{e}_{\beta}]=-C_{\alpha \beta}{}^i K_i{}^{\gamma} \bar{e}_{\gamma}.
\end{align}

\subsubsection{General Expression for the Metric on the Total Space}
To avoid inserting a redundant section and get straight to our results, we again refer the reader to \cite{10.1063/1.525753} (section VII therein) for details on the steps leading to the gauge-invariant bundle metric on the total space. We just summarise the main points below. \\ 
The key properties imposed on the metric $\bar{g}$ on the total space $E$ (associated to some principal bundle $P$) are the following. We require it to be independent of the choice of trivialisation, to ensure gauge invariance. We also impose that $\bar{g}$ is such that horizontal and vertical spaces are orthogonal to each other. The most general expression for a metric satisfying these requirements, in the basis outlined above, is:\footnote{We recall that $x$ are the coordinates on the base and $y$ are the coordinates on the fibre.}
\begin{align}
\bar{g}_{MN} =
    \left(\begin{array}{cc}
g_{\mu \nu}(x)+\bar{h}_{\alpha \beta}(x, y) K_i{}^\alpha(y) K_j{}^\beta(y) A_\mu^i(x) A_\nu^j(x) & A_\mu^i(x) K_i{}^\alpha(y) \bar{h}_{\alpha \beta}(x, y) \\
\bar{h}_{\alpha \beta}(x, y) A_{\nu}^i(x) K_i{}^\beta(y) & \bar{h}_{\alpha \beta}(x, y)
\end{array}\right),
\label{eq:General_metric}
\end{align}
where $g_{\mu \nu}(x)$ is some metric on the base space, $A^i_{\mu}(x)$ is the component-form of some connection on the principal bundle $P$ and $\bar{h}_{\alpha \beta}(x,y)$ is the metric on $G/H$, in non-coordinate basis, and it is allowed to smoothly vary from fibre to fibre. To be more specific,
\begin{align}
\bar{h}_{\alpha \beta}(x,y) = \bar{h}(x)(K_{\alpha},K_{\beta})|_y.
\label{eq:Fibre_metric_def}
\end{align}
The simplest limit of the formula \ref{eq:General_metric} is the case of $H =e$, where the fibre becomes the group itself and we recover a principal bundle. We further specialise to $G=\mathrm{SU(2)}$ and $M=S^4$, so that $E=P$ correspond to the family of bundles containing the usual Hopf fibration of the seven-sphere. \\
Since we take $H$ to be trivial, the hatted indices disappear, and we have that $i,j,...=\alpha, \beta,... = 1, ..., \textrm{dim}(\mathrm{SU(2)})=3$. This implies that $K_i{}^{j}(y)=\delta_i{}^j$, and consequently the metric in \ref{eq:General_metric} takes the form:
\begin{align}
    \left(\begin{array}{cc}
g_{\mu \nu}(x)+\bar{h}_{ij}(x, y) A_\mu^i(x) A_\nu^j(x) & A_\mu^i(x)  \bar{h}_{i j}(x, y) \\
\bar{h}_{i j}(x, y) A_{\nu}^j(x)  & \bar{h}_{i j}(x, y)
\end{array}\right).
\end{align}
Note that, under these assumptions, the $K_i$'s are simply the right-invariant vector fields $e_i^{L}$, which we denote with $\bar{\sigma}_i$, to connect with the previous section.
If we set $\bar{h}_{i j}(x, y) = \delta_{ij}$, then we get:\footnote{We gloss over conventions and normalisation factors in this section. They will be discussed in detail in the next section, which is a more general set-up that includes this one as a special case.}
\begin{align}
\left(\begin{array}{cc}
g_{\mu \nu}(x)+  A_\mu^i(x) A_\nu^i(x) & A_\mu^i(x)   \\
 A_{\nu}^j(x)  & \delta_{i j}
\end{array}\right).
\label{eq:Principal_bundle_metric}
\end{align}
We only need to choose $g_{\mu \nu}$ and $A_{\mu}$ at this point. We let $g_{\mu \nu}$ be the round metric on $S^4$ given by \ref{eq:Metric_on_S4_right_inv}. As for $A^i_{\mu}$, we set it to be the potential corresponding to a $k=-1$ instanton, i.e. $A^i = -\cos^2(\frac{1}{2} \mu) \bar{\Sigma}_i$. With these choices, we recover the line element of \ref{eq:Metric_on_S7_right_inv}. \\
As we mentioned, the formalism for the case of the bundle being principal (see previous section, as well as\cite{10.1063/1.522434} and \cite{PhysRevD.13.235}, for instance) is not enough to describe a geometry on exotic spheres. The ansatz \eqref{eq:General_metric}, on the other hand, is sufficiently general to handle Milnor's bundles. We now show how to realise this in practice.

\subsection{Metric on the Gromoll--Meyer Sphere via Inverse Kaluza--Klein}
\label{subsec:metric_on_Gromoll_Meyer_chap3}

\subsubsection{Preliminaries}
Let us introduce two choices of basis for the Lie algebra of $\mathrm{so(4)}$ that will be used throughout this work. We discuss the subalgebras that naturally appear in each basis, the corresponding subgroups and the quotient spaces associated to those subgroups. These are well-known results and expressions, but we introduce them here using our notation, since they are used in the construction below. We begin by the decomposition into rotations and ``boosts''. Given the usual basis $(L_{\alpha \beta})_{\mu \nu} = \delta_{\alpha \mu} \delta_{\beta \nu} - \delta_{\alpha \nu} \delta_{\beta \mu}$, we can split the generators into 3-dimensional rotations and ``boosts'' by defining $R_i = \frac{1}{2} \epsilon_{ijk} L_{jk}$ and $B_i = L_{i4}$, respectively. We will refer to the set of generators corresponding to this choice of basis as $\{T^{RB}_I \}$ ($I=1,...,6$). The subalgebra spanned by $\{R_i \}$, once exponentiated, produces the $SO(3)$ subgroup of $SO(4)$ which leaves the point $(0,0,0,1)\in \mathbb{R}^4$ fixed. The $B_i$'s
do not close to form a subalgebra. This can be seen from the commutation relations:
\begin{align}
    [R_i, R_j] = - \epsilon_{ijk} R_k \, , \quad  [R_i, B_j] = - \epsilon_{ijk} B_k \, , \quad [B_i, B_j] = - \epsilon_{ijk} R_k \, .
\end{align}
If we quotient by the $SO(3)$ above, we obtain $SO(4)/SO(3) \simeq S^3$, where the isomorphism is given by the map $SO(4) \ni A \mapsto Az \in S^3$ (see \cite{zbMATH03194988}), with $z=(0,0,0,1)$. \\

We can define a new basis for $\mathrm{so(4)}$ by taking linear combinations of the previous generators as: $M_i = (R_i + B_i) $ and $N_i = (R_i - B_i) $.
We denote this set of generators as $\{ T^{su(2)}_I\}$. The change of basis just described illustrates the well-known Lie algebra isomorphism between $\mathrm{so(4)}$ and $\mathrm{su(2) \oplus su(2)}$. Specifically, the subset $\{ M_i \}$ spans an $\mathrm{su(2)}$ subalgebra, referred as $\mathrm{su(2)_L}$. The subset $\{ N_i \}$ spans the other $\mathrm{su(2)}$ subalgebra, referred as $\mathrm{su(2)_R}$. They exponentiate exactly to the subgroups $\mathrm{SU(2)}_L$ and $\mathrm{SU(2)}_R$ introduced in the previous section, respectively. The subalgebras explicitly read:\footnote{Note that the normalisation employed here is not the conventional one, but we find it to be more natural in the context of our construction.}
\begin{align}
    [M_i, M_j] = - 2 \epsilon_{ijk} M_k \, \quad [N_i, N_j] = -  2 \epsilon_{ijk} N_k .
\label{eq:Structure_constants_su(2)s}
\end{align}
We note that taking the quotient by one of these subgroups yields a different manifold from the previous case, i.e. $SO(4)/\mathrm{SU(2)}_{L,R} \simeq SO(3)$. The reason for this is that the two subgroups have a common $\mathbb{Z}_2$ subgroup, as mentioned in the previous section. \\

Let us now review how these different choices play a role in the geometry of the quotient space $SO(4) / SO(3) \sim S^3$. The natural basis for the Lie algebra is given by $\mathfrak{g}$ of $G$ is given by $\{R_i, B_i \}$, defined above. We let $\mathfrak{h}=\textrm{Span}(\{R_{\hat{i}} \})$, with $\hat{i},\hat{j}=1,2,3$ being the indices associated to this subalgebra. It follows that $\mathfrak{b}=\textrm{Span}(\{B_{\alpha} \})$, with the corresponding indices $\alpha, \beta=1,2,3$. This choices satisfy $C_{\hat{i} \alpha}{}^{\hat{j}} = 0$, which ensures that the algebra is reductive. \\
As opposed to the previous case with $H=e$, we now have that the $K_i$'s are not simply the $e_i^L$'s, and also that $K_{i}^{\alpha}$ is non-trivial. Regarding the former, we recall that this setting is special in the sense that the homogeneous space which we take to be the fibre turns out to be a Lie group again. By using \ref{eq:K_i_definition} we can find the $K_i$'s via a quick computation.
We perform the calculation by considering $S^3$ as embedded in $\mathbb{R}^4$ as usual ($S^3 = \{(X,Y,Z,W) \,\,\, \mathrm{ s.t. } \,\,\,  X^2 + Y^2 + Z^2 + W^2 = 1 \}$), and we find that the components of the $K_{\hat{i}}$'s are
\begin{align}
    (K_1)^{\mathrm{C}} = \left(\begin{array}{c}
0 \\
Z \\
-Y \\
0
\end{array}\right)^{\mathrm{C}} \, , \quad  (K_2)^{\mathrm{C}} = \left(\begin{array}{c}
-Z \\
0 \\
X \\
0
\end{array}\right)^{\mathrm{C}} \, , \quad  (K_3)^{\mathrm{C}} = \left(\begin{array}{c}
Y \\
-X \\
0 \\
0
\end{array}\right)^{\mathrm{C}} \end{align}
while the  components of the $K_{\alpha}$'s are given by
\begin{align}
(K_4)^{\mathrm{C}} = \left(\begin{array}{c}
W \\
0 \\
0 \\
-X
\end{array}\right)^{\mathrm{C}} \, , \quad  (K_5)^{\mathrm{C}} = \left(\begin{array}{c}
0 \\
W \\
0 \\
-Y
\end{array}\right)^{\mathrm{C}} \, , \quad  (K_6)^{\mathrm{C}} = \left(\begin{array}{c}
0 \\
0 \\
W \\
-Z
\end{array}\right)^{\mathrm{C}},
\label{eq:Ks_for_S3}
\end{align}
We thus have that (see equation \ref{eq:K_matrix_commutator}):
\begin{align}
K_{\hat{i}}{}^{\gamma} =
    \left(\begin{array}{ccc}
    \vspace{0.1cm}
0 & \frac{Z}{W} & -\frac{Y}{W}\\ \vspace{0.1cm}
-\frac{Z}{W} & 0 & \frac{X}{W}  \\
\frac{Y}{W} & - \frac{X}{W} & 0 \\
\end{array}\right)_{\hat{i}}^{\,\,\, \gamma}
\label{eq:K_rot_and_boost}
\end{align}
We need to be able to compare our results with the ones appearing in the literature (\cite{DUFF19861} specifically), and we also seek a convenient description for exotic spheres. For these reasons, we note that the basis just presented (which is not right-invariant) is not the most suitable one. This leads us to a change of basis that leads to a more suitable set-up for exotic spheres. Let us denote with $\{\mathchorus{K}_{\, \, \hat{i}'} \}$ the left-invariant vector fields on $S^3$ ($\hat{i}' = 1,2,3$), and with $\{ \mathchorus{K}_{\, \, \alpha'} \}$ the right-invariant ones ($\alpha'=4,5,6$), and it is known that each set forms a parallelisation of $S^3$. We employ the same index notation introduced in section \ref{subsec:associated_Kaluza_Klein_chap3}, and the reason for this will become apparent shortly. If again we consider the unit 3-sphere as embedded in $\R^4$ with coordinates $\{X,Y,Z,W\}$, once these vector fields are normalised, their components read:
\begin{align}
    (\mathchorus{K}_{\, \, \hat{i}'})_C =  \eta^{\hat{i}'}_{CB} X_B \, , \quad (\mathchorus{K}_{\, \, \alpha'})_C=  \bar{\eta}^{\alpha'}_{CB} X_B ,
\label{eq:Left_right_inv_vector_fields}
\end{align}
where $\eta^{\hat{i}'}_{BC}$ and $\bar{\eta}^{\alpha'}_{BC}$ the self-dual 't Hooft symbols and the anti-self-dual 't Hooft symbols, respectively.
They coincide with the $K_i$'s defined by equation \ref{eq:K_i_definition} when we choose the $\mathrm{su(2) \oplus su(2)}$ basis for $\mathrm{so(4)}$, splitting the generators as $\{ M_{\hat{i}'} \}$ and $\{ N_{\alpha'} \}$:
\begin{align}
    \mathchorus{K}_{\,\, \hat{i}'}(y)=\left.\frac{d}{d t} \bar{L}_{\exp \left(t M_{\hat{i}'}\right)}(y)\right|_{t=0} \, , \quad 
    \mathchorus{K}_{\,\, \alpha'}(y)=\left.\frac{d}{d t} \bar{L}_{\exp \left(t N_{\alpha'}\right)}(y)\right|_{t=0} .
\end{align}
Just as a quick check, we observe that they satisfy
\begin{align}
    [\mathchorus{K}_{\,\, \hat{i}'}, \mathchorus{K}_{\,\, \hat{j}'}] = 2 \epsilon_{\, \hat{i}' \hat{j}' \hat{k}'} \mathchorus{K}_{\,\, \hat{k}'} \, , \quad [\mathchorus{K}_{\,\, \alpha'}, \mathchorus{K}_{\,\, \beta'}] = 2 \epsilon_{\, \alpha' \beta' \gamma'} \mathchorus{K}_{\,\, \gamma'},
\end{align}
which is the opposite of \ref{eq:Structure_constants_su(2)s}, as we would expect from section \ref{subsec:associated_Kaluza_Klein_chap3}.
We note that
\begin{align}
    \mathchorus{K}_{\,\, 1} = K_1 + K_4 \, \quad \mathchorus{K}_{\,\, 2} = K_2 + K_5 \, \quad \mathchorus{K}_{\,\, 3} = K_3 + K_6 \, , \\ 
    \mathchorus{K}_{\,\, 4} = K_1 - K_4 \, \quad \mathchorus{K}_{\,\, 5} = K_2 - K_5 \, \quad \mathchorus{K}_{\,\, 6} = K_3 - K_6 \, .
\end{align}
It can be summarised by by
\begin{align}
    \mathchorus{K}_{\,\, i'}=M_{i'}{}^{i} K_i ,
\end{align}
with
\begin{align}
M_{i'}{}^{i} =
\left(\begin{array}{cccccc}
1 & 0 & 0 & 1 & 0 & 0 \\
0 & 1 & 0 & 0 & 1 & 0 \\
0 & 0 & 1 & 0 & 0 & 1 \\
1 & 0 & 0 & -1 & 0 & 0 \\
0 & 1 & 0 & 0 & -1 & 0 \\
0 & 0 & 1 & 0 & 0 & -1 
\end{array} \right)_{i'}^{\,\,\,\,\, i} 
\end{align}
with inverse given by
\begin{align}
    K_i = (M^{-1})_{i}{}^{i'} \mathchorus{K}_{\,\, i'},
\end{align}
so that 
\begin{align}
    M_{i'}{}^i (M^{-1})_{i}{}^{j'} = \delta_{i'}^{j'} \quad \textrm{and} \quad (M^{-1})_{i}{}^{i'} M_{i'}{}^{j} = \delta^{j}_{i}.
\end{align}
The same holds for the basis at the origin clearly, i.e. the $T^{su(2)}$'s and $T^{RB}$'s are related by $T^{su(2)}_{ i'}=M_{i'}{}^{i} T^{RB}_i$. 
Now, this means that
\begin{align}
    \bar{h}_{\alpha' \beta'} = \bar{h}(\mathchorus{K}_{\,\, \alpha'}, \mathchorus{K}_{\,\, \beta'}) = \bar{h}(M_{\alpha ' }{}^{i} K_{i}, M_{\beta ' }{}^{j} K_{j}) = M_{\alpha ' }{}^{i} M_{\beta ' }{}^{j} \bar{h}(K_{i}, K_{j}) = \nonumber \\ M_{\alpha ' }{}^{i} K_i{}^{\alpha} M_{\beta ' }{}^{j} K_j{}^{\beta}  \bar{h}_{\alpha \beta} = W_{\alpha'}{}^{\alpha} W_{\beta'}{}^{\beta}  \bar{h}_{\alpha \beta}, 
\end{align}
where we have implicitly defined $W_{\alpha'}{}^{\alpha} = M_{\alpha ' }{}^{i} K_i{}^{\alpha}$, with inverse $(W^{-1})_{\alpha}{}^{\alpha'}$\footnote{Note that, even though $K_i{}^{\alpha}$ does not have an inverse, its contraction with $ M_{\alpha ' }{}^{i}$, which turns it into a square matrix, does.}.
Analogously, we have that $A_{\mu}=A_{\mu}^{i'} T^{su(2)}_{i'} = A_{\mu}^{i} T_{i}^{RB}$ gives
\begin{align}
    A_{\mu}^{i} = A_{\mu}^{i'} M_{i'}{}^i . 
\end{align}
Finally, 
\begin{align}
     M_{i'}{}^{i} K_i =  M_{i'}{}^{i} K_i{}^{\alpha} K_{\alpha}  =\mathchorus{K}_{\,\, i'} = \mathchorus{K}_{\,\, i'}{}^{\alpha'} \mathchorus{K}_{\,\, \alpha'} = \mathchorus{K}_{\,\, i'}{}^{\alpha'} M_{\alpha ' }{}^{i} K_{i} =  \mathchorus{K}_{\,\, i'}{}^{\alpha'} M_{\alpha ' }{}^{i} K_i{}^{\alpha} K_{\alpha},
\end{align}
from which we infer
\begin{align}
    \mathchorus{K}_{\,\,i'}{}^{\alpha'} =   M_{i'}{}^{i} K_i{}^{\alpha} (W^{-1})_{\alpha}{}^{\alpha'} .
\end{align}
An identical reasoning can be applied to obtain $K_i{}^{\alpha}$ from $\mathchorus{K}_{\,\,i'}{}^{\alpha'}$. One can check that, using \ref{eq:K_rot_and_boost} together with $K_{\alpha}{}^{\gamma} = \delta_{\alpha}{}^{\gamma}$, one obtains
\begin{align}
    \mathchorus{K}_{\,\,\hat{i}'}{}^{\alpha'} =
\left(
\begin{array}{ccc} \vspace{0.1 cm}
 1-2 \left(W^2+X^2\right) & -2 (W Z+X Y) & 2 W Y-2 X Z \\ \vspace{0.1cm}
 2 (W Z-X Y) & 1-2 \left(W^2+Y^2\right) & -2 (W X+Y Z) \\
 -2 (W Y+X Z) & 2 W X-2 Y Z & 2 \left(X^2+Y^2\right)-1 
\end{array}
\right)_{\hat{i}'}^{\,\,\,\, \alpha '} ,
\label{eq:K_in_left_right_basis}
\end{align}
with $\mathchorus{K}_{\,\,\gamma '}{}^{\alpha'} = \delta_{\gamma '}{}^{\alpha'}$, as expected. This result will be needed in what follows. For now, we point out that each row of the above matrix can be thought as a map from $S^3$ embedded in $\R^4$ to $S^2$ embedded in $\R^3$. Specifically, each row is a realisation of the Hopf map $p$ that defines the Hopf fibration $S^1 \hookrightarrow S^3 \xrightarrow{p} S^2$. \\
As we mentioned, the right/left-invariant basis is the most natural one to work with when dealing with Milnor's bundles. For this reason, we work in the left/right-invariant basis for the rest of the section. Under this assumption, we now drop the primes in the indices to ease the notation.

\subsubsection{Usual Sphere (again)}
Armed with the results obtained in the previous two sections, we are almost in the position to recover the metric \ref{eq:Metric_on_S7_right_inv} in this new setting. We define everything in the basis of left/right-invariant objects, as just mentioned, and drop the primed indices under this assumption.\\
The components of the dual one-forms to \ref{eq:Left_right_inv_vector_fields} read (see also Section \ref{sec:Invariant_objects_appendix} for more details):
\begin{align}
    (\bar{\omega}_{ \hat{i}})_C =  \eta^{\hat{i}}_{CB} X_B \, , \quad (\bar{\omega}_{ \alpha})_C=  \bar{\eta}^{\alpha}_{CB} X_B .
    \label{eq:Dual_forms_omega}
\end{align}
Let us now make the connection between this set-up and the formalism of \cite{DUFF19861}, pointing out a few subtle differences. Firstly, we have that the norm of the $\bar{\omega}_{ \alpha}$'s is half the norm of the $\bar{\Sigma}_\alpha$'s ($\alpha = 1,2,3$). The $\bar{\omega}_{ \alpha}$'s are uniformly scaled by a factor of $1/2$ so that $|\bar{\omega}_{ \alpha}| = 1/2 \,  |\bar{\Sigma}_{\alpha}| $ for any $\alpha$.
With this consideration in mind, let us choose $\bar{h} = 4 \bar{\omega}_{\alpha} \bar{\omega}_{\alpha}$, so that, according to \ref{eq:Fibre_metric_def}, we obtain $\bar{h}_{\alpha \beta}= 4 \delta_{\alpha \beta}$.
For the metric on the base, we let
\begin{align}
    ds^2(S^4)= d\mu^2 + \sin^2\mu ( \bar{\Omega}_{\alpha} \bar{\Omega}_{\alpha} ),
\label{eq:Metric_on_S4_right_inv_2}
\end{align}
where $\bar{\Omega}_{\alpha}$ is another set of Maurer-Cartan forms identical to \ref{eq:Dual_forms_omega}.
Regarding the gauge field, we recall that our definition of generators for the algebra was unconventional (see equation \ref{eq:Structure_constants_su(2)s}). With this choice, we have that an anti-instanton in the $\mathrm{su(2)}$ subalgebra labelled by Greek indices is described by $A^{\alpha}= -\cos^2(\frac{\mu}{2}) \bar{\omega}_{\alpha}$. We let the gauge field living in the other $\mathrm{su(2)}$ be trivial, i.e. $A^{\hat{i}}=0$.\\
Now, by recalling that $\mathchorus{K}_{\,\, \alpha}{}^{\gamma}= \delta_{\alpha}{}^{\gamma}$ (see \ref{eq:K_matrix_commutator} and the following comments), we can plug all these quantities in equation \ref{eq:General_metric}. By doing this, we find that the metric on the $(0,1)$ Milnor's bundle that we just outlined coincides with \ref{eq:Metric_on_S7_right_inv}, as expected.

\subsubsection{The Double Instanton}
In order to construct the Gromoll-Meyer exotic sphere, we need a transition function $f_{h,l}$ with $h=2$, $l=-1$, or anything homotopic to this (see \cite{nuimeprn10073}). Translating this in the physics language, we need an anti-instanton in one $\mathrm{su(2)}$ factor and a double instanton, i.e. $k=2$, in the other factor. While for the first one there exists a simple a explicit expression involving all the moduli, the same is not true for the second one. A closed form in terms of all the moduli exists, but it is quite complicated. We will now take a small detour to clarify this point. The most common approach to finding the general solution to the (anti-)self-duality equations, which is also the one that first appeared in the literature (see Section \ref{subsec:instantons_chap2}), starts by taking the following ansatz for the gauge field:
\begin{align}
    A_{\mu} = \sigma_{\mu \nu } \partial_{\nu} ln \rho.
\end{align}
Then, it is shown that $\rho$ must satisfy $\frac{1}{\rho} \Box \rho = 0$. The general solution is found to be:\footnote{To be precise, this misses four moduli, see \cite{PhysRevD.15.1642}.}
\begin{align}
    \rho = 1 + \sum_{i=1}^k \frac{\lambda_i^2}{(x - a_i)^2},
\end{align}
and it has winding number $k$. However, there is a subtlety when $a_i = a_j$ and $i \neq j$. It is easy to see that, for the case of $k=2$, if $a_1 = a_2$, then we obtain the a $k=1$ instanton with size squared given by $\lambda_1^2 + \lambda_2^2$. This is noted in \cite{RevModPhys.51.461}, for instance, and this seemingly singular point in the moduli space is the reason why we were not able to obtain an $SO(4)$-invariant solution for the $k=2$ gauge potential.\footnote{Note that the most elegant formalism for this computation is the ADHM construction, that can be translated into explicit expressions for the gauge potential and field strength. This is work in progress. We thank Professor Berman and Professor Travaglini for discussions on this point.} With this subtle point in mind, we present here the expression for the gauge field describing two instantons ($k=2$) with the same size:
\begin{align}
    A^a_{\mu} = -2\frac{(x-a)^2 (x-b)^2}{(x-a)^2 (x-b)^2 + \rho^2 [(x-a)^2 + (x-b)^2]} \eta^a_{\mu \nu} \Bigg( \frac{\rho^2 (x-a)_{\nu}}{(x-a)^4} + \frac{\rho^2 (x-b)_{\nu}}{(x-b)^4}
     \Bigg).
     \label{eq:Diinstanton}
\end{align}
For convenience, the gauge field here is given in the \textit{singular gauge}, while the one in \ref{eq:Metric_and_instanton_in_stereographic} is in the \textit{regular gauge}. We will come back to this double-instanton, both in singular and regular gauges, with a more appropriate quaternionic formalism, in Section \ref{subsec:k2_chap3}. Finally, note that we also chose the sizes of the two instantons to be the same here in order to keep our expression general without making it too cumbersome.

\subsubsection{Exotic Spheres}
Let us now move to the general case involving both $\mathrm{su(2)}$ components of the connection, keeping the same choices for $g_{\mu \nu}(x)$ and $\bar{h}_{\alpha \beta}(y)$. \\
Given a general connection $A_{\mu}^i$ on $P$, the metric in \ref{eq:General_metric} will include contributions from the $\mathchorus{K}_{\, \, \hat{i}}{}^{\gamma}(y)$.
In this case we will need the most general form of the metric on $E$, given by
\begin{align}
    \left(\begin{array}{cc}
g_{\mu \nu}(x) + 4\delta_{\alpha \beta} K_i^\alpha(y) K_j^\beta(y) A_\mu^i(x) A_\nu^j(x)   \,\,\,
&  4A_\mu^{\beta}(x) 
 + 4A_\mu^{\hat{i}}(x) \mathchorus{K}_{\,\,\hat{i}}{}^\beta(y)  \\
 4A_{\nu}^{\alpha}(x) + 4A_{\nu}^{\hat{i}}(x) \mathchorus{K}_{\,\,\hat{i}} {}^\alpha(y) & 4\delta_{\alpha \beta}
\end{array}\right),
\label{eq:Expansion_metric}
\end{align}
where, in the off-diagonal entries, we simply expanded $ A_{\nu}^i(x) \mathchorus{K}_{\,\,i}{}^{\alpha}$. For completeness, the expansion of $\delta_{\alpha \beta} K_i^\alpha(y) K_j^\beta(y) A_\mu^i(x) A_\nu^j(x) $ reads
\begin{align}
    A^{\alpha}_{\mu}(x) A^{\alpha}_{\nu}(x) + A^{\alpha}_{\mu}(x) \mathchorus{K}_{\,\,\hat{j}}{}^{\alpha} A^{\hat{j}}_{\nu}(x) + A^{\hat{i}}_{\mu}(x) \mathchorus{K}_{\,\,\hat{i}}{}^{\beta} A^{\beta}_{\nu}(x)  +  \mathchorus{K}_{\,\,\hat{i}}{}^\alpha(y) \mathchorus{K}_{\, \, \hat{j}}{}^\beta(y) A_\mu^{\hat{i}}(x) A_\nu^{\hat{j}}(x).
    \label{eq:Expansion_first}
\end{align}
In the previous case, by setting $A^{\hat{i}}=0$, we eliminated the two extra contributions in the off-diagonal terms of \ref{eq:Expansion_metric} and the last three terms in \ref{eq:Expansion_first}. While, in this case, we set the gauge field to be a combination of an anti-instanton (\ref{eq:Instanton_gauge}) and a double instanton (\ref{eq:Diinstanton}):
\begin{align}
    &A^{\alpha}_{\mu}= - \bar{\eta}^{\alpha}_{\mu \nu} \frac{x_{\nu}}{x^2 +1} \, , \label{eq:Single_instanton_explicit} \\ 
     &A^{\hat{i}}_{\mu} =  -\frac{(x-a)^2 (x-b)^2}{(x-a)^2 (x-b)^2 + \rho^2 [(x-a)^2 + (x-b)^2]} \eta^{\hat{i}}_{\mu \nu} \Bigg( \frac{\rho^2 (x-a)_{\nu}}{(x-a)^4} + \frac{\rho^2 (x-b)_{\nu}}{(x-b)^4}
     \Bigg). 
    \label{eq:Double_instanton_explicit}
\end{align}
This corresponds to Milnor's bundle with transition maps with winding numbers $(2,-1)$, which is an exotic sphere. To make the construction fully explicit, we should provide the matrix $\mathchorus{K}_{\,\, \hat{i}}{}^{\alpha}$ in some coordinates. We do that by choosing the Euler angles, with the non-standard conventions explained in Section \ref{sec:Invariant_objects_appendix}, yields:
\begin{align}
    \mathchorus{K}_{\,\, \hat{i}' }{}^{\alpha'} = \left(\begin{array}{ccc}
    \vspace{0.1cm}
       \cos\alpha \cos \beta \cos \gamma - \sin \alpha \sin \gamma \,\, & \,\, -\cos \beta \cos \gamma \sin \alpha - \cos \alpha \sin \gamma \,\,  &  \,\, \cos \gamma \sin \beta \\ \vspace{0.1cm}
       \cos \gamma \sin \alpha + \cos \alpha \cos \beta \sin \gamma \,\,  & \,\, \cos \alpha \cos \gamma - \cos \beta \sin \alpha \sin \gamma \,\, & \,\, \sin \beta \sin \gamma \\
       \cos \alpha \sin \beta & \sin \alpha \sin \beta & -\cos \beta 
    \end{array} \right)_{\hat{i}'}^{\,\,\,\,\, \alpha'}  
\end{align}
Now, let us make a few remarks about the metric just presented. With the choices outlined above, this is the simplest bundle Riemannian metric on the Gromoll-Meyer sphere, in that it is written as a slight generalisation of the metric in \cite{DUFF19861}. However, this is far from being the most general bundle metric on such exotic sphere. We can generalise three ingredients in our construction as follows. Firstly, we can make a different choice for the metric on the base $S^4$. Secondly, we can consider a different metric on the fibre $S^3$, as long as it is left-invariant (see \cite{10.1063/1.525753}). And, finally, we can employ a wider class of gauge fields then the (anti)-self-dual ones. We restricted to instanton ansatzes for the $|k|=1$ and $|k|=2$ components of the $\mathrm{so(4)}$ connection, but this assumption can be relaxed. However, as described in the next section, the choice of instanton connections does play a role when the exotic geometry appears in a physical theory.

\subsection{Traces of Exoticness in Four Dimensions}
\label{subsec:Exotic_physics_chap3}
As we just argued, by setting $M=S^4$, $G=SO(4)$ and $H=SO(3)$, we obtain exactly Milnor's bundles. The family of their total spaces contains the standard seven-sphere and many exotic ones. In this section, we present an instance of how two geometries associated to inequivalent differentiable structures might appear as solutions to a physical theory. To do so, we simply follow Kaluza--Klein's prescription, in our non-abelian setting. We obtain a four-dimensional theory by substituting the metric ansatz \ref{eq:General_metric} into the action of seven-dimensional gravity, with cosmological constant, and integrating over the fibre. To avoid complications, we consider the case where $\bar{h}_{\alpha \beta}$ has no $x$-dependence, as it is done in \cite{DUFF198490}, for instance. Solutions for the more general case, which includes the $x$-dependence, are currently being studied, together with their higher-dimensional interpretations in terms of D-branes. \\
The dimensional reduction outlined above yields:
\begin{align}
\begin{aligned}
\int_E d^4 x \, d^3 y  \, \bar{g}^{1 / 2} ( \bar{R} - 2 \Lambda) = V_{G/H} \int_M d^4 x \sqrt{g}  \Big(  R_M  - \frac{1}{4} g^{\mu \nu} g^{\rho \sigma} \lambda_{i j} F_{\mu \rho}{ }^i F_{\nu \sigma}{ }^j  + R_{\, G/H} -2 \Lambda \Big).
 \label{eq:Action_reduced}
 \end{aligned}
\end{align}
This result can be read off from the one in \cite{10.1063/1.525753}, by ignoring the terms coming from the $x$-dependence of the fibre metric, and we refer to section VIII therein for the intermediate steps in the computation of the reduced Ricci scalar. Using the same notation, we have defined $R_M$, $R_{G/H}$ and $V_{G/H}$ as the Ricci scalar of the base, the Ricci scalar of the fibre and the volume of the fibre, respectively. Regarding the definition of $\lambda_{ij}$, we take a small detour. It is given by:
\begin{align}
    \lambda_{ij} = \frac{1}{V_{G/H}} \int_{G/H} d^3 y \,[\bar{h}]^{1/2} \,\, \bar{h}_{\alpha \beta} K_{i}{}^{\alpha}{}(y) K_{j}{}^{\beta}{} (y) = \frac{1}{V_{G/H}} \int_{G/H} d^3 y \,[\bar{h}]^{1/2} \,\, K_{i}{}^{\alpha}{}(y) K_{j}{}^{\alpha}{} (y),
\end{align}
where $ V_{G/H} = \int_{G/H} d^3 y \,[\bar{h}]^{1/2}$. To avoid a proliferation of factors of two, we have chosen $\bar{h}_{\alpha \beta} = \delta_{\alpha \beta}$, differently from what was earlier in this section.\footnote{Note that this is consistent with scaling the generators by a factor of $1/2$ compared to the previous discussions. Schematically, if $M,N \xrightarrow{} 1/2 M, 1/2 N$, then we have that $\mathchorus{K}_{\,\,i} \xrightarrow{} 1/2 \mathchorus{K}_{\,\,i}$, while $\mathchorus{K}_{\,\, \hat{i}}{}^{\alpha}$ stays the same. Hence, the dual forms to the $\mathchorus{K}_{\,\, \alpha}$'s are scaled by a factor of two, and so are the gauge fields $A^{\alpha}$, giving the standard expression for the anti-instanton.}
As we pointed out (see comments after equation \ref{eq:K_matrix_commutator}), when $i = \gamma$, we have that $K_{\gamma}{}^{\alpha}{} = \delta_{\gamma}{}^{\alpha}{} $, and hence $\lambda_{\alpha \beta} = \delta_{\alpha \beta}$.  For the other cases, we need to examine $K_{\hat{i}}{}^{\alpha}{}$. They are the coefficients for the change of basis between the right-invariant and the left-invariant vector fields on $S^3$. To make the symmetries manifest, and make simplifications easier, we again think of $S^3$ as embedded in $\mathbb{R}^4 = \{ (X,Y,Z,W) | X,Y,Z,W \in \mathbb{R} \}$. Then, we have that (see \ref{eq:K_in_left_right_basis}):
\begin{align}
    \begin{aligned}
        K_1 = (-W^2 - X^2 +Y^2 + Z^2 , \, -2XY -2WZ , \, 2WY -2XZ) , \\
         K_2 = (-2XY + 2WZ, \, -W^2 + X^2 - Y^2 + Z^2, \, -2WX -2YZ) , \\
         K_3 = (-2WY -2XZ, \, 2WX - 2YZ, \, -W^2 + X^2 + Y^2 - Z^2 ).
    \end{aligned}
\end{align}
As we mentioned, they are all different realisations of the Hopf map from $S^3$ to $S^2$, which is curious. We see immediately that the integral of any single component of $K_{\hat{i}}{}^{\alpha}{}$ over the three-sphere vanishes, due to symmetry. This implies $\lambda_{\alpha \hat{i}} = 0$. For the same reason, $\lambda_{\hat{i} \hat{j}} = 0$ for $\hat{i} \neq \hat{j}$. The only case where we get a non-zero integral is when $\hat{i} = \hat{j}$, where we have that $K_{\hat{i}}{}^{\alpha}{} K_{\hat{i}}{}^{\alpha}{} = 1$ (no sum over $\hat{i}$), so that $\lambda_{\hat{i} \hat{i}}=1$. \\
Given that $\lambda_{ij} = \delta_{ij}$, the dimensional reduction yields Einstein--Yang--Mills action (with cosmological constant given by the Ricci scalar of the fibre), which agrees with the analogous cases examined in \cite{DUFF198490} and \cite{10.1063/1.522434}.
Hence, the dynamics of the theory is described by the standard equations of this system. Varying with respect to the metric yields: 
\begin{align}
    R_{\mu \nu} - \frac{1}{2} g_{\mu \nu} R - \frac{1}{2} g_{\mu \nu} R_{S^3} + g_{\mu \nu} \Lambda =  \frac{1}{2} \lambda_{ij} \Big[ g^{\rho \sigma} F_{\mu \rho}^i F_{\nu \sigma}^j - \frac{1}{4} g_{\mu \nu} F_{\rho \sigma}^i F^{\rho\sigma \, j} \Big] ,
    \label{eq:EoM_metric}
\end{align} 
while we do not perform the explicit variation with respect to the gauge fields because we take the Bogmonly' bound shortcut.
Firstly, we let $g_{\mu \nu}$ be the metric on $S^4$ with radius $R$, which in stereographic coordinates reads:
\begin{align}
    g_{\mu \nu} = \frac{4 R^4}{(R^2 + x^2) ^2} \delta_{\mu \nu},
\label{eq:Metric_on_S4_stereo}
\end{align}
with the determinant being $g = \big(\frac{4 R^4}{(R^2 + x^2) ^2} \big)^4$. Ricci curvature tensor, scalar curvature and Einstein tensor read:
\begin{align}
    R_{\mu \nu} = \frac{12 R^2}{(R^2 + x^2) ^2} \delta_{\mu \nu} \, ,  \quad R_{S^4} = \frac{12}{R^2} \, \, \, \implies \, \, \, G_{\mu \nu} = - \frac{12 R^2}{(R^2 + x^2) ^2} \delta_{\mu \nu} = - \frac{3}{R^2} g_{\mu \nu} \, ,
    \label{eq:Ricci_and_Einstein}
\end{align}
respectively. Then, as anticipated, we consider the self-duality equation:
\begin{align}
    F_{\mu \nu} = \frac{1}{2 \sqrt{g}} \epsilon^{\rho \sigma \tau \omega }  g_{\mu \rho} g_{\nu \sigma} F_{\tau \omega},
    \label{eq:Self_duality_curved}
\end{align}
where $\epsilon^{\rho \sigma \tau \omega } $ is the Levi-Civita symbol. The overall scaling factor in \ref{eq:Metric_on_S4_stereo} of the two contracted metrics cancels with the square root of the inverse determinant, so that \ref{eq:Self_duality_curved} reduces to the standard self-duality in $\mathbb{R}^4$, whose solutions are the well known instantons.
Hence, we can set the gauge field to be an anti-instanton:
\begin{align}
    \left( \begin{array}{c}
    A^{\alpha}_\mu \\
    A^{\hat{i}}_\mu
    \end{array} \right)
    = \left(\begin{array}{c}
    2 \bar{\eta}^{\alpha}_{\mu \nu} \frac{ x^\nu}{x^2 + R^2} \\
    0
    \end{array} \right).
\end{align}
This choice recovers the seven-dimensional ordinary sphere. Correspondingly, the choice associated to the Gromoll-Meyer sphere reads:
 \begin{align}
    \left( \begin{array}{c}
    A^{\alpha}_\mu \\
    A^{\hat{i}}_\mu
    \end{array} \right)
    =\left(\begin{array}{c}
    \mathrm{equation} \,\, \ref{eq:Single_instanton_explicit} \\
\mathrm{equation} \,\, \ref{eq:Double_instanton_explicit}
    \end{array} \right) .
\end{align}
By the usual argument, since these gauge fields satisfy the self-duality equation \ref{eq:Self_duality_curved}, then they automatically satisfy their equations of motion (see \cite{Oh:2011nv}, for example). Moreover, they give a vanishing energy-momentum tensor.
Hence, for both choices, the right-hand side of equation \ref{eq:EoM_metric} vanishes. Then, by using \ref{eq:Ricci_and_Einstein}, we find that $\Lambda = \frac{3}{R^2} + \frac{R_{S^3}}{2} $, which completes our solution.  \\
As mentioned above, we leave the discussion of the higher-dimensional field equations for a forthcoming article. However, let us make just a quick comment about the consistency of this dimensional reduction. The $y$-dependence in the 4-d equations of motion, which is the source of inconsistency pointed out in \cite{DUFF198490}, does not affect the solution. The reason for this is our choice of a four-dimensional Einstein space and an instanton gauge field, which ensures that both sides of equation \ref{eq:EoM_metric} vanish (this should be compared with equation (3) in \cite{DUFF198490}). Finally, we also note that the above article contains an argument for the existence of a consistent ansatz when the fibre is itself a non-abelian group manifold, which holds in our case due to the diffeomorphism $S^3 \simeq \mathrm{SU(2)}$.

\section{Interlude}
\label{sec:Interlude_chap3}

\subsection{Summary of Results so Far}
\label{subsec:Summary_of_results_chap3}
In the first part of this chapter, we discussed the construction of a metric on the Gromoll--Meyer sphere (one of Milnor's bundles) following an approach inspired by Kaluza--Klein theories. We begun by introducing the notion of differentiable structure and of exotic manifold. Then, we moved to the description of fibre bundles and Kaluza--Klein geometry, through a series of formalisms and examples of increasing complexity. We first reviewed Kaluza and Klein's original proposal for space-time dimensional reduction, and its interpretation as a mathematical prescription for constructing Riemannian metrics on the total space of abelian fibre bundles. The latter formalism was exemplified with the case of the Hopf fibration. We accompanied this example with a discussion on lens spaces, the lower-dimensional analogues of exotic spheres. \\
We then move to principal non-abelian bundles, and the correspondent framework of non-abelian Kaluza--Klein theory. After providing both the physical and the mathematical pictures, we presented an incarnation of such a machinery through the application to $S^7$ viewed as a quaternionic Hopf fibration $S^3 \hookrightarrow S^7 \xrightarrow{} S^4$. As discussed in the previous chapter, this involved the use of an instanton as a $\mathrm{SU(2)}$-valued gauge field specifying the connection on the bundle. Finally, we moved to the case of non-abelian non-principal bundles, by reviewing the formalism of \cite{10.1063/1.525753} for bundles with homogeneous fibre, which specifies the explicit expression of the most general metric on the total space compatible with the bundle structure. In the trivial limit of the fibre being diffeomorphic to the structure group, the formalism recovers the usual non-abelian Kaluza--Klein results (valid for principal bundles). After explicitly describing the construction of exotic spheres as total spaces of bundles with homogeneous fibres, we applied the formalism to Milnor's bundle characterised by the pair of winding numbers $(2,-1)$, which carries an inequivalent differentiable structure. In this case, we find that the associated metric is a straightforward modification of the round metric on $S^7$. This time, in addition to a BPST instanton, a ``double instanton'' appears as a second $\mathrm{SU(2)}$ connection contributing to the construction. Simplicity is the key feature of the metric expression that we derived, since it does not have any special properties (it is not Einstein, for instance). By simplicity, we mean that it arises naturally in the context of Kaluza--Klein formalism and its expression in coordinates is remarkably similar to the round metric on the ordinary $S^7$. Finding a metric with special properties in the space of all metrics compatible with the bundle structure is the subject of the second part of this chapter. After deriving the explicit form of the metric on the Gromoll--Meyer sphere, we studied the dimensional reduction of gravity (with cosmological constant) for manifolds belonging to the family of Milnor's bundles. Via a non-abelian Kaluza--Klein mechanism, with integration over the $S^3$ fibre, we obtained four-dimensional Einstein--Yang--Mills theory. We found explicit solutions for both manifolds of the exotic pair considered, which differ by the winding numbers of the instantons involved, showing how inequivalent differentiable structures in seven-dimensions descend to different solutions in the four-dimensional theory. To explore the full range of possible solutions - not only to the reduced theory, but also to the higher-dimensional field equations - it is necessary to use explicit expressions for $k=2$ connections with all the moduli. More on this can be found in the next chapter. As we mentioned in the introduction, the results presented so far are only partially novel, in that similar ideas appeared briefly in \cite{FREUND1985263} and \cite{YAMAGISHI198447}. However, neither the approach described here, nor the derivation of the explicit expressions are something that we could find in the existing literature. \\
This brings us to some final comments on the first five sections of this chapter. The metric that we derived can be significant in the understanding of exotic spheres' geometries and decisive in the search for new supergravity solutions only if one has analytical control on the coordinate expression. The crucial ingredient that makes this a non-trivial task is the double-instanton, which manifests a smaller symmetry and larger moduli space compared to the standard BPST instanton. As we show in the second part of the chapter, the study of this object, through in a suitable quaternionic framework, is the key for advancing the understanding of Kaluza--Klein geometries on the Gromoll--Meyer sphere.

\subsection{Summary of the Geometric Picture}
\label{subsec:Summary_of_the_geometric_picture_chap3}
Let us summarise, very briefly, the formalism that has been developed so far, for associated $S^3$--bundles over $S^4$. For the Kaluza--Klein ansatz, the choice of metric on $S^4$ is completely unconstrained. The metric on $S^3$ should be left-invariant (we considered the bi-invariant one for simplicity). The (differential--)topological information on the bundle is encoded in the choice of connection/gauge field. Assuming that we restrict to objects which are either self-dual or anti-self-dual, then each bundle in the discrete classification, based on $h,l$, is in one-to-one correspondence with a choice of multi--instanton configuration. Of course, one should also take into account the redundancy due to the various orientation-reversing diffeomorphisms, summarised in \ref{subsec:Milnor_construction_chap3}. Concretely, the prescription is:
 \begin{align}
    t_{h,l} = u^h v u^l 
     \iff \left(\begin{array}{c}
    A^{\alpha}_{\mu} \,\,\mathrm{has} \,\, \mathrm{instanton} \,\, \mathrm{number} \,\, h \\
A^{\hat{i}}_{\mu} \,\,\mathrm{has} \,\, \mathrm{instanton} \,\, \mathrm{number} \,\, -l
    \end{array} \right) \, .
    \label{eq:Prescription}
\end{align}
Note that the rule above applies for the specific choices of coordinates and generators that were discussed in Chapter \ref{chap:2}, and employed throughout the previous sections.
We comment on this prescription again in the second part of the chapter, but with a formalism naturally associated to Milnor's construction: that of quaternions.

\section{Quaternions, instantons and spheres}
\label{sec:Quat_inst_and_spheres_chap3}
We now summarise the quaternionic notation that will be used throughout the rest of the chapter. We show how geometric quantities of physical interest can be recast in terms of quaternionic-valued objects by focussing on $\mathrm{SU(2)}$ (multi-)instantons on $S^4$ and the vielbein of $S^3$, $S^4$.\footnote{See Section \ref{sec:VielbeinFormalism} for details on the vielbein formalism, which will be extensively employed for the rest of this chapter.} These are key ingredients of the Kaluza--Klein construction presented in \cite{Gherardini:2023uyx}, but their representation provided here is more natural and efficient for computations. We end this section by briefly discussing Milnor bundles.

\subsection{'t Hooft notation vs. quaternions}
\label{subsec:'tHooft_notation_and_quaternions_chap3}
Although not frequently employed in the traditional physics literature, instantons admit a very elegant description in terms of quaternions and quaternionic-valued forms. In the following, we review the description of the well-known BPST instanton from \cite{BELAVIN197585, tHooft:1976snw} in terms of quaternions - see \cite{Atiyah1979}, for instance. The representation that will be used throughout the chapter is summarised by the following choice of basis:
\begin{align}
      \boldsymbol{e}_{\mu} =  ( I, - i \Vec{\sigma}) \, , \quad \bar{ \boldsymbol{e}}_{\mu} = ( I, i \Vec{\sigma}) \,  .
     \label{eq:Defn_e_0}
\end{align}
where $\Vec{\sigma}$ are the Pauli matrices and $I$ is the $2 \times 2$ identity matrix (according to Section \ref{sec:Notation_and_conventions_ch1}). In this section only, we use bold symbols to denote quaternionic objects, to avoid any confusion.  With these definitions, we have the isomorphism with quaternions given by the map:
\begin{align}
    \boldsymbol{e}_{0} \xrightarrow{} \boldsymbol{1} \, , \quad
     \boldsymbol{e}_{1} \xrightarrow{} \boldsymbol{i} \, , \quad   \boldsymbol{e}_{2} \xrightarrow{} \boldsymbol{j} \, , \quad   \boldsymbol{e}_{3} \xrightarrow{} \boldsymbol{k}   \, ,
\end{align}
and the definition of $\bar{\boldsymbol{e}}_{c}$ is consistent with quaternionic conjugation. Let us denote quaternions with $\HH$, unit quaternions with $\HH^* = \{\boldsymbol{x}\in\HH:|\boldsymbol{x}|=1\}$ and imaginary quaternions with $\HH' = \{\boldsymbol{x}\in\HH:\mathrm{Im}(\boldsymbol{x}) = \boldsymbol{x} \iff \mathrm{Re}(\boldsymbol{x}) = 0 \}$; as usual, $\mathrm{Re}(\boldsymbol{x}) = \frac{1}{2} (\boldsymbol{x} + \bar{\boldsymbol{x}})$, $\mathrm{Im}(\boldsymbol{x}) = \frac{1}{2} (\boldsymbol{x} - \bar{\boldsymbol{x}})$ and $|\boldsymbol{x}|^2=\boldsymbol{x} \bar{\boldsymbol{x}}$. Then, the standard isomorphisms read:
\begin{align}
    \HH \simeq \mathbb{R}^4 \, , \quad  \HH^* \simeq S^3 \simeq \mathrm{SU(2)} \, , \quad \HH'  \simeq \su(2) \, .
    \label{eq:HH_isomorphisms}
\end{align}
Accordingly, coordinates $x^{m}$ on $\RR^4$ can be organised into a quaternionic object as 
$ \boldsymbol{x} = x^{m} \boldsymbol{e}_m $. 
The exterior derivative is given by 
$\mathrm{d} \boldsymbol{x} = \dd x^{m} \boldsymbol{e}_m $, 
as expected.
Then, the expression for the usual $k=1$ instanton field strength in regular gauge is
\begin{align}
    \boldsymbol{F}={\lambda^2\over(\lambda^2+|\boldsymbol{x}-\boldsymbol{\xi}|^2)^2} \dd \boldsymbol{x} \wedge \dd\bar{\boldsymbol{x}} \, ,
    \label{eq:Quaternionic_F_k=1}
\end{align}
where $\boldsymbol{\xi}$ is a constant quaternion, containing the position moduli, and the wedge product is defined by antisymmetrisation of component 1-forms $\dd x^m$ and quaternionic multiplication. To show the equivalence of this expression to the usual one, it is sufficient to realise that $\boldsymbol{e}_{[m} \bar{\boldsymbol{e}}_{n]} = {1\over2} (\boldsymbol{e}_{m} \bar{\boldsymbol{e}}_{n} - \boldsymbol{e}_{n} \bar{\boldsymbol{e}}_{m})$ is selfdual, \ie, 
\begin{align}
\boldsymbol{e}_{[\mu} \bar{\boldsymbol{e}}_{\nu]}=\frac{1}{2} \epsilon_{\mu \nu \rho \sigma} \boldsymbol{e}_{[\rho} \bar{\boldsymbol{e}}_{\sigma]}\;.
\label{eq:selfduality}
\end{align}
 The components of the form in \eqref{eq:Quaternionic_F_k=1} read:
\begin{align}
     \boldsymbol{F}_{\mu \nu} = {2\lambda^2\over(\lambda^2+|\boldsymbol{x}-\boldsymbol{\xi}|^2)^2} \boldsymbol{e}_{[\mu} \bar{\boldsymbol{e}}_{\nu]}  \, ,
\end{align}
matching the standard expression for the $\mathrm{SU(2)}$ 1-instanton, up to re-labelling of indices, as we shortly discuss.\footnote{Tensors of the form $\boldsymbol{e}_{[m} \bar{\boldsymbol{e}}_{n]}$, which might differ by permutations and minus signs, are often denoted as $\sigma_{m n}$ in the literature - see for instance \cite{Vandoren:2008xg}.} A key point is that $\boldsymbol{e}_{[\mu} \bar{\boldsymbol{e}}_{\nu]}$ is an imaginary quaternion for any $\mu, \nu$ and, as depicted in \eqref{eq:HH_isomorphisms}, imaginary quaternions form a representation of $\su(2)$ - the fundamental; this can be seen directly from \eqref{eq:Defn_e_0}. When extracting the components of the field strength in the basis $\{ \boldsymbol{e}_i \}$, $i=1,2,3$, one does not find the standard 't Hooft symbols. This is because of our choice in \eqref{eq:Defn_e_0}, where the real part was (naturally) labelled as the zeroth component. Instead, one finds the \textit{reversed} 't Hooft symbols, which read:
\begin{align}
& ^o\eta_{i \mu \nu} =\epsilon_{i \mu \nu 0} -\delta_{i \mu} \delta_{\nu 0} + \delta_{i \nu} \delta_{\mu 0} \, , \nonumber \\
& ^o\bar{\eta}_{i \mu \nu}=\epsilon_{i \mu \nu 0} + \delta_{i \mu} \delta_{\nu 0} - \delta_{i \nu} \delta_{\mu 0} \, ,
\label{eq:'tHooft_zero}
\end{align}
where $^o\eta_{i \mu \nu}$ is selfdual and $^o\bar{\eta}_{i \mu \nu}$ is anti-selfdual. The reverse 't Hooft symbols differ by the standard ones by moving the zeroth component to the fourth position, which also exchanges selfdual with anti-selfdual. Hence, the component expression of the field strength for a $k=1$ instanton in our conventions reads
\begin{align}
    F^i_{\mu \nu} = -  {2\lambda^2\over(\lambda^2+|\boldsymbol{x}-\boldsymbol{\xi}|^2)^2}  \left. ^o\eta^i_{\mu \nu} \right.  \, .
\end{align}

The associated gauge field, obeying $\boldsymbol{F}=d\boldsymbol{A}+\boldsymbol{A}\wedge \boldsymbol{A}$, is given by 
\begin{align}
\boldsymbol{A}={\Im( (\boldsymbol{x} - \boldsymbol{\xi})\dd \bar{\boldsymbol{x}})\over({\lambda^2 + |\boldsymbol{x} - \boldsymbol{\xi}|^2})}
 \;.
\label{eq:RegularA_k=1}
\end{align}
To show that the component expression also matches the classic BPST instanton of \cite{BELAVIN197585, tHooft:1976snw}, up to relabelling, one shall use the identity $(\boldsymbol{x} \bar{\boldsymbol{y}})_i =- \, ^o\eta_{i \mu \nu} x^{\mu} y^{\nu}$, where $(\cdot)_i$ indicates the $i^{\mathrm{th}}$ component of the quaternion, to find:
\begin{align}
    A^i_{\mu} = - {1\over\lambda^2+|\boldsymbol{x}-\boldsymbol{\xi}|^2}  \left. ^o\eta^i_{\mu \nu} (x^\nu - \xi^\nu) \right. \, .
\end{align}
On the other hand, to go from \eqref{eq:RegularA_k=1} to \eqref{eq:Quaternionic_F_k=1}, it is convenient to use the relations
\begin{align}
    \begin{aligned}
& -2 \operatorname{Re} \dd\boldsymbol{x} \wedge \operatorname{Im} \dd\boldsymbol{x}-\operatorname{Im} \dd\boldsymbol{x} \wedge \operatorname{Im} \dd\boldsymbol{x}=\dd\boldsymbol{x} \wedge \dd\bar{\boldsymbol{x} } \, , \\
& -4 \operatorname{Re} \dd\boldsymbol{x} \wedge \operatorname{Im}\dd \boldsymbol{x}+\dd\bar{\boldsymbol{x}} \wedge \dd\boldsymbol{x}=\dd\boldsymbol{x} \wedge \dd\bar{\boldsymbol{x}} \, .
\end{aligned}
\label{eq:Quat_identity_1}
\end{align}
We collect all of these, and other useful formulae for quaternionic computations, in Section \ref{sec:Quaternionc_chapB}. Finally, let us clarify an important point. The conventions outlined in this section, which naturally follow from the quaternionic description of instantons, differ from those used up to this point by an \textit{orientation-reversing} change of coordinates, which switches the first coordinate with the last one. The fourth component was (implicitly) associated to the real part of a quaternion throughout the previous sections, while from now this role is assigned to the first component. As we mentioned, this exchanges the (anti-)self-duality properties of two-forms.

\subsection{Background geometry}
\label{subsec:BGSection_chap3}

All bundles we will consider are constructed with the round $S^4$ as base space and the round $S^3$ as fiber.
The radius of $S^3$ will always be 1. To encode the relative size of the spheres, we (sometimes) introduce a radius $r$ for $S^4$. Most calculations are performed for $r=1$, the results can then be scaled appropriately.

Let us briefly sketch how the ``background geometry'' $S^4\times S^3$ is dealt with in quaternionic language.

As mentioned, we view $S^3\simeq \mathrm{SU(2)}$ as the space of unit quaternions $\{y\in\HH:|y|=1\}$ (note that we have dropped the bold notation). We will not bother to divide $S^3$ in coordinate patches. 
The vielbein can be seen as a 1-form taking values in $\HH'$, the imaginary quaternions; it reads
\begin{align}
\varepsilon=\dd y\bar y=-y\dd\bar y\;,
\end{align}
and $\dd s^2=\Re(\varepsilon\otimes\bar\varepsilon)$.
It fulfills the Maurer--Cartan equation $\dd\varepsilon-\varepsilon\wedge\varepsilon=0$.
The spin connection, also an imaginary $1$-form (\ie, an $\su(2)$-valued 1-form), fulfills the vanishing torsion condition
$\dd\varepsilon+\omega\wedge\varepsilon+\varepsilon\wedge\omega=0$. We thus have 
\begin{align}
\omega=-{1\over2}\varepsilon=-{1\over2}\dd y\bar y\;.
\end{align}
The curvature is $r=\dd\omega+\omega\wedge\omega=-{1\over4}\varepsilon\wedge\varepsilon$, with components
$r_{ij}{}^k=-{1\over2}\epsilon_{ij}{}^k$. Translating the index $k$ to an antisymmetric pair according to 
``$v^{ij}=-2\epsilon^{ij}{}_kv^k$'' gives $r_{ij}{}^{kl}=2\delta_{ij}^{kl}$, appropriate for a sphere with radius 1.

The isometry $SO(4)\simeq(\mathrm{SU(2)}\times \mathrm{SU(2)})/\ZZ_2$ of $S^3$ is realised as left and right action with unit quaternions:
$y\mapsto uy\bar v$. Notice that the choice of $\varepsilon$ above amounts to choosing the right-invariant Maurer--Cartan forms. We might as well have chosen the left-invariant ones $\epsilon'=\bar y\dd y=-\dd\bar y y$. The translation between them by conjugation with $y$ will be the source of explicit $y$-dependence in the Kaluza--Klein construction.

The $S^4$ is described in two patches, each excluding one pole of $S^4$. For each patch, we note that $\RR^4\simeq\HH$ and we use a coordinate $x\in\HH$, with the overlap $x'=x^{-1}$ between the patches. 
The metric for radius 1 is encoded in $\dd s^2=\Re(E\otimes\bar E)$ with the vielbein
\begin{align}
E={2\dd x\over1+|x|^2}\;.
\end{align}
The local $\so(4)\simeq\su(2)\oplus\su(2)$ acts by left and right multiplication by elements in $\HH'$.
The vanishing torsion condition then reads
$\dd E+\Omega_L\wedge E+E\wedge\Omega_R=0$, which is solved by
\begin{align}
\Omega_L&={\Im(x\dd\bar x)\over1+|x|^2}\;,\nn\\
\Omega_R&={\Im(\bar x\dd x)\over1+|x|^2}\;.
\end{align}
One finds left (selfdual) and right (anti-selfdual) curvatures
\begin{align}
R_L&={\dd x\wedge\dd\bar x\over (1+|x[^2)^2}={1\over4}E\wedge\bar E\;,\nn\\
R_R&={\dd \bar x\wedge\dd x\over (1+|x[^2)^2}={1\over4}\bar E\wedge E\;.\label{leftrightS4Riemann}
\end{align}
Their sum, translated from $\HH'\oplus\HH'$ to antisymmetric pairs of indices, is in flat indices
$R_{ab}{}^{cd}=2\delta_{ab}^{cd}$. The left and right spin connections are connections on the 1-instanton and 1-anti-instanton bundles on $S^4$ (see Section \ref{sec:KKModuli_vs_Instanton_moduli_chap3}).

In order to derive the action of the $SO(5)$ isometry on $x$, we start from the homogeneous coordinates of $\HH P^1=S^4$:
\begin{align}
Z=\left(\begin{matrix}z_1\\z_2\end{matrix}\right)\in\HH^2
\end{align}
on which the isometry group acts linearly, $Z\mapsto MZ$. $M$ fulfills $MM^\dagger=I=M^\dagger M$, and is a group element in $U(2,\HH)\simeq USp(4)\simeq Spin(5)$:
\begin{align}
M=\left(\begin{matrix}a&b\\c&d\end{matrix}\right)\;,\quad |a|^2+|b|^2=1=|c|^2+|d|^2\;,\quad a\bar c+b\bar d=0\;.
\label{so5matrix}
\end{align}
Note that the relations imply $|a|^2=|d|^2$, $|b|^2=|c|^2$.
$S^4=\HH P^1$ is obtained from the homogeneous coordinates $Z$ as $\HH^2/\HH=\HH^2/(\mathrm{SU(2)}\times\RR_+)$, where the orbits are generated as $Z\mapsto Z\alpha$, $\alpha\in\HH$, which commutes with left multiplication by $M$.
In the patch where $z_2\neq0$, we can choose a representative 
\begin{align}
Z=\left(\begin{matrix}x\\1\end{matrix}\right)\;,
\end{align}
leading to a quaternionic M\"obius transformation
\begin{align}
\left(\begin{matrix}x\\1\end{matrix}\right)
\mapsto\left(\begin{matrix}ax+b\\cx+d\end{matrix}\right)
\approx\left(\begin{matrix}(ax+b)(cx+d)^{-1}\\1\end{matrix}\right)\;.
\end{align}
The linearly realised $SO(4)$ subgroup is described by diagonal matrices with $|a|=1=|d|$, and $x\mapsto ax\bar d$. While this transformation leaves the form of the round metric on $S^4$ invariant, the same is not true for the expressions of $k=1,2$ instantons. Hence, the $SO(5)$ isometries of the base act non-trivially on the moduli space of the instantons. This, in turn, implies that the moduli space of the instanton is not the moduli space of the Kaluza--Klein metric, since the action of the isometry group $SO(5)$ needs to be quotiented out. This point will be discussed in more detail in Section \ref{sec:KKModuli_vs_Instanton_moduli_chap3}, and it is a key observation in order to identify special points in the moduli space of instantons.

\subsection{Left and right bundles and exotic spheres}
\label{subsec:left_and_right_bundles_chap3}
$\mathrm{SU(2)}$ instantons on $S^4$ are characterised by the instanton number\footnote{There may be sign differences due to conventions across the literature, due to \eg\ definition of dualisation. In our conventions, selfdual solutions according to eq. \eqref{eq:selfduality} have positive instanton number.}
\begin{align}
k=-{1\over4\pi^2}\int_{S^4}\Re(F\wedge F)\;.\label{eq:InstantonNumber}
\end{align}
Evaluating this integral, we need to use the two patches of $S^4$. Let the gauge transformation on the overlap be $g$, so that 
$A'=g\dd g^{-1}+gAg^{-1}$, $F'=gFg^{-1}$. The integral can then be written as 
\begin{align}
k=&-{1\over4\pi^2}\int_{S^3}\Re(A\wedge\dd A+{2\over3}A\wedge A\wedge A)\nn\\
&+{1\over4\pi^2}\int_{S^3}\Re(A'\wedge\dd A'+{2\over3}A'\wedge A'\wedge A')\;,
\label{kCS}
\end{align}
where the coboundary $S^3$ is contained in both patches. A standard choice is the unit sphere, $|x|=1$. Using the gauge transformation,
this turns into
\begin{align}
k={1\over12\pi^2}\int_{S^3}\Re(g^{-1}\dd g)^3\;.
\end{align}
This is minus the winding number of $g$ on $S^3$. Take \eg\ $g=x$ ($|x|=1$).
We can think of $e=g^{-1}\dd g$ as a vielbein on $S^3$ of radius 1 (see Section \ref{subsec:BGSection_chap3}).
Then, $\Re(g^{-1}\dd g)^3=-\epsilon_{ijk}\dd x^\mu\wedge\dd x^\nu\wedge\dd x^\rho e_\mu{}^ie_\nu{}^je_\rho{}^k$, and
$*\Re(g^{-1}\dd g)^3=-6$. The integral becomes $k={1\over12\pi^2}\times(-6)\times2\pi^2=-1$. 

For a selfdual $F$, \eqref{eq:InstantonNumber} becomes $k={1\over8\pi^2}\int_{S^4}d^4x\sqrt gg^{\mu \rho}g^{\nu \sigma}F_{\mu \nu}{}^iF_{\rho \sigma}{}^i
=\int_{S^4}d^4x\sqrt g\II$. We refer to $\II={1\over8\pi^2}g^{\mu \rho}g^{\nu \sigma}F_{\mu \nu}{}^iF_{\rho \sigma}{}^i$ as the instanton scalar, and $\sqrt g\II$ as the instanton density.

When constructing $S^3$ bundles over $S^4$, there are two $\mathrm{SU(2)}$'s present, acting on the unit quaternion $y$ parametrising $S^3$ by left and right multiplication. Both can be twisted on the overlap as above, leading to instantons for each $\mathrm{SU(2)}$. 
These bundles are so called Milnor bundles \cite{10.2307/1969983}, with overlaps
\begin{align}
x'&=x^{-1}\;,\nn\\
y'&=e^{-m}ye^{n}\;,\label{Overlap}
\end{align}
where $y\in\HH$, $|y|=1$ parametrises $S^3$, $x\in\HH$ parametrises $\RR^4$, and $e={x\over|x|}$.
With this definition of the integer winding numbers, $m$ and $n$ coincide with the instanton numbers of the left and right $\mathrm{SU(2)}$, respectively. Note that this labelling differs from the original one employed by Milnor, where the two integers ($h=-m$ and $l=n$) correspond to the powers of the quaternions, and there is no minus sign involved.

There are a priori two copies of $SO(4)\simeq (\mathrm{SU(2)}\times \mathrm{SU(2)})/\ZZ_2$, where the $\mathrm{SU(2)}$'s act by left and right multiplication by unit quaternions on $x$ and $y$:
\begin{align}
x&\mapsto\alpha x\bar\beta\;,\nn\\
y&\mapsto\gamma y\bar\delta\;.
\end{align}
A selfdual 2-form with basis elements $\dd x\wedge \dd \bar x$ transforms only under $\mathrm{SU(2)}_\alpha$:
$\dd x\wedge \dd \bar x\mapsto\alpha \dd x\wedge \dd \bar x\bar\alpha$, and anti-selfdual $\dd \bar x\wedge \dd x$ under $\mathrm{SU(2)}_\beta$.
$m>0$ in eq. \eqref{Overlap}, and also $n>0$, corresponds to selfdual instantons, and $m,n<0$ to anti-selfdual instantons.
Conjugation of $x$ interchanges $\mathrm{SU(2)}_\alpha\leftrightarrow \mathrm{SU(2)}_\beta$, and conjugation of $y$ interchanges
$\mathrm{SU(2)}_\gamma\leftrightarrow \mathrm{SU(2)}_\delta$. 
Thus, from eq. \eqref{Overlap}, $x$-conjugation gives $(m,n)\mapsto(-m,-n)$, instantons are interchanged with anti-instantons,
while $y$-conjugation gives $(m,n)\mapsto(n,m)$. 

Milnor showed in \cite{10.2307/1969983}, via Morse theory, that when $-m + n = 1$, the total space of the  bundle is homeomorphic to the topological $7$-sphere; $-m + n = -1$ also guarantees the existence of a homeomorphism, by the same argument or just by realising that $m,n \xrightarrow{} -m,-n$ is an orientation reversing isomorphism of vector bundles (this is the $x$-conjugation mentioned above, see \cite{McEnroe2016MILNORSCO} for a detailed account). Moreover, when $(m+n)^2 \neq 1(\bmod 7)$, the total space cannot be diffeomorphic to the ordinary $S^7$ - which is obtained as $m=1$, $n=0$. Hence, when both conditions are met, we are in the presence of an exotic sphere.
The simplest such case is $m=2$, $n=1$; we will investigate its geometry in some detail.\footnote{These windings are opposite to those appearing in \cite{Gherardini:2023uyx}. This does not really make a difference, since the two choices are related simply by an $x$-conjugation, or change of orientation.}

The left- and right-twisted bundles with instanton numbers $(m,n)$ can be obtained from the (principal) $(S^3\times S^3)$-bundle with
\begin{align}
x'&=x^{-1}\;,\nn\\
y'&=e^{-m}y\;,\\
z'&=e^{-n}z\;\nn
\end{align}
by modding out $(y,z)\approx(y\bar\delta,z\bar\delta)$. Choose $(y,1)$ as a representative. Then,
$(y',z')=(e^{-m}y,e^{-n})\approx(e^{-m}ye^{n},1)$, and the Milnor bundle is obtained.

Let us start from a metric $ds^2=\Re(E\otimes\bar E+\varepsilon\otimes\bar\varepsilon+\varphi\otimes\bar\varphi)$, where $E$ is the quaternionic vielbein on $S^4$, and $\varepsilon$ and $\varphi$ are imaginary quaternionic vielbeins on the two $S^3$'s, with
\begin{align}
\varepsilon&=a(\dd y\bar y+A)\;,\nn\\
\varphi&=b(\dd z\bar z+B)\;.
\end{align}
$a$ and $b$ are the radii of the $S^3$'s, and $A$ and $B$ are $\mathrm{SU(2)}$ connections (\ie, $A=\dd x^\mu A_\mu{}^i(x)e_i$ etc.) with instanton numbers $m$ and $n$. It is then straightforward to calculate the distance between orbits. It is encoded in the new vielbein $\tilde\varepsilon$ on $S^3$ parametrised by $y$:
\begin{align}
\tilde\varepsilon={ab\over\sqrt{a^2+b^2}}(\dd y\bar y+A-yB\bar y)\;.
\end{align}
Taking $a=b=\sqrt2$ gives the vielbein we will use in the following, with unit radius for $S^3$.

\section{The Kaluza--Klein metric and its curvature}
\label{sec:Curvature_chap3}
In this section, we present the Kaluza--Klein metric in the quaternionic notation just outlined, by making the connection with component expressions of Section \ref{subsec:metric_on_Gromoll_Meyer_chap3} (and \cite{Gherardini:2023uyx}) explicit. Moreover, we comment on the role of some special diffeomorphisms of the total space, which will be considered in the next sections. Then, we perform the calculation of the Riemann tensor, Ricci curvature and Ricci scalar, finding an agreement with existing results in the literature.

\subsection{Metric and isometries}
\label{subsec:Isometries_chap3}
Let the metric on the total space of the bundle be $ds^2=E^a\otimes E^a+\varepsilon^i\otimes\varepsilon^i$ ($a=0\ldots3$, $i=1\ldots3$), where we write 
$\varepsilon=\varepsilon^ie_i$ as a 1-form in $\HH'$, the imaginary quaternions:
\begin{align}
E^a&=\dd x^\mu E_\mu{}^a \, , \nn \\
\varepsilon&=\epsilonzero+A-yB\bar y\;.
\label{eq:Vielbein_quaternions}
\end{align}
Following the notation of Section \ref{subsec:BGSection_chap3}, we use $E_\mu{}^a(x)$ to denote the vielbein on an $\RR^4$ patch of $S^4$, and $\epsilonzero$ for the vielbein on the round $S^3$ ($\epsilonzero = \dd y\bar{y}$, $|y|=1$). 
$A$ and $B$ are connections for the left and right $\mathrm{SU(2)}$ isometries on $S^3$.

Let us now briefly show the equivalence of this ansatz with the one of \cite{Gherardini:2023uyx}. The components of the metric therein are given by:
\begin{align}
    \left(\begin{array}{cc}
g_{\mu \nu}(x) + 4\delta_{i j} \mathchorus{K}_{\,\, I}{}^i(y) \mathchorus{K}_{\,\, J}{}^j(y) A_\mu^I(x) A_\nu^J(x)   \,\,\,
&  4A_\mu^j(x) 
 + 4A_\mu^{\hat{i}}(x) \mathchorus{K}_{\,\,\hat{i}}{}^j(y)  \\
 4A_{\nu}^{i}(x) + 4A_{\nu}^{\hat{i}}(x) \mathchorus{K}_{\,\,\hat{i}} {}^i(y) & 4\delta_{i j}
\end{array}\right),
\label{eq:Expansion_metric_1}
\end{align}
where we have re-labelled some indices to make the notation consistent with the choices above, so that $I = ( i, \hat{i} ) = 1,\dots, 6$ refer to the two $su(2)$ components of the Lie algebra $so(4)$; accordingly, $A_{m}^{i}$ are the components of $A$ and $A_m^{\hat{i}}$ are components of $B$. The factors of $4$ are simply due to an unconventional choice of the generators of $\mathrm{SU(2)}$ (see \cite{Gherardini:2023uyx} for more details), and the reader is referred to \cite{10.1063/1.525753} for a thorough derivation of the general ansatz. It is worth noting that, upon the choice a of bi-invariant metric on the fibre, which identifies the right- and left-invariant vector fields with the Killing vectors $\xi_{I}{}^{\tau}$ ($\tau$ being the curved index on $S^3$, with coordinates $z^{\tau}$), then one can re-write the ansatz as:
\begin{align}
    \mathrm{d} s^2=\left(E_{\mu}^{a} \mathrm{d} x^{\mu}\right)^2+\left(\varepsilon_\tau{}^i \mathrm{d} z^\tau- \varepsilon_{\tau}{}^i \, \xi_{I}{}^\tau A_{\mu}^{I} \mathrm{d} x^{\mu}\right)^2 \, .
    \label{eq:Classic_KK_ansatz}
\end{align}
This is a more common expression for the Kaluza--Klein ansatz within the physics literature (see \cite{DUFF198490, Bailin:1987jd, Salam:1981xd}, for instance, and Section \ref{sec:Bundle_adapted_vs_coordinate_adapted} for a discussion of coordinate adapted vs bundle-adapted bases). Before discussing its isometries, let us quickly return to \eqref{eq:Expansion_metric_1} to expose its equivalence with \eqref{eq:Vielbein_quaternions}. In \cite{Gherardini:2023uyx}, the unit $S^3$ was embedded in $\RR^4$ as $\{ (X, Y, Z, W ) : X^2 + Y^2 + Z^2 + W^2 = 1 \}$, which yields $\mathchorus{K}_{\,\, i }{}^{j} = \delta_{i }{}^{j}$ and
\begin{align}
    \mathchorus{K}_{\,\,\hat{i}}{}^{i} =
\left(
\begin{array}{ccc} \vspace{0.1 cm}
 1-2 \left(W^2+X^2\right) & -2 (W Z+X Y) & 2 W Y-2 X Z \\ \vspace{0.1cm}
 2 (W Z-X Y) & 1-2 \left(W^2+Y^2\right) & -2 (W X+Y Z) \\
 -2 (W Y+X Z) & 2 W X-2 Y Z & 2 \left(X^2+Y^2\right)-1 
\end{array}
\right)_{\hat{i}}^{\,\,\,\, i } .
\label{eq:K_in_left_right_basis}
\end{align}
With the identification $y=(-W,X,Y,Z)_c \, e_c$, one finds that $(y B \bar{y})^{i} = -A_\mu^{\hat{i}}(x) \mathchorus{K}_{\,\,\hat{i}}{}^i \mathrm{d}x^\mu$, which proves the equivalence. Note the efficiency of the quaternionic notation, where the whole matrix \eqref{eq:K_in_left_right_basis} is encoded by the conjugation by $y$ in eq. \eqref{eq:Vielbein_quaternions}.

Let us now turn our attention to the diffeomorphisms of the total space of the bundles that we are considering. Some of them are particularly relevant within the Kaluza--Klein construction: they are the isometries of the base and the base-dependent isometries of the fibre. We start by discussing the former in the specific context of our investigation. 

Isometries of the base play a role in the ``inverse'' construction that we are focussing on, where they determine one (or more) natural choice(s) of connection on the bundle. Concretely, the round metric on the base manifold $S^4$ is invariant under $SO(5)$ transformations, which were reviewed in Section \ref{subsec:BGSection_chap3}. These transformations, however, do not necessarily leave the gauge field unchanged. Hence, all of those gauge field configurations that are related by $SO(5)$ transformations should be identified for our purposes, since plugging them into the Kaluza--Klein ansatz just produces diffeomorphic metrics on the total space. As discussed in Section \ref{sec:KKModuli_vs_Instanton_moduli_chap3}, the case of $k=1$ instantons is special, in that there exist a choice of moduli which is fixed point of the $SO(5)$ action. This is therefore a reasonable choice for the connection on the bundle, which is always made in all the constructions of the round metric on $S^7$ treated as a quaternionic Hopf fibration. For the $k=2$ case, things are a bit more subtle, since there is no fixed point. This is also discussed in the next Section.

Isometries of the fibre, on the other hand, have been discussed thoroughly in the literature, and we just quickly review them here. To do that, it is convenient to consider the metric \eqref{eq:Classic_KK_ansatz}. Then, non-Abelian gauge transformations arise by considering the effect on the components $\bar{g}_{\mu \tau}$ of the infinitesimal isometry of the fibre metric $g_{\tau \omega }$, with $x$-dependent parameters:
\begin{align}
    z^{\tau} \rightarrow z^{\tau}+\xi_{I}^{\tau}(y) \theta^{I}(x) \, .
\end{align}
One finds that:
\begin{align}
    A_{\mu}^{I} \rightarrow A^{'}{}_{\mu}^{I}=A_{\mu}^{I}+\partial_{\mu} \theta^{I}+C_{IJK} \theta^{J} A_\mu^{K} \, 
\end{align}
where $C_{IJK}$ are the structure constants of the algebra associated to the isometries of the fibre, \ie, $\so(4)\simeq\su(2)\oplus\su(2)$ for us (since the fibre is $S^3$). Hence, base-dependent isometries of the fibre effectively implement gauge transformations on the connection of the bundle, as one would expect from the Kaluza--Klein ansatz.

\subsection{Bundle vielbein, connection and curvature}
\label{subsec:BundleGeometry_chap3}
We want to find the spin connections, and then the curvature, associated with $\eqref{eq:Vielbein_quaternions}$. 
Let us divide the $\so(7)$ spin connection in three parts, depending on the index structure, schematically
\begin{align}
\left(\begin{matrix}\Omega&-\nu^\transpose\\ \nu&\omega\end{matrix}\right)\;.
\end{align}
Let us begin with $\omega$, the $\so(3)$ spin connection on $S^3$. 
It is convenient to represent it as a 1-form in $\HH'$.
It is
\begin{align}
\omega=\omegazero+{1\over2}(A+yB\bar y)\;,
\end{align}
where $\omegazero$ is the connection on the round $S^3$ defined in Section \ref{subsec:BGSection_chap3}.
Note the different relative sign of $A$ and $B$ compared to $\varepsilon$.
It is then straightforward to verify that
\begin{align}
d\varepsilon+\omega \wedge \varepsilon + \varepsilon \wedge \omega
=F-yG\bar{y}\equiv\sF\;,
\end{align}
with $F=dA+A\wedge A$, $G=dB+B\wedge B$. This comes from an interplay between terms with different signs:
\begin{align}
d\varepsilon+\omega\wedge\varepsilon+\varepsilon\wedge\omega
&=d\epsilonzero+\omegazero\wedge\epsilonzero+\epsilonzero\wedge\omegazero\nn\\
&\qquad+dA-ydB\bar y-\dd y\wedge By+yB\wedge \dd \bar y\nn\\
&\qquad+\omegazero\wedge(A-yB\bar y)+(A-yB\bar y)\wedge\omegazero\nn\\
&\qquad+{1\over2}(A+yB\bar y)\wedge\epsilonzero+{1\over2}\epsilonzero\wedge(A+yB\bar y)\\
&\qquad+{1\over2}(A+yB\bar y)\wedge(A-yB\bar y)+{1\over2}(A-yB\bar y)\wedge(A+yB\bar y)\nn\\
&=dA+A\wedge A-y(dB+B\wedge B)\bar y\;,\nn
\end{align}
where we have used $\dd y=\epsilonzero y$, $\dd \bar y=-\bar y\epsilonzero$ on the second line and
$\omegazero=-{1\over2}\epsilonzero$ on the third line.
Similar statements relate the spin connection on $S^3$ to the gauge connections. Let $X(x),Y(x)\in\HH'$, and
let $Z=X-yY\bar y$, $\tilde Z=X+yY\bar y$. Then,
\begin{align}
D^{(\omega)}Z&=dZ+[\omega,Z] \label{DZeq}\\
&=D^{(A)}X-yD^{(B)}Y\bar y-{1\over2}[\varepsilon,X+yY\bar y]\\
&\equiv DZ-{1\over2}[\varepsilon,\tilde Z]\nn
\end{align}
 by a similar calculation.

The remaining parts of the spin connection are
\begin{align}
\nu_{ia}&={1\over2}\imath_a\sF^i\;,\nn\\
\Omega_{ab}&=\Omegazero_{ab}-{1\over2}\sF_{ab}{}^i\varepsilon^i\;,
\end{align}
where $dE^a+\Omegazero{}^a{}_b\wedge E^b=0$.

The corresponding three parts of the Riemann tensor, decomposed as
\begin{align}
\left(\begin{matrix}R&-\varrho^\transpose\\ \varrho&r\end{matrix}\right)\;,
\end{align}
are
\begin{align}
R&=d\Omega+\Omega\wedge\Omega-\nu^\transpose\wedge\nu\;,\nn\\
\varrho&=d\nu+\nu\wedge\Omega+\omega\wedge\nu=D^{(\Omega,\omega)}\nu\;,\\
r&=d\omega+\omega\wedge\omega-\nu\wedge\nu^\transpose\;.\nn
\end{align}

In the resulting expressions, it is good to keep all $\dd y$'s expressed by $\varepsilon$, in order to be able to read off the flat components of the Riemann tensor.  A good check is that the components obtained this way are gauge covariant.
\begin{align}
r&=-{1\over4}\varepsilon\wedge\varepsilon+{1\over2}\tilde\sF+{1\over4}\imath_a\sF\wedge\imath_a\sF\;.
\end{align}
where $\tilde\sF=F+yG\bar y$
(still expressed as a 2-form in $\HH'$). 

Expressing also $\varrho$ as a 2-form $\varrho_a$ in $\HH'$,
\begin{align}
\varrho_a={1\over2}(d\imath_a\sF+\Omega_{ab}\imath_b\sF+\omega\wedge\imath_a\sF+\imath_a\sF\wedge\omega)\;,
\end{align}
we can use eq. \eqref{DZeq} to get the result
\begin{align}
\varrho_a={1\over2}(D^{(\Omegazero,A)}\imath_aF-yD^{(\Omegazero,B)}\imath_aG\bar y)
-{1\over4}\sF_{ab}{}^j\varepsilon^j\wedge\imath_b\sF
-{1\over4}\varepsilon\wedge\imath_a\tilde\sF-{1\over4}\imath_a\tilde\sF\wedge\varepsilon\;.
\end{align} 
The final result is checked for the window (\yng(2,2)) symmetry, and is, in flat components:
\begin{align}
R_{ab,cd}&=\Rzero_{ab,cd}-{1\over2}\sF_{ab}{}^i\sF_{cd}{}^i+{1\over2}\sF_{a[c}{}^i\sF_{d]b}{}^i\;,\nn\\
R_{ab,ci}&=-{1\over2}D_c\sF_{ab}{}^i\;,\nn\\
R_{ab,ij}&=-\epsilon_{ijk}\tilde\sF_{ab}{}^k-{1\over2}\sF_{[a}{}^{ci}\sF_{b]c}{}^j\;,\nn\\
R_{ai,bj}&=-{1\over2}\epsilon_{ijk}\tilde\sF_{ab}{}^k+{1\over4}\sF_{a}{}^{cj}\sF_{bc}{}^i\;,
\label{RiemannTensor}\\
R_{ai,jk}&=0\;,\nn\\
R_{ij,kl}&=2\delta_{ij}^{kl}\;.\nn
\end{align}
The covariant derivative is with $\Omegazero$, $A$ and $B$ (and thus does not feel the $y$'s in $\sF$).
(We have reverted to the notation $R$ for all components.)

The Ricci tensor obtained from this Riemann tensor is
\begin{align}
R_{ab}&=\Rzero_{ab}-{1\over2}\sF_a{}^{ci}\sF_{bc}{}^i\;,\nn\\
R_{ai}&={1\over2}D^b\sF_{ab}{}^i=0\;, \label{eq:Ricci_general_components}
\\
R_{ij}&=2\delta_{ij}+{1\over4}\sF^{abi}\sF_{ab}{}^j\;.\nn
\end{align}

We will always keep the radius of $S^3$ to $1$. The relative size of $S^4$ and $S^3$ is encoded in the radius of $S^4$. 
Geometrical quantities are obtained by scaling the radius 1 results to radius $r$. 
Then, it is clear that \eg\ the part $R_{ab}$ of the Ricci tensor as well as $\sF_{ab}{}^i$, both with flat indices, scale as $r^{-2}$. 

We can check that the 1-instanton (see Section \ref{subsec:'tHooft_notation_and_quaternions_chap3} and next section) reproduces the round and squashed $S^7$. Let the $S^4$ have the round metric with radius $r$. Then, $\Rzero_{ab}={3\over r^2}\delta_{ab}$. Also, let $\sF=F$. A 1-instanton of unit size has 
\begin{align}
F={\dd x\wedge \dd \bar x\over(1+|x|^2)^2}={1\over4r^2}E\wedge\bar E\;,
\end{align}
so that $F_{ab}{}^i=-{1\over2r^2}\Re(e_a\bar e_be_i)$, which gives
$F_a{}^{ci}F_{bc}{}^i={3\over4r^4}\delta_{ab}$, $F^{abi}F_{ab}{}^j={1\over r^4}\delta_{ij}$.
The non-vanishing parts of the Ricci tensor are
\begin{align}
R_{ab}&=({3\over r^2}-{3\over8r^4})\delta_{ab}\;,\nn\\
R_{ij}&=(2+{1\over4r^4})\delta_{ij}\;.
\end{align}
The metric is Einstein for $r={1\over2}$ and $r={\sqrt5\over2}$, with $R_{AB}=k\delta_{AB}$, $k=6$ and ${54\over25}$ respectively. The former case is the round $S^7$ with radius 1, and the latter the squashed $S^7$. It can be checked that, in the case $r={1\over2}$, the expressions in eq. \eqref{RiemannTensor} give $R_{AB,CD}=2\delta_{AB}^{CD}$, where $A=(a,i)$, so the sectional curvature is identically 1.

\section{Instanton moduli and Kaluza--Klein moduli}
\label{sec:KKModuli_vs_Instanton_moduli_chap3}

In this section, we comment on the symmetries and moduli spaces of the various geometric quantities that appear in the metric ansatz just presented. This involves deriving the (regular) form of the $k=2$ instanton in our quaternionic language.
In what follows, we only consider bundles over a round $S^4$, which has isometry $SO(5)$.
This isometry will typically be broken by the presence of gauge connections $A$ and $B$. Additionally, the isometry of the round $S^3$ may be broken, partially or entirely.
Instanton solutions are parametrised by locus and size moduli (and relative $\mathrm{SU(2)}$ orientation moduli for $k>1$, which we do not consider); these, however, do not coincide with the moduli of the space of geometric solutions.
As we already mentioned, if an instanton solution breaks part of $SO(5)$, the corresponding generators will transform the solution to other solutions. Since the ``geometric'' or ``Kaluza--Klein'' moduli should be counted modulo diffeomorphisms, the action of $SO(5)$ should be divided out, yielding a parameter space which is much smaller than the instanton moduli space. All of this is described in details below, together with the discussion of special choices in the moduli space.

Finally, in addition to instanton moduli, we also introduce a geometric modulus in the form of the radius of the base $S^4$. 

\subsection{Instanton solutions and moduli}
\label{subsec:Intaton_and_moduli_chap3}

The $k$-instanton moduli space 
is the space of selfdual ($k>0$) or anti-selfdual ($k<0$) solutions with instanton number $k$. 
Note that the $\RR^4$ patches of $S^4$ are conformal to flat $\RR^4$. Dualisation of forms of degree ${d\over2}$ in $d$ dimensions only depends on the conformal class of the metric, so selfduality is the same on the round $S^4$ as on $\RR^4$. 
For instanton number $k>0$ the moduli space has dimension $8k-3$. It can be parametrised by $k$ loci, or ``centra'', $k$ (real) sizes and $k-1$ relative $\mathrm{SU(2)}$ orientations, in total $4k+k+3(k-1)=8k-3$. Unlike instantons on $\RR^4$, where the moduli space has dimension $8k$, the overall orientation is a gauge parameter. The orientations may be combined with the sizes in quaternionic parameters, whose modulus is the size and whose ``phase'' is the orientation.

The most general method for finding instanton solutions (in any gauge) is the ADHM construction \cite{ATIYAH1978185}.
A somewhat simpler method, which does not capture the orientation moduli, is the method of harmonic functions \cite{PhysRevD.15.1642}.
We will not consider orientation moduli, the presence of which alters solutions significantly, so this method is in principle sufficient. 
It however has the drawback that connections and field strengths are given in ``singular gauge''.
Mathematically speaking, a singular gauge is not good, specifically it involves singularities (for the gauge connection and field strength) in each patch. Roughly speaking, in our previous terminology, the expression for $F'$ is used in the patch containing $x=0$. If one calculates the instanton number as in eq. \eqref{kCS}, the Chern--Simons integral can instead be localised close to the singularities.
In order to arrive at a ``regular gauge'', a ``singular gauge transformation'' has to be applied.
Even if the singular gauge is mathematically unsound, the expressions involved turn out to be somewhat simpler than the regular ones. The actual (regular) field strength can then be encoded in a singular one, together with the transformation that removes the singularity. For more details on the latter, see the following Subsections.

The construction from a harmonic function is straightforward. Let $\phi(x)$ be a harmonic function on 
$\RR^4\backslash\{a_1,\ldots,a_k\}$ with flat metric. Then a connection
\begin{align}
A=-{1\over2}\*_\mu\log\phi\,\Im(\bar e_\mu\dd x)
\end{align}
has a selfdual field strength
\begin{align}
F=-{1\over8}\bar e_\mu\dd x\wedge\dd\bar x e_\nu(\*_\mu\*_\nu\log\phi+\*_\mu\log\phi\,\*_\nu\log\phi)\;.
\end{align}
When the calculation is performed using quaternions, the crucial identity is (with $\*=e_\mu\*_\mu$)
$\*\bar\*\log\phi+(\*\log\phi)(\bar\*\log\phi)=0$. 
For a $k$-instanton, the harmonic function can be taken as
\begin{align}
\phi=1+\sum_{i=1}^k{\lambda_i^2\over|x-a_i|^2}\;,
\end{align}
where $\lambda_i$ are size moduli and $a_i$ location moduli (all different). This captures $5k$ of the $8k-3$ moduli on $S^4$.
The solutions are singular at $x=a_i$.

\subsection{$k=1$}
\label{subsec:k1inst_chap3}

From the harmonic function $\phi=1+{\lambda^2\over|x-a|^2}$, we obtain the connection
\begin{align}
A={\lambda^2\Im(\bar x_a\dd x)\over|x_a|^2(\lambda^2+|x_a|^2)}\;,
\end{align} 
with $x_a=x-a$, and the field strength
\begin{align}
F={\lambda^2\bar x_a\dd x\wedge\dd\bar x x_a\over|x_a|^2(\lambda^2+|x_a|^2)^2}\;.
\end{align}
Clearly, the singularity at $x=a$ is an angular discontinuity in $F$ (but $A$, and hence the bundle metric, has a stronger singularity), which can be removed by a ``singular gauge transformation'' with parameter $g={x_a\over|x_a|}$. The regular connection and field strength are
\begin{align}
A'&=g\dd g^{-1}+gAg^{-1}={\Im(x_a\dd\bar x)\over\lambda^2+|x_a|^2}\;,\nn\\
F'&=gFg^{-1}={\lambda^2\dd x\wedge\dd\bar x\over(\lambda^2+|x_a|^2)^2}\; ,\label{k1solution}
\end{align}
as presented in Section \ref{subsec:'tHooft_notation_and_quaternions_chap3}.

How does the isometry $SO(5)$ act on the moduli of instantons? Consider the field strength $F'$ as above,
with $\lambda\in\RR$ size modulus and $\xi\in\HH$ location moduli.
Under $SO(5)$ as in eq. \eqref{so5matrix},
\begin{align}
x&\mapsto(ax+b)(cx+d)^{-1}\;,\label{MobiusEq}\\
\dd x&\mapsto|cx+d|^{-2}(a-b\bar x)\dd x(cx+d)^{-1}\;,\nn
\end{align}
where the second transformation is obtained after a short calculation using the conditions on the matrix $M$.
Thus,
\begin{align}
\dd x\wedge \dd \bar x\mapsto|cx+d|^{-6}(a-b\bar x)\dd x\wedge \dd \bar x\overline{(a-b\bar x)}
=|cx+d|^{-4}u\dd x\wedge \dd \bar x\bar u\;,\label{dxdxtransformation}
\end{align}
where $u={a-b\bar x\over|cx+d|}$, which is a unit quaternion. We work with solutions modulo gauge transformations, so $u$ can be discarded.
We also need the transformation of the function in front in eq. \eqref{k1solution}, which becomes
\begin{align}
{\lambda^2\over(\lambda^2+|x-\xi|^2)^2}
\mapsto|cx+d|^4{\lambda^2\over (\lambda^2|cx+d|^2+|(a-\xi c)x+b-\xi d|^2)^2}
\end{align}
Rewriting this as $|cx+d|^4{\lambda'^2\over(\lambda'^2+|x-\xi'|^2)^2}$ involves one non-trivial check, that the same result for $\lambda'$ is obtained in the denominator and in the overall factor, so that one stays in the same class of $2$-forms, eq. \eqref{k1solution}. The factors of $|cx+d|$ are cancelled against those in eq. \eqref{dxdxtransformation}. The result is
\begin{align}
\xi\mapsto\xi'&=-{(a-\xi c)^{-1}(b-\xi d)+{\lambda^2\bar cd\over|a-\xi c|^2}\over1+{\lambda^2|c|^2\over|a-\xi c|^2}}\;,\nn\\
\lambda\mapsto\lambda'&={\lambda\over|a-\xi c|^2+\lambda^2|c|^2}\;.
\end{align}

Both $\xi'$ and $\lambda'$ in general depend on both $\xi$ and $\lambda$.  
The size modulus is not a scalar, and the location moduli do not transform with a simple M\"obius transformation, but one modified by $\lambda$. 
Transformations with $b=c=0$ act as expected, $\xi\mapsto\bar a\xi d$, $\lambda\mapsto\lambda$.
An instanton centered at $x=0$ transforms to
\begin{align}
\xi'&=-{\bar a b+\lambda^2\bar c d\over |a|^2+\lambda^2|c|^2}\;,\\
\lambda'&={\lambda\over  |a|^2+\lambda^2|c|^2}\;.\nn
\end{align}
A size $1$ instanton at $\xi=0$ is invariant (it is like a ``constant function'', being completely delocalised).
For any size modulus, one may always use an isometry to move a 1-instanton to be centered at $x=0$.

These considerations were based on the transformation of the field strength. 
It is quite instructive to elaborate on the transformation of $\dd x$ by itself.
A little calculation yields the transformation property
\begin{align}
{\bar x \dd x\over1+|x|^2}\mapsto(cx+d){\bar x \dd x\over1+|x|^2}(cx+d)^{-1}+(cx+d)\dd(cx+d)^{-1}\;.
\end{align}
This explains more or less directly the appearance of a gauge transformation of the connection (the one discarded above).

It is informative to examine the transformations under an ``inversion'', with the matrix $A$ having $a=d=0$. It sends the origin to infinity (so one needs to use the other patch, with coordinate $y=x^{-1}$).
We can view such a transformation as the limit of an element
\begin{align}
A_\beta={1\over\sqrt{1+|\beta|^2}}\left(\begin{matrix}1&\beta\\\bar\beta&-1\end{matrix}\right)
\end{align}
as $\beta\rightarrow\infty$. Let us take $\beta\in\RR$. Then, according to eq. \eqref{MobiusEq}, $x\mapsto x^{-1}$ (which is the coordinate transformation to the other patch).
If we let $A_\beta$ act on the moduli parameters $\xi=0$ and $\lambda$, however, the result is
\begin{align}
\xi'&={(\lambda^2-1)\beta\over1+\lambda^2|\beta|^2}\;,\nn\\
\lambda'&={\lambda(1+|\beta|^2)\over1+\lambda^2|\beta|^2}\;,
\end{align}
and the limit $\beta\rightarrow\infty$ is well defined. It agrees with the field strength \eqref{k1solution} with $\xi=0$ transformed to the other patch with the appropriate gauge transformation reflecting $k=1$:
With $x'=x^{-1}$, $\dd x=-x'^{-1}\dd x'x'^{-1}$, and
\begin{align}
F={\lambda^2\over(\lambda^2+|x'|^{-2})^2}x'^{-1}\dd x'x'^{-1}\wedge \bar x'^{-1}\dd \bar x'\bar x'^{-1}
={\lambda^{-2}\over(\lambda^{-2}+|x'|^2)^2}{\bar x'\over|x'|}\dd x'\wedge \dd \bar x'{x'\over|x'|}\;.
\end{align}
which again  has $\xi=0$ but size $\lambda'={1\over\lambda}$. An instanton centered at infinity is also centered at $0$, but with the inverse size.

The instanton scalar (here calculated when the center is at $x=0$) is 
\begin{align}
\II={1\over 8\pi^2}{(1+|x|^2)^4\over16}{4\lambda^4\times12\over(\lambda^2+|x|^2)^4}
={3\lambda^4\over8\pi^2}\Bigl({1+|x|^2\over\lambda^2+|x|^2}\Bigr)^4 \, ,
\label{eq:Inst_dens_k=1}
\end{align}
where the middle factor is $1/\sqrt g$ and the number $12$ comes from 
$\Re(e_{[a}\bar e_{b]}e^i)\Re(e_{[a}\bar e_{b]}e^i)=12$, see below.
The integral is of course $\int_{S^4}d^4x\sqrt g\II=1$.
For $\lambda=1$, $\II$ is constant over $S^4$.

In conclusion, we can always choose the locus to $0$. Then, the Kaluza--Klein moduli space only contains the size parameter $\lambda$, and only $\lambda\leq1$ (or $\lambda\geq1$). 
In the construction of the exotic $S^7$, it will be taken to $\lambda=1$ when centered at $x=0$, which is the only solution that does not break $SO(5)$.

\subsection{$k=2$}
\label{subsec:k2_chap3}

Let $x_a=x-a$, $x_b=x-b$. The singular gauge connection for a $k=2$ instanton, obtained from the harmonic function
$f=1+{\lambda_a^2\over|x_a|^2}+{\lambda_b^2\over |x_b|^2}$, is
\begin{align}
A={1\over1+{\lambda_a^2\over|x_a|^2}+{\lambda_b^2\over |x_b|^2}}\left(
{\lambda_a^2\Im(\bar x_a\dd x)\over|x_a|^4}+{\lambda_b^2\Im(\bar x_b\dd x)\over|x_b|^4}\right) \;.
\label{SingularA}
\end{align}
The field strength $F=dA+A\wedge A$ is easiest calculated in singular gauge, and then transformed to the regular one.
Given the form of eq. \eqref{SingularA}, it is clear that it will involve factors $\Re\omega$ and $\Im\omega$, where
$\omega=\bar x\dd x$ (with $x$ replaced by $x_a$ or $x_b$). One then uses identities like
\begin{align}
\Re\omega\wedge\Im\omega&=-{1\over4}\omega\wedge\bar\omega+{1\over4}\bar\omega\wedge\omega\;,\nn\\
\Im\omega\wedge\Im\omega&=-{1\over2}\omega\wedge\bar\omega-{1\over2}\bar\omega\wedge\omega\; ,
\end{align}
which are rearrangements of \eqref{eq:Quat_identity_1}, to arrive at the result
\begin{align}
F&={1\over(1+{\lambda_a^2\over|x_a|^2}+{\lambda_b^2\over|x_b|^2})^2}	
\left[{\lambda_a^2\over|x_a|^6}(1+{\lambda_b^2\over|x_b|^2})\bar x_a\dd x\wedge \dd \bar xx_a\right.\nn\\
&\qquad\left.+{\lambda_b^2\over|x_b|^6}(1+{\lambda_a^2\over|x_a|^2})\bar x_b\dd x\wedge \dd \bar xx_b
-{\lambda_a^2\lambda_b^2\over|x_a|^4|x_b|^4}(\bar x_a\dd x\wedge \dd \bar xx_b+\bar x_b\dd x\wedge \dd \bar xx_a)\right]\;,\\
&={1\over(\lambda_b^2|x_a|^2+\lambda_a^2|x_b|^2+|x_a|^2|x_b|^2)^2}
\left[\lambda_a^2|x_b|^2(\lambda_b^2+|x_b|^2){\bar x_a\dd x\wedge \dd \bar xx_a\over|x_a|^2}\right.\nn\\
&\qquad\biggl.+\lambda_b^2|x_a|^2(\lambda_a^2+|x_a|^2){\bar x_b\dd x\wedge \dd \bar xx_b\over|x_b|^2}
-\lambda_a^2\lambda_b^2(\bar x_a\dd x\wedge \dd \bar xx_b+\bar x_b\dd x\wedge \dd \bar xx_a)\biggr]\;.\nn
\end{align}
This explicitly displays selfduality with respect to a metric conformal to the flat metric on $\RR^4$, since all terms contain the selfdual $\dd x\wedge \dd \bar x$. It is clear that the singularities of $F$ at $x=a$ and $x=b$ are angular discontinuities. It can be checked
that $F'=gFg^{-1}$ is regular.
Terms in the scalar curvature and Ricci tensor are conveniently calculated in the singular gauge. Even though this corresponds to using coordinates with coordinate singularities at $x=a$ and $x=b$, the terms appearing in $R_{ab}$ are not affected by the gauge/coordinate transformation. The expressions for $R_{ij}$ need to be transformed to regular gauge.

The angular discontinuities in $F$ may be removed by the following ``singular gauge transformation'' \cite{Giambiagi:1977yg}.
Let $z={x_a\over|x_a|^2}-{x_b\over|x_b|^2}=\bar x_a^{-1}-\bar x_b^{-1}$. 
Then, 
\begin{align}
dz&=-\bar x_a^{-1}\dd \bar x\bar x_a^{-1}+\bar x_b^{-1}\dd \bar x\bar x_b^{-1}
=-{x_a\dd \bar xx_a\over|x_a|^4}+{x_b\dd \bar xx_b\over|x_b|^4}\;,\nn\\
|z|^2&={|a-b|^2\over|x_a|^2|x_b|^2}\;.
\end{align}
We want to make a gauge transformation with $g={z\over|z|}$. Then,
$A'=gdg^{-1}+gAg^{-1}$. We have
\begin{align}
gdg^{-1}&={\Im(z\dd \bar z)\over|z]^2}
=-{|x_b|^2\over |a-b|^2}{\Im(\dd x\bar x_a)\over|x_a|^2}-{|x_a|^2\over |a-b|^2}{\Im(\dd x\bar x_b)\over|x_b|^2}\nn\\
&+{1\over|a-b|^2}{\Im(x_a\bar x_b\dd x\bar x_b)\over|x_b|^2}
+{1\over|a-b|^2}{\Im(x_b\bar x_a\dd x\bar x_a)\over|x_a|^2}\;.\label{gdginveq}
\end{align}
Note that the divergence in the first term behaves as $-{\Im(\dd x\bar x_a)\over|x_a|^2}$ when $x\rightarrow a$, and similarly for the second term around $x=b$. Note also that the third and fourth terms are finite but not continuous at
$x=b$ and $x=a$, respectively. 
The other term in $A'$ is
\begin{align}
&gAg^{-1}=|z|^{-2}zA\bar z	\nn\\
&\quad={|x_a|^2|x_b|^2\over|a-b|^2}{1\over1+{\lambda_a^2\over|x_a|^2}+{\lambda_b^2\over |x_b|^2}}
   (\bar x_a^{-1}-\bar x_b^{-1})
   \left({\lambda_a^2\Im(\bar x_a\dd x)\over|x_a|^4}+{\lambda_b^2\Im(\bar x_b\dd x)\over|x_b|^4}\right)
   (x_a^{-1}-x_b^{-1})\nn\\
&\quad={|x_a|^2|x_b|^2\over|a-b|^2}{1\over1+{\lambda_a^2\over|x_a|^2}+{\lambda_b^2\over |x_b|^2}}\\
&\qquad\times\left[\lambda_a^2\left(
{\Im(\dd x\bar x_a)\over|x_a|^6}-{\Im(\dd x\bar x_b)\over|x_a|^4|x_b|^2}
+{\Im(x_b\bar x_a\dd x\bar x_b)\over|x_a|^4|x_b|^4}-{\Im(x_b\bar x_a\dd x\bar x_a)\over|x_a|^6|x_b|^2}\right)\right.\nn\\
&\left.\qquad\quad+\lambda_b^2\left({\Im(\dd x\bar x_b)\over|x_b|^6}-{\Im(\dd x\bar x_a)\over|x_a|^2|x_b|^4}
+{\Im(x_a\bar x_b\dd x\bar x_a)\over|x_a|^4|x_b|^4}-{\Im(x_a\bar x_b\dd x\bar x_b)\over|x_a|^2|x_b|^6}
\right)\right]\;.\nn
\end{align}
Number the terms (1)-(8) according to the position in the last parenthesis.
Terms (2), (3), (6) and (7) are regular at $x=a$ and $x=b$. The terms (1) and (4) are singular at $x=a$ and (5) and (8) at $x=b$.
At $x\approx a$, the behaviour of the singular terms (1) and (4) is
\begin{align}
(gAg^{-1})_{(1)+(4)}\approx
{\Im(\dd x\bar x_a)\over|x_a|^2}-{\Im((a-b)\bar x_a\dd x\bar x_a)\over|a-b|^2|x_a|^2}\;,
\end{align}
which cancels the behaviour of the first and fourth terms in $gdg^{-1}$, eq. \eqref{gdginveq}. In the same way, terms 
(5) and (8) cancel the singular behaviour of the second and third terms in $gdg^{-1}$.

The result is regular. It could of course be rewritten in a manifestly regular way, but we have no need for that expression.
If we examine the behaviour of $A'$ as $|x|\rightarrow\infty$, we find that the leading term comes entirely from 
$gdg^{-1}$ and is
\begin{align}
A'={\Im(x\dd \bar x)\over|x|^2}+{\Im(x(\bar a-\bar b)x\dd \bar x(a-b)\bar x)\over|a-b|^2|x|^4}+O(|x|^{-2})\;.
\end{align}

If we choose a frame where $a-b$ is real, this leading term equals $hdh^{-1}$, where $h={x^2\over|x|^2}$,
displaying the correct winding.

Extending the calculation of the $SO(5)$ transformations of moduli for $k=2$ seems complicated.
There will certainly be no $SO(5)$ fixed points in the $k=2$ moduli space. However, one will clearly always be able to transform the centra to (\eg) $\pm\xi$, $\xi\in\RR$ (or some similar desired relation if the sizes are different), so that the remaining parameters are two sizes and one distance (again, disregarding internal orientation). 

We will restrict our attention to equal size parameters. What does this mean, given the lesson from $k=1$ that size parameters are not scalar? We short-circuit this question by defining the class of equal-size 2-instantons as the solutions that are obtained from those with centra $\pm a$ and equal size $\lambda$ by an $SO(5)$ transformation. Then we will have no need for the explicit form of the other solutions in the orbits under $SO(5)$.
The only transformation still needed to divide out is the inversion.

\begin{figure}
\begin{center}
\includegraphics[scale=.5]{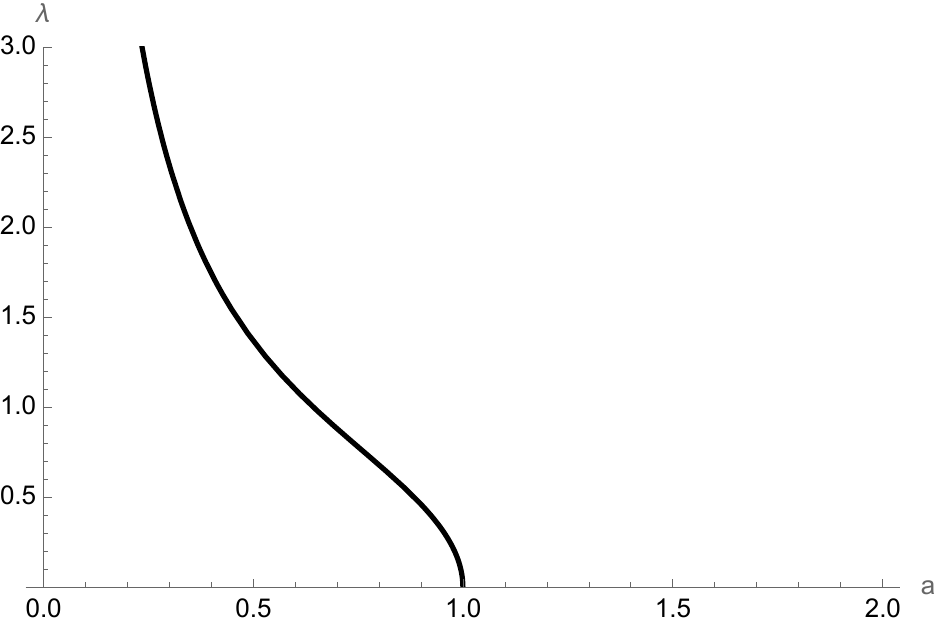}
\end{center}
\caption{The subspace of the geometric moduli space for equal size instantons. The curve separates the region of solutions with $\II(0)\geq\II(\infty)$ (left) from those with $\II(0)\leq\II(\infty)$ (right).}
\label{k2Figure}
\end{figure}

The field strength in a singular gauge is
\begin{align}
F&={\lambda^2\over(|x_+|^2|x_-|^2+\lambda^2|x_+|^2+\lambda^2|x_-|^2)^2}\nn\\
&\times\Bigl(|x_+|^2(\lambda^2+|x_+|^2){\bar x_-\dd x\wedge \dd \bar x x_-\over|x_-|^2}
+|x_-|^2(\lambda^2+|x_-|^2){\bar x_+\dd x\wedge \dd \bar x x_+\over|x_+|^2}\Bigr.    \label{k2SingularF}\\
&\qquad\Bigl.-\lambda^2(\bar x_+\dd x\wedge \dd \bar x x_-+\bar x_-\dd x\wedge \dd \bar x x_+)\Bigr)\;,\nn
\end{align}
where $x_\pm=x\pm a$. For convenience, we take $a\in\RR$ (by an $SO(4)$ rotation).
A clue about the behaviour under an inversion is obtained by looking at the prefactor, governed by the function 
\begin{align}
f_{a,\lambda}(x)=|x_+|^2|x_-|^2+\lambda^2|x_+|^2+\lambda^2|x_-|^2
\end{align}
appearing in the denominator. Under an inversion $x'=x^{-1}$, we have
\begin{align}
{1\over a^2|x|^2}f_{a,\lambda}(x)={1\over a'^2|x'|^2}f_{a',\lambda'}(x')\;,
\end{align} 
where
\begin{align}
a'&={1\over\sqrt{a^2+2\lambda^2}}\;,\nn\\
\lambda'&={\lambda\over a\sqrt{a^2+2\lambda^2}}\;.   \label{aLambdaTransf}
\end{align}
In principle, it remains to be checked that the full solution transforms like this, but it is the only possibility. 
Note that ${\lambda'\over a'}={\lambda\over a}$, so a solution with two ``well separated'' instantons remains well separated viewed from the antipode (but see below). Solutions with $a^2(a^2+2\lambda^2)=1$ are invariant under inversion.
In the geometric moduli space, it is sufficient to include sizes $0<\lambda^2\leq{1-a^4\over2a^2}$, to the left of the curve in Figure
\ref{k2Figure}.

We will now examine the behaviour of the instanton scalar $\II={1\over8\pi^2}g^{\mu \rho}g^{\nu \sigma}F_{\mu \nu}{}^iF_{\rho \sigma}^i$. 
Since, in the solution \eqref{k2SingularF}, $\dd x\wedge \dd \bar x$ is conjugated with different vectors in different terms, we need the more general identity
for arbitrary vectors $a,b,c,d$:
\begin{align}
\Re(\bar a e_{[a}\bar e_{b]}be^i)\Re(\bar ce_{[a}\bar e_{b]}de^i)=4\bigl[2(a\cdot c)(b\cdot d)+2(a\cdot d)(b\cdot c)-(a\cdot b)(c\cdot d)\bigr]\;,
\end{align}
where $(u\cdot v)=\Re(\bar u v)$ is the ordinary scalar product.
A partial result is
\begin{align}
&4\pi^2\sqrt g \II={8\lambda^4\over(|x_+|^2|x_-|^2+\lambda^2|x_+|^2+\lambda^2|x_-|^2)^4}\nn\\
&\quad\times\Bigl[3|x_+|^4(\lambda^2+|x_+|^2)^2+3|x_-|^4(\lambda^2+|x_-|^2)^2
	+4\lambda^2(2|x_+|^2|x_-|^2+(x_+\cdot x_-)^2)\nn\\
&\qquad+2(\lambda^2+|x_+|^2)(\lambda^2+|x_-|^2)(4(x_+\cdot x_-)^2-|x_+|^2|x_-|^2)        \label{PartialResult}\\
&\qquad-12\lambda^2|x_+|^2(\lambda^2+|x_+|^2)(x_+\cdot x_-)
              -12\lambda^2|x_-|^2(\lambda^2+|x_-|^2)(x_+\cdot x_-)
	\Bigr]\;.\nn
\end{align}
We then insert $|x_\pm|^2=|x|^2\pm2(a\cdot x)+a^2$, $(x_+\cdot x_-)=|x|^2-a^2$. Let us call the object within square brackets in eq. 
\eqref{PartialResult} $X_{a,\lambda}(x)$. Also, let $f_{a,\lambda}(x)=|x_+|^2|x_-|^2+\lambda^2|x_+|^2+\lambda^2|x_-|^2$. Then,
\begin{align}
X_{a,\lambda}(x)&=12|x|^8+16a^2|x|^6+8a^2(a^2-2\lambda^2)|x|^4\nn\\
&+16a^4(a^2+2\lambda^2)|x|^2+12a^4(a^2+2\lambda^2)^2       \label{NotSoPartialR}\\
&+\left(128|x|^4+64(6a^2+\lambda^2)|x|^2+128 a^2(a^2+2\lambda^2)\right)(a\cdot x)^2
+64(a\cdot x)^4\;, \nn
\end{align}
and
\begin{align}
f_{a,\lambda}(x)=|x|^4+2(a^2+\lambda^2)|x|^2+a^2(a^2+2\lambda^2)-4(a\cdot x)^2\;.
\label{eq:Ricci_equal_size_3}
\end{align}
It is then straightforward to verify the behaviour under inversion
\begin{align}
f_{a',\lambda'}(x^{-1})&=a^{-2}(a^2+2\lambda^2)^{-1}|x|^{-4}f_{a,\lambda}(x)\;,\nn\\
X_{a',\lambda'}(x^{-1})&=a^{-4}(a^2+2\lambda^2)^{-2}|x|^{-8}X_{a,\lambda}(x)\;.
\end{align}
Together with the second eq. in \eqref{aLambdaTransf} and $\sqrt g\mapsto |x|^8\sqrt g$, this shows that
$4\pi^2\II={8\lambda^4\over\sqrt g}{X_{a,\lambda}(x)\over f_{a,\lambda}(x)^4}$ is invariant under an inversion.
This is a good consistency check on the calculations leading to eq. \eqref{NotSoPartialR}.

We can now start to investigate the behaviour of $\II$ in (for example) the left region of Figure \ref{k2Figure}.
The values at $x=0$ and $x=\infty$ provide one interesting piece of  input:
\begin{align}
4\pi^2\II(0)&={6\lambda^4\over a^4(a^2+2\lambda^2)^2}\;,\nn\\
4\pi^2\II(\infty)&=6\lambda^4\;.
\end{align}
This gives the simple characterisation of the left half of the ``phase diagram'', Figure \ref{k2Figure}, that it consists of the solutions with
$\II(0)\geq\II(\infty)$.

If we consider $\II$ as a function of the two variables $|x|^2$ and $(a\cdot x)^2$, we note that $X_{a,\lambda}$ increases with increasing $(a\cdot x)^2$ for constant $|x|^2$, while $f_{a,\lambda}$ decreases. This implies that any local maximum must lie on the real line.

\begin{figure}
\begin{center}
\includegraphics[scale=.5]{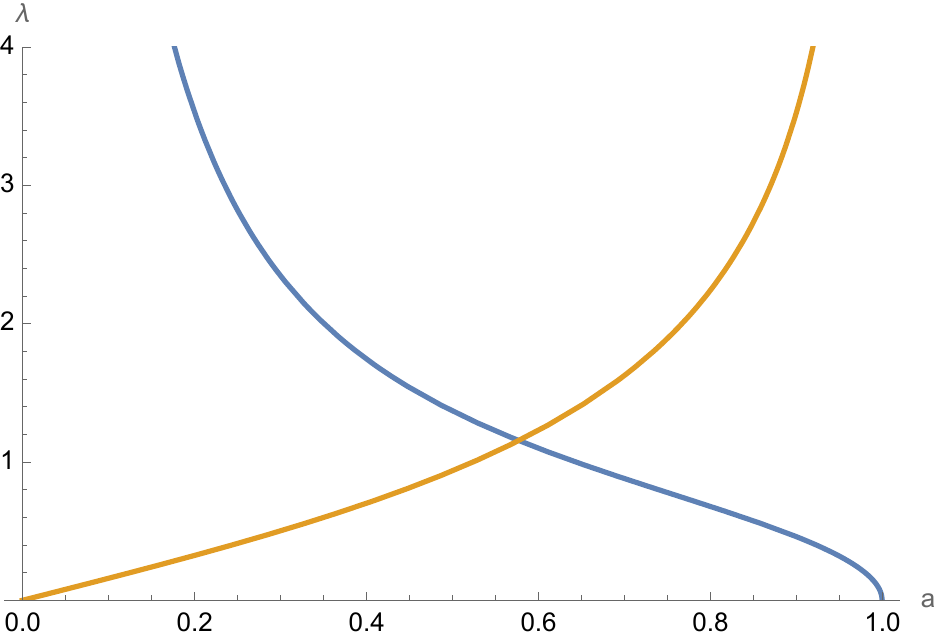}
\end{center}
\caption{The subspace of the geometric moduli space for equal size instantons, with the critical line for appearance/disappearance of twin peaks. The blue line is the same as the one in Figure \ref{k2Figure}, while the orange line divides solutions with a single peak from those that show two peaks. The intersection of the two lines occurs $a={1\over\sqrt3}$, $\lambda={2\over\sqrt3}$, which is the ``special'' point of the moduli space that we will focus on.  \label{k2Figure2}}
\end{figure}

It is straightforward to see that all partial derivatives $\*_m\II$ vanish at $x=0$.
We may ask if $x=0$ is a local maximum, minimum or a saddle point. 
It turns out that a second directional derivative orthogonal to $a$ is always (in the parameter region) negative.
The second directional derivative along  $a$ may be positive or negative.
We find it to be positive for small $\lambda$ (and small enough $a$) and negative for large $\lambda$, the critical point being
$\lambda^2={a^2(a^2+5)\over2(1-a^2)}$. This divides the region of parameter space in two parts, one where the size parameter is small, so the instantons are separated, yielding two peaks, one where the size is large enough relative to the separation, so there is only a single peak. This second critical line is included in Figure \ref{k2Figure2}. 
The two curves intersect in the ``special'' point $a={1\over\sqrt3}$, $\lambda={2\over\sqrt3}$. 
For these values of the moduli, $\II$ is constant along the great circle through the origin and $a$.
There is in fact an enhancement of isometry at this point, and it can on good grounds be considered the ``center'' of the $k=2$ moduli space.

\begin{figure}
\begin{center}
\includegraphics[scale=.22]{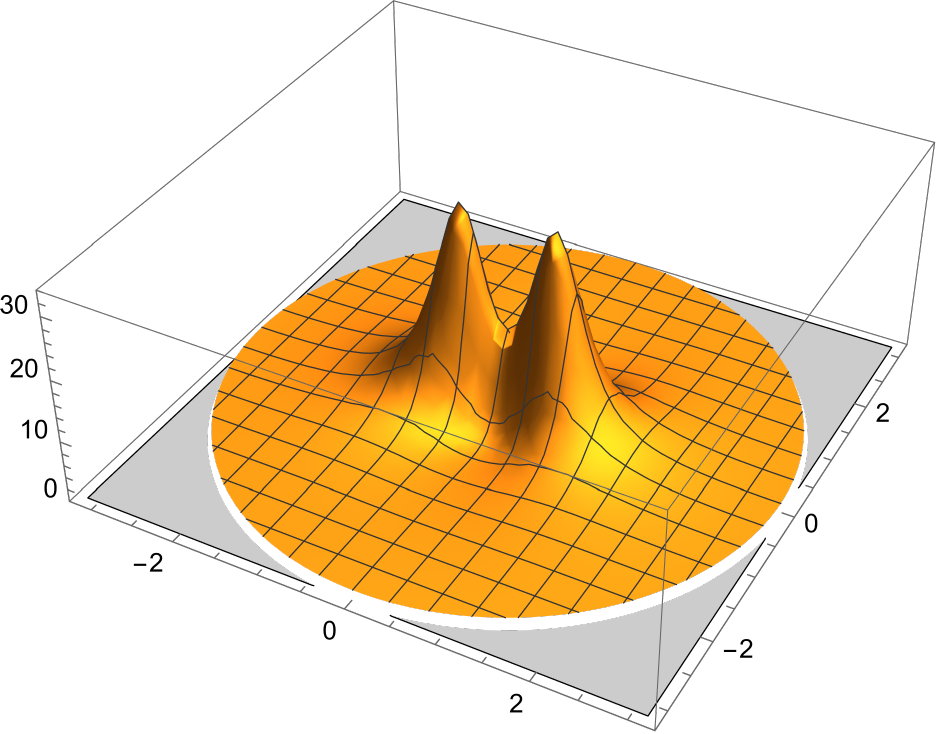}
\includegraphics[scale=.22]{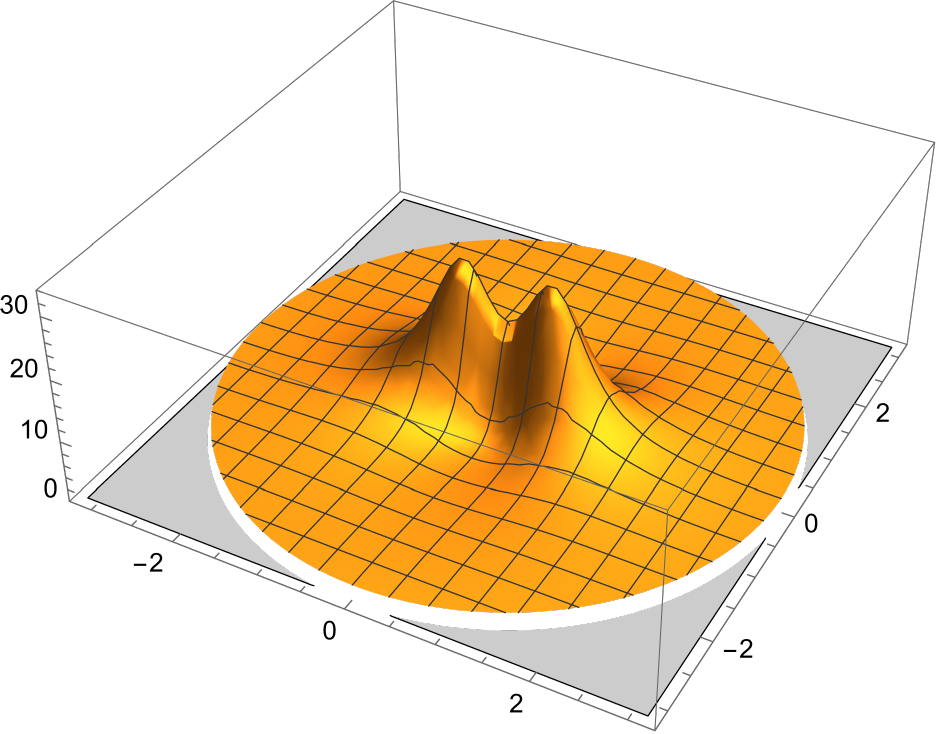}
\includegraphics[scale=.22]{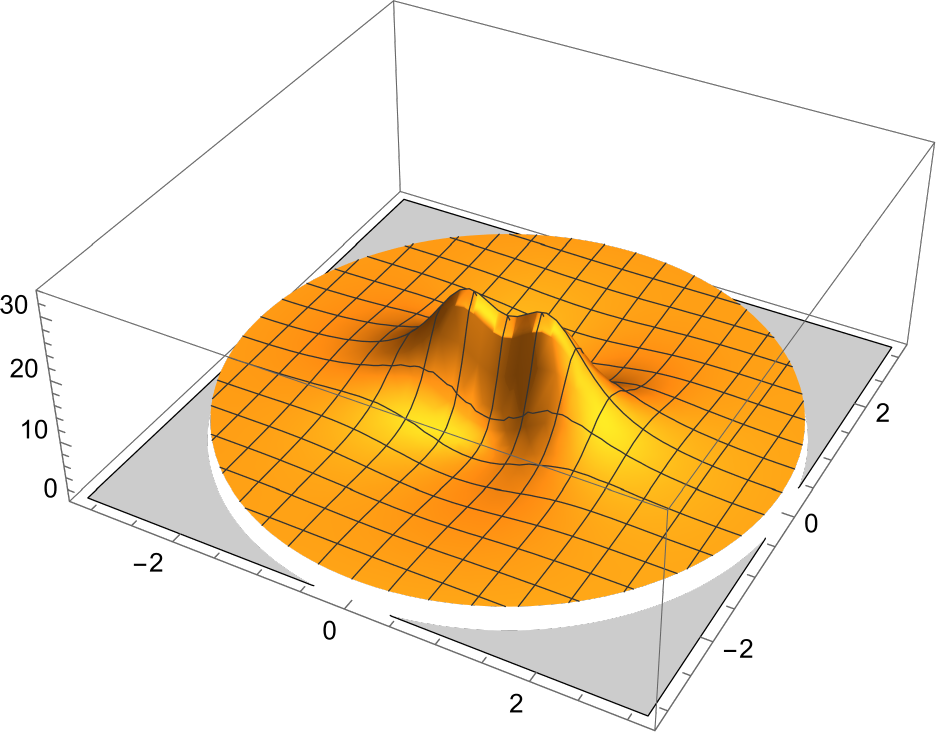}
\includegraphics[scale=.22]{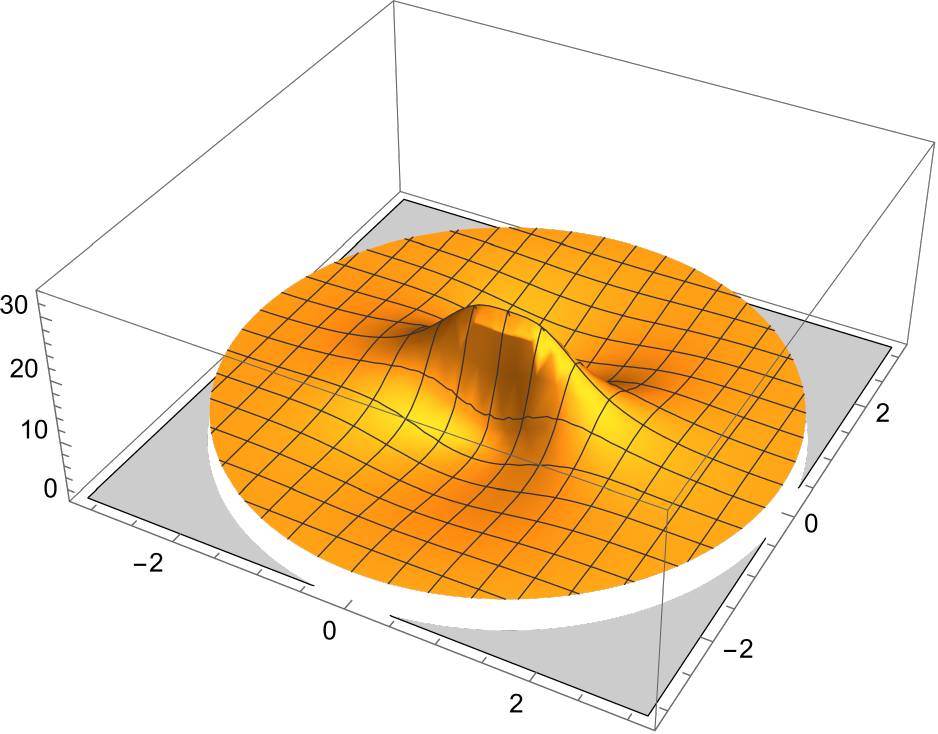}
\includegraphics[scale=.22]{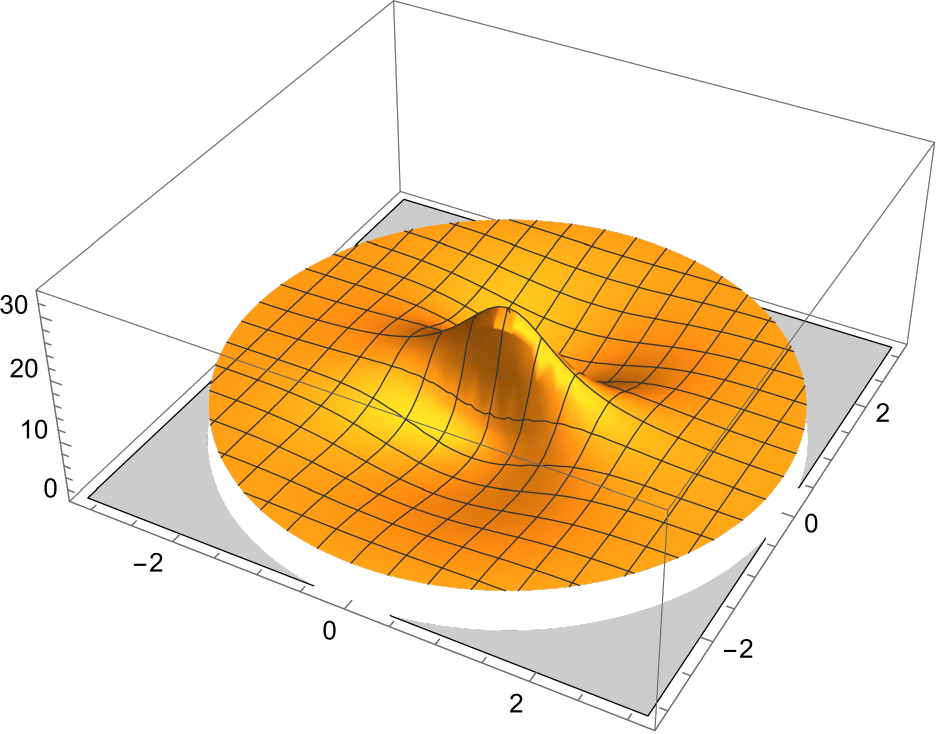}
\includegraphics[scale=.22]{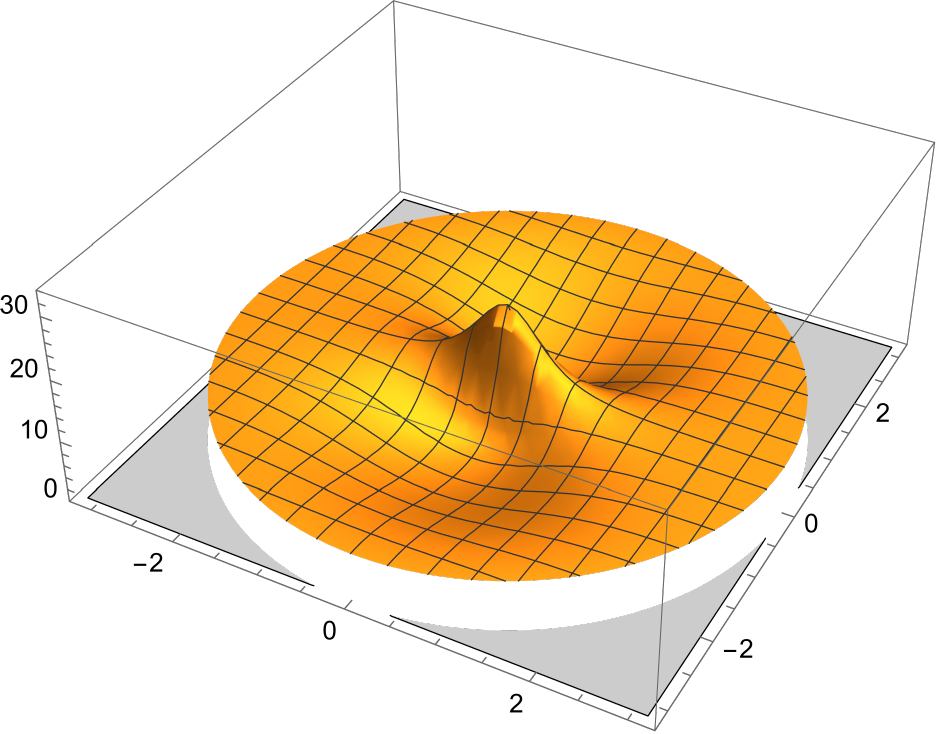}
\includegraphics[scale=.22]{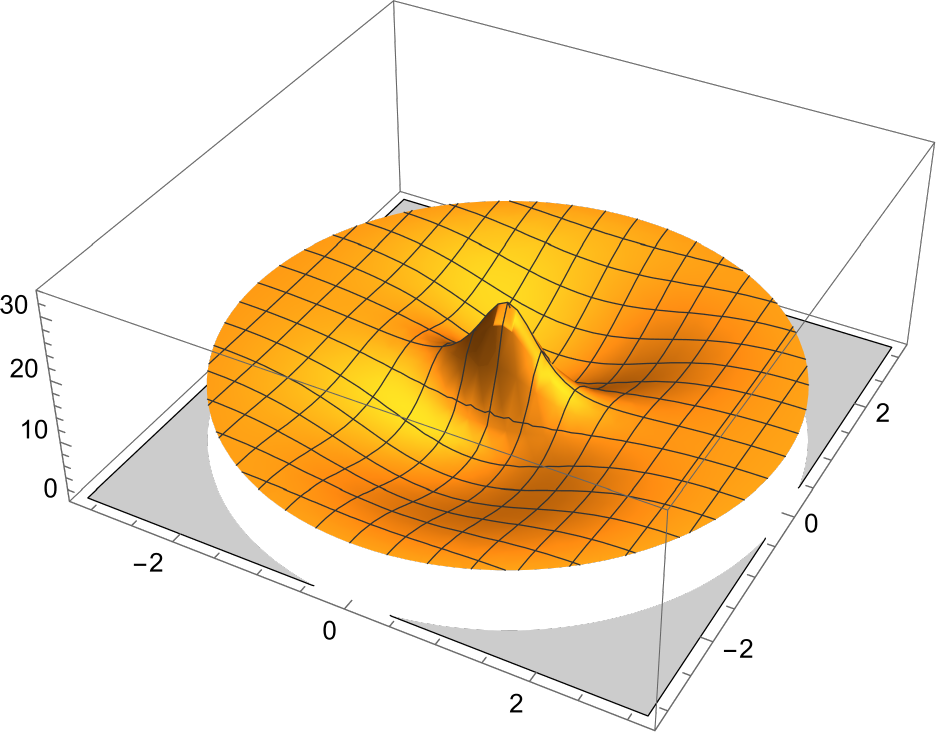}
\includegraphics[scale=.22]{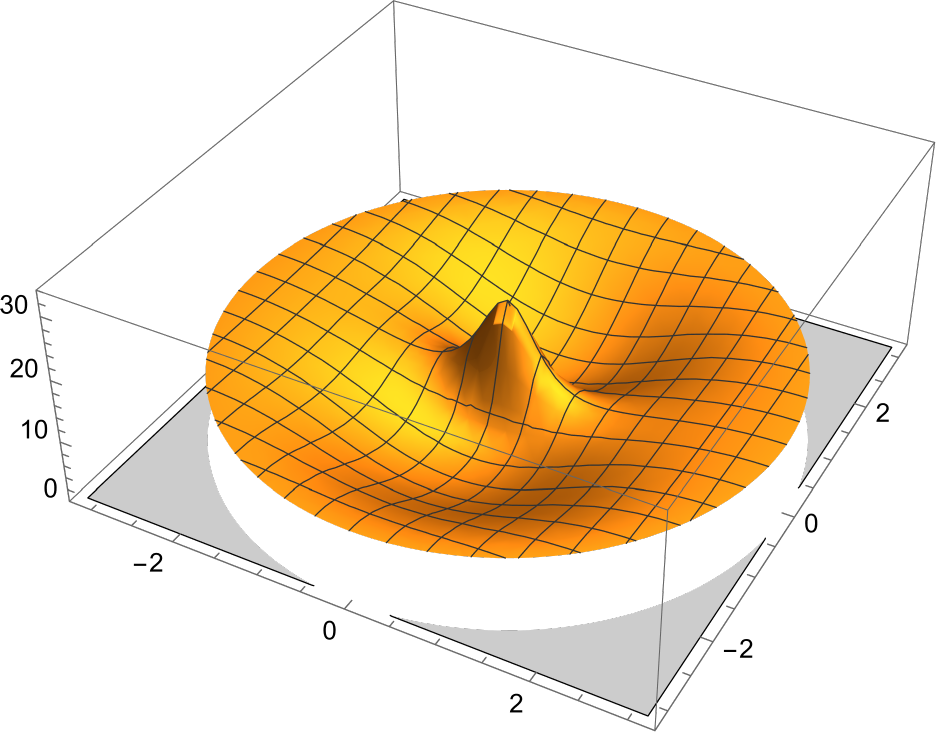}
\includegraphics[scale=.22]{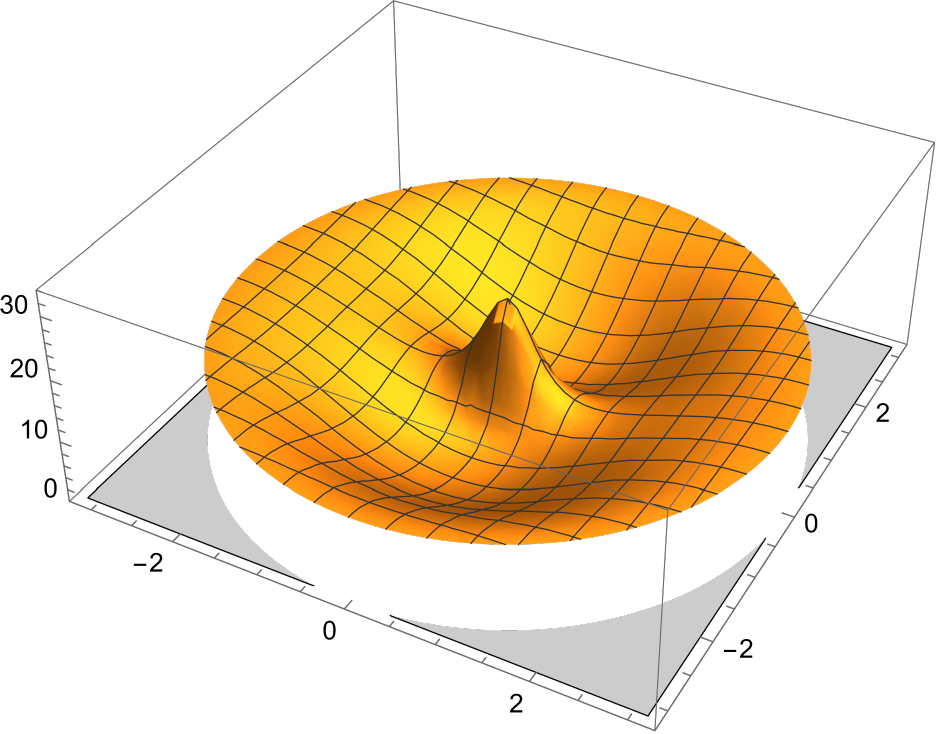}
\includegraphics[scale=.22]{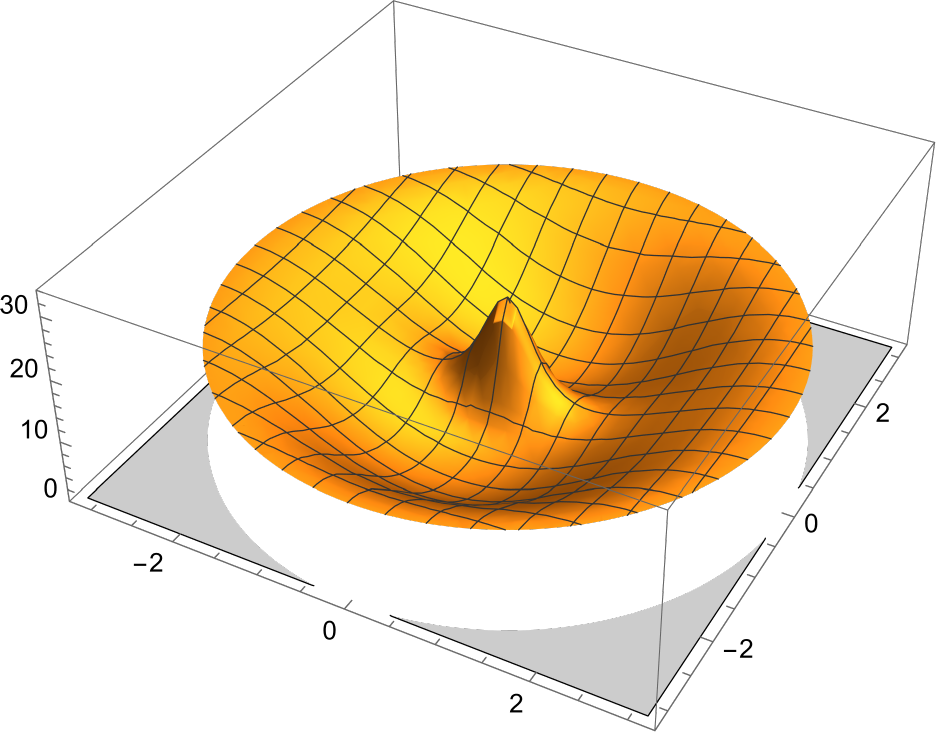}
\includegraphics[scale=.22]{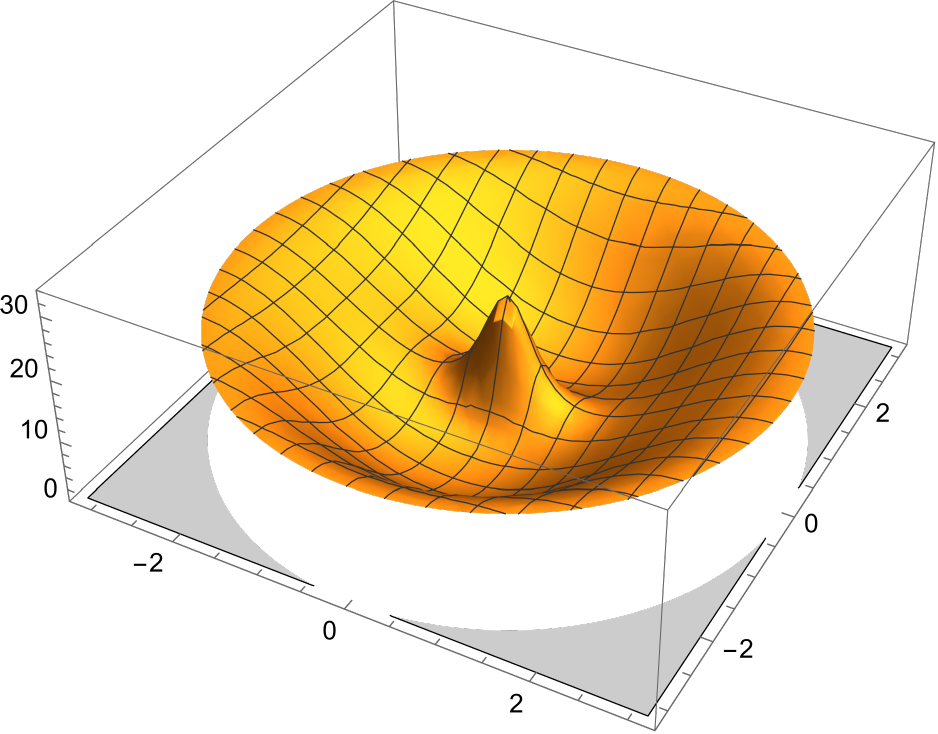}
\includegraphics[scale=.22]{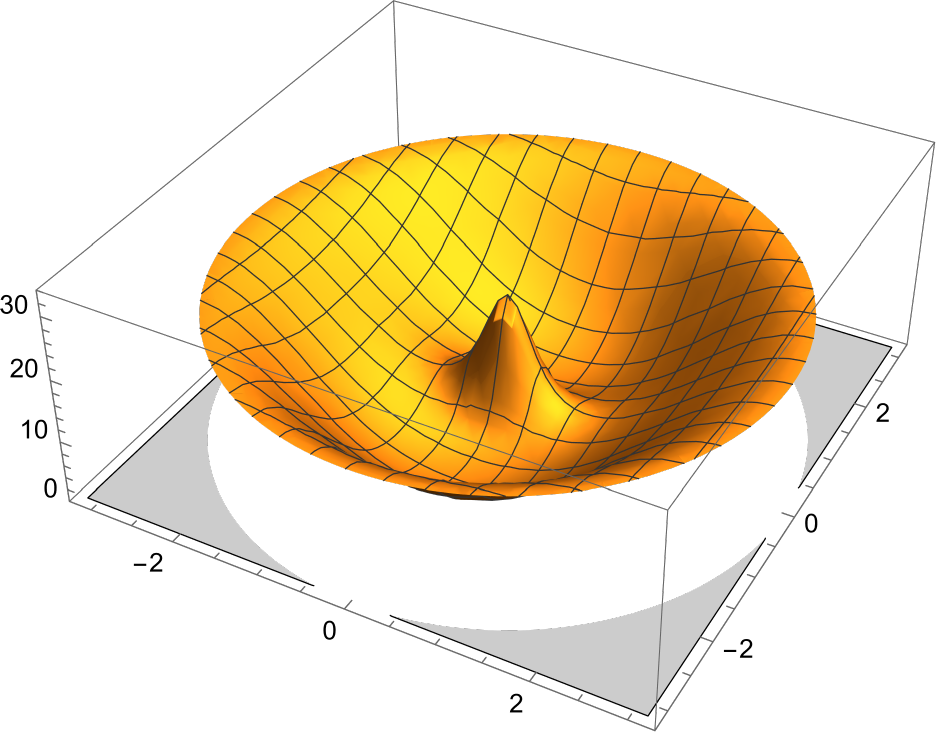}
\end{center}
\caption{Plots of $4\pi^2\II$ for $a={1\over2}$ and $\lambda^2={n\over8}$, $n=4,\dots,15$. 
The radial direction in the plot is polar angle (``$\theta$'') on the sphere.
$n=7$, the upper right plot, is on the critical line where the peaks coalesce. $n=15$, the lower right plot, is inversion-invariant.}
\label{ahalfPlots}
\end{figure}

The algebraic equation for stationary points of $\II$ at the real axis, away from $x=0$, is a cubic equation for $(\Re x)^2$. 
A careful analysis of this equation (discriminant, sum and product of roots) for all values of the parameters, gives at hand that there are no other local maxima than the ones already mentioned. The behaviour described above is illustrated in Figure \ref{ahalfPlots}, which considers multiple values of $\lambda$ for a fixed $a$, illustrating how the two peaks merge into a single one when the size becomes large enough compared to the separation, as well as the interplay between $\II(0)$ and $\II(\infty)$.

If we restrict our attention to solutions with $\II(0)\geq\II(\infty)$ (to the left of the blue curve in Figure \ref{k2Figure2}), the maximum value is $\II(0)$, as long as we are above the critical curve (orange). Below the critical curve, the maximum value is attained at the two peaks. Their precise location requires solving a cubic equation. When $\lambda\rightarrow0$, they approach $\pm a$, and their height diverges as ${3\over2}(1+a^2)^4\lambda^{-4}$.

\subsection{The special point in the $k=2$ moduli space}
\label{subsec:Special_point_k=2_chap3}

\begin{figure}
\begin{center}
\includegraphics[scale=.6]{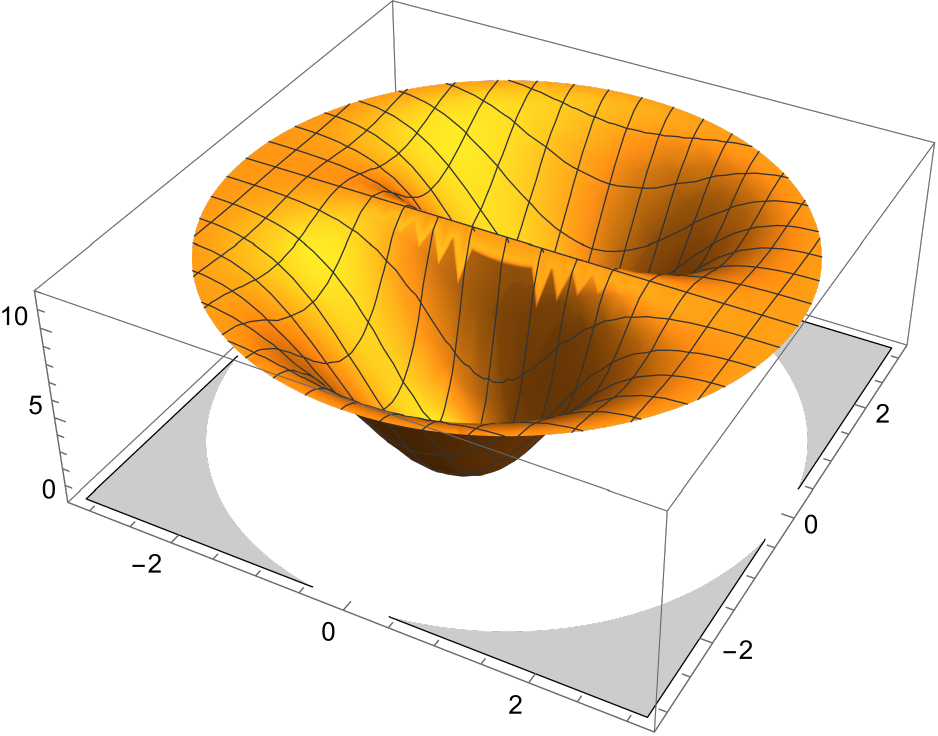}
\end{center}
\caption{Plot of $4\pi^2\II$ for $a={1\over\sqrt3}$, $\lambda={2\over\sqrt3}$. 
}
\label{specialPlot}
\end{figure}

At the special point $a={1\over\sqrt3}$, $\lambda={2\over\sqrt3}$, the  functions $f$ and $X$ can be rewritten as
\begin{align}
f_{{1\over\sqrt3},{2\over\sqrt3}}(x)&=(1+|x|^2)^2+{4\over3}|\Im x|^2\;,\\
X_{{1\over\sqrt3},{2\over\sqrt3}}(x)&=12\bigl((1+|x|^2)^4-{32\over9}(1+|x|^2)^2|\Im x|^2+{16\over27}|\Im x|^4\bigr)\;.\nn
\end{align}
The instanton scalar becomes
\begin{align}
4\pi^2\II={32\over3}{1-{32\over9}{|\Im x|^2\over (1+|x|^2)^2}+{16\over27}{|\Im x|^4\over (1+|x|^2)^4}
\over(1+{4\over3}{|\Im x|^2\over (1+|x|^2)^2})^4 } \, .
\label{SpecialI}
\end{align}
$\II$ is constant on surfaces $|\Im x|={\rho\over 2}(1+|x|^2)$. This is the stereographic image\footnote{The stereographic projection is along lines in $\RR^5=\HH\oplus\RR$ from $(0,2)$ through the point $(u,v)$: $|u|^2+(v-1)^2=1$ on a unit $S^4$ to $(2x,0)$. The factor $2$ is to obtain the standard metric $ds^2={4|\dd x|^2\over(1+|x|^2)^2}$.}
 of the space $|\Im u|=\rho$ in the unit $S^4$. The parameter $\rho$ lies in the interval $0\leq \rho\leq 1$. For $0<\rho<1$ this is 
 $S^2_\rho\times S^1_{\scaleto{\sqrt{1-\rho^2}}{8pt}}$, where the subscripts indicate radius.
 For $\rho=0$ it degenerates to $S^1$ (the compactified real line $\Im x=0$), and for $\rho=1$ to $S^2$ ($\Re x=0$, $|\Im x|=1$); this is shown in Figure \ref{specialPlot}, where only two dimension are depicted. As a consequence, the $S^2$ is represented by $S^0=\{\pm1\}$, and what looks like two minima is actually a $2$-sphere of minima.

 The maximum of $\II$ is attained at $\Im x=0$ (and $|x|=\infty$), with $4\pi^2\II={32\over3}$, and the minimum at $\Re x=0$, $|\Im x|=1$, with
 $4\pi^2\II={1\over2}$.
 
 As a check of normalisation, we can perform the integration of the instanton density for the special solution. Using the slicing in 
 $S^2_\rho\times S^1_{\scaleto{\sqrt{1-\rho^2}}{8pt}}$, we get the integration measure 
 $d^4x\sqrt g=dV_{S^4}={d\rho\over\sqrt{1-\rho^2}}dV_{S^1_{\scaleto{\sqrt{1-\rho^2}}{6pt}}}dV_{S^2_\rho}$.
 For a function which only depends on $\rho$,
 \begin{align}
 \int d^4x\sqrt g f(\rho)=8\pi^2\int_0^1 d\rho\,\rho^2f(\rho) \, ,
 \end{align}
(reproducing $\Vol(S^4)={8\pi^2\over3}$).
 Applied on $\II={8\over3\pi^2}{1-{8\over9}\rho^2+{1\over27}\rho^4\over(1+{1\over3}\rho^2)^4}$, this gives the instanton number
 \begin{align}
k= \int_{S^4}d^4x\sqrt g\II= 8\pi^2\times{8\over3\pi^2}\times{3\over32}=2\;.\label{SpecialK}
 \end{align}
 The integrals corresponding to  the three terms in the numerator each also contains a contribution to the last factor in eq. \eqref{SpecialK} which is a rational number times $\pi\sqrt3$. These cancel in the sum, providing a strong consistency check.

\section{Metric of maximal isometry and its Ricci tensor }
\label{sec:CalcRicciSec}
In this section, we put together all of the results accumulated so far. By focussing on the special points in the moduli space of the $k=1$ and $k=2$ instantons described in the previous section, we show that the resulting Kaluza--Klein metric has the maximal isometry, \ie, $SO(3) \times O(2)$ \cite{10.2307/1971078}, and establish bounds on the radius of the base $S^4$ to ensure positivity of the Ricci tensor.

\subsection{Special point and symmetry enhancement}
\label{subsec:special_point_enhancement_chap3}

The (or, a) regular field strength $F'$ is obtained by applying a singular gauge transformation as $F'=gFg^{-1}$, where 
$g={\bar x_a^{-1}-\bar x_b^{-1}\over|\bar x_a^{-1}-\bar x_b^{-1}|}$. Focussing on the special values of the moduli discussed in Section \ref{subsec:Special_point_k=2_chap3}, a calculation yields:
\begin{align}
F'&={4/3\over((1+|x|^2)^2+{4\over3}|\Im x|^2)^2}\label{regk2sol}\\
&\times\left(Q_0 \dd x\wedge \dd \bar x
+Q_1 ({\Im x\over|\Im x|} \dd x\wedge \dd \bar x-\dd x\wedge \dd \bar x{\Im x\over|\Im x|})
+Q_2{\Im x\over|\Im x|}\dd x\wedge \dd \bar x {\Im x\over|\Im x|}\right)\;,\nn
\end{align}
where
\begin{align}
Q_0&=2(1+|x|^2)^2-{2\over 3}(5+3|x|^2)|\Im x|^2\;,\nn\\
Q_1&=2(1+|x|^2)\Re x|\Im x|\;,\\
Q_2&=-{2\over3}(1+3|x|^2)|\Im x|^2\;.\nn
\end{align}
The $SO(3)$ symmetry is manifest in this expression, which is evidently invariant under rotations of the imaginary part of $x$ preserving its norm. It is however more convenient to use an ``orthogonal'' set $\{\omega_I\}_{I=1}^3$ for the selfdual $\su(2)$-valued 2-forms:
\begin{align}
\omega_1&={1\over8}\Bigl({\Im x\over|\Im x|} \dd x\wedge \dd \bar x-\dd x\wedge \dd \bar x{\Im x\over|\Im x|}\Bigr)\;,\nn\\
\omega_2&={1\over8}\Bigl(\dd x\wedge \dd \bar x+{\Im x\over|\Im x|}\dd x\wedge \dd \bar x {\Im x\over|\Im x|}\Bigr)\;,\\
\omega_3&={1\over8}\Bigl(\dd x\wedge \dd \bar x-{\Im x\over|\Im x|}\dd x\wedge \dd \bar x {\Im x\over|\Im x|}\Bigr)\;.\nn
\end{align}
Then, defining $M_{IJ}{}^{ij}=\sqrt gg^{\mu \rho}g^{\nu \sigma}\omega_{I \mu \nu}{}^i\omega_{J \rho \sigma}{}^j$, $M_{IJ}^{ij}=0$ for $I\neq J$, and
$M_{11}{}^{ij}=M_{22}{}^{ij}=\delta^{ij}-{x^ix^j\over|\Im x|^2}\equiv \Pperp{ij}$, $M_{33}{}^{ij}={x^ix^j\over|\Im x|^2}\equiv \Ppar^{ij}$. These are the projection operators on imaginary quaternions ortogonal and parallel to $\Im x$, respectively.
A M\"obius transformation $x\mapsto(x+\beta)(1-\beta x)^{-1}$, $\beta\in\RR$, preserves ${\Im x\over1+|x|^2}$
(and thus ${\Im x\over|\Im x|}$). 
This means that this $SO(2)$ rotation 
leaves these projection operators invariant. 
It acts on $\dd x\wedge \dd \bar x$ as
\begin{align}
\dd x\wedge \dd \bar x&\mapsto (1+\beta^2)|1-\beta x|^{-4}{1-\beta x\over|1-\beta x|}\dd x\wedge \dd \bar x{1-\beta\bar x\over|1-\beta x|}\;,\nn\\
{\dd x\wedge \dd \bar x\over(1+|x|^2)^2}&\mapsto {1-\beta x\over|1-\beta x|}{\dd x\wedge \dd \bar x\over(1+|x|^2)^2}{1-\beta\bar x\over|1-\beta x|}\;.
\end{align}
The conjugation with ${1-\beta x\over|1-\beta x|}$ commutes with $\Im x$.
Thus, an $SO(2)$ rotation induces an $\mathrm{SU(2)}$ gauge transformation. An element $h_\varphi=\cos\varphi+{\Im x\over|\Im x|}\sin\varphi$ transforms $\omega_I$ as 
$\omega_I\mapsto h_\varphi\omega_I\bar h_\varphi=(R_\varphi)_I{}^J\omega_J$, with
\begin{align}
R_\varphi=\left(\begin{matrix}\cos2\varphi&\sin2\varphi&0\\-\sin2\varphi&\cos2\varphi&0\\0&0&1\end{matrix}\right)\;.
\end{align}
This implies that $F$ is invariant modulo a gauge transformation, and $SO(2)$ is an iso\-metry. Together with the $\ZZ_2$ transformation $\Re x\mapsto -\Re x$, we obtain an $O(2)$.

Expressing $F'$ in the new basis, 
\begin{align}
F¨'={4/3\over((1+|x|^2)^2+{4\over3}|\Im x|^2)^2}q^I\omega_I\;,
\end{align}
where
\begin{align}
(q^1)^2+(q^2)^2&=64(1+|x|^2)^4\Bigl(1-4{|\Im x|^2\over(1+|x|^2)^2}\Bigr)\;,\nn\\
q^3&=8(1+|x|^2)^2\Bigl(1-{4\over3}{|\Im x|^2\over(1+|x|^2)^2}\Bigr)\;.
\end{align}
This shows that $Y^{ij}={1\over2}g^{mp}g^{nq}F'_{mn}{}^iF'_{pq}{}^j$ is invariant under $SO(2)$ and takes the form
\begin{align}
Y^{ij}&={32\over9}\Bigl(1+{4\over3}{|\Im x|^2\over (1+|x|^2)^2}\Bigr)^{-4}\nn\\
&\times\left[\Bigl(1-4{|\Im x|^2\over (1+|x|^2)^2}\Bigr)\delta^{ij}
+{4\over3}{|\Im x|^2\over (1+|x|^2)^2}\Bigl(1+{4\over 3}{|\Im x|^2\over(1+|x|^2)^2}\Bigr){x^ix^j\over|\Im x|^2}
\right]\\
&={32\over9}(1+{\rho^2\over3})^{-4}\left[(1-\rho^2)\delta^{ij}+{\rho^2\over3}(1+{\rho^2\over3}){x^ix^j\over|\Im x|^2}
\right]\;.\nn
\end{align}
Taking the trace gives back eq. \eqref{SpecialI}.
The factor $1-4{|\Im x|^2\over (1+|x|^2)^2}=1-\rho^2$ is positive semidefinite, it has its global minimum $0$ at $\Re x=0$, $|\Im x|=1$. 
$Y^{ij}$ is a positive definite combination of the projection operators $P^{ij}_{/\!/}={x^ix^j\over|\Im x|^2}$ and $P^{ij}_\perp=\delta^{ij}-{x^ix^j\over|\Im x|^2}$, with coefficients that are invariant under the ``extra'' $SO(2)$.

\subsection{The exotic sphere with maximal isometry}
\label{subsec:maximal_isometry_metric_chap3}

The bundle vielbein is given as in Section \ref{subsec:BundleGeometry_chap3}, with 
$\sF=F-yG\bar y$, where $G$ is the regular $SO(5)$-symmetric $k=1$ solution \eqref{k1solution} with $\lambda=1$ and $a=0$ (we drop the primes), and $F$ is the regular $k=2$ solution at the special point, eq. \eqref{regk2sol}.
Notice that there is no freedom in relative positioning on $S^4$ of the $k=1$ and $k=2$ solutions, when the $k=1$ instanton is at the symmetric values of the moduli. 

The product space $S^4\times S^3$ of course has isometry $SO(5)\times SO(4)$.
Any instanton solution for the left/right $\mathrm{SU(2)}$ of $S^3$ links the left/right isometry of $S^3$ to one of the $\mathrm{SU(2)}$'s in 
$(\mathrm{SU(2)}\times \mathrm{SU(2)})/\ZZ_2\simeq SO(4)\subset SO(5)$ of $S^4$ through the 't Hooft symbols. 
The maximally symmetric $k=1$ solution has isometry $SO(5)\times SO(3)$, which at the particular value of the relative radius giving the round $S^7$ gets enhanced to $SO(8)$. The maximally symmetric $k=2$ solution has isometry $SO(3)\times SO(3)\times O(2)$.
The bundle at hand, with both left and right instantons, thus breaks all $S^3$ isometry. The remaining isometry is
$SO(3)\times O(2)$.

\subsection{Ricci tensor and its bounds}
\label{subsec:Ricci_tensor_and_bounds_chap3}

In order to calculate the (field strength)${}^2$ contributions to the components of the Ricci tensor, let components of a field strength be real proportional to
\begin{align}
f(\alpha,\beta)={1\over2}(\bar \alpha\dd x\wedge \dd \bar x\beta+\bar \beta\dd x\wedge\dd \bar x\alpha)\;,
\end{align}
$\alpha,\beta\in\HH$,
and define the map from the tensor product of $\HH'$-valued selfdual 2-forms to $\vee^2\HH'$: $\varrho^{ij}(f,g)=\sqrt gg^{\mu \rho}g^{\nu \sigma}f_{\mu \nu}{}^{(i}g_{\rho \sigma}{}^{j)}$. Then,
\begin{align}
\varrho^{ij}(f(\alpha,\beta),f(\gamma,\delta))=-8\Re(e^{(i}\bar \alpha\gamma e^{j)}\bar \delta\beta)-8\Re(e^{(i}\bar \alpha\delta e^{j)}\bar \gamma\beta)\;.
\end{align}
If the field strength is conjugated by $u$, $|u|=1$, $F\mapsto uF\bar u$, we instead need to calculate
$\varrho^{ij}(f(\alpha\bar u,\beta\bar u),f(\gamma\bar u,\delta \bar u))$. This is equivalent to conjugating the $e^i$'s by $\bar u$, and is in general different from $\varrho^{ij}(f(\alpha,\beta),f(\gamma,\delta))$, unless at least 3 of the quaternions $\alpha,\beta,\gamma,\delta$ are real proportional to each other, 
making $\varrho^{ij}(f(\alpha,\beta),f(\gamma,\delta))$ proportional to $\delta^{ij}$.

When constructing the contribution from $\sF^2$ to the components $R_{ij}$ of the Ricci tensor, they will all be of the above form. Contraction with $\delta^{ij}$ gives the terms in the contribution to $R_{ii}$, but also to $R_{aa}$. We can then observe, than since the field strength is selfdual, the $\sF^2$ contribution to $R_{ab}$ is automatically proportional to $\delta_{ab}$. So, calculating 
$\varrho^{ij}(f(\alpha,\beta),f(\gamma,\delta))$ for the various terms in $\sF$ gives all information needed for the whole Ricci tensor.

We choose, in $\sF=F-yG\bar y$, to let $F$ be the 2-instanton solution. When the centra $a,b\in\RR$, the only (non-real) quaternion appearing multiplying $\dd x\wedge\dd \bar x$ from the left or right is ${\Im x\over|\Im x|}$, which is abbreviated as $I$ below. $G$ is the taken to be the 1-instanton solution, which is conjugated by $y$. The relevant $\varrho^{ij}$'s can be calculated as:
\begin{align}
\varrho^{ij}(f(1,1),f(1,1))&=16\delta^{ij}\;,\nn\\
\varrho^{ij}(f(1,1),f(1,I))&=0\;,\nn\\
\varrho^{ij}(f(1,1),f(I,I))&=-16\delta^{ij}+32I^iI^j\;,\nn\\
\varrho^{ij}(f(1,I),f(1,I))&=16(\delta^{ij}-I^iI^j)\;,\nn\\
\varrho^{ij}(f(1,I),f(I,I))&=0\;,\\
\varrho^{ij}(f(I,I),f(I,I))&=16\delta^{ij}\;,\nn\\
\varrho^{ij}(f(1,1),f(y,y))&=16\delta^{ij}(1-2|\Im y|^2)+32y^iy^j\;,\nn\\
\varrho^{ij}(f(1,I),f(y,y))&=32y_0(\delta^{ij}(y\cdot I)-y^{(i}I^{j)})+32\epsilon^{(i}{}_{kl}y^{j)}y^kI^l\;,\nn\\
\varrho^{ij}(f(I,I),f(y,y))&=-16\delta^{ij}(1-2|\Im y|^2)+32I^iI^j(1-2|\Im y|^2)\nn\\
&\qquad -32y^iy^j+64(y\cdot I)y^{(i}I^{j)}
+64y_0\epsilon^{(i}{}_{kl}I^{j)}I^ky^l\;,\nn\\
\varrho^{ij}(f(y,y),f(y,y))&=16\delta^{ij}\;.\nn
\end{align}
The first six can be obtained from the following three by letting $y=1$ or $y=I$.
It is convenient to use the linear combinations
\begin{align}
\omega_1&=-{1\over4}f(1,I)={1\over8}\Bigl(I \dd x\wedge \dd \bar x-\dd x\wedge \dd \bar xI\Bigr)\;,\nn\\
\omega_2&={1\over8}(f(1,1)-f(I,I))={1\over8}\Bigl(\dd x\wedge \dd \bar x+I\dd x\wedge \dd \bar x I\Bigr)\;,\\
\omega_3&={1\over8}(f(1,1)+f(I,I))={1\over8}\Bigl(\dd x\wedge \dd \bar x-I\dd x\wedge \dd \bar x I\Bigr)\;.\nn
\end{align}
in the expansion of the 2-instanton field strength $F$. They fulfill
$\varrho^{ij}(\omega_I,\omega_J)=0$ for $I\neq J$ and
$\varrho^{ij}(\omega_1,\omega_1)=\varrho^{ij}(\omega_2,\omega_2)=\delta^{ij}-I^iI^j=\Pperp^{ij}$,
$\varrho^{ij}(\omega_3,\omega_3)=I^iI^j=\Ppar^{ij}$, which are projection matrices on the imaginary quaternions orthogonal and parallel to $I$, respectively. 

We now let $\sF=F-yG\bar y$, with 
$F=(1+[x|^2)^{-2}q^I\omega_I$ and $G={1\over4}(1+|x]^2)^{-2}\gamma f(1,1)$.	

Then,
\begin{align}
(1+|x|^2)^4\varrho^{ij}(\sF,\sF)
&=\gamma^2\delta^{ij}-{1\over2}\gamma q^I\varrho(\omega_I,f(y,y))+((q^1)^2+(q^2)^2)\Pperp^{ij}+(q^3)^2\Ppar^{ij}\;,
\end{align}
so, with $S^4$ having unit radius,
\begin{align}
g^{\mu \rho}g^{\nu \sigma}\sF_{\mu \nu}{}^i\sF_{\rho \sigma}{}^j
&={1\over16}\bigl[\gamma^2\delta^{ij}-{1\over2}\gamma q^I\varrho^{ij}(\omega_I,f(y,y))\nn\\
&\quad+((q^1)^2+(q^2)^2)\Pperp^{ij}+(q^3)^2\Ppar^{ij}\bigr]\;.\label{rhoFF}
\end{align}
The mixed term is somewhat complicated. The three symmetric $(3\times3)$-matrices $\varrho(\omega_I,f(y,y))$ have entries that are functions on $S^3$. 
Let $y=\xi+\eta I+\zeta J$ in a local quaternionic basis\footnote{Notice that this basis is local both on $S^4$ ($\Im x$ defines the $I$ direction) and on $S^3$ ($\Im y$ then defines the $IJ$ plane), and in general not used for anything but local algebraic considerations.} $(1,I,J,K)$, where $J={y-(y\cdot I)I\over|y-(y\cdot I)I|}$ and $K=IJ$ (the basis degenerates if $\Im y$ is parallel to $I$, but that case is easy to treat). 
The coefficients obey $\xi^2+\eta^2+\zeta^2=1$.
Then,
\begin{align}
{1\over4}f(y,y)&=-2\xi\eta\omega_1+(\xi^2-\eta^2)\omega_2+(\xi^2+\eta^2)\omega_3\nn\\
&+{1\over2}\xi\zeta f(1,J)+{1\over2} \eta\zeta f(I,J)+{1\over4}\zeta^2f(J,J)\;.
\end{align}
We can now calculate the three matrices $M_I^{ij}={1\over4}\varrho(\omega_I,f(y,y))$ occurring in the mixed term. In the $IJK$ basis they are
\begin{align}
M_1&=\left(\begin{matrix}\qquad0\qquad&\xi\zeta&\eta\zeta\\\xi\zeta&\quad-2\xi\eta\quad&\zeta^2
			\\\eta\zeta&\zeta^2&\quad-2\xi\eta\quad\end{matrix}\right)\;,\nn\\
M_2&=\left(\begin{matrix}\qquad0\qquad&-\eta\zeta&-\xi\zeta\\-\eta\zeta&\xi^2-\eta^2+\zeta^2&\qquad0\qquad\\
			-\xi\zeta&\qquad0\qquad&\quad \xi^2-\eta^2-\zeta^2\quad\end{matrix}\right)\;,\\
M_3&=\left(\begin{matrix}\quad \xi^2+\eta^2-\zeta^2&-\eta\zeta&\xi\zeta\\
			-\eta\zeta&\qquad0\qquad&\qquad0\qquad\\\xi\zeta&0&0\end{matrix}\right)\;.\nn
\end{align}
It turns out that all eigenvalues of all three matrices lie in the interval $[-1,1]$ everywhere on $S^3$.
Inserting in eq. \eqref{rhoFF},
\begin{align}
Y^{ij}&\equiv{1\over2}g^{\mu \rho}g^{\nu \sigma}\sF_{\mu \nu}{}^i\sF_{\rho \sigma}{}^j\nn\\
&={1\over32}\bigl(\gamma^2\delta^{ij}-2\gamma q^I M_I^{ij}
+((q^1)^2+(q^2)^2)\Pperp^{ij}+(q^3)^2\Ppar^{ij}\bigr)\;.
\label{FFfunc}
\end{align}
All the $y$-dependence is in the matrices $M_I$. 

We are interested in finding bounds of the eigenvalues of this matrix.
In principle, this can be done by solving the cubic equations for the eigenvalues and study their dependence on $y$ and on the components $q^I$ (which depend on $x$). In practise, we only want to solve those cubic equations that reduce to quadratic ones.
We also want to use some properties of the solution at the special point in the $k=2$ moduli space.

We saw that we could freely rotate between $\omega_1$ and $\omega_2$. We use that freedom to set $q^1=0$ (this is a gauge choice; the contribution to $R_{ij}$ changes under gauge transformations).
It turns out that it is practical to use the basis $M_\pm=M_3\pm M_2$ (which means going back to $f(1,1)$ and $f(I,I)$).
Eq. \eqref{FFfunc} then becomes
\begin{align}
Y^{ij}={1\over32}\bigl(\gamma^2\delta^{ij}-2\gamma q^+ M_+^{ij}-2\gamma q^- M_-^{ij}
+(q^++q^-)^2\Ppar+(q^+-q^-)^2\Pperp   \bigr)\;.
\label{FFfunc1}
\end{align}
The eigenvalue structure of $M_\pm$ is quite simple.
The eigenvalues of $M_+$ are $\{-1+2\xi^2,-1+2\xi^2,1\}$ with eigenvectors
$\{ (0,0,1),(\zeta,\eta,0),(-\eta,\zeta,0) \}$
The eigenvalues of $M_-$ are $\{-1+2\eta^2,-1+2\eta^2,1\}$ with eigenvectors
$\{ (0,1,0),(-\zeta,0,\xi),(\xi,0,\zeta) \}$.
In order to give a lower bound on eq. \eqref{FFfunc1}, we want to complete the square to absorb negative terms. This only needs to be done for positive eigenvalues of $M_\pm$. Three regions of $S^3$ need to be considered:
\begin{itemize}
\item $\xi^2\leq{1\over2}$, $\eta^2\leq{1\over2}$. Both $M_+$ and $M_-$ have $1$ as the only positive eigenvalue.
\item $\xi^2> {1\over2}$. All eigenvalues of $M_+$ are positive, $1$ is the only positive eigenvalue of $M_-$.
\item $\eta^2>{1\over2}$. All eigenvalues of $M_-$ are positive, $1$ is the only positive eigenvalue of $M_+$.
\end{itemize}

Denote the projections on the eigenvalue $1$ subspaces of $M_\pm$ as $\Pi_\pm$.
In the first region, we write
\begin{align}
32Y&=(\gamma-q^+\Pi_+-q^-\Pi_-)^2-2\gamma q^+(M_+-\Pi_+)-2\gamma q^-(M_--\Pi_-)\nn\\
&\quad-(q^+\Pi_++q^-\Pi_-)^2
+(q^++q^-)^2\Ppar+(q^+-q^-)^2\Pperp\;.
\end{align}
All terms on the first line are non-negative, as are the terms $(q^+)^2(1-\Pi_+)+(q^-)^2(1-\Pi_-)$, so we have
\begin{align}
32Y\geq q^+q^-(-2+4\Ppar-\Pi_+\Pi_--\Pi_-\Pi_+)\label{qpqmlimit}
\end{align}
(inequality between matrices meaning contracted with any vector as $v^\transpose Yv$).
Solving for the eigenvalues of this matrix involves a ``hard''  cubic equation. 
Instead we discard the positive term with $\Ppar$. The maximal eigenvalue of $\Pi_+\Pi_-+\Pi_-\Pi_+$ takes the maximum value $2$ in the region 
(when $\zeta=0$, \ie, at $(\xi,\eta,\zeta)=({1\over\sqrt2},{1\over\sqrt2},0)$), so $32Y\geq-4q^+q^-$.
For the special solution, as a function of $\rho={2|\Im x|\over1+|x|^2}$, $q^+q^-$ takes its maximal value $4$ at $\rho=1$, so the result from the first region is
\begin{align}
Y\geq-{1\over2}\;.\label{Ylimit}
\end{align}
The same procedure in the other two regions yields the same limit, as does completing the square with the whole matrix
$q^+M_++q^-M_-$, disregarding eigenvalue signs.

If we apply this to the Ricci tensor on the exotic $S^7$, and let $S^4$ have radius $r$,
\begin{align}
R_{ij}=2\delta_{ij}+{1\over2r^4}Y_{ij}\;.
\end{align} 
Inserting the limit \eqref{Ylimit} shows that $R_{ij}$ is positive definite when $r^4>{1\over8}$ ($r\gtrapprox0.5946$).

In order to get the contribution to $R_{ab}$, we use 
$\sF_a{}^{ci}\sF_{bc}^i={1\over4}\delta_{ab}\sF^{cdi}\sF_{cd}{}^i={1\over2}\delta_{ab}Y^{ii}$ (the product of two selfdual 2-forms does not contain a traceless symmetric tensor). We immediately get
\begin{align}
32Y^{ii}=3\gamma^2-2\gamma q^+(4\xi^2-1)-2\gamma q^-(4\eta^2-1)
+(q^++q^-)^2+2(q^+-q^-)^2\;.
\end{align}
The $\gamma^2$ term from $G^2$ is proportional to the 1-instanton density and the $q^2$ terms from $F^2$ to the 2-instanton density. The mixed $\gamma q$ terms from $FG$ have average\footnote{On a unit $S^n$, the average value of the square of a coordinate in the embedding $\RR^{n+1}$ is ${1\over n+1}$.} 0 over $S^3$, so
${1\over8\pi^2}\int_{S^4}d^4x\sqrt gg^{\mu \rho}g^{\nu \sigma}\sF_{\mu \nu}{}^i\sF_{\rho \sigma}{}^i=3$.
For the special solution, the maximal value is attained at $\Im x=0$ and $\xi=\eta=0$ (\ie, $\Re y=0$, $\Re(\bar xy)=0$), and is
$Y^{ii}_{\hbox{\tiny max}}={3\over2}+{8\over3}+{32\over3}={89\over6}$, the three terms representing $G^2$, $FG$ and $F^2$, respectively.
For $S^4$ of radius $r$, 
\begin{align}
R_{ab}=\delta_{ab}\bigl({3\over r^2}-{1\over4r^4}Y^{ii}\bigr)\geq\delta_{ab}\bigl({3\over r^2}-{89\over24r^4}\bigr)\;.
\label{eq:Bound_R_ab}
\end{align}
$R_{ab}$ is positive definite when $r^2>{89\over72}$ ($r\gtrapprox1.112$). This limit is stronger than the one obtained from positivity of $R_{ij}$. We end this section with a remark concerning some relevant invariants
associated with the Gromoll--Meyer sphere: the Pontryagin class and the instanton numbers, as described in \cite{McEnroe2016MILNORSCO}, for instance. We note that these invariants are, of course, the same for all members in the class of metrics considered above (i.e., independent of moduli); but a description of how they influence the curvature remains elusive.

\section{Energy conditions}
\label{sec:Energy_conditions_chap3}

In this section, we finally come to applying our results in a physical setting, to determine whether a static exotic sphere solution defines a physically acceptable space-time. To do so, we consider some of the energy conditions. Assuming a mostly plus signature, so that $v^A v_A < 0$ defines a \textit{time-like} vector $v^A$, the most famous ones read (\cite{Curiel_2017,Martin-Moruno:2017exc}):
\begin{itemize}
    \item Weak Energy Condition (WEC): $G_{AB} v^A v^B \geq 0 $ for $v^A$ time-like.
     \item Strong Energy Condition (SEC): $(T_{AB} - \frac{1}{2} T g_{AB}) v^A v^B \geq 0 \iff R_{AB} v^A v^B \geq 0$ for $v^A$ time-like.
     \item Null Energy Condition (NEC): $T_{AB} k^A k^B \geq 0 \iff G_{AB} k^A k^B \geq 0$ for $k^A$ null.
     \item Dominant Energy Condition (DEC): $ G_{AB} v^A v^B \geq 0$ for $v^A$ time-like and $-G^A{}_B v^B$ is causal.
\end{itemize}
In the above, $T_{AB}$ is the stress-energy tensor, which equals (modulo Einstein gravitational constant) the Einstein tensor $G_{AB}$. The SEC and WEC both imply the NEC, while there in general is no implication between them. 

For a static space-time without any warp factor, a sufficient condition to satisfy the SEC is to have a spatial manifold with non-negative Ricci curvature (see \cite{Econditions}, for instance). This, however, is automatically true provided that the bound after \eqref{eq:Bound_R_ab} is met. 
For a space-time with scalar curvature $R\geq0$, $G_{AB}v^Av^B=R_{AB}v^Av^B-{1\over2}Rv^2\geq R_{AB}v^Av^B$ if $v^2\leq0$, so SEC implies WEC.

The dominant energy condition states that, for any future-pointing vector $v$ with $v^2\leq0$, the vector $w^A=-G^A{}_Bv^B$ also satisfies the same condition. For a static space-time with non-negative Ricci tensor this is also automatically satisfied. 
Namely, the vector $w$ becomes $w^0={1\over2}Rv^0$, $w^\mu={1\over2}Rv^\mu-R^\mu{}_\nu v^\nu$, and
\begin{align}
w^2={1\over4}R^2v^2-(RR_{\mu \nu}-R_{\mu \rho}R^\rho{}_\nu)v^\mu v^\nu\;.
\end{align}
The matrix $RR_{\mu \nu}-R_{\mu \rho}R^\rho{}_\nu$ is easily seen to have non-negative eigenvalues if $R_{mn}$ has non-negative eigenvalues, so
$w^2\leq{1\over4}R^2v^2\leq0$. Hence, we see that, provided $r \geq \frac{89}{72}$, all four physical energy conditions are met: weak, strong, null and dominant.

\section{Summary and Outlook}
\label{sec:Conclusions_chap3}

In this chapter, we have focused on metrics of the Kaluza--Klein type defined on the Gromoll--Meyer sphere, which is one of the exotic $7$-spheres. The first half of the chapter was dedicated to introducing the Kaluza--Klein framework and its two-fold nature; we emphasized its interpretation as a dimensional-reduction of physical theories and its mathematical presentation as a tool for describing metrics on the total space of fibre bundles. We reviewed the formalism focusing on increasing levels of complexity, from abelian principal fibre bundles to non-abelian non-principal ones. We focused on one example for each step of generalisation, with the last one being the Gromoll--Meyer sphere, realised according to Milnor's construction. By virtue of this construction (as an associated $SO(4)$-bundle), its Kaluza--Klein ansatz involves two $\mathrm{SU(2)}$ connections, which have instanton number of absolute value $1$ and $2$, respectively. The known expressions for these gauge fields were used to provide an explicit coordinate form of the Kaluza--Klein metric on the Gromoll--Meyer sphere and to comment on the associated Kaluza--Klein reduction from $7$ dimensions to $4$. A more detailed summary of the first part of the chapter is provided Section \ref{sec:Interlude_chap3}, and let us now focus on the second part. The ingredients involved in the Kaluza--Klein constructions are: round metrics on $S^4$ (the base space) and $S^3$ (the fibre), and $k=1,2$ $\mathrm{SU(2)}$ instanton gauge fields. Consistently with Milnor's original construction, the last five sections used quaternionic-valued objects for describing the geometric quantities mentioned above. This constitutes a significant computational advantage: through quaternionic algebra and calculus, we carried out a number of calculations that could almost certainly not be performed using component notation. For consistency, we determined the Ricci tensor associated with our Kaluza--Klein ansatz, finding perfect agreement with the literature. In order to understand the properties of the metric, we performed a detailed study of its most characteristic ingredient: the $k=2$ instanton gauge field. Starting from the original ansatz in \cite{PhysRevD.15.1642}, we computed its field strength and applied the regularising gauge transformation proposed in \cite{Giambiagi:1977yg}; we note that the above steps are straightforward and well-known for the $k=1$ (BPST) case, but much more non-trivial when the charge is doubled. Moreover, we studied the relation between the instantons' moduli space and the Kaluza--Klein metric's moduli space, and found that only a quotient of the former contributes to the latter. Specifically, one should identify all the instantons' configuration which are related via an $SO(5)$ transformation, \ie, via an isometry of the base. This motivated a special choice for the instantons' moduli, which resulted in the corresponding Kaluza--Klein metric having the maximal isometry group: $SO(3) \times O(2)$. It is natural to ask about the possible link between the construction discussed in this thesis, for some choice of the $S^4$ radius, and the one proposed by Gromoll and Meyer in \cite{10.2307/1971078}. We were also able to establish a bound on the radius of the base space $S^4$, $r$, which ensures a positive Ricci tensor: $ r > \frac{89}{72}$. 
When the inequality above is met, the $8$-dimensional space-time whose spatial manifold is an exotic sphere satisfies the strong, weak, null and dominant energy conditions. 
This provides the first and most natural application of our closed-form formulae for the Gromoll--Meyer sphere's metric and curvature to deduce some physical properties of this geometry in the context of gravity. The explicit expressions that were derived in this chapter could be leveraged to find solutions of higher-dimensional theories with the geometry exotic spheres, as we discuss below in more detail.

From a mathematical point of view, the relatively simple expression for the Riemann tensor, in particular the concrete expressions for the special point in the $k=2$ instanton moduli space, should facilitate an extensive investigation of the behaviour of the sectional curvature. It is known (\cite{nuimeprn10073}) that the Gromoll--Meyer sphere allows metrics with almost everywhere positive sectional curvature. It would be interesting to perform such an investigation for the sectional curvature of the bundle metric at the special points in the $k=1$ and $k=2$ moduli spaces, with the $S^4$ radius $r$ still as a free parameter. Other relevant questions begging for answers concern the geodesic structure of our metric: cut loci and Wiedersehen property, for instance (see \cite{Duran2001}). We leave all these questions for future investigation. Moreover, it would be interesting to repeat the same study for a larger portion of the $k=2$ instanton's moduli space. Generalising it to arbitrary positions is the first step, and including the gauge orientation would exhaust the whole moduli space. At that point, it would be interesting to examine the condition for the metric to be Einstein, and possibly prove a non-existence theorem in case such a condition cannot be satisfied. We also note that we limited ourselves to just one of the 27 exotic differentiable structures on $S^7$, to avoid cumbersome expressions. We believe that studying the results of this construction for the remaining exotic spheres would give valuable insights on such manifolds, as well as the Berger space, which was recently shown in \cite{10.2307/40067878} to be diffeomorphic to the total space of an $S^3$-bundle over $S^4$. \\
Finally, from the physics side, it would be natural to use the explicit results that we derived as a starting point for constructing solutions to supergravity theories in dimensions $7$ or higher, supported by appropriate fluxes. Concretely, two directions come to mind. The first one consists of extending this work to finding exotic sphere solutions in $7$-dimensional supergravity. A second interesting option would be to try and construct Freund-Rubin-like solutions to $11$-dimensional supergravity, with an exotic sphere as the internal manifold  (note that such solutions with non-Einstein metrics on the internal space exist, see \cite{POPE1985352}, for instance).

\chapter{Exotic Structures, Discontinuities \\
and General Relativity \,\,\,\,\,\,\,\,\,\,\,\,\,\,\,\,\,\,\,\,\,\,\,\,\,\,}
\label{chap:4}
This chapter summarises some facts about the differential-geometric and differentia-topological properties of exotic spheres and other exotic manifolds. Moreover, it discusses some features of the homeomorphic maps between ordinary spheres and exotic ones, together with some possible interpretations in the context of general relativity.

\section{Introduction, Overview and Structure}
General relativity, often referred as the most elegant theory ever conceived, has its mathematical roots in differential geometry, admitting a very rigorous formulation in terms of manifolds and metrics on them. 
This allowed to settle formally many questions on various aspects of the theory, while many more remain open. A key role in the formalism of GR is played by coordinates. The horizon singularity in the Schwarzschild metric is the prototypical example of how crucial it is to understand that local coordinates, which only give partial information on the manifold, need to be patched together consistently to describe a global object (\cite{Kruskal:1960, MTW}). As it is well known, choosing different sets of coordinates to describe the same spacetime should not affect the physics, which  is how the diffeomorphism invariance comes about. The study of these coordinates' choices, unlike the neighbouring layers of structure (topology and geometry), does not have a name.\footnote{Although its existence is expressed through the adjective \textit{differential}, which is often placed before ``topology'' and ``geometry''.} One reason for this might be that, in many simple cases, there is no room for manoeuvrer between the topological and differentiable structures, and they essentially coincide. As we have seen, this is not the case for exotic manifolds. The absence of a diffeomorphism shows that two exotic manifolds cannot be viewed as different descriptions of the same spacetime; but rather, they represent two, irreconcilable, models of the universe. Possibly because most of the framework of general relativity was developed before the discovery of exotic spaces, gauging the possible implications of differentiable structures on the physical picture is not immediate.\footnote{Some works have gone as far as formulating general relativity without invoking a differentiable structures, such as \cite{samann2024brief}.} Some attempts at doing so within general relativity were initiated by Brans (see \cite{Brans:1992mj,Brans1994a,Brans1994b}) and are currently being pursued by Asselmayer--Maluga and Krol (see \cite{AsselmeyerKrol2012,AsselmeyerKrol2014,AsselmeyerKrol2018}).\footnote{For a wider overview on the appearance of exotic differentiable structures in theoretical physics, see Section \ref{sec:Intro_chap3}.} The second part of this chapter provides another proposal for how to interpret the role of inequivalent differentiable structures from the gravitational point of view, by inspecting some features of homeomorphic maps between exotic $7$--spheres and ordinary ones. Before that, in order to motivate and contextualise the investigation, a broad overview on exotic spheres and exotic manifolds is presented in the first part of the chapter; specifically, we discuss some features of the Gromoll--Meyer sphere that are not mentioned in the previous chapter, briefly comment on other exotic $7$--spheres, and recall some facts about exotic differentiable structures more in general. Then, we set the focus on the Gromoll--Meyer sphere once again, this time under the lens of differential topology. Since the key object in general relativity and supergravity is the metric tensor, the pragmatic reader might be satisfied with the results of the previous chapter, which provides a very detailed description of the geometry with maximal isometry. However, there is a set of more qualitative questions that are not fully answered by focusing on the Kaluza--Klein geometry of exotic spheres, such as the following ones. \\
What does an exotic sphere ``look like'', and how does it differ from an ordinary sphere? \\
How does the difference between two inequivalent differentiable structures manifest itself? \\
Does the ``exoticness'' property have some quantifiable consequences in terms of gravity? \\
This chapter is also devoted to provide partial answers to such questions, by examining what features forbid the uplift of homeomorphisms between two exotic manifolds to diffeomorphisms; in spirit, this investigation might be considered analogous to some of the discussions in \cite{Rohm:1988yz} about topological defects. \\

The structure of this chapter is the following. Section \ref{sec:Facts_about_exot_spheres_chap4} provides some additional results about exotic $7$--spheres different from the Gromoll--Meyer one, and two discussions aimed at unveiling how and why the Gromoll--Meyer sphere differs from a homogeneous space. \\
In Section \ref{sec:Facts_about_general_exotic_chap4}, we discuss some other exotic manifolds in various dimensions, and briefly comment on the relation between smooth functions and differentiable structures. \\
Section \ref{sec:Homeo_chap4} examines two possible realisations of a homeomorphism between the ordinary $7$--sphere and the Gromoll--Meyer one, with the aim of investigating the obstruction preventing the map from being a diffeomorphism; this manifests as a discontinuity in the Jacobian. \\
Section \ref{sec:Changing_diff_struc_chap4} proposes the use of such a map to achieve a \textit{transport of (differentiable) structure}, and comments on its implications for the geometrical objects. \\
We offer a summary of the themes discussed and some promising future directions in Section \ref{sec:Conclusions_chap4}.

\section{Facts about Exotic $7$--Spheres}
\label{sec:Facts_about_exot_spheres_chap4}
This section contains a collection of facts concerning all exotic $7$--spheres, which are stated without a detailed derivation. It also focuses on some features of the Gromoll--Meyer sphere that make it \textit{not} a quotient manifold, but not far from it either.

\subsection{Classification}
\label{subsec:Classification_chap4}

Ever since the appearance of \cite{10.2307/1969983}, the complete catalogue of exotic $7$--spheres has been a central test-bed for high-dimensional differential topology. Kervaire and Milnor endowed the set of oriented homotopy $7$-spheres with the connected-sum operation, proving that it forms a finite abelian group $\Theta_{7}\cong\mathbb Z_{28}$ and identifying an explicit Milnor sphere as a generator (\cite{KervaireMilnor1963}). A key to distinguishing the $28$ smooth structures is the Eells–Kuiper $\mu$-invariant: for any closed spin $7$-manifold $M$, $\mu(M)\in\mathbb Z_{28}$ is defined in terms of the first Pontrjagin number and the signature of a bounding $8$-manifold, and Eells and Kuiper proved in \cite{EellsKuiper62} that $\mu$ is a diffeomorphism invariant which gives a bijection $\Theta_{7}\!\xrightarrow{\;\cong\;}\mathbb Z_{28}$.  In geometric terms, orientation reversal sends $\mu$ to $-\mu$ in $\mathbb{Z}_{28}$. Because $0$ and $14$ are their own negatives, the remaining $26$ classes form $13$ distinct pairs, yielding exactly $15$ unoriented diffeomorphism classes of homotopy $7$-spheres ($1$ standard and $14$ exotic). Shortly thereafter Brieskorn showed in \cite{Brieskorn1966} that all classes in $\Theta_{7}$ can also be realised as links of isolated hypersurface singularities—today called Brieskorn spheres—thereby connecting the smooth classification with complex-algebraic singularity theory.  As we have discussed at length, a different but equally explicit realisation, following Milnor's original paper \cite{10.2307/1969983}, arises from $S^{3}$-bundles over $S^{4}$; for this construction, Crowley and Escher determined exactly which integer pairs $(m,n)$ yield exotic spheres and gave a full diffeomorphism classification of these bundles, recovering $15$ of the $28$ smooth structures and explaining how the remaining classes are obtained by connected sum (\cite{CROWLEY2003363}). Modern work has subsumed the exotic case into a uniform picture of all closed smooth $2$-connected $7$-manifolds: the invariant $\mu$ has been generalised to a ``global'' Eells–Kuiper invariant $\tilde{\mu}$ defined for every spin $7$-manifold and, together with algebraic data of $H^{4}$ and its torsion linking form, proved that $\tilde{\mu}$ completes the diffeomorphism classification in \cite{CrowleyNordstrom2019}; when the manifold is a homotopy sphere this result reduces to the classical $\mu$ and hence re-derives $\Theta_{7}\cong\mathbb Z_{28}$. Taken together, these results show that, regardless of what construction is considered (Brieskorn link, or connected sum), the smooth structures on the topological $7$--sphere are exhausted by $28$ possibilities, and that the Eells–Kuiper invariant (classical or generalised) is the decisive complete invariant in dimension $7$.

\subsection{Exotic Spheres are \textit{not} Quotients}
\label{subsec:Not_quotients_chap4}
One interesting property of exotic spheres is that they cannot be obtained as quotient manifolds, i.e.~$G/H$ (\cite{GroveZiller00}). There is, moreover, a more concrete way of showing how obstructions arise when trying to build an exotic sphere via standard quotient constructions, such as the one leading to the squashed $7$-sphere. 
In \cite{CASTELLANI1984394}, it is shown how $S^7$ can be obtained by taking the quotient $SO(5) / SO(3)$, where the $SO(3)$ is embedded as one of the two subgroups that are present in $SO(5)$. The resulting manifold admits both the usual sphere geometry, but also the ``squashed'' geometry with a smaller amount of symmetry. \\
To be rigorous, one should work with double covers (i.e.~simply connected groups), and consider $\textrm{Spin}(5)/\mathrm{SU(2)}$ or $Sp(2)/Sp(1)$, which can be found in \cite{https://doi.org/10.48550/arxiv.2111.13221} (Section 2.2.1), \cite{Bais:1983wc} and \cite{PhysRevLett.50.2043} .\footnote{For a complete list of all the realisations of $S^7$ as a quotient manifold, see \cite{Coquereaux:1983kj}. The manifold $SO(5)/SO(3)$ is a \textit{Stiefel manifold} (\cite{Nikonorov2004CompactHE}). It can also be found in Appendix C of \cite{CASTELLANI1984429}. For metrics on it, should refer to references therein, and also to \cite{ABBASSI2010131}. Note that another manifold which is realised as the quotient of $SO(5)$ by (a special embedding of) $SO(3)$ is the Berger space - see \cite{ball2020associative}. Very explicit realisations of quotient manifolds arising from different embeddings of $SO(3)$ in $SO(5)$ can also be found in Appendix B of \cite{CASTELLANI1984429}. As an aside, the Berger space was shown to be diffeomorphic to the total space of a $S^3$ bundle over $S^4$ in \cite{10.2307/40067878}.} These distinctions are often (pragmatically) overlooked in the physics literature since they are irrelevant at the level of algebras. Let us consider the ``standard'' $\textrm{Spin}(4)$ subgroup of $\textrm{Spin}(5)$. Since $\textrm{Spin}(4) = \mathrm{SU(2)} \times \mathrm{SU(2)}$, then the quotient space $\textrm{Spin}(5)/\mathrm{SU(2)}$ is defined by the embedding of $\mathrm{SU(2)}$, which is the map $\mathrm{SU(2)} \xrightarrow{} \mathrm{SU(2)} \times \mathrm{SU(2)}$. This map is characterised by the winding numbers $p,q$, just as for the case of $U(1) \xrightarrow{} U(1) \times U(1)$. The choice $(p,q) = (0,1)$ gives the standard $S^7$, as described in \cite{CASTELLANI1984394}. However, different embeddings will, in general, produce different spaces. And, as we mentioned, such spaces are labelled by $\pi_3 ( \mathrm{SU(2)} \times \mathrm{SU(2)}) = \pi_3 (S^3 \times S^3) = \pi_3 (SO(4))$, which is the same quantity that characterises the $S^3$ bundles over $S^4$; in that setting, the winding $(0,1)$ also corresponds to the standard $S^7$. In the light of these considerations, consider trying to realise these spaces with the same technique described in \cite{CASTELLANI1984394}. To do that, one needs to understand how the winding is encoded at the level of the algebra. \\
Let us start with the usual $U(1)$ case. The map 
\begin{align}
    W_p: U(1) &\xrightarrow{} U(1) \nonumber \\
    e^{i \theta} &\mapsto W_p( e^{i \theta}) =  e^{ip \theta}
\end{align}
has winding number $p$. It can be viewed at the level of the algebra as the map between generators:
\begin{align}
    w_p: \mathfrak{u}(1) &\xrightarrow{} \mathfrak{u}(1) \nonumber \\
    \theta &\mapsto w_p( \theta) =  p\theta.
\end{align}
Now, the question is how the same can be done for $\mathrm{SU(2)}$ (the fact that it can be done follows from $\mathrm{SU(2)}$ being simply connected). A general $g$ of $\mathrm{SU(2)}$ can be written as 
\begin{align}
    g=u_4 \mathbf{1}_2+i\left(u_1 \sigma_1+u_2 \sigma_2+u_3 \sigma_3\right) \equiv u_4 \mathbf{1}_2+i \vec{u} \cdot \vec{\sigma},
\end{align}
with $u_1^2 + u_2^2 + u_3^2 + u_4^2 =1$ and $\sigma_i$ are the Pauli matrices. If we let $u_4 = \cos(\theta/2)$, then it follows that $\sin(\theta / 2)= |\vec{u}|$. Also defining $n_i=\frac{-u_i}{|\vec{u}|}$ (i=1,2,3) we obtain:
\begin{align}
    \left(\cos \frac{\theta}{2}\right) \mathbf{1}_2-i\left(\sin \frac{\theta}{2}\right) \vec{n} \cdot \vec{\sigma}=\mathrm{e}^{-i \theta \vec{n} \cdot \vec{\sigma} / 2},
    \label{eq:SU(2)_vs_su(2)_element}
\end{align}
where the identity (which is crucial) follows from the fact that $(\vec{n}\cdot \vec{u})^2 = \mathbf{1}_2$. \\
We can define the map $W_p$ with winding number $p$:
\begin{align}
    W_p: \mathrm{SU(2)} &\xrightarrow{} \mathrm{SU(2)} \nonumber \\
    g= \left(\cos \frac{\theta}{2}\right) \mathbf{1}_2-i\left(\sin \frac{\theta}{2}\right) \vec{n} \cdot \vec{\sigma} &\mapsto W_p(g) = \left(\cos \frac{p\theta}{2}\right) \mathbf{1}_2-i\left(\sin \frac{p\theta}{2}\right) \vec{n} \cdot \vec{\sigma},
\end{align}
which clearly ``wraps around'' $\mathrm{SU(2)}$ $p$ times, i.e.~each point of the target $SU(2)$ is the image of $p$ points. The equivalent map at the level of algebras (the equivalence emerges from \ref{eq:SU(2)_vs_su(2)_element}) is:
\begin{align}
     w_p: \mathfrak{su}(2) &\xrightarrow{} \mathfrak{su}(2) \nonumber \\
    \mathfrak{g}= \frac{\theta}{2} \vec{n} \cdot \vec{\sigma} &\mapsto w_p(\mathfrak{g}) = \frac{p\theta}{2} \vec{n} \cdot \vec{\sigma}.
\end{align}
The above steps show how to encode $g^n$ for $g\in \mathrm{SU(2)}$ at the level of the algebra. What goes wrong, however, is that there are no homomorphisms $H: \mathrm{SU(2)} \xrightarrow{} \mathrm{SU(2)}$ with homotopy class bigger than one (see \cite{osti_6662393}). It also easy to see that $(g h)^n \neq g^n h^n $ since $\mathrm{SU(2)}$ is non-abelian. Hence, one cannot embed $\mathrm{SU(2)}$ into $\mathrm{SU(2)} \times \mathrm{SU(2)} $ with arbitrary winding numbers, but the only allowed non-trivial choices are $(0,1)$, $(1,0)$ and $(1,1)$. Note that, however, this is not true for the abelian case of $U(1) \xrightarrow{} U(1)$, where homomorphisms of any winding are possible, which give rise to lens spaces in the context of the Hopf construction (see Section \ref{sec:lens_spaces_chap3}). Note that, of course, this not a proof, but just an instance of ``what goes wrong'' when trying to construct an exotic sphere as a quotient manifold.

\subsection{An Exotic Sphere as a Bi-quotient: the Gromoll--Meyer Construction}
\label{subsec:Martin_chap4}
As we just discussed, exotic $7-$spheres cannot be obtained as quotient spaces. However, they can be constructed as \textit{bi-quotient} manifolds, as
Gromoll and Meyer shown in \cite{10.2307/1971078}. This is a summary of their construction, presented in a very explicit and detailed fashion, based on Martin Cederwall's notes (\cite{Martin}).

First, let us consider the $10$-dimensional compact group $\USp(4)\simeq\Spin(5)$ (often referred as $\Sp(2)$ in the mathematical literature ). It consists of $2\times2$ $\HH$-valued matrices ($\HH$ is the algebra of quaternions):
\begin{align}
U=\left(\begin{matrix}a&b\\ c&d\end{matrix}\right)\;,
\end{align}
such that $UU^\dagger=I$, where $\dagger$ stands for transpose and $\HH$-conjugation.

This group manifold is an $S^3\times S^3$ bundle over $S^4$, \ie, a principal $\SU(2)\times\SU(2)$ bundle, as we are about to show.
We can parametrise the group in terms of $x\in \RR^4$ and $y,z\in S^3$, with $\RR^4$ represented by $\HH$ and $S^3$ by unit elements 
in $\HH$ (\ie, $y\bar y=1=z\bar z$). Then,
\begin{align}
U=\invsq x\left(\begin{matrix}y&xz\\ -\bar xy&z\end{matrix}\right)
\qquad\hbox{or}\qquad
U=\invsq{x'}\left(\begin{matrix}\bar x'y'&z'\\ -y'&x'z'\end{matrix}\right)\;.
\label{UPatches}
\end{align}
The first form applies when $a$ and $d$ are non-zero, the second when $b$ and $c$ are non-zero ($|a|=|d|$ and $|b|=|c|$ everywhere).
Equalling the two expressions on the overlap gives the transition functions
\begin{align}
x'&=x^{-1}\;,\nn\\
y'&=e^{-1}y\;,\\
z'&=ez\;,\nn
\label{USp4overlap}
\end{align}
where $e={x\over|x|}$.
We see that the two $\RR^4$'s are patched to $S^4$ and that the transitions of the $S^3$'s are those of instanton bundles with instanton numbers $-1$ and $1$. If we were to consider the corresponding principal $SO(4)$--bundle, then one would find that the transition functions read (see Section \ref{subsec:Milnor_construction_chap3} and specifically equation \eqref{eq:From_SU(2)_times_SU(2)_to_SO(4)}):
\begin{align}
    (x, z \, y^{-1 }) \mapsto (x^{-1} , e (z \, y^{-1} ) e ) \, , 
\end{align}
whose total space is $SO(5)$ (see \cite{Rigas1978}), as expected.

There is a natural, and isometric for the Cartan--Killing metric, action of $\SU(2)\times\SU(2)\subset\USp(4)$ by diagonal elements,
\begin{align}
U\mapsto\left(\begin{matrix}\gamma&0\\ 0&\delta\end{matrix}\right)U\left(\begin{matrix}\bar\alpha&0\\ 0&\bar\beta\end{matrix}\right)\;,
\end{align}
where $\alpha,\beta,\gamma,\delta$ are unit quaternions.
The action on the coordinates is
\begin{align}
\begin{matrix}
x\mapsto\gamma x\bar\delta\hfill\\
y\mapsto\gamma y\bar\alpha\hfill\\
z\mapsto\delta z\bar\beta\hfill
\end{matrix}
\qquad\qquad
\begin{matrix}
x'\mapsto\delta x'\bar\gamma\hfill\\
y'\mapsto\delta y'\bar\alpha\hfill\\
z'\mapsto\gamma z'\bar\beta\hfill
\end{matrix}
\label{su24transf}
\end{align}

Subgroups $\SU(2)\subset\SU(2)^4$ can be used to obtain $S^7$'s as quotients of $\USp(4)$.
Choosing (for example) the subgroup defined by $\alpha=\gamma=\delta=1$ yields the standard $S^7$, while
the choice $\beta=1$, $\alpha=\gamma=\delta$ yields the exotic Gromoll--Meyer $S^7$.
From the transformations \eqref{su24transf} it is clear that there are no fixed points in either of these cases:
in the first case $z\mapsto z\bar\beta$, $z'\mapsto z'\bar\beta$, and in the second one
$z\mapsto\alpha z$, $z'\mapsto\alpha z'$.\footnote{Note that one could equivalently choose the diagonal of one of the left $\SU(2)$'s and both right $\SU(2)$'s. This is however not convenient with the parametrisation \eqref{UPatches}, the identification $x'=x^{-1}$ is then destroyed.}
The standard $S^7$ is thus obtained as a right coset $\USp(4)/\SU(2)$, while the exotic one is a bi-quotient, involving both left and right group action.

Before taking quotients, consider the Cartan--Killing metric on $\USp(4)$. It is left- and right-invariant, and so invariant under
$\SU(2)^4$. (One may consider deformations that preserve this structure.)
It is given as 
\begin{align}
ds^2=\tr(dUdU^\dagger)\;.
\end{align}
Expressing it in terms of the parametrisation \eqref{UPatches} yields, after a short calculation,
an expression $ds^2=e\bar e+\varepsilon\bar\varepsilon+\varphi\bar\varphi$, where $e,\varepsilon,\varphi$ are vielbein components, 1-forms in $\HH$ ($\varepsilon$ and $\varphi$ are imaginary), given by
\begin{align}
e&={2dx\over1+|x|^2}\;,\nn\\
\varepsilon&=dy\bar y+A\;,\\
\varphi&=dz\bar z+B\;,\nn
\end{align}
where the $\SU(2)\times\SU(2)$ instanton gauge connections $A$ and $B$ on $S^4$ are
\begin{align}
A&={\Im(xd\bar x)\over1+|x|^2}\;\nn\\
B&={\Im(\bar xdx)\over1+|x|^2}\;.
\end{align}
We note that the first parts of $\varepsilon$ and $\varphi$ are the right-invariant Maurer--Cartan forms on the $S^3$'s. The presence of the instantons identify the left $\SU(2)$'s on the $S^3$'s with the left and right $\SU(2)$ on $\RR^4$, in accordance with the transformations
\eqref{su24transf}. The field strengths are straightforwardly obtained,
\begin{align}
F=dA+A\wedge A&={dx\wedge d\bar x\over(1+|x]^2)^2}\;,\nn\\
G=dB+B\wedge B&={d\bar x\wedge dx\over(1+|x]^2)^2}\;.
\end{align}
They are manifestly selfdual/anti-selfdual, and agree with the standard form of a single instanton on $S^4$ with center at $x=0$ and radius $1$. 
(Deformations respecting $\SU(2)^4$ can be obtained for example by changing the radii of the $S^3$'s relative to the one of $S^4$. All of these are $1$ here.)

Let us now turn to quotients, and first practice on the easy case leading to the standard $S^7$.
As already mentioned, it corresponds to ``elimination'' of $z$ by the $\SU(2)$ with parameter $\beta$ in
eq. \eqref{su24transf}. The coordinates $x$ and $y$ (and in the second patch $x'$ and $y'$) are inert, \ie, remain the same on an orbit.
A representative can be taken as
\begin{align}
U=\invsq x\left(\begin{matrix}y&x\\ -\bar xy&1\end{matrix}\right)
\qquad\hbox{or}\qquad
U=\invsq{x'}\left(\begin{matrix}\bar x'y'&1'\\ -y'&x'\end{matrix}\right)\;.
\label{OrbitRepr}
\end{align}
Due to the $B$ connection, $dx$ is not orthogonal to the orbit. One needs to add $dz=-Bz$, so that the resulting displacement is orthogonal to $\varphi$, and thereby orthogonal to the orbit. The metric on the quotient space is then given by $ds^2=e\bar e+\varepsilon\bar\varepsilon$, with the same transitions for the $x$ and $y$ coordinates and the $A$ connection as before.
This is of course the result from a standard coset construction.

The corresponding construction in the Gromoll--Meyer case looks much more complicated.
The orbit is now given as
\begin{align}
\begin{matrix}
x\mapsto\alpha x\bar\alpha\hfill\\
y\mapsto\alpha y\bar\alpha\hfill\\
z\mapsto\alpha z\hfill
\end{matrix}
\qquad\qquad
\begin{matrix}
x'\mapsto\alpha x'\bar\alpha\hfill\\
y'\mapsto\alpha y'\bar\alpha\hfill\\
z'\mapsto\alpha z'\hfill
\end{matrix}
\label{su24transfGM}
\end{align}
Representatives on the orbit are again given by eq. \eqref{OrbitRepr}.
However, setting these equal (modulo an $\alpha$-transformation) on the overlap requires the $U$ of the first patch to be transformed with $\alpha=e^{-1}={\bar x\over|x]}$. Given the transformations \eqref{su24transfGM} and the original overlap equation
\eqref{USp4overlap}, this gives
\begin{align}
x'&=x^{-1}\;,\nn\\
y'&=e^{-2}ye\;.
\label{GMoverlap}
\end{align}
This is the overlap of the simplest exotic $S^7$ in the Milnor construction.

It is interesting to consider the metric inherited from the Cartan--Killing metric on $\USp(4)$ (or some deformation); in particular, with the aim of comparing it with the Kaluza--Klein ansatz discussed in the previous chapter. The latter procedure provides metrics on Milnor bundles, but it is not clear that they are be diffeomorphic to the ones obtained from the quotient construction. For instance, it is legitimate to ask whether the quotient construction yields a recognisable $2$-instanton connection. One fact that can be quickly inferred is that the resulting metric will have at least an $\SO(3)$ isometry, deriving from the right action on $z$, $z\mapsto z\bar\beta$. On the representatives, 
$x\mapsto\beta x\bar\beta$, $y\mapsto\beta y\bar\beta$, under which the real parts are invariant.

A set of tangent vectors to the orbit 
is $(dx,dy,dz)=([e_i,x],[e_i,y],e_iz)=v_i$ for $e_i$ the imaginary unit quaternions.
The resulting flat components are
\begin{align}
e(v_i)&={2[e_i,x]\over1+|x|^2}\;,\nn\\
\varepsilon(v_i)&=[e_i,y]\bar y+{x[e_i,\bar x]\over 1+|x|^2}\;,\\
\varphi(v_i)&=e_i+{\bar x[e_i,x]\over 1+|x|^2}\;.\nn
\label{TangentVectors}
\end{align}
The principle is to find vectors $dx+\ldots$, $dy+\ldots$, where the ellipses denote objects along the orbit, chosen so that the resulting vectors are orthogonal to the orbit, and calculate their metric.
Given a vector $u$, its projection orthogonal to the orbit is
\begin{align}
\tilde u=u-(M^{-1})^{ij}\langle u,v_i\rangle v_j\;,
\end{align} 
where $M_{ij}=\langle v_i,v_j\rangle$, and the scalar product on the quotient is 
\begin{align}
g(u,u')=\langle\tilde u,\tilde u'\rangle
=\langle u,u'\rangle-\langle u,v_i\rangle(M^{-1})^{ij}\langle v_j,u'\rangle\;.
\end{align}
The procedure produces cumbersome results, so let us just focus on one ingredient, for illustrative purposes, i.e.~the metric for the $S^3$ parametrised by $y$ when $\Im x=0$. Then all $x$-dependence in
\eqref{TangentVectors} goes away. Let $\vec y=\Im y$.
We get 
\begin{align}
\langle v_i,v_j\rangle=\Re([e_i,y]\overline{[e_j,y]})=\delta_{ij}(1+4|\vec y|^2)-4y_iy_j\;,
\end{align}
 and
\begin{align}
(M^{-1})^{ij}={1\over1+4|\vec y|^2}(\delta^{ij}+4y^iy^j)\;.
\end{align}
We also have $\langle dy,v_i\rangle=\Re(dy[e_i,\bar y])=-2\varepsilon_{ijk}y_jdy_k$.
This results in
\begin{align}
ds^2=|dy|^2-{4\over1+4|\vec y|^2}(|\vec y|^2|d\vec y|^2-(\vec y\cdot d\vec y)^2)\;.
\end{align}
The metric becomes dependent on the polar angle $\theta$. If we parametrise $S^3$ as
$y=\cos\theta+\eta\sin\theta$, where $\eta$ is a unit imaginary quaternion parametrising $S^2$, the metric is
\begin{align}
ds^2=d\theta^2+{\sin^2\theta\over1+4\sin^2\theta}|d\eta|^2\;,
\end{align}
with obvious $\SO(3)$ isometry, and being ``squashed'' compared to the round metric. 

In summary, this investigation (based on \cite{Martin}), illustrates how the Gromoll--Meyer construction nicely matches Milnor's one from a topological-differential point of view; at the same time, it shows how recovering the geometric features from the double-quotient construction is far from trivial.

\subsection{Some more Facts about Exotic Spheres}
\label{subsec:More_facts_chap4}

\subsubsection{Different Realisations}
As mentioned in \ref{subsec:Classification_chap4}, exotic $7$--spheres admit different realisations, in addition to Milnor's original construction \cite{10.2307/1969983}. Let us now elaborate on two of those: the first one being relevant for our current investigations, presented in this chapter, and the second one being instrumental for some further investigations described in Section \ref{sec:Conclusions_chap4}.

A prolific method for realising exotic $7$--spheres comes from \emph{twisted spheres}: given an orientation-preserving diffeomorphism $f\colon S^{\,n-1}\!\to S^{\,n-1}$, gluing two $n$-discs along~$f$ produces the manifold $\Sigma_f=D^{n}\cup_{f}D^{n}$, always homeomorphic to $S^{n}$ but carrying a smooth structure which might be exotic. In fact, every oriented exotic $7$-sphere is realised as some $\Sigma_f$, and the set of isotopy classes of gluing maps is isomorphic to the group $\Theta_{7}\cong\mathbb{Z}_{28}$ (\cite{Milnor1965hCobordism, Smale1962}). Diffeomorphisms $f: S^6 \xrightarrow{} S^6$ which yield exotic spheres are, by definition, topologically isotopic to the identity but not smoothly isotopic. They are sometimes referred as \textit{exotic diffeomorphisms}, and some explicit examples are discussed in Section \ref{subsec:Other_homeos_chap4}.

A second, seemingly quite different, avenue arises from \emph{Brieskorn spheres}. They were mentioned in Section \ref{subsec:Classification_chap4}, but let us provide a bit more details on such a construction. Under suitable arithmetic conditions on integers $a_0,\dots,a_k\ge 2$, the link of the isolated complex hypersurface singularity $z_0^{\,a_0}+\dots+z_k^{\,a_k}=0$ in $\mathbb{C}^{k+1}$ is the smooth manifold 
\begin{align}
\Sigma(a_0,\dots,a_k)=\bigl\{z_0^{\,a_0}+\dots+z_k^{\,a_k}=0,\;\|z\|=1\bigr\}\subset S^{\,2k+1}(1).
\end{align}
Brieskorn showed that many such links are exotic spheres~\cite{Brieskorn66}. The real dimension of these links is $2k-1$. Consequently, when $k=4$ they live in dimension $7$, and suitable choices of exponents realise every element of $\Theta_{7}$. For instance, Brieskorn proved that the classic family $\Sigma(2,2,2,3,6j-1)$ corresponds to the class $j \in \mathbb{Z}_{28}$. Milnor’s study of the Milnor fibration~\cite{Milnor68} reveals rich geometric structures on these manifolds, and plumbing computations align their $\mu$-invariants with those of twisted spheres.

The fact that every exotic $7$-sphere can be described both by clutching two discs and as the link of a singularity bridges differential topology and algebraic geometry.  Translating a gluing map into weighted-homogeneous exponents encodes differential-topological data in algebraic terms, while the Eells–Kuiper invariant provides a common yardstick for distinguishing the resulting smooth structures.  This interplay continues to inspire new techniques for studying exotic smooth structures in higher dimensions.

\subsubsection{Curvature Properties}

Not long after Milnor’s seminal paper, attention has partially shifted from mere existence to the differential–geometric properties of exotic spheres. A first benchmark is positive \emph{scalar} curvature.  Hitchin used the $\alpha$–invariant of spin manifolds to exhibit exotic spheres in dimensions $8k+1$ and $8k+2$ that cannot carry such metrics, showing that the smooth structure alone may obstruct curvature conditions (\cite{Hitchin74}).  Conversely, Gromov–Lawson’s surgery theory in \cite{GromovLawson80} and the subsequent classification by Stolz in \cite{Stolz92} imply that every exotic sphere bounding a parallelisable manifold (the subgroup $bP_{n+1}\subset\Theta_n$) does admit metrics of positive scalar curvature; in particular this covers all exotic $7$-spheres.

Strengthening to positive \emph{Ricci} curvature, Wraith established a surgery theorem that proved Ricci-positive metrics on all homotopy spheres in $bP_{n+1}$, including every oriented exotic $7$-sphere (\cite{Wraith97}).  Brieskorn links turned out to be especially fruitful: Boyer, Galicki and Kollár produced Sasaki–Einstein structures on each such sphere, giving infinitely many inequivalent Einstein (hence Ricci-positive) metrics in dimensions $7$, $11$ and $15$ and confirming that all $28$ oriented diffeomorphism classes in dimension $7$ admit Einstein metrics, see \cite{BoyerGalickiKollar05}.

For \emph{sectional} curvature the picture is more restrictive.  Gromoll and Meyer wrote down a metric of non-negative sectional curvature on a specific exotic $7$-sphere in \cite{GromollMeyer74}.  Petersen and Wilhelm later showed that the same sphere supports a metric whose sectional curvature is everywhere strictly positive (\cite{PetersenWilhelm08}); however, an unfixable gap was subsequently found in their proof. To date, and to the author's knowledge, no exotic sphere in dimension 7 is known to admit a metric with strictly positive sectional curvature. Nonetheless, techniques of cohomogeneity-one actions greatly enlarged the stock of non-negatively curved examples: Grove and Ziller constructed such metrics on ten out of the fourteen unoriented exotic classes, demonstrating that non-negative curvature is far more common than strict positivity, as discussed in \cite{GroveZiller00}.  Whether \emph{every} exotic sphere admits positive sectional curvature is still open.

Taken together, these results show that exotic smooth structures seldom prevent favourable curvature in the weaker scalar or Ricci senses, yet they can pose formidable obstacles to positive sectional curvature.  Closing this gap—either by finding new positively curved examples or by proving definitive obstructions—remains one of the central challenges in high-dimensional Riemannian geometry.

\section{Facts about Generic Exotic Differentiable Structures}
\label{sec:Facts_about_general_exotic_chap4}
In this section, we ``zoom out'' from exotic spheres, and provide a (far from exhaustive) overview on exotic manifolds and exotic differentiable structures more in general.

\subsection{Other Interesting Exotic Manifolds}
\label{subsec:Other_interesting_chap4}
The discovery of exotic spheres showed that topological and differentiable categories can diverge even for the simplest closed manifolds.  Yet spheres are only the tip of the iceberg: many other spaces support unexpected smooth structures, each illuminating a different facet of high-dimensional topology, gauge theory or geometric analysis.

Among all examples, nothing is stranger than an \emph{exotic} $\mathbb{R}^{4}$. Freedman proved that any closed simply–connected topological $4$-manifold is determined by its intersection form, and in particular that there is a unique topological $\mathbb{R}^{4}$ - see \cite{Freedman82}.  Donaldson’s gauge-theoretic constraints on smooth intersection forms, however, imply that some of those topological models cannot be smoothed in the standard way, as shown in \cite{Donaldson83}. In \cite{Gompf95}, Gompf combined handle calculus with Casson handles to build explicit ``small’’ exotic $\mathbb{R}^{4}$’s that embed smoothly in the standard $\mathbb{R}^{4}$ and ``large’’ ones that do not.  There are uncountably many pairwise non-diffeomorphic versions, and every smooth, simply connected, open $4$-manifold contains at least one such exotic $\mathbb{R}^{4}$ as an open subset~\cite{BizacaGompf96}.  These pathologies occur only in dimension~$4$, making exotic Euclidean space a laboratory where Donaldson–Seiberg–Witten theory meets Casson-handle wildness, with potential ramifications in quantum gravity and low-energy gauge-field models.

Exotic phenomena are not limited to open manifolds: there exist closed $4$-manifolds that are homeomorphic but not diffeomorphic to familiar complex surfaces.  Dolgachev constructed the first simply-connected complex surface, now called a \emph{Dolgachev surface}, that is homeomorphic to the elliptic surface $E(1)=\mathbb{C}P^{2}\#9\overline{\mathbb{C}P}{}^{2}$ yet not diffeomorphic to it (\cite{Dolgachev81}).  Fintushel and Stern’s knot-surgery technique later produced infinite families of pairwise exotic copies of many rational and elliptic surfaces by excising a torus neighbourhood and regluing via the complement of a knot in \cite{FintushelStern98}. These constructions are ``interesting’’ because the resulting manifolds carry the exact same intersection form (and are thus homeomorphic), yet are distinguished by their Seiberg–Witten invariants, which are modified by the Alexander polynomial of the knot; they demonstrate that $4$-dimensional differential topology is rich enough to encode knot theory inside seemingly rigid complex surfaces.

Contractible $4$-manifolds furnish another source of exoticity.  Mazur’s original example showed that a manifold can be contractible but have boundary a non-trivial homology $3$-sphere~\cite{Mazur61}.  Akbulut later introduced the notion of a \emph{cork}: a compact contractible $4$-manifold whose boundary contains an involution extending to the interior only after altering the smooth structure of the ambient space~\cite{Akbulut91}.  Akbulut and Matveyev proved that \emph{every} pair of simply connected, closed, exotic $4$-manifolds differs by twisting a single cork~\cite{AkbulutMatveyev97}.  Corks are “interesting’’ because they reduce complicated questions about exotic smooth structures to local modifications inside a topologically trivial core, giving an operational handle on otherwise intangible smooth phenomena.

Outside dimension four, smoothing theory predicts much tamer behaviour, yet exotic smooth structures still arise on some aspherical manifolds.  For tori, the classical theorem of Moise rules out exotic structures in dimensions~$\le 3$, but for $n\ge5$ there are tori that are homeomorphic yet not diffeomorphic to the standard $\mathbb{T}^{n}$. Hsiang and Wall first discovered such exotic structures using surgery theory, and Farrell–Jones later employed controlled topology and hyperbolisation techniques to build such examples~\cite{FarrellJones89}.  These “exotic tori’’ show that even the archetype of a flat manifold can admit non-standard smoothness when the dimension is high enough. Crucially, by Bieberbach's theorem, an exotic torus cannot admit a strictly flat metric, underscoring how a change in smooth structure can act as a rigid obstruction to standard Riemannian geometries.

Higher-dimensional manifolds with exceptional holonomy supply a final showcase.  Joyce constructed compact $7$-manifolds with holonomy $\mathrm{G}_{2}$ by resolving quotients of $\mathbb{T}^{7}$ and produced families that are homeomorphic yet distinguished by their $\mathrm{G}_{2}$ structures~\cite{Joyce00}.  More recently, Crowley, Goette and Nordström introduced an analytic $\nu$-invariant that separates many of Joyce’s examples which had previously been topologically indistinguishable~\cite{CrowleyNordstrom18}. These manifolds captivate both geometers and physicists: in M-theory a change of smooth structure on a $\mathrm{G}_{2}$ background can alter the spectrum of effective field theories, making exotic $\mathrm{G}_{2}$’s a bridge between pure mathematics and string phenomenology.

Taken together, exotic $\mathbb{R}^{4}$’s, Dolgachev and knot-surgery surfaces, corks, exotic tori, and $\mathrm{G}_{2}$ manifolds illustrate that unusual smooth structures pervade manifold theory well beyond the realm of spheres.  Each class is “interesting’’ for a different reason—whether it be uniqueness of dimension, interplay with gauge theory, local generation of global exoticity, or curvature-theoretic surprises. Last but not least, there is a final exotic manifold which is worth mentioning: the exotic $4-$sphere. They were not listed among the other cases for one reason: their existence is still uncertain, and, if proved, it would solve the long-standing smooth Poincare conjecture in dimension 4 (\cite{doi:10.1142/9789812772107_0004}).

\subsection{Scalar Fields and Functions}
\label{subsec:Scalar_fields_chap4}
Given some manifold $\mathcal{M}$, there is not many simple definitions that one can make without an atlas. Functions, i.e.~maps $\mathcal{M} \xrightarrow{} \mathbb{R}$ and curves, i.e.~maps $ \mathbb{R} \xrightarrow {}\mathcal{M} $, are two of them. Both of them appear copiously in general relativity and analogous theories; curves give a trajectory inside spacetime, and functions are nothing but scalar fields. It is interesting to note that functions/scalar fields are intimately linked to differentiable structure. In fact, they ``specify'' the differentiable structure, in a sense that is clarified below. Let us first recall that a topological function (a function defined on a topological manifold) might be smooth with respect to one choice of atlas, and not with respect to another one. It is straightforward to construct a one-dimensional example by considering the atlases described in Section \ref{sec:Exotic_diff_struct_chap3}. In essence, this feature boils down to the fact that each chart is only defined with a \textit{homeomorphism} between (a piece of) the manifold and $\mathbb{R}^n$. It also hints at an interesting theorem, which is that two atlases (on the same manifold) are compatible if and only if they determine the same set of smooth functions (see \cite{1717244}). But, for the same reason outlined in Section \ref{sec:Exotic_diff_struct_chap3}, the resolution to this seemingly troublesome fact comes by considering that each function can be pulled back via the diffeomorphism that relates the two manifolds. This shows that the discontinuous nature of the function is in fact an artefact of a ``weird'' but harmless atlas choice, which can be put in smooth one-to-one correspondence with the ``nice'' atlas. This is, of course, assuming the existence of such a diffeomorphism. A pair of exotic manifolds, by definition, does not have this feature; therefore, smooth functions in one manifold cannot always be pulled back to smooth functions in the other manifold. Given the close relation between scalars and atlases, it is curious to note that one of the fathers of scalar-tensor theories, Brans (\cite{Brans:2008zz,brans1997gravitytenaciousscalarfield}), is also one of the pioneers in gauging the implications of exotic differentiable structures in physics. Nothing but a funny coincidence.

\section{Homeomorphisms}
\label{sec:Homeo_chap4}
In this section, we present a few homeomorphisms between one exotic sphere, the Gromoll--Meyer one, and the ordinary 7-sphere. While the homeomorphism in the twisted sphere picture is a straightforward application of Alexander's trick, the map that we obtain within Milnor's construction is a modification of a little-known approach due to Tamura. The relevance of this study is two-fold. Firstly, it provides some intuition on what mathematical features are behind the smooth incompatibility between two exotic manifolds, which might help developing some constructive results on the generation of new exotic manifolds. Secondly, it is relevant for concretely gauging the implications of exotic differentiable structures in the context of GR, which is the subject  of the next section.

\subsection{Alexander's Trick}
\label{subsec:Alexander_chap4}
As mentioned in Section \ref{subsec:More_facts_chap4}, consider an orientation–preserving diffeomorphism
\begin{align}
f\colon S^{6}\;\longrightarrow\;S^{6} \, ,
\end{align}
which is smooth, but not smoothly isotopic to the identity. Then, gluing two copies of the $7$--disc along their boundary according to $f$ produces the
twisted sphere
\begin{align}
\Sigma^{7}_{f}=D^{7}\;\cup_{\,f}\;D^{7} \,.
\end{align}
Such a manifold inherits a smooth structure making
it a closed, smooth $7$--manifold. If $f$ is the identity, then one obtains the usual $7$--sphere, where each disc is a hemisphere:
\begin{align}
     S^7 \simeq \Sigma^7_{Id} = D^{7}\;\cup_{\,Id}\;D^{7} \,.
\end{align}
Let us now construct a homeomorphism between a generic twisted sphere and $S^7$, via the so-called \textit{Alexander's trick}; the idea is to provide a way to extend the diffeomorphism $f: S^6 \to S^6$ on the boundary to a homeomorphism $\tilde{f}: D^7 \to D^7$ of the entire disk. Alexander's trick achieves this by ``coning-off'' the map from the boundary. Using polar coordinates $(t,v)$ for a point in the disk, where $t \in [0,1]$ is the radius and $v \in S^6$ is the direction, the extension is defined by $\tilde{f}(tv) = t f(v)$. This map is a homeomorphism that agrees with $f$ on the boundary (where $t=1$). By combining the identity map on the first disk with this cone map $\tilde{f}$ on the second (note that they agree on the boundary), we obtain a well-defined global homeomorphism $H: S^7 \to \Sigma^7_f$, proving they are topologically identical. The only locus where the map is not smooth is at the origin of the disc ($t=0$), creating a ``conical singularity''. One can even get rid of this, by replacing the linear scaling factor $t$ with a carefully chosen smooth function $\beta: [0,1] \to [0,1]$ such that $\beta(0)=0$, $\beta(1)=1$, $\beta'(t)>0$ for $t\in(0,1]$ and all of its derivatives vanish at $t=0$. The new extension, $\tilde{f}_{\text{smooth}}(tv) = \beta(t)f(v)$, is now smooth everywhere, including the origin. This specific example is a manifestation of a more general result (see \cite{Lance+2000+73+104}, Corollary 2.2): for $n>4$, any two homotopy $n$--spheres
are homeomorphic by a map which is a diffeomorphism except perhaps at
a single point. The singularity at that point might be quite wild, and we now make a few comments about this. Let
\begin{align}
H(t,v)=\alpha(t)\,f(v) \, ,
\end{align}
where
\(\alpha:[0,1]\to[0,1]\) can be set equal to the bump function $\beta(t)$ described above, to $t$, or to any arbitrary function between $0$ and $1$.

Then, the differential (in polar coordinates $(t,v)$) takes the schematic block–diagonal form
\begin{align}
DH_{\text{polar}}(t,v)=
\begin{pmatrix}
\alpha'(t) & 0\\
0 & D f(v)
\end{pmatrix} \, ,
\end{align}
where \(D f(v)\) is the differential of the map $f$ on $S^6$. To understand the true geometric singularity at the origin, however, one must convert this into the Cartesian Jacobian matrix $DH_{\text{Cart}}(x)$ for $x=tv$, whose determinant scales with the geometric volume as:
\begin{align}
\det DH_{\text{Cart}}(x)=\left(\frac{\alpha(t)}{t}\right)^6 \alpha'(t)\det Df(v) \, .
\end{align}
The singularity at the Cartesian origin $x=0$ can therefore manifest itself in two possible ways. For $\alpha(t) = t$, the map is Lipschitz but not $C^1$; the Cartesian Jacobian matrix depends purely on the direction $v=x/|x|$ of approach, resulting in a discontinuity since it does not have a well-defined limit at the origin. For $\alpha(t) = \beta(t)$ (with the flat properties described above), all derivatives vanish at $t=0$. The Cartesian Jacobian matrix smoothly approaches the zero matrix as $x \to 0$, meaning it becomes completely degenerate, i.e.~non-invertible, at the origin. Both singularities are quite severe. One of the reasons for this is that Alexander's trick is somewhat a ``brute force'' tool; it does not depend in any way on the details of the exotic diffeomorphism $f$. In order to construct homeomorphisms with milder singularities, one might consider exploiting the details of the exotic diffeomorphisms; which is possible for the case $S^6 \xrightarrow{} S^6$, as a few concrete expressions have been worked out. If one lets $S^6=\left\{(p, w) \in \mathbb{H} \times \mathbb{H}:|p|^2+|w|^2=1, \mathrm{Re}(p)=0\right\}$, then an exotic diffeomorphism reads (\cite{Duran2001,7ceb639f-641d-3fa7-9bc3-1b419c5ba656}):
\begin{align}
    \sigma(p, w)= \begin{cases}\left(\frac{1}{\left(1+p^2\right)^2} \bar{w} \mathrm{e}^{-\pi p} w p \bar{w} \mathrm{e}^{\pi p} w, \frac{1}{\left(1+p^2\right)} \bar{w} \mathrm{e}^{-\pi p} w \mathrm{e}^{\pi p} w\right), & w \neq 0 \, , \\ (p, 0), & w=0 \, ,\end{cases} 
\end{align}
where
\begin{align}
e^{\pi p}\;=\;\cos\!\bigl(\pi|p|\bigr)\;+\;\frac{p}{|p|}\,\sin\!\bigl(\pi|p|\bigr) \, .
\end{align}
It is to be noted that since $p$ is a purely imaginary quaternion, then $|w|^2 = 1 - |p|^2 = 1 +p^2$. It is possible, however, to deform such a map into one only involving first and second powers of the quaternions, according to \cite{DURAN2009206}. These leads to the exotic diffeomorphism:
\begin{align}
R(p,w)=\left(
\frac{(1+4p^{2}+wp\bar{w})\,p\,(1+4p^{2}-wp\bar{w})}
     {(1+4p^{2})^{2}-|w|^{4}p^{2}},
\;
\frac{(1+4p^{2}+wp\bar{w})\,w\,(1+4p^{2}-wp\bar{w})}
     {(1+4p^{2})^{2}-|w|^{4}p^{2}}
\right) \, .
\end{align}
It is reasonable to wonder whether one could get some control over the obstruction by considering either of these maps explicitly and investigating possible continuous but non-differentiable deformations or glueings. This leads to consider the properties of
$\pi_{0}\bigl(\Diff^{+}(S^{6})\bigr)$, which is still not completely understood; some relevant known properties and fact can be found in Section \ref{subsec:More_definitions_and_theorems}. Understanding the intricate relation between exotic diffeomorphisms of $S^6$ and topological maps is a promising avenue for understanding the possible ``defects'' associated with homeomorphisms between exotic spaces. This is work in progress, and we now review another approach to tackling the same question.

\subsection{Tamura's Map}
\label{subsec:Tamura_chap4}

\subsubsection{The Original Construction}
\begin{figure}
\begin{subfigure}{0.49\textwidth}  
        \centering
        \includegraphics[width=0.98\textwidth]{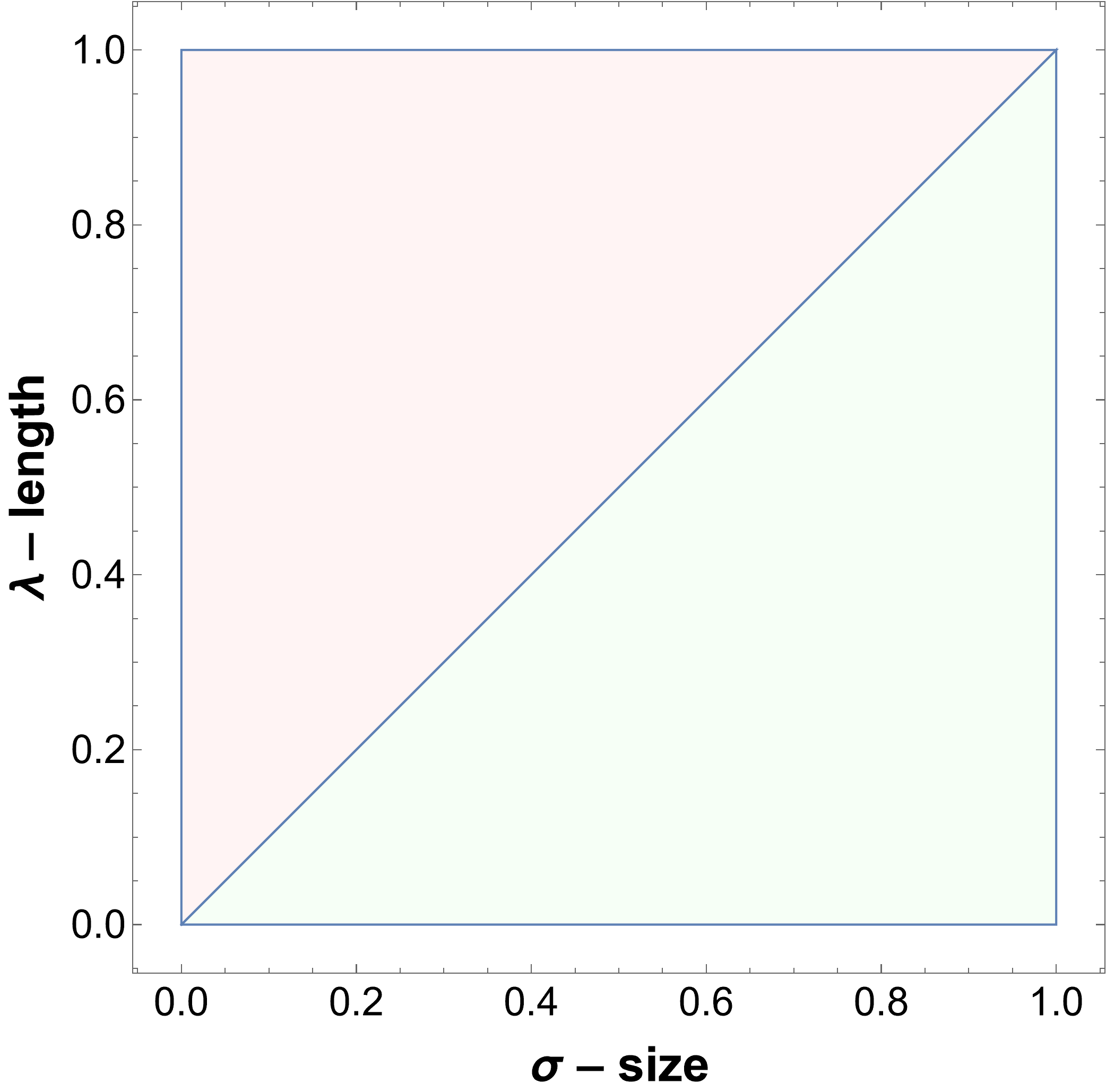}
        \caption{}
    \end{subfigure} 
    \hfill
    \begin{subfigure}{0.49\textwidth}  
        \centering
        \includegraphics[width=0.98\textwidth]{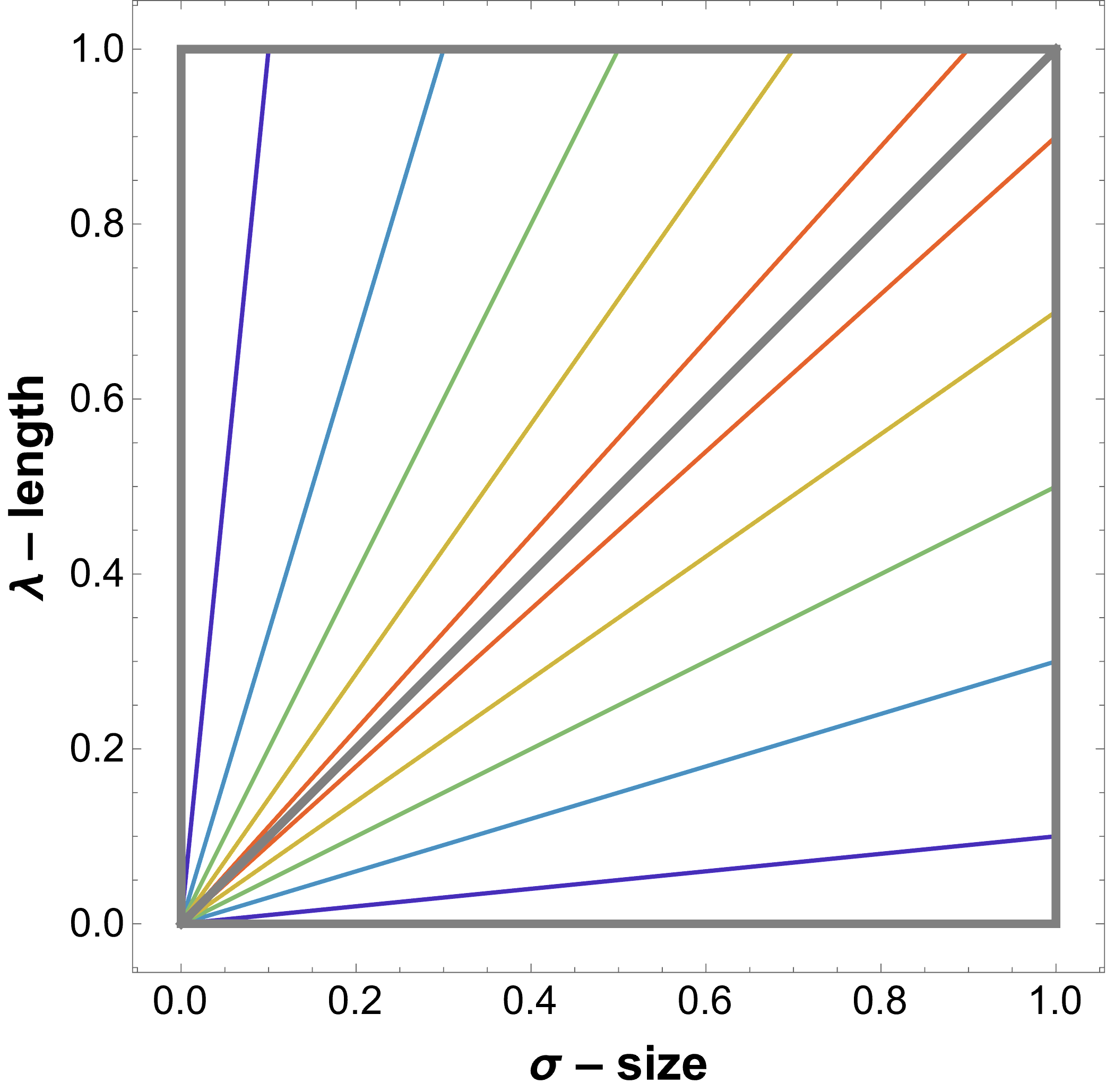}
        \caption{} 
    \end{subfigure} 
    \caption{The left plot shows how the parametrisation in \eqref{eq:Paramatrisation_easy} splits the $\sigma$-$\lambda$ square into two regions; one corresponding to $l(a,[b]_s) $, and one to $l([a]_s,b) $. The right plot contains a more detailed illustration of the lines corresponding to \eqref{eq:Paramatrisation_easy}. Purple corresponds to $s \sim 0$ (the lines coincide with the edges of the box at $s = 0$) and red to $s \sim 1$ (the two families of lines meet there, becoming the diagonal of the square).}
    \label{fig:two_regions}
\end{figure}

Two years after Milnor's discovery/invention of exotic sphere, Tamura provided the first explicit homeomorphism between an ordinary sphere and an exotic one (see \cite{10.2969/jmsj/01010029}). Although the expression is arguably quite cumbersome, the advantage of this map is that it can help making manifest the \textit{nature} of the obstruction that prevents a map between two inequivalent differentiable structures from being uplifted to a diffeomorphism.
There are a few quantities to be introduced in order to present the original construction following \cite{10.2969/jmsj/01010029} closely. Then, we proceed to demystify and clarify some of its features. Note that the construction presented in this section will appear in a more detailed form in \cite{TanNewPaper}.\\
Let $E^4$ be the interior of the three-sphere defined by unit quaternions, i.e. $E^4 = \{ q \,\, \mathrm{s.t.} \,\, ||q||^2 <1 \}$ and $\partial E^4 = S^3 = \{ q \,\, \mathrm{s.t.} \,\, ||q||^2 =1 \}$. On $S^4$, we label the North pole as $x_1$ and the south pole as $x_2$, and define $V = S^4 - x_1$, $V' = S^4 - x_2$. Then, we construct Milnor's bundles as follows. We define the two bundle charts as $p^{-1}(V) \simeq E^4 \times S^3$ and $p^{-1}(V^{-1}) \simeq E^4 \times S^3$. The points $(u,v)$ and $\left((1-\|u\|) u /\|u\|, u^{m+n} v u^{-m} /\|u\|^n\right)^{\prime}$, each defined in its own coordinate system, define the same point on the manifold.\footnote{Note that the labelling of the quaternions' exponents is different from Milnor's original construction; this alternative convention was discussed in Section \ref{subsec:Milnor_construction_chap3}.} These patches, together with the equivalence relation just specified, define Milnor's bundles. Let the corresponding total space be denoted by $B_{m,n}$, when referring to it as a topological manifold, and $M_{m,n}$, when it is endowed with the natural differentiable structure associated with the above equivalence relation. \\
The construction just outlined, which is the one used by Tamura, differs by Milnor's one in two simple ways: the choice of atlas on $\mathbb{S}^4$ and the choice on how to label the ``twisting'' of the bundle. Regarding the former, Tamura's coordinates are obtained by combining the standard stereographic projection with a transformation of the form:
\begin{align}
    \phi: \R^4 &\xrightarrow{} B^4 \, \nonumber \\
    y &\mapsto \phi(y) = \frac{y}{ 1 + ||y||} \, .
\end{align}
Regarding the latter, it corresponds to a different choice of generators for $\pi_3 (SO(4))$, and it is reviewed in Section \ref{subsec:Milnor_construction_chap3}.
Now let us define some curves.
\begin{itemize}
    \item Let $a \in S^3$ be a unit quaternion with unit norm. Then, given $0 \leq t \leq 1$, we define $ta \in \bar{E}^4$ as $[a]_t$. It follows that $\{ [a]_t \,\, s.t. \,\, 0 \leq t < 1 \} = E^4$.
    \item Let $S^2$ be defined by unit quaternions $u$ s.t. $\mathrm{Re}(u) = 0$. Let $b$ be a point on $S^2$, and let $\overline{(1,b,-1)}$ be the arc from -1 to 1 in $S^3$, passing through $b$. Then, we define $b_t$ as the point on that arc which is distant $\pi t$ from -1. This implies that $\{ b_t \,\, s.t. \,\, 0 \leq t \leq 1 \} = S^3$. 
\end{itemize}
Some more definitions are needed. We define $S^3_1 = p^{-1}(x_1) \subset p^{-1}(V')$. Moreover, let $E^4_1 = \left\{\left([a]_t, 1\right) ; a \in S^3, 0 \leq t<1\right\} \subset p^{-1}(V)$. Then, since $([a]_t , 1)$ and $((1-t)[a]_t / t, a^n )'$ are identified, we can associate any point $a$ on the boundary of $E^4_1$ with the point $a^n \in S^3_1$. Let us consider not $p^{-1}(V) - E^4_1$, and construct the following decomposition. Given $0 \leq s \leq 1$, we can define the following curves:
\begin{align}
   l(a,[b]_s) = \{([a]_t, b_{st} ) \, , \,\, 0 \leq t < 1 \}\, , \nonumber \\
    l([a]_s ,b) = \{( [a]_{st} , b_t ) \, , \,\, 0 \leq t < 1 \}\, .
    \label{eq:Paramatrisation_easy}
\end{align}
The set of all these curves covers $p^{-1}(V) - E^4_1$. We can also define their closures in $B_{m,n}$, which we denote as $\bar{l}(a,[b]_s)$ and $\bar{l}([a]_s ,b)$, respectively. Then, we have that the set of all these curves covers $B_{m,n}$. Also, $\bar{l}(a,[b]_s) - l(a,[b]_s)$ and $\bar{l}([a]_s ,b) - l([a]_s ,b)$ are contained in $E^4_1 \cup S^3_1$. \\
It is clear that this decomposition of $B_{m,n}$ into $1$-dimensional curves relies on two quaternions (the unit quaternion $a$ and the imaginary unit quaternion $b$), and two scalar parameters, which are: the ``size'' parameter which multiplies $a$ and the ``length'' parameter which specifies the position of the imaginary unit quaternion passing through $b$ along the great circle of $S^3$. These parameters, \textit{size} and \textit{length}, span the range $[ 0, 1 ]$, and we will denote them by $\sigma$ and $\lambda$, respectively. Any set of curves which covers $B_{m,n}$ must also cover this whole square region in the $\sigma$-$\lambda$ plane. The choice of parametrisation of $l(a,[b]_s)$ and $l([a]_s ,b)$, represented in Figure \ref{fig:two_regions}a and \ref{fig:two_regions}b, is one possibility. 
To obtain a set of curves which correspond to circles of longitude, however, a different parametrisation of the $\sigma$-$\lambda$ plane is needed. Let us now specialise to the case $n=1$. In Milnor's notation, this corresponds to $h+l = 1$, i.e.~the condition which guarantees topological equivalence to the 7-sphere. For this special case, in \cite{10.2969/jmsj/01010029}, two sets of curves spanning the whole manifold and only meeting at the poles $(0,1)$ and $(0,-1)$ are constructed by Tamura as follows:
\begin{align}
    \begin{aligned} & 0 \leq t \leq 1 / 4, \quad f_m\left([a]_s, b\right)(t)=\left(l\left([a]_s, b\right), t\right) \quad(s \neq 1), \\ & 1 / 4 \leq t \leq 1, \quad f_m\left([a]_s, b\right)(t)=\left(l\left([a]_{\frac{4}{3}s(1-t)}, b\right), t\right) \quad(s \neq 1), \\ & 0 \leq t \leq 1 / 4, \quad f_m\left(a,[b]_s\right)(t)=\left(l\left(a,[b]_s\right), t\right), \\ & 1 / 4 \leq t \leq 1 / 2, \quad f_m\left(a,[b]_s\right)(t)=\left(l\left(a,[b]_s\right), \frac{1}{4}+3(1-s)\left(t-\frac{1}{4}\right)\right), \\ & 1 / 2 \leq t \leq 3 / 4, \quad f_m\left(a,[b]_0\right)(t)=\left(l\left(a,[b]_0\right), 1\right), \\ & f_m\left(a,[b]_s\right)(t)=\left(l\left(a^{m+1} b_s b_{+}^{-1}{ }_{4(1-s)\left(t-\frac{1}{2}\right)} a^{-m},\right.\right. \\ & \left.\left.\left[\left(a^{m+1} b_s b_{s+4(1-s)\left(t-\frac{1}{2}\right)}^{-1} a^{-m}\right)^{-m} a^m b a^{-m}\left(a^{m+1} b_s b_{s+4(1-s)\left(t-\frac{1}{2}\right.}^{-1}\right) a^{-m}\right)^m\right]_{s+4(1-s)\left(t-\frac{1}{2}\right)}\right), \\ & \left.1-\frac{3}{4} s\right) \quad(s \neq 0), \\ & 3 / 4 \leq t \leq 1, \quad f_m\left(a,[b]_0\right)(t)=\left(l\left([-a]_{4(1-t)}, b\right), 1\right), \\ & f_m\left(a,[b]_s\right)(t)=\left(l\left(\left[a^{m+1} b_s a^{-m}\right]_{4(1-t)},\left(a^{m+1} b_s a^{-m}\right)^{-m} a^m b a^{-m}\left(a^{m+1} b_s a^{-m}\right)^m\right),\right. \\ & \left.1-\frac{3}{4} s+3 s\left(t-\frac{3}{4}\right)\right) \\ & (s \neq 0) \text {. } \\ & \end{aligned}
\label{eq:Tamura_circles_general}
\end{align}
This piece-wise construction can be made more compact by defining the two sets of curves:
\begin{align}
    \begin{aligned}
& L^{(m)}\left(a,[b]_s\right)=\left\{f_m\left(a,[b]_s\right)(t) ; 0 \leqq t \leqq 1\right\} \, , \\
& L^{(m)}\left([a]_s, b\right)=\left\{f_m\left([a]_s, b\right)(t) ; 0 \leqq t \leqq 1\right\} \, .
\end{aligned}
\end{align}
As claimed in \cite{10.2969/jmsj/01010029}, these curves show how $B_{m,1}$ is topologically a 7-sphere; because they cover the manifold, do not have common points except the two ``poles'' $(0,1)$ and $(0,-1)$, and they continuously depend on $a,b$.\footnote{Note that the curve $s=1$ is excluded from $L([a]_s, b)$, because it is actually given by $L(a, [b]_1)$.} Moreover, for the case $m=0$, it is claimed that these curves can be deformed to the usual circles of longitude of $M_{0,1}$, i.e.~the usual $7-$sphere. From the map \eqref{eq:Tamura_circles_general}, a homeomorphism between $B_{0,1}$ and $B_{m,1}$ can be constructed as:
\begin{align}
    \begin{aligned}
& g_m\left(l\left([a]_s, b\right), t\right)=\left(l\left([a]_s, b\right), t\right) \quad(0 \leq t \leq 1 / 4), \\
& g_m\left(l\left([a]_{\frac{4}{3}} s_{(1-t)}, b\right), t\right)=\left(l\left([a]_{\frac{4}{3} s(1-t)}, b\right), t\right) \quad(1 / 4 \leq t \leq 1), \\
& g_m\left(l\left(a,[b]_s\right), t\right)=\left(l\left(a,[b]_s\right), t\right) \quad(0 \leq t \leq 1 / 4), \\
& g_m\left(l\left(a,[b]_s\right), \frac{1}{4}+3(1-s)\left(t-\frac{1}{4}\right)\right)=\left(l\left(a,[b]_s\right), \frac{1}{4}+3(1-s)\left(t-\frac{1}{4}\right)\right) \\
& (1 / 4 \leq t \leq 1 / 2), \\
& g_m\left(l\left(a b_s b_{s+4(1-s)\left(t-\frac{1}{2}\right)}^{-1},[b]_{s+4(1-s)\left(t-\frac{1}{2}\right)}\right), 1-\frac{3}{4} s\right) \\
& =\left(l \left(a^{m+1} b_s b_{s+4(1-s)\left(t-\frac{1}{2}\right)}^{-1} a^{-m},\right.\right. \\
& {\left[\left(a^{m+1} b_s b_{s+4(1-s)\left(t-\frac{1}{2}\right)}^{-1} a^{-m}\right)^{-m} a^m b a^{-m}\left(a^{m+1} b_s b_{s+4(1-s)\left(t-\frac{1}{2}\right)}^{-1} a^{-m}\right)^m\right]_{s+4(1-s)\left(t-\frac{1}{2}\right)},} \\
& \left.1-\frac{3}{4} s\right) \\
& (s \neq 0,1 / 2 \leq t \leq 3 / 4), \\
& \left.g_m\left(l[-a]_{4(1-t)}, b\right), 1\right)=\left(l\left([-a]_{4(1-t)}, b\right), 1\right) \\
& (3 / 4 \leq t \leq 1), \\
& g_m\left(l\left(\left[a b_s\right]_{4(1-t)}, b\right), 1-\frac{3}{4} s+3 s\left(t-\frac{3}{4}\right)\right) \\
& =\left(l\left(\left[a^{m+1} b_s a^{-m}\right]_{4(1-t)},\left(a^{m+1} b_s a^{-m}\right)^{-m} a^m b a^{-m}\left(a^{m+1} b_s a^{-m}\right)^m\right),\right. \\
& \left.1-\frac{3}{4} s+3 s\left(t-\frac{3}{4}\right)\right) \\
& (s \neq 0,3 / 4 \leq t \leq 1) .
\end{aligned}
\label{eq:Tamura_homeomorphism_genral}
\end{align}
The reader might be quite unhappy with the presentation of this map, because many aspects of its construction are quite obscure. Let us provide some natural questions.
\begin{enumerate}
    \item What motivates this choice of parametrisation of the ``size'' of the first quaternion and ``length'' of the second one, in this map? 
    \item Why both $a$ and $b$ get ``twisted'' in some regions of the map?
    \item Why is the twist present just in $L(a, [b]_s)$ and not in $L([a]_s, b)$? 
    \item We claimed (following \cite{10.2969/jmsj/01010029}) that two curves have no points in common except $(0,-1)$ and $(0,1)$. However, this seems not to be the case for $f_m([a]_s, b)(t)$ with $s=0$. In that case, it looks like the curves corresponding to different $a$'s and same $b$'s coincide. Why?
\end{enumerate}
These are answered in the next subsection.

\subsubsection{Circles of Longitude of the Ordinary Sphere}
\begin{figure} 
        \centering
        \includegraphics[width=0.48\textwidth]{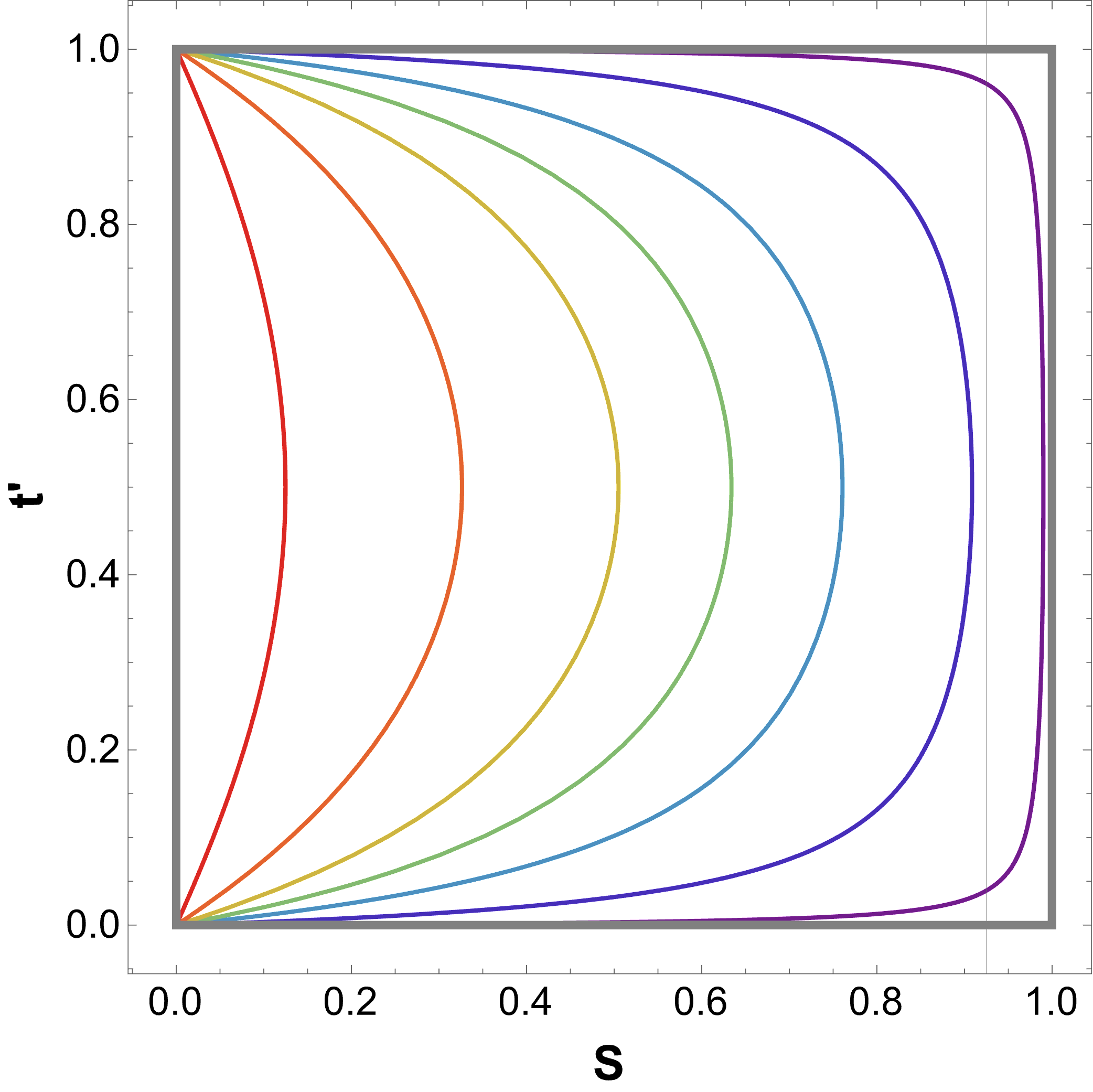}
        \caption{A plot of \eqref{eq:Smooth_parametrisation_standard_circles}, for different values of $s$: from $\sim 0$ (purple) to $\sim 1$ (red).}
        \label{fig:Smooth}

\end{figure}

The first logical step is to understand this map for the case of the ordinary sphere, i.e.~$m=0$. In this case, as we mentioned, it should be possible to deform these curves into the standard circles of longitude of the $7-$sphere realised as a quaternionic Hopf fibration. Let us review them.
If we let $S^7$ be defined by quaternions $z_0$ and $z_1$ whose norm squared add up to one, then the natural set of circles of longitude, connecting $(0,-1)$ and $(0,1)$ is given by:
\begin{align}
    z_0 = u \sin t \, , \quad  z_1 = \cos t + v \sin t \, ,
    \label{eq:Circles_Embedded}
\end{align}
where $||u||^2 + ||v||^2 = 1$ and $\text{Re}(v) = 0$. It follows from the quaternionic Hopf fibration (Section \ref{sec:ordinary_sphere_chap3}) that these curves in a local trivialisation read:
\begin{align}
    A = u \sin t (\cos t + v \sin t )^{-1} \, , \quad \, B = \frac{\cos t + v \sin t }{||\cos t + v \sin t ||} \, .
    \label{eq:Cirles_of_long}
\end{align}
By definition, $B$ is a unit quaternion, so we can identify it with $v'_{t'} = \cos t' + v' \sin t' $, where $v'$ is a unit imaginary quaternion. This leads to the identifications:
\begin{align}
    v' = v /||v|| \, , \quad \cos t'=\frac{\cos t}{\sqrt{\cos ^2 t+|v|^2 \sin ^2 t}}, \quad \sin t'=\frac{|v| \sin t}{\sqrt{\cos ^2 t+|v|^2 \sin ^2 t}} \, .
\end{align}
This yields $\tan t' =\frac{|v| \sin t}{\cos t}=|A| \tan t$, which gives 
\begin{align}
    t' =\arctan (|v| \tan t) \, , \quad  t =\arctan (\frac{1}{|v|} \tan t') \, .
    \label{eq:t_and_t'}
\end{align}
To be precise - and this is an important subtlety - one should write this as $t' = atan2(|v|\sin t, \cos t)$ or $t = atan2(\frac{1}{|v|}\sin t', \cos t')$ to specify which quadrant.
We have that $||u|| = \sqrt{1 - |v|^2}$, and we define $u' = u/||u|| = u / \sqrt{1 - |v|^2}$.
Then, the circles of longitude read:
\begin{align}
    A =  \frac{\sqrt{1 - |v|^2}}{\sqrt{\cos^2 t + |v|^2 \sin^2 t}} \,  \sin(\arctan (\frac{1}{|v|} \tan t')) \, u' v'_{-t'} \, , \quad B = v'_{t'} \, . 
    \label{eq:Circles_of_longitude_smooth_t'}
\end{align}
The map using the disc coordinates read:
\begin{align}
A =  \frac{ \frac{\sqrt{1 - |v|^2}}{\sqrt{\cos^2 t + |v|^2 \sin^2 t}} \,  \sin(\arctan (\frac{1}{|v|} \tan t'))}{1+ | \frac{\sqrt{1 - |v|^2}}{\sqrt{\cos^2 t + |v|^2 \sin^2 t}} \,  \sin(\arctan (\frac{1}{|v|} \tan t')) | } \, u' v'_{-t'} \, , \quad B = v'_{t'} \, .
\end{align}
By denoting $|v|$ with $s$, we can define 
\begin{align}
    S(t',s) =  \frac{ \frac{\sqrt{1 - s^2}}{\sqrt{\cos^2 t + s^2 \sin^2 t}} \,  \sin(\arctan (\frac{1}{s} \tan t'))}{1+ | \frac{\sqrt{1 - s^2}}{\sqrt{\cos^2 t + s^2 \sin^2 t}} \,  \sin(\arctan (\frac{1}{s} \tan t')) | } \, ,
    \label{eq:Smooth_parametrisation_standard_circles}
\end{align}
where $t$ carries an implicit dependence on $t'$ according to \eqref{eq:t_and_t'}. The behaviour of such a function, treating $t'$ as the independent variable (on the vertical axis) and $s$ as a parameter, is depicted in Figure \ref{fig:Smooth}. The reason for the unconventional choice of axes will become clear shortly. 

Before turning to Tamura's map and compare it to the circles of longitude just described, let us study this map in the other patch. According to Section \ref{subsec:quaternionic_hopf_chap3}, the coordinates of the second patch are obtained as $\tilde{A} = z_1 z_0^{-1}$ and $\tilde{B} = z_0 / ||z_0||$, so that the change of coordinates reads:
\begin{align}
    A' = 1/ A \, , \quad B' = \frac{A B}{ ||A||} \, . 
\end{align}
If one plugs \eqref{eq:Circles_Embedded}, the result is:
\begin{align}
    A' =  (\cos t + v \sin t ) u^{-1} (\sin t)^{-1} \, , \quad \, B' = \frac{ u \sin t}{|| u \sin t ||} = \frac{u}{||u||} \textrm{sgn}(\sin(t))\, . 
\end{align}
The ``origin'' in the other patch sits at $z_1 = 0$. This is achieved for $\cos t = 0$ and $v = 0 $. At those values, $B' = u$, so that the position on the fibre still depends on the quaternion specifying the circle of longitude.

\begin{figure}
\begin{subfigure}{0.49\textwidth} 
        \centering
        \includegraphics[width=0.98\textwidth]{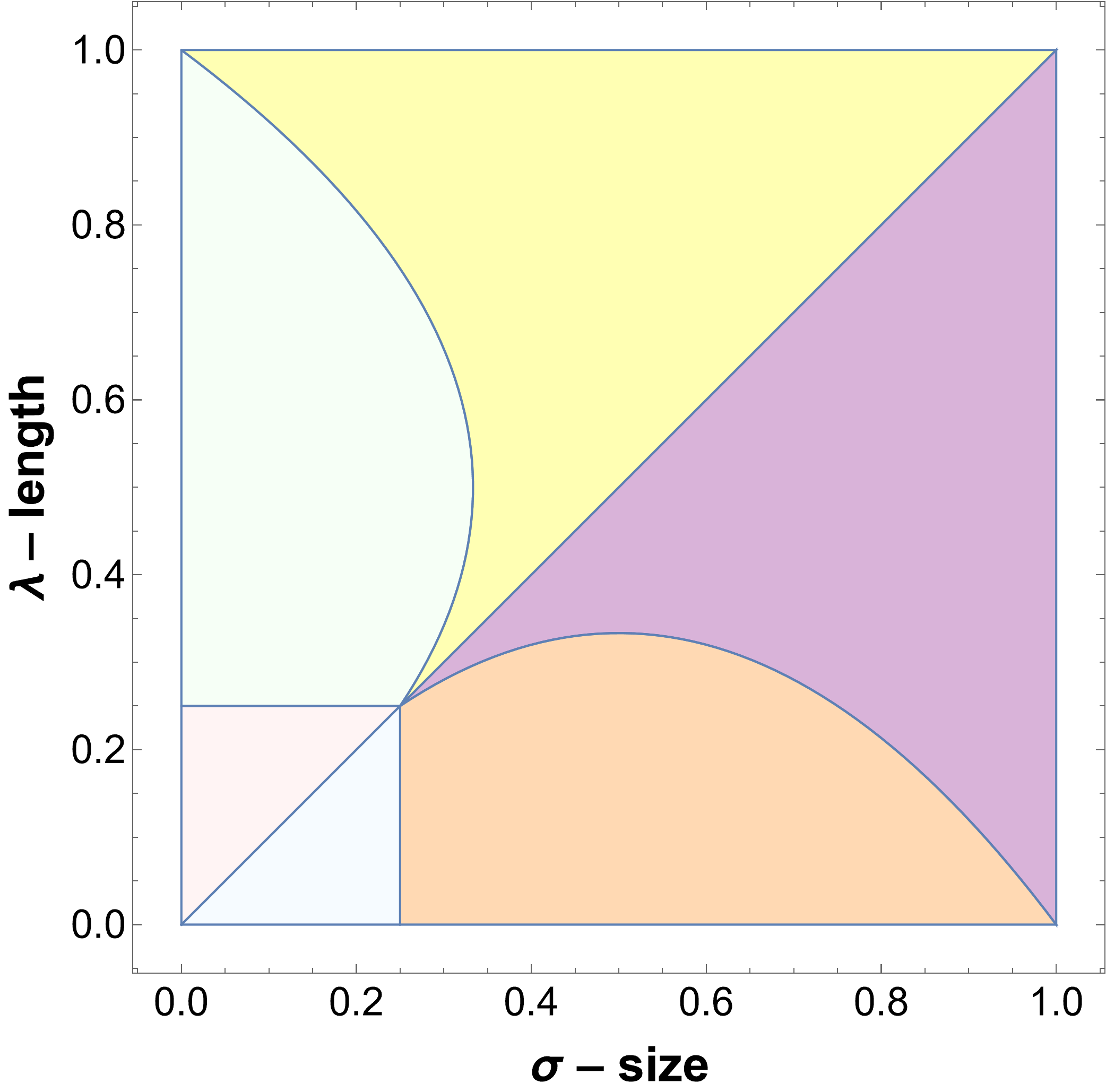}
        \caption{}
    \end{subfigure} 
    \hfill
    \begin{subfigure}{0.49\textwidth}  
        \centering
        \includegraphics[width=0.98\textwidth]{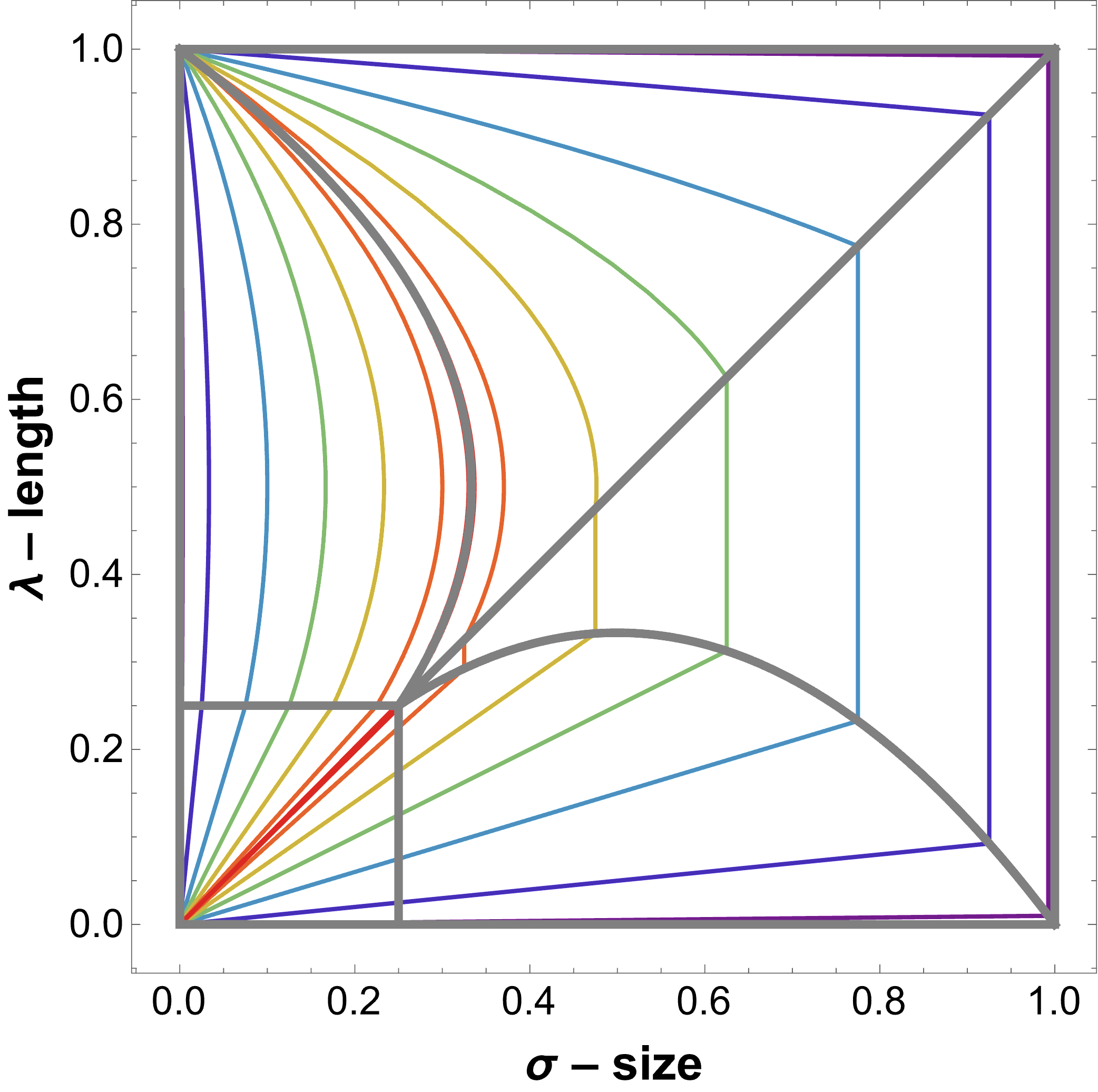} 
        \caption{}
    \end{subfigure} 
    \caption{The left plot shows the $6$ regions into which the $\sigma$-$\lambda$ plane is divided according to Tamura's expression for the circles of longitude in \eqref{eq:Tamura_circles_S7}. The pink and green regions (those that touch the $\sigma = 0$ axis) correspond to $l\left([a]_s, b\right)$, and the other four to $l\left(a, [b]_s\right)$. On the right plot, the lines are displayed for different choices of $s$: red means $s \sim 1$, while purple is $s \sim 0$.}
    \label{fig:Six_regions}
\end{figure}

Let us now return to Tamura's map for the ordinary sphere (i.e.~$m=0$), which reads:
\begin{align}
    \begin{aligned} & 0 \leq t \leq 1 / 4, \quad f_m\left([a]_s, b\right)(t)=\left(l\left([a]_s, b\right), t\right) \quad(s \neq 1), \\ & 1 / 4 \leq t \leq 1, \quad f_m\left([a]_s, b\right)(t)=\left(l\left([a]_{\frac{4}{3}s(1-t)}, b\right), t\right) \quad(s \neq 1), \\ & 0 \leq t \leq 1 / 4, \quad f_m\left(a,[b]_s\right)(t)=\left(l\left(a,[b]_s\right), t\right), \\ & 1 / 4 \leq t \leq 1 / 2, \quad f_m\left(a,[b]_s\right)(t)=\left(l\left(a,[b]_s\right), \frac{1}{4}+3(1-s)\left(t-\frac{1}{4}\right)\right), \\ & 1 / 2 \leq t \leq 3 / 4, \quad f_m\left(a,[b]_0\right)(t)=\left(l\left(a,[b]_0\right), 1\right), \\ & f_m\left(a,[b]_s\right)(t)=\left(l\left(a \, b_s \, b^{-1}_{s+4(1-s)\left(t-\frac{1}{2}\right)} ,\right.\right. \\ & \left.\left[ b \right]_{s+4(1-s)\left(t-\frac{1}{2}\right)} \right), \\ & \left.1-\frac{3}{4} s\right) \quad(s \neq 0), \\  & 3 / 4 \leq t \leq 1, \quad f_m\left(a,[b]_0\right)(t)=\left(l\left([-a]_{4(1-t)}, b\right), 1\right), \\ & f_m\left(a,[b]_s\right)(t)=\left(l\left(\left[a \, b_s \right]_{4(1-t)},b\right),\right. \\ & \left.1-\frac{3}{4} s+3 s\left(t-\frac{3}{4}\right)\right) \\ & (s \neq 0) \text {. } \\ & \end{aligned}
    \label{eq:Tamura_circles_S7}
\end{align}
By comparing this with the standard (smooth) circles of longitude just defined, we address the four questions posed above, and therefore reveal the reason behind some apparently obscure features of Tamura's map. Let us begin with the choice of parametrisation employed in \eqref{eq:Tamura_circles_S7} (as well as \eqref{eq:Tamura_circles_general}). As shown in Figure \ref{fig:Six_regions}a, the map is defined piece-wise according to a split of the $\lambda - \sigma$ (length-size) plane into $6$ regions. The two regions that contain the vertical axis $\sigma = 0$ (coloured in light pink and green) correspond to the two pieces of $f_m([a]_s,b)(t)$. The other four regions are associated to $f_m(a,[b]_s)(t)$, in the order given by starting at the origin and going anti-clock-wise. These regions cover the whole $\lambda - \sigma$ square. The finer picture is given in Figure \ref{fig:Six_regions}b, where we plot the curves in the $\lambda - \sigma$ plane that are implicitly defined in \eqref{eq:Tamura_circles_S7}. They are treated as functions of $t$, and plotted for a range of values of the parameter $s$. The boundaries of the various regions are highlighted in gray. According to Figure \ref{fig:Six_regions}, the curves which entirely lie in the two regions on the left correspond to $f_m([a]_s,b)(t)$, while those that are on the right belong to $f_m(a,[b]_s)(t)$.
The similarity between this set of piece-wise functions and that depicted in Figure \ref{fig:Smooth} is evident. In fact, one is simply the discrete version of the other, naturally associated with the partition of the $\lambda - \sigma$ plane shown in Figure \ref{fig:Six_regions} a. Hence, $s$ in \eqref{eq:Tamura_circles_general} and \eqref{eq:Tamura_circles_S7} can be naturally identified with $|v|$, and $t$ in \eqref{eq:Tamura_circles_general} with $t'$. This answers the first question of the list, but it raises another one: why should one consider this partition in the first place? The answer is tightly related to that of questions 2. and 3. from the previous subsection, i.e.~it is related to the ``twist'' in Tamura's map. 

The (sub)set of curves $f_m([a]_s,b)(t)$ only contains points whose such that the size of $a$ is at most $1/3$; this means that the curves entirely lie in the first chart. Conversely, the set $f_m(a,[b]_s)(t)$ contains a curve ($s=0$) which passes through the pole \textit{not} contained in the first chart. Hence, the definition of $f_m([a]_s,b)(t)$ must be consistent and compatible with the transition functions of the bundle, which is why the ``twist'' is present in the generic map \eqref{eq:Tamura_circles_general}, and it depends on $m$. On the other hand, the curves $f_m([a]_s,b)(t)$ are ``insensitive'' to the global structure of the manifold, and therefore can be chosen to have a direct product structure. The reason behind this choice is that the general map \eqref{eq:Tamura_circles_general} contains terms of the form $a^{m+1} b a^{-m}$, which become ill-defined at $a=0$. Hence, the region of the $\lambda - \sigma$ plane near $\sigma = 0$ must ``shielded'' from the twist. The curves $f_m([a]_s,b)(t)$, which are defined exactly on a neighbourhood of $\sigma = 0$, do not contain the twist for this reason. This answers the second and third questions of the list, as well as explaining the need for partitioning the $\lambda - \sigma$ into separate regions.

Let us address the last question now. In \cite{10.2969/jmsj/01010029} it is claimed that two curves have no points in common except $(0,-1)$ and $(0,1)$, while this seems not to be the case for $f_m([a]_s, b)(t)$ with $s=0$, where different $a$'s correspond to the same point. It follows from the discussion above, that such a degeneracy is just the result of the quaternion $A$ into a ``size'' parameter and the unitary part. Hence, it is just an artifact of the parametrisation, and the curves are indeed disjoint except for the poles. 

After carefully examining the similarity between \eqref{eq:Tamura_circles_S7} and the standard circles of longitude \eqref{eq:Circles_of_longitude_smooth_t'}, we can neatly summarise Tamura's map for $S^7$ in a single line as
\begin{align}
        (a b_{f(s,t)} {S(t,s)},b_{t}) \, ,
\end{align}
where $f(s,t)$ is zero on the regions corresponding to $f_m([a]_s, b)(t)$ (i.e.~the two ones that touch the $\sigma = 0$ axis in Figure \ref{fig:Six_regions}a) and non-zero in those corresponding to $f_m(a, [b]_s)(t)$ (i.e.~the remaining four ones in Figure \ref{fig:Six_regions}b). This feature is not required for the case $m=0$, as we discussed, but it is necessary for having well-defined circles of longitude when $m>1$, since terms of the form $a^{m+1} b a^{-m}$ appear. The homeomorphism \eqref{eq:Tamura_homeomorphism_genral} is based on mapping circles of one bundle to circles of longitude of the other, and therefore the same behaviour of $f(s,t)$ is imposed for any value of $m$.

\subsubsection{Circles of Longitude of the Gromoll--Meyer Sphere}
Once the structure of Tamura's circles of longitude is clarified, then making them well-defined on the Gromoll--Meyer sphere amounts to replacing $a b_{f(s,t)} S(t,s)$ with $a^2 b_{f(s,t)} a^{-1}  S(t,s)$ and $b_t$ with $((a^2 b_{f(s,t)} a^{-1})a b_t a^{-1} (a^2 b_{f(s,t)} a^{-1}))_t$. This ``twist'', prescribed by Tamura's map, respects the patching of the bundle. The corresponding map from the $7$-sphere to the Gromoll--Meyer sphere is obtained by mapping circles of longitude to circles of longitude, as in \eqref{eq:Tamura_homeomorphism_genral}. The specific form of the obstruction to the differentiability of this map lies in the choice of the functions $f(s,t)$ and $S(t,s)$. We now construct it explicitly, but we do not use Tamura's coordinates and parametrisation; instead we work with coordinates analogous to the ones of the quaternionic Hopf fibration. As we have shown, for the ordinary sphere in these coordinates, the circles of longitude read:
\begin{align}
    A = u \sin t (\cos t + v \sin t )^{-1} \, , \quad \, B = \frac{\cos t + v \sin t }{||\cos t + v \sin t ||} \, ,
    \label{eq:Circles_of_longitude}
\end{align}
where $(t = \pi /2 , \, v = 0)$ and $(t = 3\pi /2 , \, v = 0)$ are excluded from the chart. To ease the notation, let us define $\cos t + v \sin t =: \mathrm{cs}(v, t)$. According to Section \ref{subsec:quaternionic_hopf_chap3}, the coordinates of the second patch are obtained as $A' = z_1 z_0^{-1}$ and $B' = z_0 / ||z_0||$, so that the change of coordinates reads:
\begin{align}
    A' = 1/ A \, , \quad B' = \frac{A B}{ ||A||} \, . 
\end{align}
If one plugs \eqref{eq:Circles_of_longitude}, the result is:
\begin{align}
    A' =  \mathrm{cs}(v, t) u^{-1} (\sin t)^{-1} \, , \quad \, B' = \frac{ u \sin t}{|| u \sin t ||} = \frac{u}{||u||} \textrm{sgn}(\sin(t)) \, .
    \label{eq:Circles_of_long_S7}
\end{align}
Here, it is the loci $u=0$ and $t = 0, \pi$ which are excluded.

Before moving to the ``twisted'' circles of longitude, let us first review the transition functions of  the Gromoll--Meyer sphere in these coordinates. Using tildes to avoid confusion with the ordinary $S^7$, we have that the change of coordinates between the two patches reads:
\begin{align}
    \tilde{A} \mapsto \tilde{A}' = 1 / \tilde{A} , \, \quad \tilde{B} \mapsto \tilde{B}' = \frac{\tilde{A}^2 \tilde{B} \tilde{A}^{-1}}{||\tilde{A}||} \, . 
\end{align}
Suppose now that we were to set $\tilde{A}$ and $\tilde{B}$ equal to the same expressions in \eqref{eq:Cirles_of_long}. Then, one would find $\tilde{B}' = u \, \mathrm{cs}(v, t)^{-1} \frac{u}{||u||}\, \mathrm{cs}(v, t) \, u^{-1}$, which does \textit{not} have a well defined limit at $(t = \pi /2 , \, v = 0)$ or $(t = 3 \pi /2 , \, v = 0)$. \\
This is just a confirmation of the fact that the circles of longitude need to be modified in order to account for the non-trivial twist which is intrinsic to exotic spheres. Based on Tamura's map, the choice for these ``twisted'' circles is:
\begin{align}
    &\tilde{A} = u^2 \sin(t) \, \mathrm{cs}(v, g)^{-1} u^{-1} \, , \nonumber \\ &\tilde{B} = u  \, \mathrm{cs}(v, g) \, u^{-1} \, \frac{\, \mathrm{cs}(v, t) \,}{||\, \mathrm{cs}(v, t) \,||} \,   u \, \mathrm{cs}(v, g)^{-1} u^{-1} \, ,
\end{align}
where $g = g(t, ||v||)$. A priori, these expressions are not defined at: $u =0$; $(t=\pi/2, v=0)$ and $(g=\pi/2, v=0)$; $(t=3\pi/2, v=0)$ and $(g=3\pi/2, v=0)$. However, by requiring that
\begin{align}
g(t,||v||) = 0 \quad \textrm{for} \quad ||v|| > 1 - \varepsilon \quad \quad \textrm{(constraint 1)} \, ,
\end{align}
then everything is well-defined at $u = 0$; this is the first constraint we encounter. The loci $(g=\pi/2, v=0)$ and $(g=3 \pi/2, v=0)$ are also new features, compared to the standard sphere case. A second constraint, associated with it, emerges from the global structure of the exotic sphere; in fact, the advantage of this ansatz is manifest when moving to the other patch:
\begin{align}
    \tilde{A}' = u \sin(t)^{-1} \, \mathrm{cs}(v, g) \,  u^{-2} \, , \nonumber \\ \tilde{B}' = \mathrm{sgn}(\sin(t)) \frac{u^2}{||u||} \, \mathrm{cs}(v, g)^{-1}\, \mathrm{cs}(v, t) \, \frac{|| \, \mathrm{cs}(v, g) \, ||}{||\, \mathrm{cs}(v, t) \,||} u^{-1} \, .
\end{align}
The potentially harmful loci here are $u = 0$, $t = 0, \pi$ (same as before), alongside with $(t=\pi/2, v=0)$ and $(g=\pi/2, v=0)$, and same for $3 \pi /2$. $\tilde{B}'$ has a well defined limit at the last two locations if one imposes 
\begin{align}
  g(\pi /2 , ||v||=0) = \pi /2 + n \pi  \, , \quad g(3\pi /2 , ||v||=0) = 3 \pi /2 + m \pi \quad\textrm{(constraint 2)} \, ,  
\end{align} 
where $n,m$ are arbitrary integers; this gives the second constraint on $g$. Note that, as part of this constraint, $g$ should attain the values $\pi / 2 + n \pi$ \textit{only} at those points.
Finding a function $g$ with these properties is straightforward. Let us assume that we choose a such a $g$. Then, the most stringent constraint, comes from considering the first derivatives of $\tilde{B}'$ at the points $(g=t=\pi /2 , ||v||=0)$ and $(g=t=3\pi /2 , ||v||=0)$. A necessary condition for the derivatives to have a unique limit is:
\begin{align}
    g_t\left(\pi /2 , 0\right)=1, \quad   g_t\left( 3 \pi /2 , 0\right)=1  \quad \textrm{(constraint 3)} \, ,
\end{align}
where $g_t$ denote the derivative of $g$ with respect to $t$. These constraints (together with the implicit one that $g$ should be periodic over $2 \pi$) are irreconcilable. The detailed proof can be found below; while the heuristics behind it are the following. \\
Constraint 1 imposes that $g = 0$ for all t when $||v||=1$; assuming that $g_{||v||}$ is a continuous family of maps, the constraint implies that the curve 
\begin{align}
    g_{1} :[0, 2 \pi] &\xrightarrow{} S^1 \subset \mathbb{R}^2 \nonumber \\
    t &\mapsto( 1,  0 ) \, ,
\end{align}
is just constant, having trivial homotopy. Therefore,  then for a fixed $w<1$, $g_{w}:[0, 2 \pi] \xrightarrow{} S^1 $ cannot loop around the circle; since this would imply a change of homotopy. So the curve $g_{w}$ has to ``turn back on itself'' somewhere. But this feature is not compatible with the derivatives requirement at $t = \pi / 2$ and $t = 3 \pi / 2$ (constraint 3), if one wants to only have zeroes of $\cos(g)$ at $t = \pi /2$ and $t = 3 \pi /2$ only (constraint 2). \\
What does this imply? It is unavoidable that the derivatives of $\tilde{B}'$ develop a discontinuity at $(t= 3 \pi /2, ||v|| = 0)$, i.e.~on the three-sphere defined by $||u||^2 = 1$. \\
A current work in progress is to write down in a sensible way ``how'' the derivatives are discontinuous on the $S^3$ above; in other words, along which directions the derivatives give different limits. The proof that this foliation cannot be smoothed further than $C^0$ goes as follows. \\
Let \(h(t) := g(t,0)\). To analyze the derivative, we consider the continuous real-valued lift of \(h\) to \(\mathbb{R}\). 
Constraint 1 imposes that \(g(t, s) = 0\) for \(s > 1-\varepsilon\), which means the boundary curve has a winding number of zero. By the continuity of \(g\) with respect to \(||v||\), the winding number of \(h(t)\) must also be identically zero. Therefore, its real-valued lift strictly satisfies:
\begin{align}
    h(0) = h(2\pi) \, .
    \label{eq:winding_zero}
\end{align}
Because \(h(0) = h(2\pi)\), we can uniquely extend \(h\) to a \(2\pi\)-periodic continuous function on all of \(\mathbb{R}\), satisfying \(\tilde{h}(t+2\pi) = \tilde{h}(t)\) everywhere.

Constraint 2 dictates that \(\cos(\tilde{h}(t)) = 0\) if and only if \(t \equiv \pi/2 \pmod{\pi}\). In the real-valued lift, this means \(\tilde{h}(t)\) intersects the set \(S = \{\frac{\pi}{2} + n\pi \mid n \in \mathbb{Z}\}\) \textit{only} at these points.

Consider the open interval \(I_1 = (\pi/2, 3\pi/2)\). Since \(\tilde{h}(t)\) is continuous and cannot intersect \(S\) on this interval, \(\tilde{h}(I_1)\) must be entirely contained between two adjacent elements of \(S\). Thus, there exists a fixed integer \(k\) such that for all \(t \in I_1\):
\begin{align}
    k\pi - \frac{\pi}{2} < \tilde{h}(t) < k\pi + \frac{\pi}{2} \, .
\end{align}
By continuity, the limits at the endpoints must evaluate to the boundaries of this interval. 
Now we apply Constraint 3, which requires \(\tilde{h}'(\pi/2) = 1\) and \(\tilde{h}'(3\pi/2) = 1\). 
Because the derivative at \(\pi/2\) is strictly positive, \(\tilde{h}(t)\) is strictly increasing as it enters \(I_1\). It must therefore enter from the lower bound:
\begin{align}
    \tilde{h}(\pi/2) = k\pi - \frac{\pi}{2} \, . \label{eq:lower_bound_1}
\end{align}
Similarly, because the derivative at \(3\pi/2\) is strictly positive, \(\tilde{h}(t)\) is strictly increasing as it exits \(I_1\). It must therefore exit through the upper bound:
\begin{align}
    \tilde{h}(3\pi/2) = k\pi + \frac{\pi}{2} \, . \label{eq:upper_bound_1}
\end{align}
Subtracting \eqref{eq:lower_bound_1} from \eqref{eq:upper_bound_1} yields a strict geometric requirement:
\begin{align}
    \tilde{h}(3\pi/2) - \tilde{h}(\pi/2) = \pi \, . \label{eq:variation_1}
\end{align}

We can apply the exact same logic to the adjacent interval \(I_2 = (3\pi/2, 5\pi/2)\). Because \(\tilde{h}\) does not intersect \(S\) on \(I_2\), and because \(\tilde{h}'(3\pi/2) = 1\) and \(\tilde{h}'(5\pi/2) = \tilde{h}'(\pi/2) = 1\), \(\tilde{h}\) must again enter \(I_2\) from a lower bound and exit through an upper bound. This yields:
\begin{align}
    \tilde{h}(5\pi/2) - \tilde{h}(3\pi/2) = \pi \, . \label{eq:variation_2}
\end{align}

Adding \eqref{eq:variation_1} and \eqref{eq:variation_2} gives the total variation of \(\tilde{h}\) over one full period:
\begin{align}
    \tilde{h}(5\pi/2) - \tilde{h}(\pi/2) = 2\pi \, .
\end{align}
However, because \(\tilde{h}\) is \(2\pi\)-periodic as established by \eqref{eq:winding_zero}, we must have \(\tilde{h}(5\pi/2) = \tilde{h}(\pi/2)\). This implies the contradiction.

Therefore, no such continuous function \(g(t, ||v||)\) can exist. As a consequence, a discontinuity in the derivatives on the three-sphere defined by \(||u||^2 = 1\) is an unavoidable topological feature of this modified chart transition.

\subsection{Other Options}
\label{subsec:Other_homeos_chap4}
We would like to emphasize  that the two constructions just reviewed are not the only possibilities for realising a homeomorphism between the ordinary $7$--sphere and the Gromoll--Meyer one. 

In \cite{7ceb639f-641d-3fa7-9bc3-1b419c5ba656}, for instance (see pages 5039-5040), they propose a construction for obtaining a homeomorphism which is a diffeomorphism everywhere but at a point (one of the poles). Another option is to build a homeomorphism by considering the flow associated with the Morse function used by Milnor (\cite{10.2307/1969983, McEnroe2016MILNORSCO}). Finally, the realisation as Brieskorn spheres is also a potentially fruitful avenue, which could help understanding some features of the smooth ``defects'' through the study of Brieskorn links.

We do not comment further on this here, but we just note that many other closed-form expressions for homeomorphisms can be derived with reasonable effort. However, to our knowledge, a systematic study on the properties of the obstruction which forbids the uplift of the map to a diffeomorphism is still missing.

\section{Changing Differentiable Structure}
\label{sec:Changing_diff_struc_chap4}
In this section, we discuss the notion of ``change of differentiable structure'', proposing to interpret it as a \textit{global} change of coordinates. We emphasize how the non-differentiability is a main feature of this transformation and we argue that, despite the inherent discontinuities, it is possible for a change of differentiable structure to map a continuous metric to a continuous metric. Discussing direct and concrete applications to general relativity, however, comes at the cost of abandoning the well-established smooth category; we enter in the regime of distributional and generalised calculus. We emphasise that this section is, in order, speculative, mostly qualitative and incomplete. The aim of the following discussion is simply to present a set of (educated) guesses for how inequivalent differentiable structures might affect the geometry of spacetime.\\

Any map from an exotic (differentiable) manifold to its ``standard'' counterpart cannot be smooth with a smooth inverse, by definition. A natural question is to quantify the degree of non-smoothness of such a map. In other words, for which $k$ the map is $C^k$. If this $k$ could be made arbitrary large, then exotic spaces would essentially be physically indistinguishable from their standard counterparts. However, due to a theorem of Whitney, $k=0$ is the only option. This implies that the derivatives of the homeomorphism necessarily contain a discontinuity. Hence, given a smooth metric on one of the two manifolds, trying to pull-back, in the appropriate generalised or distributional sense, produces a discontinuous metric on the other manifold, in general. Metrics with discontinuities in their components have been discussed in a number of contexts and admit a formulation within GR, see for instance \cite{C_K_Raju_1982, Gemelli:2007tj}, and even Israel's junction conditions have been generalised to the discontinuous case (\cite{Thakur1998}). Despite this, it can be argued on physical grounds that the most relevant metrics are at least continuous (\cite{Israel1966, BarrabesIsrael1991, GerochTraschen1987, Taub1980}). Among the few exceptions, there is Penrose's pp-wave metric (\cite{Penrose:1972xrn}), which contains delta functions at the location of the wave:
\begin{align}
    \mathrm{d} s^2=f(x, y) \delta(u) \mathrm{d} u^2-\mathrm{d} u \mathrm{~d} v+\mathrm{d} x^2+\mathrm{d} y^2 \, , 
    \label{eq:Penrose_pp}
\end{align}
Penrose also provided an argument to relate such an expression to a continuous representation of the same metric, through a (formal) ``discontinuous change of coordinates''. Further study on this matter was presented in \cite{Aichelburg_1996, Aichelburg_1997}, where in the latter the authors interpret the discontinuous change of coordinates as a change of differentiable structure; a few years later, the same transformation was reviewed in a distributional sense and referred to as a ``generalized coordinate transformation'' in \cite{Kunzinger:1998xw}. Since then, further studies on transformations of similar nature and metrics of low regularity appeared from various members of the University of Vienna: \cite{Erlacher:2010ts, Steinbauer_2006, S_mann_2024, PhysRevD.100.024040}; they are aimed at developing the appropriate technical formalism based on distributional and generalised functions. \\
While the underlying mathematical framework developed in the papers above is very relevant to this work, the physical motivation differs. Those studies were considering the process of starting with a specific discontinuous metric and making it regular through a discontinuous change of coordinates. In our case, we begin with a non-smooth change of coordinates (the discontinuity is somewhat milder, since the map is $C^0$), and study the implications of this transformation on generic metrics.

Another set of works which is naturally connected to the current investigation is the one in \cite{Reintjes_2015, reintjes2017spacetimelocallyinertialpoints, Reintjes_2019, reintjes2020how, Reintjes:2022rmh, Reintjes_2023, reintjes2024essentialregularitysingularconnections}.
Some of the premises of these works are based on the fact that a smooth change of coordinates cannot remove discontinuities present in the metric or its derivatives, and therefore one needs to consider a non-smooth transformation to remove them. The key insight which lies at the core of our analysis is that transforming continuous metric by a discontinuous Jacobian, does not necessarily yield a discontinuous metric.\footnote{For other works concerning rigorous analysis of discontinuities arising in general relativity, see \cite{KunzingerSteinbauer2002, SteinbauerVickers2006, Chru_ciel_1998,Sbierski:2015nta,a965bffe-de90-3a7a-a652-4c0d2147a13b}.} \\
In other words, they employ discontinuous Jacobians to cancel discontinuities in the metric; we instead focus on the case where the discontinuities of the Jacobian cancel out among themselves in the transformation of the metric. Let us spell this out in more details, with the simplest possible example - too simple to arise directly from a homeomorphism between two exotic manifolds, but still very illustrative. Consider the pull-back of the metric, which reads:
\begin{align}
    g'(y) = J(y) g(y(x)) J(y)^T \, ,
\end{align}
where all objects are matrices and juxtaposition stands for matrix multiplication. Suppose that one wants to transform this under a homeomorphism relating two exotic manifolds. By definition, such a map fails to be differentiable, so let us suppose that this failure is localised on a hypersurface denoted by $y_0$. The details of $y_0$ are quite crucial, in that not much is known about which hypersurfaces the obstruction can be localised on. For the present discussion, we assume that it is a co-dimension $1$ locus, for the sake of simplicity. The following argument, however, can be adapted to other hypersurfaces. For $g'(y)$ to remain continuous on $y_0$ (where $J(y)$ is discontinuous), consider the left and right limits of $J$ as $y \rightarrow y_0$ :
\begin{align}
J\left(y_0^{-}\right), \quad J\left(y_0^{+}\right)
\end{align}

Since $g(y(x))$ is continuous, let $g_0=g\left(y_0\right)$.
We want:
\begin{align}
\lim _{y \rightarrow y_0^{-}} J(y) g(y) J(y)^T=\lim _{y \rightarrow y_0^{+}} J(y) g(y) J(y)^T .
\end{align}

With continuity of $g$, this reduces to:
\begin{align}
J\left(y_0^{-}\right) g_0 J\left(y_0^{-}\right)^T=J\left(y_0^{+}\right) g_0 J\left(y_0^{+}\right)^T .
\end{align}

Now suppose $J\left(y_0^{+}\right)$ and $J\left(y_0^{-}\right)$ differ by some invertible matrix $Q$, i.e.
\begin{align}
J\left(y_0^{+}\right)=J\left(y_0^{-}\right) Q
\end{align}
We substitute this into the continuity condition:
\begin{align}
J\left(y_0^{-}\right) g_0 J\left(y_0^{-}\right)^T=\left(J\left(y_0^{-}\right) Q\right) g_0\left(Q^T J\left(y_0^{-}\right)^T\right)
\end{align}

Multiplying on the left by $J\left(x_0^{-}\right)^{-1}$ and on the right by $\left(J\left(x_0^{-}\right)^{-1}\right)^T$, we simplify this requirement to:
\begin{align}
g_0=Q g_0 Q^T
\end{align}

This means the discontinuity ``jumps'' within the set of transformations that leave $g_0$ invariant under congruence. In other words, $Q$ must belong to the group:
\begin{align}
G_{g_0}=\left\{Q \in G L(n) \mid Q g_0 Q^T=g_0\right\} .
\end{align}

The group $G_{B_0}$ can be viewed as the automorphism group of the bilinear form defined by $g_0$. If $g_0$ is, for example, the identity matrix, then $G_{g_0}$ is the orthogonal group $O(n)$ (if we consider the real case).

Thus, if the discontinuity in $J$ occurs so that $J\left(y_0^{+}\right)$ differs from $J\left(y_0^{-}\right)$ by an element of the group $G_{g_0}$, then $g'(y)=J(y) g(y(x)) J(y)^T$ remains continuous at $y_0$. Clearly, unless the metric is Einstein, the corresponding Ricci tensor will develop a singularity, bringing us into the realm of shock waves - see \cite{Alcubierre_1997}.

\section{Summary and Outlook}
\label{sec:Conclusions_chap4}

In this chapter, we have surveyed various aspects of exotic $7$--spheres and exotic manifolds. We have presented a representation-theoretic obstruction to building the Gromoll--Meyer sphere as a quotient space, analogously to what is done for the ordinary sphere, and discussed in detail its realisation as a double quotient. We have also reviewed the classification of exotic $7$--spheres and commented on their known curvature properties. Then, we have moved to more general exotic manifolds, listing a number of interesting examples and expanding on the role of scalar fields in the definition of a differentiable structure. After this broad overview, we focused on a very specific aspect, present in any pair of exotic manifolds: the homeomorphic map proving their topological equivalence. By definition, and according to Whitney's theorem, the Jacobian of such a map contains a discontinuity or a degeneracy. Not much is known on the explicit form of such a ``defect'', which obstructs the uplift of the map to a diffeomorphism. For this reason, we presented two explicit realisations of homeomorphisms between the Gromoll--Meyer sphere and the ordinary one. One is obtained through the standard Alexander's trick, while the other appeared in a paper published shortly after Milnor's results (\cite{10.2969/jmsj/01010029}). We end by commenting on how such a homeomorphism could be identified with a continuous (but not differentiable) change of coordinates, and the effects of this on transformation on the geometry. This procedure only makes sense in the realm of generalised functions and distributional calculus, and it can be identified as \textit{global} change of coordinates, as opposed to \textit{local} ones that are common in general relativity. We emphasize that the material presented in the second part of the chapter is not sufficient for establishing a definite result on the role of differentiable structures in general relativity. The reason for this is mainly the absence of a systematic study concerning the obstructions present in the homeomorphisms between exotic manifolds. By presenting some explicit realisations, however, we hope to expose some features that might be common to a wider set of maps. Moreover, the discussion about a possible mathematical and physical interpretation of such transformations might help identifying some of the most interesting questions about this uncharted territory.

In addition to what was discussed in this chapter, we believe that it would also be worth investigating the concept of ``dynamical change of differentiable structure'', to be compared with the analogous change of topology in general relativity (\cite{Geroch1967Topology, Penrose1972Techniques,HawkingEllis1973,Gibbons:2011dh,Horowitz:1991fr}).
Another aspect that might require further inspection concerns what happens to Einstein's equations under a change of differentiable structure, i.e.~a non-differentiable change of coordinates. They are not invariant under such transformations, making this question quite non-trivial to answer(\cite{Westman_2009}, footnote 10).

\chapter{Machine Learning, Einstein Metrics \\ 
and AInstein \,\,\,\,\,\,\,\,\,\,\,\,\,\,\,\,\,\,\,\,\,\,\,\,\,\,\,\,\,\,\,\,\,\,\,\,\,\,\,\,\,\,\,\,\,\,\,\,\,\,\,\,\,\,\,\,\,\,}
\label{chap:5}
This chapter reviews a recent numerical method for approximating Riemannian Einstein metrics on arbitrary manifolds, based on machine learning (\cite{hirst2025ainsteinnumericaleinsteinmetrics}). We present its application to the case of ordinary spheres and discuss the route to generalising it to exotic ones.

\section{Introduction, Overview and Structure} 
Finding Einstein metrics on a given manifold has been a central problem in differential geometry for decades. 
An Einstein metric is defined by the condition $Ric(g)=\lambda g$, where $Ric$ is the Ricci curvature tensor, $g$ is the Riemannian metric, and $\lambda$ is a constant. 
These metrics play a prominent role in differential geometry and they are ubiquitous in theoretical physics since they solve Einstein's equations with a cosmological constant. 

Although finding Einstein metrics has been an active area of research for over a century, the field remains vibrant due to numerous unresolved questions. 
Some of them concern the existence (or non-existence) of Einstein metrics on various manifolds, whilst others concern finding appropriate closed-form expressions for those metrics in the cases where their existence has been proven non-constructively. 
Regarding the former type of question, one of the most well-known open problems is whether $S^2 \times S^2$ admits a Ricci-flat (or more generally non-standard Einstein) metric \cite{Besse:1987pua}. 
Similarly, the question of whether $S^n$, with $n>3$, admits non-round Einstein metrics with positive or zero Ricci curvature continues to be a major challenge \cite{Berger_2003}. 
Concerning the search for concrete description of metrics which are known to exist, the Calabi-Yau case stands out as the most prominent example \cite{Yau1978OnTR}, but there are many other analogous scenarios, like exotic $7$-spheres \cite{boyer2003einstein, boyer2004einstein}, as we mentioned in Chapter \ref{chap:3}. In addition to its relevance in differential geometry, finding a new Einstein metric numerically represents an impactful result in theoretical physics as well, with immediate applications in ordinary general relativity or higher-dimensional models, like Kaluza--Klein theories or supergravities. However, the problems mentioned above exemplify how difficult it is to either construct analytic solutions or prove they cannot exist, in the cases where there is little (if any) isometry involved.

This difficulty in verifying existence, and explicitly constructing metrics, has motivated efforts from physicists and mathematicians to explore and develop new methods which are computational in nature. 
Numerical approaches are crucial tools for generating results where analytic techniques are infeasible. 
Already many excellent works have developed numerical schemes to solve Einstein's equations in a variety of scenarios \cite{Figueras:2012xj, Pretorius:2005gq, Dias:2015rxy, Clough_2015, Lehner:2010pn, Dias:2015pda, Chesler:2013lia, Chaurasia:2025}, as far as construction of  black hole/string solutions \cite{Wiseman:2002zc, Kudoh:2003ki, Headrick:2009pv}.
However, these numerical approaches are subject to a curse of dimensionality, where scaling to higher dimensions and more parameters leads to an insurmountable demand for data.

It is here the recent revolution in novel methods of computation statistics can be capitalised on, where copious successes have been seen with application of these techniques across academic fields; techniques of \textit{machine learning}.
In recent years, the first applications of machine learning to numerically approximate metrics have occurred, for complex geometries relevant to string theory, holography, and numerical relativity. 
The most popular compactification spaces for string theory are Calabi-Yau manifolds, where there has been many exceptional works approximating their metrics with machine learning \cite{Ashmore:2019wzb, Douglas:2020hpv, Anderson:2020hux, Jejjala:2020wcc, Larfors:2021pbb, Ashmore:2021ohf, Larfors:2022nep, Berglund:2022gvm, Gerdes:2022nzr, Hendi:2024yin, Ek:2024fgd, Butbaia:2024xgj, Mirjanic:2024gek}; among these works are some very nice packages \cite{Gerdes:2022nzr, Butbaia:2024xgj}, notably \texttt{cymetric} \cite{Larfors:2021pbb} which we take structural inspiration from.
Other exemplary works numerically solving Einstein's equations for specific manifolds in restricted settings include \cite{deluca2024, Li:2023, chen2024}, where machine learning methods support their approaches.\footnote{Other exciting applications of machine learning within mathematical physics can be found in \cite{manningcoe2025grokkingvslearningfeatures,Armstrong-Williams:2024nzy,Berglund:2024reu,Costantino:2024joa,Hirst:2024abn,Hirst:2023kdl,Aggarwal:2023swe,Chen:2023whk,Berglund:2023ztk,He:2023csq,Cheung:2022itk,Chen:2022jwd,Bao:2022rup,Dechant:2022ccf,Hirst:2022qqr,Arias-Tamargo:2022qgb,Berman:2021mcw,Bao:2021ofk,Bao:2021ohf,Bao:2021olg,Bao:2021auj,Bao:2021vxt,Bao:2020nbi,He:2020eva,Bao:2020sqg, capuozzo2024machinelearningtoricduality, fivefolds,Niarchos:2023lot,Berman:2023rqb,Seong:2023njx,Halverson:2019tkf,Loges:2021hvn,Loges:2022mao,Chen:2020dxg,Kantor:2021kbx,Kantor:2021jpz,Kantor:2022epi,Abel:2014xta,Bies:2020gvf,Krippendorf:2021uxu,Constantin:2021for,Abel:2021rrj,Abel:2023zwg,Dubey:2023dvu, CONSTANTIN2025116778,Gao:2021xbs,coates2022machine,Coates2023,coates2023machine,Manko:2022zfz,Choi:2023rqg, Douglas:2024pmn, swirszcz2025advancinggeometryaimultiagent}. They include works on amoeba, branes configurations, Calabi--Yau manifolds in various dimensions,  conformal theories, (string and non-string) phenomenology, $G_2$ manifolds and polytopes. The age of application of machine learning to unveil new physical and mathematical understanding is, alluring, just at its beginning.}

In this chapter, we summarise a novel semi-supervised machine learning approach  to approximate general Einstein metrics on a broad class of manifolds. The code repository for the package can be found at: \url{https://github.com/xand-stapleton/ainstein}. It is written in \texttt{Python 3} and built on \texttt{TensorFlow} (\cite{tensorflow2015whitepaper}). We structure the chapter as follows.
Section \ref{sec:background_chap5} is devoted to introducing the relevant notions from differential geometry and machine learning. The general structure behind the \textit{AInstein} neural network is also presented. \\
In Section \ref{sec:results_chap5}, we demonstrate its potential by focusing on the specific case of spheres in dimensions $2,3,4,5$, with the aim of shedding light on longstanding open problems, providing new perspectives for analysis, and stimulating further research into the numerical and analytical aspects of Riemannian Einstein geometry. After this successful validation of our method, we aim at applying to other settings with larger relevance in theoretical physics, by looking for black hole solutions and moving to Lorentzian signature.

\section{Background}
\label{sec:background_chap5}
In this section, we introduce the differential geometry and machine learning background underlying the \textit{AInstein} package.
\subsection{Differential Geometry}
\label{subsec:bkg_dg_chap5}
When performing analytic calculations, the use of coordinates in expressions carries some disadvantages. To mention two, it often requires working with cumbersome formulae and can hide the global nature of the objects being described. When tackling a problem via numerical approximation techniques, the situation changes: there is no other choice than to implement coordinate expressions. This prompts the question of which coordinates shall be used to cover the manifolds considered in this work, i.e.~$n$-dimensional spheres. One of the most natural choices consists of the standard stereographic projection atlas. However, since its coordinates span $\mathbb{R}^n$ entirely, sampling and visualising a whole patch becomes non-trivial. For this reason, we use a modified version of the above, where the stereographic projection from $S^n$ to $\mathbb{R}^n$ is followed by a mapping of $\mathbb{R}^n$ to $B^n$, the $n$-dimensional unit open ball. 

Consider the usual stereographic atlas, as defined in Section \ref{subsec:Working_definitions_chapA1}.
Then, we compose this with an additional transformation, mapping the stereographic coordinates to the so-called \textit{ball coordinates}, $\phi_i: \mathbb{R}^n \xrightarrow{} B^n$, reads
\begin{equation}
\begin{split}
    \phi_1(X_1, X_2, \cdots, X_n) & = \left( \frac{X_1}{1+\sqrt{1+|X|^2}}, \frac{X_2}{1+\sqrt{1+|X|^2}} , \cdots , \frac{X_n}{1+\sqrt{1+|X|^2}} \right) \\
    & =: \left(x_1, x_2 , \cdots, x_n \right) \, ,
\end{split}
\end{equation}
where $|X|^2 = X_1^2 + X_2^2 + \cdots + X_n^2$; and similarly for $\phi_2$, which defines coordinates $(\tilde{x}_1, \tilde{x}_2, \cdots, \tilde{x}_n)$ for the second patch. The two ball patches are related by the coordinate transformation
\begin{align}
    \tau(x_1, x_2, \cdots , x_n) = \frac{|x| - 1 }{|x| (|x| + 1)} (x_1 , x_2 , \cdots , x_n) = (\tilde{x}_1, \tilde{x}_2, \cdots , \tilde{x}_n ) \, ,
    \label{eq:Change_of_coords}
\end{align}
whilst the entries of the corresponding Jacobian matrix read
\begin{align}
\label{eqn:jacobian}
    J_{ij} = \delta_{ij} \frac{|x|-1}{|x|(1+|x|)} + x_i x_j \frac{1 + 2 |x| - |x|^2 }{ |x|^3 (1 + |x|)^2} \, ,
\end{align}
for $i = 1, 2, \cdots, n$.
It follows from the dependence of the prefactor of \eqref{eq:Change_of_coords} on $|x|$ that co-dimension $1$ spheres centred at the origin in one patch get mapped into co-dimension $1$ spheres in the other. The two radii will be related by the following identity
\begin{align}\label{eq:radii_patchchange}
    r_{\tilde{x}}^2 = \big( \frac{1-r_x}{1+ r_x}\big)^2 \, . 
\end{align}
The \textit{mid-point} radius, which we define to be the radius of the co-dimension $1$ sphere which is mapped to itself under the change of coordinates between the two ball patches, is given by $r_m = \sqrt{2} - 1$.
Consequently, considering the set of points in the ball with radius up to $r_m + \varepsilon$ for both patches is sufficient to have a non-trivial overlap region, and therefore cover the whole manifold. 

On spheres in general dimension, $S^n$, the Einstein equation with positive constant $R_{ij} = \lambda g_{ij}$ for $\lambda = 1$ is solved by the round metric; which in ball coordinates reads
\begin{align}\label{eq:roundmetric}
g_{i j} = \frac{16 (1 - |x|^2)^2 }{(1 + |x|^2)^4} \delta_{i j}  +  \frac{64  }{(1 + |x|^2)^4} x_i x_j \, ,
\end{align}
for both ball patches. The metric is in fact invariant under the change of coordinates between the two patches. For clarity, let us remind our conventions for the Christoffel symbols and the Ricci tensor in components (according to Section \ref{eq:mn_conventions}):\footnote{To be precise, we train the neural network to predict the vielbein (see Section \ref{sec:VielbeinFormalism}), rather than the metric, for convenience. We find this more natural since it lowers the dimension of the output. However, the final stage of the pipeline constructs the metric from the vielbein (according to the Cholesky decomposition - see next section), and the computation of the Ricci tensor is carried out with the standard formulae according to \eqref{eq:Ricci_and_Christoffel}.}
\begin{align}
\label{eq:Ricci_and_Christoffel}
\begin{aligned}
\Gamma_{i j}^k & :=\frac{1}{2} g^{k l}\left(\partial_i g_{j l}+\partial_j g_{i l}-\partial_l g_{i j}\right) \, , \\
R_{j k} & :=\partial_i \Gamma_{j k}^i-\partial_j \Gamma_{k i}^i+\Gamma_{i p}^i \Gamma_{j k}^p-\Gamma_{j p}^i \Gamma_{i k}^p \, .
\end{aligned}
\end{align}
This chapter focuses on the Einstein condition above, which can be written globally as $Ric(g) = \lambda g$, for spheres $S^n$ with $n = 2,3,4,5$. Since the Ricci tensor is invariant under conformal scaling, we can restrict to $\lambda \in \{+1, 0, -1\}$ without loss of generality (\cite{Besse:1987pua}). While dimensions $2,3$ are a safe arena to corroborate our method since the metrics are completely classified, dimensions $4,5$ host long-standing open questions regarding the existence of Ricci-flat metrics on spheres for $\lambda = 0$. \\

\subsection{Machine Learning}\label{subsec:bkg_ml_chap5}

\subsubsection{An Introduction to Neural Networks}
Before proceeding to the discussion of the concrete implementation of ``AInstein'', let us review th pivotal concept of a \textit{neural network}; we do so by following \cite{capuozzo2024machinelearningtoricduality} closely. As the name suggests, a neural network is a computational model inspired by the structure and functioning of biological neurons. The primary goal of neural networks is to approximate complex, often non-linear mappings between input and output data through layers of interconnected nodes, or neurons. Mathematically, a neural network can be described as a composition of functions, where each layer represents one such function and the output of one layer becomes the input to the next.

Consider a neural network composed of $ L $ layers. Given an input vector $ x_0 \in \mathbb{R}^{n_0} $, where $n_i$ is the dimension of the $i^\text{th}$ layer, the output of the $ \ell^\text{th}$ layer is typically denoted as $ x_\ell $, and is computed recursively as
\begin{equation*}
x_\ell = \sigma(W_\ell x_{\ell-1} + b_\ell),
\end{equation*}
where $ W_\ell \in \mathbb{R}^{n_\ell \times n_{\ell-1}} $ represents the weight matrix, $ b_\ell \in \mathbb{R}^{n_\ell} $ is the bias vector, and $ \sigma $ is the \textit{activation function} applied element-wise. The function $ \sigma $ introduces non-linearity into the network, a critical feature which allows neural networks to model non-linear mappings.

The network's final layer, denoted $ x_L $, provides the model's prediction. Neural networks are typically trained in a supervised fashion, whereby the objective is to adjust the parameters $ \{W_\ell, b_\ell\}_{\ell=1}^L $ so as to minimise a predefined \textit{loss function} $ \mathcal{L}(x_L, y) $, where $ y $ represents the true target values. Optimisation is achieved through backpropagation, which computes the gradient of the loss function with respect to each parameter via the chain rule, followed by a gradient-based optimisation method such as stochastic gradient descent.

Training is often performed using \textit{mini-batch gradient descent}, whereby the dataset is divided into batches. We will also adopt this approach. If the total dataset has $ N $ samples and the chosen batch size is $ B $, the training data is partitioned into $ N/B $ batches, and the network parameters are updated after processing each batch. This strikes a balance between full-batch gradient descent and stochastic gradient descent. 

While batches and layers can be specified by a discrete number, i.e. their \textit{size}, the choice of activation function consists in picking one out of an uncountable infinity of non-linear functions. Some popular choices are: ReLU (Rectified Linear Unit), Leaky ReLU, linear, softmax, sigmoid, GELU. A key point to note is that activation functions are often non-smooth, and this feature is behind the astonishing approximating power of neural networks in many circumstances. For our application, however, we aim at approximating a \textit{smooth} Riemannian metric, and it is therefore crucial to ensure that the neural network only involves smooth operations. Accordingly, we used GELU as the activation function, as we discuss below. 

We end this section by commenting on the simplest and most commonly used type of neural network: the \textit{fully connected neural network}. In this type of architecture, each neuron in a given layer is connected to every neuron in the subsequent layer, hence the name. Such a structure ensures that the information from one layer is propagated entirely to the next, and the network is completely specified by its layers and their activation functions. As we now describe, AInstein is built as a (combination of) fully connected neural netowrk(s).

\subsubsection{An Introduction to the AInstein Network}

This section outlines the overall structure of the ``AInstein'' neural network, the regimes in which it may be trained, and the losses which encode the constraints necessary for the model to output a sensible Einstein metric.

The AInstein model is trained to predict the components of the metric $g_{\mu\nu}$ satisfying $R_{\mu\nu} = \lambda g_{\mu\nu}$ given a pair of points in two patches over a given domain. 

Without loss of generality, let $X_\text{Patch 1}$ and $X_\text{Patch 2}$ be a pair of datasets constituting $N$ points of dimension $n$ represented by $n$-tuples from patches $X_\text{Patch 1}$ and $X_\text{Patch 2}$, such that
\begin{align}
X_\text{Patch 1} &:= \{ x_j = (x^0_j, \ldots, x^n_j) \; | \; j \in 0, \ldots, N \} \, , \\
X_\text{Patch 2} &:= \{ \tilde x_j = (\tilde x^0_j, \ldots, \tilde x^n_j) \; | \; j \in 0, \ldots, N \} \, ,
\end{align}
where $j$ indexes the elements of the dataset. Points in $X_\text{Patch 2}$ are related to those in $X_\text{Patch 1}$ by a transition function $T$ such that $\tilde x^i_j = T(x_j^0, \ldots, x_j^n)$. For the specific case of spheres, the map $T$ is identified with $\tau$ in \eqref{eq:Change_of_coords}. Prior to training the network, points in patch 1 are randomly sampled according to the scheme specified in Section \ref{app:sampling} in order to generate a set of training data.

\begin{figure}[h]
\centering
\resizebox{0.5\textwidth}{!}{\begin{tikzpicture}[
    roundnode/.style={circle, draw=black, thick, minimum size=1cm},
    squarednode/.style={rectangle, draw=black, thick, minimum size=1cm, anchor=center},
    ->, thick]

    \node[roundnode] (A1) {$x_{\text{Patch 1}}^i$};
    \node[below=0.4cm of A1] (dots1) {$\vdots$};
    \node[roundnode, below=0.4cm of dots1] (A2) {$x_{\text{Patch 1}}^d$};
    
    \node[draw, rectangle, minimum height=6cm, minimum width=1.8cm, anchor=center] (Abox) at ($(A1)!0.5!(A2)$) {}; 
    
    \node[squarednode, right=3.5cm of Abox, yshift=1.5cm] (H1) {$\{H^{(l)}_{\text{Patch 1}}\}$};

    \coordinate (MidH1) at ($(Abox.east) + (2,1.5)$);
    
    \draw (Abox.east) -- (MidH1) -- (H1.west);

    \node[squarednode, right=1.9cm of Abox, yshift=-1.5cm] (T2) {$T_{\text{Patch 2}}$};

    \draw (Abox.east) -- (T2.west);

    \node[squarednode, right=2.5cm of T2, anchor=center] (H2) {$\{H^{(l)}_{\text{Patch 2}}\}$};
    
    \draw (T2.east) -- (H2.west);
    
    \coordinate (Hmid) at ($(H1)!0.5!(H2)$);
    
    \node[squarednode, right=3.5cm of Hmid] (Concat) {Concat};

    \coordinate (MidH1Concat) at ($(H1.east) + (1.5,0)$);

    \draw (H1.east) -- (MidH1Concat) -- (Concat.west);

    \draw (H2.east) -- (Concat.west);

    \node[roundnode, right=1.5cm of Concat] (O) {O};

    \draw (Concat.east) -- (O.west);

\end{tikzpicture}}
\caption{Overview sketch of the AInstein architecture. Here, $T$ is a patch transition function layer which converts the points in patch 1 to their equivalents in patch 2, $\{H^{(l)}_\text{Patch $p$}\}$ a set of hidden layers with non-linear activations for each patch, and `Concat' a concatenation layer, then followed by a Cholesky transform on the output of the hidden states in the pipeline, producing the metrics on both patches.}
\label{fig:machine_learning_arch}
\end{figure}
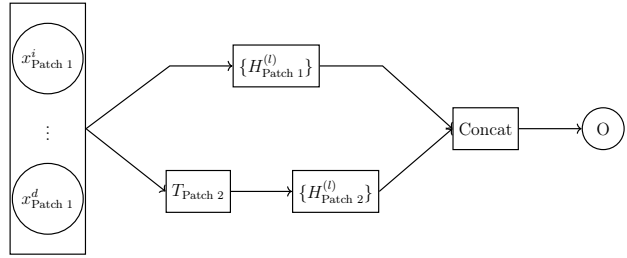

As an architecture, we choose a modified multi-layer perceptron network (MLP). In general, an MLP may be defined recursively layer-by-layer, 
\begin{align}
\begin{aligned}
\phi^{(1)}_i &= b^{(1)}_i + w^{(1)}_{ij}x_j \\
h^{(1)}_i &= \sigma(\phi^{(1)}_i) \\
&\vdots \\
\phi^{(l)}_i &= b^{(l)}_i + w^{(l)}_{ij} h^{(l-1)}_j \\
h^{(l)}_i &= \sigma(\phi^{(l)}_i) \, ,
\end{aligned}
\end{align}
where $\phi^{(l)}_i$ is the output of the $l$-th hidden layer pre-activation, and $h^{(l)}_i$ the output of the subsequent activation function used to introduce non-linearity in the network.

Let $\mathcal N_\text{AInstein}$ be a concatenation of a pair of sub-networks $\mathcal N_\text{Patch 1}, \mathcal N_\text{Patch 2}$ each taking input points $x_i$ on their respective patch. More specifically, one may write $\mathcal N_\text{AInstein}$ as
\begin{align}
\mathcal N^{\theta_1, \theta_2}_\text{AInstein} := \mathcal N^{\theta_1}_\text{Patch 1} \oplus \mathcal N^{\theta_2}_\text{Patch 2} \, ,
\end{align}
where $\mathcal N^{\theta_1}_\text{Patch 1}$ and $\mathcal{N}^{\theta_2}_\text{Patch 2}$ are the neural networks, parametrised\footnote{The parameters in this case are the set of all weights and biases $w^{(\ell)}$, $b^{(\ell)} \ \forall \ell$.} by variables $\theta_1$ and $\theta_2$, and learn the metric in patches 1 and 2 respectively. For notational simplicity, we choose to henceforth suppress the explicit dependence on $\theta_1$, $\theta_2$, and define $\mathcal N^{\theta_1}_\text{Patch 1} \oplus \mathcal N^{\theta_2}_\text{Patch 2}$ to act such that,
\begin{align}
(\phi_i^{(l)})_\text{AInstein}(x_j, \tilde x_j):= (\phi_i^{(l)})_\text{Patch 1}(x_j) \oplus (\phi_i^{(l)})_\text{Patch 2}(\tilde x_j)\\
(h_i^{(l)})_\text{AInstein}(x_j, \tilde x_j) := (h_i^{(l)})_\text{Patch 1} (x_j) \oplus (h_i^{(l)})_\text{Patch 2}(\tilde x_j),
\end{align}
for $x_j \in X_\text{Patch 1}$, $\tilde x_j \in X_\text{Patch 2}$.

Assuming the network has depth $D$, we denote the output of the final layer of each subnetwork by
\begin{equation}
    \phi^D_\text{Patch 1}(x) := (g_{ij})_x^\text{Patch 1} \quad \phi^D_\text{Patch 2}(\tilde x):= (g_{ij})^\text{Patch 2}_{\tilde x},
\end{equation}
where it is understood that the patch label is associated with both the (sub)network and the data on which it acts.

The full model provides an output $\mathcal N_\text{AInstein}(x, \tilde x) = (g_{ij})_x^\text{Patch 1} \oplus (g_{ij})_{\tilde{x}}^\text{Patch 2}$., where this predicted metric is evaluated at points $(x, \tilde x)$ from each patch respectively. The sub-networks are trained simultaneously subject to the loss function defined in equation \eqref{eq:whole_loss}. The architecture is depicted in Figure \ref{fig:machine_learning_arch}.

The specified model, $\mathcal N_\text{AInstein}$, may be trained in two regimes: the \textit{supervised} regime and the \textit{semi-supervised} regime. The main contribution of this work arises from training the model in the semi-supervised regime subject to the losses presented in Section \ref{ssub:losses}. To enhance training convergence, rather than initializing the network's weights from a random configuration, one can leverage the identical architecture shared between the supervised and unsupervised models. Specifically, the initialization can be derived from the parameters obtained by training the supervised model on a known function. 

\subsubsection{Supervised Reference Models}
In the supervised regime of training the $\mathcal{N}_\text{AInstein}$ architecture, the outputs of the function are known in the training data.
Therefore for every input point $x$ the output metric is known for that point $g_{ij}(x)$, such that the training seeks to minimise a mean squared error loss between the known metric components at each point and the components predicted by the model.

The known Einstein metric on the sphere we consider is the round metric, for $\lambda = +1$, as defined in \eqref{eq:roundmetric}.
By training the same architecture in a supervised manner, using explicitly computed round metric components as output, the architecture is trained to model this round metric.
This is important as the test loss scores on this trained metric set an important baseline for comparison, with full knowledge of the output metric values for the manifold points, how well can an Einstein metric be modelled with the allocated computational resources.
These loss scores are hence reported alongside the semi-supervised test losses, dictating the loss order which indicates the architecture has learnt a metric function which truly exists.

Further to a supervised training of the round metric, the supervised architecture can be used to design intelligent starting points for the model.
With random initialisation of the parameters $(\theta_1,\theta_2)$ the initial function represented by $\mathcal{N}_\text{AInstein}$ is far from smooth which leads to a blow up of Einstein loss values, however if we could pick parameter values which represented a smoother function the loss order would initialise within a computable range and encourage sensible learning.
To do this we choose to train a supervised model to predict the identity function in each patch ($g_{ij}=\delta_{ij}$), an ansatz which is completely flat and hence also smooth, despite substantially violating the overlap condition.
This is trained again with a mean squared error loss, now with network outputs which match $\delta_{ij}$ for \textit{every} input point.
After training, the parameters $(\theta_1,\theta_2)$ are saved, and used to initialise the $\mathcal{N}_\text{AInstein}$ function in the semi-supervised training (as well as the supervised training of the round metric to ensure fair comparison).

\subsubsection{Semi-Supervised Loss Components}
\label{ssub:losses}

As with all deep learning tasks, we must supply the network with a loss function to use during training. This loss acts on both subnetworks simultaneously, and contains a set of designed loss components, from which we consider their weighted sum, which is minimised where the output and $\mathcal{N}_\text{AInstein}$ function represents a sensible Einstein metric.

The loss may be written as
%
\begin{equation}
\label{eq:whole_loss}
\begin{split}
\mathcal{L}_\text{AInstein}[\theta_1, \theta_2](g_x^\text{Patch 1}, g_{\tilde{x}}^\text{Patch 2}) 
&:= f_1 \bigg( \mathcal{L}^\text{Einstein}_\text{Patch 1}[\theta_1] (g_x^\text{Patch 1}) + \mathcal{L}^\text{Einstein}_\text{Patch 2}[\theta_2] (g_{\tilde{x}}^\text{Patch 2}) \bigg) \\
& + f_2 \bigg(\mathcal{L}^\text{Overlap}[\theta_1, \theta_2](g_x^\text{Patch 1},g_{\tilde{x}}^\text{Patch 2})\bigg) \\
& + f_3 \bigg(\mathcal{L}^\text{Finiteness}_\text{Patch 1}[\theta_1](g_x^\text{Patch 1}) + \mathcal{L}^\text{Finiteness}_\text{Patch 2}[\theta_2](g_{\tilde{x}}^\text{Patch 2})\bigg) \;,
\end{split}
\end{equation}
where $f_i$ are the respective loss term multipliers, specifying the relative importance of the loss components; practically we used $(f_1,f_2,f_3) = (1,10,1)$.
Each loss component implicitly contains a filter, which weights the contribution of points depending on which part of the patches are most important to that loss, this improves the global metric learning and additionally improves numerical stability; more information is provided in Section \ref{app:filters}. We now describe these loss components in detail.

\paragraph{Einstein loss}
To satisfy the Einstein condition of the solution, we impose the following loss term:
\begin{equation}
    \mathcal L^\text{Einstein}_\text{Patch $p$}[\theta_p](g_x^\text{Patch $p$}) := ||\lambda (g_{ij})_x^\text{Patch $p$} -  (R_{ij})_{x}^\text{Patch $p$}|| \, ,
    \label{eq:Einstein_loss}
\end{equation}
where $p \in \{1, 2 \}$, $\lambda$ is the Einstein constant, $R_{ij}$ is the \textit{Ricci tensor}, and $||\cdot||$ represents the Euclidean 2-norm. By inspection, it is evident the Einstein loss term penalises metrics which deviate far from $\lambda R_{ij}$ evaluated at point $x$. 
This loss term is weighted according to the point's radial coordinate, as described in Section \ref{app:filters}, to prioritise points in the patch region used in defining the global metric model.

\paragraph{Overlap loss}
This loss component enforces the gluing condition of the patches, ensuring the metric evaluated on points in one patch is consistent with the companion metric evaluated on equivalent points in the other.

Concretely, let $x_j \in X_\text{Patch 1}$ possess an associated point\footnote{Here we consider that $x_j$ and $\tilde x_j$ are finite.} $\tilde x_j \in X_\text{Patch 2}$ related by $\tilde x_j = T(x_j)$, where $T$ is an appropriate transition function. Equivalently, the metric is related by the Jacobian matrix $J$, meaning one may write an overlap loss as
\begin{equation}
\begin{split}
   \mathcal L^{\text{Overlap}}[\theta_1, \theta_2](g^\text{Patch 1}_{x},g^\text{Patch 2}_{\tilde{x}}) := \ & ||(g_{ij})_x^\text{Patch 1} - J_{ki} (g_{kl})^\text{Patch 2}_{\tilde x} J_{lj}|| \\
    + \ & ||J_{ki}(g_{kl})_x^\text{Patch 1}J_{lj} -  (g_{ij})^\text{Patch 2}_{\tilde x}|| \;,
\end{split}
\end{equation}
where $J$ is the analytically known Jacobian matrix corresponding to the change of coordinates between the two patches. For the case of spheres, it is given by \eqref{eqn:jacobian}, and is equal to its inverse. 
This loss term is also weighted according to the point's radial coordinate, as described in Section \ref{app:filters}, but in a different way to prioritise points in the overlap region.

\paragraph{Finiteness loss}
In order to ensure finiteness, and discourage the machine learning algorithm from approaching the ``zero-metric'' ($g_{ij} \sim 0$), we introduce a loss which takes the form:\footnote{By inspection, we found that the neural network tended to minimise the loss \eqref{eq:Einstein_loss} by simultaneously making the components of the metric and of the Ricci tensor smaller and smaller. This is clearly a numerical artifact: the violation of the Einstein condition should be small compared to the components of the metric, not just in absolute terms.}

\begin{equation}
\begin{split}
\mathcal L^\text{Finiteness}_\text{Patch p}[\theta_p](g^\text{Patch p}_{x}) &:= 1 +\left(h \, e^{-\left(\frac{F - c_f}{w_f}\right)^{t_f}}- h \right)^2\\ &+ \left(\frac{F - (c_f + w_f )}{ s } \right) \cdot \frac{1+\tanh \left(\frac{F - (c_f + w_f )}{2}\right)}{2} \\ 
& + \left(\frac{-F + (c_f - w_f)}{s}\right) \cdot \frac{1+\tanh \left(\frac{-F + (c_f - w_f )}{2}\right)}{2} \, ,
\end{split}
\end{equation}
for parameters $(F,h,c_f,w_f,t_f,s)$. Here $F = \sum_{i,j} ||(g_{ij})^\text{Patch p}_x||$ is the sum of the absolute value of all the components of the metric; $h$ controls the ``height'' of the well with centre $c_f$, and width $w_f$. Moreover, $t_f$ controls how vertical the walls are, and $s$ determines the gradient of the ``slopes'' which emanate from the well. A plot of this filter, as used in the finiteness loss is shown in Figure \ref{fig:filters_f}. The motivation for this filter is to avoid the components of the predicted metric getting arbitrarily close to zero. Since it involves the sum of the components of the predicted metric, we supplement it with a dimension-dependent normalisation.

\paragraph{Filters}
Many of the loss components, if implemented na\"{i}vely for the case of spheres, result in training behaviours which are unpredictable and numerically unstable. This is because of the pathological behaviour of the metric and the Jacobian as one approaches the boundary of the unit ball (see \eqref{eqn:jacobian} and \eqref{eq:roundmetric}). As such, we introduce a set of loss `filters' which smoothly\footnote{It is important each filter is smooth and differentiable to enable derivatives to be taken for back-propagation.} suppress each loss component's contribution depending on the location of the point being evaluated in the patch.

To explain our choices in more detail, let us consider the usual stereographic projection (or simple modifications of it); both patches cover the whole sphere with the exception of their associated pole. Since the overlap consists of the whole sphere excluding the poles, in the coordinates of each patch, the overlap region is the whole of $\mathbb{R}^n$ (or $B^n$, if working with ball coordinates) excluding the origin. If the patches are made smaller, the overlap region shrinks. As discussed in Section \ref{subsec:bkg_dg_chap5} (see \eqref{eq:radii_patchchange} and the following comments), we can choose our charts to consist of the $n$-ball with radius $r_m +  \varepsilon$, and the corresponding overlap region is the annulus between $\frac{1 - r_m -  \varepsilon}{1 + r_m +  \varepsilon}$ and $r_m +  \varepsilon$. This is what is implemented in our code, with the choice of $\varepsilon$ being one of the hyperparameters. With these charts, one needs to introduce a filter that devalues contributions from points whose radius is larger than $r_m +  \varepsilon$ when evaluating the Einstein condition; because they are not contained within the patch. When calculating the overlap loss, another filter should devalue points outside of the annulus overlap region described above.
These filters are described further in Section \ref{app:filters}.


\subsubsection{Global Test Loss}\label{ssub:global_loss}
As described in Section \ref{ssub:losses}, the final global model of the trained metrics are restricted to patches of radii $r_m + \varepsilon$, using an overlap region of radii $\in [\frac{1 - r_m -  \varepsilon}{1 + r_m +  \varepsilon}, r_m +  \varepsilon]$.
Where the training loss used includes weighted contributions from \textit{all} points (which improves the learning), our final testing evaluation is restricted to just these patch regions required for global definition.

The training filters are hence converted into hard cutoffs, and the finiteness loss which has non-geometric motivation is ignored.
Hence, the global test loss, as reported in the results of Section \ref{sec:results_chap5}, is defined 
\begin{equation}\label{eq:global_loss}
\begin{split}
\mathcal{L}_\text{Global}[\theta_1, \theta_2](g_x^\text{Patch 1}, g_{\tilde{x}}^\text{Patch 2}) 
&:= f_1 \bigg( \mathcal{L}^\text{Einstein}_\text{Patch 1}[\theta_1] (g_x^\text{Patch 1} \; \big| \; ||x|| < r_m+\varepsilon) \\
& \quad + \mathcal{L}^\text{Einstein}_\text{Patch 2}[\theta_2] (g_{\tilde{x}}^\text{Patch 2} \; \big| \; ||\tilde{x}|| < r_m+\varepsilon) \bigg) \\
+ f_2 \bigg(\mathcal{L}^\text{Overlap}[\theta_1, \theta_2] & (g_x^\text{Patch 1},g_{\tilde{x}}^\text{Patch 2} \; \big| \; ||x|| \in \bigg[\frac{1 - r_m -  \varepsilon}{1 + r_m +  \varepsilon}, r_m +  \varepsilon \bigg])\bigg) \;,
\end{split}
\end{equation}
where $||\cdot||$ indicates the 2-norm of the input point, which equals its radial coordinate.

\section{Results}
\label{sec:results_chap5}
To train the network and obtain Einstein metrics, data must first be generated.
To match the architecture style described in Section \ref{subsec:bkg_ml_chap5}, where the input is a point's coordinates in one patch and the output is the metric vielbeins for all patches, the data need only be generated for the first patch.

The patches are represented in ball coordinates such that for the sphere $S^n$ each patch is a unit $B^n$, which we parameterise by $n$ Euclidean coordinates evaluating in the range $x_i \in (-1,1)$.
The patch is sampled using a modified beta distribution, designed to prioritise the patch overlap region and minimise numerical instabilities; more details are given in Section \ref{app:sampling}, including exemplary plots of the distributions in 2d.

For training, the number of points sampled were $(10^4, 10^4, 10^5, 10^5)$ for dimensions $(2,3,4,5)$ respectively; consistently resampled across all runs and Einstein constants, where testing used $10^4$ independently sampled points.
Traditionally, exponential increases in the sampling size is desired as data dimension increases, which makes the displayed results for higher dimensions all the more impressive. 
Consequently, we would also expect performances to improve further with more training data.

Once the patch data has been sampled, the NN architecture is initialised. 
For the hyperparameters introduced in Section \ref{subsec:bkg_ml_chap5} and listed in Section \ref{app:hyperparams}, the model parameters are set such that the metrics are identity matrices. 
To do this a supervised model with the same architecture is first trained on independent inputs sampled equivalently, paired with output vielbeins which produce the identity matrix for every point in both patches.
Four networks of this form are trained for each of the four considered dimensions, and their parameters are used to initiate each model of that respective dimension in the subsequent learning.
These start points\footnote{Preliminary investigations used random initialisations for the model parameters, but since the metrics they represented were so far from being smooth the Einstein loss condition blew up, obstructing sensible learning and often exceeding the floating point memory limit.} are by nature smooth, and violate the Einstein equation in each patch to an order comparable with the metric components, but are exceptionally far from satisfying the overlap gluing conditions between the patches, hence representing \textit{non-geometric} starting points.
It is worth emphasising here that these identity function start point are completely independent of the problem, or any knowledge of solutions, they can be quickly and cheaply defined for any dimension and proved surprisingly effective. 

With the data sampled, and architectures initialised, 10 independent runs were performed for each investigation (over varying dimension and Einstein constant), and final performances were evaluated with the Global test loss described in Section \ref{ssub:global_loss}.
However, an additional means of assessing the test loss measures was also devised.
Since the round metric is known to exist as a solution for $\lambda = +1$ in all dimensions, and the explicit metric form can be computed for any input patch point using \eqref{eq:roundmetric}, a supervised model can be trained to explicitly model this metric.
This is done by training the same architecture, also initialised with the same pre-trained identity functions, with MSE loss on input-output pairs of the point coordinates in patch 1 and the round metric vielbein coordinates for that point in both patches.
These were trained with the same hyperparameters for each dimension, had Global test loss scores equivalently independently evaluated, and provide an important comparison baseline for the main investigation test loss scores.
These baselines represent the feasible limit of solutions to the Einstein equations from these techniques with the compute resources provided. 

\subsection{Local Einstein Geometries}
As a warm-up, to test the effectiveness of the Einstein loss, and the code functionality, we begin with a single patch.
By working in a single patch without boundary conditions, the architecture is being trained to find Einstein metrics on a space which is topologically equivalent to $\mathbb{R}^n$.
The solutions to the Einstein equations in the cases of $\lambda \in \{-1,0,+1\}$ are known, and represent spherical, flat, and hyperbolic spaces, often expressed with trigonometric functions, and which here would be restricted to the ball patch (hence `local').

The data is sampled in the same way, except the NN metric architecture is set up with only a single patch subnetwork, outputting the metric vielbein for the input patch alone.
Since there is only one patch, the overlap loss is redundant, and hence ignored.
This leaves the Einstein loss and finiteness loss as the only terms in the training loss, where each is now only for the single patch.
The multiplier weightings of these two losses are set as equal to mirror the behaviour in \eqref{eq:whole_loss} for the full training loss, and the global test loss has only a single contributing term: the Einstein loss for the patch.

\begin{figure}[!t]
    \centering
    \begin{subfigure}{0.32\textwidth}
        \centering
        \includegraphics[width=0.98\textwidth]{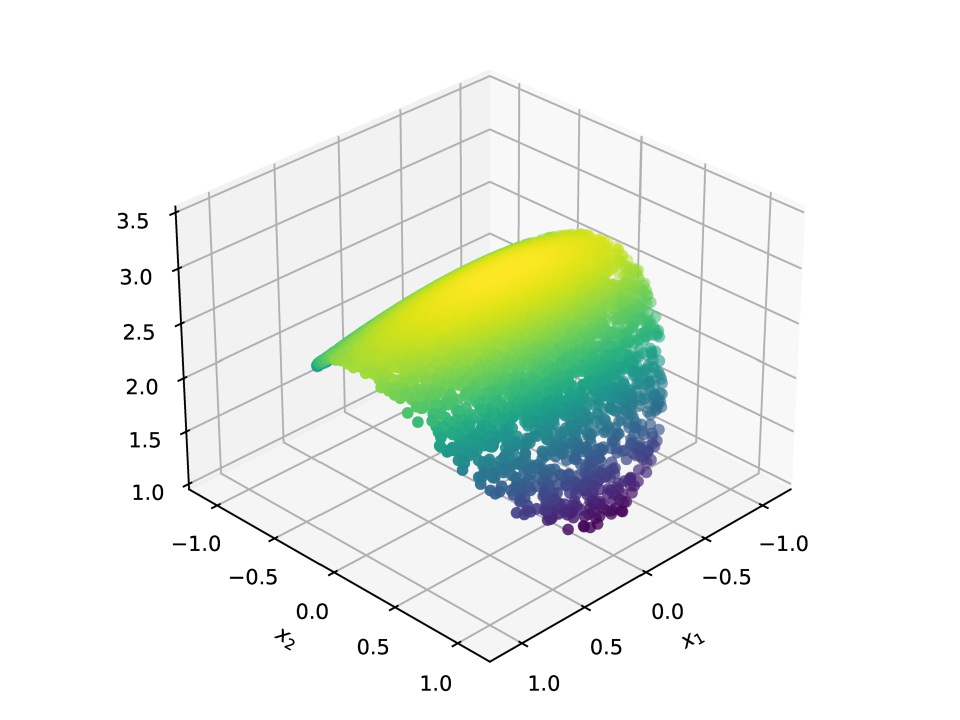}
        \caption{$g_{00}$ ($\lambda = +1$)}
    \end{subfigure} 
    \begin{subfigure}{0.32\textwidth}
        \centering
        \includegraphics[width=0.98\textwidth]{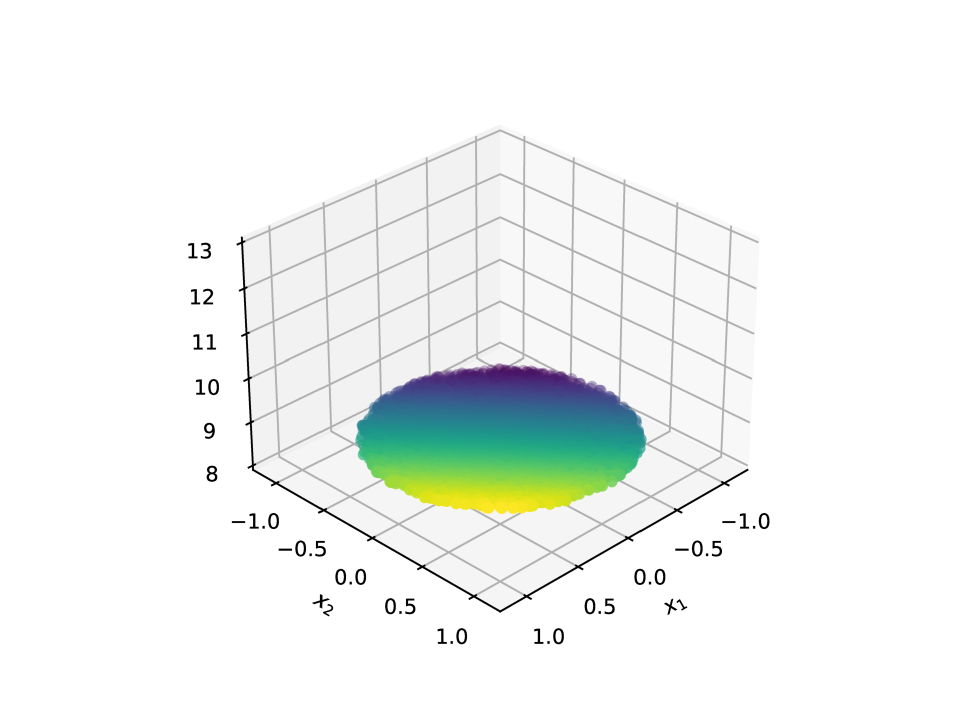}
        \caption{$g_{00}$ ($\lambda = 0$)}
    \end{subfigure}
    \begin{subfigure}{0.32\textwidth}
        \centering
        \includegraphics[width=0.98\textwidth]{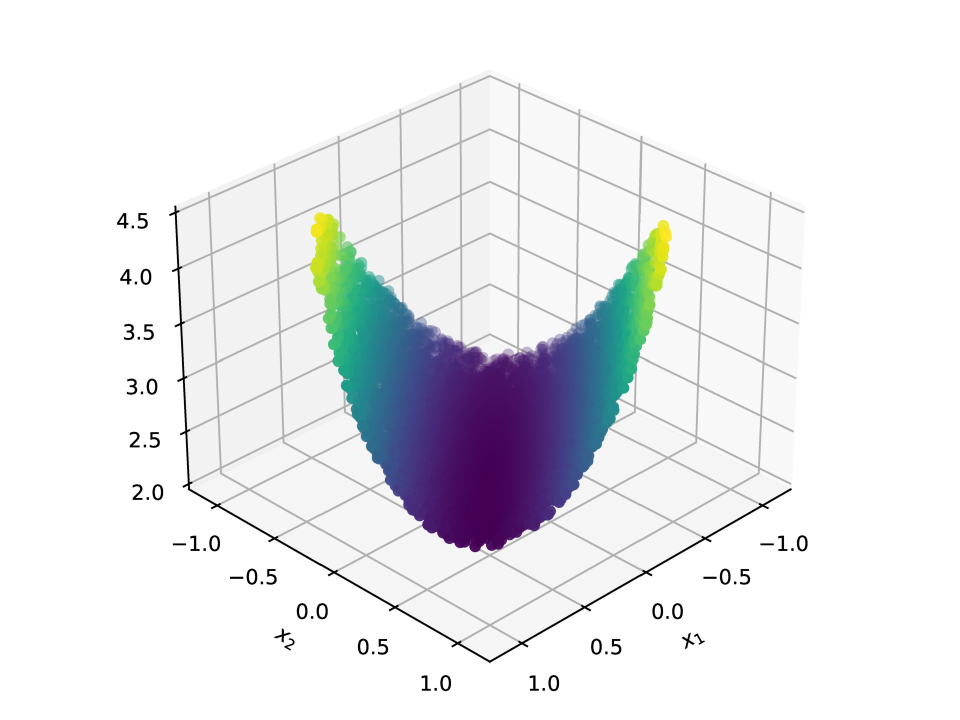}
        \caption{$g_{00}$ ($\lambda = -1$)}
    \end{subfigure}\\
    \begin{subfigure}{0.32\textwidth}
        \centering
        \includegraphics[width=0.98\textwidth]{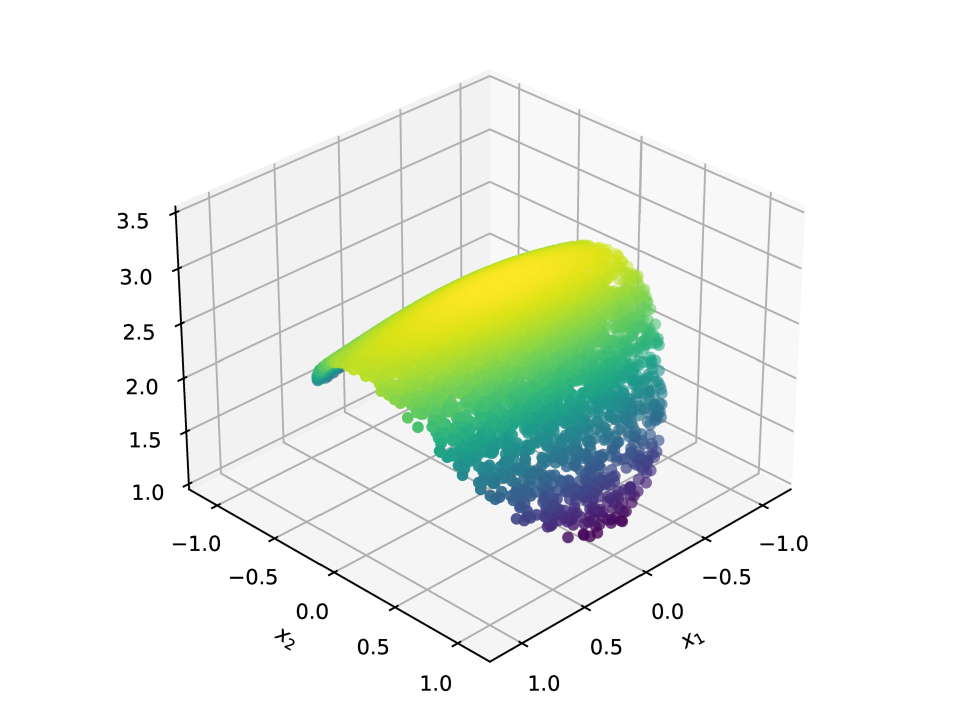}
        \caption{$R_{00}$ ($\lambda = +1$)}
    \end{subfigure} 
    \begin{subfigure}{0.32\textwidth}
        \centering
        \includegraphics[width=0.98\textwidth]{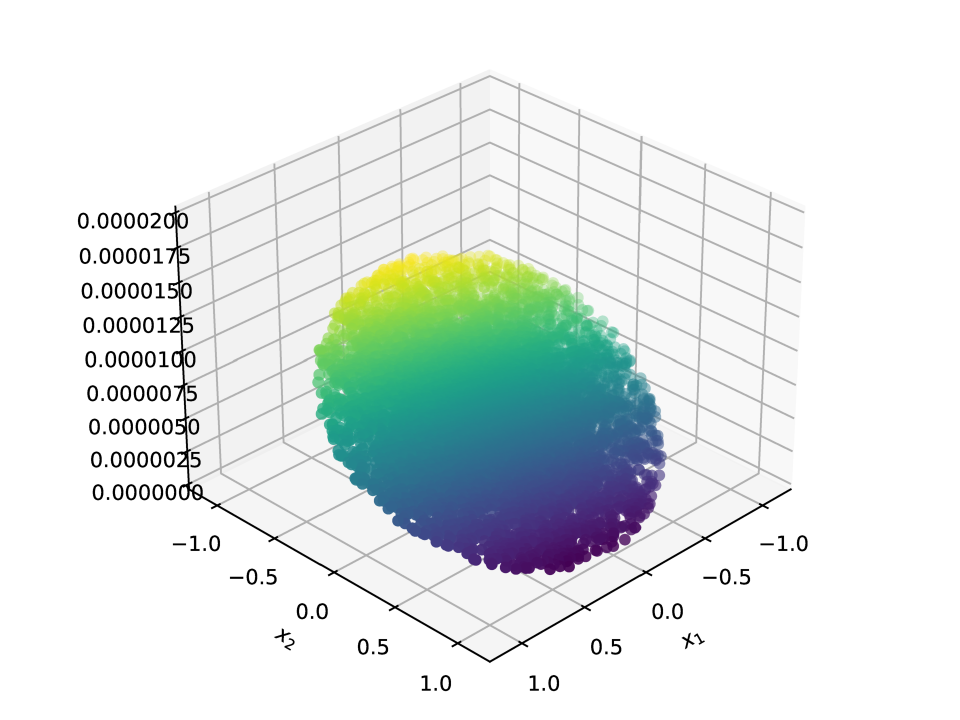}
        \caption{$R_{00}$ ($\lambda = 0$)}
    \end{subfigure}
    \begin{subfigure}{0.32\textwidth}
        \centering
        \includegraphics[width=0.98\textwidth]{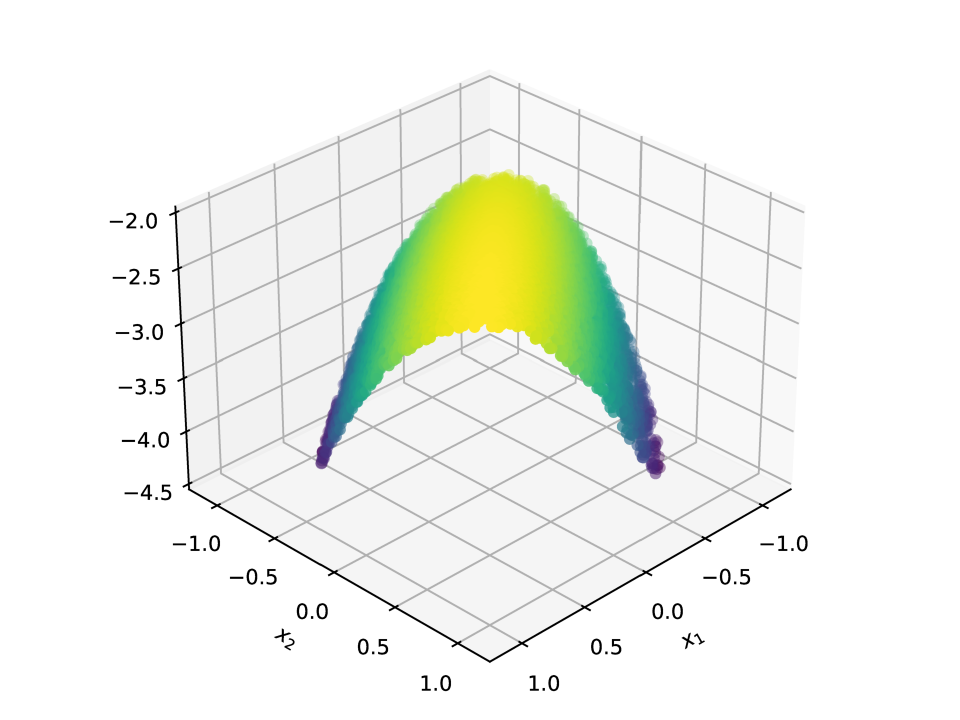}
        \caption{$R_{00}$ ($\lambda = -1$)}
    \end{subfigure}
    \caption{Visualisations of the $(0,0)$ components of the learnt metrics and their respective Ricci tensors, in 2d on a single patch. These metrics solve the Einstein equation with Einstein constants of $\lambda \in \{+1, 0, -1\}$ respectively. We emphasise the $R_{00}$ $(\lambda = 0)$ scale is $\sim 10^{-5}$, indicating Ricci-flat.}
    \label{fig:vis_2d1p}
\end{figure}

Training with the same hyperparameters, as stated in Section \ref{app:hyperparams}, 10 runs for each $\lambda$ value were performed for a 2d ball patch, starting from the same identity initialisation\footnote{Since the architecture has changed by removing one subnetwork for the second patch, technically a new 1-patch version of the supervised identity function was trained to be used for initialisation.}.
The trained metrics were evaluated on independent sample sets, and test Einstein losses computed, reported in Table \ref{tab:local_losses}.
Visualisation of the $(0,0)$ components for a single run are shown in Figure \ref{fig:vis_2d1p}, were the other components had similar behaviour.

\begin{table}[!t]
\centering
\begin{tabular}{|c|ccc|}
\hline
\multirow{2}{*}{\begin{tabular}[c]{@{}c@{}}Loss\\ Component\end{tabular}} & \multicolumn{3}{c|}{Einstein Constant $\lambda$}                            \\ \cline{2-4} 
& \multicolumn{1}{c|}{$+1$} & \multicolumn{1}{c|}{$0$} & $-1$ \\ \hline
Einstein & \multicolumn{1}{c|}{0.038 $\pm$ 0.016}    & \multicolumn{1}{c|}{0.000 $\pm$ 0.000}    &   0.025 $\pm$ 0.017  \\ \hline
\end{tabular}
\caption{Global test loss results averaged over 10 runs, for NN approximations of Einstein metrics with the respective curvatures on single patches in 2d; note overlap loss not applicable, so the global loss's only contribution is from the Einstein loss. All losses are reported to 3 decimal places with standard deviations across their 10 runs.}
\label{tab:local_losses}
\end{table}

The losses in Table \ref{tab:local_losses} are all very low, significantly $<1$. 
The $\lambda=0$ case is especially low since the identity initialisation already satisfies this Einstein condition, but the others which are initialised not satisfying the condition modify their metrics to satisfy the condition well, reaching similar performance scores up to error.
The visualisations on Figure \ref{fig:vis_2d1p} are especially insightful, the shapes show trigonometric-like behaviour, matching the expected style.
The computed Ricci tensors on the test data for $\lambda=+1$ look identical to the metric, for $\lambda=0$ are near-identically 0 throughout the patch, and for $\lambda=-1$ is the negation of the metric.

These performances validate this machine learning approach nicely, and set up scope for development to more non-trivial manifolds, with boundary conditions or further patches -- the latter we focus on now.

\subsection{Global Einstein Geometries on Spheres}
Extending the setup to a more non-trivial manifold, one wishes to consider multiple patches satisfying a gluing condition on their overlap.
In this work, we do this by considering spheres, $S^n$, covered by an atlas with 2 patches, as described in Section \ref{subsec:bkg_dg_chap5}.
The gluing condition, associated to the transition function of the atlas, is defined over the patches with a weighting that prioritises an overlap region for radii $\sim r_m$, and is packaged within an overlap loss term.
This is coupled with the Einstein loss and the finiteness loss defined for both patches in the full training loss, according to Section \ref{subsec:bkg_ml_chap5}.

In performing these investigations, again 10 runs were trained for each investigation spanning the Einstein constants $\lambda \in \{+1,0,-1\}$ and dimensions $\{2,3,4,5\}$. 
The architectures were initialised using the parameters from a supervised pre-trained identity function for each patch, and run with hyperparameters as specified in Section \ref{app:hyperparams}.
The trained metrics were evaluated using the Global test loss, which had only Einstein and overlap contributions as described in \eqref{eq:global_loss}.
The Einstein contribution was computed on each patch for test points within a restricted radii of $r_m+0.1$, and the overlap contribution was computed for the test points with radii in the range $[\frac{1-(r_m+0.1)}{1+(r_m+0.1)}, r_m+0.1]$, which selects the same points for both patches.
We emphasise that a restriction of $r_m + \varepsilon$ for $0 < \varepsilon << 1$ is sufficient to give a global description of the manifolds, but to ensure sufficient test data for each loss this was expanded to an upper width given by $\varepsilon = 0.1$.
The proportion of test points in each patch and the overlap region was remarkably consistent across each runs metric testing for all $\lambda$ and dimension.
The average proportions of test points in the (restricted patch 1, restricted patch 2, overlap region) were $(0.594, 0.594, 0.188)$, which can be multiplied by $10^4$ to get the number of points contributing to each loss term. 

The average Global test losses are shown in Table \ref{tab:global_test_losses}, with a breakdown into the sublosses in Section \ref{app:extra_results}.
In addition to reporting Global test losses for the considered dimensions and $\lambda$ values run with the semi-supervised architecture, results are also reported for supervised models trained to model the analytic round metric defined in \eqref{eq:roundmetric} which solves the Einstein equations with $\lambda=+1$.
The supervised Global test loss scores set a threshold for learning a true metric, as we know the round metric to exist in all dimensions.

\begin{table}[!t]
\centering
\begin{tabular}{|c|ccc!{\vrule width 1.5pt}c|}
\hline
\multirow{2}{*}{Dimension} & \multicolumn{3}{c!{\vrule width 1.5pt}}{Einstein Constant $\lambda$} & \multirow{2}{*}{\begin{tabular}[c]{@{}c@{}}Supervised\\ $\lambda = +1$\end{tabular}} \\ \cline{2-4} 
& \multicolumn{1}{c|}{$+1$} & \multicolumn{1}{c|}{$0$} & $-1$ &  \\ \hline
2                          & \multicolumn{1}{c|}{0.083 $\pm$ 0.023} & \multicolumn{1}{c|}{2.881 $\pm$ 0.113} & \multicolumn{1}{c!{\vrule width 1.5pt}}{4.364 $\pm$ 0.093} & 0.096 $\pm$ 0.013 \\ \hline
3                          & \multicolumn{1}{c|}{0.151 $\pm$ 0.027} & \multicolumn{1}{c|}{5.560 $\pm$ 0.160} & 8.641 $\pm$ 0.183 & 0.195 $\pm$ 0.020 \\ \hline
4                          & \multicolumn{1}{c|}{0.150 $\pm$ 0.018} & \multicolumn{1}{c|}{8.494 $\pm$ 0.121} & 14.928 $\pm$ 1.317 & 0.248 $\pm$ 0.024 \\ \hline
5                          & \multicolumn{1}{c|}{0.244 $\pm$ 0.039} & \multicolumn{1}{c|}{10.810 $\pm$ 0.185} & 18.798 $\pm$ 2.024 & 0.518 $\pm$ 0.063 \\ \hline
\end{tabular}
\caption{Global test loss results averaged over 10 runs, for NN approximations of Einstein metrics with the respective curvatures on spheres in dimensions 2-5 (2-patches). For comparison, the right-hand column shows the respective global test losses for the \textit{supervised} NN model approximation of the analytic round metric (which satisfies the Einstein equation for $\lambda = +1$). All losses are reported with standard deviations across their 10 runs.} 
\label{tab:global_test_losses}
\end{table}

Interpreting the losses, one can see in each dimension for the case of $\lambda=+1$ the semi-supervised architecture has learnt to approximate an Einstein metric exceptionally well.
Where in the supervised case the output is explicitly known, in the semi-supervised case the only conditions informing the learning are the values of the Einstein and other losses, and the model starts from an identity initialisation which is far from satisfying the Einstein and overlap conditions (training loss often starts $>10^4$).
It is therefore exceptionally impressive that the model can learn to approximate these $\lambda=+1$ Einstein metrics so well, even exceeding the performance scores of the supervised model\footnote{Despite the supervised model's training being informed by the exact metric values at each training datapoint, the Ricci tensor is so highly sensitive that the semi-supervised architecture can better learn the metric, even without the explicit knowledge of its values.}.

Where existence of Einstein metrics with $\lambda=+1$ is known and proven for spheres $S^n$ in any dimension, those with $\lambda = 0,-1$ are forbidden in dimensions $2,3$ (\cite{Besse:1987pua}).
The runs where the model attempts to find a metric with $\lambda = 0,-1$ in those dimensions satisfyingly fail: all losses are large ($>1$, and over an order of magnitude above that of the supervised model), and these set the opposing loss score baselines for comparison where a metric does not exist.
Of greater significance are the $\lambda=0$ and $\lambda=-1$ cases for higher dimensions; especially the former, from a physics perspective. Specifically, the existence of Ricci-flat metrics on $S^{4,5}$ is an open problem which excitingly this machine learning approach can provide a new numerical perspective on.
Therefore, of new insight are the results for 4d \& 5d, which are not conclusive\footnote{One may comment that 10 runs is not particularly many for finding a likely obscure metric, we add here that $\sim 50$ more runs were performed for the Ricci-flat search in further hope of finding suitable metrics, all with similar performance scores; and we plan to continue submitting runs in search of evidence for their existence.}, with losses of order $10$, and thus much larger than the supervised model losses.
These results hence provide new numerical evidence \textit{against} this longstanding open problem of Ricci-flat metric existence on the spheres $S^4$ and $S^5$, and no examples of Einstein metric with negative Einstein constant are found either.

\begin{figure}[H]
    \centering
    \begin{subfigure}{0.24\textwidth}
        \centering
        \includegraphics[width=0.98\textwidth]{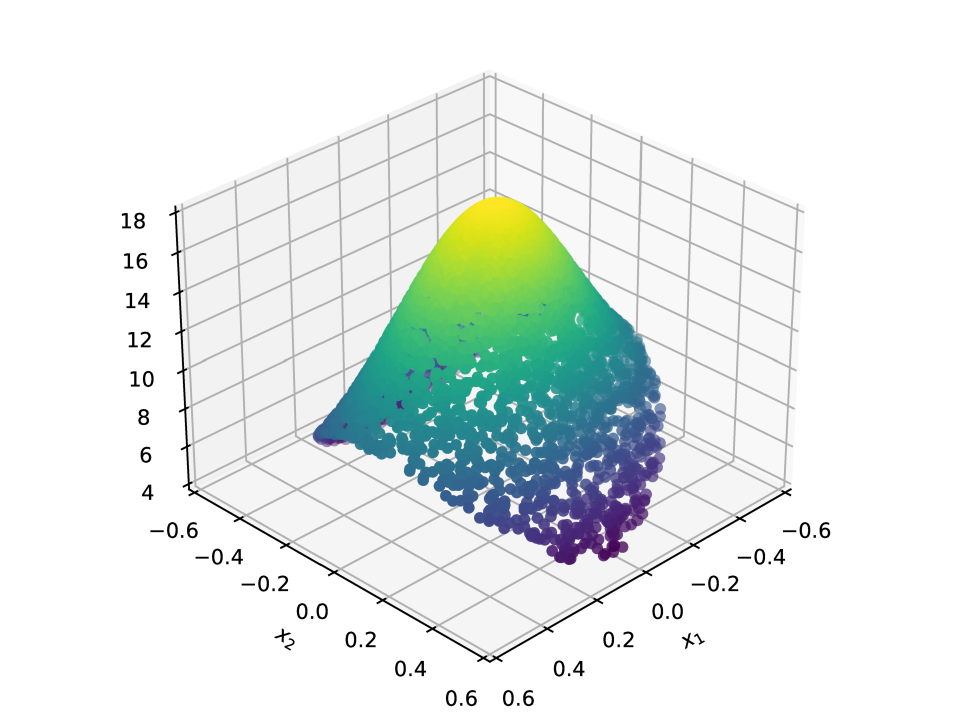}
        \caption{$g_{00}$ Patch 1}
        \label{fig:vis_2dpos_g001}
    \end{subfigure} 
    \begin{subfigure}{0.24\textwidth}
        \centering
        \includegraphics[width=0.98\textwidth]{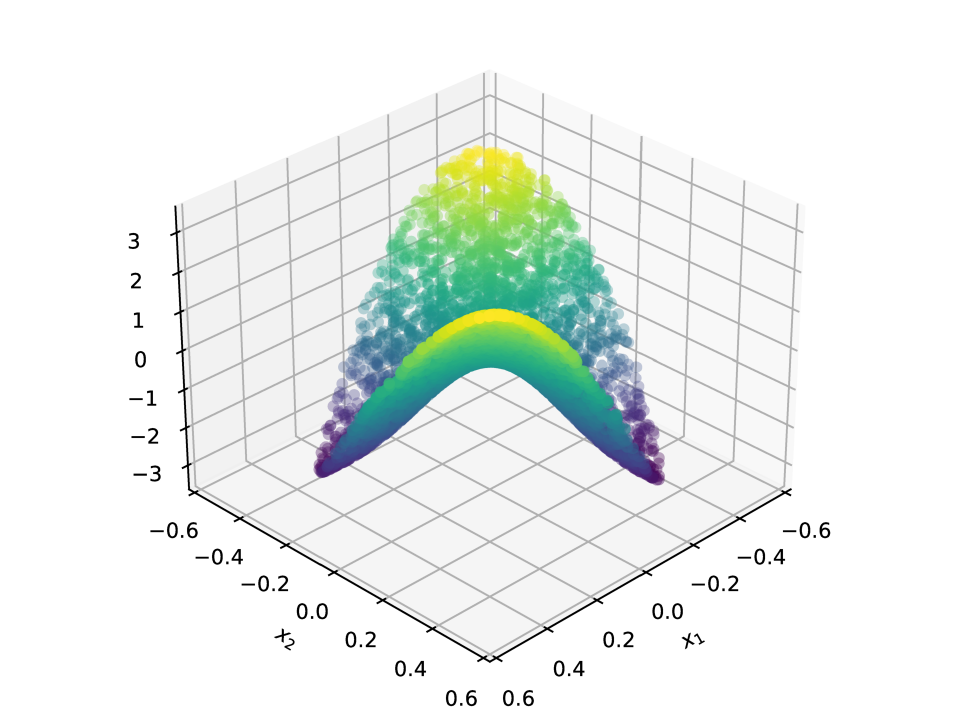}
        \caption{$g_{01}$ Patch 1}
    \end{subfigure} 
    \begin{subfigure}{0.24\textwidth}
        \centering
        \includegraphics[width=0.98\textwidth]{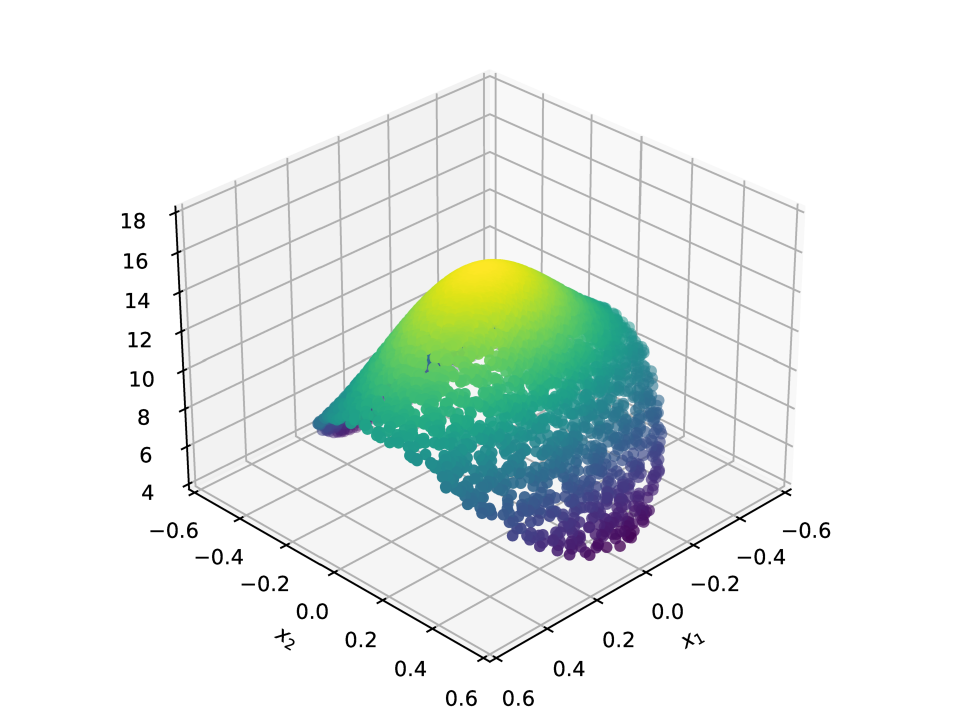}
        \caption{$g_{00}$ Patch 2}
    \end{subfigure} 
    \begin{subfigure}{0.24\textwidth}
        \centering
        \includegraphics[width=0.98\textwidth]{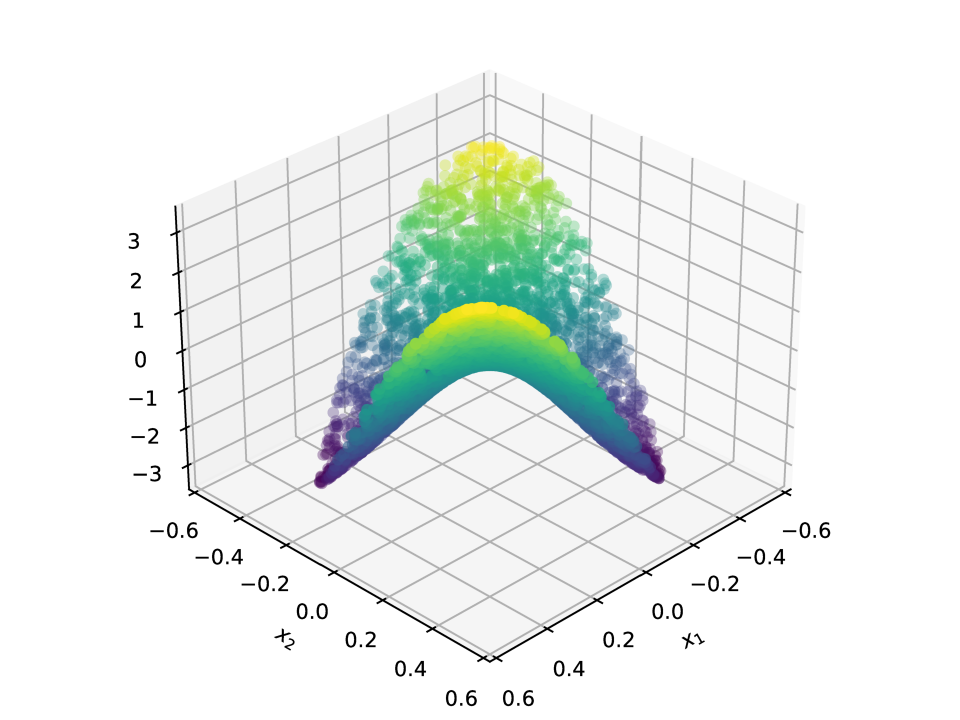}
        \caption{$g_{01}$ Patch 2}
    \end{subfigure}\\
    \begin{subfigure}{0.24\textwidth}
        \centering
        \includegraphics[width=0.98\textwidth]{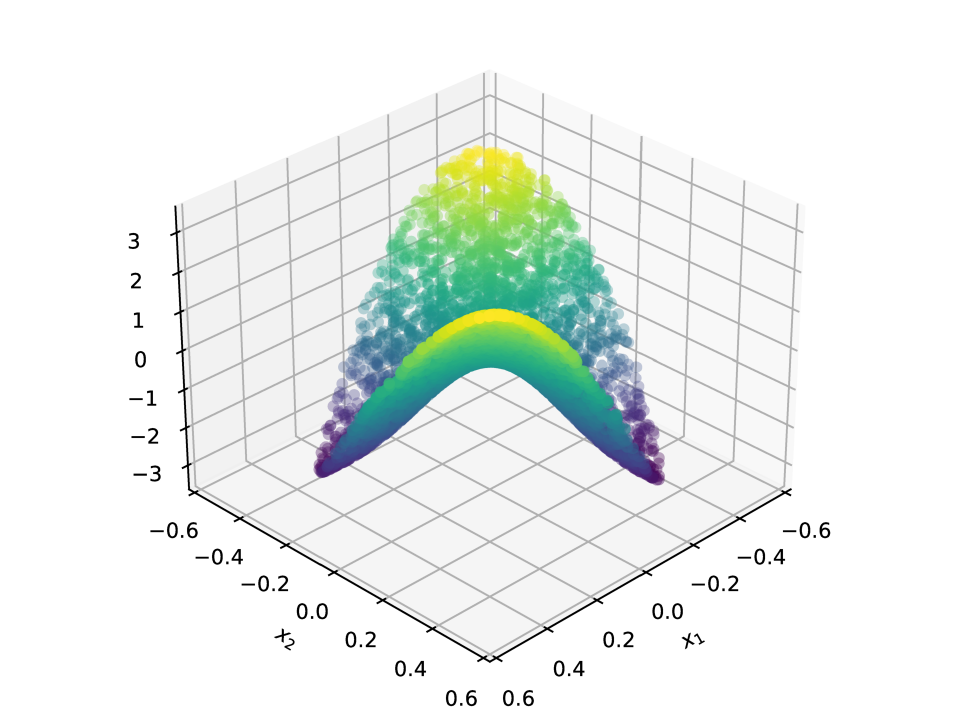}
        \caption{$g_{10}$ Patch 1}
    \end{subfigure} 
    \begin{subfigure}{0.24\textwidth}
        \centering
        \includegraphics[width=0.98\textwidth]{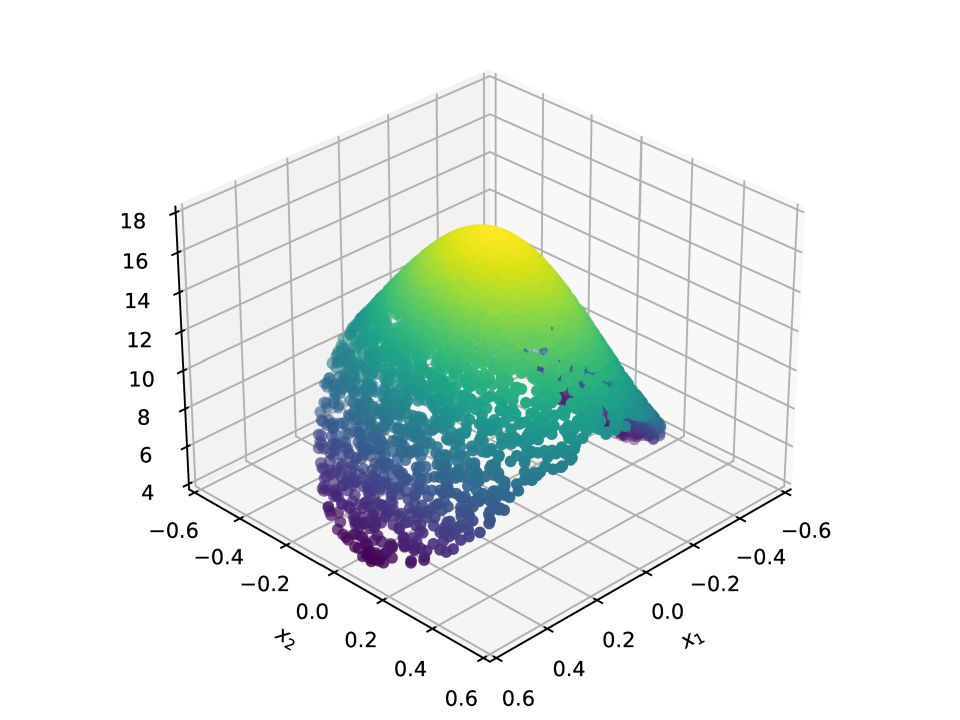}
        \caption{$g_{11}$ Patch 1}
    \end{subfigure} 
    \begin{subfigure}{0.24\textwidth}
        \centering
        \includegraphics[width=0.98\textwidth]{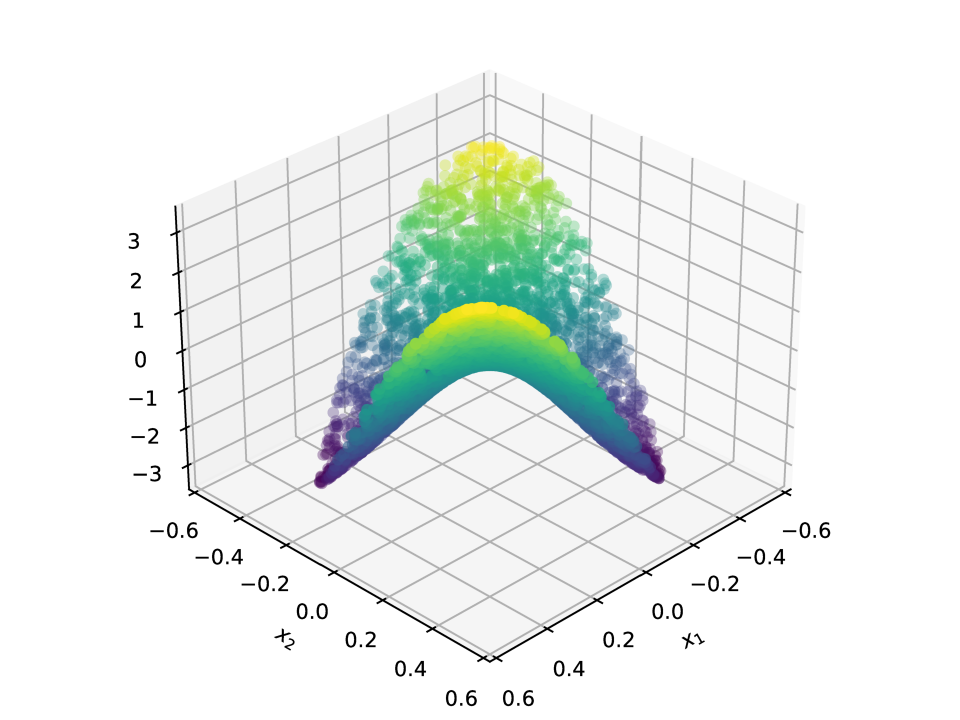}
        \caption{$g_{10}$ Patch 2}
    \end{subfigure} 
    \begin{subfigure}{0.24\textwidth}
        \centering
        \includegraphics[width=0.98\textwidth]{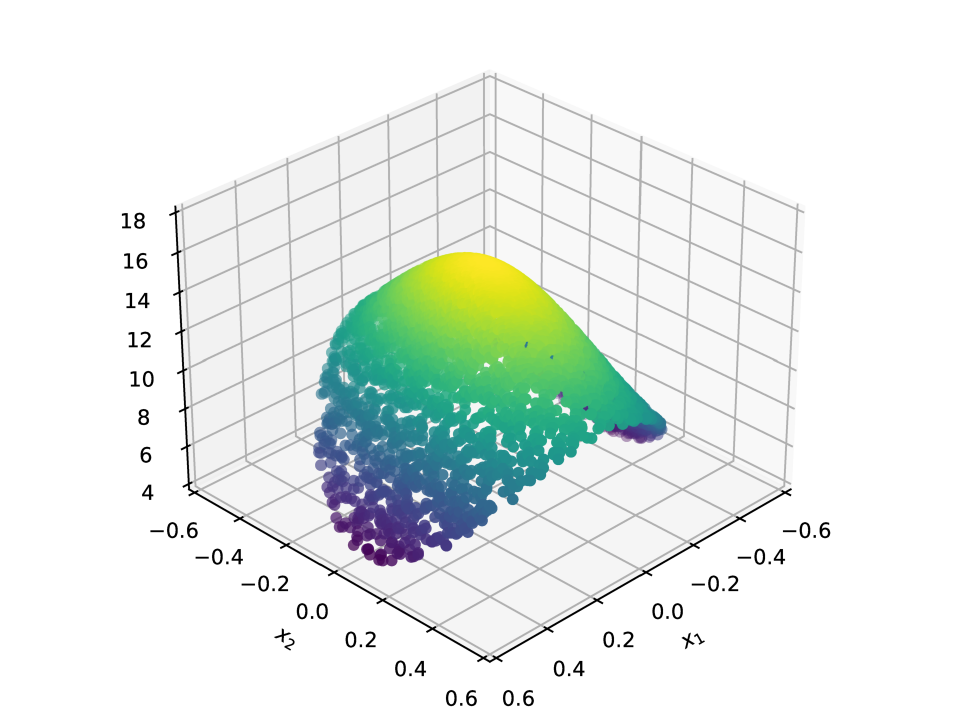}
        \caption{$g_{11}$ Patch 2}
    \end{subfigure}
    \caption{Visualisations of the learnt metrics, $g_{ij}$, in 2d, on the 2 patches, trained with positive Einstein constant (such that $R_{ij} = g_{ij}$).}
    \label{fig:vis_2dpos_g}
\end{figure}

\begin{figure}[H]
    \centering
    \begin{subfigure}{0.24\textwidth}
        \centering
        \includegraphics[width=0.98\textwidth]{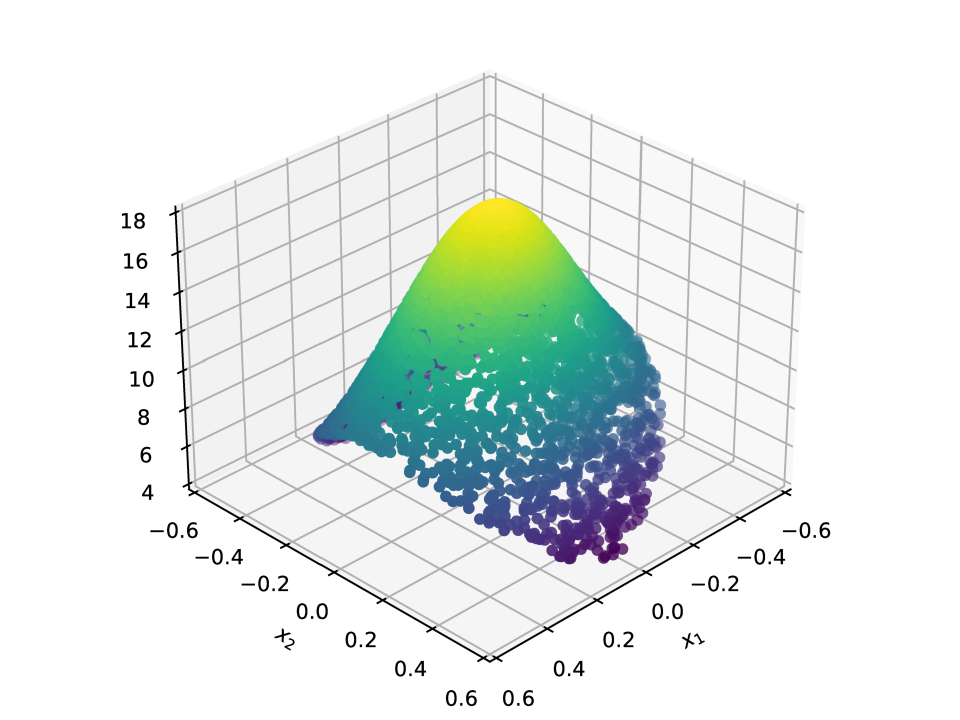}
        \caption{$R_{00}$ Patch 1}
    \end{subfigure} 
    \begin{subfigure}{0.24\textwidth}
        \centering
        \includegraphics[width=0.98\textwidth]{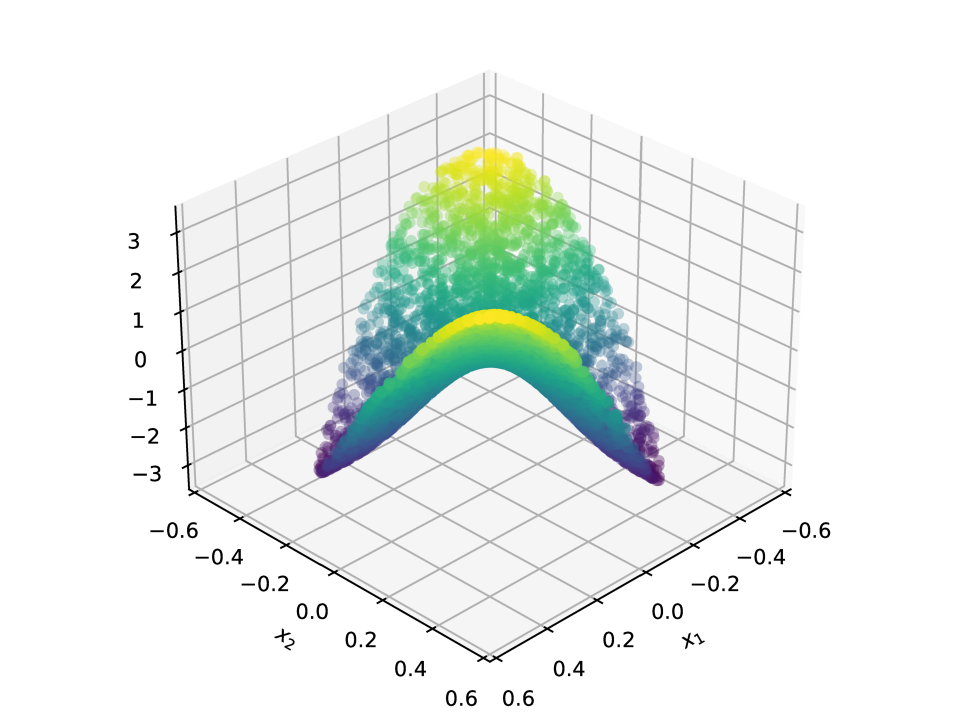}
        \caption{$R_{01}$ Patch 1}
    \end{subfigure} 
    \begin{subfigure}{0.24\textwidth}
        \centering
        \includegraphics[width=0.98\textwidth]{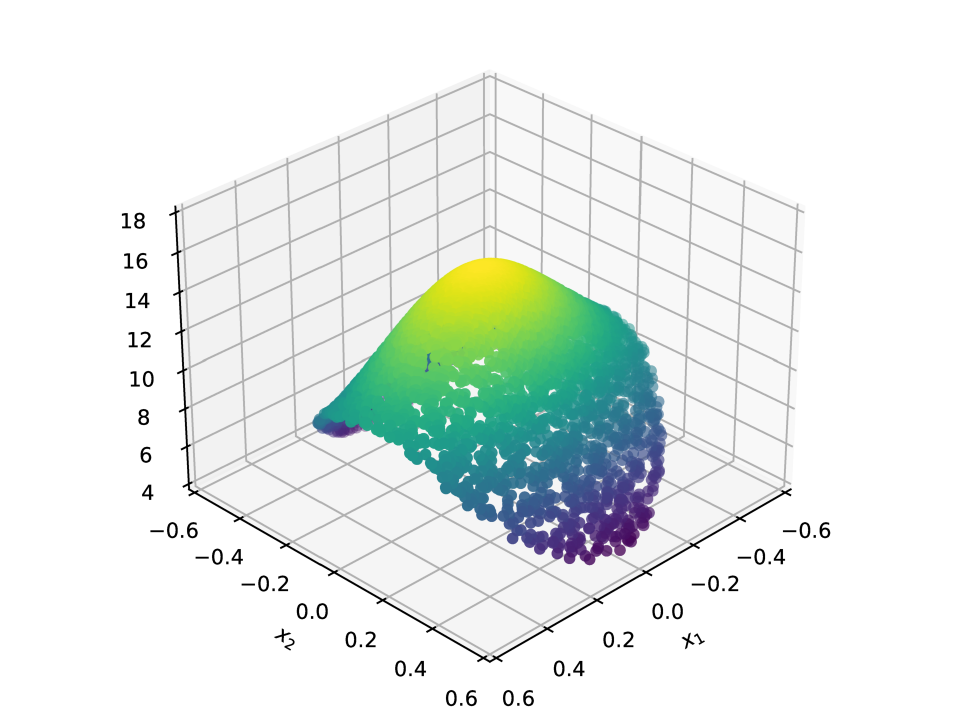}
        \caption{$R_{00}$ Patch 2}
    \end{subfigure} 
    \begin{subfigure}{0.24\textwidth}
        \centering
        \includegraphics[width=0.98\textwidth]{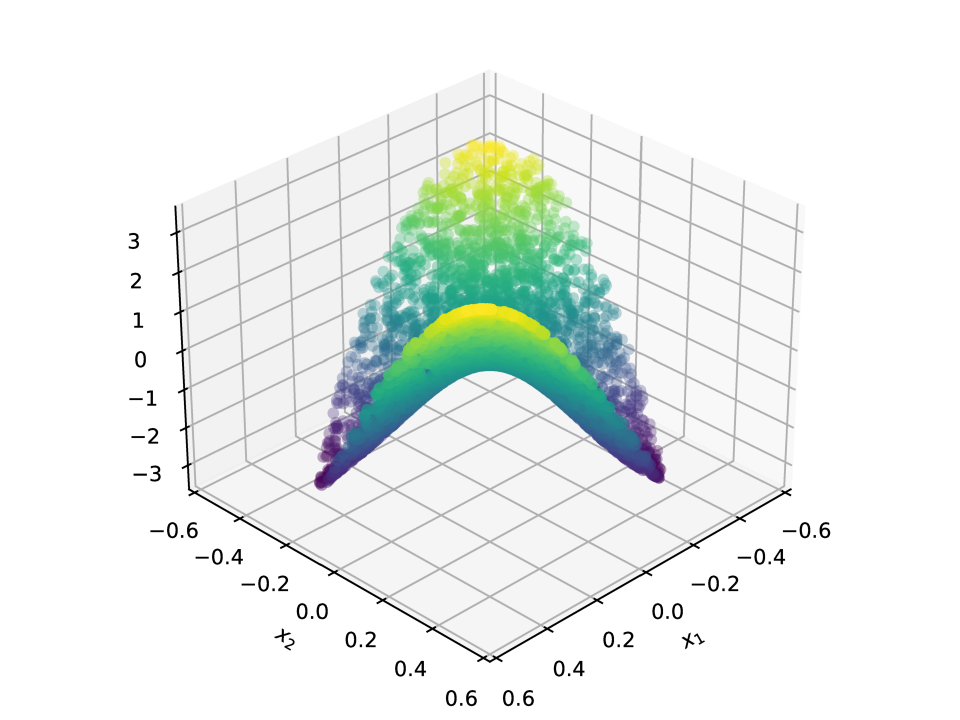}
        \caption{$R_{01}$ Patch 2}
    \end{subfigure}\\
    \begin{subfigure}{0.24\textwidth}
        \centering
        \includegraphics[width=0.98\textwidth]{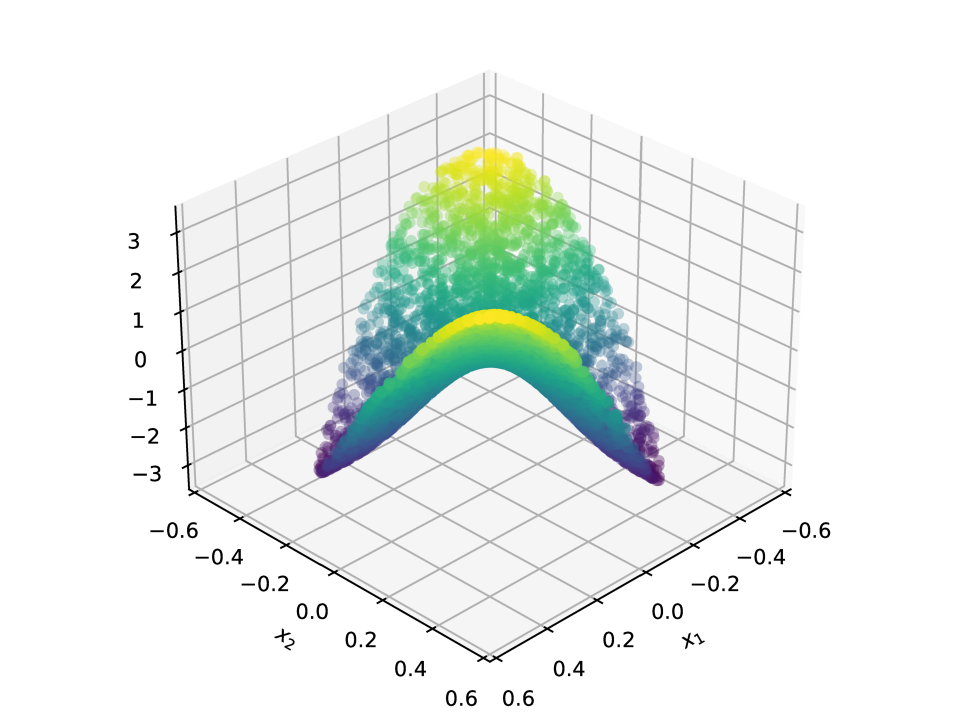}
        \caption{$R_{10}$ Patch 1}
    \end{subfigure} 
    \begin{subfigure}{0.24\textwidth}
        \centering
        \includegraphics[width=0.98\textwidth]{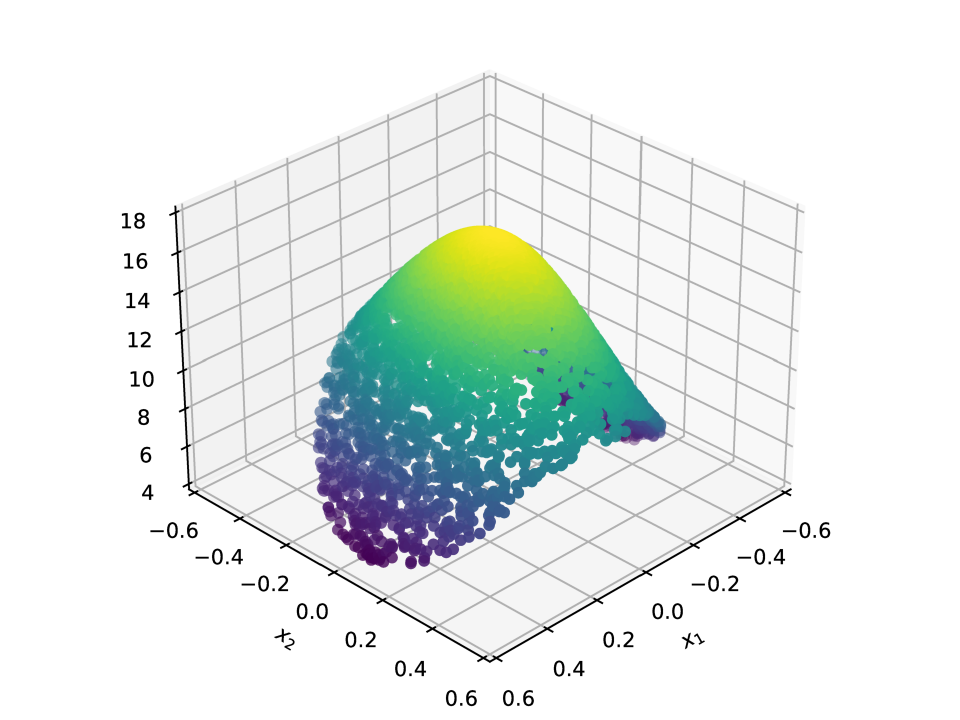}
        \caption{$R_{11}$ Patch 1}
    \end{subfigure} 
    \begin{subfigure}{0.24\textwidth}
        \centering
        \includegraphics[width=0.98\textwidth]{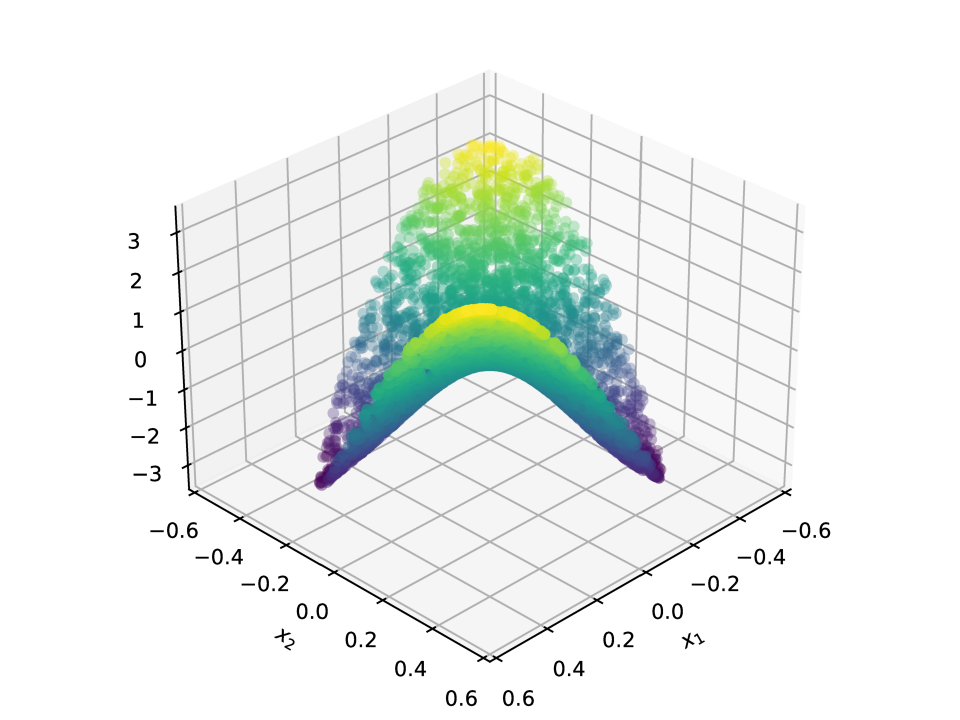}
        \caption{$R_{10}$ Patch 2}
    \end{subfigure} 
    \begin{subfigure}{0.24\textwidth}
        \centering
        \includegraphics[width=0.98\textwidth]{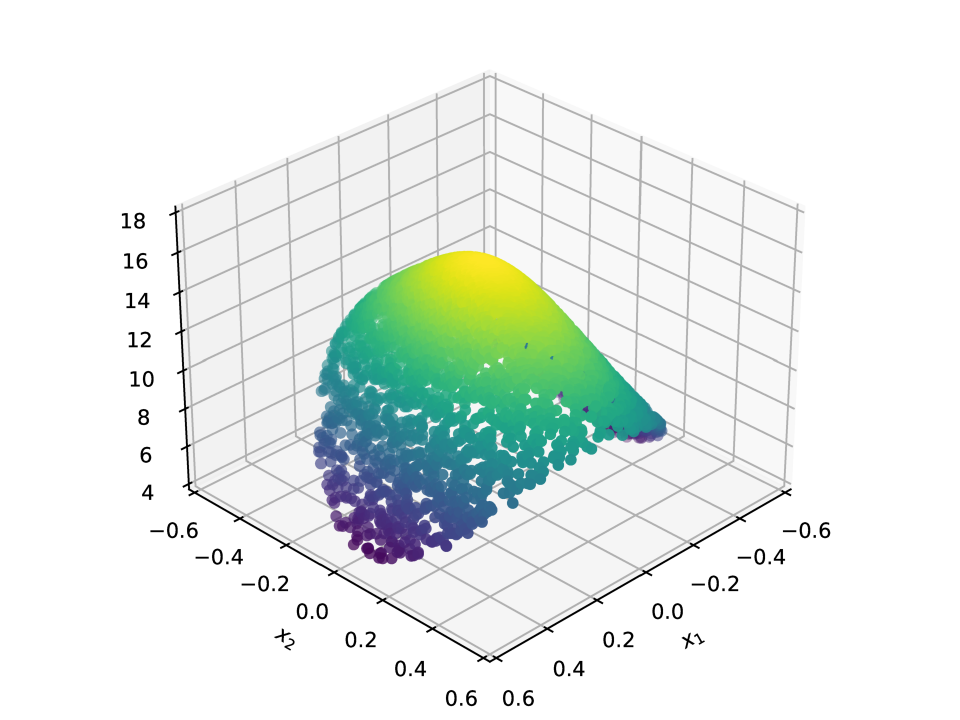}
        \caption{$R_{11}$ Patch 2}
    \end{subfigure}
    \caption{Visualisations of the Ricci tensors, $R_{ij}$, of the learnt metrics in 2d, on the 2 patches, trained with positive Einstein constant (such that $R_{ij} = g_{ij}$).}
    \label{fig:vis_2dpos_R}
\end{figure}

\subsubsection{Visualisations}
To make tangible the metric learning provided by this package and the respective semi-supervised models, we present here visualisations of an example run of the 2d $\lambda=+1$ investigation.
Figure \ref{fig:vis_2dpos_g} shows the 4 metric components ($g_{ij}$) in both patches, whilst Figure \ref{fig:vis_2dpos_R} shows the 4 respective Ricci components ($R_{ij}$) in both patches also.
The plot data uses the same test data, with the same patch restriction to radii $r_m+0.1 \sim 0.51$ to reflect the required patch and overlap elements for building the global manifold.
We emphasise that the behaviour was consistent across the 10 runs, and note that equivalent visualisations for the 2d $\lambda \in \{0,-1\}$ investigations are shown in Section \ref{app:extra_vis}.

Since the $\lambda=+1$ investigation involves solving the Einstein equation $R_{ij}=g_{ij}$, one expects a solution to have identical metric and Ricci components over the patch, these Figures \ref{fig:vis_2dpos_g} \& \ref{fig:vis_2dpos_R} demonstrate this especially well, with matching components between metric and Ricci, equally good in both patches.
These visualisations corroborate the strong learning of the $\lambda=+1$ Einstein metrics, confirming that the low losses observed for $\lambda=+1$ in Table \ref{tab:global_test_losses} do truly represent good Einstein metrics\footnote{We add that visualisations were also generated in higher-dimensions, using 2d sections of the patches, and matching was equivalently good.}.

\begin{figure}[!t]
    \centering
    \begin{subfigure}{0.48\textwidth}
        \centering
        \includegraphics[width=0.98\textwidth]{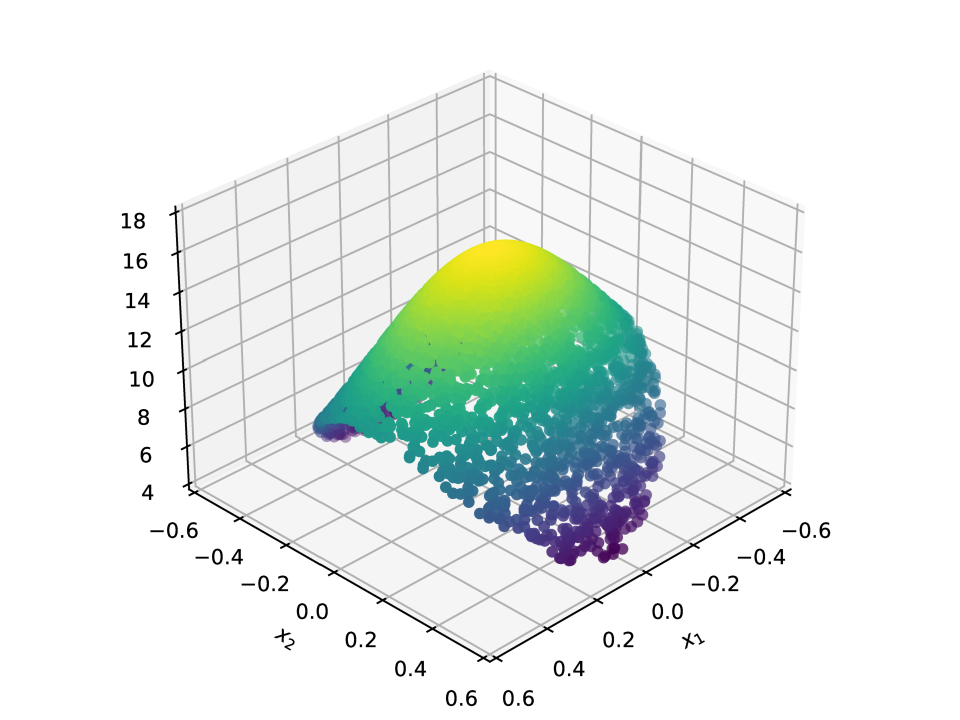}
        \caption{$g_{00}$ Analytic}
    \end{subfigure} 
    \begin{subfigure}{0.48\textwidth}
        \centering
        \includegraphics[width=0.98\textwidth]{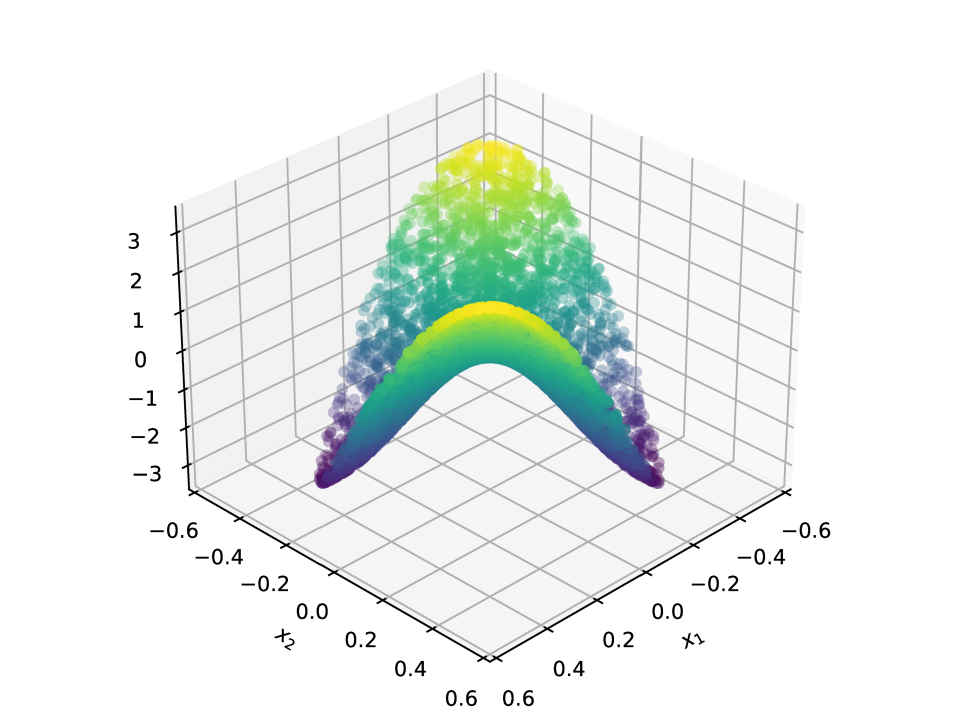}
        \caption{$g_{01}$ Analytic}
    \end{subfigure}\\
    \begin{subfigure}{0.48\textwidth}
        \centering
        \includegraphics[width=0.98\textwidth]{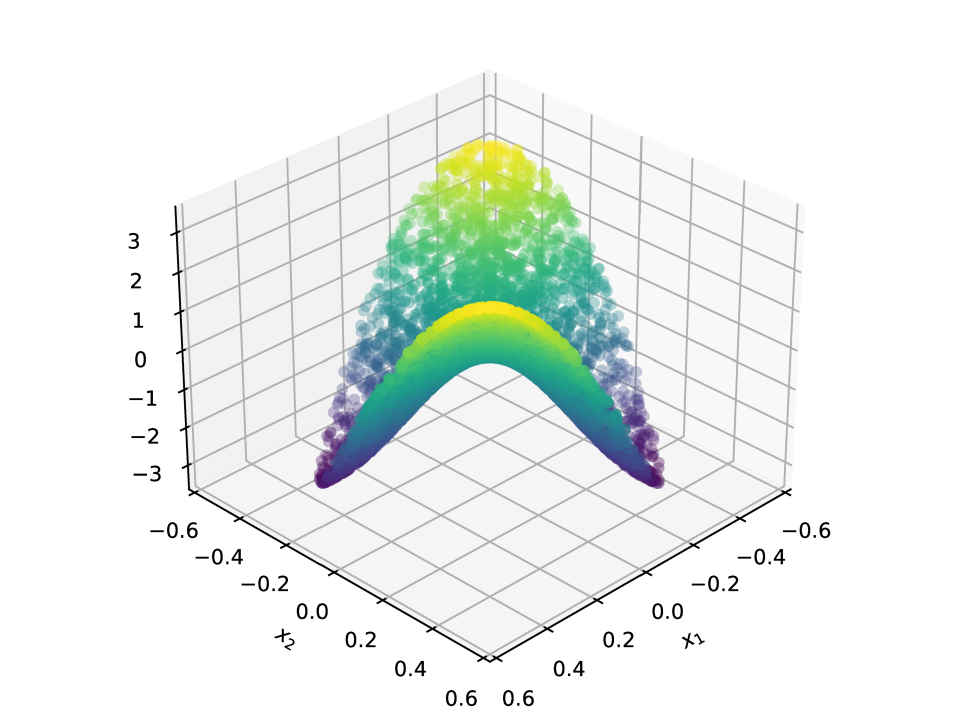}
        \caption{$g_{10}$ Analytic}
    \end{subfigure} 
    \begin{subfigure}{0.48\textwidth}
        \centering
        \includegraphics[width=0.98\textwidth]{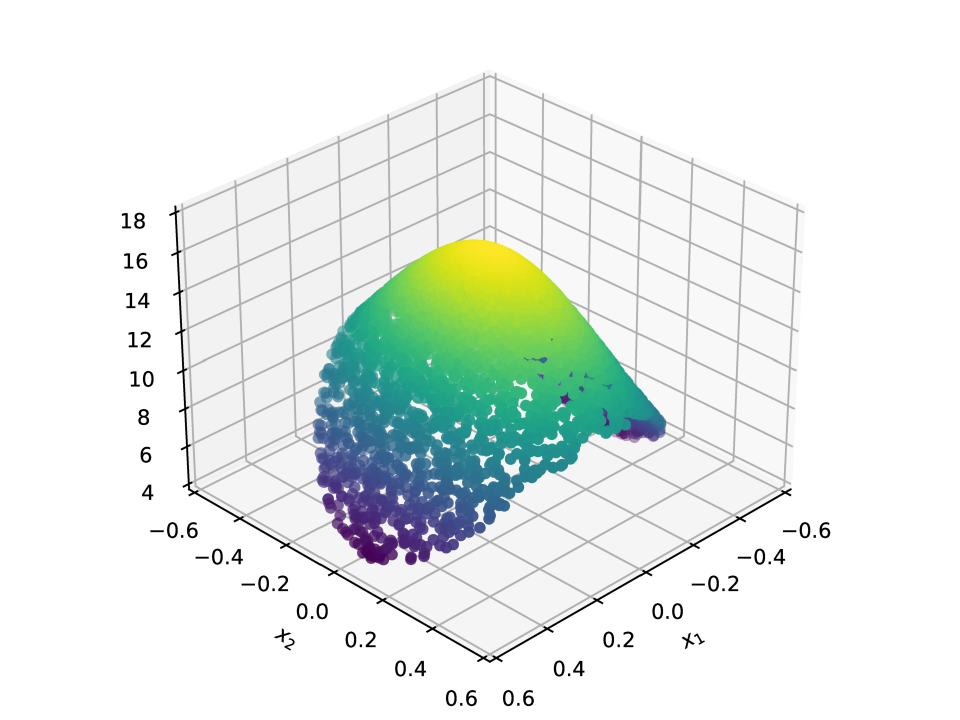}
        \caption{$g_{11}$ Analytic}
    \end{subfigure} 
    \caption{Visualisations of the analytic round metric, $g_{ij}$, in 2d on a ball patch. This metric solves the Einstein metric equation with positive Einstein constant ($R_{ij} = g_{ij}$), such that each metric component $g_{ij}$ equals its equivalent Ricci component $R_{ij}$.}
    \label{fig:vis_analytic}
\end{figure}

As a final comparison, in Figure \ref{fig:vis_analytic}, the metric components of the analytic round metric of \eqref{eq:roundmetric} are computed and plotted in the same visualisation style.
This metric is the same in both patches, and these plotted metric values were computed in the same way as the outputs used for the training of the supervised models whose test scores are shown in Table \ref{tab:global_test_losses}.
Of note is that these visualisations are strikingly similar to those in Figure \ref{fig:vis_2dpos_g}, indicating that the 2d $\lambda=+1$ model learnt by the semi-supervised model is this known analytic round metric, yet learned \textit{better} without the knowledge of the metric values, relying only on solving the Einstein equation directly.

\section{Current Work: Lens Spaces}
\label{sec:Lens_chap5}
At the time of writing this thesis, the AInstein code has been generalised to handle manifold structures that involve more than $2$ patches. This, in principle, allows to consider all possible manifolds - although some constructions might be unfeasible in practice. \\
One case that is currently being tested is that of lens spaces, which were discussed in Section \ref{sec:lens_spaces_chap3}. As we discussed above, a precise knowledge of the explicit atlas and its transition functions is needed for this numerical scheme. This was partially discussed in Section \ref{subsec:Kaluza_Klein_on_lens_chap3}, where we presented one of the change of coordinates associated with the Hopf bundle, for $\sigma_1 = \sigma_2 = \alpha$:
\begin{align}
    (X, u) \mapsto (Y, u') = (1/\bar{X}, \frac{\cos (\Theta)}{1-\sin(\Theta)} ) \, ,
    \label{eq:Tra_alpha_alpha}
\end{align}
where $\Theta_{\alpha}$ is given by $\Theta_ = k \atantwo(x_2 , x_1) + \atantwo( \frac{u^2 -1}{u^2 + 1} , \frac{2 u }{u^2 + 1 }) $. Just to expand on this expression, let us briefly comment on its underlying maps. First, we are mapping $u$, the original coordinate, back to the circle embedded in $R^2$, which yields $(x,y)=(\frac{2u}{u^2 + 1},\frac{u^2 - 1}{u^2 + 1})$. Then, we are extracting the angle of that point, via $\atantwo( \frac{u^2 -1}{u^2 + 1} , \frac{2 u }{u^2 + 1 })$, and adding the usual ``twist'' $k \atantwo(x_2 , x_1)$. Finally, we are mapping the resulting angle, $\Theta$, to a stereographic coordinate, via $\frac{\cos (\Theta)}{1-\sin(\Theta)}$.
Note that $\atantwo(x_2 , x_1)$ just gives the angle of $X$ with respect to the horizontal axis. Also note that the inverse of the above transformation, i.e.~from $U_{B \alpha}$ to $U_{A \alpha}$, reads the same but with $ \atantwo(x_2 , x_1) \xrightarrow[]{} -  \atantwo(x_2 , x_1)$, as expected. \\
When $\sigma_1 = \alpha$ and $\sigma_2 = \beta$, there is an extra change of coordinates on the fibre, and the overall transformation reads:
\begin{align}
    (X, u) \mapsto (Y, u') = (1/X, \frac{\cos (\Theta)}{1 +\sin(\Theta)} ) \, ,
    \label{eq:Tra_alpha_beta}
\end{align}
where we have used the fact that $\frac{1 - \sin(\Theta)}{\cos (\Theta)} = \frac{\cos (\Theta)}{1 +\sin(\Theta)} $, and $\Theta = k \atantwo(x_2 , x_1) + \atantwo( \frac{u^2 -1}{u^2 + 1} , \frac{2 u }{u^2 + 1 })$, as before. \\
If $\sigma_1 = \beta$ and $\sigma_2 = \beta$, then the change of coordinates reads:
\begin{align}
    (X, u) \mapsto (Y, u') = (1/X, \frac{\cos (\Theta)}{1+\sin(\Theta)} ) \, ,
     \label{eq:Tra_beta_beta}
\end{align}
with $\Theta$ now given by $\Theta = k \atantwo(x_2 , x_1) + \atantwo( \frac{-u^2 +1}{u^2 + 1} , \frac{2 u }{u^2 + 1 })$. Finally, if $\sigma_1 = \beta$ and $\sigma_2 = \alpha$, one has:
\begin{align}
    (X, u) \mapsto (Y, u') = (1/X, \frac{\cos (\Theta)}{1-\sin(\Theta)} ) \, ,
    \label{eq:Tra_beta_alpha}
\end{align}
where $\Theta = k \atantwo(x_2 , x_1) + \atantwo( \frac{-u^2 +1}{u^2 + 1} , \frac{2 u }{u^2 + 1 })$.

The Jacobian associated with $\sigma_1 = \alpha$ and $\sigma_2 = \alpha$, i.e.~\eqref{eq:Tra_alpha_alpha}, is given by:
\begin{align}
    \left(
\begin{array}{ccc}
 \frac{-x_1^2+x_2^2}{\left(x_1^2+x_2^2\right)^2} & -\frac{2 x_1 x_2}{\left(x_1^2+x_2^2\right)^2} & 0 \\
 -\frac{2 x_1 x_2}{\left(x_1^2+x_2^2\right)^2} & \frac{(x_1-x_2) (x_1+x_2)}{\left(x_1^2+x_2^2\right)^2} & 0 \\
 \frac{k x_2}{\left(x_1^2+x_2^2\right) D_{\alpha \alpha} } & -\frac{k x_1}{\left(x_1^2+x_2^2\right) D_{\alpha \alpha} } & -\frac{2}{\left(1+u^2\right) } \\
\end{array}
\right) \,
\end{align}
where $D_{\alpha \alpha} = \left(\sin \left(\atantwo(u^2-1, 2 u)+ k \atantwo(x_2, x_1)\right)-1\right)$.

The Jacobian associated with $\sigma_1 = \alpha$ and $\sigma_2 = \beta$ \eqref{eq:Tra_alpha_beta} reads:
\begin{align}
    \left(\begin{array}{ccc}
\frac{-x_1^2+x_2^2}{\left(x_1^2+x_2^2\right)^2} & -\frac{2 x_1 x_2}{\left(x_1^2+x_2^2\right)^2} & 0 \\
-\frac{2 x_1 x_2}{\left(x_1^2+x_2^2\right)^2} & \frac{(x_1-x_2)(x_1+x_2)}{\left(x_1^2+x_2^2\right)^2} & 0 \\
\frac{k x_2}{\left(x_1^2+x_2^2\right) D_{\alpha \beta} } & -\frac{k x_1}{\left(x_1^2+x_2^2\right) D_{\alpha \beta} } & -\frac{2}{\left(1+u^2\right)D_{\alpha \beta} }
\end{array}\right) \,
\end{align}
where $D_{\alpha \beta} = \left(1+\sin\left(\atantwo\left(u^2 - 1, 2 u\right)+k \atantwo(x_2, x_1)\right)\right)$.

The Jacobian associated with $\sigma_1 = \beta$ and $\sigma_2 = \beta$ \eqref{eq:Tra_beta_beta} reads:
\begin{align}
    \left(\begin{array}{ccc}
\frac{-x_1^2+x_2^2}{\left(x_1^2+x_2^2\right)^2} & -\frac{2 x_1 x_2}{\left(x_1^2+x_2^2\right)^2} & 0 \\
-\frac{2 x_1 x_2}{\left(x_1^2+x_2^2\right)^2} & \frac{(x_1-x_2)(x_1+x_2)}{\left(x_1^2+x_2^2\right)^2} & 0 \\
\frac{k x_2}{\left(x_1^2+x_2^2\right)D_{\beta \beta}} & -\frac{ k x_1}{\left(x_1^2+x_2^2\right)D_{\beta \beta}} & -\frac{2}{\left(1+u^2\right)}
\end{array}\right) \,
\end{align}
where $D_{\beta \beta} = \left(1+\sin\left(\atantwo\left(1 - u^2 , 2 u \right)+ k \atantwo(x_2, x_1)\right)\right)$.

The Jacobian associated with $\sigma_1 = \beta$ and $\sigma_2 = \alpha$ \eqref{eq:Tra_beta_alpha} reads:
\begin{align}
    \left(\begin{array}{ccc}
\frac{-x_1^2+x_2^2}{\left(x_1^2+x_2^2\right)^2} & -\frac{2 x_1 x_2}{\left(x_1^2+x_2^2\right)^2} & 0 \\
-\frac{2 x_1 x_2}{\left(x_1^2+x_2^2\right)^2} & \frac{(x_1-x_2)(x_1+x_2)}{\left(x_1^2+x_2^2\right)^2} & 0 \\
\frac{k x_2}{\left(x_1^2+x_2^2\right)D_{\beta \alpha}} & -\frac{k x_1}{\left(x_1^2+x_2^2\right)D_{\beta \alpha}} & -\frac{2}{\left(1+u^2\right)D_{\beta \alpha}}
\end{array}\right) \, ,
\end{align}
where $D_{\beta \alpha} = \left(\sin\left(\atantwo\left(1 - u^2 , 2 u \right)+ k \atantwo(x_2, x_1)\right) - 1 \right)$.
This provides a complete description of the atlas and transition functions, which can be used to implement a loss in the spirit of that described above, including multiple overlaps.

\section{Summary and Outlook}
In this chapter, we introduced a numerical scheme, based on semi-supervised machine learning, which approximates Einstein metrics on arbitrary manifolds. Results in this work restricted investigations to spheres of various dimensions, as a source of open questions regarding the existence of Einstein metrics. 

We presented an architecture which mimics the patching structure of a manifold, consisting of two parallel subnetworks. The input data are the coordinates of points in one patch; they are fed directly to the first sub-network, and they are transformed into coordinates of the second patch before being fed to the second subnetwork. Then, each subnetwork predicts the components of the metric, which we label as $g^1$ and $g^2$. The first loss component computes the Einstein condition for each patch independently, as $|\lambda g^{1,2} - Ric(g^{1,2})|$.
The second loss component ensures the correct transformation property of the metric under a change of coordinates; this is, schematically, $J^T g^1 J = g^2$, where $J$ is the Jacobian of the change of coordinates between the two patches. Such loss is evaluated to prioritise points belonging to the overlap region of the two patches. 
Finally, an artificial component is also added to the loss function to prevent the convergence to metrics with very low entries.

As mentioned, we applied our method to the case of spheres in dimension $2,3,4,5$, which admit a natural description in terms of two patches. While essentially all geometric properties have been fully understood in the former two dimensions, many questions are open in the latter two dimensions, especially regarding the existence of Ricci-flat metrics. 

For all of our runs, we initialise the neural network with a non-geometric configuration, where the metric is flat in both patches - therefore violating the patching condition. This is done in order not to introduce any bias in the process. Our findings show that the semi-supervised model trained with $\lambda = +1$ is able to converge to the round metric on $S^{2,3,4,5}$, consistently and with absolute errors in the order of $10^{-1}$ for both the Einstein condition and the overlap condition. To confirm that the output metric coincides with the maximally symmetric round one, we perform a qualitative as well as a quantitative verification. The former one consists of inspecting (sections of) the output for the various components, and comparing it with the analytic prediction; this is reported in many of the plots. The latter one is provided by comparing the performance of the semi-supervised model with a fully supervised model trained to approximate the exact analytic form of the round metric, for the same amount of data and training epochs. We find that the semi-supervised results always outperform the supervised ones, corroborating the convergence properties of our method. When applied to the cases $\lambda = 0, -1$ in dimensions $2$ and $3$, the error increases consistently by at least one order of magnitude. This is in accordance with known results which disprove the existence of Einstein metrics with zero or negative constant on $S^{2,3}$. The results concerning $S^{4,5}$ are analogous, with a marginal increase in the error across all values of $\lambda$. Since the method does not rely on any analytic assumption regarding symmetry or Killing vectors, these results provide numerical evidence towards the non-existence of Einstein metrics on $S^{4,5}$, which is a long-standing open problem in differential geometry.

The advantages of our method compared to traditional algorithms are numerous. First of all, the stochastic nature of neural networks allows for a more dynamical exploration of the landscape of metrics. Moreover, we observe an exceptionally good scaling of the number of samples with the manifold's dimension. Instead of the traditional $D^n$ associated with finite-difference methods, we find that almost no scaling is required for our purposes. As another key advantage, the neural network architecture can be adapted to predict not just one metric, but a family of them, which would allow exploration of moduli spaces of metrics. The simplest scenario for testing this feature is the case of $T^2$. Finally, the general construction of our code allows application to manifolds which are described by more than two patches; these are hard to deal with if one uses current algorithms.

In addition to tackling questions regarding the existence of certain metrics, our method could also be used to find numerical approximations for metrics lacking an analytic description\footnote{To be fully rigorous, we can foresee coupling this scheme to computer-assisted proofs' techniques (see \cite{808365d6fe2f449e8be7d40295302da1} for instance).}. In this light, it is our intention to apply it to the case of exotic $7$-spheres, for which existence results have been proven in \cite{boyer2003einstein, boyer2004einstein}, and recent progresses in the understanding of their geometry have been presented in \cite{Gherardini:2023uyx, berman2024curvatureexotic7sphere}. It is exactly for this purpose that we are currently testing lens spaces, because of their structural similarity with exotic spheres, as commented in Chapter \ref{chap:3}. Going to higher dimensions would, of course, result in more expensive computations. For this reason, we plan on further developing our code in two directions. On one side, making it suitable for GPU's. On the other side, we are planning to implement soon is the fixing of the diffeomorphism freedom at the level of the loss function, following the prescription outlined in \cite{Figueras:2012xj}. Finally, we are currently working on applying our method to problems of great relevance in theoretical physics. We are investigating Ricci-flat solutions with the Euclidean Schwarzschild's topology, and modifying the code to include Lorentzian-signature metrics.



\chapter{Conclusions}
This thesis is a study of exotic spheres through a theoretical physicist's toolkit. The previous chapters contained a plethora of techniques, analytical as well as numerical, that can be employed to investigate many unknown and mysterious properties of exotic spheres. Some of these methods, mainly gauge theoretic and differential geometric, led to new results which deepened our understanding of these manifolds. Some other approaches, based on differential topology and machine learning, are currently being developed with the hope of being effective tools to answer a number of open questions. Overall, we hope to have convinced the reader that exotic spheres offer a plethora of interesting mathematical challenges; while those concerning the metric tensor have a clear and straightforward application in the context of supergravity, those to do with the differential topology properties that make exotic spheres ``exotic'' currently seek satisfying physical interpretation. We now provide a detailed and technical summary of what outlined above.

The first chapter of this thesis discussed some epistemological aspects of this work and introduced exotic spheres from an historical perspective. It clarified the context into which the current research effort fits, by discussing the main mathematical results on exotic spheres and exotic manifolds in general, together with the few applications of these studies in theoretical physics. The theorems concerning manifolds carrying inequivalent differentiable structures largely outnumber the discussions on their possible interpretation within the context of general relativity or string theory. For this reason, we were able to list all of the latter, which include gravitational path integrals, global gravitational anomalies, supergravity compactifications and cosmology. 

In the second chapter, some of the main themes were introduced: fibre bundles, gauge and Yang--Mills theory, self-duality, and instantons. These topics are presented parallelly from the mathematical  perspective and from a theoretical physics viewpoint. While most of the chapter deals with reviewing standard results, the last section treats a recent study on a modified version of self-duality, called twisted self-duality (\cite{Berman:2022dpj}). The emergence of these equations is discussed, and solutions both in Lorentzian and Euclidean settings are derived. Their geometric interpretation is also provided, revealing a close connection with the type fibre bundles among which Milnor found the first exotic spheres. A short summary and an overview of possible future directions is provided in the last section. 

The core of this thesis consists of Chapter 3. After introducing the notion of exotic differentiable structure, the Kaluza--Klein formalism is introduced, in its progressively more complicated realisations. Each section deals with the physical interpretation of Kaluza--Klein theory as a dimensional reduction tool as well as the so-called ``inverse Kaluza--Klein'' prescription, which is a common tool among mathematicians too. First, the original abelian Kaluza--Klein ansatz is discussed, under both perspectives. As an example of a Kaluza--Klein geometry on an abelian bundle, the case of lens spaces is discussed in detail. These manifolds are the lower-dimensional analogue of exotic spheres, where quaternions are replaced by complex numbers. Then, the Kaluza--Klein machinery for non-abelian principal bundles is presented, and exemplified through the case of $S^7$, viewed as a quaternionic Hopf fibration. Finally, the case of associated (i.e.~non principal) non-abelian bundles is discussed. It is shown how this formalism is readily applicable to exotic spheres since, according to Milnor's original construction, they are obtained as (non-principal) $S^3$ bundles over $S^4$. The study is then focused to a specific exotic sphere, the Gromoll--Meyer one. According to the Kaluza--Klein ansatz for associated bundles, the only aspect of the geometry which is not widely studied in the literature is the ``double instanton'', i.e.~a self-dual $\text{SU(2)}$ connection whose winding number equals two. This object is discussed (in singular gauge), and then an explicit coordinate expression for a Kaluza--Klein metric on the Gromoll--Meyer sphere is presented, in components. This concludes the first half of the chapter, based on the results in \cite{Gherardini:2023uyx}. The second half, based on \cite{berman2024curvatureexotic7sphere}, provides a series of additional calculations, which lead to the main results of the chapter. The feasibility of such derivations heavily relies on the use of quaternionic algebra and calculus. The starting point is a rephrasing of the Gromoll--Meyer sphere's Kaluza--Klein ansatz in terms of quaternionic-valued forms. The vielbein admits an elegant description, much more concise than the component expression, allowing for a neat derivation of its Riemann tensor, Ricci tensor and scalar curvature. After having found an expression of the ``double instanton'' in regular gauge, the interplay between different moduli spaces in this problem is discussed. Then, a special point, with enhanced symmetry, is identified in the moduli space of the ``double instanton''. With this choice, it is shown that the resulting metric carries the maximal isometry allowed for any exotic sphere. Moreover, an assessment of some gravitational properties of such a geometry is carried out. By focusing on an eight dimensional Lorentzian space-time, whose space-like part is described by the metric above, it is found that the most common energy conditions (weak, strong and null) are satisfied when the radius of the base manifold $S^4$ is bigger than a certain value. The chapter ends with a brief summary of results and a discussion of the future directions. 

Chapter 4 is somewhat different from the others, in that it is completely original and not based on any existing paper. In the first part, a list of facts on exotic spheres and exotic manifolds in general is presented. This series of miscellaneous results, some of which are accompanied by detailed derivations, was included to provide the reader with an overview of the current knowledge on the topic. It is not a complete or exhaustive list, especially because it focuses on those aspects that are mainly relevant to a physicist: facts about geometry, concrete realisations of some exotic manifolds and relation of (inequivalent) differentiable structures in the formalism of general relativity. The second part of the chapter is aimed at answering the following question: how does the differentiable structure of an exotic sphere differ from the one of an ordinary sphere? The most natural way to quantify this difference is to consider homeomorphic maps from one topological manifold to the other and study how they fail to be differentiable. Two examples of such maps are presented, one following a standard recipe (the Alexander's trick), and the other one revisiting a construction from 1958 (\cite{10.2969/jmsj/01010029}) which relies on a foliation of the exotic sphere. Each of them offers insights into the nature of the ``defects'' preventing the uplift of the homeomorphic map to a diffeomorphic one. We justify the study of these maps and defects from a physics perspective too, by proposing a ``change of differentiable structure'' mechanism that, under certain assumptions, yield physically acceptable metrics. The chapter ends with a discussion on further possible investigations on exotic space-times in general relativity. \\

Chapter 5 deals with a different approach to the problem of understanding exotic spheres' geometries, based on machine learning. The method presented is based on \cite{hirst2025ainsteinnumericaleinsteinmetrics}, and is currently at an early stage; its generalisation to handle exotic spheres is currently in progress. This numerical scheme is based on neural networks, which are introduced at the beginning of the chapter. The idea behind it is to model a Riemannian metric on a given manifold with a (smooth) neural network, and impose the global patching structure as well as the Einstein condition through a loss function. The specific implementation presented in \cite{hirst2025ainsteinnumericaleinsteinmetrics} (a package called AInstein) is reviewed, and the application of the method to the case of (ordinary) spheres in dimensions $2,3,4,5$ is presented. The presence of open questions regarding the existence of Einstein metrics with zero or negative Einstein constant on $S^4, S^5$ motivated this choice. It is shown how the algorithm is able to recover the usual round metric on all spheres, but no new metrics are found by our investigations. An overview on ongoing work to generalise this scheme to exotic spheres, passing through lens spaces, is provided. The final section of this chapter discusses the advantages of AInstein compared to traditional methods, and is outlines all the possible generalisations of this method. 

There are many more foreseeable investigations, some of which are a natural continuation of those presented in this thesis, to better understand exotic spheres and exotic manifolds, alongside their implications in physics. The possible future steps in this direction were discussed at the end of each chapter. Many of them naturally lend themselves to applications of techniques from theoretical physics.

\newpage

\section*{Acknowledgements}
Firstly, I would like to thank David Berman. He gave me the once-in-a-lifetime opportunity of doing a PhD in theoretical physics; he brought me into contact with to state-of-the-art research in supergravity and exceptional field theory; he introduced me to the existence of exotic spheres and to the use of machine learning for physics. He taught me that intelligence goes beyond cleverness; that creativity can emerge even in the most constrained mathematical framework; that theoretical physics is a \textit{land of ideas}, more than calculations.
I would like to express my gratitude to Martin Cederwall. He showed me that, if one takes the time to recognise and appreciate the beautiful structures underlying many physics problems, then the result often follows smoothly and without effort. During our interactions, I could see what it means to treat mathematics as a language, rather than a tool; and what it means to speak it fluently. I am very grateful to Leonardo Cavenaghi, for his continuous interest in my work and his willingness to show me the ways of differential geometry; every discussion with him has been a source of academic inspiration. 
I am very grateful to Daniele Angella, for his academic guidance throughout these years. It is thanks to him that I could appreciate how fruitful a collaboration between researchers in different fields can be. Moreover, his kindness, patience and optimism, have been inspirational to me in several circumstances.
 
I would like to thanks Costis Papageorgakis for his precious help during these years. When I needed some advice, he has always shared some of his valuable knowledge and intuition with me.

Finally, I would like to thank Theo Kreouzis. I worked with him for one semester, and his kind, calm and composed attitude I have deeply admired. I look up to him, as an academic, but more importantly, as a person. 

These are the ``senior'' people that influenced me the most during the past four years, academically and beyond, for the reasons explained above. While working closely and/or interacting frequently with them, I came to realise that they all, in different ways, serve as examples for the person that I would like to become. \\

Among the ``less senior'' people, I would like to express my gratitude to Ed Hirst. He has introduced me to machine learning in the first place. Working with him has always been, and it still is, an absolute pleasure. How could I forget the other extraordinary member of ``exotric''? I am very grateful to Alex Stapleton, for his help with computers in general, our insightful academic discussions  and our stimulating non-academic ones. 

I would like to deeply thank Mattia Cesaro, for our numerous conversations. From physics to the most disparate topics; from Prague, to Potsdam, to Turin; I have enjoyed each of our interactions, and look forward to many more.

I am very grateful to Jonas Henkel, for his openness to ideas from physics and for his attempts to guide me inside the field of homogeneous spaces. As time passed, we found more and more points of contact, including machine learning, which I hope to be the premises for a long series of interactions.

I would like to thank Rashid Alawadi, for sharing with me his insights on fibre bundles, and for embarking on a completely new project with me.  \\

I would like to thank Ilka Agricola, Chris Hull,  Volker Branding, Gary Shiu and Carlos Nunez. At various stages of my PhD, they have shown trust and confidence in my abilities. I would also like to express my gratitude to the anonymous referee of my first JHEP paper, which was a turning point of the PhD. \\

Let me now turn to my friends who actively shaped my path within physics. Some of them have also been co-authors of some papers, but I believe it is more appropriate to list them here, since in some cases our friendship was born a long time before any academic collaboration. I would like to thank Emanuele, Emma and Pietro (in alphabetical order); eight years have passed, and we have always been together. My gratitude also goes to Andrea, whom I have known for many years before moving to London, and to Anantya, who has always been on my side since we both moved to this city. I am grateful to (again, in alphabetical order) Benjamin, Giorgio, Julian, Mitchell, Tomás for my frequent interactions with them, concerning physics and life. \\

And now, my friends outside of physics. Amongst those from Florence, I would like to thank Federico, Andrea, Alessandro, Marco and Martina, for always been there, even when we did not see each other for months; the distance has not weakened our friendships. For the same reason, I would like to thank my old-time friends from other parts of Italy, especially Bianca and Giovanni. Last but not least, those from London. I am grateful to: the rest of the Legendary 5 (Henry, Lollo and Pit), Pocock (Filo, Nicco and Mattew), the rest of the TDT (Lapo and Luca), Pitone, Mazza, Sam, Ale, Fede, Carlo, Gugli, Fabio, and my good friends from Queen Mary (Giacomo, Luce, Lucia, Matte, Sara, Sara; again, in alphabetical order!).\\

My gratitude also goes to my family; to my grandparents, aunts and uncles, and cousins Adele, Diego, Emanuela, Giulia, Giulia, Raphael and Silvia (in alphabetical order). \\

Finally, I also wish to thank all who contributed constructively to the papers presented in this thesis, and were not mentioned above. For \cite{Berman:2022dpj}, I would like to thank Malcom Perry and Chris White for discussions and comments on the manuscript and in particular Rod Halburd for early discussions on self-duality and dimensional reduction. Concerning \cite{Gherardini:2023uyx}, I am grateful Professor Johnson, Professor Schleich and Professor Derdzinski for their support during the early stages of this project. For \cite{berman2024curvatureexotic7sphere}, I would like to thank the organisers of ``Mathematical Supergravity'' at UNED, Friederik Valach and Miguel Pino Carmona for useful discussions during the preliminary phases of the project. For \cite{hirst2025ainsteinnumericaleinsteinmetrics}, I wish to thank Michael Douglas, Fabian Ruehle and Tomás Silva for their helpful comments during the ``Mathematics and Machine Learning Program'' at Harvard University, as well as Toby Wiseman for his insights on how this method compares to traditional algorithms. \\

My PhD was supported by the Science and Technology Facilities Council (STFC) Consolidated Grants ST/T000686/1 ``Amplitudes, Strings \& Duality'' and ST/X00063X/1 ``Amplitudes, Strings \& Duality'', and I also acknowledge support from Pierre Andurand over the course of the final months. The research of \cite{hirst2025ainsteinnumericaleinsteinmetrics} utilised Queen Mary's Apocrita HPC facility \cite{apocrita}, supported by QMUL Research-IT. \\

Amalia, Irene and Jacopo, thank you.

\appendix

\chapter{Mathematical Definitions and Theorems}
\label{chap:Mathematica_definitions_and_theorems}

\section{Differential Topology}
In this section, we collect a number of definitions (some more abstract than others) and theorems. It should not be thought as a comprehensive review of the differential-topological framework underlying our work, but rather as a list of facts to be aware of, since they are implicitly assumed throughout the main body.

\subsection{Some Foundational Definitions}

\begin{definition}[Topological manifold]
  Let \(n\in\mathbb{N}\).
  An \emph{\(n\)-dimensional topological manifold} is a Hausdorff, second-countable\footnote{Note that some authors do not include second-countable in their definition.}
  topological space \(M\) that is \emph{locally Euclidean of dimension \(n\)}:
  every point \(p\in M\) has an open neighbourhood \(U\subset M\) homeomorphic
  to an open subset of \(\mathbb{R}^{n}\).
\end{definition}

\medskip
\begin{definition}[Topological atlas] 
  Let \(M\) be an \(n\)-dimensional topological manifold.
  By local Euclideanness, for each \(p\in M\) choose a neighbourhood
  \(U_{p}\ni p\) and a homeomorphism  
  \(
     \varphi_{p}:U_{p}\longrightarrow\varphi_{p}(U_{p})\subset\mathbb{R}^{n}.
  \)
  The family of pairs \(\{(U_{p},\varphi_{p})\}_{p\in M}\) satisfies the
  three conditions below and therefore constitutes a \emph{topological atlas}.
  Second-countability lets us shrink this to a \emph{countable} sub-atlas if
  desired, while the axiom of choice guarantees we can make the selections. As a result of this, we define a \emph{topological atlas} on \(M\) is any collection of \textit{topological charts}
  \begin{align}
     \mathcal{A}=\{(U_\alpha,\varphi_\alpha)\}_{\alpha\in A}
  \end{align}
  such that
  \begin{enumerate}
    \item[\textnormal{(i)}] \(\displaystyle\bigcup_{\alpha\in A} U_\alpha = M\)
          (the charts cover \(M\));
    \item[\textnormal{(ii)}] each \(\varphi_\alpha:U_\alpha\to\mathbb{R}^n\) is a
          homeomorphism onto an \emph{open} subset of \(\mathbb{R}^n\);
    \item[\textnormal{(iii)}] whenever \(U_{\alpha\beta}:=U_\alpha\cap U_\beta\neq\varnothing\),
          the transition map
          \begin{align}
             \varphi_\beta\circ\varphi_\alpha^{-1}:
             \varphi_\alpha(U_{\alpha\beta}) \;\longrightarrow\;
             \varphi_\beta(U_{\alpha\beta})
          \end{align}
          is a homeomorphism between open subsets of \(\mathbb{R}^n\).
  \end{enumerate}
\end{definition}

\begin{definition}[Homeomorphism]
Let $X$ and $Y$ be topological spaces.  
A map $f\colon X \to Y$ is a \emph{homeomorphism} if it is bijective, continuous, and its inverse 
$f^{-1}\colon Y \to X$ is also continuous.  
When such an $f$ exists we say that $X$ and $Y$ are \emph{homeomorphic}; they are identical from the
purely topological point of view.
\end{definition}

\begin{remark}
  Suppose \(M\) and \(N\) are \(n\)-dimensional topological manifolds with topological
  atlases \(\mathcal{A}_M=\{(U_\alpha,\varphi_\alpha)\}\) and
  \(\mathcal{A}_N=\{(V_\beta,\psi_\beta)\}\).
  A bijection \(f:M\to N\) is a homeomorphism (in the sense above) if and only if the following equivalent condition holds:
    for every topological chart \((U_\alpha,\varphi_\alpha)\in\mathcal{A}_M\) and every topological chart
    \((V_\beta,\psi_\beta)\in\mathcal{A}_N\) with \(f(U_\alpha)\subset V_\beta\),
    the coordinate expression
    \begin{align}
        \psi_\beta\;\circ\;f\;\circ\;\varphi_\alpha^{-1}:
        \;\varphi_\alpha(U_\alpha)\longrightarrow\psi_\beta(V_\beta)
    \end{align}
    is a homeomorphism between open subsets of \(\mathbb{R}^n\).
  In other words, continuity of \(f\) and \(f^{-1}\) can be
  verified \emph{locally}, and \(f\) preserves the atlas structure by
  sending coordinate patches to coordinate patches via compatible
  homeomorphisms.
\end{remark}

\begin{definition}[Differentiable (or smooth) manifold]
An \emph{$n$-dimensional differentiable manifold} is a second-countable, Hausdorff topological space $M$
equipped with a collection $\mathcal{A}=\{(U_{\alpha},\varphi_{\alpha})\}$, called an \emph{atlas},
where each $U_{\alpha}\subset M$ is open, all $U_{\alpha}$'s cover $M$ and
\begin{align}
\varphi_{\alpha}\colon U_{\alpha}\;\xrightarrow{\;\;\cong\;\;}\; \varphi_{\alpha}(U_{\alpha})\subset\mathbb{R}^{n}
\end{align}
is a homeomorphism onto its image.  
For any overlapping charts,
$(U_{\alpha},\varphi_{\alpha})$ and $(U_{\beta},\varphi_{\beta})$, the \emph{transition maps}
\begin{align}
\varphi_{\beta}\circ\varphi_{\alpha}^{-1}
\;:\;
\varphi_{\alpha}(U_{\alpha}\cap U_{\beta}) \;\longrightarrow\;
\varphi_{\beta}(U_{\alpha}\cap U_{\beta})
\end{align}
are required to be $C^{\infty}$ diffeomorphisms.  
\end{definition}

\begin{definition}[Diffeomorphism]\label{def:diffeomorphism}
  Let \(M\) and \(N\) be \(n\)-dimensional smooth manifolds with atlases
  \(\mathcal{A}_{M}=\{(U_{\alpha},\varphi_{\alpha})\}\) and
  \(\mathcal{A}_{N}=\{(V_{\beta},\psi_{\beta})\}\).
  A map \(F\colon M\to N\) is a \textit{diffeomorphism} if
  \begin{enumerate}
    \item \(F\) is bijective;
    \item for every pair of charts with \(F(U_{\alpha})\subset V_{\beta}\) the
      coordinate representative
      \begin{align}
        \psi_{\beta} \circ F \circ \varphi_{\alpha}^{-1} \;:\;
        \varphi_{\alpha}(U_{\alpha}) \longrightarrow \psi_{\beta}(V_{\beta})
      \end{align}
      is a \(C^{\infty}\) map between open subsets of \(\mathbb{R}^{n}\);
    \item the inverse map \(F^{-1}\colon N\to M\) is also \(C^{\infty}\) (equivalently, 
          each coordinate representative above is a \(C^{\infty}\) bijection
          whose inverse is \(C^{\infty}\)).
  \end{enumerate}
  In particular, a diffeomorphism is automatically a homeomorphism of the
  underlying topological spaces, but with the stronger requirement that both
  directions respect the smooth structure in every system of local coordinates.
\end{definition}

\begin{remark}[$C^k$ manifold and $C^k$ maps]
The same objects and maps can be defined with a lower level of regularity by just substituting $C^{\infty}$ with $C^k$ in the definitions above.
\end{remark}

\begin{definition}[Equivalence relation]\label{def:eqrel}
Let \(M\) be a smooth (\(C^{\infty}\)) manifold.  
A binary relation \(\sim\subset M\times M\) is an \emph{equivalence relation} if for all \(x,y,z\in M\)
\begin{align}
\text{(i) } x\sim x,\qquad
\text{(ii) } x\sim y \;\Longrightarrow\; y\sim x,\qquad
\text{(iii) } x\sim y\text{ and }y\sim z \;\Longrightarrow\; x\sim z.
\end{align}
For \(x\in M\) the \emph{equivalence class} of \(x\) is
\begin{align}
[x]\,\coloneqq\,\{\,y\in M \mid y\sim x\,\}\subset M.
\end{align}
\end{definition}

\begin{definition}[Quotient]\label{def:quotient}
Given an equivalence relation \(\sim\) on \(M\), the \emph{quotient set}
\begin{align}
M/\!\!\sim \;=\;\bigl\{\, [x] \mid x\in M \bigr\}
\end{align}
is endowed with the \emph{quotient topology}: a subset \(U\subset M/\!\!\sim\) is open iff its pre–image under the canonical projection
\begin{align}
\pi\colon M\longrightarrow M/\!\!\sim,\qquad \pi(x)=[x],
\end{align}
is open in \(M\).
\end{definition}

\begin{proposition}[Smooth quotient manifold]\label{prop:smooth-quotient}
The topological space \(M/\!\!\sim\) carries a unique smooth-manifold structure making \(\pi\) a smooth submersion \emph{iff}
\begin{enumerate}
  \item each class \([x]\) is an embedded submanifold of \(M\);
  \item the family \(\{[x]\}_{x\in M}\) is \emph{regular}: around every \(p\in M\) there exists a chart in which all classes cut out slices of constant dimension, equivalently \(\pi\) is locally a submersion;
  \item \(M/\!\!\sim\) is Hausdorff and second–countable in the quotient topology.
\end{enumerate}
In that case \(\pi\) is a surjective smooth submersion, and \(C^{\infty}(M/\!\!\sim,N)\simeq\{f\colon M\!\to\! N \mid f\text{ smooth and constant on each }[x]\}\) for any manifold \(N\).
\end{proposition}

\begin{remark}
    As a bonus, we provide a (tentative) definition of a manifold which does not require the global topological space to begin with, but it builds on the local structure of the manifold. This is to justify the ``bottom-up'' approach which appears in Section \ref{sec:lens_spaces_chap3}.
\end{remark}

\begin{definition}[Local Manifold]
    A weak n-manifold is the result of smoothly gluing a number of open sets in $\mathbb{R}^n$, in the following way.
    Let $U^{'}_i$ be open sets of $\mathbb{R}^n$, and $\mathcal{W} = \cup_i U^{'}_i$ be the disjoint union of those. Then, we define $\mathcal{M}_W$ as $\mathcal{W}/\sim$, where the equivalence relation $\sim $ defined on points of $\mathcal{W}$ is such that:
    \begin{itemize}
        \item The only non-trivial identifications involve points belonging to different $U_i$'s ($U_i \ni x \sim y \in U_y\implies i \neq j$), it is smooth ($ U_i \ni x \sim y \in U_j$ iff $y=f_{ij}(x)$, with $f_{ij}(x)$ that maps open sets into open sets, it is $C^{\infty}$ and has a $C^{\infty}$ inverse) and it satisfies the cocycle conditions
        ($f_{ii}=\mathrm{id}$, $f_{ji}=f_{ij}^{-1}$, and
        $f_{ik}=f_{jk}\!\circ f_{ij}$ wherever all three are defined).
        \item The quotient map $\mathcal{W} \xrightarrow[]{} \mathcal{M}_W$ is open.
        \item $D=\{(x, y) \in\mathcal{W} \times \mathcal{W} \mid x \sim y\}$ is closed in $\mathcal{W} \times \mathcal{W}$.
        
    \end{itemize}
\end{definition}

\begin{remark}
    The second condition is sufficient (although not necessary) for second countability, assuming that $\mathcal{W}$ is itself second-countable; while the third condition is necessary and sufficient for Hausdorff.
\end{remark}

\textbf{Example}. \\
Let us summarise the stereographic Atlas for $S^2$ (which we define as $\{ (x,y,z) | x^2 + y^2 + z^2 =1 \}$).
The open covering is given by 
    $\{U_1, U_2 \}$, where $ U_1=\{ S^2 - \textrm{South Pole} \}$ and $U_2=\{ S^2 - \textrm{North Pole} \}$. The two maps to $\R^2$ are given by:
\begin{align}
        \psi_1: U_1 &\rightarrow \R^2 \nonumber \\
        (x,y,z) &\mapsto \psi_1   (x, y, z)= \frac{1}{1 + z } \big(x,y\big)  \eqdef \big ( X , Y \big) , \nonumber \\
        \psi_2: U_2 &\rightarrow \R^n \nonumber \\
         (x,y,z) &\mapsto \psi_1  (x,y,z)= \frac{1}{1 - z } \big(x,y\big)\eqdef \big ( X' ,Y' \big)  .
\end{align}
Thus, for completeness, the stereographic Atlas is given by the two charts $\{(U_1,\psi_1),(U_2 , \psi_2) \}$. The transition function is given by:
\begin{align}
    X'=\frac{X}{X^2 + Y^2} \quad \quad Y'=\frac{Y}{X^2 + Y^2},
\end{align}
which are $C^\infty$ since the poles are excluded from the overlap. This is extended to an arbitrary $n$-dimensional sphere in the next section. \\
Equivalently, one might construct $S^2$ without resorting to a surface embedded in $\mathbb{R}^3$, by declaring that $\mathcal{W} = \mathbb{R}_A^2 \cup \mathbb{R}_B^2 $, and that $ \mathbb{R}_A^2  \ni (X,Y) \sim (X',Y') \in \mathbb{R}_B^2 $ if and only if $(X', Y') = f(X,Y) = ( \frac{X}{X^2 + Y^2} , \frac{Y}{X^2 + Y^2} )$, with $f$ being defined everywhere but at the origin. The (set of) function(s) $f$ and the associated quotient map satisfy all the properties required by the local manifold definition, making this a local manifold realisation of $S^2$.

\begin{definition}[Homotopy group]
Fix a pointed topological space $(X,x_{0})$.  
For $n\ge 1$, the \emph{$n$-th homotopy group} of $X$ is
\begin{align}
\pi_{n}(X,x_{0}) \;=\; [\,S^{n},X\,]_{\!*},
\end{align}
the set of homotopy classes of continuous maps
$g\colon S^{n}\to X$ that send a chosen basepoint of the sphere to $x_{0}$,
where homotopies are required to keep the basepoint fixed.  
For $n\ge 2$, $\pi_{n}(X,x_{0})$ is abelian; the group operation is given by
concatenation of maps along a chosen equatorial decomposition of $S^{n}$.
\end{definition}

\begin{definition}[Homotopy]\label{def:homotopy}
Let \(M\) and \(N\) be smooth (\(C^{\infty}\)) manifolds and let
\begin{align}
f_0,f_1: M \longrightarrow N
\end{align}
be continuous maps (or smooth maps, if one desires a \emph{smooth homotopy}).  
A \emph{homotopy} from \(f_0\) to \(f_1\) is a continuous map
\begin{align}
H \;:\; M \times [0,1] \;\longrightarrow\; N
\end{align}
such that \(H(x,0)=f_0(x)\) and \(H(x,1)=f_1(x)\) for every \(x\in M\).  
If \(H\) is required to be smooth as a map of manifolds-with-boundary, one speaks of a \emph{smooth homotopy}.  
We write \(f_0 \simeq f_1\) when such a (smooth) homotopy exists.
\end{definition}

\begin{definition}[Homotopy class]\label{def:homotopy-class}
Let \(X\) and \(Y\) be smooth manifolds and let
\begin{align}
C^{\infty}(X,Y)\;=\;\{\,f:X\!\longrightarrow\!Y \mid f\text{ smooth}\,\}.
\end{align}
Declare two maps \(f,g\in C^{\infty}(X,Y)\) \emph{equivalent} if there exists a
smooth homotopy \(H:X\times[0,1]\!\longrightarrow\!Y\) with
\(H(\,\cdot,0)=f\) and \(H(\,\cdot,1)=g\).
The corresponding equivalence class of \(f\) is denoted \([f]\),
and the set of all such classes is written
\begin{align}
[X,Y]_{\!C^{\infty}}
\;=\;
C^{\infty}(X,Y)\bigl/\!\!\simeq.
\end{align}
When basepoints are specified,
\begin{align}
[(X,x_0),(Y,y_0)]_{\!C^{\infty}}
\end{align}
denotes the set of \emph{based} smooth maps
\(f:(X,x_0)\!\to\!(Y,y_0)\) modulo smooth homotopies that keep the
basepoint fixed.
\end{definition}

\begin{definition}[Homotopy group - smooth]\label{def:homotopy-group}
Let \((M,m_0)\) be a pointed smooth manifold and let \(n\ge 0\).

\medskip
\noindent
The \emph{$n^{\text{th}}$ homotopy group} of \(M\) at \(m_0\) is
\begin{align}
\pi_n(M,m_0)
\;=\;
\bigl[(S^n,s_0),(M,m_0)\bigr]_{\!C^{\infty}},
\end{align}
the set of based smooth maps \(f:(S^n,s_0)\!\to\!(M,m_0)\) modulo
based smooth homotopy.  For \(n\ge1\) this set carries a natural group
structure: given representatives \(f,g\colon S^n\!\to\!M\),
their product \(f\ast g\) is defined by
\begin{align}
(f\ast g)(x)=
\begin{cases}
f\bigl(\lambda(x)\bigr), & x\in S^n_+,\\
g\bigl(\rho(x)\bigr), & x\in S^n_-,
\end{cases}
\end{align}
where \(S^n_+\) and \(S^n_-\) are the upper and lower hemispheres,
and \(\lambda,\rho:S^n_\pm\!\to\!S^n\) are smooth rescalings chosen to be 
\emph{flat} (meaning all their derivatives vanish) at the equator \(S^{n-1}\). 
This flatness condition ensures that the concatenated map \(f\ast g\) is 
smooth everywhere, yielding a well-defined element in \(C^{\infty}(S^n,M)\). 
The resulting operation on homotopy classes is associative, has
\([c_{m_0}]\) (the constant map) as identity, and each element is
invertible up to homotopy.  Moreover, \(\pi_n(M,m_0)\) is abelian for
\(n\ge2\).

\medskip
\noindent
For \(n=0\), viewing \(S^0=\{-1,1\}\) with basepoint \(s_0=1\), the set 
\(\pi_0(M,m_0) = \bigl[(S^0,1),(M,m_0)\bigr]_{\!C^{\infty}}\) naturally 
identifies with the set of path-connected components of \(M\). It is a 
pointed set with basepoint \([m_0]\), but need not be a group.
\end{definition}

\begin{remark}[Equivalence of smooth and topological homotopy groups]
\label{rem:smooth-approximation}
Because every smooth manifold \(M\) is canonically a topological space, one 
may also consider its topological homotopy group \(\pi_n(M,m_0)\) exactly as 
in Definition~1, using continuous maps modulo continuous homotopies. By the 
Smooth Approximation Theorem (often attributed to Whitney), any continuous map 
\(S^n \to M\) is continuously homotopic to a smooth map, and any two smooth 
maps that are continuously homotopic are also smoothly homotopic. 
Consequently, the natural inclusion induces a canonical bijection
\begin{align}
\bigl[(S^n,s_0),(M,m_0)\bigr]_{\!C^{\infty}} 
\;\xrightarrow{\;\sim\;}\; 
\bigl[(S^n,s_0),(M,m_0)\bigr]_{\!*}.
\end{align}
For \(n \ge 1\), one can show that this bijection respects the group 
operations. Thus, the smooth homotopy groups of a manifold coincide exactly 
with its topological homotopy groups, fully justifying the use of the same 
notation \(\pi_n(M,m_0)\) in both contexts.
\end{remark}

\subsection{Some Working Definitions}
\label{subsec:Working_definitions_chapA1}

\begin{definition}
  The \emph{$n$-sphere of radius~\(\rho>0\)}, denoted \(S^n_\rho\), is the subset of
  \(\mathbb R^{\,n+1}\) defined by the equation
  \begin{align}
     x_1^{\,2}+\dots+x_{n+1}^{\,2}=\rho^{2}.
  \end{align}
  \label{def:Sphere_embedded_rho}
\end{definition}

\begin{definition}
  The \emph{stereographic atlas} on \(S^n_\rho\) is built from the open cover
  \(\{U_1,U_2\}\) where  

  \begin{align}
     U_1 = S^n_\rho\setminus\text{(South pole)},\qquad
     U_2 = S^n_\rho\setminus\text{(North pole)},
  \end{align}
  with North and South poles at \((0,\dots,0,\pm\rho)\).

  The stereographic projections are
  \begin{align}
    \psi_1 : U_1 &\longrightarrow \mathbb R^{\,n},
    &
    (x_1,\dots,x_{n+1})
    &\;\longmapsto\;
      \frac{\rho}{\rho+x_{n+1}}\,(x_1,\dots,x_n)=:(X_1,\dots,X_n),
    \nonumber\\
    \psi_2 : U_2 &\longrightarrow \mathbb R^{\,n},
    &
    (x_1,\dots,x_{n+1})
    &\;\longmapsto\;
      \frac{\rho}{\rho-x_{n+1}}\,(x_1,\dots,x_n)=:(X'_1,\dots,X'_n).
  \end{align}
  Thus the stereographic atlas is the pair of charts
  \(\bigl\{(U_1,\psi_1),(U_2,\psi_2)\bigr\}\).
  \label{def:Stereo_atlas_rho}
\end{definition}

\begin{remark}
For completeness, the inverses of the stereographic projections are given by
\begin{align}
  \psi_1^{-1}: \R^n &\;\longrightarrow\; U_1 \nonumber\\
  \bigl(X_{1}, X_{2}, \ldots , X_{n}\bigr)
  &\;\longmapsto\;
  \Bigl(
      \frac{2\rho^{2}X_{1}}{X_i X_{i}+\rho^{2}},
      \frac{2\rho^{2}X_{2}}{X_i X_{i}+\rho^{2}},
      \ldots ,
      \frac{2\rho^{2}X_{n}}{X_i X_{i}+\rho^{2}},
      \frac{-\,X_i X_{i}+\rho^{2}}{X_i X_{i}+\rho^{2}}
  \Bigr), \nonumber\\
  \psi_2^{-1}: \R^n &\;\longrightarrow\; U_2 \nonumber\\
  \bigl(X_{1}, X_{2}, \ldots , X_{n}\bigr)
  &\;\longmapsto\;
  \Bigl(
      \frac{2\rho^{2}X_{1}}{X_i X_{i}+\rho^{2}},
      \frac{2\rho^{2}X_{2}}{X_i X_{i}+\rho^{2}},
      \ldots ,
      \frac{2\rho^{2}X_{n}}{X_i X_{i}+\rho^{2}},
      \frac{\,X_i X_{i}-\rho^{2}}{X_i X_{i}+\rho^{2}}
  \Bigr).
\end{align}
Here \(X_i X_{i}:=\sum_{i=1}^{n}X_{i}^{2}\) is the squared Euclidean norm in \(\R^{n}\).
\end{remark}

\begin{remark}[Transition map between the two stereographic charts]
  On the overlap \(U_{1}\cap U_{2}=S^{n}_{\rho}\setminus\{\text{North,\;South poles}\}\)
  the two coordinate representations are related by the smooth map
  \begin{align}
     \psi_{21}\;:=\;\psi_{2}\circ\psi_{1}^{-1}
     \;:\;
     \R^{n}\setminus\{0\}\;\longrightarrow\;\R^{n}\setminus\{0\},\qquad
     (X_{1},\dots,X_{n})
     \;\longmapsto\;
     \Bigl(
        \frac{\rho^{2}X_{1}}{X_{i}X_{i}},
        \;\dots\;,
        \frac{\rho^{2}X_{n}}{X_{i}X_{i}}
     \Bigr),
  \end{align}
  where \(X_{i}X_{i}:=\sum_{i=1}^{n}X_{i}^{2}\).
  Its inverse (the other transition map)
  \(\psi_{12}:=\psi_{1}\circ\psi_{2}^{-1}\) is obtained by the same
  formula with primed coordinates:
  \begin{align}
     (X'_{1},\dots,X'_{n})
     \;\longmapsto\;
     \Bigl(
        \frac{\rho^{2}X'_{1}}{X'_{i}X'_{i}},
        \;\dots\;,
        \frac{\rho^{2}X'_{n}}{X'_{i}X'_{i}}
     \Bigr),\qquad
     X'_{i}X'_{i}:=\sum_{i=1}^{n}(X'_{i})^{2}.
  \end{align}
\end{remark}

\begin{remark}[Jacobian components of the transition map]\label{rmk:Jacob_components}
The Jacobian matrix \(J\) with entries
\(J_{kj}=\dfrac{\partial X'_k}{\partial X_j}\) is, component-wise,
\begin{align}
     J_{kj}(X)
     \;=\;
     \frac{\rho^{2}}{(X_i X_{i})^{2}}
     \Bigl[
        (X_i X_{i})\,\delta_{kj}\;-\;2\,X_k X_j
     \Bigr]
   ,
   \qquad k,j=1,\dots ,n.
\end{align}
The same component formulas hold for the inverse transition
map~\(\psi_{12}\) upon replacing \(X_k\) by \(X'_k\).
\end{remark}

\begin{remark}[Standard stereographic transition involves conjugation]
Let 
\(\mathbb K=\mathbb C,\mathbb H,\mathbb O\)
be the complex, quaternionic and octonionic division algebras.
Write elements of \(\mathbb K\) as \(u\) and denote conjugation by
\(\bar u\). Then,
consider the usual stereographic atlas
\(\{(U_1,\psi_1),(U_2,\psi_2)\}\) on  
\(S^2_\rho,S^4_\rho,S^8_\rho\), defined as \(|u|^{2}+t^{2}=\rho^{2}\) :
\begin{align}
   \psi_{1}(u,t)=\frac{\rho\,u}{\rho+t},
   \qquad
   \psi_{2}(u,t)=\frac{\rho\,u}{\rho-t},
   \qquad
   (u,t)\in\mathbb K\times\R.
\end{align}
Set
\(Z:=\psi_{1}(u,t)\) and \(Z':=\psi_{2}(u,t)\).
A direct calculation (identical for \(\C,\HH,\OO\)) gives the transition function:
\begin{align}
      Z \;=\; \psi_{1}\circ\psi_{2}^{-1}(Z')
      \;=\; \frac{\rho^{2}}{\;\overline{Z'}\;} \, .
\end{align}
Thus, the standard atlas introduces conjugation.
\end{remark}

\begin{definition}[Complex, quaternionic, octonionic stereographic projections]
Keep the south-pole chart \((U_{2},\psi_{2})\) as above and
modify the north-pole chart by inserting a minus sign in every
imaginary coordinate:
\begin{align}
   \widehat{\psi}_{1}(u,t):=\frac{\rho\,\overline{u}}{\rho+t},
   \qquad (u,t)\in U_{1}.
\end{align}
Explicitly:
\begin{align}
\begin{array}{ll}
S^2_\rho: & x_{1}+ix_{2}\mapsto x_{1}-ix_{2},\\
S^4_\rho: & x_{1}+ix_{2}+jx_{3}+kx_{4}\mapsto
            x_{1}-ix_{2}-jx_{3}-kx_{4},\\
S^8_\rho: & x_{1}+e_{2}x_{2}+\dots+e_{8}x_{8}\mapsto
            x_{1}-e_{2}x_{2}-\dots-e_{8}x_{8}.
\end{array}
\end{align}
We refer to the atlas
\(\bigl\{(U_{1},\widehat{\psi}_{1}),(U_{2},\psi_{2})\bigr\}\)
as the \emph{complex/quaternionic/octonionic stereographic projection}, depending on the division algebra involved;
it is defined only for \(S^{2},S^{4},S^{8}\).
\end{definition}

\begin{remark}[Transition functions]
Let:
\begin{align}
   Z:=\widehat{\psi}_{1}(u,t),
   \qquad
   Z':=\psi_{2}(u,t).
\end{align}
Then, solving \(u\) and \(t\) from \(Z'\) and substituting into \(Z\) gives the transition functions
\begin{align}
   Z
   \;=\;
   \frac{\rho^{2}}{Z'} \, .
\end{align}
Thus, with the complex/quaternionic/octonionic stereographic atlas, the overlap map is the pure inversion
\(Z\mapsto1/Z'\).
\end{remark}

\begin{remark}[Jacobian in $4$-D written with ’t Hooft symbols]
For the modified atlas $\{(U_{1},\widehat{\psi}_{1}),(U_{2},\psi_{2})\}$ on
$S^{4}_{\rho}\subset\mathbb H\times\mathbb R$
(write $X=(X_{1},\dots ,X_{4})$, $X'=(X'_{1},\dots ,X'_{4})$,
$R^{2}=X'_{i}X'_{i}$ and $^o\eta_{\mu \nu}$ be the Minkowski metric $\mathrm{diag}(-1,1,1,1)$):

\begin{align}
   X_{\mu}
   =\frac{-\rho^{2}\, \, ^o\eta_{\mu \nu }\,X'_{\nu}}{R^{2}} \, ,
\end{align}

And the corresponding Jacobian reads:
\begin{align}
     J_{\mu\nu}(X'):=\frac{\partial X_{\mu}}{\partial X'_{\nu}}
   =\frac{- \rho^{2}\, \, ^o\eta_{\mu \sigma}}{R^{4}}
     \bigl[R^{2}\,\delta_{\sigma\nu}-2X'_{\sigma}X'_{\nu}\bigr].
\end{align}
\end{remark}

\begin{remark}[Orientation of the two stereographic atlases]
On the overlap \(U_{1}\cap U_{2}\) the usual stereographic transition  
\(\psi_{21}(X)=\rho^{2}X/|X|^{2}\) has Jacobian  
\begin{align}
   \det J(X)=-
             \Bigl(\tfrac{\rho^{2}}{|X|^{2}}\Bigr)^{\!n}\;<0 ,
\end{align}
so the standard stereographic atlas \(\{(U_{1},\psi_{1}),\,(U_{2},\psi_{2})\}\)
is \emph{orientation–reversing} for every dimension \(n\).

For the complex/quaternionic/octonionic atlas one obtains  
\(\widehat J=D\,J\) with
\(D=\mathrm{diag}(1,-1,\dots,-1)\) and
\(\det D=(-1)^{\,n-1}\).
Hence  
\begin{align}
   \det\widehat J(X)=(-1)^{\,n-1}\det J(X)
=(-1)^{\,n}\!
    \Bigl(\tfrac{\rho^{2}}{|X|^{2}}\Bigr)^{\!n}.
\end{align}
In the dimensions where the atlas is defined
(\(n=2,4,8\)) this determinant is \emph{positive},
so the complex/quaternionic/octonionic stereographic atlas
\(\{(U_{1},\widehat\psi_{1}),\,(U_{2},\psi_{2})\}\)
is \emph{orientation–preserving}.
\end{remark}

\subsection{More Definitions and Theorems}
\label{subsec:More_definitions_and_theorems}

\begin{theorem}[Whitney Approximation Theorem, $C^{1}\!\to\!C^{\infty}$ case \cite{whitney1934, whitney1936}]
Let \(M\) and \(N\) be smooth (\(C^{\infty}\)) manifolds, with \(M\) second-countable and without boundary.
Suppose
\begin{align}
f\colon M \longrightarrow N
\end{align}
is a map of class \(C^{1}\).
For every compact set \(K\subset M\) and every \(\varepsilon>0\), there exists a \(C^{\infty}\) map
\begin{align}
g\colon M \longrightarrow N
\end{align}
such that
\begin{align}
\sup_{x\in K} d_{N}\!\bigl(f(x),g(x)\bigr)<\varepsilon,
\quad
\sup_{x\in K}\bigl\lVert Df(x)-Dg(x)\bigr\rVert<\varepsilon.
\end{align}
Here, \(d_{N}\) is any Riemannian distance on \(N\). The second inequality is made well-defined by viewing \(N\) as smoothly embedded in some Euclidean space \(\mathbb{R}^K\) (via the Whitney Embedding Theorem), so that the differentials \(Df(x), Dg(x)\colon T_xM \to \mathbb{R}^K\) map into the same vector space. 
Consequently, every \(C^{1}\) map \(f\) is \(C^{1}\)-homotopic to a smooth map; in particular, the homotopy class of \(f\) always contains a \(C^{\infty}\) representative.
\end{theorem}

\begin{remark}
    The theorem above implies that any homeomorphism between two exotic manifolds must be $C^0$ but not $C^1$, and hence some discontinuity or degeneracy (rank drop) must appear in the Jacobian.
\end{remark}

\begin{definition}
  Let $M$ be a smooth, closed, oriented manifold.
  \begin{enumerate}
    \item $\Diff(M)$ is the \emph{group of smooth diffeomorphisms} of $M$
          endowed with the Whitney $\Sm$–topology.
    \item $\Diffp(M)$ is the subgroup of orientation-preserving elements.
    \item $\Homeo(M)$ (respectively $\Homeop(M)$) is the corresponding
          group of (orientation-preserving) homeomorphisms with the
          compact–open topology.
  \end{enumerate}
\end{definition}

\begin{definition}[Isotopy and pseudo-isotopy]
  \label{def:isotopy}
  Let $f,g\in\Diff(M)$.
  \begin{enumerate}
    \item A \emph{(smooth) isotopy} from $f$ to $g$ is a smooth map
          $F\colon M\times[0,1]\to M$ with
          $F(\,\cdot,0)=f,\;F(\,\cdot,1)=g$ and each slice
          $F_{t}:=F(\,\cdot,t)\in\Diff(M)$.
    \item A \emph{pseudo-isotopy} is a diffeomorphism
          $H\in\Diff\!\bigl(M\times[0,1]\bigr)$ that restricts to
          $\operatorname{id}$ on
          $M\times\{0\}\ \cup\ \partial M\times[0,1]$.
  \end{enumerate}
\end{definition}

\begin{remark}
    These definitions set the stage for the study the deformation properties of certain maps. This is relevant when considering the twisted sphere construction, since it involves the use of an \textit{exotic diffeomorphism}, defined below.
\end{remark}

\begin{theorem}[Cerf {\cite{Cerf70}}]
  \label{thm:cerf}
  If $V$ is simply connected and $\dim V\ge 6$, then every
  pseudo-isotopy of $V$ is isotopic (rel~$V\times\{0\}$) to the
  identity.  In particular
  \begin{align}
      \pi_{0}\bigl(\Diffp(S^{n-1})\bigr)\;\cong\;
      \Theta_{n}\qquad(n\ge 6),
  \end{align}
  where $\Theta_{n}$ is the Kervaire–Milnor group of homotopy $n$-spheres.
\end{theorem}

For $n=7$ the calculation of Kervaire and Milnor gives:

\begin{theorem}[Kervaire–Milnor {\cite{KervaireMilnor63}}]
  \label{thm:Theta7}
  $\displaystyle
    \Theta_{7}\;\cong\;\mathbb Z/28$.
  Consequently
  \begin{align}
      \pi_{0}\bigl(\Diffp(S^{6})\bigr)\;\cong\;\mathbb Z/28,
      \qquad
      \pi_{0}\bigl(\Diff(S^{6})\bigr)\;\cong\;
      \mathbb Z/28\;\times\;\mathbb Z/2.
  \end{align}
  Hence $\Diff(S^{6})$ has $56$ path-components, $28$ of which preserve
  orientation.
\end{theorem}

\begin{definition}[Exotic diffeomorphism]
  \label{def:exotic-diffeo}
  An \emph{exotic diffeomorphism} of $S^{6}$ is an
  $f\in\Diffp(S^{6})$ whose isotopy class represents a non-trivial
  element of $\Theta_{7}$.
\end{definition}

\begin{corollary}
  \label{cor:no-smooth-path}
  Let $f\in\Diffp(S^{6})$ be exotic.  There is
  \emph{no} smooth isotopy inside $\Diff(S^{6})$ from $f$ to the
  identity.  Any continuous path $t\mapsto f_{t}$ with
  $f_{0}=f,\;f_{1}=\operatorname{id}$ necessarily exits the diffeomorphism
  group (the derivative becomes singular or the inverse ceases to be
  smooth) at some $t$.
\end{corollary}

\begin{remark}
Switching to the topological category collapses almost all of the
higher smooth complexity.
\end{remark}

\begin{theorem}[Stable homeomorphism theorem; Kirby {\cite{Kirby69}},
  Kirby–Siebenmann {\cite{KirbySiebenmann77}}, Quinn {\cite{Quinn82}}]
  \label{thm:stable-homeo}
  For every $n\ge 5$
  \begin{align}
       \pi_{0}\!\bigl(\Homeop(S^{n})\bigr)=0,
       \qquad
       \pi_{0}\!\bigl(\Homeo(S^{n})\bigr)\cong\mathbb Z/2.
  \end{align}
  In particular, \emph{every} orientation-preserving self-homeomorphism
  of $S^{6}$ is topologically isotopic to the identity.
\end{theorem}

\begin{remark}
  The proof runs through the annulus conjecture and shows that any
  orientation-preserving homeomorphism of $\mathbb R^{n}$ is a finite
  product of homeomorphisms each fixing some open ball.  An Alexander
  trick applied to such a ball converts the product into a path inside
  $\Homeop(S^{n})$, yielding path-connectedness.
\end{remark}

\begin{corollary}
  \label{cor:topological-path}
  Although an exotic $f\in\Diffp(S^{6})$ is not smoothly isotopic to
  the identity, it \emph{is} topologically isotopic to the identity.
  The obstruction detected by $\Theta_{7}$ is purely smooth; it
  disappears in $\Homeop(S^{6})$.
\end{corollary}

\begin{remark}
The following table provides a summary of the results mentioned so far.

\begin{center}
\begin{tabular}{|c|c|c|}
  \hline
  Group & $\pi_{0}$ &  Consequence for paths to $\operatorname{id}$ \\ \hline\hline
  $\Diffp(S^{6})$ & $\mathbb Z/28$ &
     no smooth path for exotic classes \\ \hline
  $\Diff(S^{6})$ & $\mathbb Z/28\times\mathbb Z/2$ &
     orientation reversal also obstructed (Jacobian sign) \\ \hline
  $\Homeop(S^{6})$ & $0$ &
     every orientation-preserving map is topologically isotopic
     to $\operatorname{id}$ \\ \hline
  $\Homeo(S^{6})$ & $\mathbb Z/2$ &
     degree $\pm1$ is the only remaining obstruction \\ \hline
\end{tabular}
\end{center}

\medskip
The table emphasises that \emph{all} of the extra $28$ components of
$\Diffp(S^{6})$ arise from smooth phenomena (exotic spheres).  Once the
intermediate maps are only required to be homeomorphisms, these
obstructions vanish and a path to the identity always exists.
\end{remark}

\section{Differential Geometry}
In this section, we review some aspects of differential geometry that are particularly relevant to our work: the vielbein formalism and invariant geometry on group manifolds.

\subsection{Vielbein Formalism}
\label{sec:VielbeinFormalism}
Let $\{x^{\mu}\}$ be local coordinates on a smooth $m$–manifold $M$ with associated coordinate basis $\{\partial_{\mu}\}$ for the tangent space and $\{dx^{\mu}\}$ for the cotangent space; a \emph{non-coordinate basis} (or \emph{vielbein field}) is a smooth set of vector fields $\{\hat e_{a}\}\;(a=1,\dots ,m)$ on the same chart such that $$\hat e_{a}= \hat e_{a}{}^{\mu}\,\partial_{\mu},\qquad \hat e_{a}{}^{\mu}(x)\in GL(m,\mathbb R),$$ where the matrices $\hat e_{a}{}^{\mu}$ are called the \textit{vielbein} components. Given a pseudo-Riemannian metric $g$ of signature $(t,s)$ ($t+s=m$), one can always choose the $\hat e_{a}$ so that $$g(\hat e_{a},\hat e_{b})=\eta_{ab},\qquad\eta_{ab}= \operatorname{diag}(\underbrace{-1,\dots ,-1}_{t},\underbrace{+1,\dots ,+1}_{s}),$$ thus the metric becomes constant in this basis.  Introducing the inverse matrix $$e^{a}{}_{\mu}\equiv (\hat e_{a}{}^{\mu})^{-1}$$ we have $$e^{a}{}_{\mu}\,\hat e_{b}{}^{\mu}=\delta^{a}{}_{b},\qquad \hat e_{a}{}^{\mu}\,e^{a}{}_{\nu}= \delta^{\mu}{}_{\nu},$$ and the metric takes the useful form $$g_{\mu\nu}=e^{a}{}_{\mu}\,e^{b}{}_{\nu}\,\eta_{ab},\qquad g=\eta_{ab}\,\bar\theta^{a}\otimes\bar\theta^{b},$$ where the dual co-frame is $$\bar\theta^{a}=e^{a}{}_{\mu}\,dx^{\mu},\qquad\langle\bar\theta^{a},\hat e_{b}\rangle =\delta^{a}{}_{b}. $$  Because the basis is generally anholonomic, $$[\hat e_{a},\hat e_{b}]=c^{c}{}_{ab}\,\hat e_{c},$$ with structure functions $$c^{c}{}_{ab}=2\,\hat e_{[a}{}^{\mu}\,\partial_{\mu}\hat e_{b]}{}^{\nu}\,e^{c}{}_{\nu}\,,$$ and the determinant condition $$\det(\hat e_{a}{}^{\mu})>0$$ is imposed to preserve orientation.  Covariant differentiation in this basis is encoded by the \emph{spin-connection components} $$\nabla_{a}\hat e_{b}\equiv\nabla_{\hat e_{a}}\hat e_{b}=\omega_{ab}{}^{c}\,\hat e_{c},$$ related to the coordinate Christoffel symbols by
\begin{align}
\omega_{ab}{}^{c}= \hat e_{a}{}^{\mu}\,e^{c}{}_{\nu}\bigl(\partial_{\mu}\hat e_{b}{}^{\nu}+ \hat e_{b}{}^{\beta}\,\Gamma_{\mu\beta}{}^{\nu}\bigr),
\end{align}
and metric compatibility implies the antisymmetry $$\omega_{abc}\equiv\omega_{ab}{}^{d}\,\eta_{dc}=-\omega_{acb}\,.$$  Defining the \emph{torsion} and \emph{curvature} 2-forms by
\begin{align}
T^{a}&=\tfrac12\,T^{a}{}_{bc}\,\bar\theta^{b}\wedge\bar\theta^{c},\qquad\; T^{a}{}_{bc}= \omega^{a}{}_{cb}-\omega^{a}{}_{bc}-c^{a}{}_{bc},\\
R^{a}{}_{b}&=\tfrac12\,R^{a}{}_{bcd}\,\bar\theta^{c}\wedge\bar\theta^{d},
\end{align}
and assembling $$\omega^{a}{}_{b}\equiv\omega^{a}{}_{b c}\,\bar\theta^{c},$$ Cartan’s structure equations become
\begin{align}
T^{a}=d\bar\theta^{a}+ \omega^{a}{}_{b}\wedge\bar\theta^{b},\qquad
R^{a}{}_{b}=d\omega^{a}{}_{b}+ \omega^{a}{}_{c}\wedge\omega^{c}{}_{b},
\end{align}
whose exterior derivatives yield the familiar Bianchi identities.  Setting $T^{a}=0$ singles out the Levi-Civita connection, which can alternatively be obtained directly from $$\omega_{ab}{}^{c}=\tfrac12\bigl(c_{ab}{}^{c}-c_{a}{}^{c}{}_{b}+c_{b}{}^{c}{}_{a}\bigr).$$  The curvature components extracted from $R^{a}{}_{b}$ reproduce the coordinate Riemann tensor through $$R^{\rho}{}_{\sigma\mu\nu}=e^{a}{}_{\mu}\,e^{b}{}_{\nu}\,\hat e_{c}{}^{\rho}\,\eta_{ad}\,R^{d}{}_{b c a}\,.$$  In practice the procedure is: choose a convenient orthonormal co-frame $$\{\bar\theta^{a}\},$$ solve $$d\bar\theta^{a}+\omega^{a}{}_{b}\wedge\bar\theta^{b}=0$$ (Levi-Civita) for the spin connection, insert into the second Cartan equation to obtain curvature, and then contract indices as needed to form objects such as the Ricci tensor or Einstein tensor.

\subsection{Invariant Geometry on Group Manifolds}
\label{subsec:Invariant_geometry_abstract}

Let \(G\) be an \(n\)-dimensional Lie group with local coordinates \(y^{m}\;(m=1,\dots,n)\) chosen such that the identity element \(e\) corresponds to \(y^m=0\). The heart of a Lie group's structure is its Lie algebra, \(\mathfrak{g}\), which can be identified with the tangent space at the identity, \(\mathfrak g=T_{e}G\). The algebra represents the set of all possible "infinitesimal transformations" away from the identity. We denote by \(\{T_{a}\}_{a=1}^{n}\) a fixed basis of \(\mathfrak g\), whose elements satisfy the commutation relations that define the algebraic structure:
\begin{align}
[T_{a},T_{b}] \;=\; C^{c}{}_{ab}\,T_{c}.
\end{align}
The constants \(C^{c}{}_{ab}\) are the structure constants of the Lie algebra. The group's smooth manifold structure is intimately linked to its algebraic structure through the group multiplication. For any element \(g\in G\), the \emph{left} (\(L_{g}\)) and \emph{right} (\(R_{g}\)) translations are the fundamental diffeomorphisms of the group onto itself, defined by \(L_{g}(h)=gh\) and \(R_{g}(h)=hg\). These maps allow us to relate the geometry at the identity to the geometry at any other point on the group manifold.

\subsubsection*{Invariant vector fields, one--forms and Maurer--Cartan forms}

The concept of invariance is central to the geometry of Lie groups. We can construct vector fields that ``look the same'' at every point by using the group's own translations to propagate a vector from the identity. This is formalized by the pushforward map \((L_g)_*\), which takes a tangent vector at the identity and transports it to the tangent space at the point \(g\).

A left-invariant vector field is generated by taking a specific vector \(T_a \in \mathfrak{g}\) and left-translating it to every point \(g \in G\). A right-invariant vector field is generated similarly using right translations.
\begin{align}
\bigl( X^{L}_{a} \bigr)_{g}\;=\;(L_{g})_{*}T_{a},
\qquad
\bigl( X^{R}_{a} \bigr)_{g}\;=\;(R_{g})_{*}T_{a}
\end{align}

\noindent
In local coordinates \(y^m\), these vector fields are expressed as differential operators,
\begin{align}
X^{L}_{a}=k_{a}{}^{m}(y)\,\partial_{m},
\qquad
X^{R}_{a}=\bar k_{a}{}^{m}(y)\,\partial_{m},
\end{align}
where the coefficient functions \(k_{a}{}^{m}(y)\) and \(\bar k_{a}{}^{m}(y)\) depend on the position \(y\) on the manifold. A crucial property is that these vector fields form a Lie algebra under the standard vector field commutator that is isomorphic (for left-invariant fields) or anti-isomorphic (for right-invariant fields) to the original Lie algebra \(\mathfrak{g}\). This is reflected in the commutation relations for their coefficient functions:
\begin{align}
k_{a}{}^{m}\partial_{m}k_{b}{}^{n}-k_{b}{}^{m}\partial_{m}k_{a}{}^{n}
      = C^{c}{}_{ab}k_{c}{}^{n}, \qquad \text{which implies} \quad [X^L_a, X^L_b] = C^c_{ab} X^L_c.
\end{align}
For the right-invariant vector fields, the composition of right translations \(R_g \circ R_h = R_{hg}\) leads to a reversal in the algebra's structure, resulting in a sign flip:
\begin{align}
\bar k_{a}{}^{m}\partial_{m}\bar k_{b}{}^{n}-\bar k_{b}{}^{m}\partial_{m}\bar k_{a}{}^{n} = -C^{c}{}_{ab}\bar k_{c}{}^{n}, \qquad \text{which implies} \quad [X^R_a, X^R_b] = -C^c_{ab} X^R_c.
\end{align}
Dual to the basis of invariant vector fields, we can define a basis of invariant one-forms. These are constructed such that they form an orthonormal basis with the vector fields at every point.

The left-invariant one-forms \(\theta^a\) and right-invariant one-forms \(\bar{\theta}^a\) are defined as the dual basis to the invariant vector fields.
\begin{align}
\theta^{a} \;=\; k^{a}{}_{m}(y)\,dy^{m},
\qquad
\bar\theta^{a} \;=\; \bar k^{a}{}_{m}(y)\,dy^{m}
\end{align}
where the matrix of coefficients \(k^{a}{}_{m}\) is the inverse of \(k_{a}{}^{m}\), i.e., \(k^{a}{}_{m}k_{b}{}^{m}=\delta^{a}{}_{b}\), and similarly for \(\bar k\).
By this construction, the one-forms are dual to the vector fields, meaning \(\theta^{a}(X^{L}_{b})=\delta^{a}{}_{b}\). The key property is their invariance under the corresponding group translation, expressed via the pullback map. For any \(g \in G\), we have \(L_{g}^{*}\theta^{a}=\theta^{a}\), and \(\bar\theta^{a}\) enjoys the analogous right--invariance, \(R_{g}^{*}\bar\theta^{a}=\bar\theta^{a}\). A more abstract but powerful way to introduce these forms is through the Maurer-Cartan form, which captures the infinitesimal change of a group element \(g\) relative to itself, expressed as an element of the Lie algebra.

The left- and right-invariant Maurer-Cartan forms are \(\mathfrak{g}\)-valued one-forms defined in a coordinate-free way as:
\begin{align}
\Theta_{L}\;=\;g^{-1}dg=\theta^{a}T_{a},
\qquad
\Theta_{R}\;=\;dg\,g^{-1}=\bar\theta^{a}T_{a}
\end{align}
Here, \(dg\) represents the infinitesimal displacement from \(g\). Multiplying by \(g^{-1}\) on the left maps this displacement from the tangent space at \(g\) back to the tangent space at the identity, \(\mathfrak{g}\). The components of this \(\mathfrak{g}\)-valued form in the basis \(\{T_a\}\) are precisely the left-invariant one-forms \(\theta^a\). The algebraic structure of the group is entirely encoded in the exterior derivatives of these forms, which satisfy the celebrated \emph{Maurer--Cartan structure equations}:
\begin{align}
d\theta^{a} + \tfrac12\,C^{a}{}_{bc}\,\theta^{b}\!\wedge\!\theta^{c}=0,
\qquad
d\bar\theta^{a} - \tfrac12\,C^{a}{}_{bc}\,\bar\theta^{b}\!\wedge\!\bar\theta^{c}=0.
\end{align}

\subsubsection*{Invariant metrics and Killing vectors}

To introduce a Riemannian metric on the group manifold, we begin by defining an inner product on the tangent space at the identity, \(\mathfrak{g}\). We choose a positive--definite, symmetric inner product \(\kappa_{ab}=\kappa(T_{a},T_{b})\) on \(\mathfrak g\). This inner product can then be extended to the entire manifold by declaring that the invariant frame fields should be orthonormal at every point.

A left-invariant metric \(g^L\) is constructed by propagating the inner product \(\kappa\) over \(G\) using left translations. This is achieved by defining the metric in terms of the left-invariant one-forms. A right-invariant metric is defined analogously.
\begin{align}
g^{L} \;=\; \kappa_{ab}\,\theta^{a}\!\otimes\!\theta^{b},
\qquad
g^{R} \;=\; \kappa_{ab}\,\bar\theta^{a}\!\otimes\!\bar\theta^{b}
\end{align}

\noindent
By construction, these metrics are invariant under the respective translations: \(L_{g}^{*}g^{L}=g^{L}\) and \(R_{g}^{*}g^{R}=g^{R}\). A particularly important case arises when the metric is invariant under both left and right translations simultaneously. This occurs if the initial inner product \(\kappa\) on \(\mathfrak{g}\) is invariant under the Adjoint representation of \(G\). If \(\kappa\) is \(\mathrm{Ad}\)-invariant, then \(g^{L}=g^{R}\equiv g\), and the metric is called \emph{bi--invariant}. For compact simple Lie groups, the Killing form provides a natural choice for such an inner product. The symmetries of a Riemannian manifold are generated by Killing vector fields.

A vector field \(X\) on a Riemannian manifold \((M,g)\) is a \emph{Killing} vector field if the flow it generates consists of isometries. Infinitesimally, this means the metric is unchanged along the flow of \(X\), which is expressed by the vanishing of the Lie derivative:
\(\mathcal L_{X}g=0\).

\noindent
The invariant vector fields are the natural candidates for Killing vectors on a Lie group. Since the flow generated by a left-invariant vector field \(X^L_a\) is a family of right translations (and vice-versa), and the metrics were constructed to be invariant under these translations, it follows directly that:
\begin{align}
\mathcal L_{X^{R}_{a}}g^{L}=0, \qquad \mathcal L_{X^{L}_{a}}g^{R}=0.
\end{align}
This means the right-invariant vector fields are the Killing vectors of the left-invariant metric, and the left-invariant vector fields are the Killing vectors of the right-invariant metric. If the metric \(g\) is bi--invariant (i.e., \(g^L=g^R=g\)), then a powerful result emerges: both sets of invariant vector fields are Killing vectors for the same metric \(g\). This endows the group manifold with a large symmetry group containing both left and right translations.
\begin{align}
\mathcal L_{X^{L}_{a}}g=0 \quad \text{and} \quad \mathcal L_{X^{R}_{a}}g=0.
\end{align}

\subsection{Application of Invariant Geometry to the Kaluza--Klein metric}
\label{sec:Bundle_adapted_vs_coordinate_adapted}

Consider the total space \(P\) of a principal \(G\)--bundle over a four--manifold \(M_{4}\) and a Yang--Mills connection
\(A=A_{\mu}^{a}(x)\,dx^{\mu}\,T_{a}\).

\paragraph{Bundle--adapted (vielbein) form.}
Using the left--invariant one--forms, the $(4+n)$--dimensional metric is written
\begin{align}
ds^{2}=g_{\mu\nu}(x)\,dx^{\mu}dx^{\nu}
      +\kappa_{ab}\bigl(\theta^{a}+A_{\mu}^{a}(x)\,dx^{\mu}\bigr)
                 \bigl(\theta^{b}+A_{\nu}^{b}(x)\,dx^{\nu}\bigr).
                 \label{eq:bundle_adapt}
\end{align}
Note that for the gauge symmetries (generated by $X^L_a$) to act as isometries of this total metric, the inner product ($\kappa_{ab}$) must be ($\mathrm{Ad}$)-invariant, guaranteeing that the fiber metric is bi-invariant.

\paragraph{Coordinate form.}
Replace \(\theta^{a}=k^{a}{}_{m}(y)\,dy^{m}\); expanding (1) in the coordinate basis
\(\{dx^{\mu},dy^{m}\}\) gives
\begin{align}
\begin{aligned}
g_{\mu\nu}            &= g_{\mu\nu}(x)+\kappa_{ab}A_{\mu}^{a}A_{\nu}^{b},\\
g_{\mu m}             &= \kappa_{ab}A_{\mu}^{a}k^{b}{}_{m}(y),\\
g_{mn}                &= \kappa_{ab}k^{a}{}_{m}(y)\,k^{b}{}_{n}(y).
\end{aligned} 
\label{eq:Coord_basis_metric}
\end{align}

\noindent
Equation \eqref{eq:Coord_basis_metric} is precisely the ``coordinate--basis'' Kaluza--Klein ansatz often written
\(g_{\mu m}=A_{\mu}^{a}k^{\phantom{a}}_{a m}\)
and \(g_{mn}=k_{m}^{\ a}k_{n}^{\ b}\kappa_{ab}\).
Conversely, contracting \eqref{eq:Coord_basis_metric} with the Killing vectors $k^{m}{}_{a}$ and using $k^{a}{}_{m}k_{b}{}^{m}=\delta^{a}{}_{b}$
reconstructs \eqref{eq:bundle_adapt}.

Thus the two apparently different metrics are in fact the same geometry,
expressed either in the \emph{bundle--adapted} orthonormal basis
\(\{\theta^{a}\}\) or in the \emph{coordinate} basis \(\{dy^{m}\}\); the
dictionary
\begin{align}
\theta^{a}\;=\;k^{a}{}_{m}(y)\,dy^{m},
\qquad
k^{a}{}_{m}=\theta^{a}(\partial_{m})
\end{align}
completes the identification of all associated quantities (frames, gauge
potentials, curvatures) in the two languages.

\subsection{Lens Spaces (Most General Definition)}
\label{subsec:Lens_app}
A very detailed definition of lens spaces can be found in \cite{Watkins1990}, where it is also shown how a quotienting a manifold by some group which acts freely yields a well-defined manifold. Note that there are other definitions which can found in the literature (see for instance the Appendix of \cite{URBANTKE2003125}, which is itself based on \cite{BottTu1982}, page 243), employing analogous notation with different meanings.

To be safe, let us provide the most general and standard definitions of lens spaces, which refers to $3$ dimensions. The classical Hopf fibration discussed in Section \ref{sec:lens_spaces_chap3},
\begin{align}
S^{1}\hookrightarrow S^{3}\xrightarrow{\;h\;}S^{2} \, ,
\end{align}
can be reconstructed by gluing two trivial \(S^{1}\)-bundles over the northern and southern hemispheres of \(S^{2}\) along their common equator with the degree-one map \(e^{i\theta}\!\mapsto\!e^{i\theta}\).
Replacing this clutching map by the degree-\(n\) map \(e^{i\theta}\!\mapsto\!e^{in\theta}\) produces a principal \(S^{1}\)-bundle whose Euler (Chern) number is \(n\).  
Its total space is the 3–manifold
\begin{align}
L(n,1)\;=\;S^{3}\big/\!\bigl\langle(z_{1},z_{2})\longmapsto(e^{2\pi i/n}z_{1},\,e^{2\pi i/n}z_{2})\bigr\rangle \, .
\end{align}
This is only the first column in the table of lens spaces (\cite{MilnorStasheff1974,HatcherAT}). 
A lens space in full generality is defined, for integers \(p\ge1\) and \(q\) coprime to \(p\), by
\begin{align}
L(p,q)\;=\;S^{3}\big/\!\bigl\langle(z_{1},z_{2})\longmapsto(e^{2\pi i/p}z_{1},\,e^{2\pi iq/p}z_{2})\bigr\rangle,
\qquad\gcd(p,q)=1\;.
\end{align}
The second integer \(q\) measures the relative twisting of the two complex coordinates; geometrically it governs how the meridional and longitudinal curves on the torus orbits of the action are identified \cite{Rolfsen1976}.  
Circle bundles over \(S^{2}\) involve the \textit{diagonal} \(S^{1}\)-action, so they realise only the sub-family \(L(p,1)\).  
Allowing the generator to act with weight \(q\neq1\) breaks the principal-bundle interpretation but still yields a free \(\mathbb Z_{p}\)-action on \(S^{3}\), thereby filling out all lens spaces.

All spaces \(L(p,q)\) with the same \(p\) share the homotopy type of a \(K(\mathbb Z_{p},1)\), yet they need not be homeomorphic.  
A classical computation using Reidemeister torsion shows that
\begin{align}
L(p,q)\;\cong\;L(p,q')
\;\Longleftrightarrow\;
q'\equiv\pm q^{\pm1}\pmod p,
\end{align}
so for \(p\ge3\) several distinct manifolds can arise - see \cite{Reidemeister1935}.  
Orientation reversal changes \(q\) to \(-q\), while reversing both orientation and the generator in \(\mathbb Z_{p}\) sends \(q\) to its inverse.  
Consequently the one-parameter family coming from the generalised Hopf construction represents just a single equivalence class whenever \(p=1,2\) but only a fraction of the classes for larger \(p\).

\section{Miscellaneous}

In this section, we collect various results concerning differential geometry and differential topology, that are not instrumental for the results derived in this thesis, but extend or clarify some discussions that appear in the main text.

\subsection{One Point Compactification}
\label{subsec:One_point_compactifications}

Finite action solutions in Yang--Mills theory satisfy
\begin{align}
    \int \operatorname{tr}\left[\partial_\mu A_\nu-\partial_\nu A_\mu+\left[A_\mu, A_\nu\right]\right]^2<\infty .
\end{align}
A sufficient condition for this is that $F_{\mu \nu}$ goes to zero at infinity faster than $\frac{1}{|x|^2}$. In other words, a suitable field strength for a finite action solution must vanish at infinity. In practice, since $F_{\mu \nu} = 0 \iff A_{\mu}= g \partial_{\mu} g^{-1}$ for some group element $g$, and therefore a pure gauge configuration at infinity guarantees $F_{\mu \nu}=0$. There is an interesting way of characterising continuous functions that vanish at infinity, which is via one-point compactification. Let us restrict to the case $\R^n$. A continuous function on $R^n$ vanishes at infinity exactly if it extends to a continuous function on $R^n \cup \{ \infty \}$, and it takes the value zero at the new point. As a one-dimensional example, consider 
\begin{align}
    f: \R \xrightarrow{} \R \nonumber \\
    u \mapsto \frac{1}{1 + u^2},
\end{align}
which vanishes at infinity. Now, what is the extension over $\R \cup \{ \infty \}$. First of all we might ask what $\R \cup \{ \infty \}$ is. It's just the circle, c.f. stereographic projection:
\begin{align}
    \varphi_N(x, z)=\frac{x}{1-z}, \quad \varphi_N^{-1}(u)=\left(\frac{2 u}{u^2+1}, \frac{u^2-1}{u^2+1}\right) \, .
\end{align}
Then, the function $f$ has to be extended to a function $F: S^1 \xrightarrow{} \R$ on the circle. To achieve this, first, let $f$ be the coordinate expression of $F$ on the chart that excludes $\{\infty \}$:
\begin{align}
    f = F \circ \varphi_N^{-1}.
\end{align}
The neighbourhood of $\{\infty \}$ corresponds to the points $u\xrightarrow{} \pm \infty$, which give $f(u)=F(\varphi_N^{-1}(u)) \xrightarrow{}0$. Hence, $F=0$ makes $F$ a continuous function on $S^1$, fulfilling the condition mentioned above.

With this in mind, we observe that the condition of finiteness of the action for $SU(2)$ Yang--Mills theory on $\mathbb{R}^4$ can be translated into a topological one: instead of considering solutions on $\mathbb{R}^4$, they are naturally formulated on $S^4 = \R^4 \cup \{ \infty \}$. For a discussion of one-point compactification in the context of instantons, the reader is referred to \cite{RevModPhys.52.175} and \cite{Luscher:1974ez}.

\subsection{Quotient Topology}
\label{subsec:Quotient_topology}

As we pointed out in \ref{sec:lens_spaces_chap3}, the topology of the total space is pulled back via the local trivialisations. However, another natural choice of topology would be the quotient topology, since $E$ is a quotient manifold. The natural question is how are they related. \\
Let the map $P$ be defined as
\begin{align}
    P: X &\rightarrow E \nonumber \\
    p \in X &\mapsto [p].
\end{align}
Then, the quotient topology is defined by choosing the open sets of $E$ to be those $O_E \in E$ s.t. $P^{-1} ( O_E )$ is an open set of $X$.\\
Suppose that $O_E$ is an open set with the pulled-back topology, i.e. $ \phi_i^{-1} (O_E \, \cap \, \pi^{-1}(U_i)) \subset U_i \times F $ is open $\forall \, U_i$. Notice that, almost by definition, $P^{-1}(O_E) = \cup_i \, \phi_i^{-1} (O_E \, \cap \, \pi^{-1}(U_i))$. Hence, if $O_E$ is open in the pulled-back topology, then $P^{-1}(O_E)$ is open in $X$ (it is the disjoint union of open sets) and so $O_E \in E$ is open in the quotient topology. \\
Vice-versa, assume that $O_E$ is s.t. $P^{-1}(O_E)$ is an open set of $X$. Then, again due to the disjoint union topology, it follows that $X_i \cap P^{-1}(O_E) \in U_i \times F$ is open $\forall X_i$. Now, notice that $X_i = \phi^{-1} \pi^{-1} ( U_i ) $. Since $f^{-1}(A) \cap f^{-1}(B) = f^{-1} (A \cap B)$ (the preimage of the intersection of two sets is the intersection of the preimage of each set), then $X_i \cap P^{-1}(O_E) = \phi_i^{-1} ( \pi^{-1}(U_i) \cap O_E ) $.

\chapter{Calculations on Instantons, Invariant Geometry and Kaluza--Klein \,\,\,\,\,\,\,\,\,\,\,\,\,\,\,\,\,\,\,\,}
This chapter contains some auxiliary computations and identities that refer to: the BPST instanton in its different gauge forms, in the standard 't Hooft symbols description; the quaternionic formalism applied to the description of principal $SO(4)$--bundles over $S^4$; the global and coordinate expressions for invariant objects on $S^3$; some miscellaneous results on the standard and quaternionic Hopf fibration.

\section{Instantons in Components}
\label{sec:Instantons_in_Components}
In this section, we collect various calculations and results concerning the BPST instanton discussed in Chapter \ref{chap:2}. Consistently with the description presented therein, we use the conventions that are most common in the theoretical physics literature, i.e.~those that naturally give rise to the 't Hooft symbols when (anti-)self-dual objects are written in components.

\subsection{Gauge Choice}
In the conventions outlined in Chapter \ref{chap:2}, the instanton solution reads:
\begin{align}
    A_{\mu}(x)=
\frac{2\,\eta^{a}_{\mu\nu}(x-x_{0})^{\nu}}
     {(x-x_{0})^{2}+\rho^{2}}\;T_{a} \, ,
\end{align}
while the anti-instanton reads:
\begin{align}
    \tilde{A}_{\mu}(x)=
\frac{2\,\bar{\eta}^{a}_{\mu\nu}(x-x_{0})^{\nu}}
     {(x-x_{0})^{2}+\rho^{2}}\;T_{a} \, .
\end{align}
Intimately related with them, are the following group elements:
\begin{align}
    &U(x)  = i x_\mu \sigma_\mu / \sqrt{x^2}, \quad U^{-1}(x)=-i x_\mu \bar{\sigma}_\mu / \sqrt{x^2} \, \nonumber \\
    &\implies U^{-1} \partial_{\mu} U = - \bar{\sigma}_{\mu \nu} \frac{x_{\nu}}{x^2} = 2 \frac{\eta_{a \mu \nu} x_{\nu}}{x^2} \, T^a \, ,
\end{align}
and
\begin{align}
    &V(x) = -U(x)^{-1} =  i x_\mu \bar{\sigma}_\mu / \sqrt{x^2}, \quad V^{-1}(x)=-i x_\mu \sigma_\mu / \sqrt{x^2} \, \nonumber \\
    &\implies V^{-1} \partial_{\mu} V  = - \sigma_{\mu \nu} \frac{x_{\nu}}{x^2} = 2 \frac{\bar{\eta}_{a \mu \nu} x_{\nu}}{x^2} \, T^a \, .
\end{align}
It follows that 
\begin{align}
    A_\mu \stackrel{|x|^2 \rightarrow \infty}{=} U^{-1} \partial_\mu U \, , \quad  \tilde{A}_\mu \stackrel{|x|^2 \rightarrow \infty}{=} V^{-1} \partial_\mu V \, ,
\end{align}
and also that it is possible to re-write the gauge fields as:
\begin{align}
    &A_{\mu} = U(x-x_0)^{-1} \partial_{\mu} U(x-x_0) \frac{(x-x_0)^2}{(x-x_0)^2 + \rho^2 } \, , \nonumber \\
    &\tilde{A}_{\mu} = V(x-x_0)^{-1} \partial_{\mu} V(x-x_0) \frac{(x-x_0)^2}{(x-x_0)^2 + \rho^2 } \, .
\end{align}
This form makes it evident that a gauge transformation leads to the ``singular gauge'' expressions\footnote{Note that $\partial_{\mu} U  U^{-1} = -U \partial_{\mu} U^{-1} = - V^{-1} \partial_{\mu} V$.}
\begin{align}
    &A'_{\mu} = U(x-x_0) A_{\mu} U(x-x_0)^{-1} +  U(x-x_0) \partial_{\mu} U(x-x_0)^{-1} = \frac{2 \rho^2 \bar{\eta}_{a\mu \nu} (x - x_0)_{\nu}}{ (x - x_0)^2 [(x - x_0)^2 + \rho^2]} \, , \nonumber \\
    &\tilde{A}'_{\mu} = V(x-x_0) \tilde{A}_{\mu} V(x-x_0)^{-1} +  V(x-x_0) \partial_{\mu} V(x-x_0)^{-1} = \frac{2 \rho^2 \eta_{a\mu \nu} (x - x_0)_{\nu}}{ (x - x_0)^2 [(x - x_0)^2 + \rho^2]} \, .
\end{align}

\subsection{Calculation of the $k=1$ Winding}
Following \cite{vandoren2008lectures}, for a vanishing field strength at infinity (which is necessary for a finite action, and also for extending the objects from $\mathbb{R}^4$ to its one point compactification $S^4$), we require $A_\mu \xrightarrow{x \rightarrow \infty} U^{-1} \partial_\mu U$, which is used in the following calculation of the winding number. \\
The winding number is given by
\begin{align}
    k=-\frac{1}{16 \pi^2} \int \mathrm{d}^4 x \operatorname{tr}^* F_{\mu \nu} F_{\mu \nu}, 
\end{align}
whose integrand can be simplified by noticing that 
\begin{align}
    \operatorname{tr}{ }^* F_{\mu \nu} F_{\mu \nu}=2 \partial_\mu \operatorname{tr} \epsilon_{\mu \nu \rho \sigma}\left\{A_\nu \partial_\rho A_\sigma+\frac{2}{3} A_\nu A_\rho A_\sigma\right\} .
\end{align}
Because of $F_{\mu \nu} \xrightarrow{x \rightarrow \infty} 0$, we can replace $\partial_{\rho} A_{\sigma}$ with $-A_{\rho} A_{\sigma}$, and the integral becomes
\begin{align}
    k=\frac{1}{24 \pi^2} \oint_{S^3 \text { (space) }} \mathrm{d} \Omega_\mu \epsilon_{\mu \nu \rho \sigma} \operatorname{tr}\left\{\left(U^{-1} \partial_\nu U\right)\left(U^{-1} \partial_\rho U\right)\left(U^{-1} \partial_\sigma U\right)\right\} .
\end{align}
The $SU(2)$ element $U$ is defined everywhere on the 3-sphere at infinity, which is labelled as $S^3$ (space), providing a map $S^3$ (space)$\xrightarrow[]{} S^3$ (group)$=SU(2)$. The integral above counts the number of times that $S^3$ (space) is wrapped around $S^3$ (group). As a check, if we set $U=V$, where $V$ is the identity map given by
\begin{align}
    V(x)=-i x_\mu \bar{\sigma}_\mu / \sqrt{x^2},
\end{align}
then we find $k=1$, as expected.

\subsection{Some Relations in Components}
\label{subsec:Some_Relations_in_Components}
A relation that follows from the choice of basis $\sigma_{\mu}$, $\bar{\sigma}_{\mu}$ is:
\begin{align}
    \sigma_\rho \bar{\sigma}_\mu+\sigma_\mu \bar{\sigma}_\rho=2 \delta_{\rho \mu} \, ,
\end{align}
which is useful when computing the Maurer--Cartan forms $U^{-1} \partial_{\mu} U$ and $U \partial_{\mu} U^{-1}$. 

The relation
\begin{align}
    \epsilon_{a b c} \eta_{b \mu \nu} \eta_{c \rho \sigma}=\delta_{\mu \rho} \eta_{a \nu \sigma}+\delta_{\nu \sigma} \eta_{a \mu \rho}-\delta_{\mu \sigma} \eta_{a \nu \rho}-\delta_{\nu \rho} \eta_{a \mu \sigma} \, ,
\end{align}
and an identical one for $\bar{\eta}_{a \mu \nu}$, yield the key simplification when computing field strength for (anti-)instantons from their gauge fields.

Another important relation, which takes the same form for $\eta$ and $\bar{\eta}$ is:
\begin{align}
    \eta_{a \mu \nu} \eta_{a \rho \sigma}=\delta_{\mu \rho} \delta_{\nu \sigma}-\delta_{\mu \sigma} \delta_{\nu \rho}+\epsilon_{\mu \nu \rho \sigma} \, .
\end{align}
For a number of other relations concerning $\sigma_{\mu \nu}$, $\bar{\sigma}_{\mu \nu}$, $\eta_{a \mu \nu}$, $\bar{\eta}_{a \mu \nu}$, can be found in the Appendix of \cite{vandoren2008lectures}.

\section{Quaternions}
\label{sec:Quaternionc_chapB}
In this section, we collect some quaternionic identities that were used throughout this work.

With the conventions spelled out in Section \ref{sec:Notation_and_conventions_ch1}vand in Section \ref{subsec:'tHooft_notation_and_quaternions_chap3}, one finds that the key relation in the dictionary between quaternionic notation and component notation is:
\begin{align}
    \dd x \wedge \dd\bar{x} = - \, ^o\eta^i_{\mu \nu}  e_i  \dd x^\mu \wedge \dd x^\nu \, ,
\end{align}
where $^o\eta_{i \mu \nu}$ are the \textit{reversed} 't Hooft symbols given in \eqref{eq:'tHooft_zero}. Another useful formula comes from considering that a quaternion $x$ left-multiplying the conjugate of another quaternion $\bar{y}$ yields:
\begin{align}
     (x\bar y)_i =- \, ^o\eta_{i \mu \nu} x^{\mu} y^{\nu} \, ,
\end{align}
where $(\cdot)_i$ stands for the $i^{\mathrm{th}}$ component, and $i=1,2,3$.
A few other identities are easy to see from the fact that the Pauli matrices are traceless. A straightforward one is: $\mathrm{Re}(x) = \frac{1}{2} \mathrm{Tr}(x)$. Moreover, if $x$ is imaginary, then $x= - \mathrm{Re}(x e_i) e_i$.

Moreover, when calculating the field strength from the gauge field (from both $k=1$ and $k=2$ instantons), the following relations become useful:
\begin{align}
    \begin{aligned}
& -2 \operatorname{Re} \dd x \wedge \operatorname{Im} \dd x-\operatorname{Im} \dd x \wedge \operatorname{Im} \dd x=\dd x \wedge \dd\bar{x } \, , \\
& -4 \operatorname{Re} \dd x \wedge \operatorname{Im}\dd x +\dd \bar{x} \wedge \dd x=\dd x \wedge \dd \bar{x} \, .
\end{aligned}
\label{eq:Quat_identity_1_App}
\end{align}
They of course hold for any $\HH$-valued 1-form. Similarly, any identity that holds for matrices in general holds for quaternions, such as $\dd x^{-1} = x^{-1} \dd x x^{-1} $.

A word of caution: elements in $\HH$ (or $\HH'$) are used to encode vectors under the $SO(4)$ of the tangent space of the base $S^4$, but also \eg\ elements in some $\su(2)$ Lie algebra. The index-free notation is efficient, but when one needs the transformation properties, for example when taking a covariant derivative, one needs to keep track of which $\su(2)\oplus\su(2)$ acts by left and right multiplication on the element in question (or $\su(2)$ by commutation, for an element in $\HH'$). Take for example the selfdual part of the Riemann tensor on $S^4$ of eq. \eqref{leftrightS4Riemann}, $R_L={1\over4}E\wedge\bar E$. It is a 2-form taking values in the left $\su(2)_L$ of the $S^4$ tangent space, and fulfills
$D^{(\Omega_L)}R_L=0$ (and in fact even $D_\mu^{(\Omega_L)}R_L=0$). The maximally symmetric $\su(2)$ 1-instanton field strength is $F={1\over4}E\wedge\bar E$, formally the same expression. Now, however, it is a 2-form valued in the gauge Lie algebra $\su(2)_g$,
and $D^{(A)}_\mu F=0$.
Consider a more general $\su(2)_g$-valued selfdual 2-form, like $G=\bar u E\wedge\bar Eu$. This is a typical expression for terms in the $k=2$ field strength. Here, $u$ (which is $1$ for the $k=1$ $F$ above) must be thought of as a bifundamental under $\su(2)_L\oplus\su(2)_g$. A covariant derivative of $G$ becomes
\begin{align}
D_mG=D_\mu\bar uE\wedge\bar Eu+\bar uE\wedge\bar ED_\mu u\;,
\end{align}
where $D_\mu u=\*_\mu u+\Omega_{L \mu}u-uA_\mu$, of course with quaternionic multiplication. For the symmetric 1-instanton, where $u=1$ and ``$\Omega_L=A$'', this vanishes.

\section{Invariant Vectors, One-forms and Metrics on $S^3$ in Different Coordinates}
\label{sec:Invariant_objects_appendix}

In this section, we review the left/right-invariant vector fields and the Maurer-Cartan forms on $S^3 \simeq SU(2)$, both with their local and global descriptions. This is done to make a connection between different conventions and notations used in \cite{DUFF19861, Awada:1982pk,Hatsuda:2009vj,Gherardini:2023uyx} and those used throughout this dissertation.

\subsection{Global Description}
Let us first focus on $S^3$ as embedded in $\mathbb{R}^4$.

\subsubsection{``Standard'' Conventions}

The geometry of the 3-sphere can be elegantly described by identifying it with the Lie group SU(2). We consider a point on S$^3$ embedded in $\mathbb{R}^4$ with coordinates $x^\mu = (x^1, x^2, x^3, x^4)$ satisfying the constraint $x^\mu x^\mu = 1$. This point is mapped to an SU(2) group element $g$ via the Pauli matrices $\sigma_a$ as
\begin{align*}
g = x^4 I + i x^a \sigma_a,
\end{align*}
where $a \in \{1,2,3\}$ and $I$ is the $2 \times 2$ identity matrix. The inverse is $g^{-1} = g^\dagger = x^4 I - i x^a \sigma_a$. The fundamental geometric objects are the left- and right-invariant Maurer-Cartan 1-forms, which are $\mathfrak{su}(2)$-valued and defined as
\begin{align*}
\omega_L = g^{-1}dg \quad \text{and} \quad \omega_R = dg g^{-1}.
\end{align*}
We can expand these forms in a basis of the Lie algebra to find the real-valued 1-forms, $e_L^a$ and $e_R^a$, known as the left- and right-invariant vielbeins (or frame fields). A direct calculation reveals that these vielbeins can be written compactly using the self-dual ($\eta$) and anti-self-dual ($\bar{\eta}$) 't Hooft symbols.
With the standard normalization choice $T^a = \frac{\sigma^a}{2i}$, the left-invariant vielbein is given by $e_L^a = -\bar{\eta}^a_{\mu\nu} x^\nu dx^\mu$, while the right-invariant one is $e_R^a = \eta^a_{\mu\nu} x^\nu dx^\mu$. These forms satisfy the Maurer-Cartan structure equations, which encode the algebra of $\mathfrak{su}(2)$. For the left-invariant forms, the equation is
\begin{align*}
d e_L^a -\varepsilon_{abc} e_L^b \wedge e_L^c = 0,
\end{align*}
and for the right-invariant forms, it is
\begin{align*}
d e_R^a + \varepsilon_{abc} e_R^b \wedge e_R^c = 0,
\end{align*}
where the sign change reflects the anti-isomorphism between the algebra of right-invariant fields and $\mathfrak{su}(2)$. Dual to these 1-forms are the left- and right-invariant vector fields, $L_a$ and $R_a$. Their components (scaled by a factor of $2$) are given by the same 't Hooft symbol expressions:
\begin{align*}
L_a =  \frac{1}{2} \eta^a_{\mu\nu} x^\nu \partial_\mu \quad \text{and} \quad R_a = - \frac{1}{2} \bar{\eta}^a_{\mu\nu} x^\nu \partial_\mu.
\end{align*}
These vector fields generate the symmetries of the group manifold and their Lie brackets reproduce the underlying algebra. The left-invariant vector fields close to form a copy of the $\mathfrak{su}(2) \cong \mathfrak{so}(3)$ algebra, while the right-invariant fields form another, commuting copy. Their commutation relations are
\begin{align*}
[L_a, L_b] &= \varepsilon_{abc} L_c, \\
[R_a, R_b] &= -\varepsilon_{abc} R_c, \\
[L_a, R_b] &= 0.
\end{align*}
This structure reflects the fact that the symmetry group of S$^3$ is SU(2)$_L \times$SU(2)$_R \cong$ SO(4). Finally, a bi-invariant metric on S$^3$ can be constructed from either set of vielbeins. The metric is given by the sum of the squares of the vielbeins:
\begin{align*}
ds^2 = \delta_{ab} e_L^a \otimes e_L^b = \delta_{ab} e_R^a \otimes e_R^b.
\end{align*}
Substituting the expression for the left-invariant vielbein, the metric tensor components are $g_{\mu\nu} = \delta_{\mu\nu} - x_\mu x_\nu$, using one of the identities in  \ref{sec:Instantons_in_Components}, and the constraint $x^\rho x^\rho = 1$. This is the projector onto the tangent space of the unit sphere. The metric is therefore
\begin{align*}
ds^2 = (\delta_{\mu\nu} - x_\mu x_\nu)dx^\mu dx^\nu = \delta_{\mu\nu}dx^\mu dx^\nu,
\end{align*}
where the last equality holds because the constraint $x^\mu x^\mu = 1$ implies $x^\mu dx^\mu = 0$. This is precisely the standard round metric on the unit 3-sphere.

\subsubsection{Conventions of \cite{Gherardini:2023uyx}}
Let us now review the conventions of \cite{Gherardini:2023uyx} employed in Section \ref{subsec:metric_on_Gromoll_Meyer_chap3}, which are slightly different from the standard ones spelled above.\footnote{Analogous discussions can be found in \cite{10.1063/1.2358391} and \cite{https://doi.org/10.15488/12546}.} \\
Given the usual definition of $S^3$ as $ \{(X,Y,Z,W) \,\,\, \mathrm{ s.t. } \,\,\,  X^2 + Y^2 + Z^2 + W^2 = 1 \}$, let us consider the map:
\begin{align} 
    d: S^3 &\xrightarrow{} SU(2) \nonumber \\
    p = (X, Y, Z, W) &\mapsto g =
     \left( \begin{array}{cc}
    z_1 & z_2^* \\
    z_2 & - z_1^* ,
    \end{array} \right),
    \label{eq:Map_from_S3_to_SU(2)}
\end{align}
where $z_1 \defeq Z + i W \, , \,\, z_2 \defeq X + i Y
$ and $(\,)^*$ denotes complex conjugation.\footnote{Note that \eqref{eq:Map_from_S3_to_SU(2)} yields a matrix with determinant $-1$ (mapping into $U(2)$ rather than strictly $\mathrm{SU(2)}$), but that its Maurer-Cartan forms are exactly as intended.} With the (unconventional) choice of normalisation for the Maurer-Cartan forms
\begin{align}
    g^{-1}dg = -i \sigma_a \omega^a \, ,
\label{eq:MC_form_our_conventions}
    \\
    dgg^{-1} = i \sigma_a \bar{\omega}^a \, ,
\end{align}
the left-invariant one-forms read:
\begin{align}
    \omega^a = \eta^{a}_{CB} X_B \mathrm{d}X_C \, .
\end{align}
The right-invariant ones are given by:
\begin{align}
    \bar{\omega}^a = \bar{\eta}^{a}_{CB} X_B \mathrm{d}X_C \, ,
\end{align}
Note that these forms obey \textit{identical} Maurer--Cartan equations, as opposed to the ones defined in the previous section. The vector fields dual to $\omega^a$, read:
\begin{align}
    V_a = \eta^{a}_{CB} X_B \frac{\partial}{\partial X_C }.
\end{align}
They generate right-translations and obey $[V_a,V_b] = 2 \epsilon_{abc}V_c$. Similarly, the vector fields dual to $\bar{\omega}^a$, read:
\begin{align}
    \bar{V}_a = \bar{\eta}^{a}_{CB} X_B \frac{\partial}{\partial X_C },
\end{align}
generate left-translations and obey $[\bar{V}_a,\bar{V}_b] = 2 \epsilon_{abc}\bar{V}_c$. Again, the same algebra is obeyed, and it is scaled by a factor of two if compared with the one of the previous section.

\subsection{Stereographic Coordinates}

Let us focus on the same objects in stereographic coordinates
$y^{i}$ ($i=1,2,3$), obtained by projecting from the south pole
$x^{4}=-1$ onto the hyperplane $x^{4}=0$:
\begin{align}
x^{a}= \frac{2y^{a}}{1+r^{2}}, 
\qquad
x^{4}= \frac{1-r^{2}}{1+r^{2}},
\qquad
r^{2}:=y^{i}y^{i}.
\end{align}
\bigskip
\textbf{Left- and right-invariant one–forms.}\;
With $T_{a}=-\tfrac{i}{2}\sigma_{a}$ the Maurer–Cartan forms
$\omega_{L}=g^{-1}dg$ and $\omega_{R}=dg\,g^{-1}$ expand as
\begin{align}
\omega_{L}=2 e^{a}_{L}\,T_{a},
\qquad
\omega_{R}=-2 e^{a}_{R}\,T_{a},
\qquad
e^{a}_{L}=\eta^{a}_{\mu\nu}x^{\nu}dx^{\mu},
\quad
e^{a}_{R}=\bar\eta^{a}_{\mu\nu}x^{\nu}dx^{\mu}.
\end{align}
They obey the Maurer–Cartan equations
\begin{align}
de^{a}_{L}+\varepsilon_{abc}\,e^{b}_{L}\!\wedge e^{c}_{L}=0,
\qquad
de^{a}_{R}-\varepsilon_{abc}\,e^{b}_{R}\!\wedge e^{c}_{R}=0.
\end{align}

\bigskip
\textbf{Left- and right-invariant vector fields.}\;
Using the same symbols but replacing $dx^{\mu}\!\mapsto\!\partial_{\mu}$
gives
\begin{align}
L_{a}=\eta^{a}_{\mu\nu}x^{\nu}\partial_{\mu},
\qquad
R_{a}=\bar\eta^{a}_{\mu\nu}x^{\nu}\partial_{\mu}.
\end{align}

which satisfy the Lie brackets
\begin{align}
[L_{a},L_{b}]=\varepsilon_{abc}\,L_{c},
\qquad
[R_{a},R_{b}]=-\varepsilon_{abc}\,R_{c},
\qquad
[L_{a},R_{b}]=0.
\end{align}

\bigskip
\textbf{Bi-invariant metric.}\;
Declaring the coframe $(e^{a}_{L})$ (or $(e^{a}_{R})$) orthonormal,
\begin{align}
ds^{2}=\delta_{ab}\,e^{a}_{L}\!\otimes e^{b}_{L}
      =\delta_{ab}\,e^{a}_{R}\!\otimes e^{b}_{R},
\end{align}
yields
\begin{align}
ds^{2}= \frac{4}{(1+r^{2})^{2}}\;\delta_{ij}\,dy^{i}dy^{j},
\end{align}
the standard round metric on the unit 3–sphere in stereographic
coordinates.  Both frames $\{L_{a}\}$ and $\{R_{a}\}$ are orthonormal
Killing vectors of this metric, fully realising the
$\mathfrak{su}(2)_{L}\!\oplus\!\mathfrak{su}(2)_{R}\cong\mathfrak{so}(4)$
isometry algebra.

\subsection{Euler-Angle Coordinates} 

Let us now parametrise the group element by Euler angles
\begin{align}
(\alpha,\beta,\gamma)\in[0,2\pi)\times[0,\pi]\times[0,2\pi)
\end{align}
via the factorisation
\begin{align}
g(\alpha,\beta,\gamma)
   =e^{\,-\frac{i\alpha}{2}\sigma_{3}}\;
    e^{\,-\frac{i\beta}{2}\sigma_{1}}\;
    e^{\,-\frac{i\gamma}{2}\sigma_{3}}
   \;\in SU(2),
\qquad
\sigma_{a}=\text{Pauli}.
\end{align}
The inverse is $g^{-1}=g^{\dagger}$, so the left- and right-invariant
Maurer–Cartan forms are as usual
\begin{align}
\omega_{L}=g^{-1}dg=\theta^{a}T_{a},
\qquad
\omega_{R}=dg\,g^{-1}=\bar\theta^{a}T_{a},
\qquad
T_{a}=-\tfrac{i}{2}\sigma_{a}.
\end{align}

\bigskip
\textbf{Invariant one-forms (vielbeins).}
A straightforward calculation gives

\begin{align}
\begin{aligned}
\theta^{1}&=\cos\gamma\,d\beta
            +\sin\gamma\,\sin\beta\,d\alpha,\\
\theta^{2}&=-\sin\gamma\,d\beta
            +\cos\gamma\,\sin\beta\,d\alpha,\\
\theta^{3}&=d\gamma+\cos\beta\,d\alpha;
\end{aligned}
\qquad
\begin{aligned}
\bar\theta^{1}&=\cos\alpha\,d\beta   +\sin\alpha\,\sin\beta\,d\gamma,\\
\bar\theta^{2}&=\sin\alpha\,d\beta -\cos\alpha\,\sin\beta\,d\gamma,\\
\bar\theta^{3}&=d\alpha+\cos\beta\,d\gamma.
\end{aligned}
\end{align}

They obey the Maurer–Cartan structure equations

\begin{align}
d\theta^{a}+\tfrac12\varepsilon_{abc}\,\theta^{b}\wedge\theta^{c}=0,
\qquad
d\bar\theta^{a}-\tfrac12\varepsilon_{abc}\,\bar\theta^{b}\wedge\bar\theta^{c}=0,
\label{eq:Maurer_cartan_equation_Euler}
\end{align}
consistent with the conventions $C^{a}{}_{bc}=\varepsilon_{bc}{}^{a}$.

\bigskip
\textbf{Invariant vector fields.}
Demanding $\theta^{a}(L_{b})=\delta^{a}{}_{b}$ and
$\bar\theta^{a}(R_{b})=\delta^{a}{}_{b}$ yields the dual frames

\begin{align}
\begin{aligned}
L_{1}&=\cos\gamma\,\partial_{\beta}
       +\frac{\sin\gamma}{\sin\beta}\,\partial_{\alpha}
       -\sin\gamma\,\cot\beta\,\partial_{\gamma},\\
L_{2}&=-\sin\gamma\,\partial_{\beta}
       + \frac{\cos\gamma}{\sin\beta}\,\partial_{\alpha}
       - \cos\gamma\,\cot\beta\,\partial_{\gamma},\\
L_{3}&=\partial_{\gamma};
\end{aligned}
\qquad
\begin{aligned}
R_{1}&=\cos\alpha\,\partial_{\beta}
       +\frac{\sin\alpha}{\sin\beta}\,\partial_{\gamma}
       -\sin\alpha\,\cot\beta\,\partial_{\alpha},\\
R_{2}&=\sin\alpha\,\partial_{\beta}
       -\frac{\cos\alpha}{\sin\beta}\,\partial_{\gamma}
       +\cos\alpha\,\cot\beta\,\partial_{\alpha},\\
R_{3}&=\partial_{\alpha}.
\end{aligned}
\end{align}

\bigskip
\textbf{Lie-algebra relations.}
Direct commutation confirms

\begin{align}
[L_{a},L_{b}]=\varepsilon_{abc}\,L_{c},
\qquad
[R_{a},R_{b}]=-\varepsilon_{abc}\,R_{c},
\qquad
[L_{a},R_{b}]=0,
\end{align}
realising $\mathfrak{su}(2)_{L}\oplus\mathfrak{su}(2)_{R}\cong\mathfrak{so}(4)$.

\bigskip
\textbf{Bi-invariant metric.}
Declaring the coframe $(\theta^{a})$ orthonormal gives the round metric

\begin{align}
ds^{2}=\delta_{ab}\,\theta^{a}\!\otimes\!\theta^{b}
      =\delta_{ab}\,\bar\theta^{a}\!\otimes\!\bar\theta^{b}.
\end{align}
In Euler angles this reads

\begin{align}
ds^{2}= d\beta^{2}
       +\sin^{2}\beta\,d\alpha^{2}
       +(d\gamma+\cos\beta\,d\alpha)^{2},
\end{align}
the familiar Hopf-fibration form with base $S^{2}$ and fibre
$S^{1}$.  Frames $\{L_{a}\}$ and $\{R_{a}\}$ are orthonormal Killing
vectors of this metric, in accordance with Section \ref{subsec:Invariant_geometry_abstract}.

\subsection{Hopf Coordinates}

We introduce Hopf coordinates
\begin{align}
(\theta,\varphi,\psi)\in[0,\pi]\times[0,2\pi)\times[0,2\pi),
\end{align}
and write
\begin{align}
z_{1}=\cos(\theta/2)\,e^{\,i(\varphi+\psi)/2}, 
\qquad
z_{2}=\sin(\theta/2)\,e^{\,i(-\varphi+\psi)/2},
\end{align}
so that $|z_{1}|^{2}+|z_{2}|^{2}=1$.  The point $(z_{1},z_{2})\in S^{3}$
is mapped to the group element
\begin{align}
g(\theta,\varphi,\psi)=
\begin{pmatrix}
z_{1} & z_{2}\\
-\bar z_{2} & \bar z_{1}
\end{pmatrix}
=e^{\,i\psi\sigma_{3}/2}\,
 e^{\,i\theta\sigma_{2}/2}\,
 e^{\,i \varphi\sigma_{3}/2}\in SU(2).
\end{align}

With the Lie-algebra basis $T_{a}=-\tfrac{i}{2}\sigma_{a}$, the
Maurer–Cartan forms $\omega_{L}=g^{-1}dg$ and
$\omega_{R}=dg\,g^{-1}$ expand as $\omega_{L}=e^{a}_{L}T_{a}$ and
$\omega_{R}=e^{a}_{R}T_{a}$. This gives the
left- and right-invariant coframes
\begin{align}
\begin{aligned}
e^{1}_{L}&=\sin\varphi d\theta - \sin\theta\cos\varphi d\psi,\\
e^{2}_{L}&=-\cos\varphi,d\theta - \sin\theta\sin\varphi d\psi,\\
e^{3}_{L}&=-d\varphi - \cos\theta d\psi;
\end{aligned}
\qquad
\begin{aligned}
e^{1}_{R}&=-\sin\psi d\theta + \sin\theta \cos\psi d\varphi,\\
e^{2}_{R}&=-\cos\psi d\theta - \sin\theta \sin\psi d\varphi, \\
e^{3}_{R}&=-d\psi - \cos\theta d\varphi .
\end{aligned}
\end{align}

These forms obey the Maurer–Cartan structure equations
\begin{align}
de^{a}_{L}+\tfrac12\varepsilon_{abc}\,e^{b}_{L}\wedge e^{c}_{L}=0,
\qquad
de^{a}_{R}-\tfrac12\varepsilon_{abc}\,e^{b}_{R}\wedge e^{c}_{R}=0,
\end{align}
with structure constants $C^{c}{}_{ab}=\varepsilon_{ab}{}^{c}$.

Requiring $e^{a}_{L}(L_{b})=\delta^{a}{}_{b}$ and
$e^{a}_{R}(R_{b})=\delta^{a}{}_{b}$ yields the dual vector fields
\begin{align}
\begin{aligned}
L_{1}&=\sin\varphi \partial_{\theta} + \cot\theta \cos\varphi \partial_{\varphi} - \frac{\cos\varphi}{\sin\theta} \partial_{\psi},\\
L_{2}&=-\cos\varphi,\partial_{\theta} + \cot\theta \sin\varphi \partial_{\varphi} - \frac{\sin\varphi}{\sin\theta} \partial_{\psi},\\
L_{3}&=-\partial_{\varphi};
\end{aligned}
\qquad
\begin{aligned}
R_{1}&=-\sin\psi \partial_{\theta} + \frac{\cos\psi}{\sin\theta} \partial_{\varphi} - \cot\theta \cos\psi \partial_{\psi},\\
R_{2}&=-\cos\psi \partial_{\theta} - \frac{\sin\psi}{\sin\theta} \partial_{\varphi} + \cot\theta \sin\psi \partial_{\psi},\\
R_{3}&=-\partial_{\psi}.
\end{aligned}
\end{align}

A direct computation confirms the Lie brackets
\begin{align}
[L_{a},L_{b}]=\varepsilon_{abc}\,L_{c},
\qquad
[R_{a},R_{b}]=-\varepsilon_{abc}\,R_{c},
\qquad
[L_{a},R_{b}]=0,
\end{align}
realising the isometry algebra
$\mathfrak{su}(2)_{L}\oplus\mathfrak{su}(2)_{R}\cong\mathfrak{so}(4)$.

Declaring the coframe $(e^{a}_{L})$ (or equivalently $(e^{a}_{R})$)
orthonormal produces the bi-invariant metric
\begin{align}
ds^{2}=\delta_{ab}\,e^{a}_{L}\otimes e^{b}_{L}
      =\delta_{ab}\,e^{a}_{R}\otimes e^{b}_{R}
      =d\theta^{2}
       +\sin^{2}\theta\,d\varphi^{2}
       +(d\psi+\cos\theta\,d\varphi)^{2}.
\end{align}
This is the standard round metric on $S^3$ (of radius $2$) written in Hopf
coordinates; the one-forms above are orthonormal, and the vector
fields are Killing, with appropriate commutation relations.

\subsection{Conventions of \cite{Gherardini:2023uyx}}
Let us return to \eqref{eq:Map_from_S3_to_SU(2)} once again, to present a local description of the same invariant geometry. A possible choice of Euler angles parametrising the group elements is: 
\begin{align}
    g=\left(\begin{array}{cc} \vspace{0.1cm}
e^{i \frac{\alpha+\gamma}{2}} \cos \left(\frac{\beta}{2}\right) & -e^{i \frac{\alpha-\gamma}{2}} \sin \left(\frac{\beta}{2}\right) \\
-e^{-i \frac{\alpha-\gamma}{2}} \sin \left(\frac{\beta}{2}\right) & -e^{-i \frac{\alpha+\gamma}{2}} \cos \left(\frac{\beta}{2}\right)
\end{array}\right),
\end{align}
i.e. 
\begin{align}
\begin{array}{ll}\vspace{0.3cm}
Z=\cos \left(\frac{\alpha+\gamma}{2}\right) \cos \left(\frac{\beta}{2}\right), & W=\sin \left(\frac{\alpha+\gamma}{2}\right) \cos \left(\frac{\beta}{2}\right), \\
X=-\cos \left(\frac{\alpha-\gamma}{2}\right) \sin \left(\frac{\beta}{2}\right), & Y=\sin \left(\frac{\alpha-\gamma}{2}\right) \sin \left(\frac{\beta}{2}\right) .
\label{eq:Angular_coords_defn}
\end{array}
\end{align}
Using the same normalisation, we obtain:
\begin{align}
    \begin{aligned}
\omega^1 & = \frac{1}{2} (\cos \gamma \sin \beta \mathrm{d} \alpha - \sin \gamma \mathrm{d} \beta), \\
\omega^2 & =\frac{1}{2} (\sin \beta \sin \gamma \mathrm{d} \alpha + \cos \gamma \mathrm{d} \beta), \\
\omega^3 & =\frac{1}{2} ( - \cos \beta \mathrm{d} \alpha - \mathrm{d} \gamma),
\end{aligned}
\end{align}
and 
\begin{align}
\begin{aligned}
& \bar{\omega}^1 = \frac{1}{2} (\sin \alpha \mathrm{d} \beta - \cos \alpha \sin \beta \mathrm{d} \gamma), \\
& \bar{\omega}^2=\frac{1}{2}(\cos \alpha \mathrm{d} \beta + \sin \alpha \sin \beta \mathrm{d} \gamma), \\
& \bar{\omega}^3 = \frac{1}{2} (\mathrm{d} \alpha + \cos \beta \mathrm{d} \gamma),
\end{aligned}
\end{align}
where the usual ranges are $ 0 \leq \beta \leq \pi  , 0 \leq \gamma \leq 2\pi, 0 \leq \alpha \leq 4 \pi$. It is interesting to note that $w^{1,2,3} = - \sigma_{x,y,z}$, where $\sigma_{x,y,z}$ are the three one-forms employed by Eguchi and Hanson in the construction of their gravitational instanton (see \cite{EGUCHI197982}). They obey:
\begin{align}
    d \omega^a = - \epsilon_{abc} \,\omega^b \wedge \omega^c \, , \label{eq:Left_MC_eqn}\\
    d \bar{\omega}^a = - \epsilon_{abc} \, \bar{\omega}^b \wedge \bar{\omega}^c. \label{Right_MC_eqn}
\end{align}
We can use the above forms to write the metric on the unit sphere as:
\begin{align}
    g_{S^3} = \omega^a \otimes \omega^a = \bar{\omega}^a \otimes \bar{\omega}^a = (\mathrm{d} \alpha \otimes \mathrm{d} \alpha+\mathrm{d} \beta \otimes \mathrm{d} \beta+\mathrm{d} \gamma \otimes \mathrm{d} \gamma+2 \cos \beta \mathrm{d} \alpha \otimes \mathrm{d} \gamma).
\end{align}

The dual vector fields to $\omega^a$ are:
\begin{align}
    \begin{aligned}
V_1 & =2 (\frac{\cos \gamma}{\sin \beta} \frac{\partial}{\partial \alpha} - \sin \gamma \frac{\partial}{\partial \beta} - \cot \beta \cos \gamma \frac{\partial}{\partial \gamma}), \\
V_2 & =2 (\frac{\sin \gamma}{\sin \beta} \frac{\partial}{\partial \alpha} + \cos \gamma \frac{\partial}{\partial \beta} - \cot \beta \sin \gamma \frac{\partial}{\partial \gamma}), \\
V_3 & = -2 \frac{\partial}{\partial \gamma},
\end{aligned}
\end{align}
while the vector fields dual to $\bar{\omega}^a$ are given by
\begin{align}
    \begin{aligned}
& \bar{V}_1=2 (\cos \alpha \cot \beta \frac{\partial}{\partial \alpha} + \sin \alpha \frac{\partial}{\partial \beta} - \frac{\cos \alpha}{\sin \beta} \frac{\partial}{\partial \gamma}), \\
& \bar{V}_2=2 (-\sin \alpha \cot \beta \frac{\partial}{\partial \alpha} + \cos \alpha \frac{\partial}{\partial \beta} + \frac{\sin \alpha}{\sin \beta} \frac{\partial}{\partial \gamma}), \\
& \bar{V}_3= 2 \frac{\partial}{\partial \alpha} .
\end{aligned}
\end{align}
Once we identify $V_a$ with $\mathchorus{K}_{\,\, \hat{i}'}$ and $\bar{V}_a$ with $\mathchorus{K}_{\,\, \alpha'}$, these expressions allow us to calculate the matrix $\mathchorus{K}_{\,\, \hat{i}'}{}^{\alpha'}$ in coordinates.

\subsection{Instanton in Stereographic Coordinates and Angular Coordinates on $S^4$}
\label{sec:Appendix_instanton}
The connection for the standard $k=1$ instanton  with unit size centered at the origin, in regular gauge, reads (see \cite{vandoren2008lectures}):
\begin{align}
   A_{\mu}^a = \Big(\frac{1}{x^2 + 1} \Big) 2  \eta^{a}_{\mu \nu} x^\nu.
    \label{eq:Standard_instanton_Vandoren}
\end{align}
This expression is implicitly assuming that the generators satisfy the conventional algebra $[T_a, T_b] = \epsilon_{abc} T_c$. Clearly, for any other choice of generators (equation \ref{eq:Structure_constants_su(2)s}, for example), the components are scaled accordingly. \\ 
In \cite{DUFF19861}, however, the $k=1$ instanton is presented in angular coordinates, i.e.~equation \ref{eq:Instanton_gauge}. We schematically show the connection between the two expressions. As we stressed several times, one should think of equation \ref{eq:Standard_instanton_Vandoren} not as a field living on $\R^4$, but as the coordinate expression of a connection on $S^4$. In other words, the $\{x^{\mu} \}$ are stereographic coordinates - see \ref{subsec:Working_definitions_chapA1} for explicit formulae.
The coordinates used by \cite{DUFF19861}, on the other hand, can be read off from the metric \ref{eq:Metric_on_S4_right_inv}. They are a mixture of standard spherical coordinates and Euler angles. If we let the four sphere be embedded in $\mathbb{R}^5$ as $a_1^2 + \cdots + a_5^2 = 1$, then, in one patch, these coordinates read read\:\footnote{Note that, clearly, the metric is insensitive to exchanging labels on the coordinates. In particular, any choice of $a_{1,...,4}$ gives an explicit isomorphism between $S^3$ and $SU(2)$ as presented above.}
\begin{align}
    a_5=\cos \mu \, \, \, , \quad a_4 = \sin \mu \cos u \cos \frac{\theta}{2} \,\,\, , \quad a_3 = \sin \mu \sin u \cos \frac{\theta}{2} \,\,\, , \nonumber \\
    a_2 = -\sin \mu \cos v \sin \frac{\theta}{2} \,\,\, , \quad a_1 = \sin \mu \sin v \sin \frac{\theta}{2},
    \label{eq_Mixed_coords_Duff}
\end{align}
where $u= (\phi + \psi)/2$, $v =(\phi - \psi)/2$ and  $ 0 \leq \theta \leq \pi  , 0 \leq \phi \leq 2\pi, 0 \leq \psi \leq 4 \pi$.
By the inverse stereographic projection, given a set of stereographic coordinates $x_i$, one obtains
\begin{align}
    x_i x_i + 1 = \frac{a_1^2 +a_2^2 +a_3^2 +a_4^2 }{(1+a_5)^2} +1 = \frac{2}{1+a_5}  = \frac{1}{\cos^2(\mu/2)}.
\end{align}
This shows that the scalar factor in equation \ref{eq:Instanton_gauge} matches the one in equation \ref{eq:Standard_instanton_Vandoren}. Regarding the equivalence between $\eta^{a}_{\mu \nu} x_S^\nu$ and the left-invariant form $\Sigma_i$, one can follow the same calculation presented in the previous section. We can use equations \ref{eq:Stereo_proj_S} and \ref{eq_Mixed_coords_Duff} to obtain a change of coordinates analogous to equation \ref{eq:Angular_coords_defn}, with an additional scaling depending on $\mu$. Then, the only difference is that an extra term, due to such scaling and proportional to the identity, will appear in the right-hand side of equation \ref{eq:MC_form_our_conventions}.

\subsection{Left-invariant vs Right-invariant}
\label{Sec:Left_inv_vs_right_inv}
The usual Fubini-Study metric reads
\begin{align}
    \mathrm{d} s^2=\left(1+\bar{q}_k q_k\right)^{-1} \mathrm{~d} \bar{q}_i \mathrm{~d} q_i-\left(1+\bar{q}_k q_k\right)^{-2} \bar{q}_i \mathrm{~d} q_i \mathrm{~d} \bar{q}_j q_j,
\end{align}
where $q_i$ are two quaternionic coordinates, and $\bar{(\cdot)}$ denotes conjugation.
With the parametrisation
\begin{align}
    q_1=\tan \chi \cos \left(\frac{1}{2} \mu\right) U, \quad q_2=\tan \chi \sin \left(\frac{1}{2} \mu\right) V,
\end{align}
where $U,V$ are unit quaternions so that $U \bar{U} = V \bar{V} = 1$, we can obtain a more familiar form of the metric. To do this, we first note that
\begin{align}
    2 U^{-1} d U=i \sigma_1+j \sigma_2+k \sigma_3, 2 \quad 2V^{-1} d V=i \Sigma_1+j \Sigma_2+k \Sigma_3,
    \label{eq:Left_inv_forms_for_Fubini-Study}
\end{align}
with $d \sigma_i = - \frac{1}{2} \epsilon_{ijk} \sigma_{j} \wedge \sigma_{k}$ and $d \Sigma_i = - \frac{1}{2} \epsilon_{ijk} \Sigma_{j} \wedge \Sigma_{k}$.
Then, we obtain
\begin{align}
    d s^2=d \chi^2+\frac{1}{4} \sin ^2 \chi\left[d \mu^2+\frac{1}{4} \sin ^2 \mu \omega_i^2+\frac{1}{4} \cos ^2 \chi\left(\nu_i+\cos \mu \omega_i\right)^2\right],
\end{align}
where $\nu_i = \sigma_i + \Sigma_i$ and $\omega_i = \sigma_i - \Sigma_i$.\\
Now, right-invariant one-forms can be defined analogously to \ref{eq:Left_inv_forms_for_Fubini-Study}:
\begin{align}
      2 d U U^{-1}=-i \tilde{\sigma}_1+-j \tilde{\sigma}_2+-k \tilde{\sigma}_3, \quad 2 d V  V^{-1} =-i \tilde{\Sigma}_1+-j \tilde{\Sigma}_2+-k \tilde{\Sigma}_3. 
\end{align}
Then, we have that again $d \tilde{\sigma_i} = - \frac{1}{2} \epsilon_{ijk} \tilde{\sigma}_{j} \wedge \tilde{\sigma}_{k}$ and $d \tilde{\Sigma}_i = - \frac{1}{2} \epsilon_{ijk} \tilde{\Sigma}_{j} \wedge \tilde{\Sigma}_{k}$, which holds with our conventions (see \ref{eq:Left_MC_eqn} and \ref{Right_MC_eqn}, up to normalisation).
Let us now consider the metric
\begin{align}
     \mathrm{d} s^2=\left(1+\bar{q}_k q_k\right)^{-1} \mathrm{~d}q_i \mathrm{~d}  \bar{q}_i-\left(1+\bar{q}_k q_k\right)^{-2} q_i \mathrm{~d}  \bar{q}_i \mathrm{~d} q_j  \bar{q}_j,
\end{align}
where the order of multiplication has been reversed. Then, all the steps that led to the result above still hold if we put tildes on $\sigma_i$ and $\Sigma_i$\footnote{In this section, we choose to distinguish right-invariant forms by using tildes because bars are used to denote conjugation. In the main text, right-invariant forms will be denoted by bars, since there is no risk of confusion there.}.
Hence, we obtain
\begin{align}
     d s^2=d \chi^2+\frac{1}{4} \sin ^2 \chi\left[d \mu^2+\frac{1}{4} \sin ^2 \mu \tilde{\omega}_i^2+\frac{1}{4} \cos ^2 \chi\left(\tilde{\nu}_i+\cos \mu \tilde{\omega}_i\right)^2\right],
\end{align}
where $\tilde{\nu}_i$ and $\tilde{\omega}_i$ are defined analogously to before.

\chapter{Details on Neural Networks \\  
and ``AInstein'' \,\,\,\,\,\,\,\,\,\,\,\,\,\,\,\,\,\,\,\,\,\,\,\,\,\,\,\,\,\,\,\,}
\label{app:extra_results}

This chapter expands on the results of Section \ref{sec:results_chap5}, displaying further breakdown of the test losses as the performance measures of the learning, as well as further example visualisations for 2d runs with other values of $\lambda$.

\section{Losses}\label{app:extra_losses}
The results in Table \ref{tab:global_test_losses} display the Global test losses, averaged over the 10 runs with standard deviations, for each of the investigations performed.
The Global test loss has 3 components, the Einstein loss in each of the 2 patches, and the overlap loss, calculated with respective multiplier weightings as described in Section \ref{subsec:bkg_ml_chap5}.

In Table \ref{tab:full_loss_results}, the average values of the constituent losses used in computing each Global test loss across the investigations are shown, again with standard deviations over the 10 runs.

\begin{table}[!t]
\centering
\begin{tabular}{|cc|ccc!{\vrule width 1.5pt}c|}
\hline
\multicolumn{2}{|c|}{\multirow{2}{*}{Dimension}}       & \multicolumn{3}{c!{\vrule width 1.5pt}}{Einstein Constant $\lambda$}  & \multirow{2}{*}{\begin{tabular}[c]{@{}c@{}}Supervised\\ $\lambda = +1$\end{tabular}} \\ \cline{3-5} 
\multicolumn{2}{|c|}{}                                 & \multicolumn{1}{c|}{$+1$}         & \multicolumn{1}{c|}{$0$}         & $-1$    &     \\ \hline
\Xhline{2\arrayrulewidth}
\multicolumn{1}{|c|}{\multirow{4}{*}{2}} & Global      & \multicolumn{1}{c|}{0.083 $\pm$ 0.023} & \multicolumn{1}{c|}{2.881 $\pm$ 0.113} & \multicolumn{1}{c!{\vrule width 1.5pt}}{4.364 $\pm$ 0.093} & 0.096 $\pm$ 0.013 \\ \cline{2-6} 
\multicolumn{1}{|c|}{}                   & Einstein patch 1 & \multicolumn{1}{c|}{0.077 $\pm$ 0.032} & \multicolumn{1}{c|}{11.992 $\pm$ 0.522} & \multicolumn{1}{c!{\vrule width 1.5pt}}{19.728 $\pm$ 0.772} & 0.219 $\pm$ 0.034 \\ \cline{2-6} 
\multicolumn{1}{|c|}{}                   & Einstein patch 2 & \multicolumn{1}{c|}{0.073 $\pm$ 0.021} & \multicolumn{1}{c|}{12.391 $\pm$ 0.674} & \multicolumn{1}{c!{\vrule width 1.5pt}}{19.596 $\pm$ 0.341} & 0.198 $\pm$ 0.034 \\ \cline{2-6} 
\multicolumn{1}{|c|}{}                   & Overlap     & \multicolumn{1}{c|}{0.076 $\pm$ 0.021} & \multicolumn{1}{c|}{0.731 $\pm$ 0.030} & \multicolumn{1}{c!{\vrule width 1.5pt}}{0.868 $\pm$ 0.019} & 0.064 $\pm$ 0.013 \\ \hline
\Xhline{2\arrayrulewidth}
\multicolumn{1}{|c|}{\multirow{4}{*}{3}} & Global      & \multicolumn{1}{c|}{0.151 $\pm$ 0.027} & \multicolumn{1}{c|}{5.560 $\pm$ 0.160} & \multicolumn{1}{c!{\vrule width 1.5pt}}{8.641 $\pm$ 0.183} & 0.195 $\pm$ 0.020 \\ \cline{2-6} 
\multicolumn{1}{|c|}{}                   & Einstein patch 1 & \multicolumn{1}{c|}{0.217 $\pm$ 0.052} & \multicolumn{1}{c|}{25.631 $\pm$ 1.019} & \multicolumn{1}{c!{\vrule width 1.5pt}}{41.246 $\pm$ 1.392} & 0.434 $\pm$ 0.058 \\ \cline{2-6} 
\multicolumn{1}{|c|}{}                   & Einstein patch 2 & \multicolumn{1}{c|}{0.188 $\pm$ 0.053} & \multicolumn{1}{c|}{25.444 $\pm$ 0.838} & \multicolumn{1}{c!{\vrule width 1.5pt}}{42.160 $\pm$ 1.042} & 0.439 $\pm$ 0.059 \\ \cline{2-6} 
\multicolumn{1}{|c|}{}                   & Overlap     & \multicolumn{1}{c|}{0.126 $\pm$ 0.021} & \multicolumn{1}{c|}{1.008 $\pm$ 0.018} & \multicolumn{1}{c!{\vrule width 1.5pt}}{1.164 $\pm$ 0.021} & 0.127 $\pm$ 0.018 \\ \hline
\Xhline{2\arrayrulewidth}
\multicolumn{1}{|c|}{\multirow{4}{*}{4}} & Global      & \multicolumn{1}{c|}{0.150 $\pm$ 0.018} & \multicolumn{1}{c|}{8.494 $\pm$ 0.121} & \multicolumn{1}{c!{\vrule width 1.5pt}}{14.928 $\pm$ 1.317} & 0.248 $\pm$ 0.024 \\ \cline{2-6} 
\multicolumn{1}{|c|}{}                   & Einstein patch 1 & \multicolumn{1}{c|}{0.343 $\pm$ 0.070} & \multicolumn{1}{c|}{40.827 $\pm$ 0.939} & \multicolumn{1}{c!{\vrule width 1.5pt}}{74.663 $\pm$ 4.943} & 0.640 $\pm$ 0.092 \\ \cline{2-6} 
\multicolumn{1}{|c|}{}                   & Einstein patch 2 & \multicolumn{1}{c|}{0.303 $\pm$ 0.051} & \multicolumn{1}{c|}{41.170 $\pm$ 1.059} & \multicolumn{1}{c!{\vrule width 1.5pt}}{74.845 $\pm$ 3.626} & 0.603 $\pm$ 0.043 \\ \cline{2-6} 
\multicolumn{1}{|c|}{}                   & Overlap     & \multicolumn{1}{c|}{0.100 $\pm$ 0.012} & \multicolumn{1}{c|}{1.144 $\pm$ 0.081} & \multicolumn{1}{c!{\vrule width 1.5pt}}{1.470 $\pm$ 0.700} & 0.148 $\pm$ 0.018 \\ \hline
\Xhline{2\arrayrulewidth}
\multicolumn{1}{|c|}{\multirow{4}{*}{5}} & Global      & \multicolumn{1}{c|}{0.244 $\pm$ 0.039} & \multicolumn{1}{c|}{10.810 $\pm$ 0.185} & \multicolumn{1}{c!{\vrule width 1.5pt}}{18.798 $\pm$ 2.024} & 0.518 $\pm$ 0.063 \\ \cline{2-6} 
\multicolumn{1}{|c|}{}                   & Einstein patch 1 & \multicolumn{1}{c|}{0.615 $\pm$ 0.132} & \multicolumn{1}{c|}{53.410 $\pm$ 1.641} & \multicolumn{1}{c!{\vrule width 1.5pt}}{97.398 $\pm$ 10.361} & 2.032 $\pm$ 0.291 \\ \cline{2-6} 
\multicolumn{1}{|c|}{}                   & Einstein patch 2 & \multicolumn{1}{c|}{0.595 $\pm$ 0.181} & \multicolumn{1}{c|}{54.186 $\pm$ 1.487} & \multicolumn{1}{c!{\vrule width 1.5pt}}{97.198 $\pm$ 12.189} & 1.552 $\pm$ 0.356 \\ \cline{2-6} 
\multicolumn{1}{|c|}{}                   & Overlap     & \multicolumn{1}{c|}{0.148 $\pm$ 0.022} & \multicolumn{1}{c|}{1.131 $\pm$ 0.066} & \multicolumn{1}{c!{\vrule width 1.5pt}}{1.218 $\pm$ 0.077} & 0.211 $\pm$ 0.016 \\ \hline
\end{tabular}
\caption{Global test loss results, with decompositions into the constituent sublosses: Einstein loss patch 1, Einstein loss patch 2, Overlap loss; averaged over 10 runs. Losses computed for NN approximations of Einstein metrics with the respective curvatures on spheres in dimensions 2-5 (2-patches). For comparison, the right-hand column shows the respective global test losses for the \textit{supervised} NN model approximation of the analytic round metric (which satisfies the Einstein equation for $\lambda = +1$). All losses are reported with standard deviations across the 10 runs in each case.}
\label{tab:full_loss_results}
\end{table}

One can see that the overlap loss is naturally lower, which is a good sign of consistency, since the patching condition is essential for ensuring the global metric definition is consistent; this is what motivated the higher multiplier weighting of this overlap loss component.
The Einstein losses within each investigation are approximately equal between the 2 patches, supporting the symmetric treatment of the patches.
Furthermore, the $\lambda=+1$ investigations all have low values across the loss components, particularly with both Einstein losses $<1$.
Conversely, the Einstein losses in the $\lambda \in \{0,-1\}$ investigations are all much higher, demonstrating further the geometric obstruction to learning Einstein metric's with these Einstein constants in these dimensions.

\section{Visualisations}\label{app:extra_vis}
\begin{figure}[!t]
    \centering
    \begin{subfigure}{0.24\textwidth}
        \centering
        \includegraphics[width=0.98\textwidth]{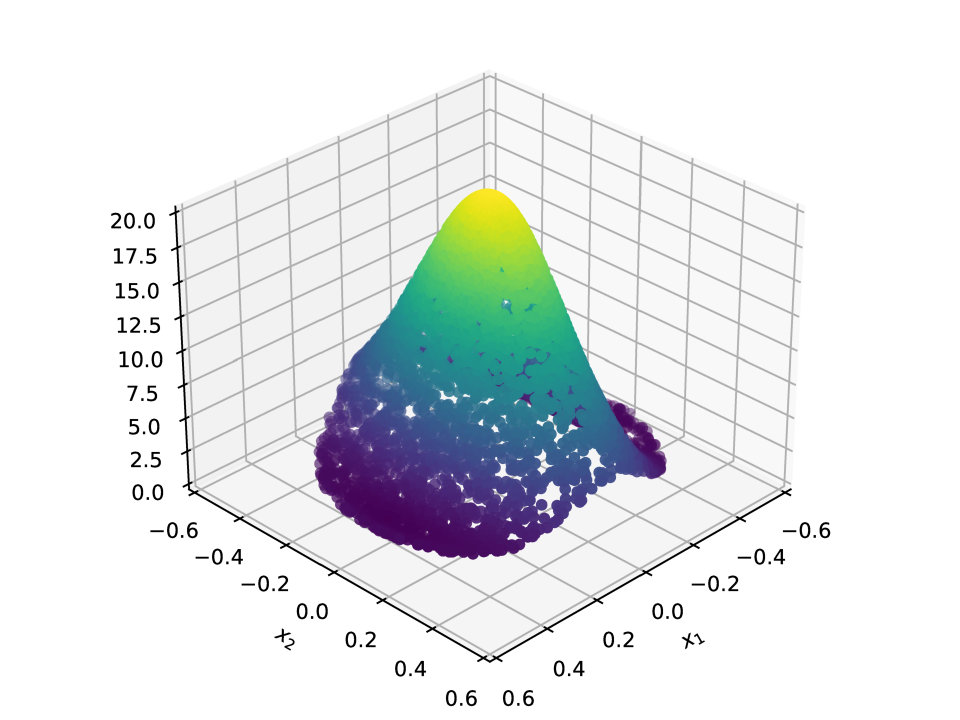}
        \caption{$g_{00}$ Patch 1}
        \label{fig:vis_2d0_g001}
    \end{subfigure} 
    \begin{subfigure}{0.24\textwidth}
        \centering
        \includegraphics[width=0.98\textwidth]{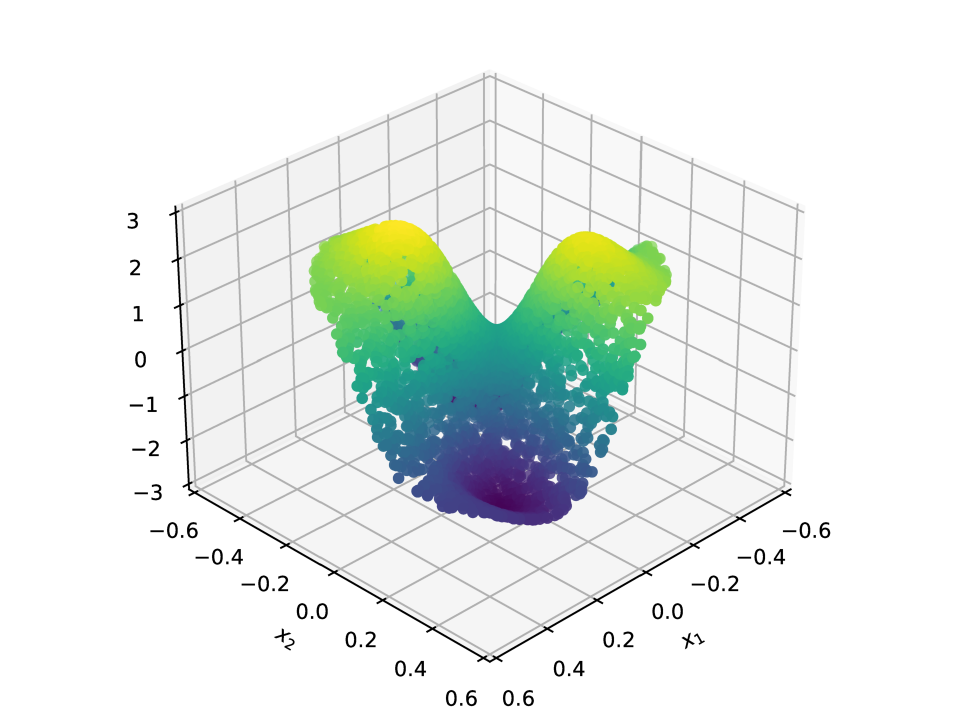}
        \caption{$g_{01}$ Patch 1}
    \end{subfigure} 
    \begin{subfigure}{0.24\textwidth}
        \centering
        \includegraphics[width=0.98\textwidth]{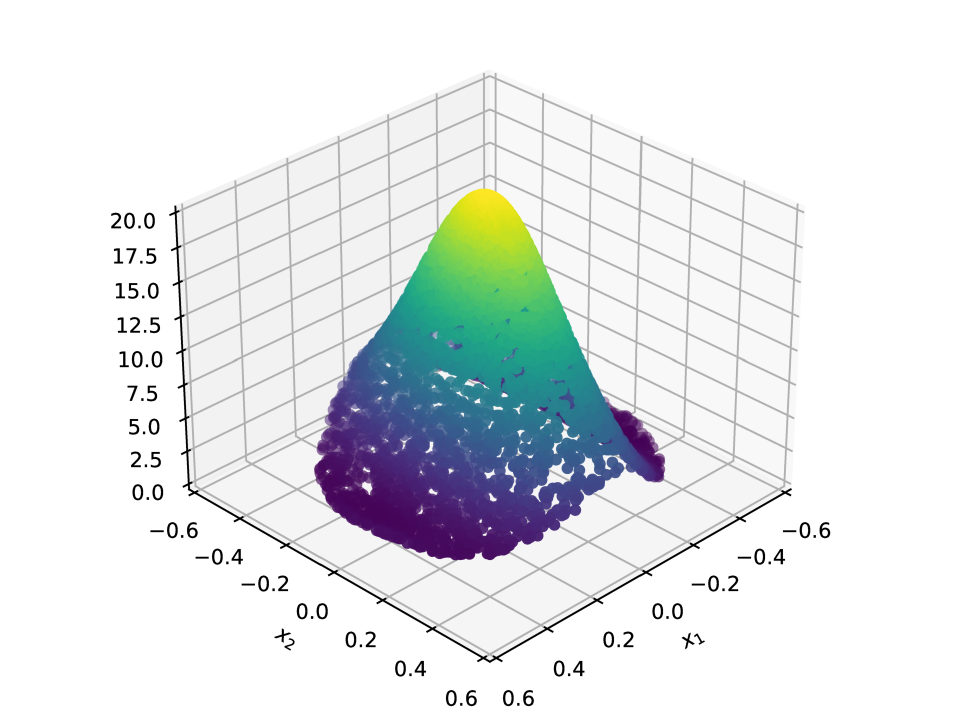}
        \caption{$g_{00}$ Patch 2}
    \end{subfigure} 
    \begin{subfigure}{0.24\textwidth}
        \centering
        \includegraphics[width=0.98\textwidth]{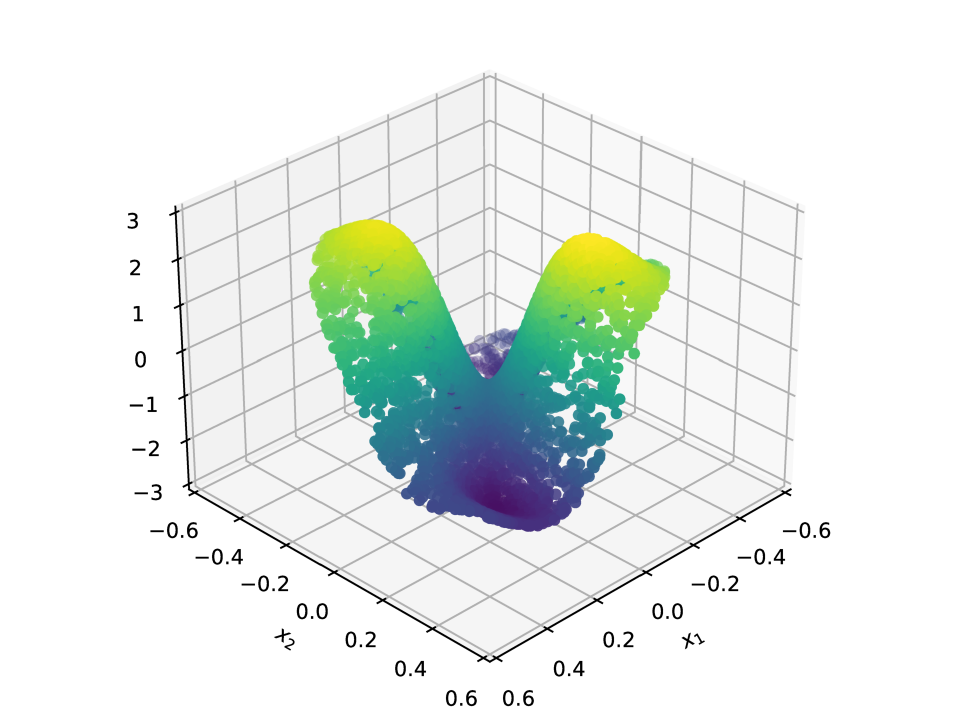}
        \caption{$g_{01}$ Patch 2}
    \end{subfigure}\\
    \begin{subfigure}{0.24\textwidth}
        \centering
        \includegraphics[width=0.98\textwidth]{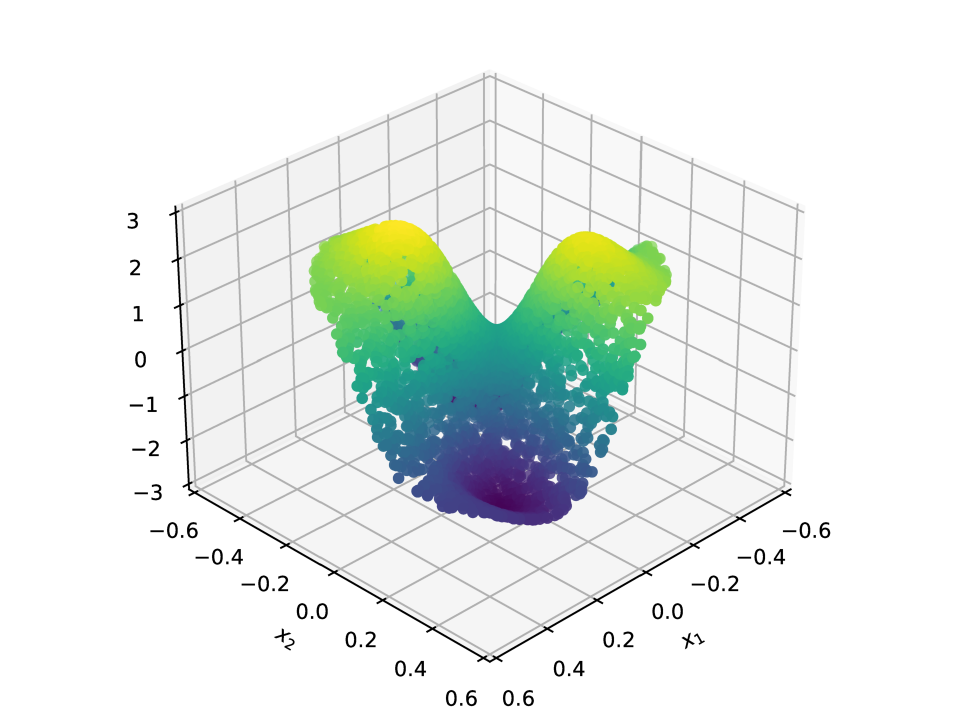}
        \caption{$g_{10}$ Patch 1}
    \end{subfigure} 
    \begin{subfigure}{0.24\textwidth}
        \centering
        \includegraphics[width=0.98\textwidth]{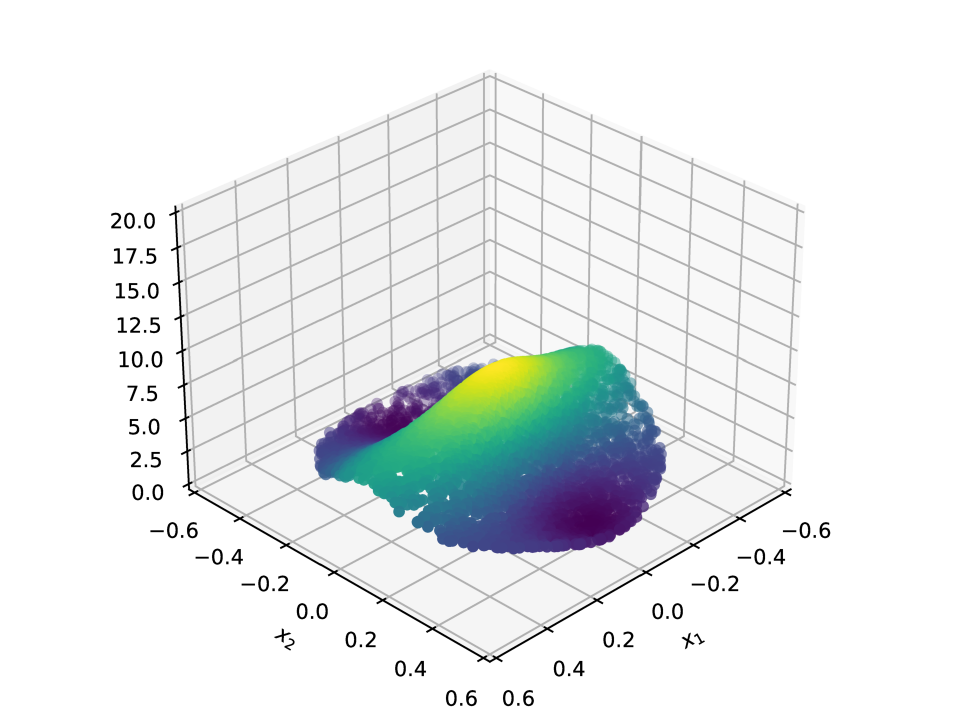}
        \caption{$g_{11}$ Patch 1}
    \end{subfigure} 
    \begin{subfigure}{0.24\textwidth}
        \centering
        \includegraphics[width=0.98\textwidth]{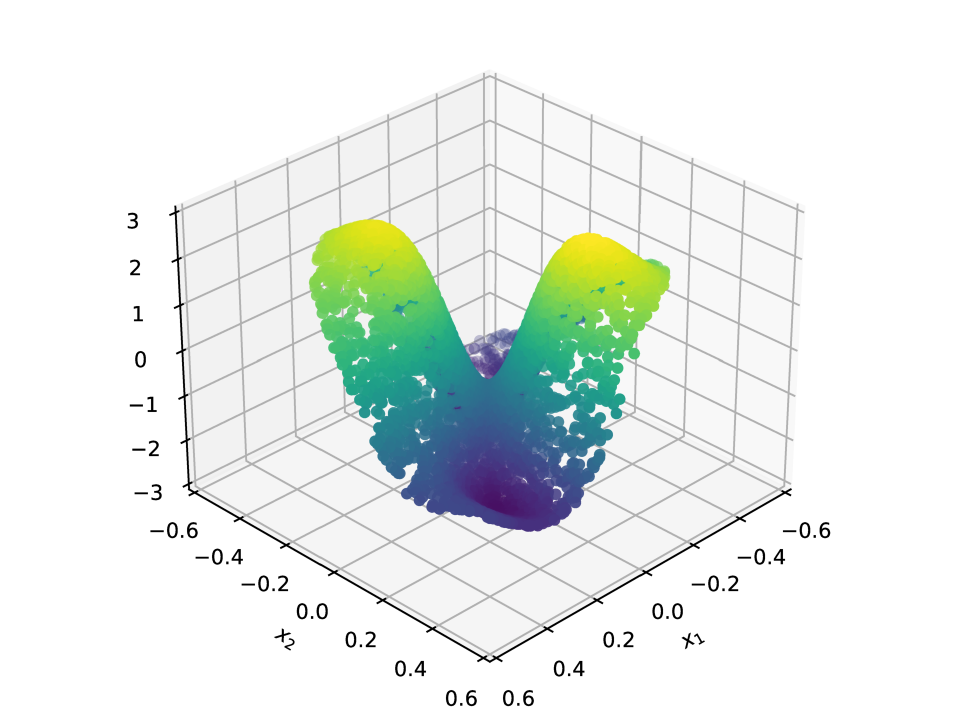}
        \caption{$g_{10}$ Patch 2}
    \end{subfigure} 
    \begin{subfigure}{0.24\textwidth}
        \centering
        \includegraphics[width=0.98\textwidth]{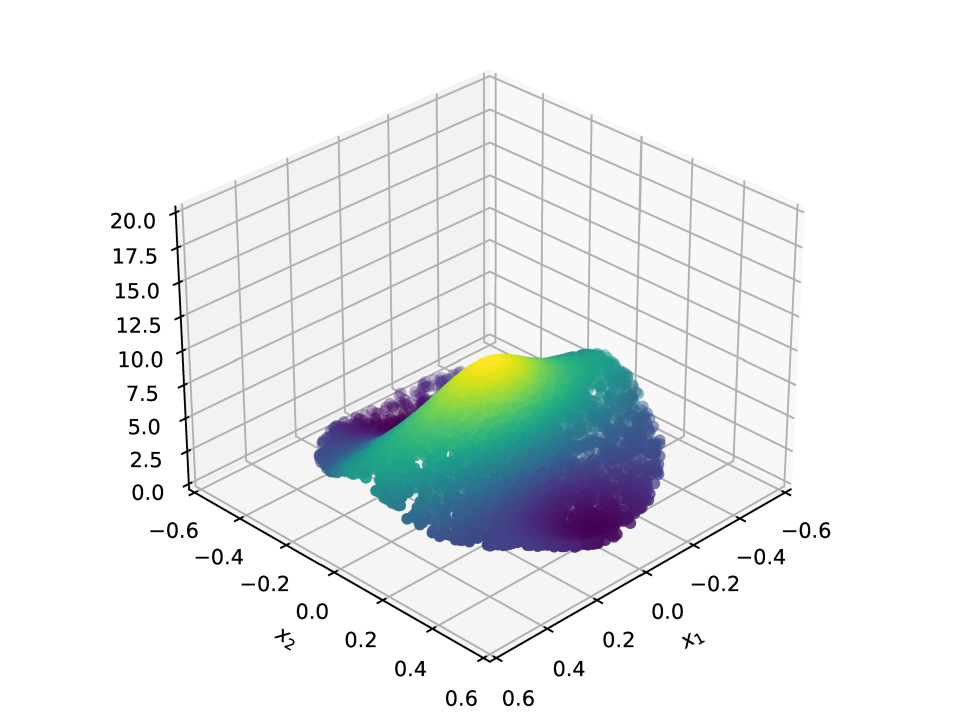}
        \caption{$g_{11}$ Patch 2}
    \end{subfigure}
    \caption{Visualisations of the learnt metrics, $g_{ij}$, in 2d, on the 2 patches, trained with zero Einstein constant (such that $R_{ij} = 0$), and the metric's goal is to be Ricci-flat.}
    \label{fig:vis_2d0_g}
\end{figure}

\begin{figure}[!t]
    \centering
    \begin{subfigure}{0.24\textwidth}
        \centering
        \includegraphics[width=0.98\textwidth]{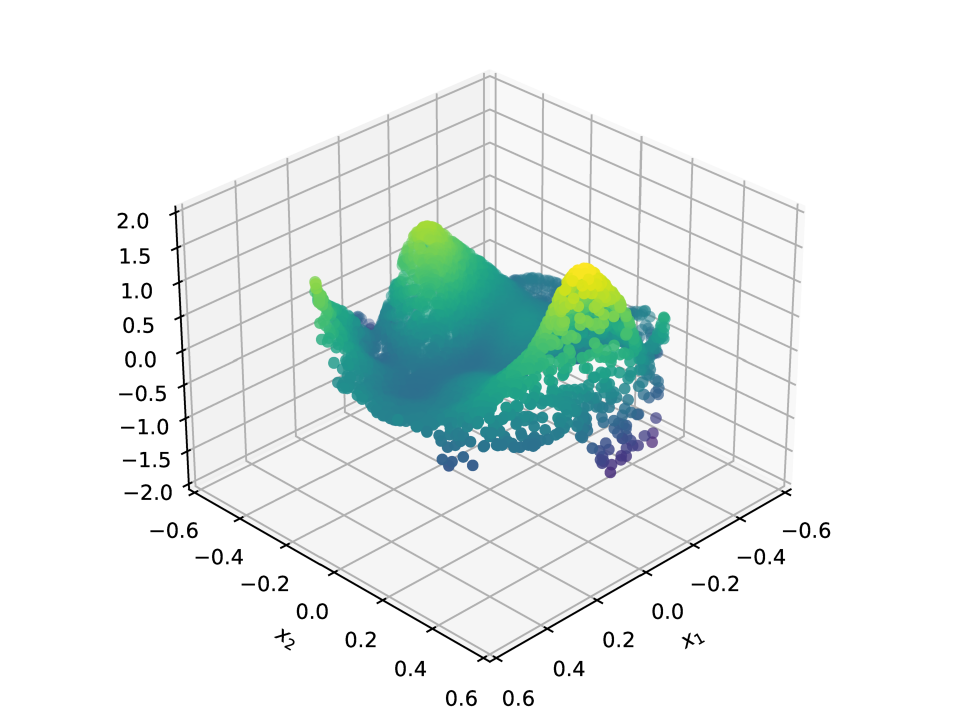}
        \caption{$R_{00}$ Patch 1}
    \end{subfigure} 
    \begin{subfigure}{0.24\textwidth}
        \centering
        \includegraphics[width=0.98\textwidth]{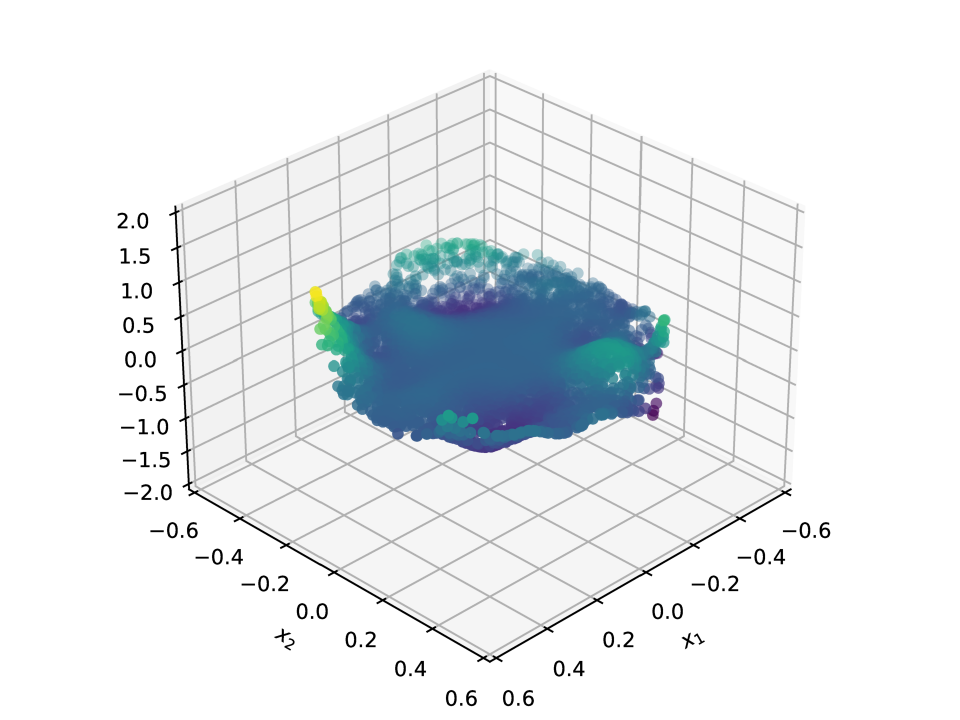}
        \caption{$R_{01}$ Patch 1}
    \end{subfigure} 
    \begin{subfigure}{0.24\textwidth}
        \centering
        \includegraphics[width=0.98\textwidth]{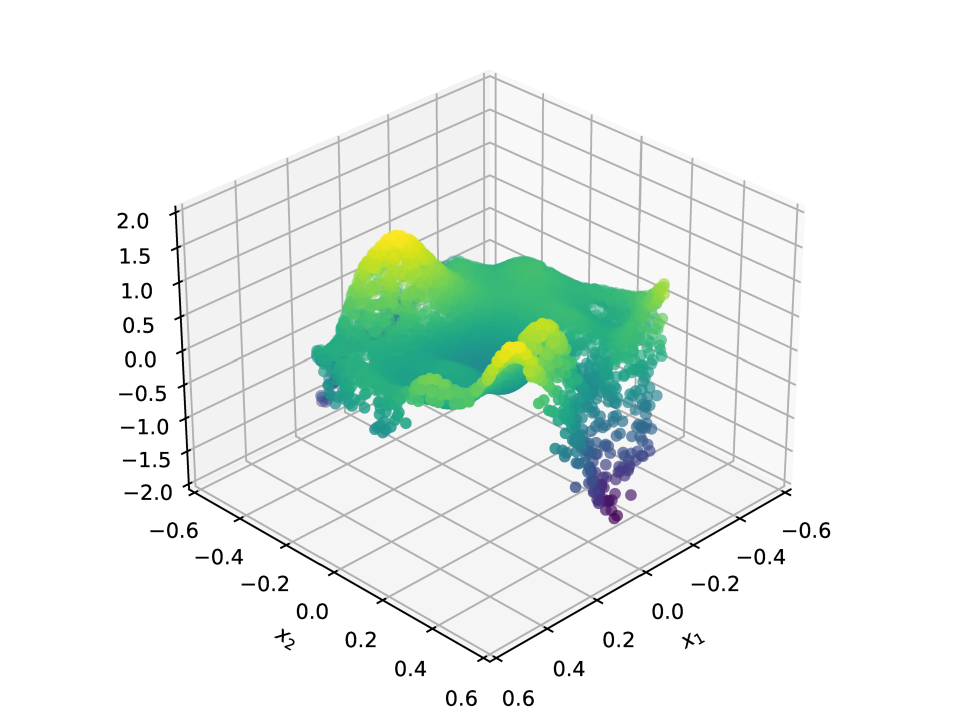}
        \caption{$R_{00}$ Patch 2}
    \end{subfigure} 
    \begin{subfigure}{0.24\textwidth}
        \centering
        \includegraphics[width=0.98\textwidth]{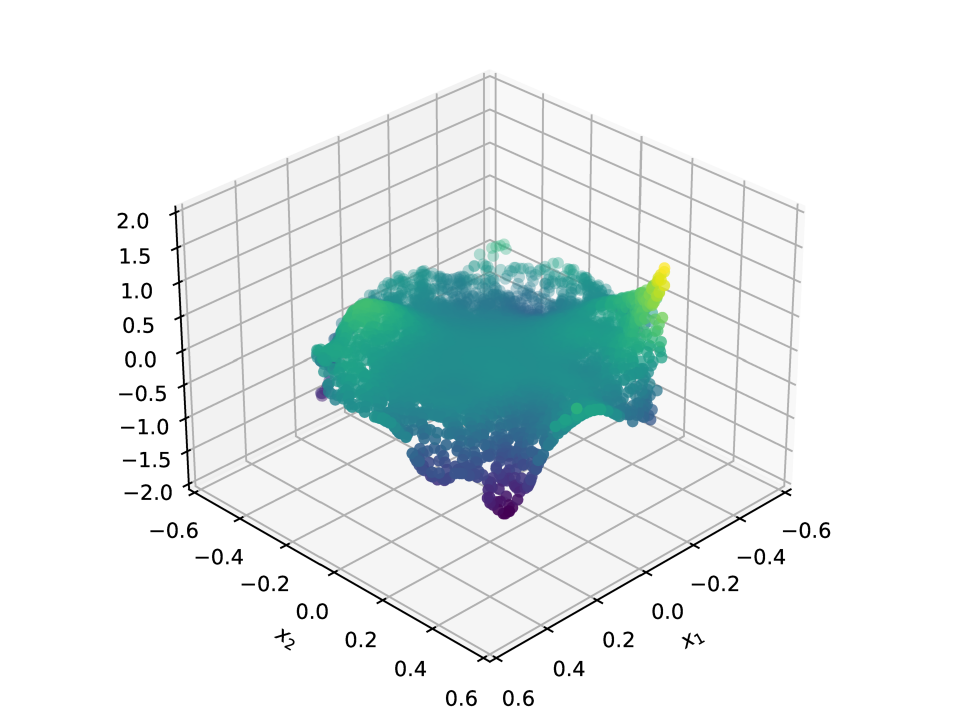}
        \caption{$R_{01}$ Patch 2}
    \end{subfigure}\\
    \begin{subfigure}{0.24\textwidth}
        \centering
        \includegraphics[width=0.98\textwidth]{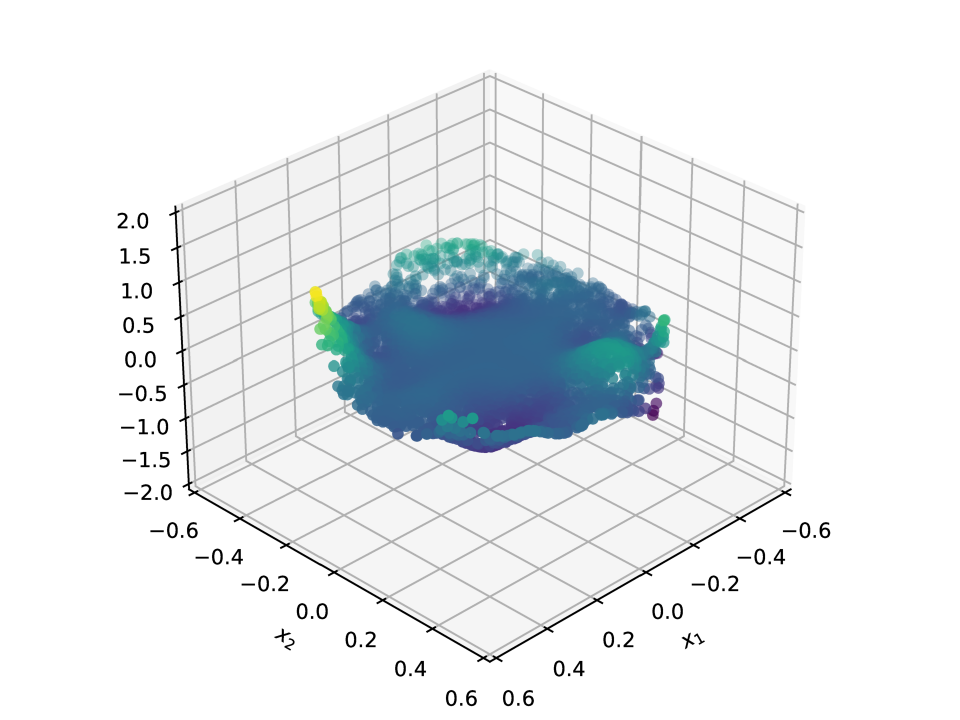}
        \caption{$R_{10}$ Patch 1}
    \end{subfigure} 
    \begin{subfigure}{0.24\textwidth}
        \centering
        \includegraphics[width=0.98\textwidth]{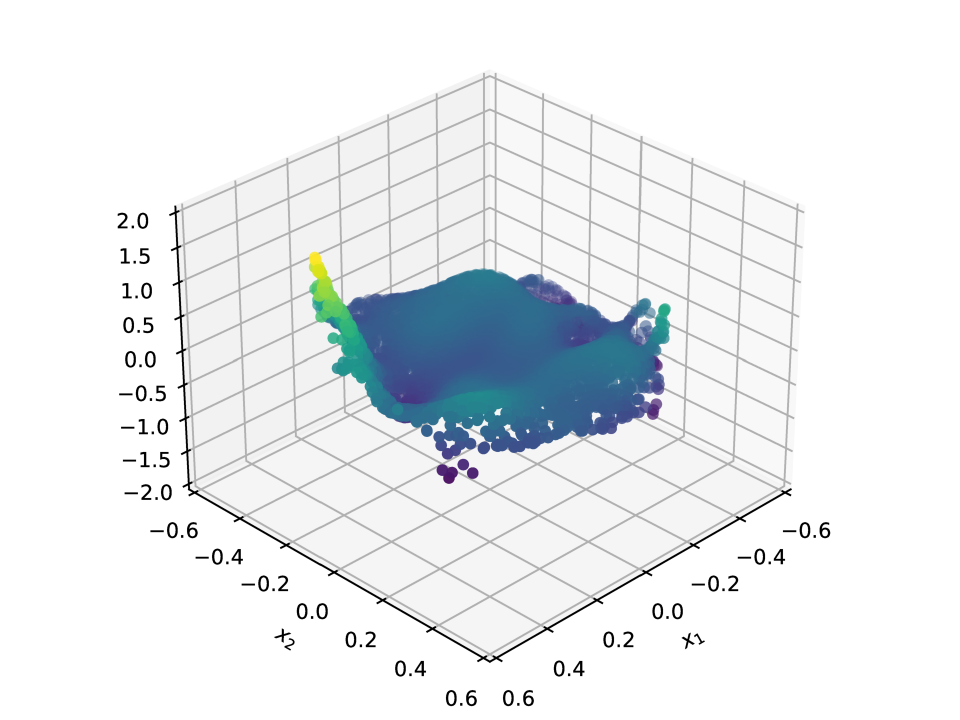}
        \caption{$R_{11}$ Patch 1}
    \end{subfigure} 
    \begin{subfigure}{0.24\textwidth}
        \centering
        \includegraphics[width=0.98\textwidth]{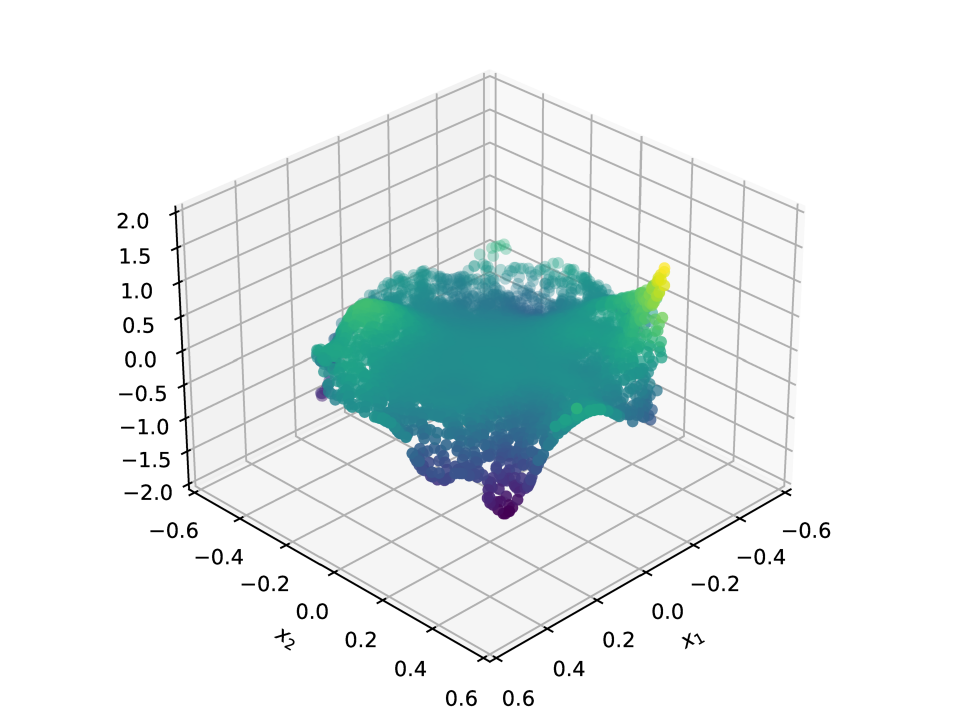}
        \caption{$R_{10}$ Patch 2}
    \end{subfigure} 
    \begin{subfigure}{0.24\textwidth}
        \centering
        \includegraphics[width=0.98\textwidth]{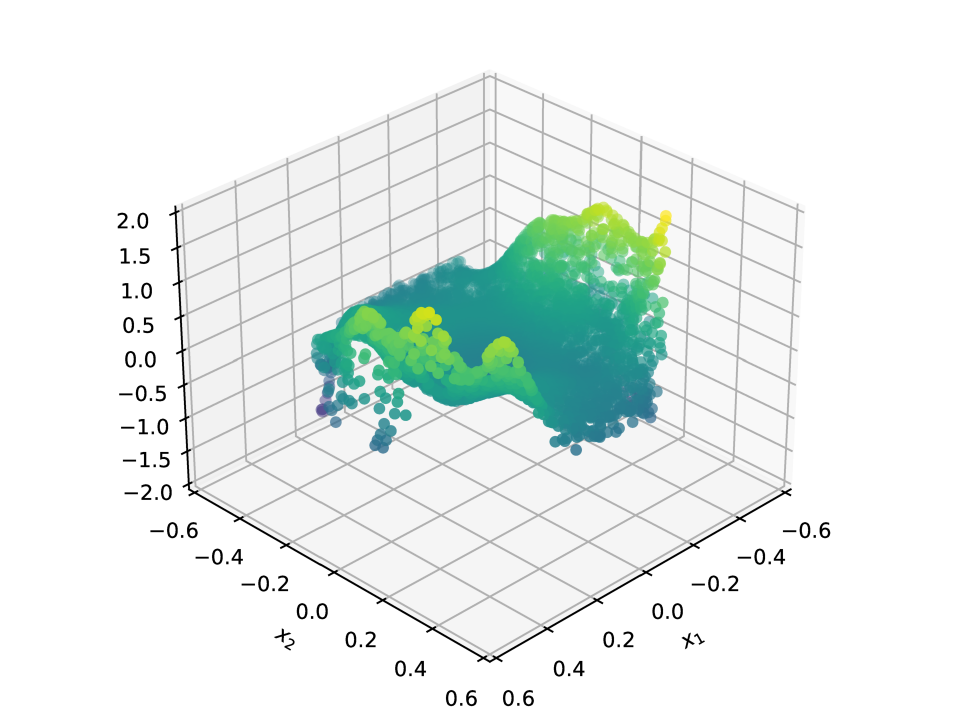}
        \caption{$R_{11}$ Patch 2}
    \end{subfigure}
    \caption{Visualisations of the Ricci tensors, $R_{ij}$, of the learnt metrics in 2d, on the 2 patches, trained for zero Einstein constant (such that $R_{ij} = 0$), and the metric's goal is to be Ricci-flat.}
    \label{fig:vis_2d0_R}
\end{figure}

\begin{figure}[!t]
    \centering
    \begin{subfigure}{0.24\textwidth}
        \centering
        \includegraphics[width=0.98\textwidth]{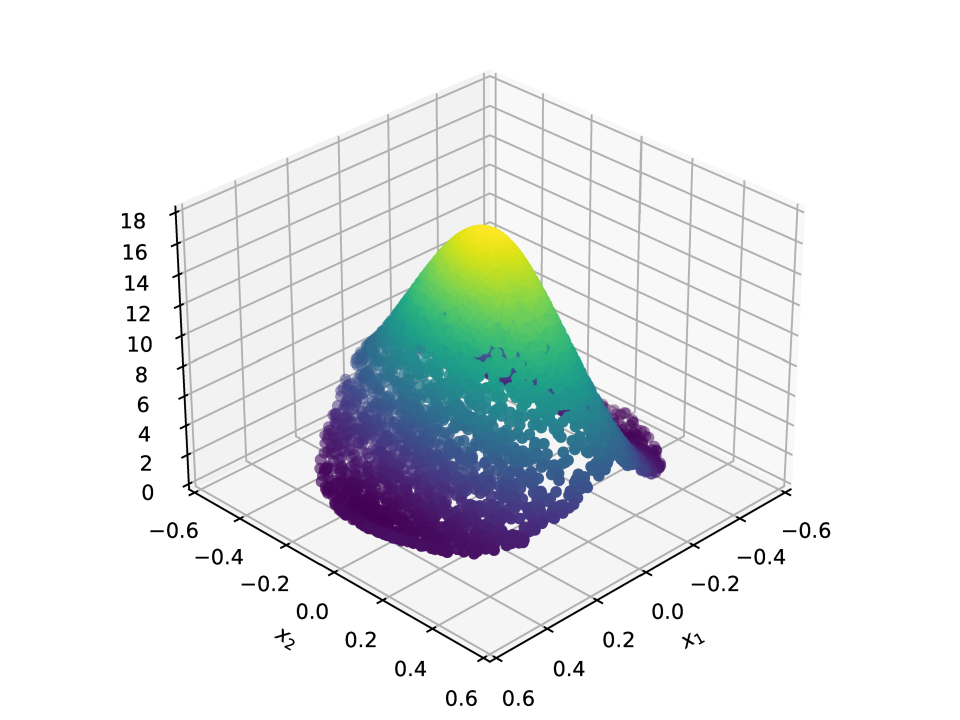}
        \caption{$g_{00}$ Patch 1}
        \label{fig:vis_2dneg_g001}
    \end{subfigure} 
    \begin{subfigure}{0.24\textwidth}
        \centering
        \includegraphics[width=0.98\textwidth]{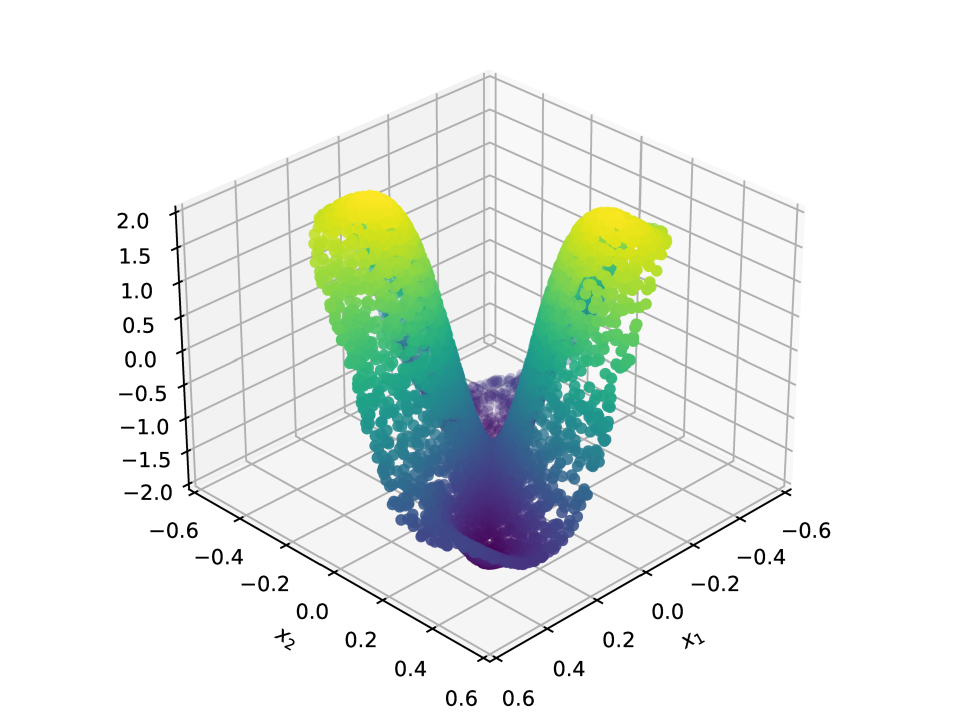}
        \caption{$g_{01}$ Patch 1}
    \end{subfigure} 
    \begin{subfigure}{0.24\textwidth}
        \centering
        \includegraphics[width=0.98\textwidth]{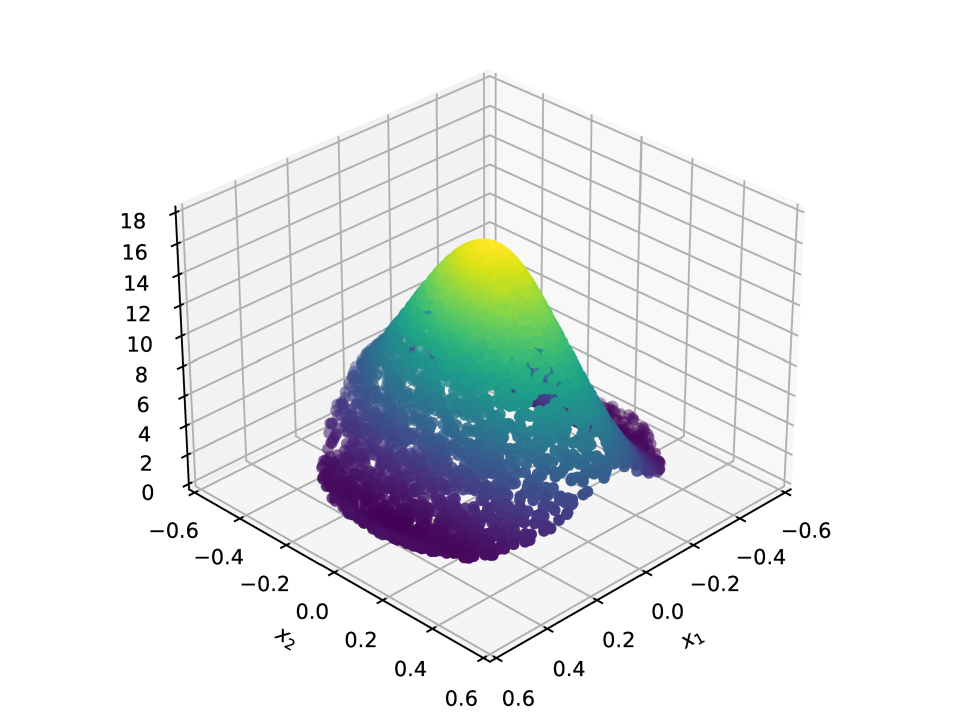}
        \caption{$g_{00}$ Patch 2}
    \end{subfigure} 
    \begin{subfigure}{0.24\textwidth}
        \centering
        \includegraphics[width=0.98\textwidth]{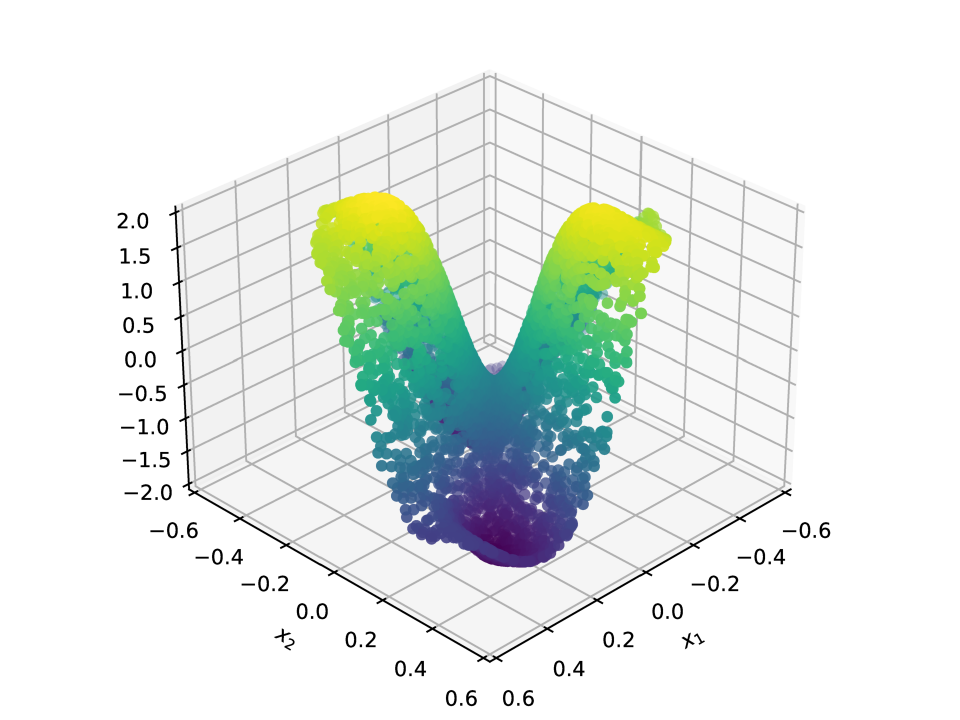}
        \caption{$g_{01}$ Patch 2}
    \end{subfigure}\\
    \begin{subfigure}{0.24\textwidth}
        \centering
        \includegraphics[width=0.98\textwidth]{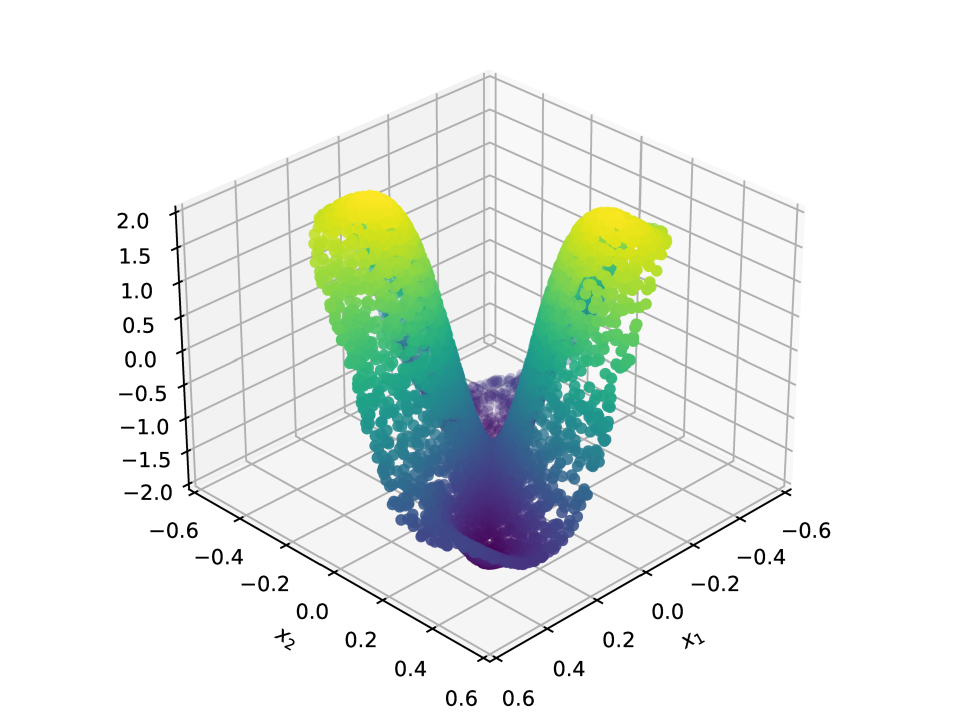}
        \caption{$g_{10}$ Patch 1}
    \end{subfigure} 
    \begin{subfigure}{0.24\textwidth}
        \centering
        \includegraphics[width=0.98\textwidth]{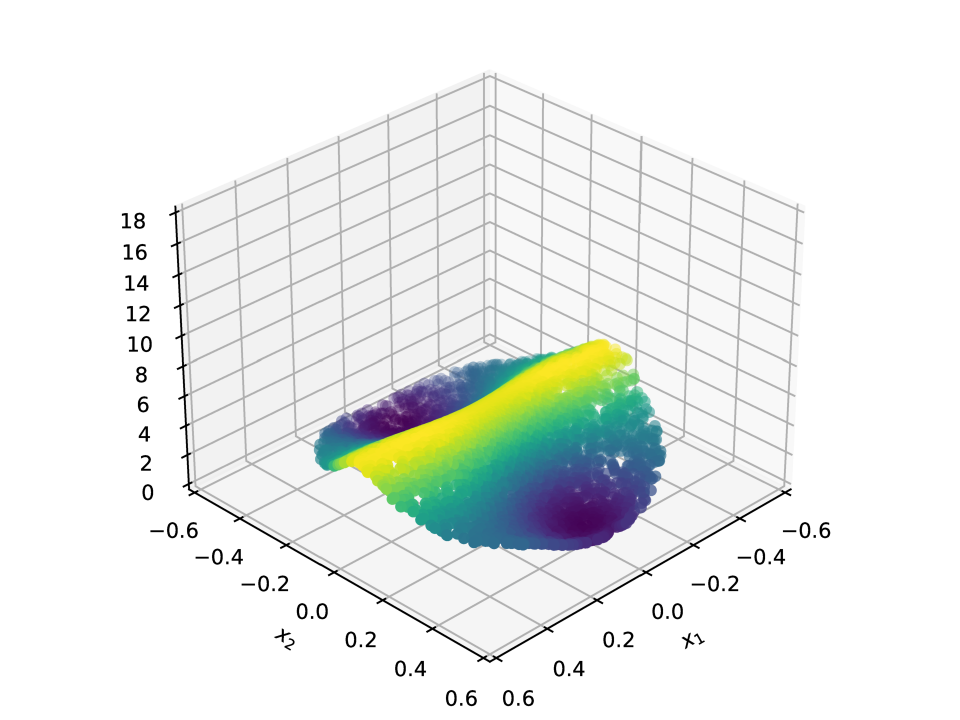}
        \caption{$g_{11}$ Patch 1}
    \end{subfigure} 
    \begin{subfigure}{0.24\textwidth}
        \centering
        \includegraphics[width=0.98\textwidth]{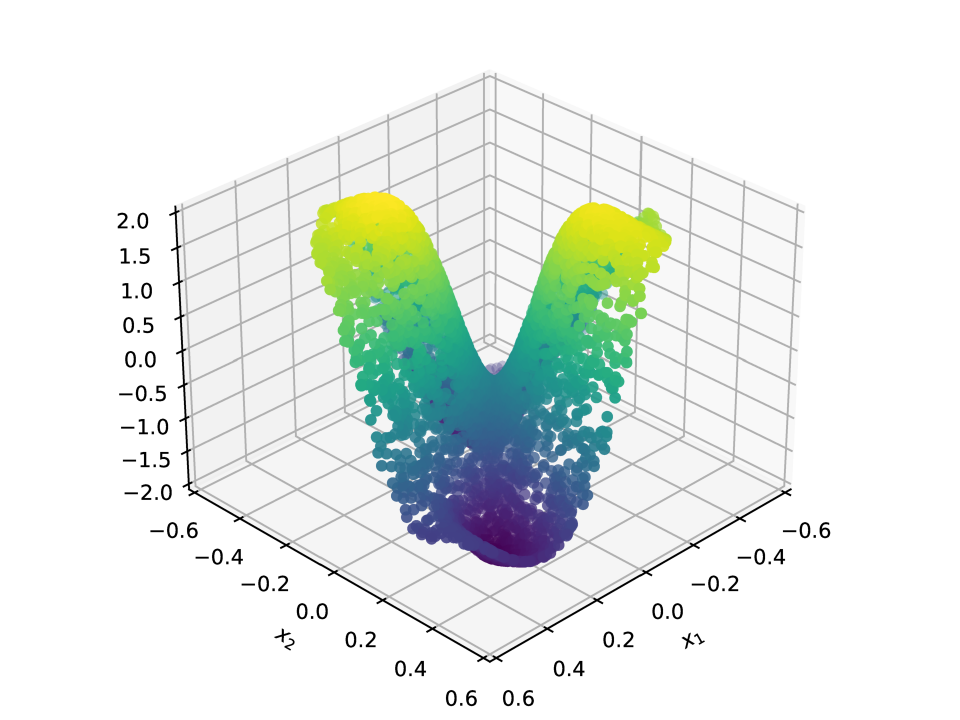}
        \caption{$g_{10}$ Patch 2}
    \end{subfigure} 
    \begin{subfigure}{0.24\textwidth}
        \centering
        \includegraphics[width=0.98\textwidth]{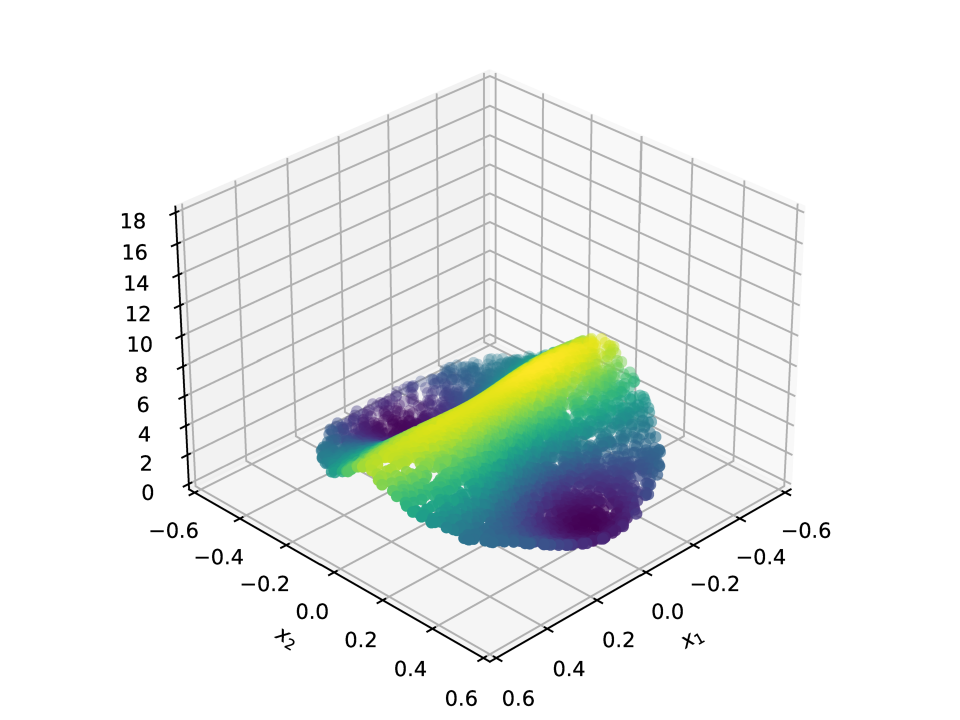}
        \caption{$g_{11}$ Patch 2}
    \end{subfigure}
    \caption{Visualisations of the learnt metrics, $g_{ij}$, in 2d, on the 2 patches, trained with negative Einstein constant (such that $R_{ij} = -g_{ij}$).}
    \label{fig:vis_2dneg_g}
\end{figure}

\begin{figure}[!t]
    \centering
    \begin{subfigure}{0.24\textwidth}
        \centering
        \includegraphics[width=0.98\textwidth]{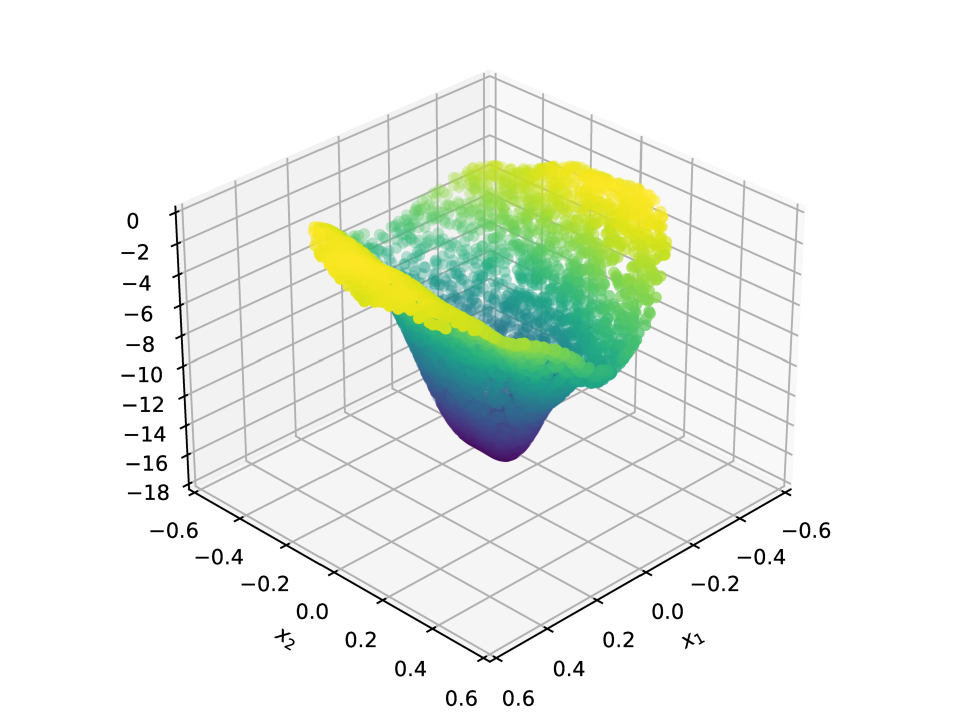}
        \caption{$R_{00}$ Patch 1}
    \end{subfigure} 
    \begin{subfigure}{0.24\textwidth}
        \centering
        \includegraphics[width=0.98\textwidth]{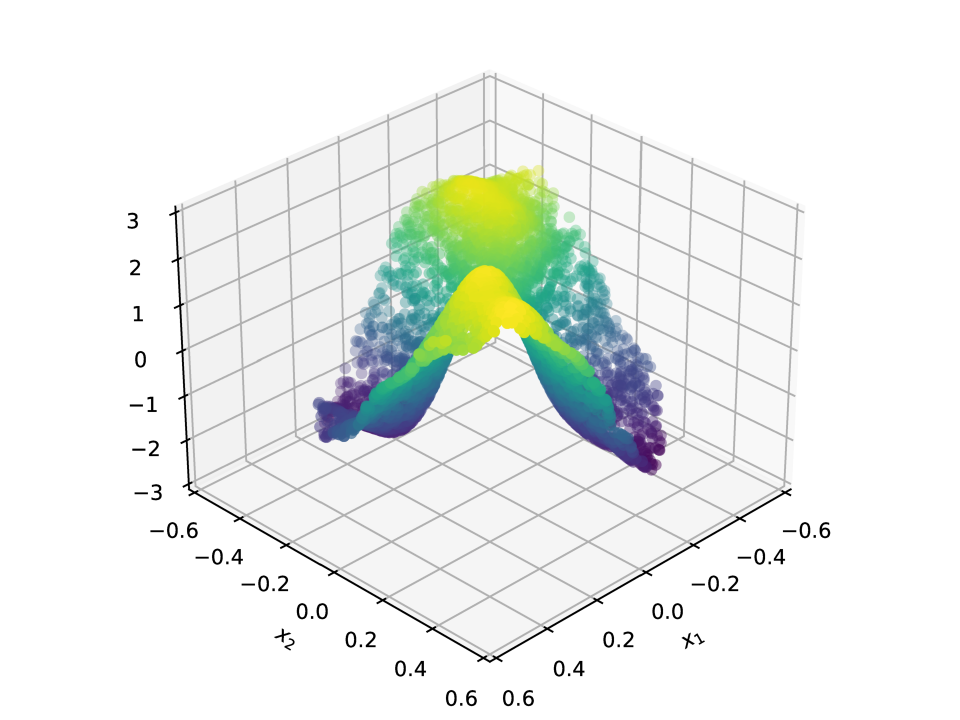}
        \caption{$R_{01}$ Patch 1}
    \end{subfigure} 
    \begin{subfigure}{0.24\textwidth}
        \centering
        \includegraphics[width=0.98\textwidth]{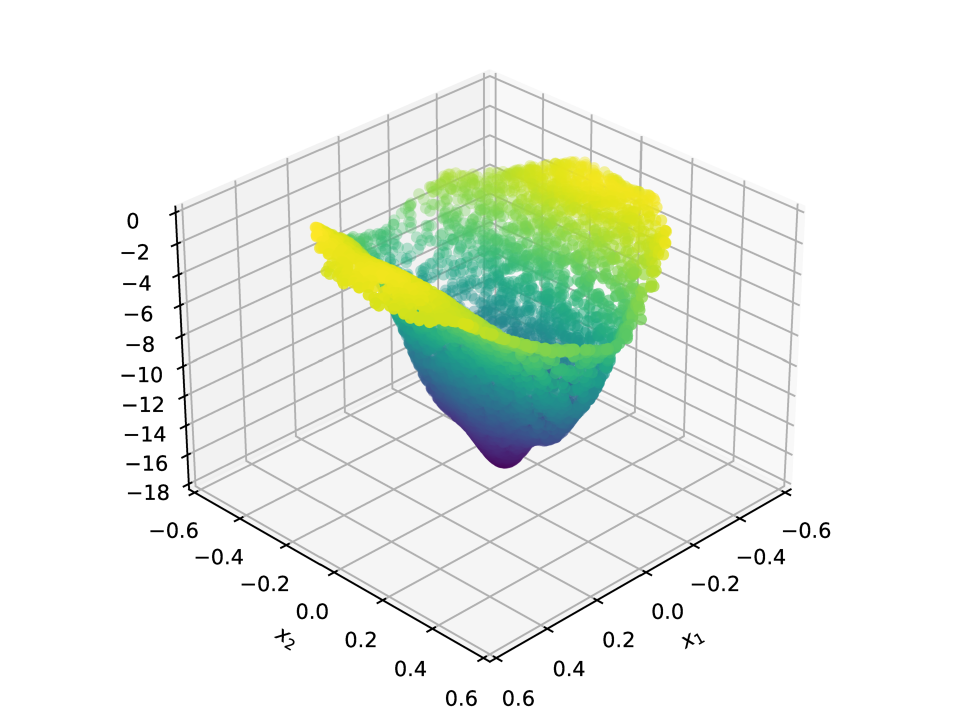}
        \caption{$R_{00}$ Patch 2}
    \end{subfigure} 
    \begin{subfigure}{0.24\textwidth}
        \centering
        \includegraphics[width=0.98\textwidth]{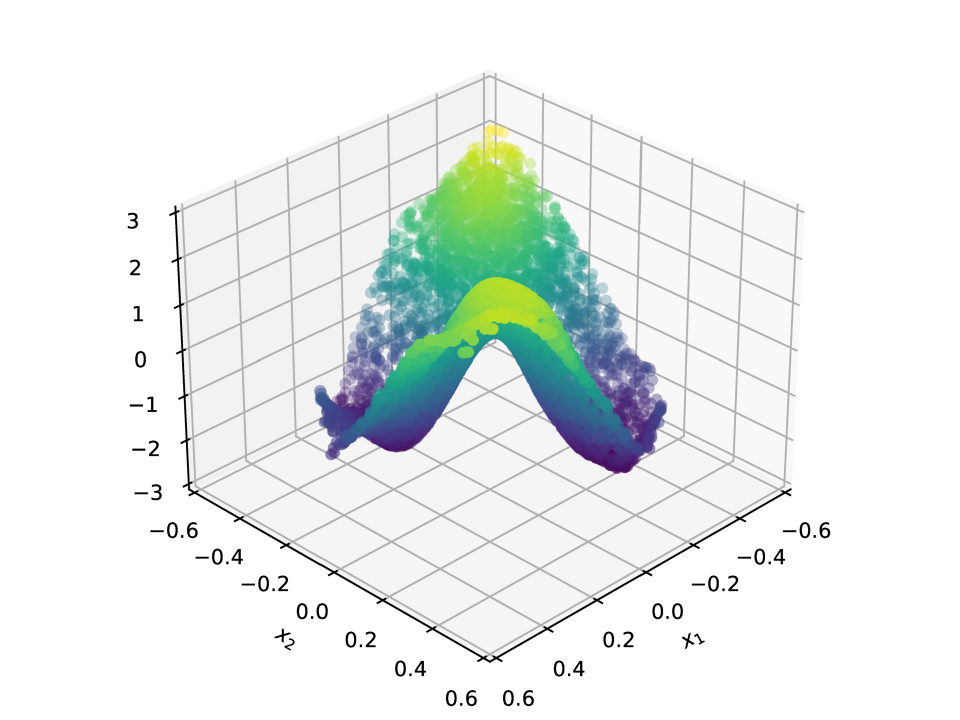}
        \caption{$R_{01}$ Patch 2}
    \end{subfigure}\\
    \begin{subfigure}{0.24\textwidth}
        \centering
        \includegraphics[width=0.98\textwidth]{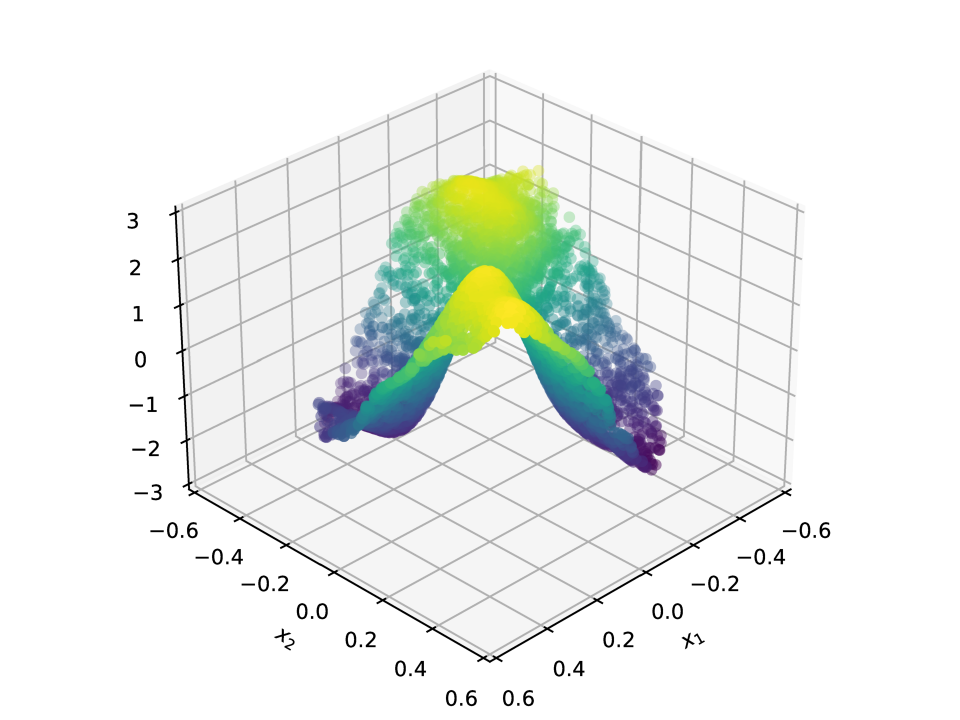}
        \caption{$R_{10}$ Patch 1}
    \end{subfigure} 
    \begin{subfigure}{0.24\textwidth}
        \centering
        \includegraphics[width=0.98\textwidth]{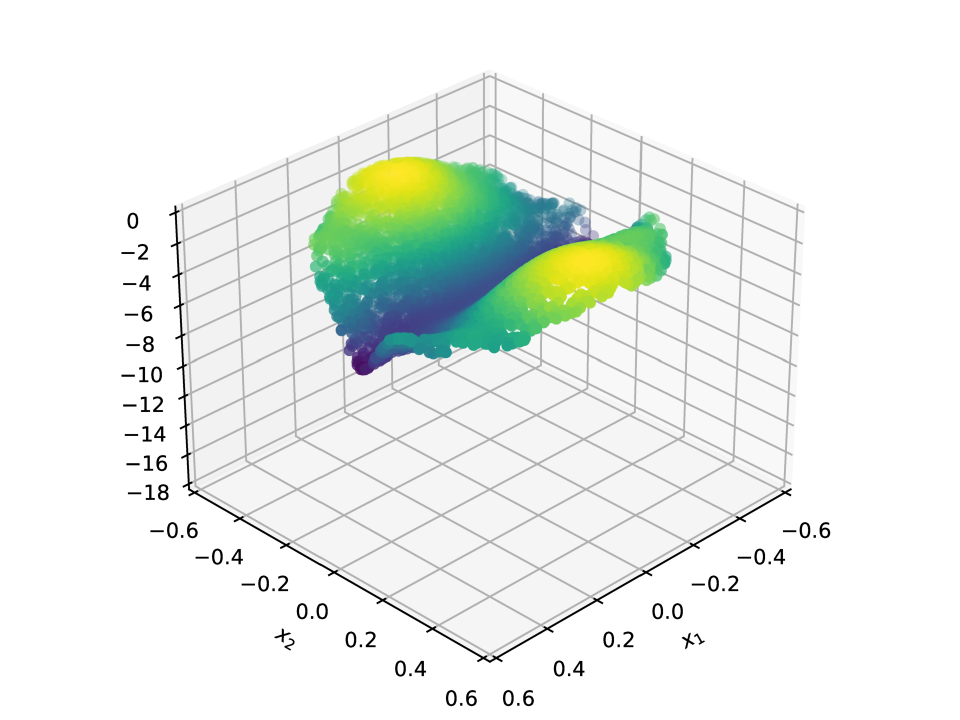}
        \caption{$R_{11}$ Patch 1}
    \end{subfigure} 
    \begin{subfigure}{0.24\textwidth}
        \centering
        \includegraphics[width=0.98\textwidth]{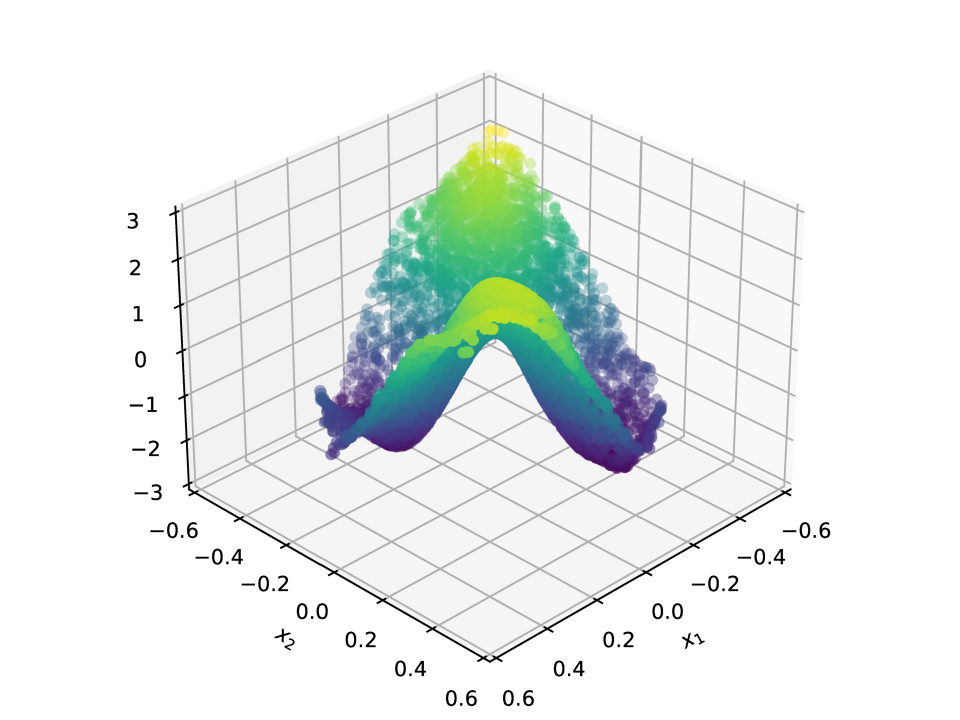}
        \caption{$R_{10}$ Patch 2}
    \end{subfigure} 
    \begin{subfigure}{0.24\textwidth}
        \centering
        \includegraphics[width=0.98\textwidth]{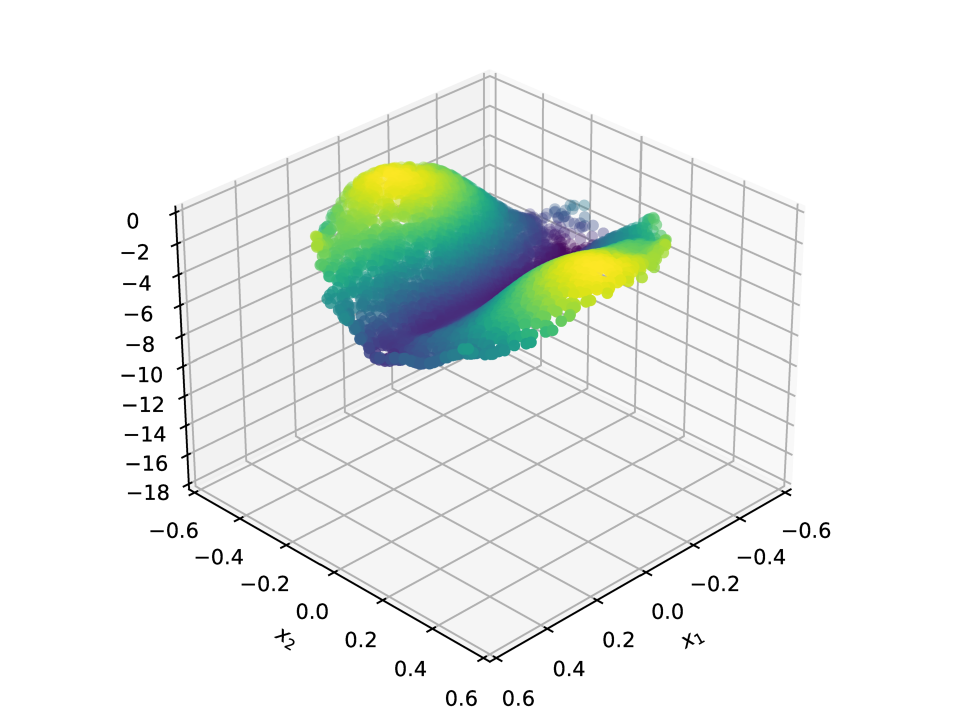}
        \caption{$R_{11}$ Patch 2}
    \end{subfigure}
    \caption{Visualisations of the Ricci tensors, $R_{ij}$, of the learnt metrics in 2d, on the 2 patches, trained for negative Einstein constant (such that $R_{ij} = -g_{ij}$).}
    \label{fig:vis_2dneg_R}
\end{figure}

To extend the visual interpretation of the metric learning, as shown in Figures \ref{fig:vis_2dpos_g} \& \ref{fig:vis_2dpos_R}, here equivalent plots are shown for example runs from the 2d $\lambda \in \{0,-1\}$ investigations.
For $\lambda=0$, the metric components in both patches are shown in Figure \ref{fig:vis_2d0_g}, whilst the equivalent Ricci tensor components are shown in Figure \ref{fig:vis_2d0_R}.
Then for $\lambda = -1$, the metric components in both patches are shown in Figure \ref{fig:vis_2dneg_g}, whilst the equivalent Ricci tensor components are shown in Figure \ref{fig:vis_2dneg_R}. 

For $\lambda=0$ the model is clearly trying to set all the Ricci tensor components to 0, however it fails with clear instabilities it cannot avoid due to the geometric obstruction to existence of Ricci-flat metrics.
Whereas for $\lambda=-1$ the respective Ricci components look somewhat like inversions of the metric components as the model tries to match these Ricci components to the negative values of the metric.
However, again there are clear instabilities around the patch centre, and at the edges of the plotting restriction where the overlap region is defined, where the model expectedly cannot overcome these geometric obstructions.

A final comment, is that the shape of the components looks somewhat similar between the $\lambda$ values, for example with conical-like shapes for the $(0,0)$ components.
Upon further inspection of these components one can start to see the differing curvatures.
In Figure \ref{fig:vis_2dpos_g001} (for $\lambda=+1$) the cone outline from the centre along the $x_1$ axes the outline starts to curve up, whereas in Figure \ref{fig:vis_2d0_g001} (for $\lambda=0$) the outline is quite flat, and finally in Figure \ref{fig:vis_2dneg_g001} (for $\lambda=-1$) the outline curves downwards.
These opposing visual curvatures match the expected behaviour, and demonstrate the subtlety in the learning of Einstein metrics via this highly non-linear and extremely sensitive Einstein equation.

\section{Details of Manifold Sampling}\label{app:sampling}
The boundary of the open ball patches represent the infinite limits of the stereographic real plane and where the sphere projections break down, unsurprisingly it is here that the greatest numerical instabilities are seen.
Conversely, points near the ball centre in patch 1 map to near the boundary in patch 2, and thus optimal sampling to avoid instabilities skews generation to the parts of the patch away from these extremities.
Additionally, since the patch gluing conditions require each patch only up the $r_m + \varepsilon$, to ensure consistent gluing at the overlap points should be dense near this midpoint.

From these motivations, the ball sampling procedure used polar coordinates for the patch, implementing a modified beta function for the radii, and sampled the angles uniformly; then transforming into the Euclidean coordinate inputs.
The beta function skews sampling to prioritise radii near to $r_m$; and to ensure the patches are sampled symmetrically, half the requested number of samples are generated using the same beta function for patch 2 and are transformed back to patch 1.
The general beta function is defined by the distribution
\begin{equation}\label{eq:beta_fn}
    f(r;\alpha,\beta) := \frac{r^{\alpha - 1}(1-r)^{\beta - 1}}{\int_0^1 t^{\alpha - 1}(1 - t)^{\beta - 1}dt}\;,
\end{equation}
for $r$ the sampled variable in the domain $[0,1]$, for us the radius of the sampled point in polar coordinates of the ball patch, and the parameters $\alpha,\beta>0$ control the distribution shape.

The mean of this distribution is $\frac{\alpha}{\alpha + \beta}$, therefore to encourage sampling to be symmetric between the patches we set this mean to equal the radial midpoint $r_m = \sqrt{2}-1$; such that rearranging sets $\beta = \alpha(\frac{1}{r_m}-1) \sim 1.41 \alpha$.
However, despite the sample mean now being symmetric under the patch change, the rate of sampling density change is still not symmetric.
Therefore to rectify this, half the sampled radii are transformed using \eqref{eq:radii_patchchange}, such that the full list of sampled radii are symmetric under the patch change and both patches are then sampled equivalently.
The value of $\alpha$ then determines how skewed the distribution is, when $\alpha=\beta=1$ the numerator of \eqref{eq:beta_fn} becomes 1 and the distribution is uniform; for testing samples we take the near uniform limit with $\alpha=1$ and $\beta$ defined as above.
In the limit $\alpha << 1$ the distribution skews to prioritise the bounds of the $[0,1]$ interval, whilst the $\alpha >> 1$ limit prioritises the middle of the interval.
The latter is desired to optimise overlap and avoid numerical instability, hence after some heuristic experimentation a value of $\alpha=4$ was selected for the training samples.

\begin{figure}[!t]
    \centering
    \begin{subfigure}{0.32\textwidth}
        \centering
        \includegraphics[width=0.98\textwidth]{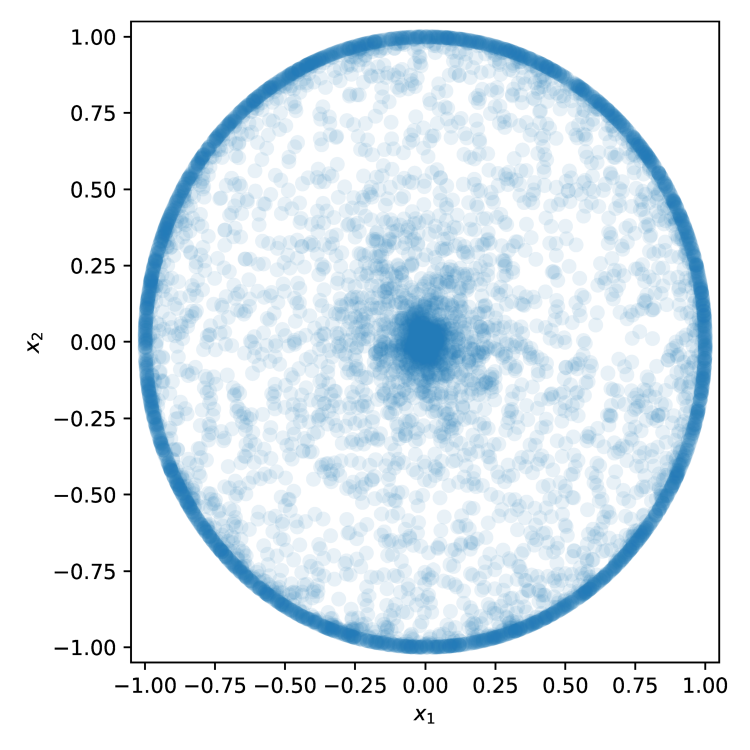}
        \caption{$\alpha=0.1$}
    \end{subfigure} 
    \begin{subfigure}{0.32\textwidth}
        \centering
        \includegraphics[width=0.98\textwidth]{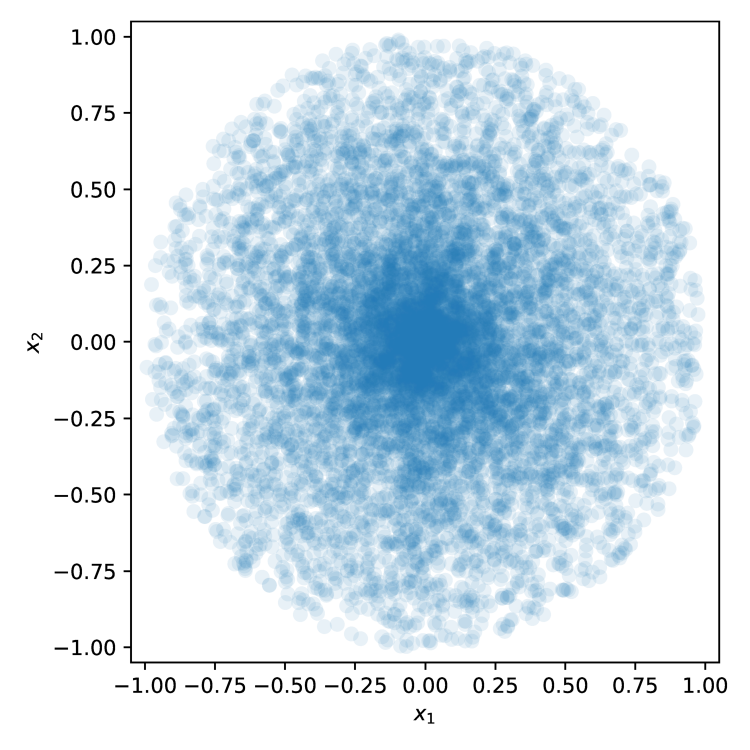}
        \caption{$\alpha=1$}
    \end{subfigure}
    \begin{subfigure}{0.32\textwidth}
        \centering
        \includegraphics[width=0.98\textwidth]{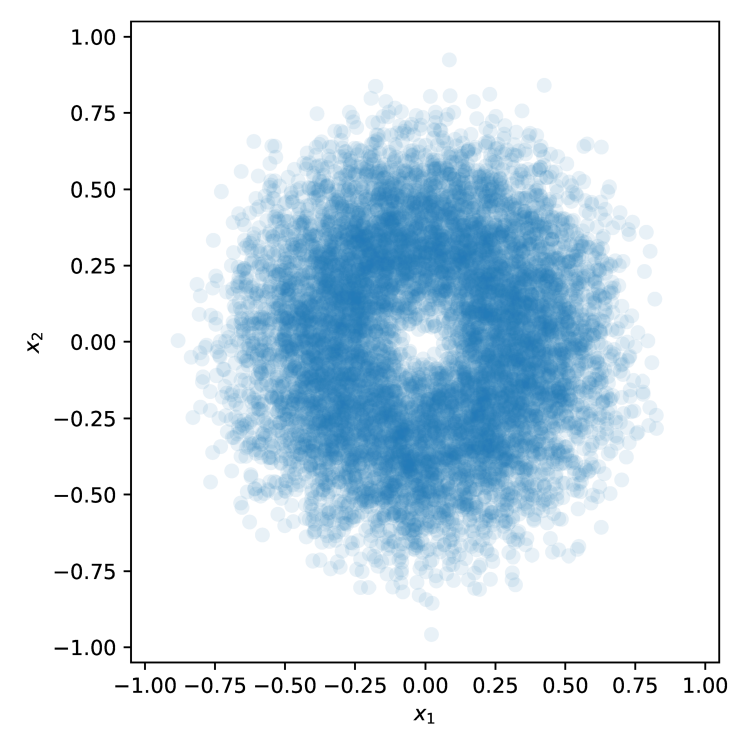}
        \caption{$\alpha=4$}
    \end{subfigure} 
    \caption{Point samples in a 2d ball patch using the modified Beta function sampling scheme. The scheme sets the $\beta$ value to centre sampling at $r_m$, and explicitly symmetrises such that these points in the other patch have the same distribution. Plots show the behaviour for varying $\alpha$.}
    \label{fig:sampling}
\end{figure}

To illustrate how the sampling in a patch varies with $\alpha$, Figure \ref{fig:sampling} shows a single patch sampled with $\alpha \in \{0.1, 1, 4\}$, due to the symmetric nature of the scheme the other patch sampling distribution looks identical.
The sampling code is highly vectorised to ensure hyper-efficient sample generation, and is released with the AInstein codebase. We include a Jupyter \cite{jupyter} notebook with interactive visualisations for varying $\alpha$.
We emphasise that $\alpha=4$ was used for training data, and $\alpha=1$ was used for testing data.

\section{Details of Data Filters}\label{app:filters}
In order to vary priority of sample points in various loss components filters were designed to apply appropriate weightings based on the sample point radii. 
Two filters were designed and used in the final model, as mentioned in Section \ref{subsec:bkg_ml_chap5}, and are detailed here.

\subsection{Radial filter in the Einstein loss}
The radial filter in the Einstein loss is of the form
\begin{align}
    e^{-(\frac{|x|- c_e}{w_e})^{t_e}} \, ,
\end{align}
with parameters $(t_e, c_e, w_e)$, where $t_e$ is even. This is a Gaussian-shaped object, where $t_e$ controls how steep the edges are. Very large $t_e$ yields a very good approximation of the rectangular function. $c_e$ is the centre of the Gaussian, and is set it to be zero for simplicity in this case, since we are not concerned with negative values of the radius. $w_e$ controls the width, which therefore determines what portion of the ball is taken into account for this loss. A plot of this filter, with an illustration of what feature each parameter controls, is shown in Figure \ref{fig:filters_e}. The plot refers exactly to the parameters which were used to collect our results.

\subsection{Radial filter in the overlap loss}
\begin{figure}[!t]
    \centering
    \begin{subfigure}{0.32\textwidth}  
        \centering
        \includegraphics[width=0.98\textwidth]{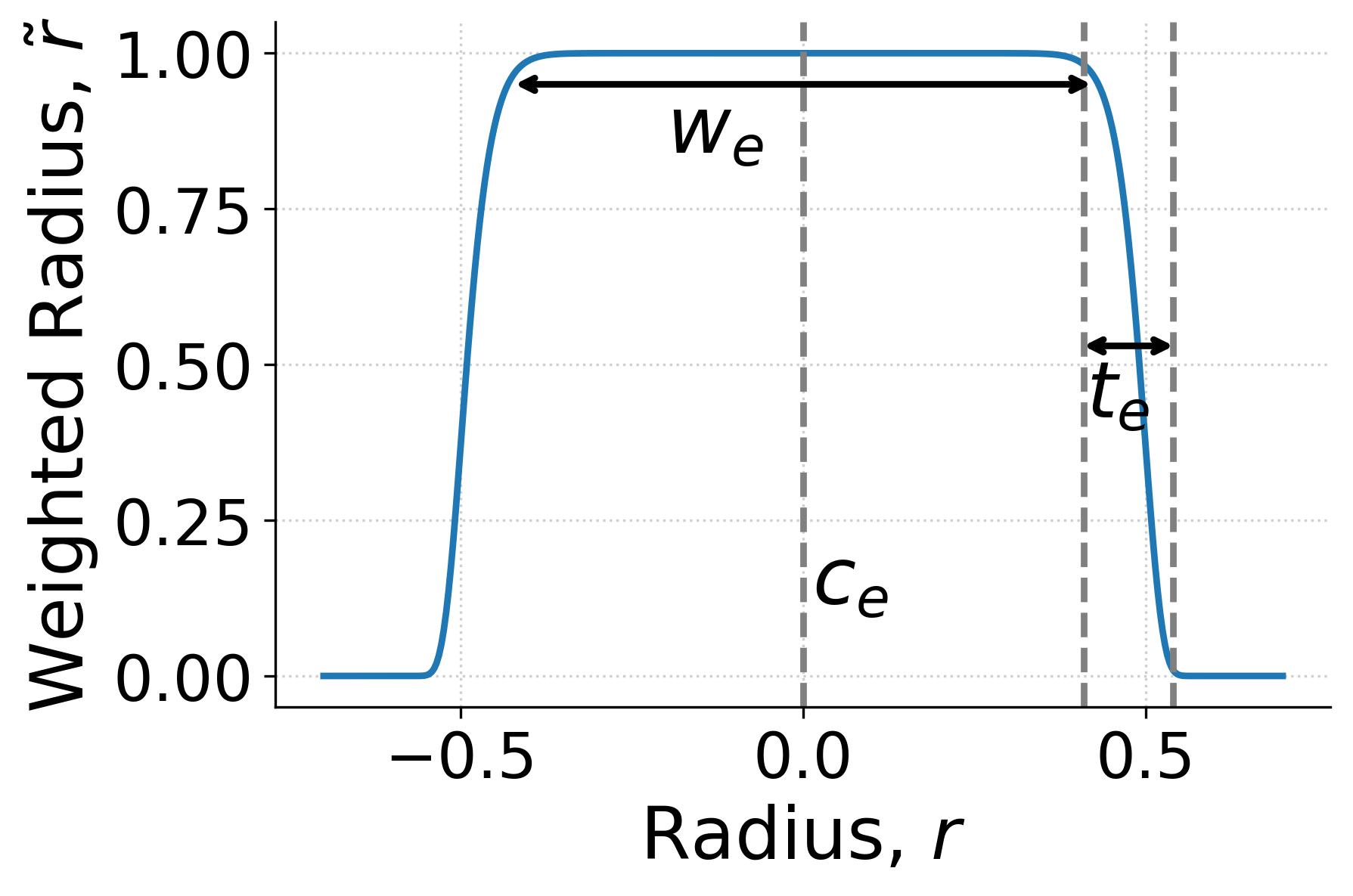}
        \caption{Einstein radial filter}
        \label{fig:filters_e}
    \end{subfigure} 
    \hfill
    \begin{subfigure}{0.32\textwidth}  
        \centering
        \includegraphics[width=0.98\textwidth]{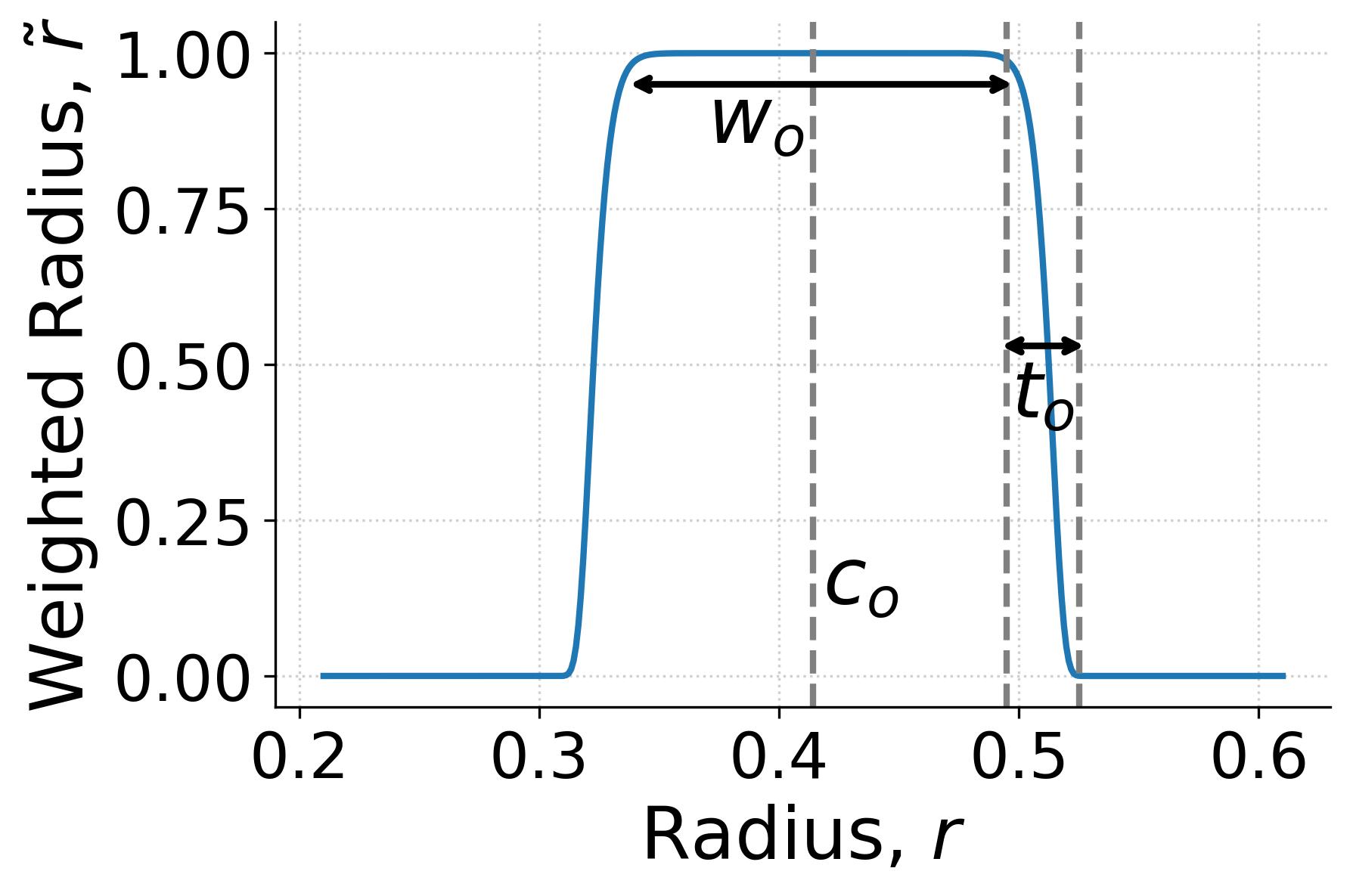}
        \caption{Overlap radial filter} 
        \label{fig:filters_o}
    \end{subfigure} 
    \hfill
    \begin{subfigure}{0.32\textwidth} 
        \centering
        \includegraphics[width=0.98\textwidth]{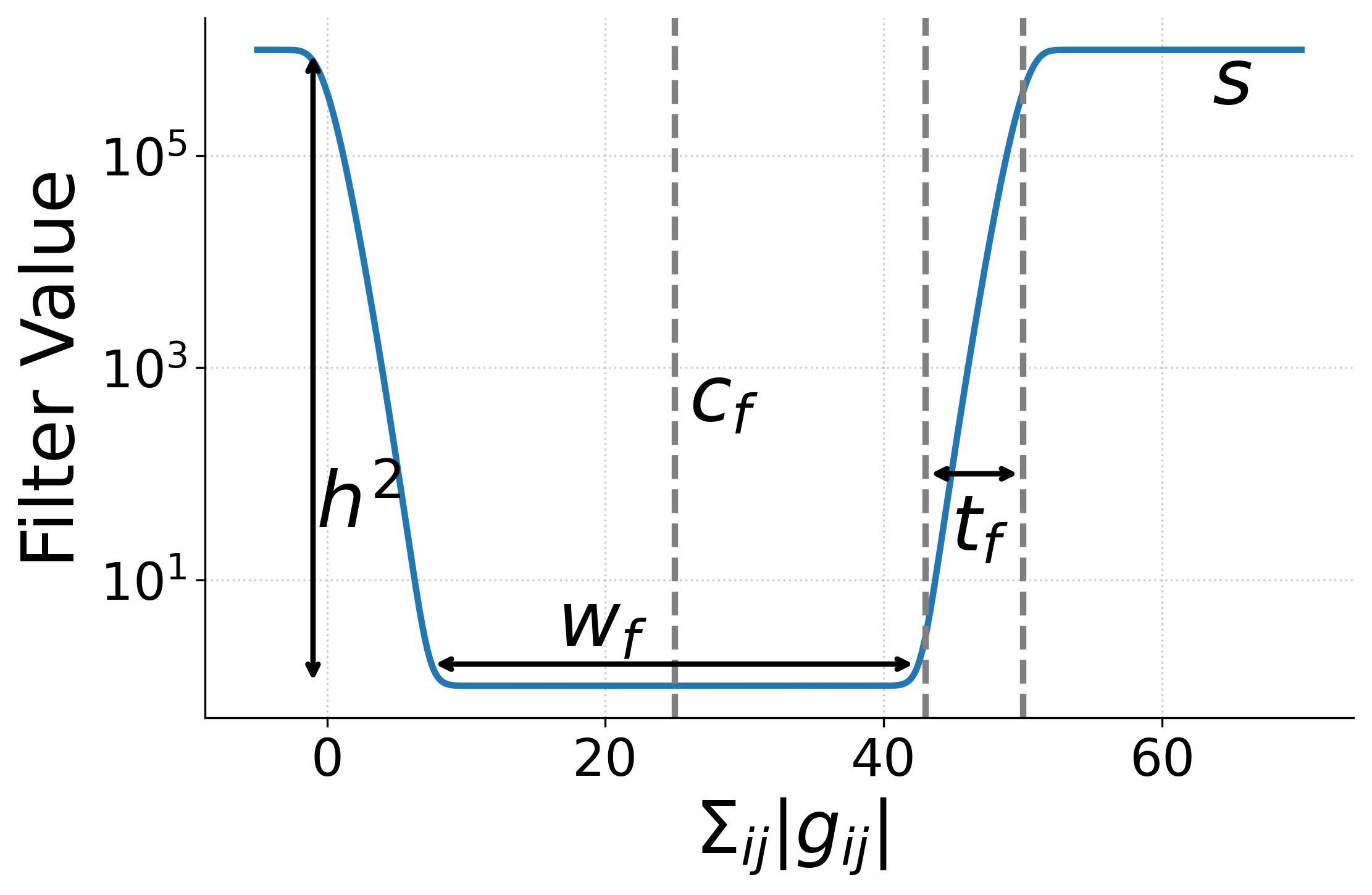}
        \caption{Finiteness filter} 
        \label{fig:filters_f}
    \end{subfigure} 
    \caption{Plots of the filter functions used in the loss components for (a) $\mathcal{L}^{\text{Einstein}}$ and (b)  $\mathcal{L}^{\text{Overlap}}$. As well as the finiteness loss function (c) defining $\mathcal{L}^{\text{Finiteness}}$.}
    \label{fig:filters}
\end{figure}

The radial filter used in the overlap loss has the same form, but it involves different choices of parameters:
\begin{align}
    e^{-(\frac{|x|-c_o}{w_o})^{t_o}} \, ,
\end{align}
now labelled $(t_o, c_o, w_o)$. As before, $c_o$ controls the centre of the Gaussian-like curve, $w_o$ its width and $t_o$ how vertical the walls are. In this case, however, the filter should isolate the overlap region (i.e. the annulus between $\frac{1 - (r_m +  \varepsilon)}{1 + (r_m +  \varepsilon)}$ and $r_m +  \varepsilon$), while setting to zero the other regions of the ball. A plot of the specific filter used in our runs is shown in Figure \ref{fig:filters_o}.

\section{Neural Network Hyperparameters}
\label{app:hyperparams}

\begin{table}[h!]
    \centering
    \begin{tabular}{|c|c|}
        \hline
        \textbf{Hyperparameter} & \textbf{Value} \\ \hline
        Training epochs & 500 \\ \hline
        Training samples & 10k (2D, 3D), 100k (4D, 5D) \\ \hline
        Batch size & 100 \\ \hline
        Learning rate (max,min) & (0.005, 0.001) \\ \hline
        Learning rate schedule & Cosine \\ \hline
        Optimizer & Adam (\cite{kingma2017adammethodstochasticoptimization}) \\ \hline
        Patch submodel layers & 3 Dense layers \\ \hline
        Neurons per layer & 64 \\ \hline
        Activation function & GELU \\ \hline
        Biases & On \\ \hline
    \end{tabular}
    \caption{Hyperparameters for the Einstein metric machine learning model training.}
    \label{tab:hyperparams}
\end{table}

In order to arrive at the set of hyperparameters stated above, we performed an extensive sweep for the 2d model with the experiment management tool Weights and Biases \cite{wandb}. We release the code for this feature with the package, such that it may be readily utilised by those possessing an API key.

In this work, the activation function $\sigma(x)$ is chosen to be the Gaussian Error Linear Unit (GELU)\footnote{Indeed one may choose $\sigma(x) = \text{ReLU}(x)$ here, however the constant behaviour for $x \leq 0$ leads to numerical instability for this use-case; derivatives of the network must be taken to calculate the Ricci tensor.}.

Likewise, for the finiteness loss filter we use the following hyperparameter choices:
\begin{table}[ht]
    \centering
    \begin{tabular}{|c|c|}
        \hline
        \textbf{Filter parameter} & \textbf{Value} \\ \hline
        $h$ & 1000 \\ \hline
        $c_f$ & 25 \\ \hline
        $w_f$ & 25 \\ \hline
        $t_f$ & 20 \\ \hline
        $s$ & 0.2 \\ \hline
    \end{tabular}
    \caption{Parameters for the finiteness loss filter.}
    \label{tab:filters}
\end{table}

\bibliographystyle{ieeetr}
\bibliography{Bibliography}{}

@article{Schleich_1999,
	doi = {10.1088/0264-9381/16/7/319},
  
	url = {https://doi.org/10.1088%2F0264-9381%2F16%2F7%2F319},
  
	year = 1999,
	month = {jan},
  
	publisher = {{IOP} Publishing},
  
	volume = {16},
  
	number = {7},
  
	pages = {2447--2469},
  
	author = {Kristin Schleich and Donald Witt},
  
	title = {Exotic spaces in quantum gravity: I. Euclidean quantum gravity in seven dimensions},
  
	journal = {Classical and Quantum Gravity}
}

@article{CASTELLANI1984429,
title = {A classification of compactifying solutions for d =11 supergravity},
journal = {Nuclear Physics B},
volume = {241},
number = {2},
pages = {429-462},
year = {1984},
issn = {0550-3213},
doi = {https://doi.org/10.1016/0550-3213(84)90055-5},
url = {https://www.sciencedirect.com/science/article/pii/0550321384900555},
author = {L. Castellani and L.J. Romans and N.P. Warner}
}

@article{DUFF198490,
title = "{On the consistency of the Kaluza--Klein ansatz}",
journal = {Physics Letters B},
volume = {149},
number = {1},
pages = {90-94},
year = {1984},
issn = {0370-2693},
doi = {https://doi.org/10.1016/0370-2693(84)91558-2},
url = {https://www.sciencedirect.com/science/article/pii/0370269384915582},
author = {M.J. Duff and B.E.W. Nilsson and C.N. Pope and N.P. Warner}
}

@article{CASTELLANI1984394,
title = {Symmetries of coset spaces and Kaluza-Klein supergravity},
journal = {Annals of Physics},
volume = {157},
number = {2},
pages = {394-407},
year = {1984},
issn = {0003-4916},
doi = {https://doi.org/10.1016/0003-4916(84)90066-6},
url = {https://www.sciencedirect.com/science/article/pii/0003491684900666},
author = {L Castellani and L.J Romans and N.P Warner}
}

@article{10.1063/1.525753,
    author = {Percacci, R. and Randjbar‐Daemi, S.},
    title = "{Kaluza–Klein theories on bundles with homogeneous fibers. I}",
    journal = {Journal of Mathematical Physics},
    volume = {24},
    number = {4},
    pages = {807-814},
    year = {1983},
    month = {04},
    issn = {0022-2488},
    doi = {10.1063/1.525753},
    url = {https://doi.org/10.1063/1.525753},
}

@article{10.2307/1969983,
 ISSN = {0003486X},
 URL = {http://www.jstor.org/stable/1969983},
 author = {John Milnor},
 journal = {Annals of Mathematics},
 number = {2},
 pages = {399--405},
 publisher = {Annals of Mathematics},
 title = {On Manifolds Homeomorphic to the 7-Sphere},
 urldate = {2023-05-12},
 volume = {64},
 year = {1956}
}

@article{FREUND1985263,
title = {Higher-dimensional unification},
journal = {Physica D: Nonlinear Phenomena},
volume = {15},
number = {1},
pages = {263-269},
year = {1985},
issn = {0167-2789},
doi = {https://doi.org/10.1016/0167-2789(85)90170-8},
url = {https://www.sciencedirect.com/science/article/pii/0167278985901708},
author = {Peter G.O. Freund}
}

@article{YAMAGISHI198447,
title = {Supergravity on seven-dimensional homotopy spheres},
journal = {Physics Letters B},
volume = {134},
number = {1},
pages = {47-50},
year = {1984},
issn = {0370-2693},
doi = {https://doi.org/10.1016/0370-2693(84)90981-X},
url = {https://www.sciencedirect.com/science/article/pii/037026938490981X},
author = {Kengo Yamagishi}
}

@article{Witten:1985xe,
    author = "Witten, Edward",
    editor = "Salam, A. and Sezgin, E.",
    title = "{Global gravitational anomalies}",
    reportNumber = "PRINT-85-0246 (PRINCETON)",
    doi = "10.1007/BF01212448",
    journal = "Commun. Math. Phys.",
    volume = "100",
    pages = "197",
    year = "1985"
}

@article{10.1063/1.529078,
    author = {Baadhio, Randy A. and Lee, Pascal},
    title = "{On the global gravitational instanton and soliton that are homotopy spheres}",
    journal = {Journal of Mathematical Physics},
    volume = {32},
    number = {10},
    pages = {2869-2874},
    year = {1991},
    month = {10},
    issn = {0022-2488},
    doi = {10.1063/1.529078},
    url = {https://doi.org/10.1063/1.529078},
}

@article{Brans:1992mj,
    author = "Brans, Carl H. and Randall, Duane",
    title = "{Exotic differentiable structures and general relativity}",
    eprint = "gr-qc/9212003",
    archivePrefix = "arXiv",
    reportNumber = "LOYOLA-B9202",
    doi = "10.1007/BF00758828",
    journal = "Gen. Rel. Grav.",
    volume = "25",
    pages = "205",
    year = "1993"
}

@book{book,
author = {Asselmeyer-Maluga, Torsten and Brans, Carl},
year = {2007},
month = {01},
pages = {},
title = {Exotic Smoothness and Physics},
isbn = {978-981-02-4195-7},
doi = {10.1142/4323}
}

@Book{zbMATH03194988,
 Author = {Kobayashi, Shoshichi and Nomizu, Katsumi},
 Title = {Foundations of differential geometry. {I}},
 FSeries = {Interscience Tracts in Pure and Applied Mathematics},
 Series = {Intersci. Tracts Pure Appl. Math.},
 Volume = {15},
 Year = {1963},
 Publisher = {Interscience Publishers, New York, NY},
 Language = {English},
 zbMATH = {3194988},
 Zbl = {0119.37502}
}

@article{DUFF19861,
title = {Kaluza-Klein supergravity},
journal = {Physics Reports},
volume = {130},
number = {1},
pages = {1-142},
year = {1986},
issn = {0370-1573},
doi = {https://doi.org/10.1016/0370-1573(86)90163-8},
url = {https://www.sciencedirect.com/science/article/pii/0370157386901638},
author = {M.J. Duff and B.E.W. Nilsson and C.N. Pope}
}

@misc{boyer2004einstein,
      title={Einstein Metrics on Spheres}, 
      author={Charles P. Boyer and Krzysztof Galicki and János Kollár},
      year={2004},
      eprint={math/0309408},
      archivePrefix={arXiv},
      primaryClass={math.DG}
}

@misc{boyer2003einstein,
      title={Einstein Metrics on Exotic Spheres in Dimensions 7, 11, and 15}, 
      author={Charles P. Boyer and Krzysztof Galicki and János Kollár and Evan Thomas},
      year={2003},
      eprint={math/0311293},
      archivePrefix={arXiv},
      primaryClass={math.DG}
}

@misc{acharya1999branes,
      title={Branes at conical singularities and holography}, 
      author={BS Acharya and JM Figueroa-O'Farrill and CM Hull and B Spence},
      year={1999},
      eprint={hep-th/9808014},
      archivePrefix={arXiv},
      primaryClass={hep-th}
}

@article{Fr__1999,
	doi = {10.1016/s0370-2693(99)01296-4},
  
	url = {https://doi.org/10.1016%2Fs0370-2693%2899%2901296-4},
  
	year = 1999,
	month = {dec},
  
	publisher = {Elsevier {BV}
},
  
	volume = {471},
  
	number = {1},
  
	pages = {27--38},
  
	author = {Pietro Fr{\'{e}} and Leonardo Gualtieri and Piet Termonia},
  
	title = {The structure of multiplets in {AdS}4 and the complete Osp(3$\vert$4){\texttimes}{SU}(3) spectrum of M-theory on {AdS}4{\texttimes}N0,1,0},
  
	journal = {Physics Letters B}
}

@article{10.2307/1971078,
 ISSN = {0003486X},
 URL = {http://www.jstor.org/stable/1971078},
 author = {Detlef Gromoll and Wolfgang Meyer},
 journal = {Annals of Mathematics},
 number = {2},
 pages = {401--406},
 publisher = {Annals of Mathematics},
 title = {An Exotic Sphere With Nonnegative Sectional Curvature},
 urldate = {2023-05-14},
 volume = {100},
 year = {1974}
}

@article{10.2307/1999745,
 ISSN = {00029947},
 URL = {http://www.jstor.org/stable/1999745},
 author = {Andrzej Derdzinski and A. Rigas},
 journal = {Transactions of the American Mathematical Society},
 number = {2},
 pages = {485--493},
 publisher = {American Mathematical Society},
 title = {Unflat Connections in 3-Sphere Bundles Over S4},
 urldate = {2023-05-14},
 volume = {265},
 year = {1981}
}

@article{Duran2001,
	affiliation = {IVIC-Matemáticas},
	author = {Durán, Carlos E.},
	copyright = {Kluwer Academic Publishers},
	doi = {10.1023/A:1013163427655},
	journal = {Geometriae Dedicata},
	language = {English},
	number = {1-3},
	pages = {199-210},
	title = {Pointed Wiedersehen Metrics on Exotic Spheres and Diffeomorphisms of S6},
	volume = {88},
	year = {2001},
}

@article{DURAN2009206,
title = {Equivariant homotopy and deformations of diffeomorphisms},
journal = {Differential Geometry and its Applications},
volume = {27},
number = {2},
pages = {206-211},
year = {2009},
issn = {0926-2245},
doi = {https://doi.org/10.1016/j.difgeo.2008.06.018},
url = {https://www.sciencedirect.com/science/article/pii/S0926224508000752},
author = {C. Durán and A. Rigas}
}

@article{Rigas1978,
author = {Rigas, A.},
journal = {Mathematische Annalen},
pages = {187-194},
title = {Some Bundles of Non-Negative Curvature.},
url = {http://eudml.org/doc/163081},
volume = {232},
year = {1978},
}

@article{BOUWKNEGT201546,
title = {Spherical T-duality II: An infinity of spherical T-duals for non-principal SU(2)-bundles},
journal = {Journal of Geometry and Physics},
volume = {92},
pages = {46-54},
year = {2015},
issn = {0393-0440},
doi = {https://doi.org/10.1016/j.geomphys.2015.02.003},
url = {https://www.sciencedirect.com/science/article/pii/S0393044015000315},
author = {Peter Bouwknegt and Jarah Evslin and Varghese Mathai}
}

@article{10.2307/40067878,
 ISSN = {00029327, 10806377},
 URL = {http://www.jstor.org/stable/40067878},
 author = {Sebastian Goette and Nitu Kitchloo and Krishnan Shankar},
 journal = {American Journal of Mathematics},
 number = {2},
 pages = {395--416},
 publisher = {Johns Hopkins University Press},
 title = {Diffeomorphism Type of the Berger Space SO(5)/SO(3)},
 urldate = {2023-05-15},
 volume = {126},
 year = {2004}
}

@article{POPE1985352,
title = {An SU(4) invariant compactification of d = 11 supergravity on a stretched seven-sphere},
journal = {Physics Letters B},
volume = {150},
number = {5},
pages = {352-356},
year = {1985},
issn = {0370-2693},
doi = {https://doi.org/10.1016/0370-2693(85)90992-X},
url = {https://www.sciencedirect.com/science/article/pii/037026938590992X},
author = {C.N. Pope and N.P. Warner}
}

@article{EGUCHI197982,
title = {Self-dual solutions to euclidean gravity},
journal = {Annals of Physics},
volume = {120},
number = {1},
pages = {82-106},
year = {1979},
issn = {0003-4916},
doi = {https://doi.org/10.1016/0003-4916(79)90282-3},
url = {https://www.sciencedirect.com/science/article/pii/0003491679902823},
author = {Tohru Eguchi and Andrew J Hanson}
}

@article{10.1063/1.522434,
    author = {Cho, Y. M.},
    title = "{Higher‐dimensional unifications of gravitation and gauge theories}",
    journal = {Journal of Mathematical Physics},
    volume = {16},
    number = {10},
    pages = {2029-2035},
    year = {1975},
    month = {09},
    issn = {0022-2488},
    doi = {10.1063/1.522434},
    url = {https://doi.org/10.1063/1.522434},
    eprint = {https://pubs.aip.org/aip/jmp/article-pdf/16/10/2029/8147447/2029\_1\_online.pdf},
}

@article{PhysRevD.13.235,
  title = {Geometrical approach to local gauge and supergauge invariance: Local gauge theories and supersymmetric strings},
  author = {Chang, Lay Nam and Macrae, Kenneth I. and Mansouri, Freydoon},
  journal = {Phys. Rev. D},
  volume = {13},
  issue = {2},
  pages = {235--249},
  numpages = {0},
  year = {1976},
  month = {Jan},
  publisher = {American Physical Society},
  doi = {10.1103/PhysRevD.13.235},
  url = {https://link.aps.org/doi/10.1103/PhysRevD.13.235}
}

@article{PhysRevD.15.1642,
  title = {Conformal properties of pseudoparticle configurations},
  author = {Jackiw, R. and Nohl, C. and Rebbi, C.},
  journal = {Phys. Rev. D},
  volume = {15},
  issue = {6},
  pages = {1642--1646},
  numpages = {0},
  year = {1977},
  month = {Mar},
  publisher = {American Physical Society},
  doi = {10.1103/PhysRevD.15.1642},
  url = {https://link.aps.org/doi/10.1103/PhysRevD.15.1642}
}

@article{RevModPhys.51.461,
  title = {Classical solutions of $\mathrm{SU}(2)$ Yang---Mills theories},
  author = {Actor, Alfred},
  journal = {Rev. Mod. Phys.},
  volume = {51},
  issue = {3},
  pages = {461--525},
  numpages = {0},
  year = {1979},
  month = {Jul},
  publisher = {American Physical Society},
  doi = {10.1103/RevModPhys.51.461},
  url = {https://link.aps.org/doi/10.1103/RevModPhys.51.461}
}

@article{Asselmeyer-Maluga:2017tbn,
    author = "Asselmeyer-Maluga, Torsten and Kr\'ol, Jerzy",
    title = "{How to obtain a cosmological constant from small exotic $R^4$}",
    eprint = "1709.03314",
    archivePrefix = "arXiv",
    primaryClass = "gr-qc",
    doi = "10.1016/j.dark.2017.12.002",
    journal = "Phys. Dark Univ.",
    volume = "19",
    pages = "66--77",
    year = "2018"
}

@article{nuimeprn10073,
          volume = {45},
          number = {4},
          author = {M. Joachim and David Wraith},
           title = {Exotic Spheres and Curvature},
       publisher = {American Mathematical Society},
         journal = {Bulletin of the American Mathematical Society},
           pages = {595--616},
            year = {2008},
             url = {https://mural.maynoothuniversity.ie/10073/}
}

@article{fivefolds,
author = {Alawadhi, Rashid and Angella, Daniele and Leonardo, Andrea and Schettini Gherardini, Tancredi},
title = {Constructing and Machine Learning Calabi-Yau Five-Folds},
journal = {Fortschritte der Physik},
volume = {72},
number = {2},
pages = {2300262},
doi = {https://doi.org/10.1002/prop.202300262},
url = {https://onlinelibrary.wiley.com/doi/abs/10.1002/prop.202300262},
eprint = {https://onlinelibrary.wiley.com/doi/pdf/10.1002/prop.202300262},
year = {2024}
}

@article{Gherardini:2023uyx,
    author = "Gherardini, T. Schettini",
    title = "{Exotic spheres\textquoteright{} metrics and solutions via Kaluza--Klein techniques}",
    eprint = "2309.01703",
    archivePrefix = "arXiv",
    primaryClass = "hep-th",
    doi = "10.1007/JHEP12(2023)100",
    journal = "JHEP",
    volume = "12",
    pages = "100",
    year = "2023"
}

@article{Econditions,
    author = "Allison, D.",
    title = "{Energy conditions in standard static spacetimes}",
    doi = "10.1007/BF00759321",
    journal = "General Relativity and Gravitation",
    volume = "20",
    year = "1988"
}

@article{Coquereaux:1983kj,
    author = "Coquereaux, Robert",
    title = "{Comments about Riemannian geometry, Einstein spaces, Kaluza-Klein, 11-dimensional supergravity, and all that}",
    reportNumber = "CERN-TH-3639",
    month = "6",
    year = "1983"
}

@book{Nakahara:2003nw,
    author = "Nakahara, M.",
    title = "{Geometry, topology and physics}",
    year = "2003"
}

@inproceedings{McEnroe2016MILNORSCO,
  title={MILNOR’S CONSTRUCTION OF EXOTIC 7-SPHERES},
  author={Rachel Maxine McEnroe},
  year={2016},
  url={https://api.semanticscholar.org/CorpusID:30229066}
}

@inproceedings{Bognat2018MILNORSES,
  title={MILNOR’S EXOTIC SPHERES},
  author={Adam Bognat},
  year={2018},
  url={https://api.semanticscholar.org/CorpusID:174784366}
}

@misc{Exot_world,
      title={The Exotic World of Milnor’s Spheres}, 
      author={J. Sampietro and C. Segovia},
      year={2023}
}

@misc{vandoren2008lectures,
      title={Lectures on instantons}, 
      author={Stefan Vandoren and Peter van Nieuwenhuizen},
      year={2008},
      eprint={0802.1862},
      archivePrefix={arXiv},
      primaryClass={hep-th}
}

@article{Giambiagi:1977yg,
    author = "Giambiagi, J. J. and Rothe, K. D.",
    title = "{Regular $N$-instanton fields and singular gauge transformations}",
    reportNumber = "CERN-TH-2325",
    doi = "10.1016/0550-3213(77)90022-0",
    journal = "Nucl. Phys. B",
    volume = "129",
    pages = "111--124",
    year = "1977"
}

@inbook{Curiel_2017,
   title={A Primer on Energy Conditions},
   ISBN={9781493932108},
   ISSN={2381-5841},
   url={http://dx.doi.org/10.1007/978-1-4939-3210-8_3},
   DOI={10.1007/978-1-4939-3210-8_3},
   booktitle={Einstein Studies},
   publisher={Springer New York},
   author={Curiel, Erik},
   year={2017},
   pages={43–104} }

@article{Vandoren:2008xg,
    author = "Vandoren, Stefan and van Nieuwenhuizen, Peter",
    title = "{Lectures on instantons}",
    eprint = "0802.1862",
    archivePrefix = "arXiv",
    primaryClass = "hep-th",
    reportNumber = "ITP-UU-08-05, SPIN-08-05",
    month = "2",
    year = "2008"
}

@article{Martin-Moruno:2017exc,
    author = "Martin-Moruno, Prado and Visser, Matt",
    title = "{Classical and semi-classical energy conditions}",
    eprint = "1702.05915",
    archivePrefix = "arXiv",
    primaryClass = "gr-qc",
    doi = "10.1007/978-3-319-55182-1_9",
    journal = "Fundam. Theor. Phys.",
    volume = "189",
    pages = "193--213",
    year = "2017"
}

@article{ATIYAH1978185,
title = {Construction of instantons},
journal = {Physics Letters A},
volume = {65},
number = {3},
pages = {185-187},
year = {1978},
issn = {0375-9601},
doi = {https://doi.org/10.1016/0375-9601(78)90141-X},
url = {https://www.sciencedirect.com/science/article/pii/037596017890141X},
author = {M.F. Atiyah and N.J. Hitchin and V.G. Drinfeld and Yu.I. Manin}
}

@article{Salam:1981xd,
    author = "Salam, Abdus and Strathdee, J. A.",
    title = "{On Kaluza--Klein theory}",
    reportNumber = "IC-81-211",
    doi = "10.1016/0003-4916(82)90291-3",
    journal = "Annals Phys.",
    volume = "141",
    pages = "316--352",
    year = "1982"
}

@article{BELAVIN197585,
title = "{Pseudoparticle solutions of the Yang--Mills equations}",
journal = {Physics Letters B},
volume = {59},
number = {1},
pages = {85-87},
year = {1975},
issn = {0370-2693},
doi = {https://doi.org/10.1016/0370-2693(75)90163-X},
url = {https://www.sciencedirect.com/science/article/pii/037026937590163X},
author = {A.A. Belavin and A.M. Polyakov and A.S. Schwartz and Yu.S. Tyupkin}
}

@article{tHooft:1976snw,
    author = "'t Hooft, Gerard",
    editor = "Shifman, Mikhail A.",
    title = "{Computation of the quantum effects due to a four-dimensional pseudoparticle}",
    reportNumber = "PRINT-76-0551 (HARVARD)",
    doi = "10.1103/PhysRevD.14.3432",
    journal = "Phys. Rev. D",
    volume = "14",
    pages = "3432--3450",
    year = "1976",
    note = "[Erratum: Phys.Rev.D 18, 2199 (1978)]"
}

@article{Asselmeyer:1996bh,
    author = "Asselmeyer, T.",
    title = "{Generation of source terms in general relativity by differential structures}",
    eprint = "gr-qc/9610009",
    archivePrefix = "arXiv",
    doi = "10.1088/0264-9381/14/3/016",
    journal = "Class. Quant. Grav.",
    volume = "14",
    pages = "749--758",
    year = "1997"
}

@article{Anderson:2020hux,
archiveprefix = {arXiv}, journal = {arXiv},
author = {Anderson, Lara B. and Gerdes, Mathis and Gray, James and Krippendorf, Sven and Raghuram, Nikhil and Ruehle, Fabian},
doi = {10.1007/s00220-021-04185-2},
eprint = {2012.04656},
journal = {Commun. Math. Phys.},
number = {1},
pages = {585--634},
primaryclass = {hep-th},
title = {{Moduli-dependent Calabi-Yau and SU(3)-structure metrics from Machine Learning}},
url = {https://arxiv.org/abs/2012.04656},
volume = {386},
year = {2021}
}

@manual{apocrita,
author = {King, Thomas and
          Butcher, Simon and
          Zalewski, Lukasz},
doi = {10.5281/zenodo.438045},
month = {March},
title = {{Apocrita - High Performance Computing Cluster for Queen Mary University of London}},
url = {https://doi.org/10.5281/zenodo.438045},
year = {2017}
}

@article{Ashmore:2019wzb,
archiveprefix = {arXiv}, journal = {arXiv},
author = {Ashmore, Anthony and He, Yang-Hui},
doi = {10.1002/prop.202000068},
eprint = {1910.08605},
journal = {Fortsch. Phys.},
number = {9},
pages = {2000068},
primaryclass = {hep-th},
title = {{Machine Learning Calabi-Yau Metrics}},
url = {https://arxiv.org/abs/1910.08605},
volume = {68},
year = {2020}
}

@article{Ashmore:2021ohf,
archiveprefix = {arXiv}, journal = {arXiv},
author = {Ashmore, Anthony and Calmon, Lucille and He, Yang-Hui and Ovrut, Burt A.},
doi = {10.1142/S2810939222500034},
eprint = {2112.10872},
journal = {International Journal of Data Science in the Mathematical Sciences},
number = {1},
pages = {49--61},
primaryclass = {hep-th},
reportnumber = {LIMS-2021-018},
title = {{Calabi-Yau Metrics, Energy Functionals and Machine-Learning}},
volume = {1},
year = {2023}
}

@book{Berger_2003,
author = {Berger, Marcel},
isbn = {3 540 65317 1},
publisher = {Springer},
title = {A panoramic view of Riemannian geometry},
year = {2003}
}

@article{Berglund:2022gvm,
archiveprefix = {arXiv}, journal = {arXiv},
author = {Berglund, Per and Butbaia, Giorgi and H\"uubsch, Tristan and Jejjala, Vishnu and Mayorga Pe\~na, Dami\'an and Mishra, Challenger and Tan, Justin},
doi = {10.4310/ATMP.2023.v27.n4.a3},
eprint = {2211.09801},
journal = {Adv. Theor. Math. Phys.},
number = {4},
pages = {1107--1158},
primaryclass = {hep-th},
title = {{Machine-learned Calabi\textendash{}Yau metrics and curvature}},
volume = {27},
year = {2023}
}

@article{berman2024curvatureexotic7sphere,
archiveprefix = {arXiv}, journal = {arXiv},
author = {David S. Berman and Martin Cederwall and Tancredi Schettini Gherardini},
eprint = {2410.01909},
journal = {arXiv:2410.01909},
primaryclass = {hep-th},
title = {Curvature of an exotic 7-sphere},
url = {https://arxiv.org/abs/2410.01909},
year = {2024}
}

@book{Besse:1987pua,
address = {Berlin, Heidelberg, New York},
author = {Besse, Arthur L.},
isbn = {978-0-387-15279-0},
publisher = {Springer-Verlag},
title = {{Einstein Manifolds}},
year = {1987}
}

@article{Butbaia:2024xgj,
archiveprefix = {arXiv}, journal = {arXiv},
author = {Butbaia, Giorgi and Mayorga Pe\~na, Dami\'an and Tan, Justin and Berglund, Per and H\"ubsch, Tristan and Jejjala, Vishnu and Mishra, Challenger},
eprint = {2410.19728},
month = {10},
primaryclass = {hep-th},
title = {{cymyc -- Calabi-Yau Metrics, Yukawas, and Curvature}},
year = {2024}
}

@article{Chaurasia:2025,
author = {Chaurasia, Siddhant and others},
doi = {10.1103/PhysRevD.111.044024},
journal = {Phys. Rev. D},
number = {4},
pages = {044024},
title = {{Numerical-relativity surrogate model for hyperbolic encounters of binary black holes}},
url = {https://journals.aps.org/prd/abstract/10.1103/PhysRevD.111.044024},
volume = {111},
year = {2025}
}

@article{chen2024,
archiveprefix = {arXiv}, journal = {arXiv},
author = {Xun Chen and Mei Huang},
eprint = {2401.06417},
primaryclass = {hep-ph},
title = {Machine learning holographic black hole from lattice QCD equation of state},
url = {https://arxiv.org/abs/2401.06417},
year = {2024}
}

@article{Chesler:2013lia,
author = {Chesler, Paul M. and Yaffe, Laurence G.},
eprint = {1309.1439},
journal = {JHEP},
pages = {086},
title = {{Numerical solution of gravitational dynamics in asymptotically anti-de Sitter spacetimes}},
volume = {07},
year = {2014}
}

@article{Clough_2015,
archiveprefix = {arXiv}, journal = {arXiv},
author = {Clough, Katy and Figueras, Pau and Finkel, Hal and Kunesch, Markus and Lim, Eugene A and Tunyasuvunakool, Saran},
doi = {10.1088/0264-9381/32/24/245011},
eprint = {1503.03436},
issn = {1361-6382},
journal = {Classical and Quantum Gravity},
month = {December},
number = {24},
pages = {245011},
primaryclass = {gr-qc},
publisher = {IOP Publishing},
title = {GRChombo: Numerical relativity with adaptive mesh refinement},
url = {http://dx.doi.org/10.1088/0264-9381/32/24/245011},
volume = {32},
year = {2015}
}

@article{deluca2024,
archiveprefix = {arXiv}, journal = {arXiv},
author = {De Luca, G. Bruno},
eprint = {2501.00093},
primaryclass = {hep-th},
title = {Machine Learning Gravity Compactifications on Negatively Curved Manifolds},
url = {https://arxiv.org/abs/2501.00093},
year = {2024}
}

@article{Dias:2015pda,
author = {Dias, Oscar J.C. and Santos, Jorge E. and Way, Benson},
eprint = {1510.02804},
journal = {Class. Quant. Grav.},
pages = {133001},
title = {{Numerical Methods for Finding Stationary Gravitational Solutions}},
volume = {33},
year = {2016}
}

@article{Dias:2015rxy,
archiveprefix = {arXiv}, journal = {arXiv},
author = {Dias, Oscar J. C. and Horowitz, Gary T. and Santos, Jorge E.},
doi = {10.1088/0264-9381/32/14/145003},
eprint = {1501.06574},
journal = {Class. Quant. Grav.},
pages = {145003},
primaryclass = {hep-th},
title = {Black Resonators and the Instability of Anti-de Sitter Space},
volume = {32},
year = {2015}
}

@inproceedings{Douglas:2020hpv,
archiveprefix = {arXiv}, journal = {arXiv},
author = {Douglas, Michael R. and Lakshminarasimhan, Subramanian and Qi, Yidi},
booktitle = {Proceedings of the 2nd Mathematical and Scientific Machine Learning Conference},
eprint = {2009.14786},
pages = {223--252},
primaryclass = {hep-th},
series = {Proceedings of Machine Learning Research},
title = {{Numerical Calabi-Yau metrics from holomorphic networks}},
url = {https://proceedings.mlr.press/v145/douglas22a.html},
volume = {145},
year = {2022}
}

@article{Douglas:2024pmn,
archiveprefix = {arXiv}, journal = {arXiv},
author = {Douglas, Michael R. and Platt, Daniel and Qi, Yidi},
eprint = {2405.19402},
month = {5},
primaryclass = {math.DG},
title = {{Harmonic $1$-forms on real loci of Calabi-Yau manifolds}},
year = {2024}
}

@article{Ek:2024fgd,
archiveprefix = {arXiv}, journal = {arXiv},
author = {Ek, Carl Henrik and Kim, Oisin and Mishra, Challenger},
eprint = {2410.11284},
month = {10},
primaryclass = {hep-th},
title = {{Calabi-Yau metrics through Grassmannian learning and Donaldson's algorithm}},
year = {2024}
}

@article{Figueras:2012xj,
author = {Figueras, Pau and Lucietti, James and Wiseman, Toby},
eprint = {1104.4489},
journal = {Class. Quant. Grav.},
pages = {215018},
title = {{Ricci solitons, Ricci flow, and strongly coupled CFTs}},
volume = {28},
year = {2011}
}

@article{Gerdes:2022nzr,
archiveprefix = {arXiv}, journal = {arXiv},
author = {Gerdes, Mathis and Krippendorf, Sven},
doi = {10.1088/2632-2153/acdc84},
eprint = {2211.12520},
journal = {Mach. Learn. Sci. Tech.},
number = {2},
pages = {025031},
primaryclass = {hep-th},
title = {{CYJAX: A package for Calabi-Yau metrics with JAX}},
volume = {4},
year = {2023}
}

@article{Headrick:2009pv,
author = {Headrick, Matthew and Kitchen, Timothy and Wiseman, Toby},
eprint = {0905.1822},
journal = {Class. Quant. Grav.},
pages = {035002},
title = {{A New approach to static numerical relativity, and its application to Kaluza-Klein black holes}},
volume = {27},
year = {2010}
}

@article{Hendi:2024yin,
archiveprefix = {arXiv}, journal = {arXiv},
author = {Hendi, Yacoub and Larfors, Magdalena and Walden, Moritz},
eprint = {2407.06914},
month = {7},
primaryclass = {hep-th},
reportnumber = {UUITP-21/24},
title = {{Learning Group Invariant Calabi-Yau Metrics by Fundamental Domain Projections}},
year = {2024}
}

@article{Jejjala:2020wcc,
archiveprefix = {arXiv}, journal = {arXiv},
author = {Jejjala, Vishnu and Mayorga Pena, Damian Kaloni and Mishra, Challenger},
doi = {10.1007/JHEP08(2022)105},
eprint = {2012.15821},
journal = {JHEP},
pages = {105},
primaryclass = {hep-th},
title = {{Neural network approximations for Calabi-Yau metrics}},
volume = {08},
year = {2022}
}

@inproceedings{jupyter,
booktitle = {Positioning and Power in Academic Publishing: Players, Agents and Agendas},
editor = {Fernando Loizides and Birgit Scmidt},
title = {Jupyter Notebooks - a publishing format for reproducible computational workflows},
author = {Thomas Kluyver and Benjamin Ragan-Kelley and Fernando P{\'e}rez and Brian Granger and Matthias Bussonnier and Jonathan Frederic and Kyle Kelley and Jessica Hamrick and Jason Grout and Sylvain Corlay and Paul Ivanov and Dami{\'a}n Avila and Safia Abdalla and Carol Willing and  Jupyter development team},
publisher = {IOS Press},
address = {Netherlands},
year = {2016},
pages = {87--90},
url = {https://eprints.soton.ac.uk/403913/}
}

@article{kingma2017adammethodstochasticoptimization,
archiveprefix = {arXiv}, journal = {arXiv},
author = {Diederik P. Kingma and Jimmy Ba},
eprint = {1412.6980},
primaryclass = {cs.LG},
title = {Adam: A Method for Stochastic Optimization},
url = {https://arxiv.org/abs/1412.6980},
year = {2017}
}

@article{Kudoh:2003ki,
author = {Kudoh, Hideaki and Wiseman, Toby and Gualtieri, Leonardo},
eprint = {hep-th/0311225},
journal = {Phys. Rev. D},
pages = {104014},
title = {{Numerical linear perturbations of static black holes in higher dimensions}},
volume = {69},
year = {2004}
}

@article{Larfors:2021pbb,
archiveprefix = {arXiv}, journal = {arXiv},
author = {Larfors, Magdalena and Lukas, Andre and Ruehle, Fabian and Schneider, Robin},
eprint = {2111.01436},
month = {11},
primaryclass = {hep-th},
reportnumber = {UUITP-53/21},
title = {{Learning Size and Shape of Calabi-Yau Spaces}},
year = {2021}
}

@article{Larfors:2022nep,
archiveprefix = {arXiv}, journal = {arXiv},
author = {Larfors, Magdalena and Lukas, Andre and Ruehle, Fabian and Schneider, Robin},
doi = {10.1088/2632-2153/ac8e4e},
eprint = {2205.13408},
journal = {Mach. Learn. Sci. Tech.},
number = {3},
pages = {035014},
primaryclass = {hep-th},
reportnumber = {UUITP-25/22},
title = {{Numerical metrics for complete intersection and Kreuzer\textendash{}Skarke Calabi\textendash{}Yau manifolds}},
volume = {3},
year = {2022}
}

@article{Lehner:2010pn,
author = {Lehner, Luis and Pretorius, Frans},
eprint = {1006.5960},
journal = {Phys. Rev. Lett.},
pages = {101102},
title = {{Black Strings, Low Viscosity Fluids, and Violation of Cosmic Censorship}},
volume = {105},
year = {2010}
}

@article{Li:2023,
archiveprefix = {arXiv}, journal = {arXiv},
author = {Li, Zhi-Han and Li, Chen-Qi and Pang, Long-Gang},
eprint = {2309.07397},
journal = {arXiv preprint},
primaryclass = {gr-qc},
title = {{Solving Einstein Equations Using Deep Learning}},
url = {https://arxiv.org/abs/2309.07397},
year = {2023}
}

@article{Mirjanic:2024gek,
archiveprefix = {arXiv}, journal = {arXiv},
author = {Mirjani\'c, Viktor and Mishra, Challenger},
eprint = {2412.19778},
month = {12},
primaryclass = {hep-th},
title = {{Symbolic Approximations to Ricci-flat Metrics Via Extrinsic Symmetries of Calabi-Yau Hypersurfaces}},
year = {2024}
}

@article{Pretorius:2005gq,
archiveprefix = {arXiv}, journal = {arXiv},
author = {Pretorius, Frans},
doi = {10.1103/PhysRevLett.95.121101},
eprint = {gr-qc/0507014},
journal = {Phys. Rev. Lett.},
pages = {121101},
primaryclass = {gr-qc},
title = {Evolution of Binary Black Hole Spacetimes},
volume = {95},
year = {2005}
}

@misc{tensorflow2015whitepaper,
author = {
              Mart\'{i}n~Abadi and
                     Ashish~Agarwal and
                     Paul~Barham and
                     Eugene~Brevdo and
                     Zhifeng~Chen and
                     Craig~Citro and
                     Greg~S.~Corrado and
                     Andy~Davis and
                     Jeffrey~Dean and
                     Matthieu~Devin and
                     Sanjay~Ghemawat and
                     Ian~Goodfellow and
                     Andrew~Harp and
                     Geoffrey~Irving and
                     Michael~Isard and
                     Yangqing Jia and
                     Rafal~Jozefowicz and
                     Lukasz~Kaiser and
                     Manjunath~Kudlur and
                     Josh~Levenberg and
                     Dandelion~Man\'{e} and
                     Rajat~Monga and
                     Sherry~Moore and
                     Derek~Murray and
                     Chris~Olah and
                     Mike~Schuster and
                     Jonathon~Shlens and
                     Benoit~Steiner and
                     Ilya~Sutskever and
                     Kunal~Talwar and
                     Paul~Tucker and
                     Vincent~Vanhoucke and
                     Vijay~Vasudevan and
                     Fernanda~Vi\'{e}gas and
                     Oriol~Vinyals and
                     Pete~Warden and
                     Martin~Wattenberg and
                     Martin~Wicke and
                     Yuan~Yu and
                     Xiaoqiang~Zheng},
note = {Software available from tensorflow.org},
title = { {TensorFlow}: Large-Scale Machine Learning on Heterogeneous Systems},
url = {https://www.tensorflow.org/},
year = {2015}
}

@misc{wandb,
title = {Experiment Tracking with Weights and Biases},
year = {2020},
note = {Software available from wandb.com},
url={https://www.wandb.com/},
author = {Biewald, Lukas},
}

@article{Wiseman:2002zc,
author = {Wiseman, Toby},
eprint = {hep-th/0209051},
journal = {Class. Quant. Grav.},
pages = {1137-1176},
title = {{Static axisymmetric vacuum solutions and nonuniform black strings}},
volume = {20},
year = {2003}
}

@article{Yau1978OnTR,
author = {Shing-Tung Yau},
journal = {Communications on Pure and Applied Mathematics},
pages = {339-411},
title = {On The {R}icci Curvature of a Compact {K}ahler Manifold and the Complex {M}onge-{A}mpere Equation, {I}*},
volume = {31},
year = {1978}
}

@article{808365d6fe2f449e8be7d40295302da1,
title = "Parameterization of Invariant Manifolds for Periodic Orbits (II): A Posteriori Analysis and Computer Assisted Error Bounds",
author = "Roberto Castelli and Lessard, {Jean Philippe} and James, {Jason D.Mireles}",
year = "2017",
month = aug,
day = "22",
doi = "10.1007/s10884-017-9609-z",
language = "English",
volume = "30",
pages = "1--57",
journal = "Journal of Dynamics and Differential Equations",
issn = "1040-7294",
publisher = "Springer New York",
number = "4",
}

@article{10.1063/1.2358391,
    author = {Gerdt, V. and Horan, R. and Khvedelidze, A. and Lavelle, M. and McMullan, D. and Palii, Yu.},
    title = "{On the Hamiltonian reduction of geodesic motion on SU(3) to SU(3)/SU(2)}",
    journal = {Journal of Mathematical Physics},
    volume = {47},
    number = {11},
    pages = {112902},
    year = {2006},
    month = {11},
    issn = {0022-2488},
    doi = {10.1063/1.2358391},
    url = {https://doi.org/10.1063/1.2358391},
    eprint = {https://pubs.aip.org/aip/jmp/article-pdf/doi/10.1063/1.2358391/13781646/112902\_1\_online.pdf},
}

@article{https://doi.org/10.15488/12546,
  doi = {10.15488/12546},
  
  url = {https://www.repo.uni-hannover.de/handle/123456789/12646},
  
  author = {Kumar, Kaushlendra},
  
  language = {en},
  
  title = {Solutions of Yang–Mills theory in four-dimensional de Sitter space},
  
  publisher = {Hannover : Institutionelles Repositorium der Leibniz Universität Hannover},
  
  year = {2022},
  
  copyright = {CC BY 3.0 DE}
}

@article{Klebanov_2009,
	doi = {10.1088/1126-6708/2009/03/140},
  
	url = {https://doi.org/10.1088%2F1126-6708%2F2009%2F03%2F140},
  
	year = 2009,
	month = {mar},
  
	publisher = {Springer Science and Business Media {LLC}
},
  
	volume = {2009},
  
	number = {03},
  
	pages = {140--140},
  
	author = {Igor R Klebanov and Thomas Klose and Arvind Murugan},
  
	title = {AdS4/CFT3 -- Squashed, Stretched and Warped},
  
	journal = {Journal of High Energy Physics}
}

@article{Oh:2011nv,
    author = "Oh, John J. and Park, Chanyong and Yang, Hyun Seok",
    title = "{Yang-Mills Instantons from Gravitational Instantons}",
    eprint = "1101.1357",
    archivePrefix = "arXiv",
    primaryClass = "hep-th",
    doi = "10.1007/JHEP04(2011)087",
    journal = "JHEP",
    volume = "04",
    pages = "087",
    year = "2011"
}

@article{Awada:1982pk,
    author = "Awada, M. A. and Duff, M. J. and Pope, C. N.",
    title = "{N=8 Supergravity Breaks Down to N=1}",
    reportNumber = "ICTP/82-83/4",
    doi = "10.1103/PhysRevLett.50.294",
    journal = "Phys. Rev. Lett.",
    volume = "50",
    pages = "294",
    year = "1983"
}

@article{Aharony_2000,
	doi = {10.1016/s0370-1573(99)00083-6},
  
	url = {https://doi.org/10.1016%2Fs0370-1573%2899%2900083-6},
  
	year = 2000,
	month = {jan},
  
	publisher = {Elsevier {BV}
},
  
	volume = {323},
  
	number = {3-4},
  
	pages = {183--386},
  
	author = {Ofer Aharony and Steven S. Gubser and Juan Maldacena and Hirosi Ooguri and Yaron Oz},
  
	title = {Large N field theories, string theory and gravity},
  
	journal = {Physics Reports}
}

@article{RevModPhys.52.175,
  title = {The geometrical setting of gauge theories of the Yang-Mills type},
  author = {Daniel, M. and Viallet, C. M.},
  journal = {Rev. Mod. Phys.},
  volume = {52},
  issue = {1},
  pages = {175--197},
  numpages = {0},
  year = {1980},
  month = {Jan},
  publisher = {American Physical Society},
  doi = {10.1103/RevModPhys.52.175},
  url = {https://link.aps.org/doi/10.1103/RevModPhys.52.175}
}

@misc{https://doi.org/10.48550/arxiv.2111.13221,
  doi = {10.48550/ARXIV.2111.13221},
  
  url = {https://arxiv.org/abs/2111.13221},
  
  author = {de la Ossa, Xenia and Galdeano, Mateo},
  
  title = {Families of solutions of the heterotic G$_2$ system},
  
  publisher = {arXiv},
  
  year = {2021},
  
  copyright = {arXiv.org perpetual, non-exclusive license}
}

@article{Betounes:1992qz,
    author = "Betounes, D.",
    title = "{Kaluza-Klein geometry}",
    doi = "10.1016/0926-2245(91)90023-3",
    journal = "Differ. Geom. Appl.",
    volume = "1",
    pages = "77--88",
    year = "1992"
}

@article{10.2969/jmsj/01010029,
author = {Itiro TAMURA},
title = {{Homeomorphy classification of total spaces of sphere bundles over spheres.}},
volume = {10},
journal = {Journal of the Mathematical Society of Japan},
number = {1},
publisher = {Mathematical Society of Japan},
pages = {29 -- 43},
year = {1958},
doi = {10.2969/jmsj/01010029},
URL = {https://doi.org/10.2969/jmsj/01010029}
}

@article{CROWLEY2003363,
title = {A classification of S3-bundles over S4},
journal = {Differential Geometry and its Applications},
volume = {18},
number = {3},
pages = {363-380},
year = {2003},
issn = {0926-2245},
doi = {https://doi.org/10.1016/S0926-2245(03)00012-3},
url = {https://www.sciencedirect.com/science/article/pii/S0926224503000123},
author = {Diarmuid Crowley and Christine M. Escher}
}

@article{Nikonorov2004CompactHE,
  title={Compact Homogeneous Einstein 7-Manifolds},
  author={Yu. G. Nikonorov},
  journal={Geometriae Dedicata},
  year={2004},
  volume={109},
  pages={7-30}
}

@article{PhysRevLett.50.2043,
  title = {Spontaneous Supersymmetry Breaking by the Squashed Seven-Sphere},
  author = {Duff, M. J. and Nilsson, B. E. W. and Pope, C. N.},
  journal = {Phys. Rev. Lett.},
  volume = {50},
  issue = {26},
  pages = {2043--2046},
  numpages = {0},
  year = {1983},
  month = {Jun},
  publisher = {American Physical Society},
  doi = {10.1103/PhysRevLett.50.2043},
  url = {https://link.aps.org/doi/10.1103/PhysRevLett.50.2043}
}

@article{ABBASSI2010131,
title = {Naturality of homogeneous metrics on Stiefel manifolds SO(m+1)/SO(m-1)},
journal = {Differential Geometry and its Applications},
volume = {28},
number = {2},
pages = {131-139},
year = {2010},
issn = {0926-2245},
doi = {https://doi.org/10.1016/j.difgeo.2009.05.007},
url = {https://www.sciencedirect.com/science/article/pii/S0926224509000680},
author = {Mohamed Tahar Kadaoui Abbassi and Oldrich Kowalski}
}

@misc{ball2020associative,
      title={Associative Submanifolds of the Berger Space}, 
      author={Gavin Ball and Jesse Madnick},
      year={2020},
      eprint={2003.13169},
      archivePrefix={arXiv},
      primaryClass={math.DG}
}

@misc{samann2024brief,
      title={A brief introduction to non-regular spacetime geometry}, 
      author={Clemens Sämann},
      year={2024},
      eprint={2404.18651},
      archivePrefix={arXiv},
      primaryClass={gr-qc}
}

@article{Martin,
      journal={, Private Communication}, 
      author={Martin Cederwall},
      year={2024}
}

@article{thooftunp,
      journal={, Unpublished}, 
      author={G. 't Hooft},
      year={1976}
}

@article{TanNewPaper,
      journal={, To Appear}, 
      author={Tancredi Schettini Gherardini},
      year={2026}
}

@article{Chru_ciel_1998,
   title={Horizons Non-Differentiable on a Dense Set},
   volume={193},
   ISSN={1432-0916},
   url={http://dx.doi.org/10.1007/s002200050336},
   DOI={10.1007/s002200050336},
   number={2},
   journal={Communications in Mathematical Physics},
   publisher={Springer Science and Business Media LLC},
   author={Chruściel, Piotr T. and Galloway, Gregory J.},
   year={1998},
   month=apr, pages={449–470} }

@article{Alcubierre_1997,
   title={Appearance of coordinate shocks in hyperbolic formalisms of general relativity},
   volume={55},
   ISSN={1089-4918},
   url={http://dx.doi.org/10.1103/PhysRevD.55.5981},
   DOI={10.1103/physrevd.55.5981},
   number={10},
   journal={Physical Review D},
   publisher={American Physical Society (APS)},
   author={Alcubierre, Miguel},
   year={1997},
   month=may, pages={5981–5991} }

@article{Sbierski:2015nta,
    author = "Sbierski, Jan",
    title = "{The $C_0$-inextendibility of the Schwarzschild spacetime and the spacelike diameter in Lorentzian geometry}",
    eprint = "1507.00601",
    archivePrefix = "arXiv",
    primaryClass = "gr-qc",
    doi = "10.4310/jdg/1518490820",
    journal = "J. Diff. Geom.",
    volume = "108",
    number = "2",
    pages = "319--378",
    year = "2018"
}

@article{a965bffe-de90-3a7a-a652-4c0d2147a13b,
 ISSN = {0003486X},
 URL = {http://www.jstor.org/stable/3597235},
 author = {Mihalis Dafermos},
 journal = {Annals of Mathematics},
 number = {3},
 pages = {875--928},
 publisher = {Annals of Mathematics},
 title = {Stability and Instability of the Cauchy Horizon for the Spherically Symmetric Einstein-Maxwell-Scalar Field Equations},
 urldate = {2024-06-07},
 volume = {158},
 year = {2003}
}

@article{Westman_2009,
   title={Coordinates, observables and symmetry in relativity},
   volume={324},
   ISSN={0003-4916},
   url={http://dx.doi.org/10.1016/j.aop.2009.03.014},
   DOI={10.1016/j.aop.2009.03.014},
   number={8},
   journal={Annals of Physics},
   publisher={Elsevier BV},
   author={Westman, Hans and Sonego, Sebastiano},
   year={2009},
   month=aug, pages={1585–1611} }

@article{KervaireMilnor1963,
  author  = {Kervaire, Michel A. and Milnor, John W.},
  title   = {Groups of homotopy spheres: I},
  journal = {Annals of Mathematics},
  volume  = {77},
  number  = {3},
  year    = {1963},
  pages   = {504--537}
}

@article{Whitney1940,
  author  = {Whitney, Hassler},
  title   = {On the theory of sphere-bundles},
  journal = {Proceedings of the National Academy of Sciences of the United States of America},
  volume  = {26},
  number  = {2},
  pages   = {148--153},
  year    = {1940},
  doi     = {10.1073/pnas.26.2.148}
}

@inproceedings{Ehresmann1951,
  author    = {Ehresmann, Charles},
  title     = {Les connexions infinitésimales dans un espace fibré différentiable},
  booktitle = {Colloque de Topologie (espaces fibrés), Bruxelles 1950},
  pages     = {29--55},
  year      = {1951},
  publisher = {Georges Thone}
}

@article{Maxwell1865,
  author  = {Maxwell, James Clerk},
  title   = {A Dynamical Theory of the Electromagnetic Field},
  journal = {Philosophical Transactions of the Royal Society of London},
  volume  = {155},
  pages   = {459--512},
  year    = {1865},
  doi     = {10.1098/rstl.1865.0008}
}

@article{Fock1932,
  author  = {Fock, V.\,A.},
  title   = {Konfigurationsraum und zweite Quantelung},
  journal = {Zeitschrift f{\"u}r Physik},
  volume  = {75},
  pages   = {622--647},
  year    = {1932},
  doi     = {10.1007/BF01344458}
}

@article{Dirac1931,
  author  = {Dirac, P.},
  title   = {Quantised Singularities in the Electromagnetic Field},
  journal = {Proceedings of the Royal Society of London A},
  volume  = {133},
  pages   = {60--72},
  year    = {1931},
  doi     = {10.1098/rspa.1931.0130}
}

@article{AharonovBohm1959,
  author  = {Aharonov, Y. and Bohm, D.},
  title   = {Significance of Electromagnetic Potentials in the Quantum Theory},
  journal = {Physical Review},
  volume  = {115},
  pages   = {485--491},
  year    = {1959},
  doi     = {10.1103/PhysRev.115.485}
}

@article{YangMills1954,
  author  = {Yang, C.\,N. and Mills, R.\,L.},
  title   = {Conservation of Isotopic Spin and Isotopic Gauge Invariance},
  journal = {Physical Review},
  volume  = {96},
  pages   = {191--195},
  year    = {1954},
  doi     = {10.1103/PhysRev.96.191}
}

@article{Trautman1970,
  author  = {Trautman, Andrzej},
  title   = {Fiber bundles associated with space-time},
  journal = {Reports on Mathematical Physics},
  volume  = {1},
  number  = {1},
  pages   = {29--62},
  year    = {1970}
}

@article{WuYang1975,
  author  = {Wu, T.\,T. and Yang, C.\,N.},
  title   = {Concept of Nonintegrable Phase Factors and Global Formulation of Gauge Fields},
  journal = {Physical Review D},
  volume  = {12},
  pages   = {3845--3857},
  year    = {1975},
  doi     = {10.1103/PhysRevD.12.3845}
}

@book{Atiyah1979,
  author    = {Atiyah, Michael F.},
  title     = {Geometry of Yang--Mills Fields},
  series    = {Lezioni Fermiane},
  publisher = {Scuola Normale Superiore},
  address   = {Pisa},
  year      = {1979}
}

@article{
doi:10.1073/pnas.21.7.464,
author = {Hassler Whitney },
title = {Sphere-Spaces1},
journal = {Proceedings of the National Academy of Sciences},
volume = {21},
number = {7},
pages = {464-468},
year = {1935},
doi = {10.1073/pnas.21.7.464},
URL = {https://www.pnas.org/doi/abs/10.1073/pnas.21.7.464},
eprint = {https://www.pnas.org/doi/pdf/10.1073/pnas.21.7.464}}

@book{7a97cef4-8443-3a57-aced-2037f84b9e06,
 ISBN = {9780691005485},
 URL = {http://www.jstor.org/stable/j.ctt1bpm9t5},
 author = {NORMAN STEENROD},
 publisher = {Princeton University Press},
 title = {The Topology of Fibre Bundles. (PMS-14)},
 urldate = {2025-06-25},
 year = {1951}
}

@article{Berman:2022dpj,
    author = "Berman, David S. and Gherardini, Tancredi Schettini",
    title = "{Twisted self-duality}",
    eprint = "2208.09891",
    archivePrefix = "arXiv",
    primaryClass = "hep-th",
    doi = "10.1142/S0217751X23500859",
    journal = "Int. J. Mod. Phys. A",
    volume = "38",
    number = "15n16",
    pages = "2350085",
    year = "2023"
}

@article{Kerner1968,
author = {Kerner, Ryszard},
journal = {Annales de l'I.H.P. Physique théorique},
language = {eng},
number = {2},
pages = {143-152},
publisher = {Gauthier-Villars},
title = {Generalization of the Kaluza-Klein theory for an arbitrary non-abelian gauge group},
url = {http://eudml.org/doc/75604},
volume = {9},
year = {1968},
}

@article{10.1063/PT.3.2799,
    author = {Friedman, John L.},
    title = {Historical note on fiber bundles},
    journal = {Physics Today},
    volume = {68},
    number = {6},
    pages = {11-11},
    year = {2015},
    month = {06},
    issn = {0031-9228},
    doi = {10.1063/PT.3.2799},
    url = {https://doi.org/10.1063/PT.3.2799},
    eprint = {https://pubs.aip.org/physicstoday/article-pdf/68/6/11/10101413/11\_2\_online.pdf},
}

@article{Trautman:1977im,
    author = "Trautman, Andrzej",
    title = "{Solutions of the Maxwell and Yang-Mills Equations Associated with Hopf Fibrings}",
    reportNumber = "ITP-SB-77-24",
    doi = "10.1007/BF01811088",
    journal = "Int. J. Theor. Phys.",
    volume = "16",
    pages = "561",
    year = "1977"
}

@misc{zbMATH03272259,
 author = {DeWitt, B. S.},
 title = {Dynamical theory of groups and fields},
 year = {1965},
 language = {English},
 howpublished = {London-Glasgow: {Blackie} \& {Son} {Ltd}. {XV}, 248 p. 22 s. 6 d. (1965).},
 zbMATH = {3272259},
 Zbl = {0169.57101}
}

@article{Kaluza1921,
  author  = {Kaluza, Theodor},
  title   = {Zum Unit\"atsproblem der Physik},
  journal = {Sitzungsber. Preuss. Akad. Wiss. Berlin (Math.\ Phys.)},
  year    = {1921},
  pages   = {966--972}
}

@article{Klein1926,
  author  = {Klein, Oskar},
  title   = {Quantum Theory and Five-Dimensional Theory of Relativity},
  journal = {Zeitschrift f\"ur Physik},
  volume  = {37},
  pages   = {895--906},
  year    = {1926},
  doi     = {10.1007/BF01397481}
}

@article{bourguignon1989mathematicians,
  title={A mathematician's visit to Kaluza-Klein theory},
  author={Bourguignon, Jean-Pierre},
  journal={Rend. Sem. Mat. Univ. Politec. Torino},
  volume={47},
  number={3},
  pages={143},
  year={1989}
}

@article{Bailin:1987jd,
    author = "Bailin, D. and Love, A.",
    title = "{Kaluza--Klein theories}",
    doi = "10.1088/0034-4885/50/9/001",
    journal = "Rept. Prog. Phys.",
    volume = "50",
    pages = "1087--1170",
    year = "1987"
}

@MISC {1717244,
    TITLE = {Two atlases on a manifold $M$ are equivalent if and only if they determine the same set of smooth functions $f:M\rightarrow\mathbb{R}$},
    AUTHOR = {jl2 (https://math.stackexchange.com/users/255861/jl2)},
    HOWPUBLISHED = {Mathematics Stack Exchange},
    NOTE = {URL:https://math.stackexchange.com/q/1717244 (version: 2016-03-28)},
    EPRINT = {https://math.stackexchange.com/q/1717244},
    URL = {https://math.stackexchange.com/q/1717244}
}

@article{Brans:2008zz,
    author = "Brans, Carl H.",
    editor = "Guzman Murillo, Francisco S. and Herrera-Aguilar, Alfredo and Nucamendi, Ulises and Quiros, Israel",
    title = "{Scalar-tensor theories of gravity: Some personal history}",
    doi = "10.1063/1.3058577",
    journal = "AIP Conf. Proc.",
    volume = "1083",
    number = "1",
    pages = "34--46",
    year = "2008"
}

@misc{brans1997gravitytenaciousscalarfield,
      title={Gravity and the Tenacious Scalar Field}, 
      author={Carl H. Brans},
      year={1997},
      eprint={gr-qc/9705069},
      archivePrefix={arXiv},
      primaryClass={gr-qc},
      url={https://arxiv.org/abs/gr-qc/9705069}, 
}

@misc{hirst2025ainsteinnumericaleinsteinmetrics,
      title={AInstein: Numerical Einstein Metrics via Machine Learning}, 
      author={Edward Hirst and Tancredi Schettini Gherardini and Alexander G. Stapleton},
      year={2025},
      eprint={2502.13043},
      archivePrefix={arXiv},
      primaryClass={hep-th},
      url={https://arxiv.org/abs/2502.13043}, 
}

@article{Cerf70,
  author  = {Cerf, Jean},
  title   = {La stratification naturelle des espaces de fonctions diff\'erentiables r\'eelles et le theoreme de la pseudo isotopie},
  journal = {Publications Math\'ematiques de l'IH\'ES},
  volume  = {39},
  year    = {1970},
  pages   = {5--173}
}

@article{KervaireMilnor63,
  author  = {Kervaire, Michel and Milnor, John},
  title   = {Groups of Homotopy Spheres: I},
  journal = {Annals of Mathematics},
  volume  = {77},
  year    = {1963},
  pages   = {504--537}
}

@article{Kirby69,
  author  = {Kirby, Robion},
  title   = {Stable homeomorphisms and the annulus conjecture},
  journal = {Annals of Mathematics},
  volume  = {89},
  year    = {1969},
  pages   = {575--582}
}

@book{KirbySiebenmann77,
  author    = {Kirby, Robion and Siebenmann, Laurence},
  title     = {Foundational Essays on Topological Manifolds, Smoothings, and Triangulations},
  series    = {Annals of Mathematics Studies},
  number    = {88},
  publisher = {Princeton University Press},
  address   = {Princeton, NJ},
  year      = {1977}
}

@article{Quinn82,
  author  = {Quinn, Frank},
  title   = {Ends of maps. III: Dimensions 4 and 5},
  journal = {Journal of Differential Geometry},
  volume  = {17},
  year    = {1982},
  pages   = {503--521}
}

@article{EellsKuiper62,
  author  = {Eells, James and Kuiper, Nicolaas},
  title   = {An invariant for certain smooth manifolds},
  journal = {Annali della Scuola Normale Superiore di Pisa, Serie III},
  volume  = {16},
  year    = {1962},
  pages   = {409--438}
}

@article{Brieskorn66,
  author  = {Brieskorn, Egbert},
  title   = {Beispiele zur Differentialtopologie von Singularitäten},
  journal = {Inventiones Mathematicae},
  volume  = {2},
  year    = {1966},
  pages   = {1--14}
}

@book{Milnor68,
  author    = {Milnor, John},
  title     = {Singular Points of Complex Hypersurfaces},
  series    = {Annals of Mathematics Studies},
  number    = {61},
  publisher = {Princeton University Press},
  address   = {Princeton, NJ},
  year      = {1968}
}

@article{Hitchin74,
  author  = {Hitchin, Nigel},
  title   = {Harmonic Spinors},
  journal = {Advances in Mathematics},
  volume  = {14},
  year    = {1974},
  pages   = {1--55}
}

@article{GromovLawson80,
  author  = {Gromov, Mikhail and Lawson, H.~Blaine},
  title   = {The Classification of Simply Connected Manifolds of Positive Scalar Curvature},
  journal = {Annals of Mathematics},
  volume  = {111},
  year    = {1980},
  pages   = {423--434}
}

@article{Stolz92,
  author  = {Stolz, Stephan},
  title   = {Simply Connected Manifolds of Positive Scalar Curvature},
  journal = {Annals of Mathematics},
  volume  = {136},
  year    = {1992},
  pages   = {511--540}
}

@article{Wraith97,
  author  = {Wraith, David},
  title   = {Exotic Spheres with Positive Ricci Curvature},
  journal = {Journal of Differential Geometry},
  volume  = {45},
  year    = {1997},
  pages   = {638--649}
}

@article{BoyerGalickiKollar05,
  author  = {Boyer, Charles P. and Galicki, Krzysztof and Kollár, János},
  title   = {Einstein Metrics on Spheres},
  journal = {Annals of Mathematics},
  volume  = {162},
  year    = {2005},
  pages   = {557--580}
}

@article{GromollMeyer74,
  author  = {Gromoll, Detlef and Meyer, Wolfgang},
  title   = {An Exotic Sphere with Nonnegative Sectional Curvature},
  journal = {Annals of Mathematics},
  volume  = {100},
  year    = {1974},
  pages   = {401--406}
}

@article{PetersenWilhelm08,
  author    = {Petersen, Peter and Wilhelm, Frederick},
  title     = {An Exotic Sphere with Positive Sectional Curvature},
  journal   = {arXiv:0805.0812 [math.DG]},
  year      = {2008},
  note      = {95 pp. preprint}
}

@article{GroveZiller00,
  author  = {Grove, Karsten and Ziller, Wolfgang},
  title   = {Curvature and Symmetry of Milnor Spheres},
  journal = {Annals of Mathematics},
  volume  = {152},
  year    = {2000},
  pages   = {331--367}
}

@article{Freedman82,
  author  = {Freedman, Michael H.},
  title   = {The topology of four-dimensional manifolds},
  journal = {Journal of Differential Geometry},
  volume  = {17},
  year    = {1982},
  pages   = {357--453}
}

@article{Donaldson83,
  author  = {Donaldson, Simon K.},
  title   = {An application of gauge theory to four-dimensional topology},
  journal = {Journal of Differential Geometry},
  volume  = {18},
  year    = {1983},
  pages   = {279--315}
}

@article{Gompf95,
  author  = {Gompf, Robert E.},
  title   = {An exotic menagerie of small $\mathbb{R}^{4}$'s},
  journal = {Geometry \& Topology},
  volume  = {1},
  year    = {1997},
  pages   = {315--369}
}

@article{BizacaGompf96,
  author  = {Biza\v{c}a, \v{Z}eljko and Gompf, Robert E.},
  title   = {Elliptic surfaces and some simple exotic $\mathbb{R}^{4}$'s},
  journal = {Journal of Differential Geometry},
  volume  = {43},
  year    = {1996},
  pages   = {458--504}
}

@article{Dolgachev81,
  author  = {Dolgachev, Igor},
  title   = {Simply connected algebraic surfaces of positive index},
  journal = {Lecture Notes in Mathematics},
  volume  = {901},
  year    = {1981},
  pages   = {58--71}
}

@article{FintushelStern98,
  author  = {Fintushel, Ronald and Stern, Ronald J.},
  title   = {Knots, links, and 4-manifolds},
  journal = {Inventiones Mathematicae},
  volume  = {134},
  year    = {1998},
  pages   = {363--400}
}

@article{Mazur61,
  author  = {Mazur, Barry},
  title   = {A note on some contractible 4-manifolds},
  journal = {Annals of Mathematics},
  volume  = {73},
  year    = {1961},
  pages   = {221--228}
}

@article{Akbulut91,
  author  = {Akbulut, Selman},
  title   = {A fake compact contractible 4-manifold},
  journal = {Journal of Differential Geometry},
  volume  = {33},
  year    = {1991},
  pages   = {335--356}
}

@article{AkbulutMatveyev97,
  author  = {Akbulut, Selman and Matveyev, R.},
  title   = {A convex decomposition theorem for 4-manifolds},
  journal = {International Mathematics Research Notices},
  year    = {1998},
  number  = {7},
  pages   = {371--381}
}

@article{FarrellJones89,
  author  = {Farrell, F.~Thomas and Jones, Lowell E.},
  title   = {Exotic smooth structures on manifolds homotopy equivalent to $\mathbb{T}^{n}$},
  journal = {Journal of the American Mathematical Society},
  volume  = {2},
  year    = {1989},
  pages   = {899--910}
}

@book{Joyce00,
  author    = {Joyce, Dominic},
  title     = {Compact Manifolds with Special Holonomy},
  publisher = {Oxford University Press},
  year      = {2000}
}

@article{CrowleyNordstrom18,
  author  = {Crowley, Diarmuid and Nordström, Johannes},
  title   = {New invariants of $\mathrm{G}_{2}$-manifolds},
  journal = {Proceedings of the Royal Society A},
  volume  = {474},
  year    = {2018},
  pages   = {20170624}
}

@article{Tseytlin:1990nb,
    author = "Tseytlin, Arkady A.",
    title = "{Duality Symmetric Formulation of String World Sheet Dynamics}",
    reportNumber = "KCL-TP-1990-2",
    doi = "10.1016/0370-2693(90)91454-J",
    journal = "Phys. Lett. B",
    volume = "242",
    pages = "163--174",
    year = "1990"
}

@article{Tseytlin:1990va,
    author = "Tseytlin, Arkady A.",
    title = "{Duality symmetric closed string theory and interacting chiral scalars}",
    reportNumber = "KCL-TP-1990-3",
    doi = "10.1016/0550-3213(91)90266-Z",
    journal = "Nucl. Phys. B",
    volume = "350",
    pages = "395--440",
    year = "1991"
}

@article{Berman:2007xn,
    author = "Berman, David S. and Copland, Neil B. and Thompson, Daniel C.",
    title = "{Background Field Equations for the Duality Symmetric String}",
    eprint = "0708.2267",
    archivePrefix = "arXiv",
    primaryClass = "hep-th",
    reportNumber = "QMUL-PH-2007-15",
    doi = "10.1016/j.nuclphysb.2007.09.021",
    journal = "Nucl. Phys. B",
    volume = "791",
    pages = "175--191",
    year = "2008"
}

@article{Berman:2013eva,
    author = "Berman, David S. and Thompson, Daniel C.",
    title = "{Duality Symmetric String and M-Theory}",
    eprint = "1306.2643",
    archivePrefix = "arXiv",
    primaryClass = "hep-th",
    doi = "10.1016/j.physrep.2014.11.007",
    journal = "Phys. Rept.",
    volume = "566",
    pages = "1--60",
    year = "2014"
}

@article{Alfonsi:2021bot,
    author = "Alfonsi, Luigi and Berman, David S.",
    title = "{Double field theory and geometric quantisation}",
    eprint = "2101.12155",
    archivePrefix = "arXiv",
    primaryClass = "hep-th",
    doi = "10.1007/JHEP06(2021)059",
    journal = "JHEP",
    volume = "06",
    pages = "059",
    year = "2021"
}

@article{Berman:2012uy,
    author = "Berman, David S. and Musaev, Edvard T. and Thompson, Daniel C. and Thompson, Daniel C.",
    title = "{Duality Invariant M-theory: Gauged supergravities and Scherk-Schwarz reductions}",
    eprint = "1208.0020",
    archivePrefix = "arXiv",
    primaryClass = "hep-th",
    doi = "10.1007/JHEP10(2012)174",
    journal = "JHEP",
    volume = "10",
    pages = "174",
    year = "2012"
}

@article{Berman:2011jh,
    author = "Berman, David S. and Godazgar, Hadi and Perry, Malcolm J. and West, Peter",
    title = "{Duality Invariant Actions and Generalised Geometry}",
    eprint = "1111.0459",
    archivePrefix = "arXiv",
    primaryClass = "hep-th",
    doi = "10.1007/JHEP02(2012)108",
    journal = "JHEP",
    volume = "02",
    pages = "108",
    year = "2012"
}

@article{Berman_2020,
	doi = {10.1142/s0217751x20300148},
  
	url = {https://doi.org/10.1142%2Fs0217751x20300148},
  
	year = 2020,
	month = {oct},
  
	publisher = {World Scientific Pub Co Pte Lt},
  
	volume = {35},
  
	number = {30},
  
	pages = {2030014},
  
	author = {David S. Berman and Chris Blair},
  
	title = {The geometry, branes and applications of exceptional field theory},
  
	journal = {International Journal of Modern Physics A}
}

@misc{https://doi.org/10.48550/arxiv.1412.2768,
  doi = {10.48550/ARXIV.1412.2768},
  
  url = {https://arxiv.org/abs/1412.2768},
  
  author = {Berman, David S. and Rudolph, Felix J.},
  
  title = {Strings, Branes and the Self-dual Solutions of Exceptional Field Theory},
  
  publisher = {arXiv},
  
  year = {2014},
  
  copyright = {arXiv.org perpetual, non-exclusive license}
}

@article{Cremmer_1998,
	doi = {10.1016/s0550-3213(98)00136-9},
  
	url = {https://doi.org/10.1016%2Fs0550-3213%2898%2900136-9},
  
	year = 1998,
	month = {jul},
  
	publisher = {Elsevier {BV}
},
  
	volume = {523},
  
	number = {1-2},
  
	pages = {73--144},
  
	author = {E. Cremmer and B. Julia and H. Lü and C.N. Pope},
  
	title = {Dualisation of dualities},
  
	journal = {Nuclear Physics B}
}

@article{Cremmer_1998_2,
	doi = {10.1016/s0550-3213(98)00552-5},
  
	url = {https://doi.org/10.1016%2Fs0550-3213%2898%2900552-5},
  
	year = 1998,
	month = {dec},
  
	publisher = {Elsevier {BV}
},
  
	volume = {535},
  
	number = {1-2},
  
	pages = {242--292},
  
	author = {E. Cremmer and B. Julia and H. Lü and C.N. Pope},
  
	title = {Dualisation of dualities {II}: twisted self-duality of doubled fields and superdualities},
  
	journal = {Nuclear Physics B}
}

@article{EGUCHI1980213,
title = {Gravitation, gauge theories and differential geometry},
journal = {Physics Reports},
volume = {66},
number = {6},
pages = {213-393},
year = {1980},
issn = {0370-1573},
doi = {https://doi.org/10.1016/0370-1573(80)90130-1},
url = {https://www.sciencedirect.com/science/article/pii/0370157380901301},
author = {Tohru Eguchi and Peter B. Gilkey and Andrew J. Hanson}
}

@article{PhysRevD.14.517,
  title = {Conformal properties of a Yang-Mills pseudoparticle},
  author = {Jackiw, R. and Rebbi, C.},
  journal = {Phys. Rev. D},
  volume = {14},
  issue = {2},
  pages = {517--523},
  numpages = {0},
  year = {1976},
  month = {Jul},
  publisher = {American Physical Society},
  doi = {10.1103/PhysRevD.14.517},
  url = {https://link.aps.org/doi/10.1103/PhysRevD.14.517}
}

@misc{harada2025exactvacuumsolutionhopf,
      title={Exact vacuum solution with Hopf structure in general relativity}, 
      author={Junpei Harada},
      year={2025},
      eprint={2506.20878},
      archivePrefix={arXiv},
      primaryClass={gr-qc},
      url={https://arxiv.org/abs/2506.20878}, 
}

@article{URBANTKE2003125,
title = {The Hopf fibration—seven times in physics},
journal = {Journal of Geometry and Physics},
volume = {46},
number = {2},
pages = {125-150},
year = {2003},
issn = {0393-0440},
doi = {https://doi.org/10.1016/S0393-0440(02)00121-3},
url = {https://www.sciencedirect.com/science/article/pii/S0393044002001213},
author = {H.K. Urbantke}
}

@book{Neugebauer1957,
  author    = {Otto Neugebauer},
  title     = {The Exact Sciences in Antiquity},
  publisher = {Princeton University Press},
  edition   = {2},
  year      = {1957}
}

@book{Kline1972,
  author    = {Morris Kline},
  title     = {Mathematical Thought from Ancient to Modern Times},
  publisher = {Oxford University Press},
  year      = {1972}
}

@book{Boyer1987,
  author    = {Carl B. Boyer},
  title     = {The History of the Calculus and Its Conceptual Development},
  publisher = {Dover Publications},
  year      = {1987}
}

@book{Truesdell1960,
  author    = {Clifford Truesdell},
  title     = {The Rational Mechanics of Flexible or Elastic Bodies, 1638–1788},
  publisher = {Birkh{\"a}user},
  year      = {1960},
  note      = {Introductory notes by C.\,Truesdell}
}

@book{Duhem1969,
  author       = {Pierre Duhem},
  title        = {To Save the Phenomena\@: An Essay on the Idea of Physical Theory from Plato to Galileo},
  translator   = {Edmund Doland and Chaninah Maschler},
  introduction = {Stanley L. Jaki},
  address      = {Chicago},
  publisher    = {University of Chicago Press},
  year         = {1969},
  pages        = {152},
  isbn         = {9780226169217},
  note         = {First published in French in 1908.}
}

@book{Cohen1981,
  author       = {Cohen, I. Bernard},
  title        = {The Newtonian Revolution\@: With Illustrations of the Transformation of Scientific Ideas},
  series       = {The Wiles Lectures},
  address      = {Cambridge},
  publisher    = {Cambridge University Press},
  year         = {1981},
  isbn         = {9780521229647},
  pages        = {424},
  note         = {Argues that Newton’s Principia and related work constituted a decisive “revolution” that re-made natural philosophy into modern science.}
}

@incollection{Popper1975,
  author    = {Popper, Karl Raimund},
  title     = {The Rationality of Scientific Revolutions},
  editor    = {Harr\'e, Rom},
  booktitle = {Problems of Scientific Revolution: Progress and Obstacles to Progress in the Sciences},
  address   = {Oxford},
  publisher = {Clarendon Press},
  year      = {1975},
  pages     = {72--101},
  note      = {Herbert Spencer Lecture (1973). Popper uses the transition from pre-Copernican astronomy to Newton’s mechanics as his prime historical example of a genuine “revolution.”}
}

@book{Kuhn1962,
  author    = {Thomas S. Kuhn},
  title     = {The Structure of Scientific Revolutions},
  publisher = {University of Chicago Press},
  year      = {1962}
}

@book{Popper1934,
  author    = {Popper, Karl Raimund},
  title     = {Logik der Forschung: Zur Erkenntnistheorie der modernen Naturwissenschaft},
  series    = {Schriften zur wissenschaftlichen Weltauffassung},
  volume    = {9},
  address   = {Vienna},
  publisher = {Verlag von Julius Springer},
  year      = {1934},
  note      = {Introduces falsifiability as the demarcation criterion; English translation published as \emph{The Logic of Scientific Discovery} in 1959.}
}

@book{Popper1959,
  author    = {Karl Popper},
  title     = {The Logic of Scientific Discovery},
  publisher = {Routledge},
  year      = {1959}
}

@article{Schrodinger1926a,
  author  = {Schr{\"o}dinger, Erwin},
  title   = {Quantisierung als Eigenwertproblem (Erste Mitteilung)},
  journal = {Annalen der Physik},
  series  = {4},
  volume  = {79},
  number  = {4},
  pages   = {361--376},
  year    = {1926},
  doi     = {10.1002/andp.19263840404},
  note    = {Introduces the stationary Schr{\"o}dinger wave equation and applies it to the hydrogen atom.}
}

@article{Dirac1928,
  author  = {P. A. M. Dirac},
  title   = {The Quantum Theory of the Electron},
  journal = {Proceedings of the Royal Society A},
  volume  = {117},
  pages   = {610--624},
  year    = {1928},
  doi     = {10.1098/rspa.1928.0023}
}

@article{Weyl1918,
  author  = {Weyl, Hermann},
  title   = {Gravitation und Elektrizität},
  journal = {Sitzungsberichte der Königlich Preußischen Akademie der Wissenschaften zu Berlin},
  year    = {1918},
  pages   = {465--480}
}

@book{Eddington1946,
  author    = {Eddington, Arthur Stanley},
  title     = {Fundamental Theory},
  address   = {Cambridge},
  publisher = {Cambridge University Press},
  year      = {1946},
  note      = {Mathematically elegant attempt to derive all physical constants; predictions disagreed with later data.}
}

@article{Bohr1924,
  author  = {Bohr, Niels and Kramers, Hendrik Anthony and Slater, John Clarke},
  title   = {LXXVI. The Quantum Theory of Radiation},
  journal = {Philosophical Magazine},
  series  = {6},
  volume  = {47},
  number  = {281},
  pages   = {785--802},
  year    = {1924},
  doi     = {10.1080/14786442408565262},
  note    = {Introduces the Bohr–Kramers–Slater (BKS) statistical model of atomic radiation.}
}

@article{BotheGeiger1925,
  author  = {Bothe, Walther and Geiger, Hans},
  title   = {Experimentelles zur Theorie von Bohr, Kramers und Slater},
  journal = {Die Naturwissenschaften},
  volume  = {13},
  pages   = {440--441},
  year    = {1925},
  note    = {Coincidence experiment that demonstrated strict energy–momentum conservation, refuting the BKS theory.}
}

@article{GreenePlesser1990,
  author  = {Greene, Brian R. and Plesser, Michael R.},
  title   = {Duality in Calabi--Yau Moduli Space},
  journal = {Nuclear Physics B},
  volume  = {338},
  number  = {1},
  pages   = {15--37},
  year    = {1990},
  doi     = {10.1016/0550-3213(90)90622-K}
}

@article{CandelasEtAl1991,
  author  = {Candelas, Philip and de~la~Ossa, Xenia C. and Green, Paul S. and Parkes, Linda},
  title   = {A Pair of Calabi--Yau Manifolds as an Exactly Soluble Superconformal Theory},
  journal = {Nuclear Physics B},
  volume  = {359},
  number  = {1},
  pages   = {21--74},
  year    = {1991},
  doi     = {10.1016/0550-3213(91)90292-6}
}

@article{Witten1977,
  author  = {Witten, Edward},
  title   = {Some Exact Multi--Instanton Solutions of Classical {Y}ang--{M}ills Theory},
  journal = {Physical Review Letters},
  year    = {1977},
  volume  = {38},
  number  = {3},
  pages   = {121--124},
  doi     = {10.1103/PhysRevLett.38.121}
}

@misc{Watkins1990,
  author       = {Watkins, Matthew R.},
  title        = {A Short Survey of Lens Spaces},
  howpublished = {Undergraduate dissertation, University of Exeter},
  year         = {1990},
  note         = {Available at \url{https://empslocal.ex.ac.uk/people/staff/mrwatkin/lensspaces.pdf}
                  (accessed 30 June 2025).}
}

@article{Shimada_1957, title={Differentiable Structures on the 15-Sphere and Pontrjagin Classes of Certain Manifolds}, volume={12}, DOI={10.1017/S0027763000021942}, journal={Nagoya Mathematical Journal}, author={Shimada, Nobuo}, year={1957}, pages={59–69}}

@article{Bais:1983wc,
    author = "Bais, F. A. and Nicolai, H. and van Nieuwenhuizen, P.",
    title = "{Geometry of Coset Spaces and Massless Modes of the Squashed Seven Sphere in Supergravity}",
    reportNumber = "CERN-TH-3577",
    doi = "10.1016/0550-3213(83)90328-0",
    journal = "Nucl. Phys. B",
    volume = "228",
    pages = "333--350",
    year = "1983"
}

@book{osti_6662393,
  author       = {Helgason, S},
  title        = {Differential geometry, Lie groups, and symmetric spaces},
  url          = {https://www.osti.gov/biblio/6662393},
  place        = {United States},
  publisher    = {Academic Press, Inc.,New York, NY},
  year         = {1978},
  month        = {01}}

@article{Smale1962,
    author    = {Smale, Stephen},
    title     = {On the Structure of Manifolds},
    journal   = {American Journal of Mathematics},
    volume    = {84},
    number    = {3},
    pages     = {387--399},
    year      = {1962},
    doi       = {10.2307/2372978},
    issn      = {0002-9327}
}

@book{Milnor1965hCobordism,
    author    = {Milnor, John W.},
    title     = {Lectures on the h-Cobordism Theorem},
    publisher = {Princeton University Press},
    year      = {1965},
    address   = {Princeton, NJ},
    series    = {Notes by L. Siebenmann and J. Sondow},
    isbn      = {978-0691079969}
}

@inbook{doi:10.1142/9789812772107_0004,
author = {S. Smale},
title = {Generalized Poincaré's conjecture in dimensions greater than four},
booktitle = {Topological Library},
chapter = {},
pages = {251-268},
doi = {10.1142/9789812772107_0004},
URL = {https://www.worldscientific.com/doi/abs/10.1142/9789812772107_0004},
eprint = {https://www.worldscientific.com/doi/pdf/10.1142/9789812772107_0004}
}

@inbook{Lance+2000+73+104,
url = {https://doi.org/10.1515/9781400865192-007},
title = {Differentiable structures on manifolds},
booktitle = {Surveys on Surgery Theory, Volume 1},
author = {Timothy Lance},
publisher = {Princeton University Press},
address = {Princeton},
pages = {73--104},
doi = {doi:10.1515/9781400865192-007},
isbn = {9781400865192},
year = {2000},
lastchecked = {2025-07-02}
}

@article{Thakur1998,
  author    = {Jagannath Thakur},
  title     = {Einstein equations and the joining of discontinuous metrics},
  journal   = {Pramana -- Journal of Physics},
  volume    = {51},
  number    = {6},
  pages     = {699--710},
  year      = {1998},
  month     = dec,
  doi       = {10.1007/BF02832602},
  issn      = {0304-4289},
  publisher = {Indian Academy of Sciences}
}

@inbook{Penrose:1972xrn,
    author = "Penrose, Roger",
    editor = "O'Raifeartaigh, L.",
    title = "{The geometry of impulsive gravitational waves}",
    booktitle = "{General relativity}: {Papers in honour of J.L. Synge}",
    pages = "101--115",
    year = "1972"
}

@article{Israel1966,
  author    = {W. Israel},
  title     = {Singular hypersurfaces and thin shells in general relativity},
  journal   = {Il Nuovo Cimento B},
  volume    = {44},
  number    = {1},
  pages     = {1--14},
  year      = {1966},
  doi       = {10.1007/BF02710419}
}

@article{BarrabesIsrael1991,
  author    = {C. Barrab{\`e}s and W. Israel},
  title     = {Thin shells in general relativity and cosmology: The lightlike limit},
  journal   = {Physical Review D},
  volume    = {43},
  number    = {4},
  pages     = {1129--1142},
  year      = {1991},
  doi       = {10.1103/PhysRevD.43.1129}
}

@article{GerochTraschen1987,
  author    = {Robert Geroch and Jennie Traschen},
  title     = {Strings and other distributional sources in general relativity},
  journal   = {Physical Review D},
  volume    = {36},
  number    = {4},
  pages     = {1017--1031},
  year      = {1987},
  doi       = {10.1103/PhysRevD.36.1017}
}

@article{Taub1980,
  author    = {A. H. Taub},
  title     = {Space--times with distribution valued curvature tensors},
  journal   = {Journal of Mathematical Physics},
  volume    = {21},
  number    = {6},
  pages     = {1423--1431},
  year      = {1980},
  doi       = {10.1063/1.524568}
}

@article{SteinbauerVickers2006,
  author    = {Roland Steinbauer and James A. Vickers},
  title     = {The use of Generalised Functions and Distributions in General Relativity},
  journal   = {Classical and Quantum Gravity},
  volume    = {23},
  number    = {10},
  pages     = {R91--R114},
  year      = {2006},
  doi       = {10.1088/0264-9381/23/10/R01}
}

@article{KunzingerSteinbauer2002,
  author    = {Michael Kunzinger and Roland Steinbauer},
  title     = {Generalized pseudo-Riemannian geometry},
  journal   = {Transactions of the American Mathematical Society},
  volume    = {354},
  number    = {10},
  pages     = {4179--4199},
  year      = {2002},
  doi       = {10.1090/S0002-9947-02-03058-1}
}

@article{C_K_Raju_1982,
doi = {10.1088/0305-4470/15/6/017},
url = {https://dx.doi.org/10.1088/0305-4470/15/6/017},
year = {1982},
month = {jun},
publisher = {},
volume = {15},
number = {6},
pages = {1785},
author = {C K Raju},
title = {Junction conditions in general relativity},
journal = {Journal of Physics A: Mathematical and General}
}

@article{Gemelli:2007tj,
    author = "Gemelli, Gianluca",
    title = "{Generalized regularly discontinuous solutions of the Einstein equations}",
    eprint = "0704.0103",
    archivePrefix = "arXiv",
    primaryClass = "gr-qc",
    doi = "10.1007/s10773-007-9450-y",
    journal = "Int. J. Theor. Phys.",
    volume = "46",
    pages = "3312--3330",
    year = "2007"
}

@article{Aichelburg_1996,
   title={Symmetries of pp-waves with distributional profile},
   volume={13},
   ISSN={1361-6382},
   url={http://dx.doi.org/10.1088/0264-9381/13/4/012},
   DOI={10.1088/0264-9381/13/4/012},
   number={4},
   journal={Classical and Quantum Gravity},
   publisher={IOP Publishing},
   author={Aichelburg, Peter C and Balasin, Herbert},
   year={1996},
   month=apr, pages={723–729} }

@article{Aichelburg_1997,
   title={Generalized symmetries of impulsive gravitational waves},
   volume={14},
   ISSN={1361-6382},
   url={http://dx.doi.org/10.1088/0264-9381/14/1A/004},
   DOI={10.1088/0264-9381/14/1a/004},
   number={1A},
   journal={Classical and Quantum Gravity},
   publisher={IOP Publishing},
   author={Aichelburg, Peter C and Balasin, Herbert},
   year={1997},
   month=jan, pages={A31–A41} }

@inproceedings{Erlacher:2010ts,
    author = "Erlacher, Evelina and Grosser, Michael",
    title = "{Inversion of a 'discontinuous coordinate transformation' in general relativity}",
    booktitle = "{International Conference on Generalized Functions: GF 2009}",
    eprint = "1003.4245",
    archivePrefix = "arXiv",
    primaryClass = "math-ph",
    month = "3",
    year = "2010"
}

@article{Steinbauer_2006,
   title={The use of generalized functions and distributions in general relativity},
   volume={23},
   ISSN={1361-6382},
   url={http://dx.doi.org/10.1088/0264-9381/23/10/R01},
   DOI={10.1088/0264-9381/23/10/r01},
   number={10},
   journal={Classical and Quantum Gravity},
   publisher={IOP Publishing},
   author={Steinbauer, R and Vickers, J A},
   year={2006},
   month=apr, pages={R91–R114} }

@article{S_mann_2024,
   title={Cut-and-paste for impulsive gravitational waves with $\Lambda $: the mathematical analysis},
   volume={114},
   ISSN={1573-0530},
   url={http://dx.doi.org/10.1007/s11005-024-01804-0},
   DOI={10.1007/s11005-024-01804-0},
   number={2},
   journal={Letters in Mathematical Physics},
   publisher={Springer Science and Business Media LLC},
   author={Sämann, Clemens and Schinnerl, Benedict and Steinbauer, Roland and Švarc, Robert},
   year={2024},
   month=apr }

@article{PhysRevD.100.024040,
  title = {Cut-and-paste for impulsive gravitational waves with $\mathrm{\ensuremath{\Lambda}}$: The geometric picture},
  author = {Podolsk\'y, Ji\ifmmode \check{r}\else \v{r}\fi{}\'{\i} and S\"amann, Clemens and Steinbauer, Roland and \ifmmode \check{S}\else \v{S}\fi{}varc, Robert},
  journal = {Phys. Rev. D},
  volume = {100},
  issue = {2},
  pages = {024040},
  numpages = {8},
  year = {2019},
  month = {Jul},
  publisher = {American Physical Society},
  doi = {10.1103/PhysRevD.100.024040},
  url = {https://link.aps.org/doi/10.1103/PhysRevD.100.024040}
}

@article{Kunzinger:1998xw,
    author = "Kunzinger, Michael and Steinbauer, Roland",
    title = "{A Note on the Penrose junction conditions}",
    eprint = "gr-qc/9811007",
    archivePrefix = "arXiv",
    reportNumber = "UWTHPH-1998-57",
    doi = "10.1088/0264-9381/16/4/013",
    journal = "Class. Quant. Grav.",
    volume = "16",
    pages = "1255--1264",
    year = "1999"
}

@article{Reintjes_2015,
   title={No regularity singularities exist at points of general relativistic shock wave interaction between shocks from different characteristic families},
   volume={471},
   ISSN={1471-2946},
   url={http://dx.doi.org/10.1098/rspa.2014.0834},
   DOI={10.1098/rspa.2014.0834},
   number={2177},
   journal={Proceedings of the Royal Society A: Mathematical, Physical and Engineering Sciences},
   publisher={The Royal Society},
   author={Reintjes, Moritz and Temple, Blake},
   year={2015},
   month=may, pages={20140834} }

@misc{reintjes2017spacetimelocallyinertialpoints,
      title={Spacetime is Locally Inertial at Points of General Relativistic Shock Wave Interaction between Shocks from Different Characteristic Families}, 
      author={Moritz Reintjes},
      year={2017},
      eprint={1409.5060},
      archivePrefix={arXiv},
      primaryClass={gr-qc},
      url={https://arxiv.org/abs/1409.5060}, 
}

@article{Reintjes_2019,
   title={Shock Wave Interactions and the Riemann-Flat Condition: The Geometry Behind Metric Smoothing and the Existence of Locally Inertial Frames in General Relativity},
   volume={235},
   ISSN={1432-0673},
   url={http://dx.doi.org/10.1007/s00205-019-01456-8},
   DOI={10.1007/s00205-019-01456-8},
   number={3},
   journal={Archive for Rational Mechanics and Analysis},
   publisher={Springer Science and Business Media LLC},
   author={Reintjes, Moritz and Temple, Blake},
   year={2019},
   month=oct, pages={1873–1904} }

@article{reintjes2020how,
  title={How to smooth a crinkled map of space--time: Uhlenbeck compactness for Linfty connections and optimal regularity for general relativistic shock waves by the Reintjes--Temple equations},
  author={Reintjes, Moritz and Temple, Blake},
  journal={Proceedings of the Royal Society A: Mathematical, Physical and Engineering Sciences},
  volume={476},
  number={2241},
  pages={20200177},
  year={2020},
  publisher={The Royal Society},
  doi={10.1098/rspa.2020.0177},
  url={https://royalsocietypublishing.org/doi/10.1098/rspa.2020.0177}
}

@article{Reintjes:2022rmh,
    author = "Reintjes, Moritz and Temple, Blake",
    title = "{Optimal regularity and Uhlenbeck compactness for general relativity and Yang{\textendash}Millstheory}",
    eprint = "2202.09535",
    archivePrefix = "arXiv",
    primaryClass = "gr-qc",
    doi = "10.1098/rspa.2022.0444",
    journal = "Proc. Roy. Soc. Lond. A",
    volume = "479",
    number = "2271",
    pages = "20220444",
    year = "2023"
}

@article{Reintjes_2023,
   title={Optimal regularity and Uhlenbeck compactness for general relativity and Yang–Mills theory},
   volume={479},
   ISSN={1471-2946},
   url={http://dx.doi.org/10.1098/rspa.2022.0444},
   DOI={10.1098/rspa.2022.0444},
   number={2271},
   journal={Proceedings of the Royal Society A: Mathematical, Physical and Engineering Sciences},
   publisher={The Royal Society},
   author={Reintjes, Moritz and Temple, Blake},
   year={2023},
   month=mar }

@misc{reintjes2024essentialregularitysingularconnections,
      title={The essential regularity of singular connections in Geometry}, 
      author={Moritz Reintjes and Blake Temple},
      year={2024},
      eprint={2412.08928},
      archivePrefix={arXiv},
      primaryClass={gr-qc},
      url={https://arxiv.org/abs/2412.08928}, 
}

@article{Taubes1987,
  author    = {Clifford H. Taubes},
  title     = {Gauge theory on asymptotically periodic 4-manifolds},
  journal   = {Journal of Differential Geometry},
  year      = {1987},
  volume    = {25},
  number    = {3},
  pages     = {363--430},
  note      = {Shows existence of a continuum (uncountable) of non‑diffeomorphic exotic $ \mathbb{R}^4 $’s :contentReference[oaicite:1]{index=1}}
}

@article{Duston2011,
author = {DUSTON, CHRISTOPHER L.},
title = {EXOTIC SMOOTHNESS IN FOUR DIMENSIONS AND EUCLIDEAN QUANTUM GRAVITY},
journal = {International Journal of Geometric Methods in Modern Physics},
volume = {08},
number = {03},
pages = {459-484},
year = {2011},
doi = {10.1142/S0219887811005233},

URL = { 
    
        https://doi.org/10.1142/S0219887811005233
    
    

},
eprint = { 
    
        https://doi.org/10.1142/S0219887811005233
    
    

}
}

@Article{Duston2022,
AUTHOR = {Duston, Christopher L.},
TITLE = {Metrics on End-Periodic Manifolds as Models for Dark Matter},
JOURNAL = {Universe},
VOLUME = {8},
YEAR = {2022},
NUMBER = {3},
ARTICLE-NUMBER = {167},
URL = {https://www.mdpi.com/2218-1997/8/3/167},
ISSN = {2218-1997},
DOI = {10.3390/universe8030167}
}

@article{Rohm:1988yz,
    author = "Rohm, Ryan",
    title = "{TOPOLOGICAL DEFECTS AND DIFFERENTIAL STRUCTURES}",
    reportNumber = "CALT-68-1465",
    doi = "10.1016/0003-4916(89)90085-7",
    journal = "Annals Phys.",
    volume = "189",
    pages = "223",
    year = "1989"
}

@article{3c71f429-d750-3ab8-af97-277a9e5ed9e8,
 ISSN = {0003486X, 19398980},
 URL = {http://www.jstor.org/stable/1970349},
 author = {Wu-Chung Hsiang and Wu-Yi Hsiang},
 journal = {Annals of Mathematics},
 number = {3},
 pages = {359--369},
 publisher = {[Annals of Mathematics, Trustees of Princeton University on Behalf of the Annals of Mathematics, Mathematics Department, Princeton University]},
 title = {On Compact Subgroups of the Diffeomorphism Groups of Kervaire Spheres},
 urldate = {2025-07-05},
 volume = {85},
 year = {1967}
}

@article{Headrick:2005ch,
    author = "Headrick, Matthew and Wiseman, Toby",
    title = "{Numerical Ricci-flat metrics on K3}",
    eprint = "hep-th/0506129",
    archivePrefix = "arXiv",
    reportNumber = "HUTP-05-A0028, MIT-CTP-3647",
    doi = "10.1088/0264-9381/22/23/002",
    journal = "Class. Quant. Grav.",
    volume = "22",
    pages = "4931--4960",
    year = "2005"
}

@article{Gentle2004,
  author    = {Adrian P. Gentle},
  title     = {Regge calculus: a unique tool for numerical relativity},
  journal   = {arXiv preprint},
  year      = {2004},
}

@article{Douglas2006,
  author    = {Michael R. Douglas and Robert L. Karp and Sergio Lukic and Rene Reinbacher},
  title     = {Numerical Calabi–Yau metrics},
  journal   = {Journal of Mathematical Physics},
  year      = {2006},
  volume    = {49},
  number    = {3},
  pages     = {032302},
  note      = {Approximate Ricci–flat metrics on Calabi–Yau hypersurfaces using balanced metrics :contentReference[oaicite:1]{index=1}}
}

@article{Berman:2011gg,
  author       = {Berman, David S. and Perry, Malcolm J.},
  title        = {Generalized geometry and M theory},
  journal      = {Journal of High Energy Physics},
  volume       = {2011},
  number       = {06},
  pages        = {074},
  year         = {2011},
  doi          = {10.1007/JHEP06(2011)074},
  eprint       = {1008.1763},
  archivePrefix= {arXiv},
  primaryClass = {hep-th}
}

@article{Berman:2011so55,
  author       = {Berman, David S. and Godazgar, Hadi and Perry, Malcolm J.},
  title        = {{SO}(5,5) duality in M-theory and generalized geometry},
  journal      = {Physics Letters B},
  volume       = {700},
  pages        = {65--67},
  year         = {2011},
  doi          = {10.1016/j.physletb.2011.04.046},
  eprint       = {1103.5733},
  archivePrefix= {arXiv},
  primaryClass = {hep-th}
}

@article{HullZwiebach:2009dft,
  author       = {Hull, Chris M. and Zwiebach, Barton},
  title        = {Double Field Theory},
  journal      = {Journal of High Energy Physics},
  volume       = {2009},
  number       = {09},
  pages        = {099},
  year         = {2009},
  doi          = {10.1088/1126-6708/2009/09/099},
  eprint       = {0904.4664},
  archivePrefix= {arXiv},
  primaryClass = {hep-th}
}

@article{HohmHullZwiebach:2010bia,
  author       = {Hohm, Olaf and Hull, Chris and Zwiebach, Barton},
  title        = {Background independent action for double field theory},
  journal      = {Journal of High Energy Physics},
  volume       = {2010},
  number       = {07},
  pages        = {016},
  year         = {2010},
  doi          = {10.1007/JHEP07(2010)016},
  eprint       = {1003.5027},
  archivePrefix= {arXiv},
  primaryClass = {hep-th}
}

@article{HohmHullZwiebach:2010gmetric,
  author       = {Hohm, Olaf and Hull, Chris and Zwiebach, Barton},
  title        = {Generalized metric formulation of double field theory},
  journal      = {Journal of High Energy Physics},
  volume       = {2010},
  number       = {08},
  pages        = {008},
  year         = {2010},
  doi          = {10.1007/JHEP08(2010)008},
  eprint       = {1006.4823},
  archivePrefix= {arXiv},
  primaryClass = {hep-th}
}

@article{Hull:2007gg,
  author       = {Hull, Chris M.},
  title        = {Generalised Geometry for M-Theory},
  journal      = {Journal of High Energy Physics},
  volume       = {2007},
  number       = {07},
  pages        = {079},
  year         = {2007},
  doi          = {10.1088/1126-6708/2007/07/079},
  eprint       = {hep-th/0701203},
  archivePrefix= {arXiv},
  primaryClass = {hep-th}
}

@article{PachecoWaldram:2008egg,
  author       = {Pires Pacheco, Paulo and Waldram, Daniel},
  title        = {M-theory, exceptional generalised geometry and superpotentials},
  journal      = {Journal of High Energy Physics},
  volume       = {2008},
  number       = {09},
  pages        = {123},
  year         = {2008},
  doi          = {10.1088/1126-6708/2008/09/123},
  eprint       = {0804.1362},
  archivePrefix= {arXiv},
  primaryClass = {hep-th}
}

@article{Hillmann:2009e7,
  author       = {Hillmann, Christian},
  title        = {Generalized \{E\}\_{7(7)} coset dynamics and \{D\}=11 supergravity},
  journal      = {Journal of High Energy Physics},
  volume       = {2009},
  number       = {03},
  pages        = {135},
  year         = {2009},
  doi          = {10.1088/1126-6708/2009/03/135},
  eprint       = {0901.1581},
  archivePrefix= {arXiv},
  primaryClass = {hep-th}
}

@article{Luscher:1974ez,
    author = "Luscher, M. and Mack, G.",
    title = "{Global Conformal Invariance in Quantum Field Theory}",
    reportNumber = "PRINT-74-1569 (BERN)",
    doi = "10.1007/BF01608988",
    journal = "Commun. Math. Phys.",
    volume = "41",
    pages = "203--234",
    year = "1975"
}

@article{manningcoe2025grokkingvslearningfeatures,
      title={Grokking vs. Learning: Same Features, Different Encodings}, 
      author={Dmitry Manning-Coe and Jacopo Gliozzi and Alexander G. Stapleton and Edward Hirst and Giuseppe De Tomasi and Barry Bradlyn and David S. Berman},
      year={2025},
      eprint={2502.01739},
      archivePrefix={arXiv},
      primaryClass={cs.LG}
}

@article{Armstrong-Williams:2024nzy,
    author = "Armstrong-Williams, Kymani T. K. and Hirst, Edward and Jackson, Blake and Lee, Kyu-Hwan",
    title = "{Machine Learning Mutation-Acyclicity of Quivers}",
    eprint = "2411.04209",
    archivePrefix = "arXiv",
    primaryClass = "math.CO",
    reportNumber = "QMUL-PH-24-27",
    month = "11",
    year = "2024"
}

@article{Berglund:2024reu,
    author = "Berglund, Per and Butbaia, Giorgi and He, Yang-Hui and Heyes, Elli and Hirst, Edward and Jejjala, Vishnu",
    title = "{Generating triangulations and fibrations with reinforcement learning}",
    eprint = "2405.21017",
    archivePrefix = "arXiv",
    primaryClass = "hep-th",
    reportNumber = "QMUL-PH-24-10",
    doi = "10.1016/j.physletb.2024.139158",
    journal = "Phys. Lett. B",
    volume = "860",
    pages = "139158",
    year = "2025"
}

@article{Costantino:2024joa,
    author = "Costantino, Francesco and He, Yang-Hui and Heyes, Elli and Hirst, Edward",
    title = "{Learning 3-Manifold Triangulations}",
    eprint = "2405.09610",
    archivePrefix = "arXiv",
    primaryClass = "math.GT",
    reportNumber = "QMUL-PH-23-16",
    month = "5",
    year = "2024"
}

@article{Hirst:2024abn,
    author = "Hirst, Edward",
    title = "{Calabi-Yau Links and Machine Learning}",
    eprint = "2401.11550",
    archivePrefix = "arXiv",
    primaryClass = "hep-th",
    reportNumber = "QMUL-PH-24-01",
    doi = "10.1142/S281093922440001X",
    journal = "Int. J. Data Sci. Math. Sci.",
    volume = "02",
    number = "01",
    pages = "3--14",
    year = "2024"
}

@article{Hirst:2023kdl,
    author = "Hirst, Edward and Gherardini, Tancredi Schettini",
    title = "{Calabi-Yau four-, five-, sixfolds as Pwn hypersurfaces: Machine learning, approximation, and generation}",
    eprint = "2311.17146",
    archivePrefix = "arXiv",
    primaryClass = "hep-th",
    reportNumber = "QMUL-PH-23-25",
    doi = "10.1103/PhysRevD.109.106006",
    journal = "Phys. Rev. D",
    volume = "109",
    number = "10",
    pages = "106006",
    year = "2024"
}

@article{Aggarwal:2023swe,
    author = "Aggarwal, Daattavya and He, Yang-Hui and Heyes, Elli and Hirst, Edward and Earp, Henrique N. S\'a and Silva, Tom\'as S. R.",
    title = "{Machine learning Sasakian and G2 topology on contact Calabi-Yau 7-manifolds}",
    eprint = "2310.03064",
    archivePrefix = "arXiv",
    primaryClass = "math.DG",
    reportNumber = "QMUL-PH-23-14",
    doi = "10.1016/j.physletb.2024.138517",
    journal = "Phys. Lett. B",
    volume = "850",
    pages = "138517",
    year = "2024"
}

@article{Chen:2023whk,
    author = "Chen, Siqi and Dechant, Pierre-Philippe and He, Yang-Hui and Heyes, Elli and Hirst, Edward and Riabchenko, Dmitrii",
    title = "{Machine Learning Clifford Invariants of ADE Coxeter Elements}",
    eprint = "2310.00041",
    archivePrefix = "arXiv",
    primaryClass = "cs.LG",
    reportNumber = "QMUL-PH-23-15",
    doi = "10.1007/s00006-024-01325-y",
    journal = "Adv. Appl. Clifford Algebras",
    volume = "34",
    number = "3",
    pages = "20",
    year = "2024"
}

@article{Berglund:2023ztk,
    author = "Berglund, Per and He, Yang-Hui and Heyes, Elli and Hirst, Edward and Jejjala, Vishnu and Lukas, Andre",
    title = "{New Calabi\textendash{}Yau manifolds from genetic algorithms}",
    eprint = "2306.06159",
    archivePrefix = "arXiv",
    primaryClass = "hep-th",
    reportNumber = "QMUL-PH-23-06",
    doi = "10.1016/j.physletb.2024.138504",
    journal = "Phys. Lett. B",
    volume = "850",
    pages = "138504",
    year = "2024"
}

@inproceedings{He:2023csq,
    author = "He, Yang-Hui and Heyes, Elli and Hirst, Edward",
    editor = "Srinivasa Rao, Arni S. R. and Rao, C. R. and Krantz, Steven",
    title = "{Machine Learning in Physics and Geometry}",
    booktitle = "{Artificial Intelligence}",
    publisher = "Springer",
    eprint = "2303.12626",
    archivePrefix = "arXiv",
    primaryClass = "hep-th",
    month = "3",
    year = "2023"
}

@article{Cheung:2022itk,
    author = "Cheung, Man-Wai and Dechant, Pierre-Philippe and He, Yang-Hui and Heyes, Elli and Hirst, Edward and Li, Jian-Rong",
    title = "{Clustering cluster algebras with clusters}",
    eprint = "2212.09771",
    archivePrefix = "arXiv",
    primaryClass = "hep-th",
    reportNumber = "LIMS-2022-025",
    doi = "10.4310/ATMP.2023.v27.n3.a5",
    journal = "Adv. Theor. Math. Phys.",
    volume = "27",
    number = "3",
    pages = "797--828",
    year = "2023"
}

@article{Chen:2022jwd,
    author = "Chen, Siqi and He, Yang-Hui and Hirst, Edward and Nestor, Andrew and Zahabi, Ali",
    title = "{Mahler measuring the genetic code of amoebae}",
    eprint = "2212.06553",
    archivePrefix = "arXiv",
    primaryClass = "hep-th",
    reportNumber = "LIMS-2022-024",
    doi = "10.4310/ATMP.2023.v27.n5.a3",
    journal = "Adv. Theor. Math. Phys.",
    volume = "27",
    number = "5",
    pages = "1405--1461",
    year = "2023"
}

@article{Bao:2022rup,
    author = "Bao, Jiakang and He, Yang-Hui and Heyes, Elli and Hirst, Edward",
    title = "{Machine Learning Algebraic Geometry for Physics}",
    eprint = "2204.10334",
    journal = "Springer [accepted]",
    archivePrefix = "arXiv",
    primaryClass = "hep-th",
    reportNumber = "LIMS-2022-012",
    month = "4",
    year = "2022"
}

@article{Dechant:2022ccf,
    author = "Dechant, Pierre-Philippe and He, Yang-Hui and Heyes, Elli and Hirst, Edward",
    title = "{Cluster Algebras: Network Science and Machine Learning}",
    eprint = "2203.13847",
    archivePrefix = "arXiv",
    primaryClass = "math.CO",
    reportNumber = "LIMS-2022-011",
    doi = "10.1016/j.jaca.2023.100008",
    journal = "J. Comput. Algebra",
    volume = "8",
    year = "2023"
}

@inproceedings{Hirst:2022qqr,
    author = "Hirst, Edward",
    title = "{Machine Learning for Hilbert Series}",
    booktitle = "{Nankai Symposium on Mathematical Dialogues}: {In celebration of S.S.Chern's 110th anniversary}",
    eprint = "2203.06073",
    archivePrefix = "arXiv",
    primaryClass = "hep-th",
    month = "3",
    year = "2022"
}

@article{Arias-Tamargo:2022qgb,
    author = "Arias-Tamargo, Guillermo and He, Yang-Hui and Heyes, Elli and Hirst, Edward and Rodriguez-Gomez, Diego",
    title = "{Brain webs for brane webs}",
    eprint = "2202.05845",
    archivePrefix = "arXiv",
    primaryClass = "hep-th",
    reportNumber = "LIMS-2022-08",
    doi = "10.1016/j.physletb.2022.137376",
    journal = "Phys. Lett. B",
    volume = "833",
    pages = "137376",
    year = "2022"
}

@article{Berman:2021mcw,
    author = "Berman, David S. and He, Yang-Hui and Hirst, Edward",
    title = "{Machine learning Calabi-Yau hypersurfaces}",
    eprint = "2112.06350",
    archivePrefix = "arXiv",
    primaryClass = "hep-th",
    reportNumber = "QMUL-PH-21-55, LIMS-2021-017",
    doi = "10.1103/PhysRevD.105.066002",
    journal = "Phys. Rev. D",
    volume = "105",
    number = "6",
    pages = "066002",
    year = "2022"
}

@article{Bao:2021ofk,
    author = "Bao, Jiakang and He, Yang-Hui and Hirst, Edward and Hofscheier, Johannes and Kasprzyk, Alexander and Majumder, Suvajit",
    title = "{Polytopes and Machine Learning}",
    eprint = "2109.09602",
    archivePrefix = "arXiv",
    primaryClass = "math.CO",
    reportNumber = "LIMS-2021-011",
    doi = "10.1142/S281093922350003X",
    journal = "Math. Sci.",
    volume = "01",
    pages = "181--211",
    year = "2023"
}

@article{Bao:2021ohf,
    author = "Bao, Jiakang and Hanany, Amihay and He, Yang-Hui and Hirst, Edward",
    title = "{Some Open Questions in Quiver Gauge Theory}",
    eprint = "2108.05167",
    archivePrefix = "arXiv",
    primaryClass = "hep-th",
    reportNumber = "Imperial/TP/21/AH/04, LIMS-2021-009",
    doi = "10.22199/issn.0717-6279-5274",
    journal = " Proyecciones Journal of Mathematics",
    volume = "41",
    month = "8",
    year = "2021"
}

@article{Bao:2021olg,
    author = "Bao, Jiakang and He, Yang-Hui and Hirst, Edward",
    title = "{Neurons on amoebae}",
    eprint = "2106.03695",
    archivePrefix = "arXiv",
    primaryClass = "math.AG",
    reportNumber = "LIMS-2021-007",
    doi = "10.1016/j.jsc.2022.08.021",
    journal = "J. Symb. Comput.",
    volume = "116",
    pages = "1--38",
    year = "2023"
}

@article{Bao:2021auj,
    author = "Bao, Jiakang and He, Yang-Hui and Hirst, Edward and Hofscheier, Johannes and Kasprzyk, Alexander and Majumder, Suvajit",
    title = "{Hilbert series, machine learning, and applications to physics}",
    eprint = "2103.13436",
    archivePrefix = "arXiv",
    primaryClass = "hep-th",
    doi = "10.1016/j.physletb.2022.136966",
    journal = "Phys. Lett. B",
    volume = "827",
    pages = "136966",
    year = "2022"
}

@article{Bao:2021vxt,
    author = "Bao, Jiakang and Foda, Omar and He, Yang-Hui and Hirst, Edward and Read, James and Xiao, Yan and Yagi, Futoshi",
    title = "{Dessins d\textquoteright{}enfants, Seiberg-Witten curves and conformal blocks}",
    eprint = "2101.08843",
    archivePrefix = "arXiv",
    primaryClass = "hep-th",
    doi = "10.1007/JHEP05(2021)065",
    journal = "JHEP",
    volume = "05",
    pages = "065",
    year = "2021"
}

@article{Bao:2020nbi,
    author = "Bao, Jiakang and Franco, Sebasti\'an and He, Yang-Hui and Hirst, Edward and Musiker, Gregg and Xiao, Yan",
    title = "{Quiver Mutations, Seiberg Duality and Machine Learning}",
    eprint = "2006.10783",
    archivePrefix = "arXiv",
    primaryClass = "hep-th",
    doi = "10.1103/PhysRevD.102.086013",
    journal = "Phys. Rev. D",
    volume = "102",
    number = "8",
    pages = "086013",
    year = "2020"
}

@article{He:2020eva,
    author = "He, Yang-Hui and Hirst, Edward and Peterken, Toby",
    title = "{Machine-learning dessins d\textquoteright{}enfants: explorations via modular and Seiberg\textendash{}Witten curves}",
    eprint = "2004.05218",
    archivePrefix = "arXiv",
    primaryClass = "hep-th",
    doi = "10.1088/1751-8121/abbc4f",
    journal = "J. Phys. A",
    volume = "54",
    number = "7",
    pages = "075401",
    year = "2021"
}

@article{Bao:2020sqg,
    author = "Bao, Jiakang and He, Yang-Hui and Hirst, Edward and Pietromonaco, Stephen",
    title = "{Lectures on the Calabi-Yau Landscape}",
    eprint = "2001.01212",
    journal = "Pacific Institute for Mathematical Sciences Proceedings [accepted]",
    archivePrefix = "arXiv",
    primaryClass = "hep-th",
    month = "1",
    year = "2020"
}

@article{Niarchos:2023lot,
    author = "Niarchos, V. and Papageorgakis, C. and Richmond, P. and Stapleton, A. G. and Woolley, M.",
    title = "{Bootstrability in line-defect CFTs with improved truncation methods}",
    eprint = "2306.15730",
    archivePrefix = "arXiv",
    primaryClass = "hep-th",
    doi = "10.1103/PhysRevD.108.105027",
    journal = "Phys. Rev. D",
    volume = "108",
    number = "10",
    pages = "105027",
    year = "2023"
}

@article{Berman:2023rqb,
    author = "Berman, David S. and Klinger, Marc S. and Stapleton, Alexander G.",
    title = "{Bayesian renormalization}",
    eprint = "2305.10491",
    archivePrefix = "arXiv",
    primaryClass = "hep-th",
    doi = "10.1088/2632-2153/ad0102",
    journal = "Mach. Learn. Sci. Tech.",
    volume = "4",
    number = "4",
    pages = "045011",
    year = "2023"
}

@article{7ceb639f-641d-3fa7-9bc3-1b419c5ba656,
 ISSN = {00029947},
 URL = {http://www.jstor.org/stable/3845056},
 author = {C. E. Durán and A. Mendoza and A. Rigas},
 journal = {Transactions of the American Mathematical Society},
 number = {12},
 pages = {5025--5043},
 publisher = {American Mathematical Society},
 title = {Blakers-Massey Elements and Exotic Diffeomorphisms of S6 and $S^{14}$ via Geodesics},
 urldate = {2025-07-06},
 volume = {356},
 year = {2004}
}

@article{Kruskal:1960,
  author  = {Kruskal, Martin D.},
  title   = {Maximal extension of Schwarzschild metric},
  journal = {Physical Review},
  volume  = {119},
  number  = {5},
  pages   = {1743--1745},
  year    = {1960},
  doi     = {10.1103/PhysRev.119.1743}
}

@book{MTW,
  author    = {Misner, Charles W. and Thorne, Kip S. and Wheeler, John Archibald},
  title     = {Gravitation},
  publisher = {W.\,H. Freeman},
  address   = {San Francisco},
  year      = {1973},
  isbn      = {978-0-7167-0344-0}
}

@article{Brans1994a,
  author    = {Brans, Carl H.},
  title     = {Exotic Smoothness and Physics},
  journal   = {Journal of Mathematical Physics},
  volume    = {35},
  number    = {10},
  pages     = {5494--5506},
  year      = {1994},
  doi       = {10.1063/1.530761}
}

@article{Brans1994b,
  author    = {Brans, Carl H.},
  title     = {Localized Exotic Smoothness},
  journal   = {Classical and Quantum Gravity},
  volume    = {11},
  number    = {7},
  pages     = {1785--1792},
  year      = {1994},
  doi       = {10.1088/0264-9381/11/7/015}
}

@article{AsselmeyerKrol2012,
  author        = {Asselmeyer-Maluga, Torsten and Kr{\'o}l, Jerzy},
  title         = {Exotic Smoothness and Quantum Gravity~II: Exotic $\mathbb{R}^{4}$, Singularities and Cosmology},
  journal       = {arXiv preprint},
  year          = {2012},
  eprint        = {1112.4882},
  archivePrefix = {arXiv},
  primaryClass  = {gr-qc},
  doi           = {10.48550/arXiv.1112.4882},
  note          = {arXiv:1112.4882}
}

@article{AsselmeyerKrol2014,
  author    = {Asselmeyer-Maluga, Torsten and Kr{\'o}l, Jerzy},
  title     = {Inflation and Topological Phase Transition Driven by Exotic Smoothness},
  journal   = {Advances in High Energy Physics},
  volume    = {2014},
  pages     = {867460},
  year      = {2014},
  doi       = {10.1155/2014/867460}
}

@article{AsselmeyerKrol2018,
  author    = {Asselmeyer-Maluga, Torsten and Kr{\'o}l, Jerzy},
  title     = {How to Obtain a Cosmological Constant from Small Exotic $\mathbb{R}^{4}$},
  journal   = {Physics of the Dark Universe},
  volume    = {19},
  pages     = {66--77},
  year      = {2018},
  doi       = {10.1016/j.dark.2017.12.002}
}

@article{Brieskorn1966,
  author  = {Brieskorn, Egbert V.},
  title   = {Examples of Singular Normal Complex Spaces Which Are Topological Manifolds},
  journal = {Proceedings of the National Academy of Sciences of the USA},
  volume  = {55},
  number  = {6},
  pages   = {1395--1397},
  year    = {1966},
  doi     = {10.1073/pnas.55.6.1395}
}

@article{CrowleyNordstrom2019,
  author  = {Crowley, Diarmuid and Nordström, Johannes},
  title   = {The Classification of 2-Connected 7-Manifolds},
  journal = {Proceedings of the London Mathematical Society},
  volume  = {119},
  number  = {1},
  pages   = {1--54},
  year    = {2019},
  doi     = {10.1112/plms.12222}
}

@article{Overduin_1997,
   title={Kaluza-Klein gravity},
   volume={283},
   ISSN={0370-1573},
   url={http://dx.doi.org/10.1016/S0370-1573(96)00046-4},
   DOI={10.1016/s0370-1573(96)00046-4},
   number={5–6},
   journal={Physics Reports},
   publisher={Elsevier BV},
   author={Overduin, J.M. and Wesson, P.S.},
   year={1997},
   month=apr, pages={303–378} }

@book{Bleecker1981,
  author    = {Bleecker, David},
  title     = {Gauge Theory and Variational Principles},
  series    = {Global Analysis, Pure and Applied Series},
  publisher = {Addison--Wesley Publishing Company},
  address   = {Reading, MA},
  year      = {1981},
  isbn      = {978-0201100969}
}

@article{Whitney1934,
  author  = {Whitney, Hassler},
  title   = {Analytic Extensions of Differentiable Functions Defined in Closed Sets},
  journal = {Transactions of the American Mathematical Society},
  volume  = {36},
  number  = {1},
  pages   = {63--89},
  year    = {1934},
  doi     = {10.2307/1989692}
}

@article{Whitney1936,
  author  = {Whitney, Hassler},
  title   = {Differentiable Manifolds},
  journal = {Annals of Mathematics (2)},
  volume  = {37},
  number  = {3},
  pages   = {645--680},
  year    = {1936},
  doi     = {10.2307/1968482}
}

@book{BottTu1982,
  author    = {Raoul Bott and Loring W. Tu},
  title     = {Differential Forms in Algebraic Topology},
  series    = {Graduate Texts in Mathematics},
  volume    = {82},
  publisher = {Springer},
  address   = {New York},
  year      = {1982},
  isbn      = {978-0-387-90613-3},
  doi       = {10.1007/978-1-4757-3951-0}
}

@book{MilnorStasheff1974,
  author    = {Milnor, John W. and Stasheff, James D.},
  title     = {Characteristic Classes},
  series    = {Annals of Mathematics Studies},
  volume    = {76},
  publisher = {Princeton University Press},
  address   = {Princeton, NJ},
  year      = {1974},
  isbn      = {978-0-691-08122-5}
}

@book{HatcherAT,
  author    = {Hatcher, Allen},
  title     = {Algebraic Topology},
  publisher = {Cambridge University Press},
  address   = {Cambridge},
  year      = {2002},
  isbn      = {978-0-521-79540-1},
  note      = {Available on the author's website}
}

@book{Rolfsen1976,
  author    = {Rolfsen, Dale},
  title     = {Knots and Links},
  series    = {Mathematics Lecture Series},
  volume    = {7},
  publisher = {Publish or Perish},
  address   = {Berkeley, CA},
  year      = {1976},
  isbn      = {978-0-914098-16-4}
}

@article{Reidemeister1935,
  author  = {Reidemeister, Kurt},
  title   = {Homotopieringe und Linsenräume},
  journal = {Abhandlungen aus dem Mathematischen Seminar der Universität Hamburg},
  volume  = {11},
  pages   = {102--109},
  year    = {1935},
  doi     = {10.1007/BF02996388}
}

@misc{capuozzo2024machinelearningtoricduality,
      title={Machine Learning Toric Duality in Brane Tilings}, 
      author={Pietro Capuozzo and Tancredi Schettini Gherardini and Benjamin Suzzoni},
      year={2024},
      eprint={2409.15251},
      archivePrefix={arXiv},
      primaryClass={hep-th},
      url={https://arxiv.org/abs/2409.15251}, 
}

@article{Seong:2023njx,
    author = "Seong, Rak-Kyeong",
    title = "{Unsupervised machine learning techniques for exploring tropical coamoeba, brane tilings and Seiberg duality}",
    eprint = "2309.05702",
    archivePrefix = "arXiv",
    primaryClass = "hep-th",
    reportNumber = "UNIST-MTH-23-RS-04",
    doi = "10.1103/PhysRevD.108.106009",
    journal = "Phys. Rev. D",
    volume = "108",
    number = "10",
    pages = "106009",
    year = "2023"
}

@article{Halverson:2019tkf,
    author = "Halverson, James and Nelson, Brent and Ruehle, Fabian",
    title = "{Branes with Brains: Exploring String Vacua with Deep Reinforcement Learning}",
    eprint = "1903.11616",
    archivePrefix = "arXiv",
    primaryClass = "hep-th",
    doi = "10.1007/JHEP06(2019)003",
    journal = "JHEP",
    volume = "06",
    pages = "003",
    year = "2019"
}

@article{Loges:2021hvn,
    author = "Loges, Gregory J. and Shiu, Gary",
    title = "{Breeding Realistic D-Brane Models}",
    eprint = "2112.08391",
    archivePrefix = "arXiv",
    primaryClass = "hep-th",
    doi = "10.1002/prop.202200038",
    journal = "Fortsch. Phys.",
    volume = "70",
    number = "5",
    pages = "2200038",
    year = "2022"
}

\end{document}